\pdfoutput=1
\documentclass[12pt,a4paper,oneside]{book}

\usepackage[utf8]{inputenc}
\usepackage[T1]{fontenc}
\usepackage[main=english,brazil]{babel}

\usepackage{newtxtext}

\usepackage[
    a4paper,
    top=3cm,
    bottom=2cm,
    left=3cm,
    right=2cm
]{geometry}

\usepackage{setspace}
\usepackage{placeins}

\usepackage{amsmath}
\usepackage{amsthm}
\usepackage{newtxmath}
\usepackage{physics}
\usepackage{cancel}
\usepackage{bm}

\allowdisplaybreaks

\usepackage{xcolor}
\usepackage{graphicx}
\usepackage{float}
\usepackage{subfig}
\usepackage{svg}
\usepackage{tikz}
\usepackage{adjustbox}
\usetikzlibrary{arrows.meta,positioning,shapes.geometric}
\usetikzlibrary{arrows.meta,calc,shadows}
\tikzset{
    labelbox/.style={
        draw,
        rectangle,
        minimum width=3.2cm,
        minimum height=0.65cm,
        align=center,
        font=\bfseries
    },
    detector/.style={
        draw,
        rectangle,
        fill=black,
        minimum width=1.25cm,
        minimum height=1.10cm,
        text=white,
        font=\large
    },
    sourcecircle/.style={
        draw,
        circle,
        minimum size=1.75cm,
        line width=1.0pt,
    },
    dottedarrow/.style={
        dotted,
        line width=1.0pt,
        -{Latex[length=2.5mm]}
    },
    solidarrow/.style={
        line width=0.8pt,
        -{Latex[length=2.5mm]}
    },
    smallarrow/.style={
        line width=0.6pt,
        -{Latex[length=2.0mm]}
    }
}
\usepackage{listings}

\usepackage{fancyvrb}
\usepackage{adjustbox}
\usepackage{multicol}
\usepackage{framed}
\usepackage{soul}
\usepackage{url}
\usepackage[
  backend=biber,
  style=numeric-comp,
  sorting=none,
  maxbibnames=3,
  minbibnames=1,
  giveninits=true,
  doi=true,
  url=false,
  isbn=true
]{biblatex}
\usepackage[
    colorlinks=true,
    linkcolor=blue,
    citecolor=blue,
    urlcolor=blue
]{hyperref}

\begin{document}

\frontmatter
\onehalfspacing


\begin{titlepage}
\pagestyle{empty}

\begin{center}

{\fontsize{16}{16}\selectfont Universidade de São Paulo \\}
\vspace{0.1cm}
{\fontsize{16}{16}\selectfont Instituto de Física}

\vspace{3.3cm}

{\fontsize{22}{22}\selectfont
Explorando o diagrama de fase da matéria densa e quente com interferometria HBT
\par}

\vspace{2cm}

{\fontsize{18}{18}\selectfont Vinícius Silva Franção\par}

\vspace{2cm}

\end{center}

\leftskip 6cm
\begin{flushright}
Orientadora: Profa. Dra. Frederique Marie-Brigitte Sylvie Grassi
\end{flushright}

\vspace{0.8cm}

\par
\leftskip 6cm
\noindent
Tese de doutorado apresentada ao Instituto de Física da Universidade de São Paulo, como requisito parcial para a obtenção do título de Doutor em Ciências.
\par
\leftskip 0cm

\vskip 2cm

\noindent Banca Examinadora: \\
\noindent Profa. Dra. Frederique Marie-Brigitte Sylvie Grassi (IF/USP) \\
Prof. Dr. Matthew William Luzum (IF/USP) \\
Prof. Dr. Weiliang Qian (EEL/USP) \\
Profa. Dra. Sandra dos Santos Padula (IFT/UNESP) \\
Prof. Dr. Fernando Gonçalves Gardim (UNIFAL) \\

\vfill

\begin{center}
São Paulo \\
2026
\end{center}

\clearpage
\end{titlepage}

\thispagestyle{empty}

\vspace*{2.0cm}

\begin{center}
\textbf{\Large FICHA CATALOGRÁFICA}\\
\textbf{\large Preparada pelo Serviço de Biblioteca e Informação}\\
\textbf{\large do Instituto de Física da Universidade de São Paulo}
\end{center}

\vspace{1.0cm}

\begin{center}
\setlength{\fboxsep}{0pt}
\fbox{
\begin{minipage}[t][10.0cm][t]{0.78\textwidth}

\vspace{0.8cm}

\hspace*{1.0cm}
Franção, Vinícius Silva

\vspace{0.8cm}

\hspace*{1.0cm}
Explorando o diagrama de fase da matéria densa e quente com interferometria HBT.
São Paulo, 2026.

\vspace{0.8cm}

\hspace*{1.0cm}
Tese (Doutorado) - Universidade de São Paulo. Instituto de Física.
Depto. de Física Matemática.

\vspace{0.8cm}

\hspace*{1.0cm}
Orientadora: Profa. Dra. Frederique Marie-Brigitte Sylvie Grassi.

\vspace{0.6cm}

\hspace*{1.0cm}
Área de Concentração: Física Nuclear de Altas Energias.

\vspace{0.8cm}

\hspace*{1.0cm}
Unitermos: 1. Interferometria HBT; 2. Colisões de íons pesados;
3. Diagrama de fase da QCD; 4. Hidrodinâmica relativística;
5. Femtoscopia.

\vfill

\hspace*{0.2cm}
USP/IF/SBI-XXX/2026

\vspace{0.2cm}

\end{minipage}
}
\end{center}

\clearpage


\begin{titlepage}
\pagestyle{empty}

\begin{center}

{\fontsize{16}{16}\selectfont University of São Paulo \\}
\vspace{0.1cm}
{\fontsize{16}{16}\selectfont Physics Institute}

\vspace{3.3cm}

{\fontsize{22}{22}\selectfont
Exploring the phase diagram of dense and hot matter with HBT interferometry
\par}

\vspace{2cm}

{\fontsize{18}{18}\selectfont Vinícius Silva Franção\par}

\vspace{2cm}

\end{center}

\leftskip 6cm
\begin{flushright}
Advisor: Prof. Dr. Frederique Marie-Brigitte Sylvie Grassi
\end{flushright}

\vspace{0.8cm}

\par
\leftskip 6cm
\noindent
Doctoral thesis submitted to the Institute of Physics of the University of São Paulo as a partial requirement for obtaining the degree of Doctor of Science.
\par
\leftskip 0cm

\vskip 2cm

\noindent Examining Committee: \\
\noindent Prof. Dr. Frederique Marie-Brigitte Sylvie Grassi (IF/USP) \\
Prof. Dr. Matthew William Luzum (IF/USP) \\
Prof. Dr. Weiliang Qian (EEL/USP) \\
Prof. Dr. Sandra dos Santos Padula (IFT/UNESP) \\
Prof. Dr. Fernando Gonçalves Gardim (UNIFAL) \\

\vfill

\begin{center}
São Paulo \\
2026
\end{center}

\clearpage
\end{titlepage}
\begin{singlespace}
 $ $
\end{singlespace}
\chapter*{Agradecimentos}

Antes de qualquer coisa, é fundamental fazer alguns apontamentos: nasci, cresci e ainda vivo como um homem preto e homossexual em um país localizado na periferia do sistema capitalista, onde as boas oportunidades são reservadas apenas àqueles que pertencem às classes sociais mais abastadas e àqueles que não carregam a marca de Cã. Onde se registram as maiores taxas de homicídios contra a população LGBT+ e onde pessoas negras recebem salários menores do que pessoas brancas. Neste cenário, eu optei por uma carreira que tem sido continuamente desacreditada e desfinanciada para dar lugar ao subdesenvolvimento e à dependência tecnológica.

No Brasil, onde a maior parte da população é preta e parda, é um privilégio que permite alcançar o maior grau de formação que uma pessoa pode atingir. Em um país onde a maioria das famílias é composta por mulheres negras e seus filhos, eu tive a sorte de ter a estrutura necessária para estudar e produzir conhecimento. Antes de tudo, quero reconhecer que a possibilidade de realização deste trabalho sempre esteve condicionada à sorte que tive de estar cercado por pessoas, amigos e familiares que me apoiaram e não deixaram de acreditar em mim. 

Por isso, quero agradecer inicialmente (1) Aos meus pais, Eliezi e Carlos, por todo o carinho, apoio e sustento, e por permanecerem ao meu lado nos momentos mais difíceis, (2) À Frederique, pela orientação e por ter confiado em mim, mesmo nos momentos em que eu não confiei, (3) Ao meu querido amigo e parceiro de pesquisa Jonatan, pois, sem o seu trabalho, esta tese não teria se concretizado, (4) À minha tia Severina, que me apoiou e me incentivou ao longo de toda a minha trajetória, e à minha tia Maria, que, embora não esteja mais entre os vivos, continua me acompanhando e me guiando das estrelas.


Sou grato aos professores André, Fernando e Gabriel pelas excelentes discussões, pela colaboração ao longo deste trabalho, por destacarem tanto os meus pontos fortes quanto as minhas fragilidades e por me incentivarem a continuar aprofundando e explorando tudo o que fosse possível, sem desistir. Sou grato também aos professores Raul, Renata e Enrico pela oportunidade de debater física no mais elevado nível de excelência.

Agradeço aos colegas que ainda permanecem na pesquisa: Kevin pelo trabalho que desenvolvemos em conjunto e Diana, não só pela amizade, mas por mostrar que no simples existem complexidades que podem e devem ser exploradas. Agradeço também aos que já não estão mais na pesquisa, como Leo, Pedro e William. Vocês foram muito importantes na minha formação como físico de altas energias.

Agradeço à minha amiga Mariana, que se tornou uma excelente física médica, pelo carinho e pelo apoio incondicional.

Agradeço ao Thiago e ao Ariel por me apoiarem e permanecerem ao meu lado em todas as vezes em que precisei. 

Agradeço ainda a todos os meus amigos que estiveram comigo: Ana, Olívia, Kevin, Camila, Maricchi,  Vitor, Vagner e Beh, e a todos aqueles que eu possa ter esquecido. Agradeço também a todos que estiveram ao meu lado e que, por infortúnio do destino, já não estão mais. Cada um de vocês foi parte essencial da minha trajetória e contribuiu para transformar a maneira como eu enxergo o mundo.

Agradeço ao CEFISMA, nas figuras de Ely, Giovanni, Gabi, Triz, Elisa, Marcelo, Gabriel, Igor, Camis e Tormar. Agradeço ao pessoal da UJC/PCBR, nas figuras de Raquissa, Arak, Progídio e militantes do Baixo Matão. Todos vocês tiveram um papel essencial na minha formação política e estão entre as pessoas mais preparadas com quem tive a honra de militar. E aqui estendo meus agradecimentos às entidades estudantis do IFUSP, em especial ao HS e aos coletivos Sônia Guimarães e Prisma Gabrielle Weber, os quais tive o prazer de construir.

Agradeço aos funcionários do Instituto de Física, em especial ao Demo e ao Chico, bem como aos funcionários dos bandejões, da biblioteca e da equipe de limpeza. Estendo ainda meus agradecimentos aos funcionários do Departamento de Física-Matemática, com destaque para a Simone e a Cecília por estarem sempre à minha disposição quando precisei. Registro também minha gratidão ao pessoal da CPG, do CCIFUSP e à equipe de segurança.

Agradeço ao Laboratório Nacional de Computação Científica (LNCC) pelo acesso ao supercomputador Santos Dumont (SDumont), por meio do projeto ``Simulações em alto desempenho aplicadas à física teórica'' (SADFT), vinculado à Universidade Federal da Grande Dourados (UFGD), sob a coordenação do Prof.\ André Gianini.

Por fim, expresso meus agradecimentos à FAPESP pelo suporte financeiro a este trabalho, concedido por meio dos projetos 2018/23836-0, 2020/08937-5 e 2024/07064-9, e ao INCT-FNA pelo apoio financeiro nos eventos científicos nos quais participei.

\begin{singlespace}
 $ $
\end{singlespace}
\begin{singlespace}
\chapter*{Abstract}

\noindent We investigate the HBT radii of pions produced in Au+Au collisions at selected energies from the STAR-BES program, using an event-by-event hybrid model that combines AMPT initial conditions, 3+1D viscous hydrodynamic evolution with a QCD equation of state at finite baryon chemical potential, and hadronic transport with SMASH. The goal is to explore the source’s space-time structure and its possible connection to the structure of the QCD phase diagram. The hybrid chain was first validated by comparing identified-particle spectra and anisotropic flow, yielding a globally consistent description of the experimental data, with more evident discrepancies at the lowest energies.

From identical-pion correlation functions, we extract the HBT radii $R_o$, $R_s$, and $R_l$ and the parameter $\lambda$, obtaining a reasonable description of the magnitude and the $m_T$ dependence measured by STAR. An analysis of $R_o^2-R_s^2$ shows that this observable cannot be interpreted in isolation as a measure of the emission duration, due to geometric contributions and the source’s space-time correlations. Calculations performed directly at switching and after including only decays show that decays already systematically increase the HBT radii and reduce $R_o/R_s$, bringing it closer to the data, while full hadronic transport produces additional modifications. Using a Lévy source yields radii similar to those obtained with a Gaussian fit, but with $\alpha<2$, indicating deviations from a Gaussian shape.

Finally, we introduce an equation of state containing a critical point and a first-order transition, with and without the inclusion of critical scaling in the bulk-viscosity coefficient. Despite the modification of the hydrodynamic dynamics in the critical region, no substantial changes were observed in the spectra, anisotropic flow, or HBT radii, including when the switching and decay stages are considered separately. Within the present model and the investigated energies, we therefore do not find a clear femtoscopic signature of the critical point.
\vspace{0.5cm}

\noindent\textbf{Keywords:} HBT interferometry; heavy-ion collisions; QCD phase diagram; relativistic hydrodynamics; femtoscopy.

\end{singlespace}
\begin{singlespace}
 $ $
\end{singlespace}
\begin{singlespace}
\chapter*{Resumo}
\noindent Investigamos os raios HBT de píons produzidos em colisões Au+Au, em energias selecionadas do programa STAR-BES, por meio de um modelo híbrido evento a evento, combinando condições iniciais do AMPT, evolução hidrodinâmica viscosa 3+1-D com uma equação de estado da QCD com potencial bariônico finito e transporte hadrônico com o SMASH. O objetivo é explorar a estrutura espaço-temporal da fonte de partículas e sua possível relação com a estrutura do diagrama de fases da QCD. A cadeia híbrida foi inicialmente validada pela comparação de espectros de partículas identificadas e fluxo anisotrópico, apresentando uma descrição globalmente consistente com os dados experimentais, com discrepâncias mais evidentes nas energias mais baixas.

A partir das funções de correlação de píons idênticos, extraímos os raios HBT $R_o$, $R_s$ e $R_l$ e o parâmetro $\lambda$, obtendo uma descrição razoável da escala e da dependência em $m_T$ medidas pelo STAR. A análise de $R_o^2-R_s^2$ mostra que esse observável não pode ser interpretado isoladamente como uma medida da duração da emissão, devido às contribuições geométricas e às correlações espaço-temporais da fonte. Cálculos realizados diretamente no switching e após a inclusão apenas dos decaimentos mostram que os decaimentos já aumentam sistematicamente os raios HBT e reduzem $R_o/R_s$, aproximando-o dos dados, enquanto o transporte hadrônico completo produz modificações adicionais. A utilização de uma fonte de Lévy produz raios semelhantes aos obtidos com o ajuste Gaussiano, mas com valores de $\alpha<2$, indicando desvios da forma Gaussiana.

Por fim, introduzimos uma equação de estado contendo um ponto crítico e uma transição de primeira ordem, com e sem a inclusão de critical scaling no coeficiente de viscosidade volumétrica. Apesar da modificação da dinâmica hidrodinâmica na região crítica, não foram observadas alterações substanciais nos espectros, no fluxo anisotrópico ou nos raios HBT, inclusive quando os estágios de switching e de decaimentos são considerados separadamente. Dentro do presente modelo e das energias investigadas, portanto, não encontramos uma assinatura femtoscópica clara do ponto crítico.

\vspace{0.5cm}

\noindent\textbf{Palavras-chave:} Interferometria HBT; colisões de íons pesados; diagrama de fases da QCD; hidrodinâmica relativística; femtoscopia.

\end{singlespace}
\begin{singlespace}
 $ $
\end{singlespace}
\listoffigures

\tableofcontents

\clearpage
\newcounter{frontmatterpage}
\setcounter{frontmatterpage}{\value{page}}

\mainmatter
\setcounter{page}{\value{frontmatterpage}}
\onehalfspacing

\chapter{Exploring the Phase Diagram of QCD}

The Schrödinger equation made it possible to calculate the spectrum of the hydrogen atom in terms of bound states, replacing previous models~\cite{Schrodinger1926}. Later, the Dirac equation enabled the relativistic description of the electron, which showed that properties such as spin and the fine structure of the hydrogen atom arise naturally, including the relativistic correction to the kinetic energy, the spin–orbit coupling, and the Darwin term~\cite{Dirac1928, Darwin1928}. The solutions of the Dirac equation also predicted particles with the same mass as the electron and opposite charge, the so‑called antiparticles. This prediction was confirmed in 1932, when Carl Anderson identified the positron in cosmic‑ray tracks in a cloud chamber, establishing the first direct evidence of antimatter~\cite{Anderson1933}. This showed that subatomic matter could not be described merely as electrons orbiting nuclei, but instead consisted of a far more diverse collection of particles and underlying symmetries. The discovery of the muon in cosmic rays, initially seen as Yukawa’s meson, revealed the existence of unstable particles with no direct role in ordinary matter~\cite{NeddermeyerAnderson1937,StreetStevenson1937}. 

\begin{figure}[!htbp]
    \centering
    \includegraphics[width=\linewidth]{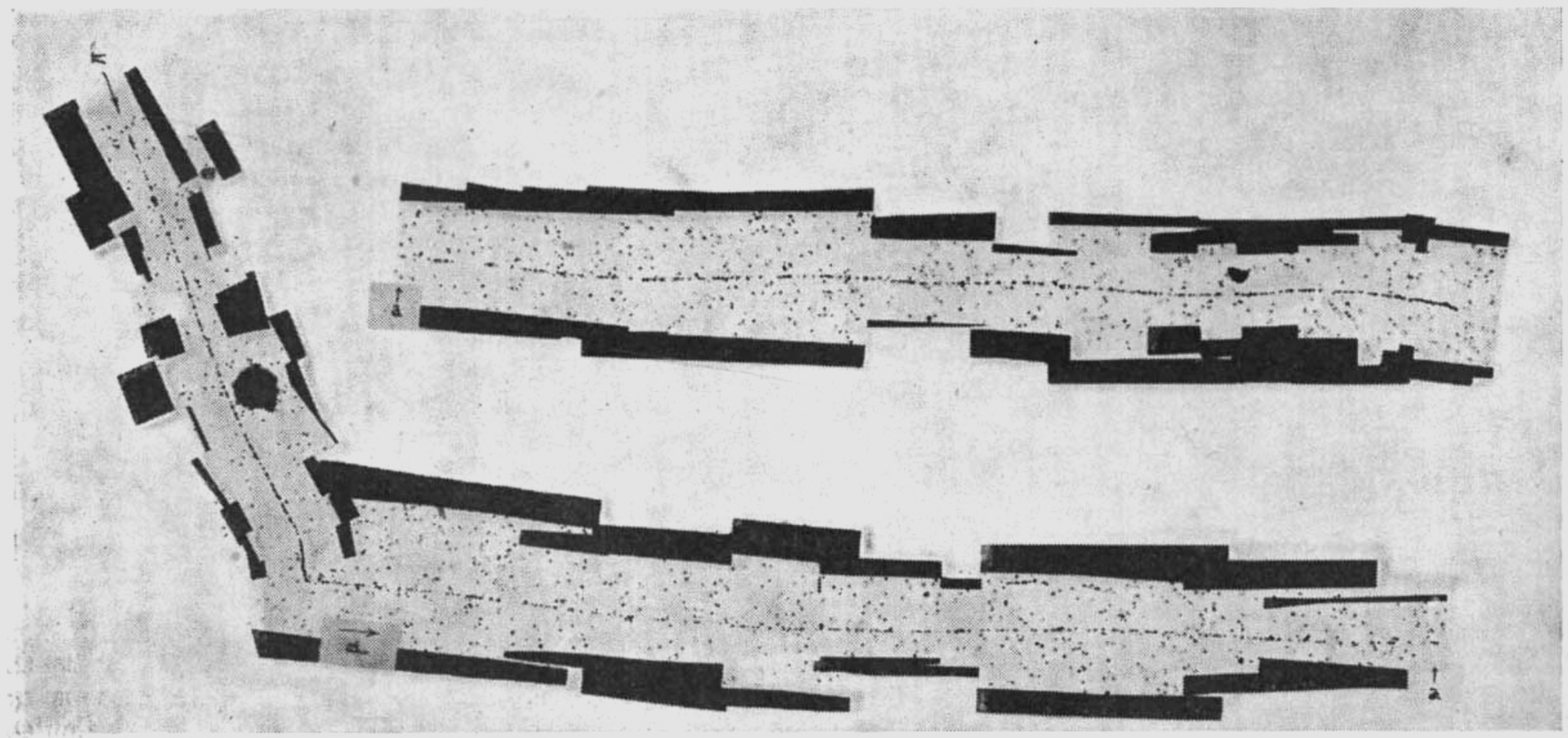}
    \caption{Microphotograph of a nuclear emulsion showing tracks from the decay of a charged pion into a muon. In this type of experiment, thick photographic plates were exposed to cosmic rays at high altitudes and then analyzed under a microscope, making it possible to identify the processes that led to the experimental observation of the pion. Taken from~\cite{Lattes1947Nature}.}
    \label{fig:plate}
\end{figure}

The experimental identification of the pion, a particle represented by the Greek letter $\pi$, involved the participation of the Brazilian physicist César Lattes, one of the leading figures in physics in Brazil, alongside Mário Schenberg, José Leite Lopes, and Jayme Tiomno. The existence of this particle had been proposed by Hideki Yukawa in 1935, when he explained the short-range nuclear force that keeps protons and neutrons bound in the atomic nucleus~\cite{Yukawa1935}. More than a decade later, using nuclear emulsions exposed to cosmic rays at high altitudes (Figure \ref{fig:plate}), Lattes, Muirhead, Occhialini, and Powell observed charged mesons and their transformation into secondary particles, providing the decisive experimental evidence for the existence of the pion~\cite{Lattes1947Nature}. 

This discovery consolidated the modern picture of hadrons as the degrees of freedom of nuclear matter and connected a theoretical prediction about nuclear forces to a direct experimental observation. Yukawa received the 1949 Nobel Prize in Physics for the prediction of mesons~\cite{NobelYukawa1949}, and Cecil Powell received the 1950 Nobel Prize for the development of the photographic method applied to nuclear processes and for the discoveries about mesons made with this method~\cite{NobelPowell1950}.

In the same year the pion was discovered, Rochester and Butler performed experiments that led to the identification of kaons and hyperons and, consequently, of strange particles~\cite{RochesterButler1947}. In the 1950s and 1960s, the appearance of a large number of mesons and baryons (the so‑called hadrons) revealed a veritable “zoo” of strongly interacting particles. There then arose the need not only to identify new particles but also to organize them into a simpler classification. This form of classification was developed in the works of Gell-Mann and Ne'eman. The quark model, initially proposed by Gell-Mann and later by Zweig, structured the organization of hadrons, giving rise to the theory known as quantum chromodynamics~\cite{GellMann1961,Neeman1961,GellMann1964,Zweig1964a,Zweig1964b}.

\section{Why Collide Particles?}

\begin{figure}[!htbp]
    \centering
    \includegraphics[width=\linewidth]{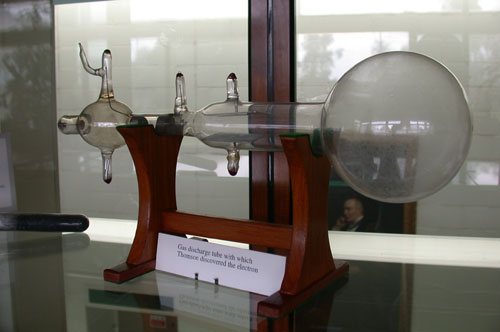}
    \caption{Cathode-ray tube used by J. J. Thomson in the study of the deflection of cathode rays by electric and magnetic fields. Taken from \cite{wismer_thomson_tube_2000}.}
    \label{fig:tubo_vacio_thompson}
\end{figure}

The transition from atomic and nuclear physics to particle physics is linked to the development of instruments to produce, control, and detect particle beams. This idea emerged at the end of the 19th century in the study of cathode rays. In 1897, while studying the deflection of these rays by electric and magnetic fields in cathode ray tubes (Figure \ref{fig:tubo_vacio_thompson}), J. J. Thomson showed that they were composed of negatively charged particles and determined their charge-to-mass ratio, thus identifying the electron~\cite{Thomson1897}.

\begin{figure}[!htbp]
    \centering
    \includegraphics[width=\linewidth]{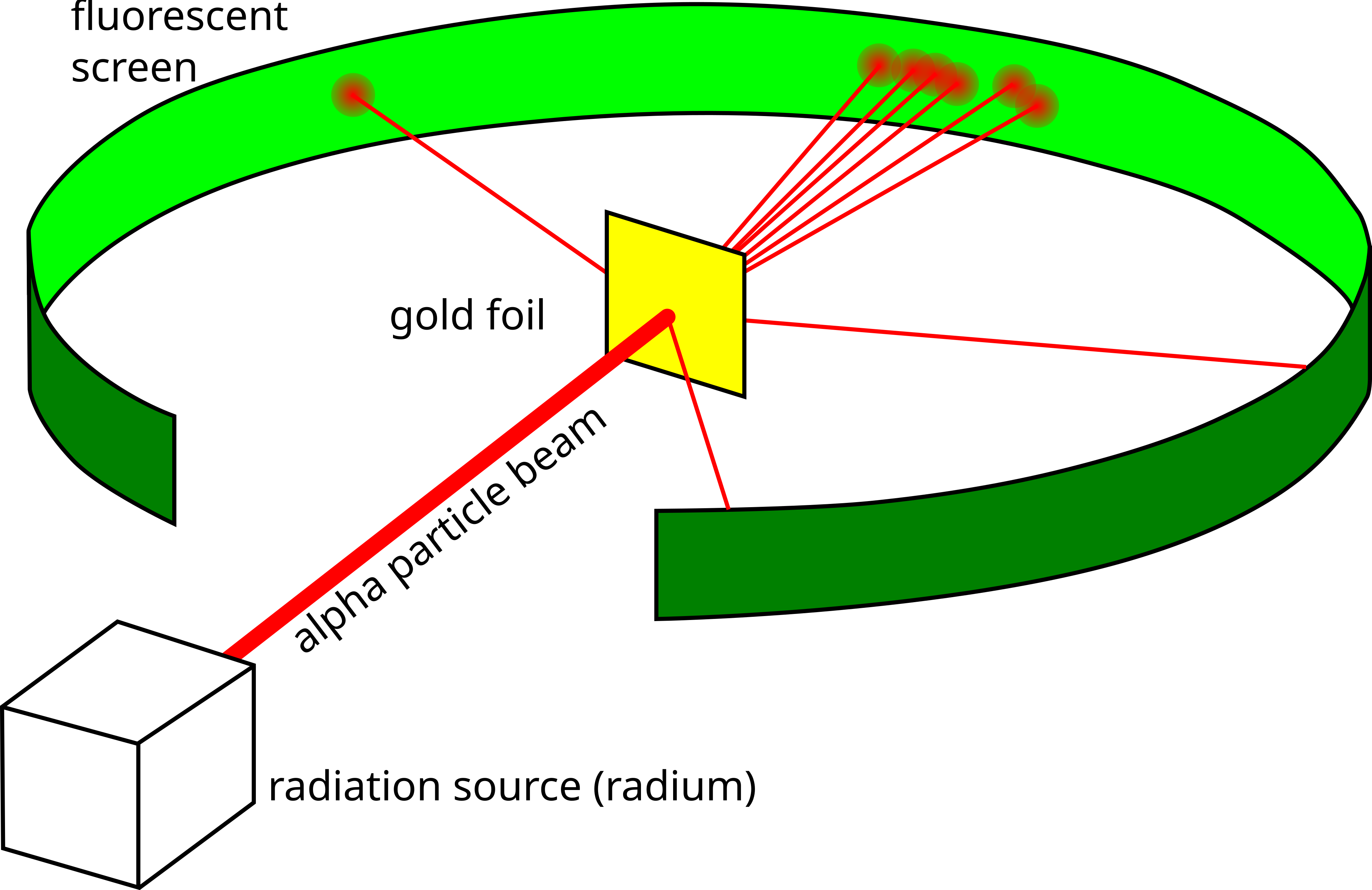}
    \caption{Schematic representation of the Geiger–Marsden experiment. $\alpha$ particles emitted by a radioactive source are collimated and directed onto a thin metal foil, while a fluorescent screen records the scattering angles. Taken from~\cite{GeigerMarsdenApparatusWikimedia}.}
    \label{fig:Geiger-Marsden_experiment}
\end{figure}

Between 1908 and 1909, Geiger and Marsden, within the context of Rutherford’s experimental program, used $\alpha$ particles --- helium nuclei --- emitted by radioactive sources as projectiles against thin metal foils (Figure~\ref{fig:Geiger-Marsden_experiment}), especially gold, reconstructing the scattering angles from the scintillations on zinc sulfide screens~\cite{GeigerMarsden1909}. The large scattering angles observed could only be understood by assuming that the positive charge and practically all the mass of the atom were concentrated in an extremely small region, which resulted in a representation of the atom very close to the modern model, in contrast with Thomson’s plum pudding model~\cite{Rutherford1911}.

These experiments consolidated the practice of investigating matter by making it interact with projectiles, that is, by accelerating them and causing them to collide with it. Initially, these projectiles came from natural sources, such as cosmic rays or radioactive decays. With the advent of accelerators, it became possible to control the incident species, its energy, its direction, and the rate at which events occurred. The work of Cockcroft and Walton in 1932 showed that artificially accelerated protons could induce the disintegration of light nuclei, making nuclear structure an object amenable to controlled investigation in the laboratory~\cite{CockcroftWalton1932a,CockcroftWalton1932b}. In the same period, the development of the cyclotron by Lawrence and Livingston introduced the repeated acceleration of ions along circular trajectories, allowing ever higher energies to be reached without the need to apply a single extremely high potential difference~\cite{LawrenceLivingston1932}. From then on, accelerators and, later, colliders came to function as dynamic microscopes of matter: by increasing the momentum of the projectiles, the wavelength associated with the probe was reduced. By increasing the energy available in the collision, the possibility of producing new degrees of freedom was opened up. It is this same logic that underlies hadronic and nuclear collision experiments used to investigate the structure of hadrons, the strong interaction, and the properties of nuclear matter under extreme conditions.

Much has been developed since then. Throughout the 20th century, accelerators and colliders were important tools for investigating fundamental interactions. Programs at CERN, Fermilab, Brookhaven, and other laboratories made it possible to produce increasingly massive particles and to create systems of matter under extreme conditions. The discovery of the vector bosons $W^\pm$ and $Z^0$ at the CERN SPS, of the top quark at the Fermilab Tevatron, and later of the Higgs boson by the ATLAS and CMS experiments at the LHC illustrates how accelerator physics has consolidated the Standard Model as the description of elementary particles and their electroweak and strong interactions~\cite{Arnison1983W,Arnison1983Z,CDF1995Top,DZero1995Top,ATLAS2012Higgs,CMS2012Higgs}.

\subsection{Collective Behavior of Strongly Interacting Matter}

Relativistic heavy-ion collisions have made it possible to investigate the collective properties of matter generated under extreme conditions of temperature and baryonic potential. The attempt to describe the collective behavior of the matter created in nuclear collisions, however, is not recent. It dates back to the 1950s, emerging as a critical response by Lev Landau to the thermodynamic treatment proposed by Enrico Fermi for particle production~\cite{aa07,aa08,aa09}. Although Fermi was a pioneer in introducing a statistical description for high-energy collisions, in his formulation, the produced system was treated as practically “frozen” before expansion could play a relevant dynamical role. Landau argued that if the interactions were strong enough to ensure local equilibrium, then the pressure generated at the initial stage could not be neglected. The system should expand before the final emission of particles, and this expansion would need to be described by relativistic hydrodynamics. The idea, initially formulated in a context far removed from that of modern heavy-ion collisions, gained renewed phenomenological relevance decades later when hydrodynamic models began to quantitatively describe collective observables, in particular, spectra and elliptic flow measured at SPS energies and, above all, in the first RHIC results~\cite{Teaney2001,Kolb2000,Huovinen2001}.

\subsection{RHIC, STAR and Beam Energy Scan}

The RHIC --- \textit{Relativistic Heavy Ion Collider} --- located at Brookhaven National Laboratory, was a heavy-ion and proton collider dedicated to the study of nuclear matter under extreme conditions, in operation from 2000 until 2026. Its initial program comprised four main experiments: BRAHMS, PHENIX, PHOBOS, and STAR. In 2005, these four experiments presented assessments of the first years of Au+Au collisions, indicating the formation of a dense medium, identified as a quark-gluon plasma~\cite{Arsene2005BRAHMS,Back2005PHOBOS,Adcox2005PHENIX,Adams2005STAR}.
\begin{figure}[!htbp]
    \centering
    \includegraphics[width=\linewidth]{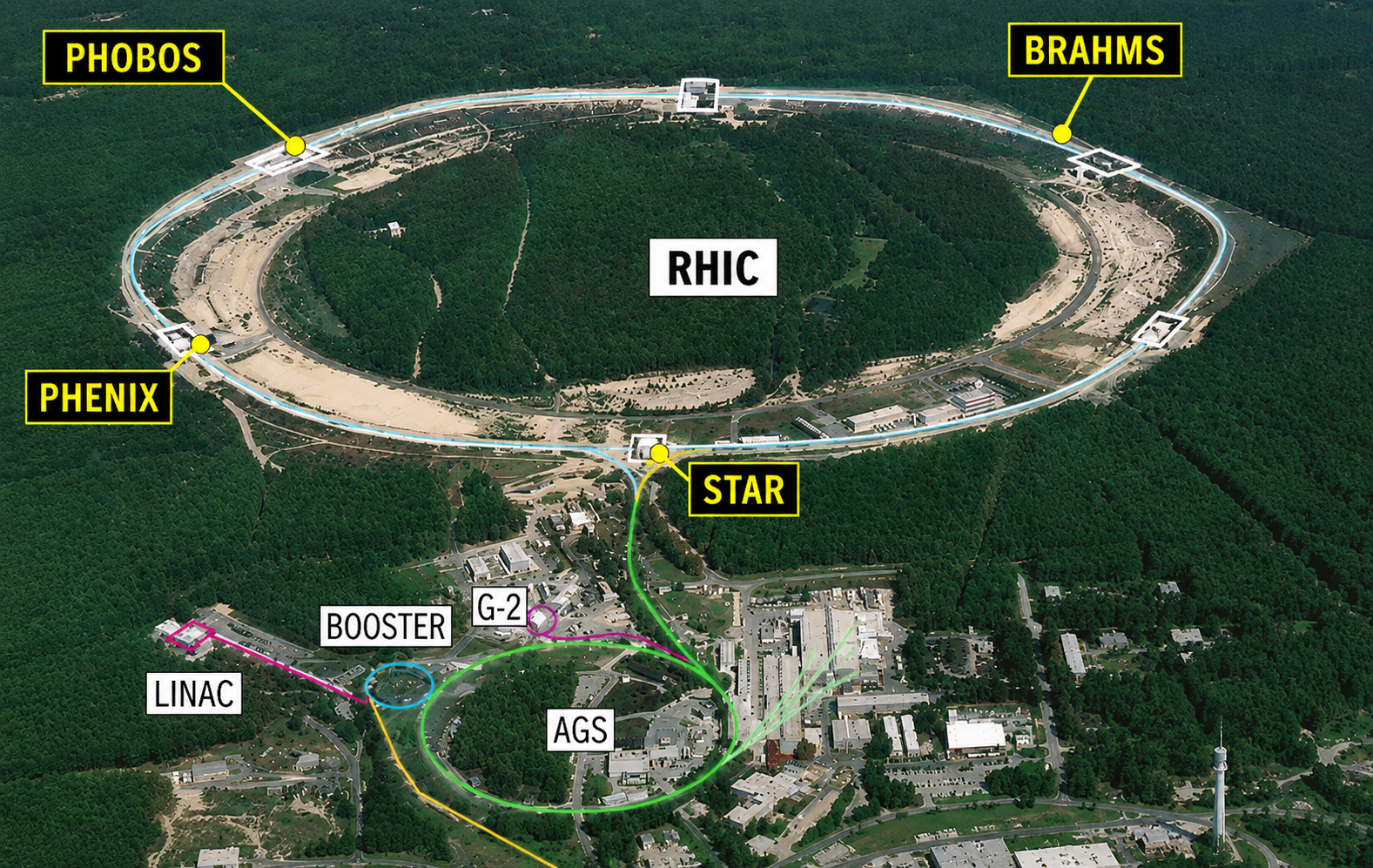}
    \caption{Schematic representation of RHIC and the locations of the four main experiments of its initial program: BRAHMS, PHENIX, PHOBOS, and STAR.}
    \label{fig:rhic}
\end{figure}

In this context, within the STAR experiment, the BES --- \textit{Beam Energy Scan} --- program  was developed, aimed at the systematic exploration of Au+Au collisions at different energies. This scan made it possible to produce systems that traverse different regimes of strongly interacting matter. In the initial stages of the collision, the matter may reach a deconfined regime, in which the relevant degrees of freedom are quarks and gluons. After expansion and cooling, the system returns to a description in terms of hadrons, which are the particles actually measured in the detectors. These two regimes belong to the same theory, QCD — Quantum Chromodynamics — which is responsible for describing the strong interaction. Therefore, studying how the matter produced in relativistic collisions arranges itself between a hadronic regime and a deconfined regime amounts to exploring QCD in a phase diagram. In this scenario, STAR-BES is one of the main experimental programs for investigating the transition between these regimes and searching for possible signs of critical structures in this region of the diagram~\cite{Caines2017,Yang2017BESII}.

\subsection{Pion Interferometry in the STAR-BES Program}

One of the ways to investigate the space-time geometry of the particle-emitting source produced in these collisions is through HBT interferometry, also known as femtoscopy. Since the microscopic evolution of the system is not directly accessible, characteristics such as the spatial size, emission time, and decoupling dynamics can be inferred from correlations between emitted particles, in particular between pairs of identical pions~\cite{GoldhaberGoldhaberLeePais1960,WiedemannHeinz1999,LisaPrattSoltzWiedemann2005}. At RHIC, the STAR experiment carried out pion interferometry measurements in Au+Au collisions at $200$ GeV~\cite{AdamsSTAR2005} and, later, within the BES program, examined how femtoscopic radii and the freeze-out eccentricity vary with the collision energy~\cite{AdamczykSTAR2015HBTBES}. These observables now serve as a rigorous test for hydrodynamic models. For a long time, achieving a simultaneous good description of the spectra, elliptic flow, and HBT radii proved problematic, a challenge that became known as the \textit{HBT puzzle}~\cite{Pratt2009ResolvingHBT,Pratt2009LongSlow}. Despite major progress in characterizing the initial conditions, implementing viscosity, handling the transition from a hydrodynamic to a hadronic description, and coupling to hadronic transport, it is still challenging to reproduce the femtoscopy data at the lower energies of the BES program~\cite{Soltz2013,PlumbergHeinz2018}. Experimental results from HBT interferometry have also been applied in phenomenological studies dedicated to the search for indications of a phase transition and a possible critical point in the QCD phase diagram~\cite{Lacey2015CEP}.

\section{Standard Model of Elementary Particle Physics}

The previous sections introduced the experimental path that leads from the identification of hadrons to the use of relativistic nuclear collisions as a tool to investigate strongly interacting matter. We now turn to the theoretical framework in which these particles and interactions are organized. In modern particle physics, this role is played by the Standard Model, formulated as a quantum field theory of elementary particles and their interactions.  
\begin{figure}[!htbp]
    \centering
    \includegraphics[width=\linewidth]{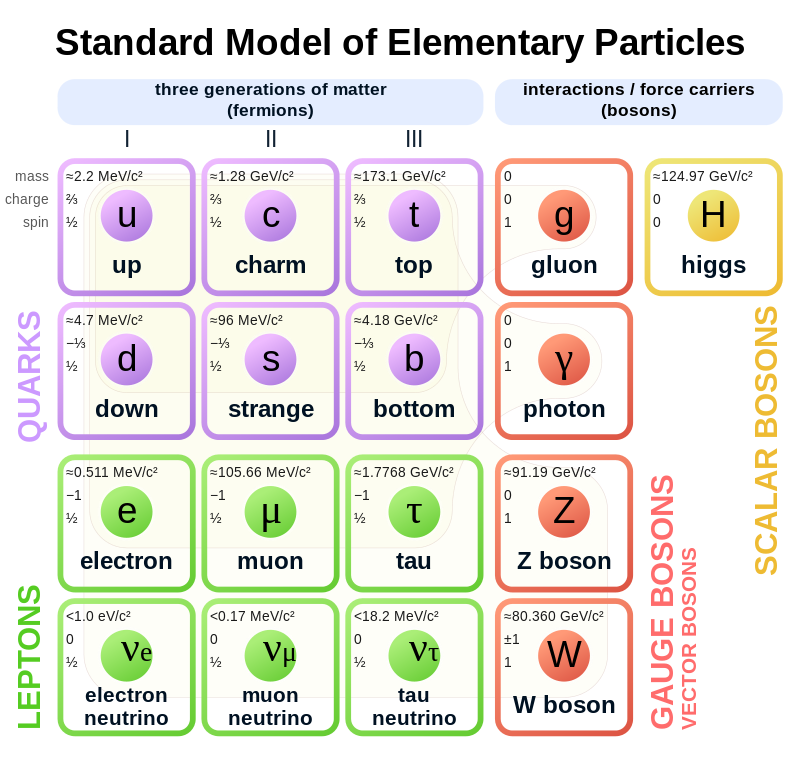}
    \caption{Elementary particles of the Standard Model, including the three generations of matter fermions, the gauge bosons associated with the fundamental interactions, and the Higgs boson. Taken from~\cite{wiki_standard_model}.}
    \label{fig:particles}
\end{figure}

In general, particles can be classified into fermions --- particles with semi-integer spin described by Fermi-Dirac statistics --- and bosons --- particles with integer spin described by Bose-Einstein statistics. Fermions include leptons and quarks, while the fundamental interactions described by the model are mediated by the so‑called gauge bosons: the photon for the electromagnetic interaction, the $W^\pm$ and $Z^0$ bosons for the weak interaction, and the gluons for the strong interaction. In addition to these degrees of freedom, the Higgs boson is associated with the mechanism of mass generation for elementary particles. In this study, we focus on the properties of strongly interacting matter created in nuclear collisions. Consequently, the pertinent part of the Standard Model is QCD, the theory that governs the strong interaction among quarks and gluons~\cite{Schwartz2014QFT}.

\subsection{Quarks and Gluons}

\begin{figure}[!htbp]
    \centering
    \includegraphics[width=\linewidth]{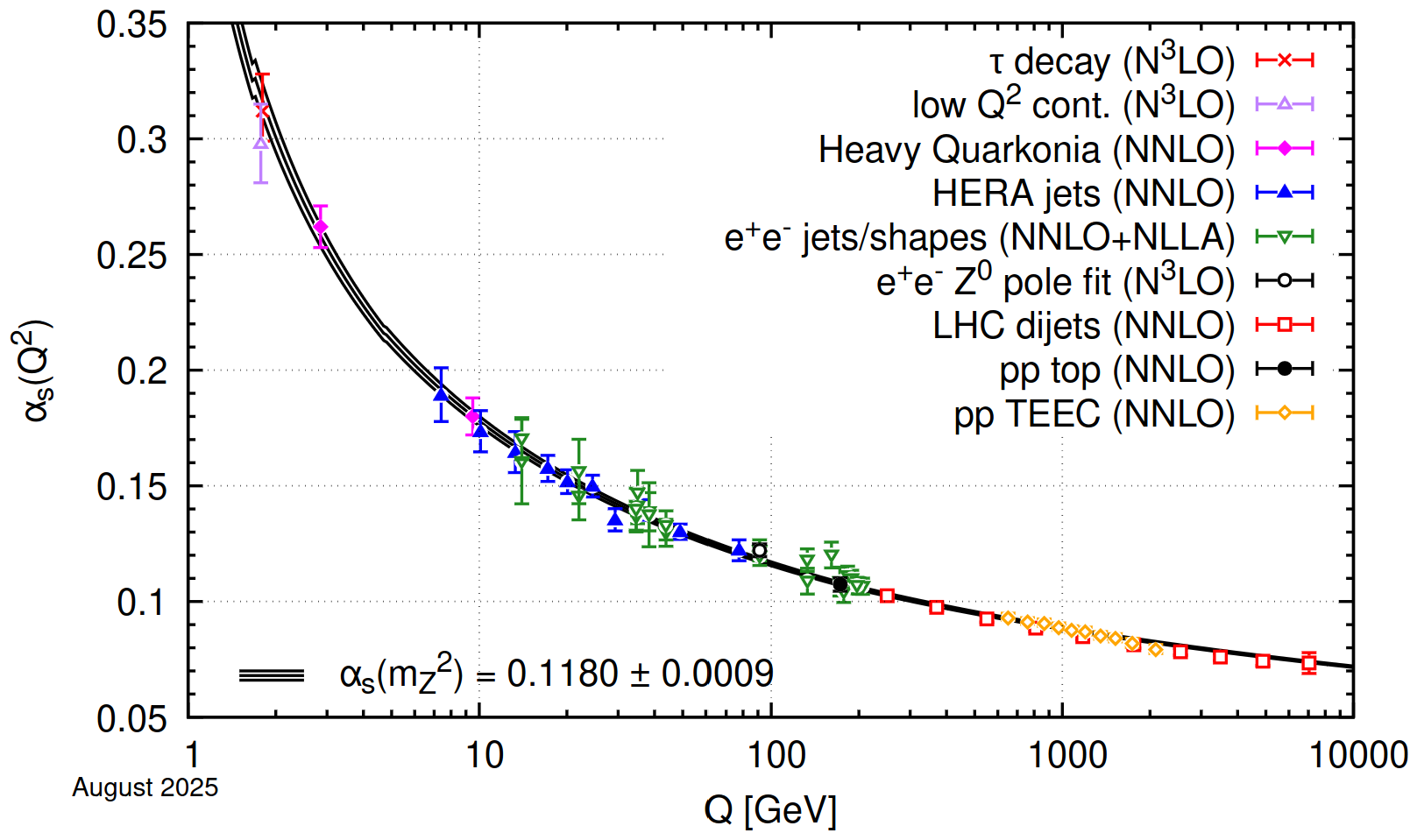}
    \caption{Summary of determinations of the strong coupling constant $\alpha_s$ as a function of the energy scale $Q$, compared with the QCD running computed at five-loop order using the PDG world average $\alpha_s(m_Z^2)=0.1180\pm0.0009$. Taken from~\cite{pdg2025}.}
    \label{fig:alpha}
\end{figure}

Quarks and gluons are the elementary degrees of freedom of QCD. Quarks carry color charge and appear in six flavors, $u$, $d$, $s$, $c$, $b$, and $t$, in addition to having fractional electric charge and baryon number. Gluons are the gauge bosons of the strong interaction and also carry color charge, which makes QCD a non-abelian theory. This feature is directly connected to two central properties of the strong interaction: asymptotic freedom and color confinement. At short distances, or equivalently at high energy scales, the effective coupling of QCD decreases (Fig. \ref{fig:alpha}), allowing quarks and gluons to be treated as almost free degrees of freedom.

At low-energy scales, on the other hand, the coupling grows, and quarks and gluons are not observed as isolated particles. The attempt to separate two quarks increases the energy stored in the color field until the production of new quark–antiquark pairs becomes energetically favorable, resulting in the formation of hadrons. This phenomenon is associated with color confinement, a property strongly supported by phenomenological evidence and by lattice QCD simulations~\cite{Schwartz2014QFT,Bazavov2014EoS,Borsanyi2014EoS}.

\subsection{Mesons, Baryons and Hadrons}

The particles observed directly in the final stages of a heavy-ion collision are not free quarks and gluons, but hadrons. Hadrons are color-neutral bound states formed by quarks and gluons, organized mainly into two classes: mesons and baryons. Mesons are composed of quark–antiquark pairs and have integer spin; therefore, they are bosons. Baryons are composed of three quarks or three antiquarks and have half-integer spin; therefore, they are fermions. Pions ($\pi^\pm$), kaons ($K^\pm$), protons ($p$, $\bar p$), neutrons $n$, and hyperons $\Lambda$ are examples of hadrons that are relevant in relativistic collisions. These particles are identified by their quantum numbers, such as electric charge (Q), baryon number (B), and strangeness (S). In particular, since these three numbers are conserved, they can be characterized in terms of conserved currents and play an important role in the evolution of matter during the stages of hydrodynamic evolution.

\subsection{QCD Phase Diagram and the Search for a Critical Point}

QCD describes both hadronic matter, in which the degrees of freedom are confined inside hadrons, and deconfined matter, in which quarks and gluons can propagate over distances larger than the typical hadronic scale. The organization of these regimes as a function of the thermodynamic conditions defines the QCD phase diagram. At low temperatures and baryon densities, matter appears in the form of hadrons. At sufficiently high temperatures, a transition to the quark–gluon plasma occurs, a deconfined phase produced in relativistic heavy-ion collisions. Lattice QCD results indicate that, for vanishing or small baryon chemical potential, this change occurs as a crossover, that is, without a singular phase transition~\cite{Aoki2006Nature,Borsanyi2014EoS,Bazavov2014EoS}. For regions of higher baryon density, however, there is still no definitive determination from first principles. In this regime, it is possible that the crossover becomes first order, ending at a critical point in the phase diagram~\cite{PhysRevLett.81.4816,stephanov2006qcdphasediagramoverview,PhysRevC.101.034901}. The search for signals of this structure is one of the main motivations of the STAR-BES program and also guides the analysis developed in this thesis~\cite{Caines2017,Yang2017BESII}.

\section{Thesis Outline}

The purpose of this thesis is to use hybrid simulations of Au+Au collisions at the beam energies covered by the STAR-BES program to investigate how the characteristics of the QCD equation of state manifest themselves in the evolution of the particle-emitting source and the resulting HBT radii.

The text is organized as follows: in Chapter 2, we present the tools and the description used in the mapping of the QCD phase diagram. In Chapter 3, we set out the standard theoretical-phenomenological model employed to describe nuclear collisions. In Chapter 4, we specify the simulation chain adopted in carrying out the numerical studies. The historical background and theoretical foundations of particle interferometry are discussed in Chapter 5 and, finally, we present and analyze the results obtained for the interferometry of pions emitted in $^{197}$Au+$^{197}$Au collisions at selected energies of the STAR BES program.

\section{Dictionary and Conventions}

Before moving to the theoretical and phenomenological developments of the following chapters, we introduce the notation, coordinate system, and kinematic conventions adopted throughout this thesis.

 We employ natural units $c = \hbar = k_B = e =  1$.  The hydrodynamic description is formulated in Milne coordinates $x^\mu = (\tau, x, y, \eta_S)$, related to Cartesian coordinates by $t = \tau \cosh \eta_S$ and $z = \tau \sinh \eta_S$, such that
\begin{align}
    \tau = \sqrt{t^2 - z^2},  \quad \eta_S = \text{atanh}\left(\frac{z}{t}\right),
\label{eq:milnecoordinates}
\end{align}
where $\tau$ plays the role of time and $\eta_S$ is called spatial-rapidity. Here, one assumes the $(+,-,-,-)$ convention for the metric. The metric in these coordinates is given by
\begin{align}
    &g_{\mu \nu} = \text{diag}(1, -1, -1, -\tau^2),\\
    &g^{\mu \nu} = \text{diag}(1, -1, -1, -\tau^{-2}),\\
    &g = \text{det}(g_{\mu \nu}) = -\tau^2,
\end{align}
and $\sqrt{-g} = \tau$. The non-zero Christoffel symbol terms are \cite{denicol2022microscopic}
\begin{align}
    &\Gamma^0_{33} = \tau, \\
    &\Gamma^{3}_{03} = \Gamma^{3}_{30} = \frac{1}{\tau} .
\end{align}
Below, we define the terms used in the hydrodynamic calculations \cite{denicol2022microscopic}.
\begin{itemize}
    \item \textbf{Covariant Derivative} is denoted as $\text d_\mu$.
    
    \item \textbf{Four-velocity} is denoted as $\text u^\mu$.
    
    \item \textbf{Comoving Derivative}, or \textbf{Total Covariant Derivative}, is denoted as $\text D = u^\mu d_ \mu$.

    \item \textbf{Expansion Scalar} is denoted as $\Theta$, and defined as $\Theta = \text d _\mu u^\mu$.
    
    \item \textbf{Covariant Levi-Civita Tensor} is denoted as $\varepsilon_{\mu \nu \alpha \beta}$ and defined as
    \begin{equation}
        \varepsilon_{\mu \nu \alpha \beta} = \sqrt{-g} \tilde\epsilon_{\mu \nu \alpha \beta},
    \end{equation}
    since $\tilde\epsilon_{\mu \nu \alpha \beta}$ is the 4-D Levi-Civita symbol with $\epsilon_{0123} = 1$.
    
    \item \textbf{Contravariant Levi-Civita Tensor} is denoted as $\varepsilon^{\mu \nu \alpha \beta}$ and defined as
    \begin{equation}
        \varepsilon^{\mu \nu \alpha \beta} = \frac{\text{sign}(g)}{\sqrt{-g}}\tilde\epsilon_{\mu \nu \alpha \beta}.
    \end{equation}
    
    \item \textbf{Projection Operator orthogonal to four-velocity} is denoted as $\Delta^{\mu \nu}$ and defined by
    \begin{align}
        \Delta^{\mu \nu} = g^{\mu \nu} - u^\mu u^\nu.
    \end{align}
    where
    \begin{align}
        &\Delta^{\mu \nu} = \Delta ^{\nu \mu},\\
        &\Delta^{\mu \nu} u_\nu = 0,\\
        &\Delta^{\mu}_{\ \mu} = d - 1
    \end{align}
    since $d$ is the space-time dimension. Here $d = 4$.
    
    \item \textbf{Double-Symmetric Projection Operator orthogonal to four-velocity} is denoted as $\Delta^{\mu \nu \alpha \beta}$ and defined by
    \begin{align}
       \Delta^{\mu \nu \alpha \beta} = \frac{1}{2}\left(\Delta^{\mu \alpha } \Delta^{\nu \beta} + \Delta^{\mu \beta}\Delta^{\nu \alpha}\right) - \frac{1}{d-1} \Delta^{\mu \nu}\Delta^{\alpha \beta}
    \end{align}
    where 
    \begin{align}
        &\Delta^{\mu \nu \alpha \beta} = \Delta^{\alpha \beta \mu \nu},\\
        &\Delta^{\mu \nu \alpha \beta}u_\mu = \Delta^{\mu \nu \alpha \beta}g_ {\mu \nu} = 0,\\
        &\Delta^{\mu \nu}_{\mu \nu} = 5.
    \end{align}

    \item \textbf{Space-Like Gradient} is denoted as $\nabla^\mu$ and defined as
    \begin{equation}
        \nabla^\mu = \Delta^{\mu \nu}\text d_\nu.
    \end{equation}

    \item \textbf{Expansion Rate} is denoted as $\theta$ and defined as $\theta = \nabla _\mu u^\mu$. Since $u_\mu D u^\mu = 0$, then $\theta = \Theta$.
    
    \item \textbf{Shear Tensor} is denoted as $\sigma^{\mu \nu}$ and defined as
    \begin{equation}
    \begin{split}
        \sigma^{\mu \nu} &= \Delta^{\mu \nu \alpha \beta}\text d_\alpha u^\beta  \\
        &= \frac{\text d^\mu u^\nu + \text d^\nu u^\mu}{2} - \frac{u^\mu \text D u^\nu + u^\nu \text D u^\mu}{2} - \frac{\Delta^{\mu \nu}}{d -1} \theta.
    \end{split}
    \end{equation}

    \item \textbf{Vorticity Tensor} is denoted as $\omega^{\mu \nu}$ and defined as
    \begin{equation}
    \begin{split}
        &\omega^{\mu \nu} = \frac{1}{2}\left(\nabla^\mu u^\nu - \nabla ^\nu u^\mu \right)\\
        &= \frac{\text d^\mu u^\nu - \text d^\nu u^\mu}{2} - \frac{u^\mu \text D u^\nu - u^\nu \text D u^\mu}{2}.
    \end{split}
    \end{equation}
    \item Given two four-vectors $A = (A^0, \vec A)$ and $B = (B^0, \vec B)$, we denote the Minkowski scalar product as $A  \cdot B = A^\mu B_\mu$ and the spatial scalar product as $\vec A \cdot \vec B = A^i B_i$.
\end{itemize}

\subsection{Kinematic Variables}

The momentum $p = (p^0, p^x, p^y, p^z)$ of a particle of mass $m$ produced in a collision can be parametrized in the following way.

We define the transverse momentum $p_T$ and the angle $\phi$
\begin{align}
    &p_T = \sqrt{(p^x)^2 + (p^y)^2},
    \label{eq:transversemomentum}\\
    &\phi = \arctan(\frac{p^y}{p^x}).
    \label{eq:transverseangle}
\end{align}
We define rapidity $y$ as
\begin{equation}
    y = \frac{1}{2}\ln(\frac{p^0+p_z}{p^0-p_z})
    \label{eq:rapidity}
\end{equation}
and it is possible to write $p^0 = m_T\cosh(y)$ and $p_z = m_T\sinh(y)$ if we define
\begin{equation}
    m_T = \sqrt{m^2 + p_T^2}
    \label{eq:transversalmass}
\end{equation}
as the transverse mass. So
\begin{equation}
    p = (m_T\cosh(y), p_T \cos(\phi), p_T\sin(\phi), m_T\sinh(y)).
\end{equation}
We also define here the pseudorapidity $\eta$ as
\begin{equation}
    \eta = \frac{1}{2}\ln(\frac{|\vec p| + p_z}{|\vec p| - p_z})
    \label{eq:pseudorapidity}.
\end{equation}


\chapter{Distributions and Equation of State for QCD Matter}
\label{chap:2}

Macroscopic descriptions of matter summarize the underlying microscopic dynamics. Although microscopic degrees of freedom interact through complex many-body processes, their collective behavior is often captured by a few thermodynamic variables: energy density, pressure, temperature, entropy density, and conserved-charge densities. Statistical mechanics links these scales by deriving macroscopic quantities from microscopic distributions and equilibrium ensembles.

This link is essential in relativistic heavy-ion collisions. After the collision, QCD matter evolves rapidly, with strong interactions, nonzero conserved charges, and a transition from deconfined quark-gluon plasma to a hadronic phase. Hydrodynamic simulations do not follow individual particles; instead, they evolve conservation equations for the energy-momentum tensor and conserved currents. To close these equations, one needs an equation of state relating pressure, temperature, chemical potentials, and densities.

The equation of state is thus a key input in hydrodynamic modeling. It determines the response to compression and expansion, the relation between energy density and pressure, the speed of sound, and the geometry of the particlization hypersurface. Because observables such as particle spectra, anisotropic flow coefficients, and femtoscopic radii depend on the medium’s space-time evolution, the equation of state directly influences how the QCD phase structure appears in final-state measurements.

This chapter introduces the kinetic-theory concepts that connect microscopic particle distributions to macroscopic hydrodynamic fields and the thermodynamic framework underlying the equation of state used here. The kinetic-theory discussion does not derive the specific equation of state employed in the simulations; rather, it defines particle currents, the energy-momentum tensor, and local equilibrium distributions, which enter the hydrodynamic formulation and the Landau matching procedure.

We then review the thermodynamic relations among pressure, energy density, entropy density, temperature, chemical potentials, and conserved-charge densities, emphasizing their role in closing the hydrodynamic equations. Finally, we discuss equations of state for QCD matter at finite baryon density, from earlier critical-point–motivated parameterizations used in models such as NexSPheRIO to the BEST equation of state adopted in this thesis.
\section{Rudiments of Kinetic Theory}

The purpose of this section is not to provide a complete theoretical treatment of kinetic theory or distribution theory. We introduce a few basic definitions that are useful for describing systems composed of many particles.

\subsection{Phase-Space Distributions}

Let us consider a system composed of many particles. At a fixed time, we define the phase-space distribution $f(x,p)$, given by
\begin{equation}
    f(x,p) = \dv{^6N}{^3x \dd^3p},
\end{equation}
where $N$ is the particle number, $\dd^3x$ is the spatial volume element, and $\dd^3p$ is the momentum-space volume element. Although $x$ and $p$ denote four-vectors, the volume elements in this definition refer only to the spatial coordinates and to the three-momentum components. With this convention, the normalization of the distribution gives the total number of particles in the system,
\begin{equation}
    \int \dd^3x \dd^3p \, f(x,p)
    =
    \int \dd^3x \dd^3p \, \dv{^6N}{^3x \dd^3p}
    =
    \int \dd N
    =
    N.
\end{equation}
Given any quantity $A=A(x,p)$, its average over phase space is
\begin{equation}
    \ev{A}
    \equiv
    \frac{\int \dd^3x \dd^3p \, A(x,p) f(x,p)}
    {\int \dd^3x \dd^3p \, f(x,p)}
    =
    \frac{\int \dd^3x \dd^3p \, A(x,p) f(x,p)}{N}.
\end{equation}
This is an average per particle. The corresponding extensive quantity is obtained by multiplying by the total particle number, namely $N\ev{A}$. One may also define marginal distributions in coordinate and momentum space,
\begin{align}
    &\frac{d^3N}{d^3x}
    \equiv
    f_X(x)
    =
    \int \dd^3p \, f(x,p),\\
    &\frac{d^3N}{d^3p}
    \equiv
    f_P(p)
    =
    \int \dd^3x \, f(x,p).
\end{align}
Here $f_X(x)$ is the spatial distribution, and $f_P(p)$ is the momentum distribution. Averages with respect to these marginal distributions can be written as
\begin{align}
    &\ev{A}_X
    =
    \frac{\int \dd^3x \, A(x) f_X(x)}{N},\\
    &\ev{A}_P
    =
    \frac{\int \dd^3p \, A(p) f_P(p)}{N},
\end{align}
with
\begin{equation}
    \int \dd^3x \, f_X(x)
    =
    \int \dd^3p \, f_P(p)
    =
    N.
\end{equation}

\subsection{Macroscopic Quantities from Phase-Space Distributions}

A system with many microscopic degrees of freedom can often be described in terms of averages of the phase-space distribution, as will become clear in what follows. The particle density in coordinate space is obtained directly from the phase-space distribution,
\begin{equation}
    n^*(x)
    =
    \dv{^3N}{^3x}
    =
    \int \dd^3p \, f(x,p).
    \label{eq:spacedensity}
\end{equation}
Analogously, the momentum-space density is given by $\tilde n^*(p)=f_P(p)$. Neither $\dd^3x$ nor $\dd^3p$ is separately Lorentz invariant. However, the measure $\dd^3p/p^0$ is Lorentz invariant \cite{Florkowski2010}. Since the total number of particles must be invariant, the quantity
\begin{equation}
    p^0 \dv{^3N}{^3p}
    =
    p^0 f_P(p)
\end{equation}
is also invariant. This motivates the definition of the invariant one-particle momentum distribution,
\begin{equation}
    \mathcal{N}(p)
    =
    p^0 f_P(p).
\end{equation}
The average energy per particle is
\begin{equation}
    \ev{E}
    =
    \frac{\int \dd^3x \dd^3p \, E(x,p) f(x,p)}{N},
\end{equation}
so that the internal energy of the system is
\begin{equation}
    U
    \equiv
    N\ev{E}
    =
    \int \dd^3x \dd^3p \, E(x,p) f(x,p).
\end{equation}
The corresponding energy density is
\begin{equation}
    \epsilon(x)
    \equiv
    \dv{^3U}{^3x}
    =
    \int \dd^3p \, E(x,p) f(x,p).
\end{equation}
For an isotropic system, the pressure is defined as
\begin{equation}
    P(x)
    =
    \frac{1}{3}
    \int \dd^3p \, \vec p \cdot \vec v \, f(x,p),
\end{equation}
where $\vec p$ is the three-momentum and $\vec v$ is the particle velocity. In the relativistic case, $\vec v=\vec p/p^0$. This expression shows that pressure is associated with the average momentum flux carried by the particles. In thermodynamics, the same quantity will later be related to the grand potential.

\subsection{Particle Four-Current and Energy-Momentum Tensor}

\begin{figure}[!htbp]
\centering
\begin{tikzpicture}[scale=2.0, every node/.style={font=\large}]

\def\cell{0.85}
\def\x0{0}
\def\y0{0}

\definecolor{energyred}{RGB}{255,90,90}
\definecolor{momorange}{RGB}{255,205,80}
\definecolor{momblue}{RGB}{90,100,255}
\definecolor{pressuregreen}{RGB}{70,235,90}

\fill[energyred, opacity=0.75]
    (\x0,\y0) rectangle ++(\cell,-\cell);

\foreach \j in {1,2,3}{
    \fill[momorange, opacity=0.75]
        (\x0+\j*\cell,\y0) rectangle ++(\cell,-\cell);
}
\foreach \i in {1,2,3}{
    \fill[momorange, opacity=0.75]
        (\x0,\y0-\i*\cell) rectangle ++(\cell,-\cell);
}

\fill[momblue, opacity=0.70]
    (\x0+\cell,\y0-\cell) rectangle ++(3*\cell,-3*\cell);

\foreach \i in {1,2,3}{
    \fill[pressuregreen, opacity=0.75]
        (\x0+\i*\cell,\y0-\i*\cell) rectangle ++(\cell,-\cell);
}

\foreach \i in {0,1,2,3}{
    \foreach \j in {0,1,2,3}{
        \node at (\x0+\j*\cell+0.5*\cell,\y0-\i*\cell-0.5*\cell)
        {$T^{\i\j}$};
    }
}

\draw[line width=1.2pt]
    (\x0-0.25,\y0+0.05) -- (\x0-0.45,\y0+0.05)
    -- (\x0-0.45,\y0-4*\cell-0.05)
    -- (\x0-0.25,\y0-4*\cell-0.05);

\draw[line width=1.2pt]
    (\x0+4*\cell+0.25,\y0+0.05) -- (\x0+4*\cell+0.45,\y0+0.05)
    -- (\x0+4*\cell+0.45,\y0-4*\cell-0.05)
    -- (\x0+4*\cell+0.25,\y0-4*\cell-0.05);

\node[energyred, align=center, font=\normalsize]
    at (\x0+0.05*\cell,\y0+1.25)
    {energy\\density};

\draw[energyred, line width=1.0pt]
    (\x0+0.45*\cell,\y0+0.65) -- (\x0+0.45*\cell,\y0+0.02);

\node[momorange, align=center, font=\normalsize]
    at (\x0+2.85*\cell,\y0+1.25)
    {momentum\\density};

\draw[momorange, line width=1.0pt]
    (\x0+2.55*\cell,\y0+0.65) -- (\x0+2.55*\cell,\y0+0.02);

\node[momblue, align=center, font=\normalsize]
    at (\x0+2.55*\cell,\y0-4*\cell-1.05)
    {momentum\\flux};

\draw[momblue, line width=1.0pt]
    (\x0+2.55*\cell,\y0-4*\cell-0.35) -- (\x0+2.55*\cell,\y0-4*\cell+0.02);

\node[pressuregreen, align=left, font=\normalsize]
    at (\x0+4*\cell+1.55,\y0-3.25*\cell)
    {pressure};

\draw[pressuregreen, line width=1.0pt]
    (\x0+4*\cell+0.05,\y0-3.5*\cell)
    -- (\x0+4*\cell+1.25,\y0-3.5*\cell);

\end{tikzpicture}
\caption{Schematic representation of the components of $T^{\mu\nu}$. The component $T^{00}$ represents the energy density, the mixed components are associated with momentum density and energy density flow, and the spatial block $T^{ij}$ describes momentum flux. For an isotropic system, the diagonal spatial components are related to the pressure.}
\label{fig:energy-momentum-tensor-components}
\end{figure}
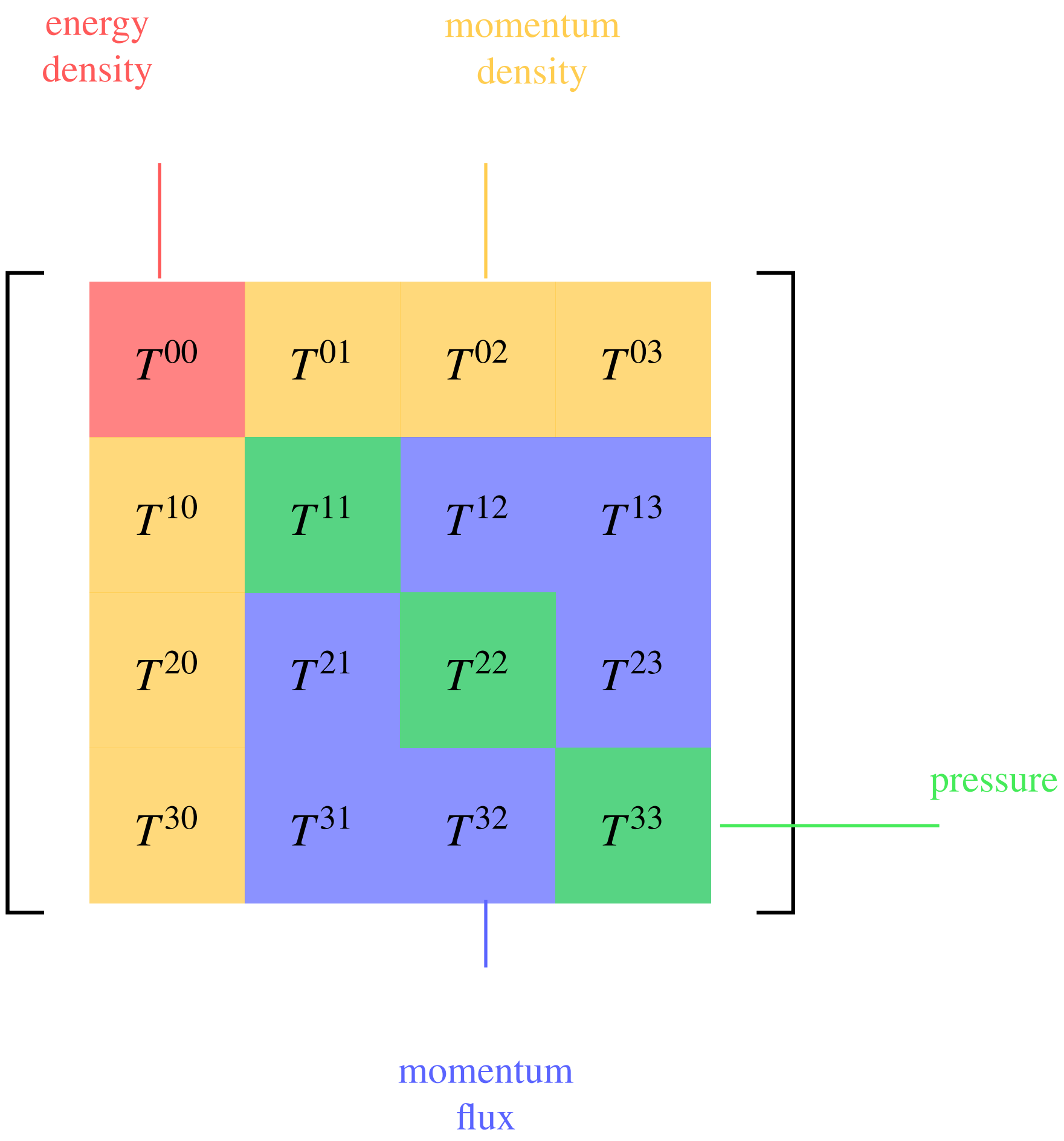

The previous definitions can be written in a covariant form. Let $u^\mu$ be the four-velocity associated with the local rest frame of the system, and let $n$ be the particle density measured in that frame. The particle four-current is defined as
\begin{equation}
    j^\mu = n u^\mu.
\end{equation}
In the local rest frame, $u^\mu=(1,0,0,0)$ and therefore $j^\mu=(n,0,0,0)$. In a general frame, the time component gives the particle density measured by that observer, while the spatial components describe the particle flux. The same object can be expressed directly in terms of the phase-space distribution. Since $\dd^3p/p^0$ is Lorentz invariant and $p^\mu$ is a four-vector, the expression \cite{groot1980relativistic, liboff2003kinetic} 
\begin{equation}
    j^\mu(x)
    =
    \int \frac{\dd^3p}{p^0} \, p^\mu f(x,p)
    \label{eq:4current}
\end{equation}
transforms as a four-vector, provided that $f(x,p)$ is a Lorentz scalar.One can readily obtain Eq. \eqref{eq:spacedensity} by examining $j^0$, which supports interpreting the integral in Eq. \eqref{eq:4current} as a four-current.. The proper density is then obtained by projection,
\begin{equation}
    n = j^\mu u_\mu.
\end{equation}
The energy-momentum tensor is constructed in an analogous way \cite{groot1980relativistic, liboff2003kinetic} ,
\begin{equation}
    T^{\mu\nu}(x)
    =
    \int \frac{\dd^3p}{p^0} \, p^\mu p^\nu f(x,p).
    \label{eq:energystresstensor}
\end{equation}
This tensor is symmetric by construction. Its components have a direct physical interpretation. The component $T^{00}$ is the energy density measured in the chosen frame,
\begin{equation}
    T^{00}
    =
    \int \dd^3p \, p^0 f(x,p),
\end{equation}
while the components $T^{0i}=T^{i0}$ are associated with energy density flux. Indeed,
\begin{equation}
    \int \dd^3x \, T^{0i}
    =
    \int \dd^3x \dd^3p \, p^i f(x,p)
    =
    N\ev{p^i}.
\end{equation}
The spatial components $T^{ij}$ describe momentum flux. For $i\neq j$, they are associated with shear stresses, while the diagonal components $T^{ii}$ describe normal stresses. In an isotropic system, these normal stresses are equal and define the pressure. Taking the trace over the spatial components gives
\begin{equation}
    \sum_i T^{ii}
    =
    \int \dd^3p \, \vec p \cdot \vec v \, f(x,p)
    =
    3P,
\end{equation}
which is consistent with the kinetic definition of pressure introduced above. Thus, the energy-momentum tensor collects in a single object the energy density, momentum density, and momentum flux of the system. This makes it one of the central quantities in any macroscopic description of relativistic matter. A schematic representation of $T^{\mu\nu}$ is shown in Fig. \ref{fig:energy-momentum-tensor-components}.

\subsection{Cooper-Frye Formula}

Consider a cloud of particles occupying a certain volume $V=V(t)$ in space, which may change with time. The total number of particles in this region, $N=N(t)$, can be quantified simply as the volume integral
\begin{equation}
    N(t) = \int_{V(t)} \dd^3r n^*(t, \Vec{r})
\end{equation}
where $n^* = n^*(\vec r,t)$ is the particle density. The Reynolds transport theorem  guarantees that the time variation of $N$ can be written as \cite{leal2007advanced}
\begin{equation}
    \dv{N}{t} = \int_{V(t)} \dd^3r\left(\frac{\partial n^*}{\partial t} + \nabla \cdot( {n^* \vec v} )\right)
\end{equation}
where $\Vec{v}$ is the velocity field of the emitted particles. We can write this equation in covariant form by making the relativistic correspondence
\begin{equation}
    n^* \to n u^0,
\end{equation}
where $n^*$ is the particle density in the laboratory frame and $n$ is the particle density in the local rest frame. We also have
\begin{equation}
    n^* \vec v \to n u^0 v^i =  n u^i.
\end{equation}
Thus, defining $u^\mu$ as the $4-$velocity field, and taking $\partial_\mu = (\partial_t, \partial_x, \partial_y, \partial_z)$, we obtain
\begin{equation}
    \dv{N}{t} = \int_{V(t)}\dd^3r \partial_\mu(nu^\mu).
\end{equation}
Integrating both sides of this equation with respect to time, with $\dd^3r \to \dd^4r =\dd t \dd^3 r $, we obtain
\begin{equation}
    N = \int_{\Delta}\dd^4r \partial_\mu(nu^\mu).
    \label{eq:particlenumberfromcurrents}
\end{equation}
where $\Delta$ is the spacetime hypervolume containing $N$. Since $\dd^4r$ is a Lorentz invariant in Cartesian coordinates, the particle number $N$ in this case is also Lorentz invariant. Although we derived the equations above using Cartesian coordinates, it is also possible to make the correspondence to generalized coordinates such that
\begin{align}
    &\int_\Delta \dd^4r \to \int_\Delta \dd^4r \sqrt{-g},\\
    &\partial _\mu (n u^\mu) \to \frac{1}{\sqrt{-g}}\partial _\mu(\sqrt{-g} n u^\mu),
\end{align}
so that
\begin{equation}
    N = \int_\Delta \dd^4r \sqrt{-g} \left( \frac{1}{\sqrt{-g}}\partial _\mu (\sqrt{-g} n u^\mu)\right)
\end{equation}
where now $r$, $u$, and $\partial_\mu$ are written in an arbitrary coordinate system, and $g$ is the determinant of the metric in this system. Now, using the $4-$dimensional version of Stokes' theorem \cite{DInvernoVickers2022}, we arrive at
\begin{equation}
    N = \int_\Sigma \dd \sigma(x) \cdot j(x),
    \label{eq:Nfromsigmaj}
\end{equation}
where $\Sigma = \partial \Delta$ is the boundary of $\Delta$. We take $(\vec r, t) \to x$, where $x$ is a four-vector parameterizing the spacetime coordinates used in the integration. Here, $d\sigma = d \sigma (x)$ is the oriented hypersurface four-element, orthogonal to the hypersurface, and we define $j^\mu (x) = nu^\mu$ as the particle four-current. The term $\dd \sigma(x)$ can then be written as \cite{Florkowski2010,Poisson2004Hypersurfaces,Gourgoulhon2012}
\begin{equation}
    d\sigma_\mu = \tilde\epsilon_{\mu  \alpha \beta \gamma} \pdv{x^\alpha}{y^1}\pdv{x^\beta}{y^2}\pdv{x^\gamma}{y^3}\dd y^1 \dd y^2 \dd y^3
    \label{eq:surfaceelement}
\end{equation}
where $(y^1,y^2,y^3)$ are the three free variables that parametrize $\Sigma$, and we write $x = x(y^1, y^2, y^3)$. We now extend the use of Eq.~\eqref{eq:Nfromsigmaj} to currents obtained from distributions, as in Eq.~\eqref{eq:4current}. In this way, the particle number $N$ becomes
\begin{equation}
    N = \int \frac{\dd^3p}{p^0} \int_\Sigma \dd \sigma(x) \cdot p f(x,p)
\end{equation}
which gives
\begin{equation}
    p^0 \dv{^3N}{^3p} = \int_\Sigma \dd \sigma(x) \cdot p f(x,p)
    \label{eq:cff}
\end{equation}
that is, an invariant distribution in momentum space. We call this equation the Cooper-Frye formula \cite{PhysRevD.10.186}. It is important because, as we will see in the theoretical collision model, the transition from the hydrodynamic description to the particle description is performed using this equation. Details of the calculation of $d\sigma$ are given in Appendix~\ref{chap:surfacelemente}.

\subsection{Boltzmann Equation}

The evolution of the phase-space distribution is governed by the Boltzmann equation,
\begin{equation}
    p^\mu \partial_\mu f(x,p)
    =
    C[f],
    \label{eq:boltzmann}
\end{equation}
where $C[f]$ is the collision term. The left-hand side describes the propagation of particles in phase space, while the collision term accounts for microscopic interactions. Although we will not solve this equation explicitly here, we include it because it underlies several key tools used throughout the field, from hadronic and partonic transport approaches to n-moment methods for deriving effective evolution equations away from equilibrium \cite{denicol2022microscopic}. In particular, taking moments of the Boltzmann equation leads to evolution equations for quantities such as particle currents and the energy-momentum tensor.

\section{Statistical Quantum Mechanics}

Classical statistical mechanics provides the tools needed to derive the thermodynamic properties of gases, including equations of state and the conditions associated with phase transitions \cite{Kardar2007}. In this section, we use an equivalent quantum-statistical formulation, which is particularly convenient for systems composed of bosons and fermions. The discussion is based on the grand-canonical ensemble, since this ensemble naturally incorporates particle exchange and chemical potentials \cite{Kardar2007,GottfriedYan2003}.

\subsection{Thermodynamic Fundamentals}

Before introducing the statistical description, let us recall some basic thermodynamic relations \cite{Kardar2007}. For a system in thermodynamic equilibrium, we denote its internal energy by $U$, its entropy by $S$, its particle number by $N$, and its volume by $V$. The differential form of the internal energy is
\begin{equation}
    \dd U = -P\dd V + T\dd S + \mu \dd N,
    \label{eq:dU1}
\end{equation}
where $P$ is the pressure, $T$ is the temperature, and $\mu$ is the chemical potential. In systems with more than one conserved quantity, the last term is generalized as
\begin{equation}
    \mu \dd N \to \sum_i \mu_i \dd N_i,
\end{equation}
where $i$ labels the conserved charges. For an extensive thermodynamic system, the Euler relation can be written as
\begin{equation}
    U = -PV + TS + \mu N.
\end{equation}
Taking the differential of this expression gives
\begin{equation}
    \dd U = -P\dd V - V\dd P + T\dd S + S\dd T + \mu \dd N + N\dd \mu.
    \label{eq:dU2}
\end{equation}
Subtracting Eq.~\eqref{eq:dU1} from Eq.~\eqref{eq:dU2}, one obtains
\begin{equation}
    V\dd P = S\dd T + N\dd \mu.
\end{equation}
This is the Gibbs--Duhem relation. Dividing by the volume, we obtain its intensive form,
\begin{equation}
    \dd P = s\dd T + n\dd \mu,
    \label{eq:gibbsduhem}
\end{equation}
where $s=S/V$ is the entropy density and $n=N/V$ is the particle number density. The energy density is defined as
\begin{equation}
    \epsilon = \frac{U}{V} = -P + Ts + \mu n.
\end{equation}
Its differential is
\begin{equation}
    \dd \epsilon = -\dd P + T\dd s + s\dd T + \mu \dd n + n\dd \mu.
\end{equation}
Using the intensive Gibbs-Duhem relation, this reduces to
\begin{equation}
    \dd \epsilon = T\dd s + \mu \dd n.
    \label{eq:gibbsduhem2}
\end{equation}
The grand potential $\Omega$ is obtained from the internal energy through a Legendre transformation from the variables $(S,N)$ to $(T,\mu)$,
\begin{equation}
    \Omega = U - TS - \mu N = -PV.
    \label{eq:grandpotential}
\end{equation}
Its differential is
\begin{equation}
    \dd \Omega = -P\dd V - S\dd T - N\dd \mu.
\end{equation}
Therefore,
\begin{align}
    &P = -\left(\pdv{\Omega}{V}\right)_{T,\mu} = -\frac{\Omega}{V},\\
    &S = -\left(\pdv{\Omega}{T}\right)_{V,\mu},\\
    &N = -\left(\pdv{\Omega}{\mu}\right)_{T,V}.
    \label{eq:thermostateequations}
\end{align}
Defining the grand-potential density $\omega\equiv\Omega/V$, one has
\begin{equation}
    \omega=-P,
\end{equation}
and
\begin{equation}
    \dd \omega = -s\dd T - n\dd \mu.
\end{equation}

The grand potential is especially useful for systems in which temperature and chemical potential are the natural thermodynamic variables. This is the case for matter with conserved charges, where chemical potentials control the corresponding charge densities. In the thermodynamic limit, the grand potential is related to the grand-canonical partition function by
\begin{equation}
    \Omega = -\frac{1}{\beta}\ln Z,
    \qquad
    \beta = \frac{1}{T},
    \label{eq:grandpotentialpartitionfunc}
\end{equation}
where $Z$ is the partition function. The Gibbs-Duhem relation in \eqref{eq:gibbsduhem2} allows to express the densities $n$ and $s$ as
\begin{align}
    &n = \frac{N}{V} = \frac{\partial P}{\partial \mu}\bigg|_T,\\
    &s = \frac{S}{V} = \frac{\partial P}{\partial T}\bigg|_\mu.
    \label{eq:derivativesgibbsduhem}
\end{align}
\subsection{Quantum Preliminaries}

Consider a thermal system composed of a single particle species. In the grand-canonical ensemble, the density matrix operator is defined as \cite{GottfriedYan2003}
\begin{equation}
   \hat \rho =
   \frac{\exp[-\beta(\hat H-\mu \hat N)]}{Z},
   \qquad
   \beta = \frac{1}{T},
   \label{matrixdensityoperatorthermo}
\end{equation}
where $\hat H$ is the Hamiltonian, $\hat N$ is the particle-number operator, and $\mu$ is the chemical potential. The partition function is fixed by the normalization condition
\begin{equation}
    \Tr \hat \rho = 1,
\end{equation}
which gives
\begin{equation}
    Z = \Tr\left\{\exp[-\beta(\hat H-\mu \hat N)]\right\}.
    \label{eq:partitionfunction}
\end{equation}
For a free quantum gas, the Hamiltonian can be written as
\begin{equation}
    \hat H = E^0 + \sum_{i,s} E_i \hat N_{i,s},
\end{equation}
where $E^0$ is the vacuum contribution, $E_i$ is the energy of the state labeled by $i$, and $s$ denotes additional quantum numbers that do not change the energy. The total particle-number operator is
\begin{equation}
    \hat N = \sum_{i,s}\hat N_{i,s},
    \label{eq:particlenumber}
\end{equation}
with
\begin{equation}
    \hat N_{i,s} = \hat a^\dagger_{i,s}\hat a_{i,s}.
    \label{eq:stateparticlenumber}
\end{equation}

\subsection{Partition Functions}

Using the occupation-number basis $\ket{\{N_{i,s}\}}$, the trace in Eq.~\eqref{eq:partitionfunction} can be written as
\begin{equation}
    Z =
    \exp(-\beta E^0)
    \prod_{i,s}
    \sum_{N_{i,s}}
    \left[
    \exp\left(\beta(\mu-E_i)\right)
    \right]^{N_{i,s}}.
\end{equation}
For bosons, the occupation number of each state can take the values $N_{i,s}=0,1,2,\ldots,\infty$. Therefore, using the geometric series,
\begin{equation}
    Z =
    \exp(-\beta E^0)
    \prod_{i,s}
    \left[
    \frac{1}{1-\exp[\beta(\mu-E_i)]}
    \right].
    \label{eq:bosonicpartition}
\end{equation}
For fermions, the occupation number is restricted to $N_{i,s}=0,1$, which gives
\begin{equation}
    Z =
    \exp(-\beta E^0)
    \prod_{i,s}
    \left[
    1+\exp[\beta(\mu-E_i)]
    \right].
    \label{eq:fermionicpartition}
\end{equation}

\subsection{Equation of State}

We now use the grand potential to obtain the thermodynamic quantities of the system. The average particle number is
\begin{equation}
    N\equiv\ev{\hat N} = \Tr\{\hat \rho \hat N\}.
\end{equation}
Using the definition of $\hat \rho$ in Eq. \eqref{matrixdensityoperatorthermo}, the relation between $\Omega$ and $Z$, and standard derivative identities, one obtains
\begin{equation}
    N = -\pdv{\Omega}{\mu}
    .
    \label{eq:avnumbpart}
\end{equation}
This agrees with Eq.~\eqref{eq:thermostateequations}. Similarly, the internal energy is obtained from
\begin{equation}
    U \equiv \ev{\hat H} = \Tr\{\hat \rho \hat H\},
\end{equation}
leading to
\begin{equation}
   U = \beta\pdv{\Omega}{\beta} + \Omega + \mu N
   .
   \label{eq:energyfromsm}
\end{equation}
The entropy is computed from the von Neumann expression
\begin{equation}
    S = -\Tr\{\hat \rho \ln \hat \rho\},
\end{equation}
which gives
\begin{equation}
   S = \beta(U-\mu N-\Omega).
   \label{eq:entropyformula}
\end{equation}
Since $\beta=1/T$, this result is equivalent to the thermodynamic identity $\Omega=U-TS-\mu N$. Thus, the derivative relations, including those from \eqref{eq:derivativesgibbsduhem}, remain valid.

\subsubsection{Bosonic Case}

For bosons, using Eqs.~\eqref{eq:bosonicpartition} and \eqref{eq:grandpotentialpartitionfunc}, the grand potential is
\begin{equation}
    \Omega =
    E^0
    +
    \frac{1}{\beta}
    \sum_{i,s}
    \ln\left(
    1-\exp[\beta(\mu-E_i)]
    \right).
\end{equation}
If the energy does not depend on the quantum number $s$, the sum over $s$ gives the degeneracy factor $g$,
\begin{equation}
    \sum_s \to g.
\end{equation}
Thus,
\begin{equation}
    \Omega =
    E^0
    +
    \frac{g}{\beta}
    \sum_i
    \ln\left(
    1-\exp[\beta(\mu-E_i)]
    \right).
\end{equation}
Using Eq.~\eqref{eq:avnumbpart}, the total particle number is
\begin{equation}
    N =
    \sum_i
    \frac{g}{\exp[\beta(E_i-\mu)]-1}.
\end{equation}
Therefore, the average occupation number of a bosonic state is
\begin{equation}
    \ev{\hat a^\dagger_{i,s}\hat a_{i,s}}
    =
    \frac{1}{\exp[\beta(E_i-\mu)]-1}.
\end{equation}
The internal energy is
\begin{equation}
    U =
    E^0
    +
    \sum_i
    \frac{gE_i}{\exp[\beta(E_i-\mu)]-1}.
\end{equation}

\subsubsection{Fermionic Case}

For fermions, using Eqs.~\eqref{eq:fermionicpartition} and \eqref{eq:grandpotentialpartitionfunc}, the grand potential is
\begin{equation}
    \Omega =
    E^0
    -
    \frac{1}{\beta}
    \sum_{i,s}
    \ln\left(
    1+\exp[\beta(\mu-E_i)]
    \right).
\end{equation}
Again, if the energy does not depend on $s$, the sum over this quantum number gives the degeneracy factor $g$,
\begin{equation}
    \Omega =
    E^0
    -
    \frac{g}{\beta}
    \sum_i
    \ln\left(
    1+\exp[\beta(\mu-E_i)]
    \right).
\end{equation}
The total particle number is
\begin{equation}
    N =
    \sum_i
    \frac{g}{\exp[\beta(E_i-\mu)]+1},
\end{equation}
and the average occupation number of a fermionic state is
\begin{equation}
    \ev{\hat a^\dagger_{i,s}\hat a_{i,s}}
    =
    \frac{1}{\exp[\beta(E_i-\mu)]+1}.
\end{equation}
The internal energy is
\begin{equation}
    U =
    E^0
    +
    \sum_i
    \frac{gE_i}{\exp[\beta(E_i-\mu)]+1}.
\end{equation}

\subsubsection{Generalization and the Parameter \texorpdfstring{$\boldsymbol{\theta}$}{theta}}

The bosonic and fermionic expressions can be written in a unified form by introducing a parameter $\theta$, with $\theta=-1$ for bosons and $\theta=+1$ for fermions. The grand potential is then
\begin{equation}
    \Omega =
    E^0
    -
    \frac{\theta g}{\beta}
    \sum_i
    \ln\left(
    1+\theta\exp[\beta(\mu-E_i)]
    \right).
\end{equation}
Defining
\begin{equation}
    h_{\theta,i}
    =
    \frac{1}{\exp[\beta(E_i-\mu)]+\theta},
\end{equation}
the particle number and internal energy become
\begin{equation}
    N =
    \sum_i g h_{\theta,i},
\end{equation}
and
\begin{equation}
    U =
    E^0
    +
    \sum_i gE_i h_{\theta,i}.
\end{equation}
For $\theta=+1$, $h_{\theta,i}$ is the Fermi–Dirac distribution, while for $\theta=-1$ it is the Bose–Einstein distribution. In the limit $\theta\to0$, one recovers the Maxwell–Boltzmann distribution for the occupation numbers \cite{Kardar2007}.

\subsubsection{The Continuum Limit}

In many physical systems, the energy spectrum must be treated as continuous. In this case, sums over discrete states are replaced by integrals over momentum space. In this way, one may write \cite{BurgessQFTNotes, Kardar2007}
\begin{equation}
    \sum_n
    \to
    \frac{V}{(2\pi)^3}
    \int \dd^3p,
\end{equation}
since $V$ is the volume of the system and the thermodynamic quantities become
\begin{align}
    &n
    =
    \frac{N}{V}
    =
    \frac{g}{(2\pi)^3}
    \int \dd^3p\, \frac{1}{\exp[\beta(E -\mu )]+\theta}
    \label{eq:partdens},\\
    &\epsilon 
    =
    \frac{U}{V}
    =
    B
    +
    \frac{g}{(2\pi)^3}
    \int \dd^3p\,  \frac{E }{\exp[\beta(E -\mu )]+\theta}, \label{eq:enerdens}\\
   &\omega 
   =
   \frac{\Omega}{V}
   =
   B
   -
   \frac{\theta g}{\beta(2\pi)^3}
   \int \dd^3p\,
   \ln\left(
   1+\theta\exp[\beta(\mu -E )]
   \right).
\end{align}
Here $B\equiv E^0/V$. Since $\omega=-P$ and using integration by parts, one may write
\begin{equation}
    P
    =
    -B
    +
    \frac{g}{(2\pi)^3}
    \int \dd^3p\,
    \frac{\vec p^{\,2}}{3E}
    \frac{1}{\exp[\beta(E -\mu )]+\theta} \label{eq:pressure}.
\end{equation}
These expressions have the same structure as the kinetic-theory definition, since we define the phase-space distribution in momentum space as
\begin{equation}
    f(x,p) = \frac{g}{(2\pi)^3}\frac{1}{\exp[\beta(E -\mu)]+\theta},
    \label{eq:distideal}
\end{equation}
since $E$, $T = 1/\beta$ and $\mu$ may vary with $(x,p)$.


\subsection{Phase Transitions}

A phase transition can be generically described as a non-analytic behavior of the thermodynamic potential, or of its derivatives, in the thermodynamic limit. These non-analyticities are associated with the distinct macroscopic forms that a substance or material may assume under different thermodynamic conditions. A simple example is water: although it is always composed of H$_2$O molecules, it can generally appear in three different phases, namely solid, liquid, and gas. These phases may be further divided into more detailed subphases depending on the thermodynamic conditions under consideration, but under ordinary temperature and pressure conditions, water is commonly described in terms of these three phases.

\begin{figure}[!htbp]
    \centering
    \adjustbox{trim=0pt 0pt 0pt 8pt,clip}{%
        \includegraphics[width=\linewidth]{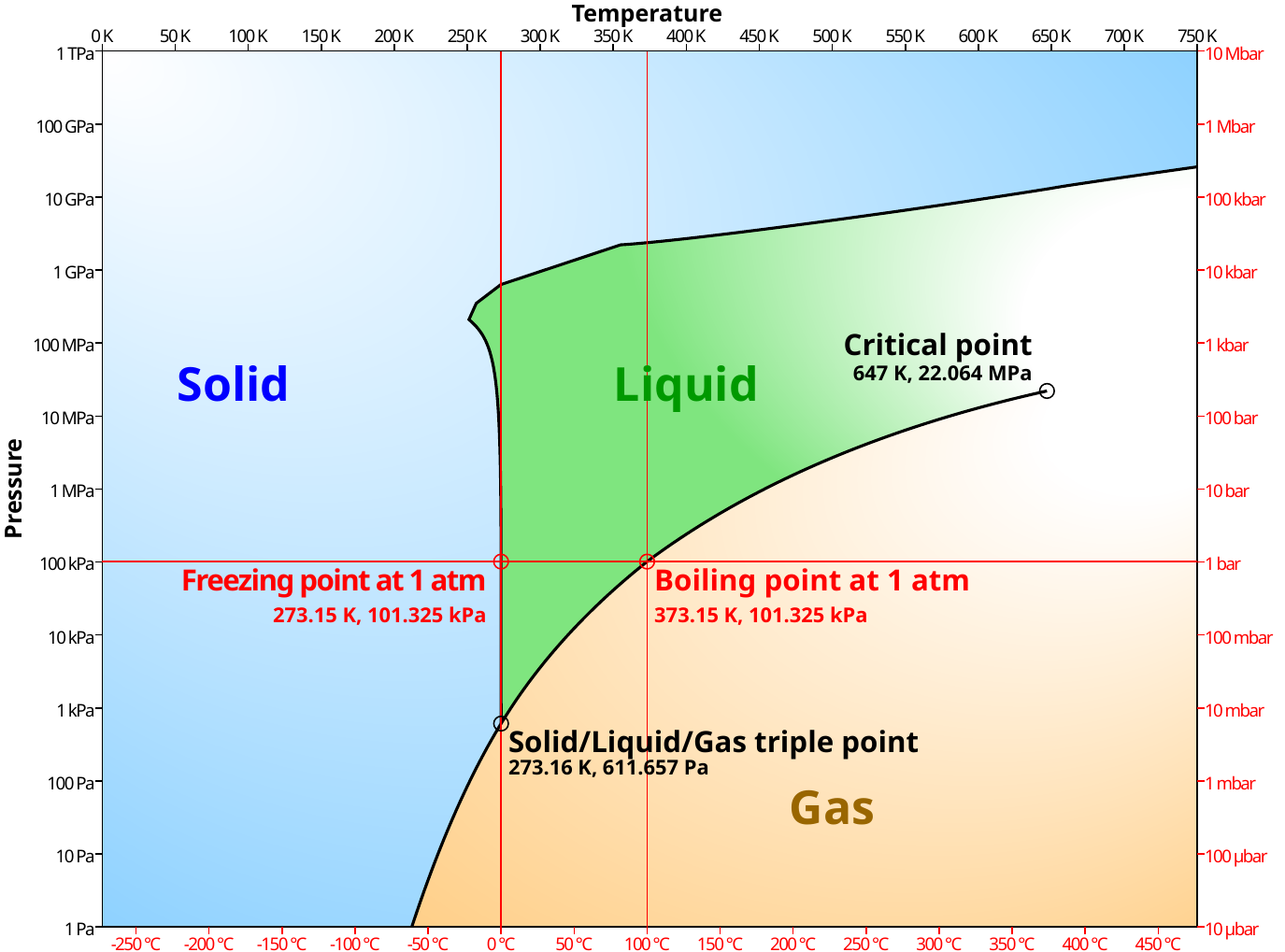}%
    }
    \caption{
    Schematic pressure--temperature phase diagram of water. The diagram shows the solid, liquid, and gaseous phases, the coexistence curves separating them, and the special locations of the triple point and the critical point. The horizontal line at atmospheric pressure indicates the usual freezing and boiling transitions of water at $0^\circ\mathrm{C}$ and $100^\circ\mathrm{C}$, respectively. Adapted from Wikimedia Commons, original work by Cmglee and simplified version by Manlleus \cite{WikimediaWaterPhaseDiagram}.
    }
    \label{fig:water_phase_diagram}
\end{figure}

At atmospheric pressure, $1~\mathrm{atm}$, water is in the solid phase for temperatures below $0^\circ\mathrm{C}$, in the liquid phase for temperatures between $0^\circ\mathrm{C}$ and $100^\circ\mathrm{C}$, and in the gaseous phase for temperatures above $100^\circ\mathrm{C}$. At $T=0^\circ\mathrm{C}\simeq 273~\mathrm{K}$ and $T=100^\circ\mathrm{C}\simeq 373~\mathrm{K}$, water undergoes what we call phase transitions. These transitions occur abruptly, with a mixed-phase region in which the two phases involved in the transition coexist. At these points, the temperature remains constant while the system absorbs or releases latent heat, changing its microscopic organization without immediately changing its temperature.

This behavior can be visualized in the phase diagram of water shown in Fig.~\ref{fig:water_phase_diagram}. In this diagram, each region of the temperature--pressure plane corresponds to a distinct thermodynamic phase. The curves separating these regions indicate the conditions under which two phases coexist in equilibrium. In particular, the horizontal line corresponding to atmospheric pressure intersects the coexistence curves at the freezing and boiling points of water discussed above. The diagram also shows that the phase structure depends not only on temperature but also on pressure: when the pressure is varied, the transition temperatures change, and special points appear, such as the triple point, where the three phases coexist, and the critical point, above which there is no longer a well-defined distinction between the liquid and gaseous phases.

\subsubsection{Classification and order parameter}

Phase transitions can be generically classified into two types: first-order and continuous transitions. A {\it first-order phase transition} (or FOPT) is characterized by a discontinuity in the first derivatives of the thermodynamic potential $\Omega$ with respect to its natural variables, such as $T$ and $\mu_B$. In this case, two distinct phases coexist along a transition line, and the passage from one phase to the other occurs with the absorption or release of latent heat. Thermodynamic quantities such as $\epsilon$, $s$, or $n_B$ may change discontinuously across the transition, while the temperature and pressure remain fixed during phase coexistence \cite{Kardar2007}. In the phase diagram of water shown in Fig.~\ref{fig:water_phase_diagram}, first-order transitions appear as coexistence lines separating distinct thermodynamic regions. For example, below the critical point, the liquid-gas line separates the liquid and gaseous phases. When crossing this line, the system changes abruptly from one phase to the other, passing through a coexistence region between liquid and vapor.

A {\it crossover}, on the other hand, is not a true phase transition in the strict sense. In this case, thermodynamic quantities change rapidly but smoothly, without any real discontinuity or singularity in the thermodynamic potential. In the water phase diagram, this can be understood by observing that, above the critical point, it is possible to pass continuously from a region with liquid-like behavior to a region with gas-like behavior without crossing any coexistence line. Nevertheless, one may define a ``pseudocritical temperature'', for instance, through the maximum of a susceptibility or through the inflection point of some thermodynamic quantity, indicating only the region where this smooth change is most pronounced.

The {\it critical end point} (CEP), or simply critical point, is the endpoint of a first-order transition line. In the case of water, it appears at the end of the liquid--gas coexistence line shown in Fig.~\ref{fig:water_phase_diagram}. As one approaches this point along the coexistence line, the difference between the liquid and gaseous phases progressively decreases, the latent heat tends to zero, and the distinction between the two phases disappears. In the thermodynamic limit, the critical point is associated with singular behavior of the thermodynamic potential, including the divergence of susceptibilities and of the correlation length. Thus, exactly at the critical point, the transition is no longer first order. It is continuous, often referred to as a second-order phase transition, and exhibits critical behavior with long-range fluctuations~\cite{Kardar2007}.

We can define here a thermodynamic quantity called the order parameter, a macroscopic quantity that distinguishes the phases of the system. In general, it assumes different values in different phases and allows one to characterize whether the transition between them occurs discontinuously, continuously, or only smoothly. In the case of the liquid-gas transition of water, for example, one possible choice is the difference between the densities of the liquid and gaseous phases. This difference is finite along the coexistence line below the critical point, but it vanishes at the critical point, where liquid and gas are no longer thermodynamically distinguishable phases.

\subsection{The Ising Model}

\begin{figure}[!htbp]
    \centering
    \includegraphics[width=0.9\linewidth]{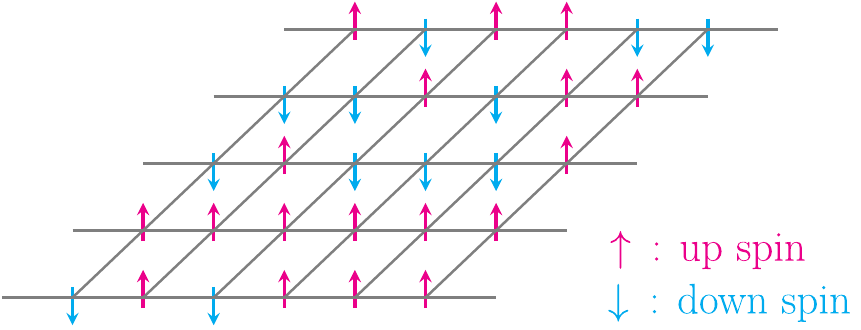}
    \caption{
    Schematic representation of the Ising model on a 2D lattice. Each site contains a discrete spin variable $s_i=\pm 1$, and neighboring spins interact through the ferromagnetic coupling in Eq.~\eqref{eq:ising_hamiltonian}. Adapted from Wikimedia Commons, original work by Ta2o~\cite{WikimediaIsingLattice}.
    }
    \label{fig:ising_lattice}
\end{figure}

The Ising model is one of the simplest models capable of describing a phase transition. Originally introduced as an effective model for ferromagnetism~\cite{Ising1925,Goldenfeld1992}, it represents a system of discretized magnetic dipoles, or ``spins'', distributed over the sites of a lattice. A 2D example can be seen in Fig.~\ref{fig:ising_lattice}. 

In the three-dimensional case, one considers a cubic lattice with $N=L^3$ sites, where each spin can assume only two values,
\begin{align}
    s_i &= \pm 1,
    \label{eq:ising_spin_values}
\end{align}
corresponding, for example, to dipoles aligned or anti-aligned with respect to a chosen direction. The energy of a microscopic configuration $\{s_i\}$ is given by the Hamiltonian
\begin{align}
    \mathcal H[\{s_i\}]
    &=
    -J\sum_{\langle i,j\rangle} s_i s_j
    -H\sum_i s_i ,
    \label{eq:ising_hamiltonian}
\end{align}
where the sum $\langle i,j\rangle$ is taken over nearest neighbors, $J$ is the coupling constant, and $H$ is the external magnetic field.

For $J>0$, configurations with neighboring spins aligned are energetically favored, characterizing ferromagnetic coupling. The second term represents the tendency of the spins to align with the external field. In thermal equilibrium, the thermodynamics of the system is determined by the canonical partition function~\cite{Goldenfeld1992,Cardy1996},
\begin{align}
    Z(T,H)
    &=
    \sum_{\{s_i\}}
    e^{-\beta \mathcal H[\{s_i\}]},
    \label{eq:ising_partition_function}
\end{align}
since there is no particle exchange. The sum in \eqref{eq:ising_partition_function} runs over all $2^N$ possible microstates. The thermodynamic potential, given by the Helmholtz free energy, is then
\begin{align}
    F(T,H)
    &=
    -\frac{1}{\beta}\ln Z(T,H).
    \label{eq:ising_helmholtz_free_energy}
\end{align}
From it, the main thermodynamic quantities can be obtained. The magnetization, which acts as the order parameter of the ferromagnetic transition, is
\begin{align}
    M(T,H)
    &=
    -\left(\frac{\partial F}{\partial H}\right)_T,
    \label{eq:ising_magnetization}
    \\
    m(T,H)
    &=
    \frac{M(T,H)}{N}.
    \label{eq:ising_magnetization_density}
\end{align}
The magnetic susceptibility and the specific heat are given by
\begin{align}
    \chi_T
    &=
    \left(\frac{\partial M}{\partial H}\right)_T,
    \label{eq:ising_susceptibility}
    \\
    C_H
    &=
    -T\left(\frac{\partial^2 F}{\partial T^2}\right)_H .
    \label{eq:ising_specific_heat}
\end{align}
The critical phenomenon emerges from the competition between energy and entropy. At low temperatures, the interaction term dominates: for $J>0$, the system minimizes its energy by aligning the spins, producing a ferromagnetic phase with spontaneous magnetization,
\begin{align}
    m &\neq 0,
    \qquad
    T<T_c,
    \qquad
    H=0.
    \label{eq:ising_ordered_phase}
\end{align}
At high temperatures, thermal fluctuations favor disordered configurations, and the average magnetization vanishes~\cite{Ising1925,Goldenfeld1992},
\begin{align}
    m &= 0,
    \qquad
    T>T_c,
    \qquad
    H=0.
    \label{eq:ising_disordered_phase}
\end{align}
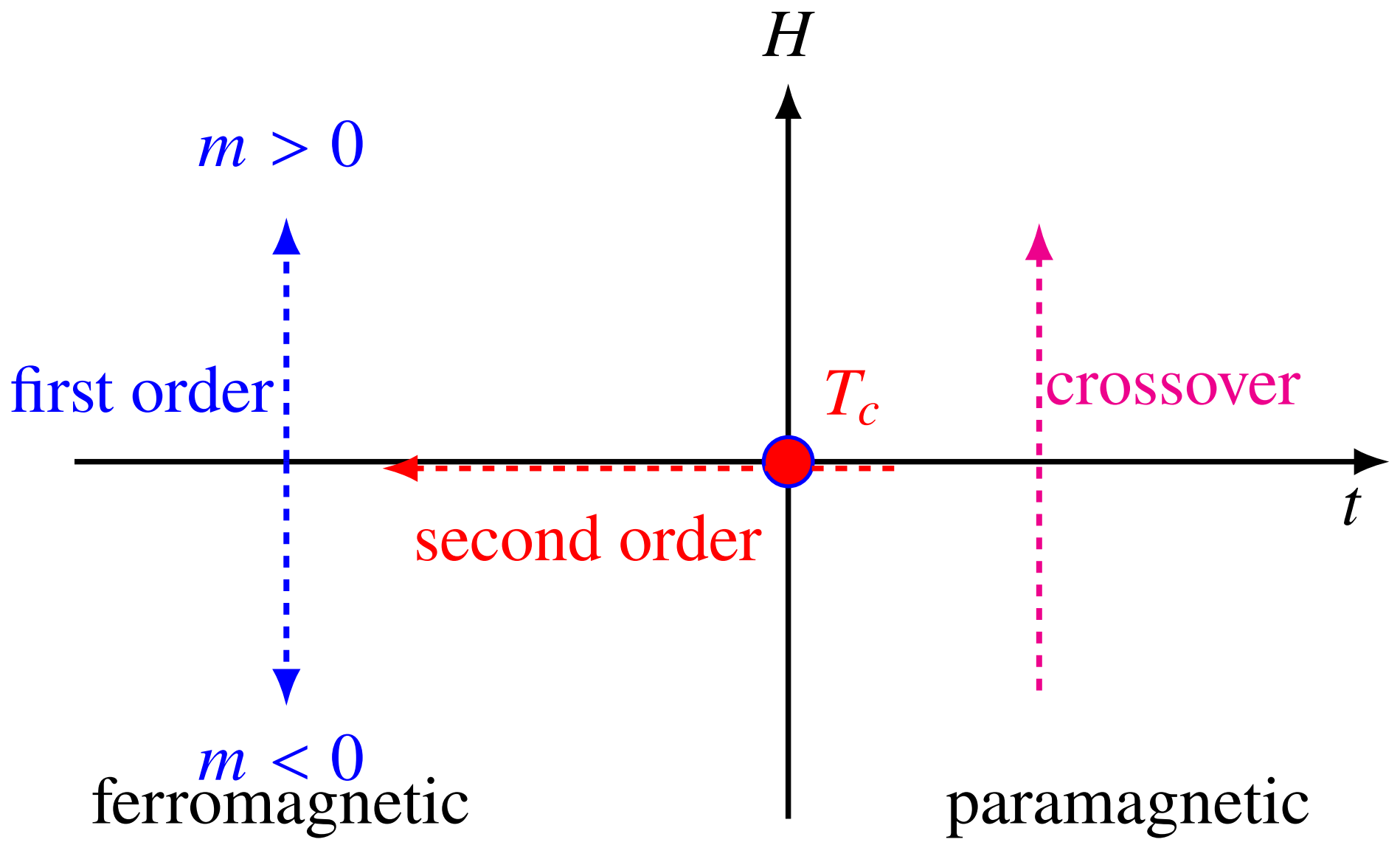
\begin{figure}[!htbp]
    \centering
    \begin{tikzpicture}[
        scale=1.0,
        >=Latex,
        every node/.style={font=\Large}
    ]

    \draw[->, line width=1.4pt] (-6.4,0) -- (5.4,0) node[below left=2pt and 2pt] {$t$};
    \draw[->, line width=1.4pt] (0,-3.2) -- (0,3.4) node[left=3pt, above=2pt] {$H$};

    \filldraw[red, draw=blue, line width=1.0pt] (0,0) circle (0.22);
    \node[red, above right=0.18cm and 0.18cm] at (0,0) {$T_c$};

    \draw[blue, dashed, line width=1.5pt, ->] (-4.5,0) -- (-4.5,2.2);
    \draw[blue, dashed, line width=1.5pt, ->] (-4.5,0) -- (-4.5,-2.2);

    \node[blue] at (-4.55,2.85) {$m>0$};
    \node[blue] at (-4.55,-2.65) {$m<0$};

    \node[blue] at (-5.8,0.65) {first order};

    \draw[red, dashed, line width=1.4pt, ->] (0.95,-0.06) -- (-3.65,-0.06);
    \node[red] at (-1.80,-0.68) {second order};

    \draw[magenta, dashed, line width=1.5pt, ->] (2.25,-2.05) -- (2.25,2.15);
    \node[magenta] at (3.45,0.65) {crossover};

    \node at (-4.55,-3.1) {ferromagnetic};
    \node at (3.05,-3.1) {paramagnetic};

    \end{tikzpicture}
    \caption{
    Schematic phase diagram of the Ising model in the plane spanned by the temperature-like variable $t=(T-T_c)/T_c$ and the external magnetic field $H$. For $t<0$ and $H=0$, the system undergoes a first-order transition between the two ferromagnetic phases with $m>0$ and $m<0$. This line terminates at the critical point $T_c$, where the transition becomes continuous. For $t>0$, the system is in the paramagnetic region and changes smoothly with $H$, corresponding to a crossover.}
    \label{fig:ising_phase_diagram_tikz}
\end{figure}
In Fig. \ref{fig:ising_phase_diagram_tikz}, an illustration of the phase diagram dynamics of the Ising model can be seen.

In the thermodynamic limit, the transition between these two phases occurs at a critical temperature $T_c$. Near $T_c$, fluctuations become correlated over increasingly large length scales, characterized by the correlation length $\xi$, which diverges as
\begin{align}
    \xi&\sim |t|^{-\nu},
    \label{eq:ising_correlation_length_scaling}
    \\
    t&\equiv\frac{T-T_c}{T_c}.
    \label{eq:ising_reduced_temperature}
\end{align}
The singularities of the thermodynamic functions near the critical point are described by critical exponents~\cite{Goldenfeld1992,Cardy1996,PelissettoVicari2002}. For $H=0$, one has, for example,
\begin{align}
    m
    &\sim (-t)^\beta,
    \qquad
    T<T_c,
    \label{eq:ising_beta_exponent}
    \\
    \chi_T
    &\sim |t|^{-\gamma},
    \label{eq:ising_gamma_exponent}
    \\
    C_H
    &\sim |t|^{-\alpha},
    \label{eq:ising_alpha_exponent}
    \\
    \xi
    &\sim |t|^{-\nu}.
    \label{eq:ising_nu_exponent}
\end{align}
On the critical isotherm $T=T_c$, the response of the magnetization to the external field is
\begin{align}
    m
    &\sim H^{1/\delta}.
    \label{eq:ising_delta_exponent}
\end{align}
Moreover, the spin-spin correlation function has the asymptotic behavior
\begin{align}
    G(r)
    &=
    \langle s(0)s(r)\rangle - \langle s\rangle^2,
    \label{eq:ising_correlation_function_definition}
    \\
    G(r)
    &\sim
    \frac{e^{-r/\xi}}{r^{d-2+\eta}},
    \label{eq:ising_correlation_function_scaling}
\end{align}
where $d=3$ for the three-dimensional Ising model. At the critical point, since $\xi\to\infty$, correlations become long ranged, and the decay is governed by a power law.

The concept of universality class arises because, sufficiently close to the critical point, the singular behavior of the system no longer depends on the microscopic details of the Hamiltonian~\cite{Cardy1996,PelissettoVicari2002}. Physically distinct systems may exhibit the same critical exponents and the same scaling laws if they share only a few general properties, such as spatial dimensionality, the symmetry of the order parameter, and the range of the interactions. In the 3D Ising model, the order parameter is the magnetization $m$, which is a scalar and has a discrete symmetry under sign inversion when the external field vanishes,
\begin{align}
    m
    &\rightarrow
    -m,
    \qquad
    H=0.
    \label{eq:ising_z2_symmetry}
\end{align}
This symmetry corresponds to the equivalence between the two possible ferromagnetic phases, with positive or negative magnetization. Therefore, three-dimensional systems with a scalar order parameter, discrete $\mathbb{Z}_2$ symmetry, and effectively short-range interactions belong to the same universality class as the 3D Ising model~\cite{PelissettoVicari2002}.

In general, this universality can be expressed through the scaling hypothesis for the singular part of the free energy per unit volume,
\begin{align}
    f_s(t,H)
    &=
    b^{-d}
    f_s\left(t b^{y_t}, H b^{y_h}\right),
    \label{eq:ising_free_energy_scaling}
\end{align}
where $b$ is a rescaling factor, $t$ is the reduced temperature defined in \eqref{eq:ising_reduced_temperature}, and $y_t$ and $y_h$ are the scaling exponents associated with temperature and the external field. The critical exponents can be related to these scaling exponents, for example, by
\begin{align}
    \nu
    &=
    \frac{1}{y_t},
    \label{eq:ising_nu_scaling_relation}
    \\
    \alpha &= 2-d\nu,\\
    \beta
    &=
    \frac{d-y_h}{y_t},
    \label{eq:ising_beta_scaling_relation}
    \\
    \gamma
    &=
    \frac{2y_h-d}{y_t},
    \label{eq:ising_gamma_scaling_relation}
    \\
    \delta
    &=
    \frac{y_h}{d-y_h},
    \label{eq:ising_delta_scaling_relation}\\
    \eta&=d+2-2y_h.
\end{align}
Thus, the critical exponents do not characterize a particular choice of lattice or microscopic parameters, but rather the critical fixed point that controls the long-distance behavior of the system.

For the 3D Ising model, typical values of the critical exponents are~\cite{PelissettoVicari2002,Hasenbusch2010,Kos2016}
\begin{align}
    \alpha
    &\simeq 0.11,
    \label{eq:ising_alpha_value}
    \\
    \beta
    &\simeq 0.326,
    \label{eq:ising_beta_value}
    \\
    \gamma
    &\simeq 1.24,
    \label{eq:ising_gamma_value}
    \\
    \delta
    &\simeq 4.79,
    \label{eq:ising_delta_value}
    \\
    \nu
    &\simeq 0.63,
    \label{eq:ising_nu_value}
    \\
    \eta
    &\simeq 0.036.
    \label{eq:ising_eta_value}
\end{align}
For this reason, the 3D Ising model is often used as a universal representative of systems with critical transitions belonging to the same universality class.

\section{Thermodynamic Descriptions of QCD Matter}

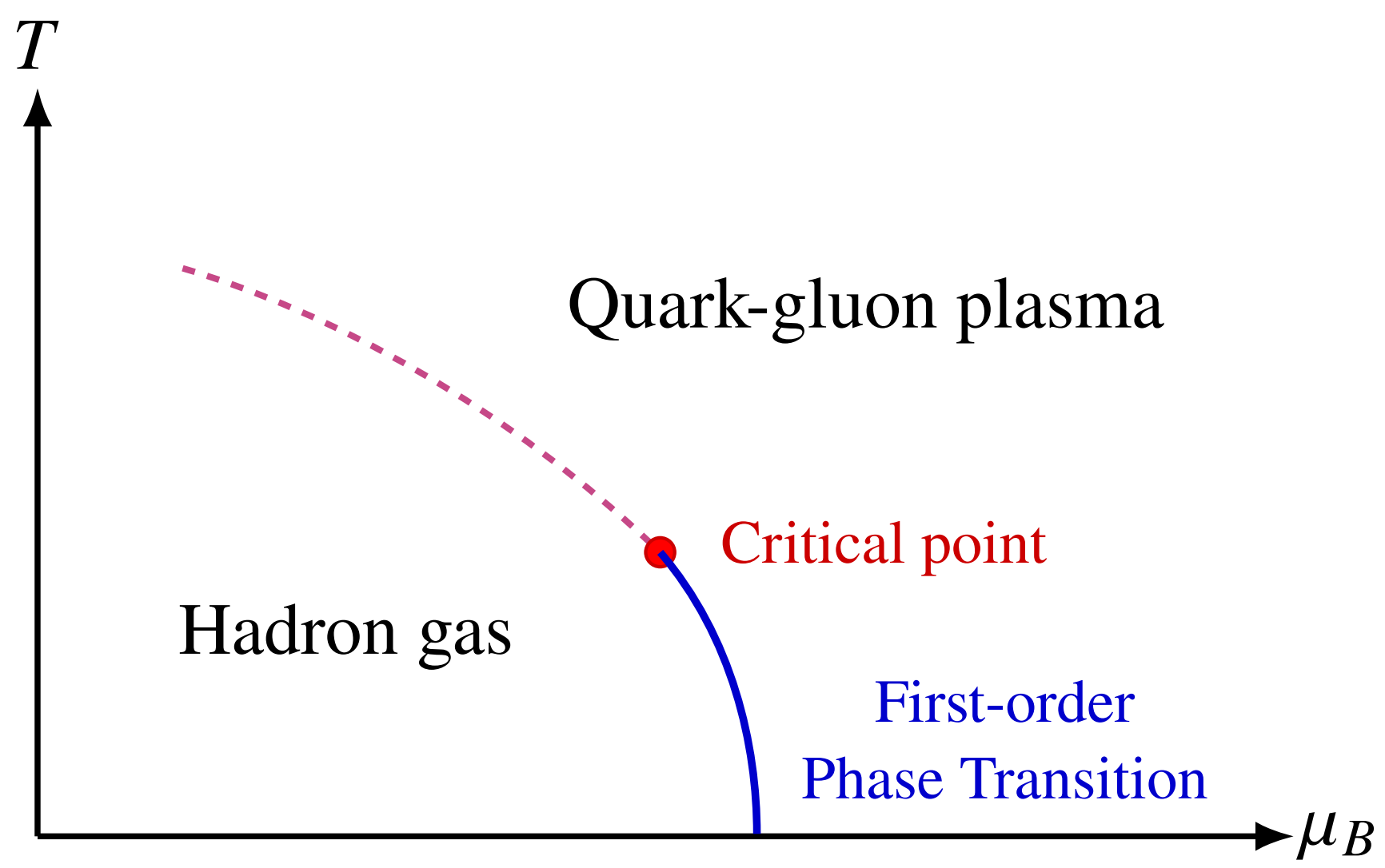
\begin{figure}[!htbp]
    \centering
    \begin{tikzpicture}[
        scale=1.,
        >=Latex,
        every node/.style={font=\large}
    ]

    \draw[->, line width=1.4pt] (0,0) -- (0,6.2);
    \draw[->, line width=1.4pt] (0,0) -- (10.4,0);

    \node[font=\bfseries\Large] at (0,6.55) {$T$};
    \node[font=\bfseries\Large] at (10.75,0) {$\mu_B$};

    \draw[magenta!80!black, dashed, line width=1.5pt]
        (1.2,4.7)
        .. controls (2.4,4.35) and (3.9,3.55) ..
        (5.15,2.35);

    \filldraw[red, draw=red!80!black, line width=0.8pt]
        (5.15,2.35) circle (0.12);

    \node[red!80!black, align=left]
        at (7,2.38)
        {Critical point};

    \draw[blue!80!black, line width=1.8pt]
        (5.15,2.35)
        .. controls (5.75,1.65) and (5.95,0.75) ..
        (5.95,0.02);

    \node[blue!80!black, align=center]
        at (8,0.8)
        {First-order\\Phase Transition};

    \node[font=\Large]
        at (6.85,4.35)
        {Quark-gluon plasma};

    \node[font=\Large]
        at (2.55,1.65)
        {Hadron gas};

    \end{tikzpicture}
    \caption{
    Schematic representation of the QCD phase diagram in the temperature--baryon-density plane. The low-temperature region is described by hadronic degrees of freedom, while the high-temperature region corresponds to the deconfined quark-gluon plasma. The blue solid line indicates a first-order phase transition, which terminates at the red critical point. The magenta dashed line represents the crossover region at lower baryon density.}
    \label{fig:qcd_phase_diagram_schematic}
\end{figure}

Strongly interacting matter described by QCD admits different descriptions depending on thermodynamic conditions. In this work, we will adopt a description composed of only two phases: one at higher temperatures, representing the QGP, and another at lower temperatures, representing the hadron gas. As we will see later, this description is obtained exactly for vanishing baryon chemical potential through lattice QCD, and effective models predict a first-order phase transition for finite baryon chemical potential $\mu_B$, which constitutes evidence for the existence of a critical point.  An illustration is shown in Fig. \ref{fig:qcd_phase_diagram_schematic}. 

In QCD matter, the conserved charges typically considered are baryon number $B$, electric charge $Q$, and strangeness $S$. The chemical potential of a particle species $i$ is
\begin{equation}
    \mu_i = B_i \mu_B + Q_i \mu_Q + S_i \mu_S ,
\end{equation}
where $B_i$, $Q_i$, and $S_i$ are its quantum numbers. In this work, we will consider only the baryon potential, so
\begin{equation}
    \mu_i \simeq B_i \mu_B ,
\end{equation}
In the hadronic phase, baryon number is carried by baryons and antibaryons, while mesons have $B_i=0$. Therefore, the net baryon density is obtained by summing the contributions of each species, weighted by its baryon number.

\subsection{MIT Bag Model}

The MIT bag model is a phenomenological description of deconfined quark matter in which quarks and gluons are treated as approximately free relativistic particles, while confinement is represented by a non-zero vacuum contribution $B$~\cite{Chodos1974MITBag,DeGrand1975Bag}. Using Eqs.~\eqref{eq:enerdens} and \eqref{eq:pressure} with the constant term, and assuming massless gluons and quarks, the degeneracies are $g_g=2\times 8=16$ for gluons and $g_q=2\times 3=6$ for each quark flavor, with the quark chemical potentials given by $\mu_q=\mu_B/3$. Under these conditions, the pressure and energy density become
\begin{align}
    &P^{\rm QGP}
    =
    g_g \frac{\pi^2}{90}T^4
    +
    \sum_q
    g_q
    \left[
        \frac{7\pi^2}{360}T^4
        +
        \frac{\mu_q^2}{12}T^2
        +
        \frac{\mu_q^4}{24\pi^2}
    \right]
    - B,
    \label{eq:mitbag_pressure}
    \\
    &\epsilon^{\rm QGP}
    =
    g_g \frac{\pi^2}{30}T^4
    +
    \sum_q
    g_q
    \left[
        \frac{7\pi^2}{120}T^4
        +
        \frac{\mu_q^2}{4}T^2
        +
        \frac{\mu_q^4}{8\pi^2}
    \right]
    + B,
    \label{eq:mitbag_energy}
    \\
    &n_B^{\rm QGP}
    =
    \frac{1}{3}
    \sum_q
    \left[
        n_q(T,\mu_q)
        -
        n_{\bar q}(T,-\mu_q)
    \right] =
    \frac{1}{3}
    \sum_q
    g_q
    \left[
        \frac{\mu_q}{6}T^2
        +
        \frac{\mu_q^3}{6\pi^2}
    \right]. \label{eq:mitbag_baryon_density}
\end{align}
At $\mu_B=0$, these equations reduce to
\begin{align}
    &P^{\rm QGP}(T,0)
    =
    g_{\rm QGP}\frac{\pi^2}{90}T^4 - B,
    \label{eq:mitbag_pressure_mu0}
    \\
    &\epsilon_{\rm QGP}(T,0)
    =
    g^{\rm QGP}\frac{\pi^2}{30}T^4 + B,
    \label{eq:mitbag_energy_mu0}
    \\
    &g_{\rm QGP}
    =
    16+\frac{7}{8}(12N_f),
    \label{eq:mitbag_degeneracy}
\end{align}
giving $g_{\rm QGP}=37$ for $N_f=2$ and $g_{\rm QGP}=47.5$ for $N_f=3$. Entropy density can be obtained from related thermodynamic expressions.

\subsection{Hadron Gas}

The hadron gas description includes the thermodynamic contributions of hadronic states. used in the calculation of thermodynamic quantities. Mesons have a baryon number $B_i=0$, while baryons have $B_i=1$ and antibaryons have $B_i=-1$. A more complete version of this description is the hadron resonance gas (HRG), in which resonances are included as additional hadron states in the thermodynamic sum. The pressure, energy density, and net baryon density of the hadron gas are then written as sums over the hadronic spectrum,
\begin{align}
    &P^{\rm HG}(T,\mu_B)
    =
    \sum_{i\in {\rm hadrons}}
    P_i(T,B_i\mu_B),
    \label{eq:hg_pressure_sum}
    \\
    &\epsilon^{\rm HG}(T,\mu_B)
    =
    \sum_{i\in {\rm hadrons}}
    \epsilon_i(T,B_i\mu_B),
    \label{eq:hg_energy_sum}
    \\
    &n_B^{\rm HG}(T,\mu_B)
    =
    \sum_{i\in{\rm hadrons}}
    B_i\, n_i(T,B_i\mu_B).
    \label{eq:hg_baryon_density_sum}
\end{align}
Here $P_i$, $\epsilon_i$, and $n_F$ are obtained from the single-species expressions in Eqs.~\eqref{eq:pressure}, \eqref{eq:enerdens}, and \eqref{eq:partdens}. Corrections to this ideal hadron-gas picture may be introduced, for example, through finite hadron-size effects, such as excluded-volume or van der Waals hard-core corrections~\cite{Hama2005Topics,Gorenstein1999ExcludedVolume}.

\subsection{Lattice-QCD Calculations}

\begin{figure}[!htbp]
    \centering
    \includegraphics[width=\linewidth]{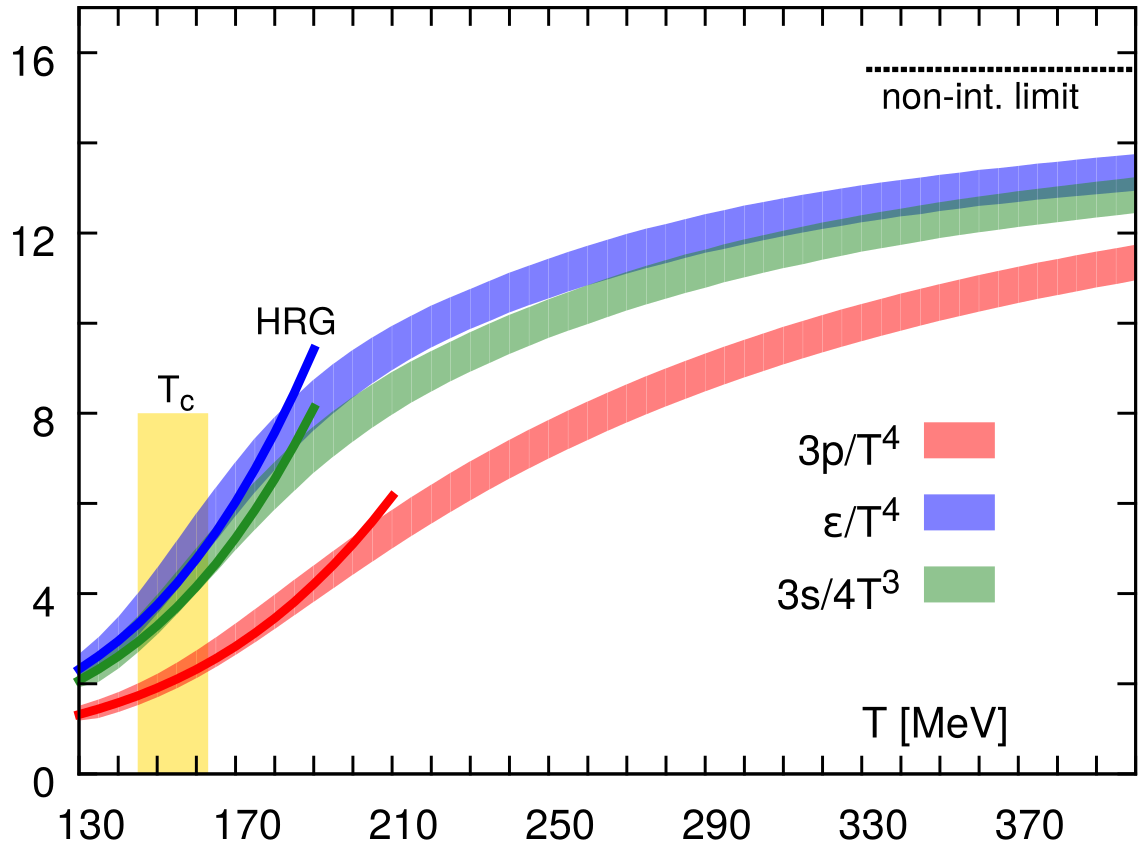}
    \caption{Lattice-QCD results (shaded curves indicating the uncertainties). The rapid increase around $T_c=(154\pm9)\,\mathrm{MeV}$ indicates the crossover from hadronic matter to a deconfined quark-gluon plasma. The dark curves show the hadron resonance gas result at low temperatures, while the dashed line indicates the non-interacting ideal-gas limit. Adapted from~\cite{Bazavov2014EoS}.
    }
    \label{fig:lattice_qcd_eos_mu0}
\end{figure}

Lattice QCD~\cite{Aoki2006Nature,Bazavov2014EoS,Borsanyi2014EoS} provides a first-principles calculation of the thermodynamics of strongly interacting matter. The theory is formulated on a discrete Euclidean spacetime lattice, and thermodynamic quantities are extracted from the partition function. At vanishing baryon chemical potential, lattice-QCD calculations with physical quark masses show that QCD does not undergo a first-order transition, but rather a smooth crossover around
\begin{align}
    T_c
    &=
    (154\pm9)\,\mathrm{MeV}.
    \label{eq:lattice_tc}
\end{align}
In this temperature region, the energy density is estimated to be
\begin{align}
    \epsilon_c
    &\simeq
    (0.18-0.5)\,\mathrm{GeV/fm^3},
    \label{eq:lattice_critical_energy_density}
\end{align}
which is of the same order but larger than the typical nuclear matter energy density,
\begin{align}
    \epsilon_{\rm nuclear}
    &\sim
    0.15\,\mathrm{GeV/fm^3}.
    \label{eq:nuclear_matter_energy_density}
\end{align}
The crossover is reflected in the rapid increase of bulk thermodynamic quantities, as shown in Fig.~\ref{fig:lattice_qcd_eos_mu0}: at low temperature, the lattice results approach the hadron resonance gas description, while at high temperature they approach the ideal quark-gluon gas limit. The deviation from this non-interacting limit near the transition indicates that nonperturbative effects remain important in the temperature range relevant for heavy-ion collisions~\cite{Aoki2006Nature,Bazavov2014EoS,Borsanyi2014EoS}.

\subsubsection{Sign Problem and Taylor Expansion}

At finite baryon chemical potential, direct lattice-QCD calculations are limited by the sign problem: for real nonzero $\mu_B$, the fermion determinant becomes complex, and the path-integral measure cannot be interpreted as a probability weight. A standard way to obtain information at small $\mu_B$ is therefore to expand the pressure around $\mu_B=0$,
\begin{align}
    \frac{P(T,\mu_B)}{T^4}
    &=
    \sum_{n=0}^{\infty}
    c_n^B(T)
    \left(
        \frac{\mu_B}{T}
    \right)^n,
    \label{eq:lattice_pressure_taylor}
    \\
    c_n^B(T)
    &=
    \frac{1}{n!}
    \left.
    \chi_n^B(T)
    \right|_{\mu_B=0},
    \label{eq:lattice_taylor_coefficients}
    \\
    \chi_n^B(T)
    &=
    \frac{\partial^n(P/T^4)}
    {\partial(\mu_B/T)^n}.
    \label{eq:lattice_baryon_susceptibilities}
\end{align}
Because of charge-conjugation symmetry, the pressure is an even function of $\mu_B$, so only even powers contribute. The coefficients in Eq.~\eqref{eq:lattice_baryon_susceptibilities} are baryon-number susceptibilities computed at $\mu_B=0$, allowing lattice information to be extended perturbatively to finite baryon density within a controlled range of $\mu_B/T$~\cite{Allton2002Taylor,Allton2003EoS,Borsanyi2021FiniteMuB,Mondal2022FiniteMuB}. This method does not remove the sign problem, but it provides the finite-density input used in modern equations of state for heavy-ion phenomenology.

\section{QCD Phase Diagram with a FOPT and a CEP}

Models of the QCD phase diagram have aimed to include phase transitions. Earlier approaches combine the MIT bag description with a hadron gas, which naturally leads to a first-order phase transition curve~\cite{hama2005topics}, and can still accommodate a phenomenological critical point~\cite{HAMA2006169}. More recent formulations include phase transitions and a critical point by exploring the universality class of the Ising model~\cite{PhysRevC.106.014909,PhysRevC.101.034901}. This choice is motivated by the conjecture that the critical behavior of QCD belongs to the same universality class as the three-dimensional Ising model, due to common symmetry properties of the order parameter~\cite{PhysRevLett.81.4816}.

In this work, we adopt the equation of state constructed by the BEST -- Beam Energy Scan Theory  -- Collaboration, which embeds a critical point and a first-order phase transition into a QCD equation of state constrained by lattice-QCD calculations at small baryon chemical potential~\cite{PhysRevC.101.034901,karthein2021strangeness, PhysRevC.107.054911}. The construction is based on mapping the universal critical behavior of the 3D Ising model onto the QCD coordinates $(T,\mu_B)$.

\subsection{Mapping Based on the Ising Model}

The three-dimensional Ising model is used to represent the universality class of the QCD critical point. In the notation introduced in the previous section, the Ising critical point is located at
\begin{equation}
    t = 0,
    \qquad
    H = 0,
\end{equation}
where $t$ is the temperature variable, $H$ is the magnetic field variable, and the magnetization $m$ is the Ising order parameter. The critical equation of state is parametrized in terms of polar variables $(R,\vartheta)$ as~\cite{PhysRevC.101.034901}
\begin{equation}
    t = R(1-\vartheta^2),
    \qquad
    H = H_0 R^{\beta\delta}\widetilde H(\vartheta),
    \label{eq:ising_parametric_variables}
\end{equation}
where $H_0$ is a normalization constant and $\widetilde H(\vartheta)$ is the scaling function associated with the magnetic-field-like direction. With this parametrization, the critical contribution to the Ising pressure is written as
\begin{equation}
    P_{\rm Ising}(R,\vartheta)
    =
    H_0 M_0 R^{2-\alpha}
    \left[
        \vartheta\,\widetilde H(\vartheta)
        -
        \mathcal G(\vartheta)
    \right],
    \label{eq:ising_pressure_parametric}
\end{equation}
where $M_0$ is a normalization constant, and $\mathcal G(\vartheta)$ is another scaling function~\cite{PhysRevC.101.034901}. This parametrization preserves the critical exponents $\alpha$, $\beta$, and $\delta$, and provides a phenomenological basis to construct the critical part of the QCD equation of state.

To relate the Ising variables to the physical QCD coordinates $(T,\mu_B)$, a linear, non-universal map is introduced:
\begin{align}
&\frac{T-T_{\rm CEP}}{T_{\rm CEP}}
=
w
\left(
    \rho\, t \sin\alpha_1
    +
    H\sin\alpha_2
\right),
\label{eq:ising_qcd_map_T}
\\
&\frac{\mu_B-\mu_{B,\rm CEP}}{T_{\rm CEP}}
=
w
\left(
    -\rho\, t \cos\alpha_1
    -
    H\cos\alpha_2
\right).
\label{eq:ising_qcd_map_muB}
\end{align}
Here $(T_{\rm CEP},\mu_{B,\rm CEP})$ is the position of the QCD critical point, $\alpha_1$ and $\alpha_2$ define the orientation of the Ising axes in the QCD phase diagram, and $w$ and $\rho$ set the size and relative scaling of the critical region.

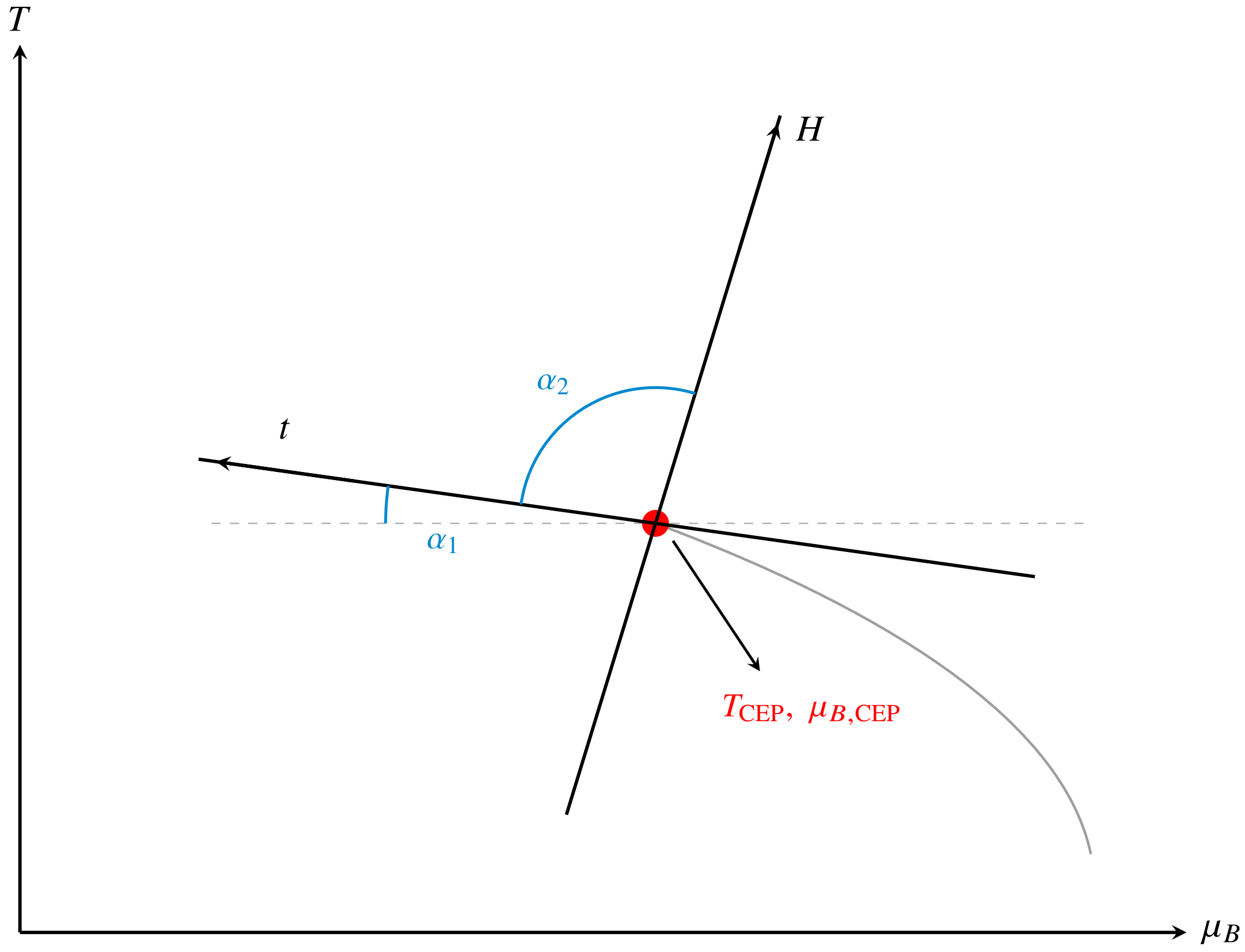
\begin{figure}[!htbp]
\centering
\begin{tikzpicture}[
    scale=2.,
    >=stealth,
    axis/.style={->, line width=1.1pt},
    mainline/.style={line width=1.1pt},
    thinmain/.style={line width=0.85pt},
    angleline/.style={cyan!70!blue, line width=0.95pt},
    labelblue/.style={cyan!70!blue, font=\small\bfseries},
    labelred/.style={red, font=\small\bfseries}
]

\draw[axis] (0,0) -- (0,5.1) node[above] {$T$};
\draw[axis] (0,0) -- (6.7,0) node[right] {$\mu_B$};

\coordinate (CP) at (3.65,2.35);

\draw[dashed, gray!70] (1.10,2.35) -- (6.15,2.35);

\draw[thinmain, gray!75]
    (CP) .. controls (4.45,2.05) and (5.95,1.40) .. (6.15,0.45);

\draw[->, black, line width=0.9pt]
    ($(CP)+(0.1,-0.10)$) -- ++(0.50,-0.75);

\fill[red] (CP) circle (2.2pt);

\node[labelred] at (4.55,1.28) {$T_{\rm CEP},\ \mu_{B,\rm CEP}$};

\draw[mainline]
    ($(CP)+(172:2.65)$) -- ($(CP)+(-8:2.20)$);

\draw[->, mainline]
    ($(CP)+(172:1.75)$) -- ($(CP)+(172:2.55)$);

\node at ($(CP)+(172:2.25)+(0.10,0.23)$) {$t$};

\draw[mainline]
    ($(CP)+(253:1.75)$) -- ($(CP)+(73:2.45)$);

\draw[->, mainline]
    ($(CP)+(73:1.65)$) -- ($(CP)+(73:2.40)$);

\node at ($(CP)+(73:2.35)+(0.2,0.02)$) {$H$};

\draw[angleline]
    ($(CP)+(180:1.55)$)
    arc[start angle=180, end angle=172, radius=1.55];

\node[labelblue] at ($(CP)+(176:1.20)+(-0.02,-0.20)$) {$\alpha_1$};

\draw[angleline]
    ($(CP)+(172:0.78)$)
    arc[start angle=172, end angle=73, radius=0.78];

\node[labelblue] at ($(CP)+(128:0.95)+(0.00,0.05)$) {$\alpha_2$};

\end{tikzpicture}
\caption{
Non-universal map from Ising variables $(t,H)$, introduced in Fig.~\ref{fig:ising_phase_diagram_tikz}, to QCD coordinates $(T,\mu_B)$. Adapted from Ref.~\cite{PhysRevC.101.034901}.
}
\label{fig:ising_qcd_map}
\end{figure}

This mapping contains six free parameters: the location of the critical point $(T_{\rm CEP},\mu_{B,\rm CEP})$, the angles $\alpha_1$ and $\alpha_2$, and the scale factors $w$ and $\rho$. 

\subsection{Pressure of the QCD Equation of State}

The QCD equation of state is constructed as the sum of a regular contribution and a critical contribution. The regular part is constrained by the Taylor expansion of lattice QCD around $\mu_B=0$, while the critical part is obtained from the Ising pressure after applying the map in Eqs.~\eqref{eq:ising_qcd_map_T} and~\eqref{eq:ising_qcd_map_muB}. In the notation used here,
\begin{equation}
    P(T,\mu_B)
    =
    P_{\rm reg}(T,\mu_B)
    +
    \mathcal I_{\rm sm}(T,\mu_B)
    P_{\rm Ising}
    \left[
        R(T,\mu_B),
        \vartheta(T,\mu_B)
    \right].
    \label{eq:qcd_pressure_best}
\end{equation}
Here $P_{\rm reg}$ represents the non-critical regular contribution originating from the Lattice-QCD pressure matched to the Hadron Gas pressure, and $\mathcal I_{\rm sm}(T,\mu_B)$ is an interpolation function that smoothly matches the Ising critical contribution to the regular QCD equation of state. Thermodynamic quantities are then obtained from the derivatives of the pressure, as in Eq. \eqref{eq:derivativesgibbsduhem}. The squared speed of sound is also computed from the thermodynamic derivatives of the equation of state as
\begin{equation}
    c_S^2 = \frac{\partial P}{\partial \epsilon}\bigg|_{\frac{s}{n_B}} 
\end{equation}
The critical contribution can significantly modify these quantities near $(T_{\rm CEP},\mu_{B,\rm CEP})$. Although the construction is formally well defined, certain parameter choices can lead to physical instabilities. In particular, small values of $w$ enhance the critical contribution and may produce negative baryon density, negative energy density, or a nonphysical speed of sound.

\subsection{Moving the Critical Point while Keeping the First-Order Line Fixed}

BEST EOS introduces a fixed first-order phase-transition curve, allowing the critical point to move along this curve while constraining the Ising-model parameters mapped onto the QCD equation of state.

The curvature of the crossover line at $\mu_B=0$ has been estimated from lattice-QCD simulations~\cite{BELLWIED2015559,PhysRevD.93.014507,PhysRevD.92.054503}: we parametrize the transition line as
\begin{equation}
T_{\rm tr}(\mu_B)
=
T_0
\left[
1
-
\kappa
\left(
\frac{\mu_B}{T_0}
\right)^2
\right]
+
\mathcal O(\mu_B^4),
\label{eq:FOPTcurve}
\end{equation}
with $T_0\simeq 155$ MeV and $\kappa\simeq 0.0149$ denoting the transition temperature and curvature at $\mu_B=0$, respectively.

Lattice-QCD results at small baryon chemical potential constrain where a critical point can appear. The radius of convergence of the pressure expansion around $\mu_B=0$ is determined by the nearest singularity in the complex $\mu_B$ plane~\cite{stephanov2006qcdphasediagramoverview}. Around the crossover temperature, lattice calculations give a radius of roughly $\mu_B/T \approx 2$~\cite{GIORDANO2021121986}. Since the dominant singularity does not lie on the real $\mu_B$ axis in the thermodynamic limit, this provides a lower bound for a possible phase transition. If a critical point exists, it should therefore lie at larger values of $\mu_B/T$.

In practice, parameter choices that place the critical point too far at large $\mu_B$ can generate instabilities in the equation of state, including negative energy density or an unphysical speed of sound. In our setup, such distant critical-point locations are therefore unsuitable. We restrict the critical-point search to the range
\begin{equation}
300~{\rm MeV}
\lesssim
\mu_{B,\rm CEP}
\lesssim
450~{\rm MeV},
\end{equation}
with the critical point moved along the transition line given by Eq.~\eqref{eq:FOPTcurve}.
\begin{figure}[!htbp]
\centering
\includegraphics[width=0.9\linewidth]{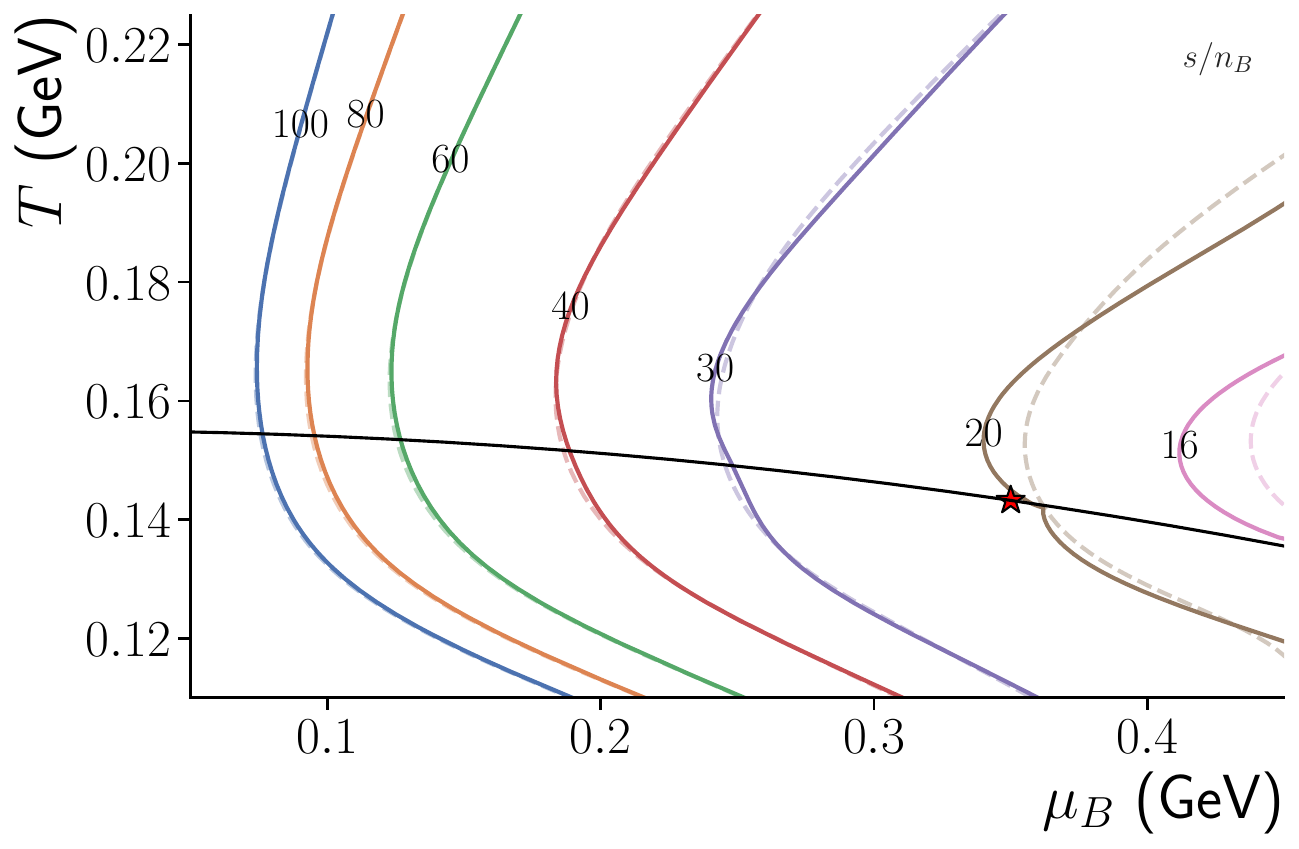}
\caption{
Isentropic trajectories with fixed $s/n_B$ in the $(T,\mu_B)$ plane are shown for both the crossover-only equation of state (dashed curves) and the equation of state with a critical point (solid curves). The critical point is at $(T_{\rm CEP},\mu_{B,\rm CEP})\simeq(143,350)$ MeV. The black curve is the transition line from Eq.~\eqref{eq:FOPTcurve}. The comparison shows how isentropic trajectories focus toward the critical region.
}
\label{fig:snb}
\end{figure}
In this work, we choose $\mu_{B,\rm CEP}=350$ MeV, following the benchmark equation of state constructed in Ref.~\cite{PhysRevC.101.034901}. From Eq.~\eqref{eq:FOPTcurve}, this corresponds to $T_{\rm CEP}\simeq 143$ MeV. This choice should not be interpreted as a prediction for the QCD critical point, whose location remains unknown, but rather as a controlled setup to investigate how critical behavior is imprinted on the thermodynamic properties of the equation of state. In Fig.~\ref{fig:snb}, we show the isentropic trajectories, $s/n_B=\mathrm{const.}$, together with the transition line and the corresponding critical-point position in the $(T,\mu_B)$ plane. The comparison between the crossover-only equation of state, shown with dashed lines, and the equation of state with a critical point, shown with solid lines, illustrates the focusing~\cite{Dore:2022qyz} of the isentropic trajectories toward the critical region. Figures~\ref{fig:thermo_quantities} and~\ref{fig:cs2_chi2} show the corresponding comparisons for the pressure, energy density, baryon density, entropy density, squared speed of sound, and second-order baryon-number susceptibility.

\begin{figure}[!htbp]
\centering
\includegraphics[width=\linewidth]{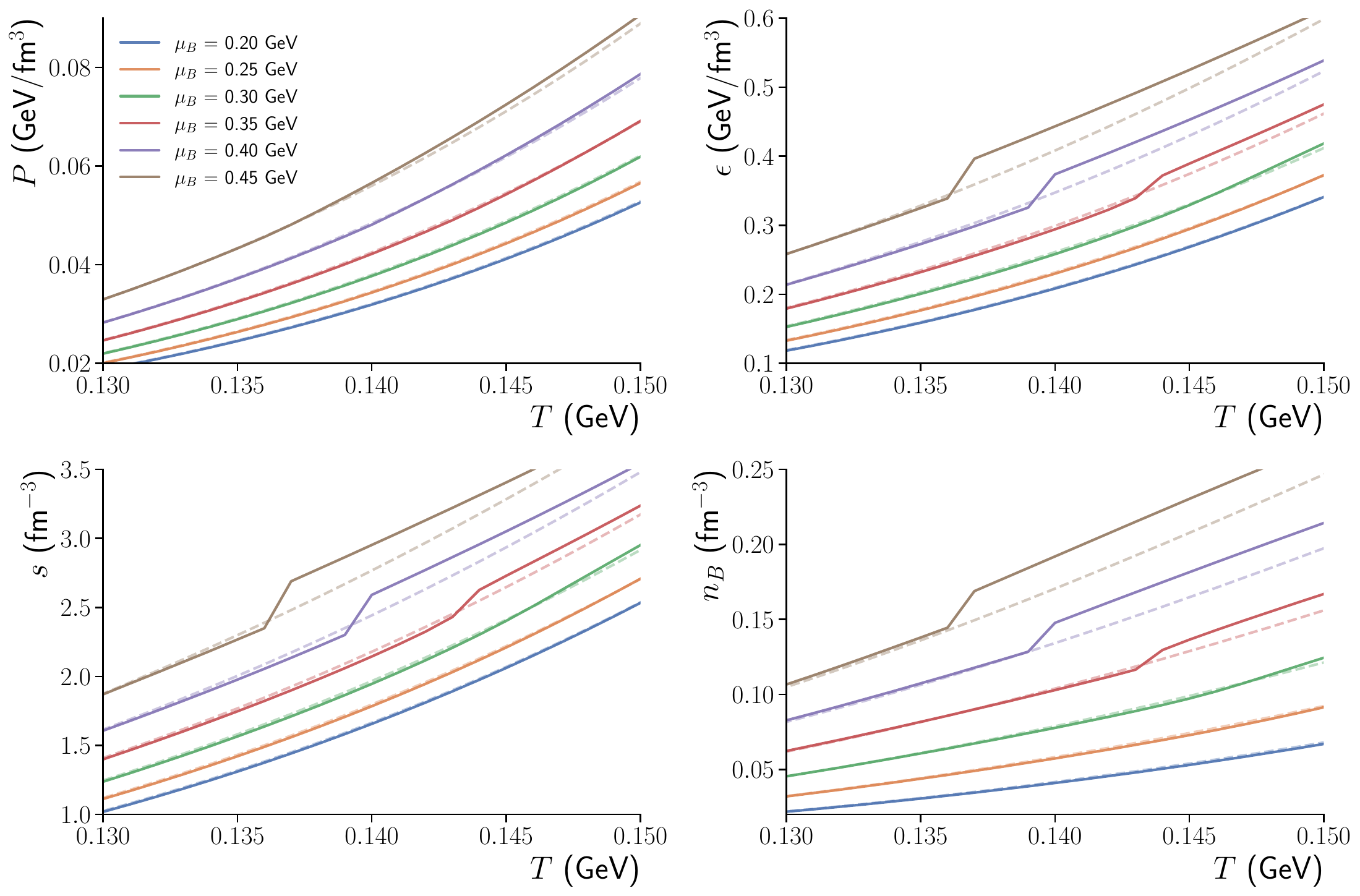}
\caption{
Thermodynamic quantities as functions of temperature at fixed baryon chemical potential $\mu_B$. The panels show energy density $\epsilon$, pressure $P$, entropy density $s$, and baryon density $n_B$. Solid lines use an equation of state with a critical point; dashed lines use a crossover-only equation of state. The comparison highlights how the critical contribution alters thermodynamic behavior near the critical region.
}
\label{fig:thermo_quantities}
\end{figure}

\begin{figure}[!htbp]
\centering
\includegraphics[width=\linewidth]{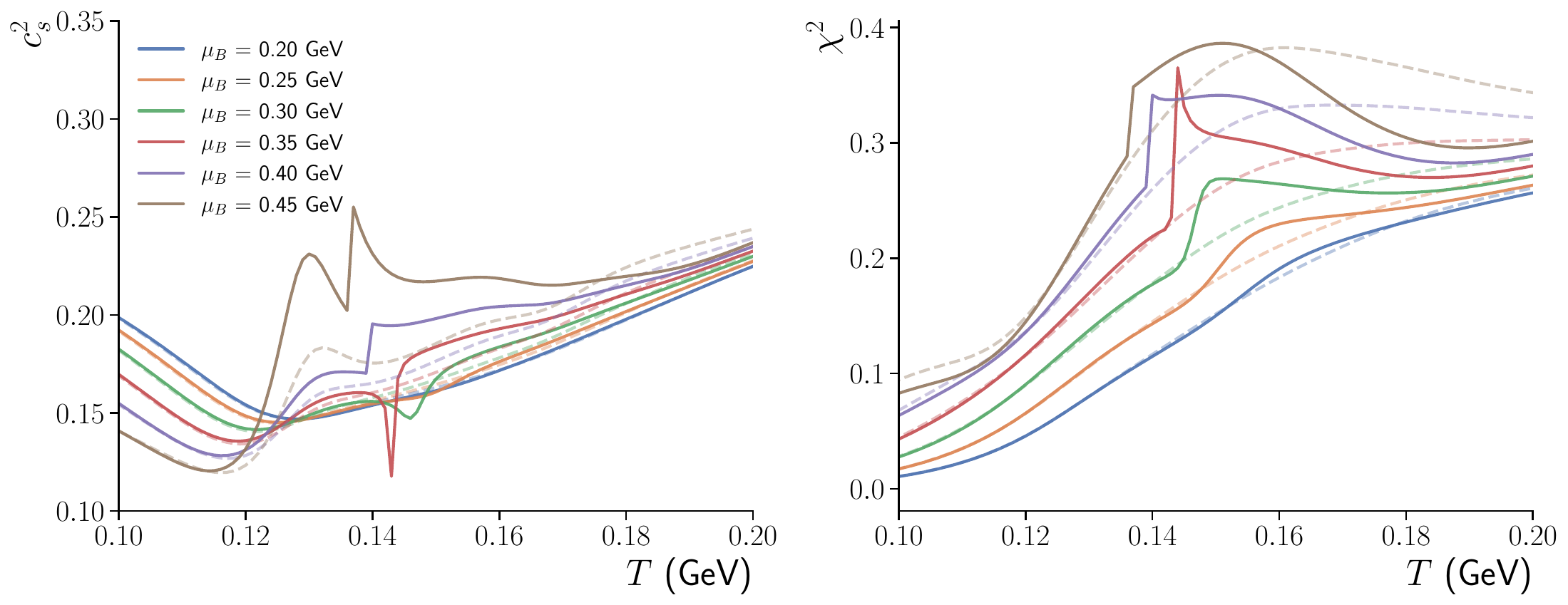}
\caption{
Squared speed of sound, $c_s^2$, and second-order baryon-number susceptibility, $\chi_2^B$, as functions of temperature for fixed values of the baryon chemical potential $\mu_B$. Solid lines correspond to the equation of state with a critical point, while dashed lines correspond to the crossover-only equation of state. These quantities are especially sensitive to the critical contribution, since the critical point modifies thermodynamic derivatives of the pressure.
}
\label{fig:cs2_chi2}
\end{figure}

\subsection{Correlation Length}

The correlation length $\xi$ can be defined as the distance measure between two particles required for them to become uncorrelated. The correlation between particle density fluctuations at two points $\vec{x}$ and $\vec{y}$ can be characterized by the correlation function

\begin{equation}
\rho(\vec{x},\vec{y}) = \langle n(\vec{x})n(\vec{y})\rangle - \langle n(\vec{x})\rangle\langle n(\vec{y})\rangle,
\end{equation}

whose spatial decay is determined by $\xi$. In simplified terms, for a homogeneous system, this dependence can be written as $\rho(\vec{x},\vec{y})\sim\exp(-|\vec{x}-\vec{y}|/\xi)$ \cite{Landau1980}. Thus, the larger the correlation length, the farther apart two points in the system can be while still exhibiting correlated fluctuations.

As the system approaches the critical point, fluctuations of the matter begin to occur on increasingly larger spatial scales. In the thermodynamic limit, exactly at the critical point, the correlation length diverges, $\xi\rightarrow\infty$, indicating that density fluctuations can remain correlated over arbitrarily large scales. This divergence of the correlation length is one of the fundamental characteristics of a second-order phase transition and is associated with the system’s universal critical behavior~\cite{doi:10.1142/S0217751X05027965}.

The BEST Collaboration’s equation of state implements the calculation of this correlation length from the critical contribution of the 3D Ising model. After mapping each point \((T,\mu_B)\) into the parametric variables \((R,\theta)\), one defines \(M(R,\theta)=M_0R^\beta\theta\) and computes
\begin{equation}
\xi(R,\theta)=a\,|M(R,\theta)|^{-\nu/\beta}\sqrt{|g(R,\theta)|},
\end{equation}
where \(a\) is a non-universal microscopic amplitude, set to \(a=1\) in the implementation, \(M_0=0.605\), \(\beta=0.326\), and \(\nu=0.63\). The scaling function \(g\) is obtained from an \(\epsilon\)-expansion near the critical point and from an asymptotic form outside that region. Here, \(\epsilon=4-d\) is the expansion parameter about the upper critical dimension \(d=4\), taking \(\epsilon=1\) for the three-dimensional Ising model. The two expressions are smoothly matched. The singular behavior of \(\xi\) is governed by the factor \(|M|^{-\nu/\beta}\), leading to its divergence at the critical point.

\section{Further Developments}

A straightforward generalization of the equation of state employed in this study is to formulate a four-dimensional BQS equation of state,
\begin{equation}
P = P(T,\mu_B,\mu_Q,\mu_S),
\end{equation}
where baryon number, electric charge, and strangeness are incorporated via their respective chemical potentials~\cite{Noronha-Hostler:2019ayj,Abuali:2025texs4d}.


A complementary approach is provided by phenomenological holographic equations of state at finite baryon density~\cite{PhysRevD.110.094006}. These models have been used to explore possible critical-point locations beyond the region directly constrained by lattice calculations, including examples at larger baryon chemical potential,
\begin{equation}
(T_{\rm CEP},\mu_{B,\rm CEP})
=
(104,589)\ {\rm MeV}
\quad \text{or} \quad
(107,571)\ {\rm MeV}.
\end{equation}
Such results provide phenomenological scenarios for finite-density QCD thermodynamics, but they should not be interpreted as first-principles lattice determinations~\cite{PhysRevD.110.094006}.

\chapter{Standard Model for a Nuclear Collision}

\begin{figure}[!htbp]
    \centering
    \includegraphics[width=0.95\linewidth]{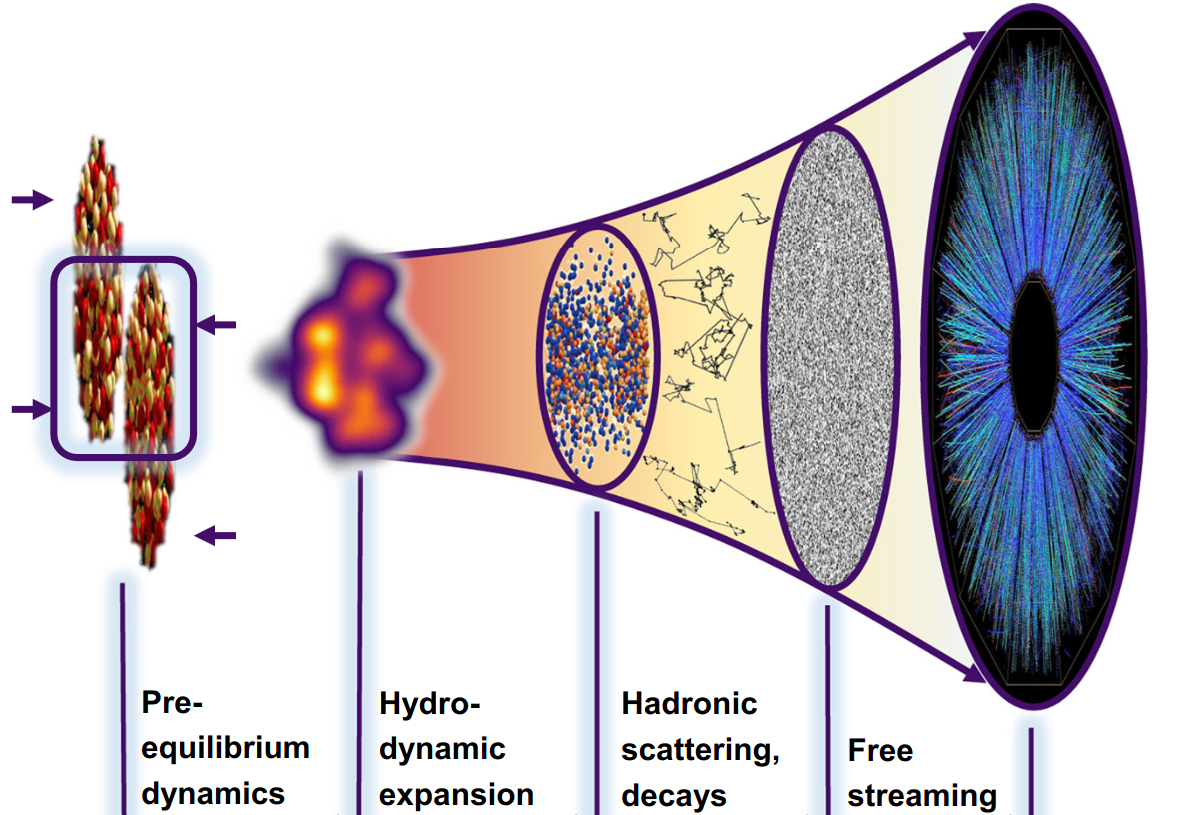}
    \caption{Collision process illustrated. Taken from \cite{kincses2025levy}.}
    \label{fig:colisao1}
\end{figure}

In an accelerator such as RHIC or the LHC, two ions from different beams collide, and the system formed is subjected to extreme conditions of high temperatures, densities, and velocities. Under these conditions, a quark–gluon plasma (QGP) is produced. This matter undergoes expansion and cooling so that the plasma’s degrees of freedom become confined into hadrons, which survive until the system is completely diluted. During this process, the system behaves like an almost perfect fluid with low viscosity. After complete dilution, the hadrons stop interacting collectively and undergo scattering processes and decay until they reach the detectors. This is the standard description of a high energy nuclear collision. An illustration of this process can be seen in Figure \ref{fig:colisao1}.

We will now present a description of each of the stages that are employed in this thesis.

\section{Collision Configuration }

\begin{figure}[!htbp]
    \centering
    \includegraphics[width=0.95\linewidth]{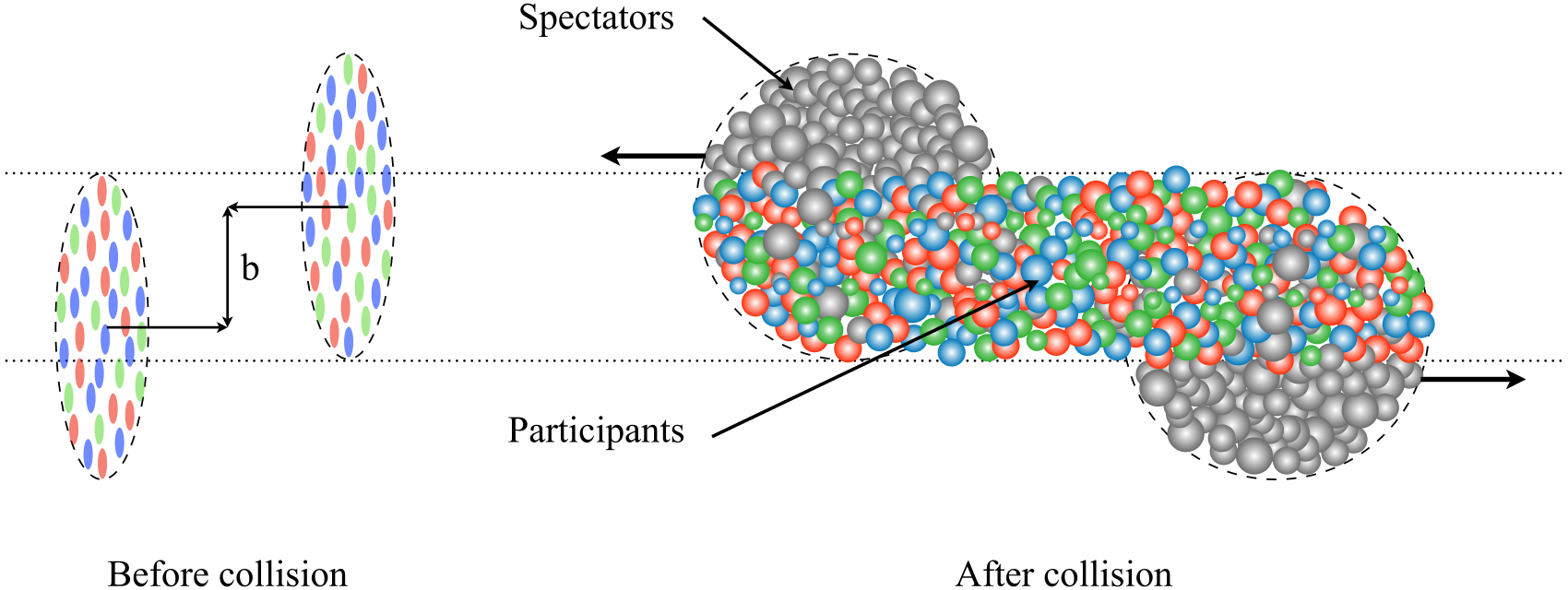}
    \caption{A schematic illustration of a heavy-ion collision, indicating the impact parameter $b$ along with the spectator nucleons and the participant nucleons. Taken from \cite{aaij2022centrality}. In the image on the right, we disregarded the contraction of the nuclei for better visualization.}
    \label{fig:colisao2}
\end{figure}

Consider a system of two nuclei undergoing a collision. The initial geometry of this interaction can be characterized in terms of nucleon–nucleon collisions, where every proton or neutron inside the projectile nuclei is regarded simply as a nucleon. Those nucleons that experience at least one inelastic interaction are referred to as participants, and the others, which do not undergo collisions or which undergo elastic collisions, are called spectators. These different components of the collision are depicted in Fig.~\ref{fig:colisao2}.

We define the impact parameter $b$ as the transverse distance between the centers of the two nuclei that collide at the instant of interaction, which determines the degree of geometrical overlap between them. We denote by $N_{\rm part}$ the number of nucleons that participate in the collision, that is, that undergo at least one inelastic collision, and by $N_{\rm coll}$ the total number of inelastic nucleon–nucleon collisions. The geometrical description of the collision that allows one to relate $b$ with $N_{\rm part}$ and $N_{\rm coll}$ is given by the so-called Glauber model \cite{Miller:2007ri}, which we will see next.

\subsection{Glauber Model}

The spatial distribution of nuclear matter is often parametrized by a Woods–Saxon distribution \cite{WoodsSaxon1954},
\begin{equation}
    \rho_A(r) =
    \rho_0
    \frac{1+w(r/R)^2}{1+\exp[(r-R)/a]} .
\end{equation}
In this parametrization, \(R\) represents the radius at half nuclear density, \(a\) controls the diffuseness of the surface, and \(w\) introduces a quadratic correction to the density in the interior of the nucleus. For \(w=0\), one recovers the usual two-parameter form of the Woods–Saxon distribution. The constant \(\rho_0\) is fixed by the normalization condition
\begin{equation}
    \int d^3r\,\rho_A(r) = A ,
\end{equation}
so that the integral of the density reproduces the number of nucleons in the nucleus under consideration. The parameter values used for some nuclei are presented in Table~\ref{tab:woods_saxon_parameters}. In this form, a spherically symmetric density is assumed. 
    
\begin{table}[htbp]
\centering
\begin{tabular}{lccccccc}
\hline
Nucleus & Symbol & Atomic number & \(A\) & \(\rho_0~[\mathrm{fm}^{-3}]\) & \(R~[\mathrm{fm}]\) & \(a~[\mathrm{fm}]\) & \(w\) \\
\hline
Copper    & Cu & 29 & 63  & 0.169 & 4.214 & 0.5860 & 0.000 \\
Gold      & Au & 79 & 197 & 0.169 & 6.380 & 0.5350 & 0.000 \\
Lead      & Pb & 82 & 208 & 0.160 & 6.624 & 0.5490 & 0.000 \\
\hline
\end{tabular}
\caption{Parameters of the Woods-Saxon nuclear density profile. Taken from \cite{deJager1974,deVries1987}.}
\label{tab:woods_saxon_parameters}
\end{table}

However, there are extensions of the parametrization that incorporate non-spherical deformations, which are relevant, for example, in isobar collisions \(^{96}\mathrm{Ru}+^{96}\mathrm{Ru}\) and \(^{96}\mathrm{Zr}+^{96}\mathrm{Zr}\) \cite{PhysRevC.110.064907}. Moreover, for light nuclei, such as in \(^{16}\mathrm{O}+^{16}\mathrm{O}\) collisions, effects associated with nontrivial nuclear configurations, such as \(\alpha\)-clustering, can also be important \cite{PhysRevLett.112.112501}.

\begin{figure}[!htbp]
\centering
\includegraphics[width=\linewidth]{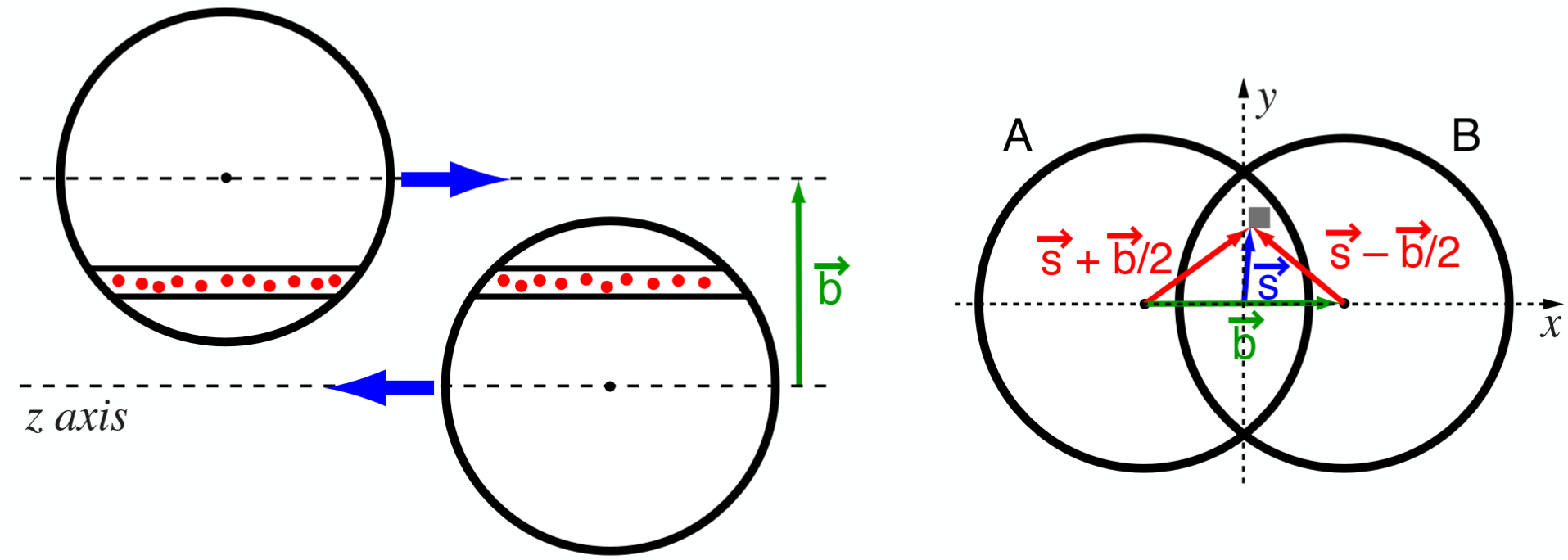}
\caption{
Longitudinal and transverse plane schematic depiction of two nuclei undergoing a collision in the Glauber framework. The dashed lines represent the beam direction, and \(\vec{b}\) corresponds to the impact parameter. The transverse coordinate \(s\) indicates the location of a nucleon tube relative to the center of one of the nuclei. Taken from \cite{BraunMunzingerSchmah2021}.}
\label{fig:glauber-side-view}
\end{figure}

The Woods-Saxon density profile provides the probability distribution used to
generate the positions of nucleons inside each nucleus. For a spherically
symmetric nucleus, the radial coordinate is sampled according to
\begin{equation}
    P(r)\,dr \propto r^2 \rho_A(r)\,dr,
\end{equation}
while the angular coordinates are sampled isotropically. After the nucleon
positions are generated, a collision with impact parameter \(\vec{b}\) is
constructed by shifting the centers of the two nuclei in the transverse plane
to \(+\vec{b}/2\) and \(-\vec{b}/2\). The corresponding optical description is expressed in terms of the nuclear
thickness function,
\begin{equation}
    T_A(\vec{s}) =
    \int dz\,\rho_A(\vec{s},z),
\end{equation}
which represents the amount of nuclear matter crossed along the longitudinal direction at a fixed transverse position \(\vec{s}\).  This construction is
illustrated schematically in Fig.~\ref{fig:glauber-side-view}, where a tube of
nucleons crosses the nucleus at a transverse distance \(s\) from its center.

For two nuclei separated by \(\vec{b}\), the nuclear overlap function is
\begin{equation}
    T_{AB}(\vec{b}) =
    \int d^2s\,
    T_A(\vec{s}+\vec{b}/2)
    T_B(\vec{s}-\vec{b}/2).
\end{equation}
It determines the mean number of binary nucleon-nucleon collisions,
\begin{equation}
    \langle N_{\rm coll}(b)\rangle
    =
    AB\sigma^{NN}_{\rm inel} T_{AB}(b),
\end{equation}
where \(\sigma^{NN}_{\rm inel}\) is the inelastic nucleon-nucleon cross
section at the collision energy. 

\subsection{Monte-Carlo Sampling}
In a Monte Carlo implementation, this continuous overlap is replaced by a
pairwise collision criterion~\cite{Miller:2007ri,Alver:2008aq}. For each nucleon \(i\) from nucleus \(A\) and
nucleon \(j\) from nucleus \(B\), one computes their transverse separation
\begin{equation}
    d_{ij}
    =
    \left|
    \vec{s}^{\,A}_i
    -
    \vec{s}^{\,B}_j
    \right|.
\end{equation}
In the black-disk approximation, the inelastic cross section is interpreted as
an effective transverse area,
\begin{equation}
    \sigma^{NN}_{\rm inel} = \pi d_{\rm max}^2 .
\end{equation}
Therefore, the pair is counted as an inelastic nucleon-nucleon collision if
\begin{equation}
    d_{ij}
    <
    d_{\rm max}
    =
    \sqrt{\frac{\sigma^{NN}_{\rm inel}}{\pi}} .
\end{equation}
The total number of pairs satisfying this condition defines \(N_{\rm coll}\) in that event. A nucleon is counted as a participant if it belongs to at least one such pair; otherwise, it is classified as a spectator. Thus, \(N_{\rm part}\) is obtained by counting all distinct projectile and target nucleons that undergo at least one inelastic collision.

\subsection{Centrality Classification}

\begin{figure}[!htbp]
\centering
\includegraphics[width=0.75\linewidth]{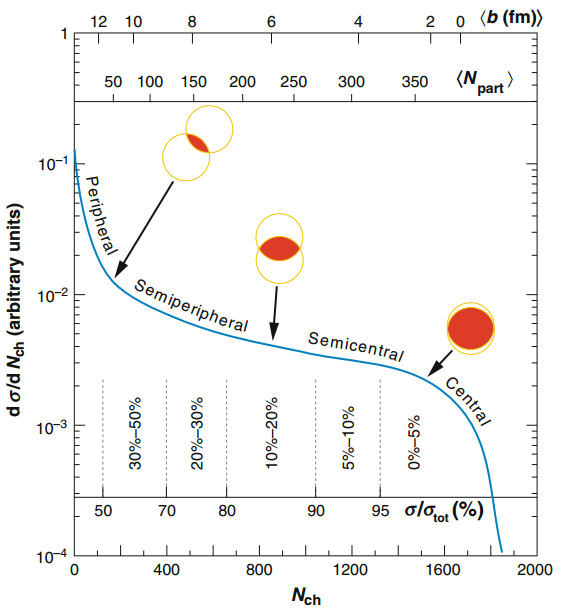}
\caption{Schematic relation between the charged-particle multiplicity \(N_{\rm ch}\), the impact parameter \(b\), and the average number of participant nucleons \(\langle N_{\rm part}\rangle\). The quantity \(\mathrm{d}\sigma/\mathrm{d}N_{\rm ch}\) represents the differential event distribution in \(N_{\rm ch}\), which is used to define centrality classes. Taken from~\cite{NSWWSTARTrigger}.}
\label{fig:centrality}
\end{figure}

Centrality is used to classify events according to the degree of geometrical overlap between the two nuclei. Since the impact parameter is not directly measurable experimentally, centrality classes are usually defined from an observable correlated with the collision geometry, such as the charged-particle multiplicity or the transverse energy deposited in a given detector acceptance. Events with the largest multiplicities correspond, on average, to the smallest impact parameters and are classified as the most central collisions, while events with smaller multiplicities correspond to more peripheral collisions. The centrality classes are then defined as percentiles of the total inelastic cross section, for example \(0\text{-}5\%\), \(5\text{-}10\%\), and so on. In the Glauber framework, these experimental classes can be related to geometrical quantities such as \(\langle b\rangle\), \(\langle N_{\rm part}\rangle\), and \(\langle N_{\rm coll}\rangle\), allowing a connection between measured event activity and the initial collision geometry~\cite{Miller:2007ri,Alver:2008aq}. We can see an illustration of this classification in Figure \ref{fig:centrality}.

\subsection{Partonic Cascade and Pre-Equilibrium}

Once the initial collision geometry is specified, the early-time dynamics must characterize how the degrees of freedom generated in the underlying nucleon–nucleon collisions evolve prior to the onset of hydrodynamics. This interval, commonly called the pre-equilibrium stage, can be described using several approaches, including free streaming \cite{Vredevoogd:2008id,Liu:2015nwa}, classical color fields \cite{McLerran:1993ni,Schenke:2012wb,Gale:2012rq}, effective kinetic theory \cite{Arnold:2002zm,Kurkela:2018wud,Kurkela:2018vqr}, or microscopic transport and partonic cascades \cite{PhysRevD.44.3501,PhysRevC.61.067901,ZHANG1998193,Lin:2004en}. In a partonic cascade framework, the system is modeled as quarks and gluons that move and undergo successive scatterings without imposing local thermal equilibrium from the outset. The result of this phase is a microscopic ensemble of partons with well-defined space-time coordinates, momenta, and conserved quantum numbers, which must then be coarse-grained into smooth fields, such as the energy–momentum tensor \(T^{\mu\nu}\) and the conserved currents, to provide initial conditions for the hydrodynamic evolution \cite{Petersen:2008dd,Oliinychenko:2015lva}.

\subsection{Hydrodynamization}

\begin{figure}[!htbp]
\centering
\includegraphics[width=\linewidth]{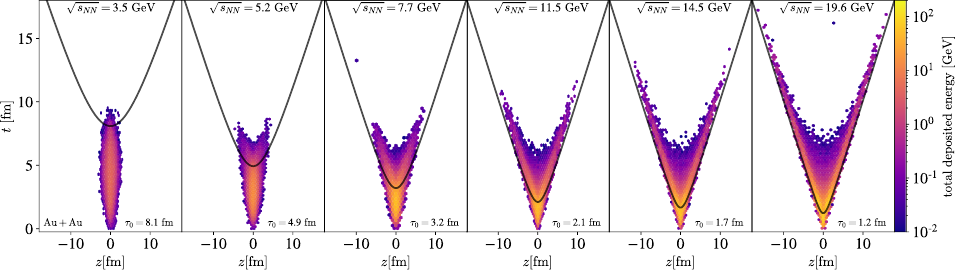}
\caption{Space-time distribution of the total deposited energy in the \(t\)-\(z\) plane for Au+Au collisions at different collision energies, obtained from SMASH-based initial conditions~\cite{w87d-ps6f}. The gray curves indicate constant proper-time hypersurfaces, \(\tau=\sqrt{t^2-z^2}=\tau_0\), used to characterize the transition from the pre-equilibrium stage to hydrodynamics. The values of \(\tau_0\) decrease with increasing \(\sqrt{s_{NN}}\), reflecting the shorter nuclear crossing time at higher collision energies. Figure produced by Renan Goés-Hirayama and provided to the author \cite{Goes-Hirayama:2026ies}.}
\label{fig:tauconst}
\end{figure}

The transition from the pre-equilibrium stage to hydrodynamics requires mapping the microscopic partonic degrees of freedom onto continuous macroscopic fields. In this work, this procedure is performed on a hypersurface of constant proper time, \(\tau=\sqrt{t^2-z^2}\), which provides a convenient approximation for the end of the pre-equilibrium evolution at collision energies above \(\sqrt{s_{NN}}=7.7~\mathrm{GeV}\) \cite{w87d-ps6f}, as shown in Fig.~\ref{fig:tauconst}.

The microscopic output of the partonic stage consists of a set of particles with definite positions and momenta. Therefore, before hydrodynamics can be initialized, this discrete information must be coarse-grained into smooth fields. Starting from the kinetic definitions of the conserved currents and the energy-momentum tensor, Eqs.~\eqref{eq:4current} and \eqref{eq:energystresstensor}, we represent the phase-space distribution on the constant-\(\tau\) hypersurface as
\begin{equation}
    f(x,p)=\sum_{i\in\mathrm{partons}}\delta(\vec p-\vec p_i)\,g(\vec x-\vec x_i).
\end{equation}
This expression encodes that each parton has a well-defined microscopic momentum \(\vec p_i\), described by a Dirac delta in momentum space, while its position is distributed over a finite spatial region via the function \(g(\vec x - \vec x_i)\). If one were to use a delta function in coordinate space as well, the resulting fields would remain singular and thus inappropriate as initial conditions for hydrodynamics. The smearing function therefore carries out the coarse-graining required to turn a discrete set of partons into continuous profiles. This construction must not be viewed as implying full local thermal equilibrium; it merely yields smooth energy, momentum, and charge densities from which the hydrodynamic variables can be determined. Inserting this distribution into the kinetic definition of the energy-momentum tensor gives
\begin{equation}
    T^{\mu\nu}(\vec x)=\sum_{i\in\mathrm{partons}}\frac{p_i^\mu p_i^\nu}{p_i^0}\,g(\vec x-\vec x_i),
\end{equation}
while the conserved currents are obtained as
\begin{equation}
    J_a^\mu(\vec x)=\sum_{i\in\mathrm{partons}}q_i^{(a)}\frac{p_i^\mu}{p_i^0}\,g(\vec x-\vec x_i),
\end{equation}
where \(q_i^{(a)}\) denotes the charge carried by parton \(i\) associated with the conserved current \(J_a^\mu\).  Once \(T^{\mu\nu}\) and the conserved currents are constructed, the local hydrodynamic fields are obtained by the corresponding matching procedure, which determines the energy density, flow velocity, and charge densities that initialize the subsequent hydrodynamic evolution \cite{Petersen:2008dd,Oliinychenko:2015lva}.

\section{Viscous Hydrodynamics}

We assume that the hot, dense matter produced can be described by a hydrodynamic evolution after the pre-equilibrium stage. Next, we will present an overview of what fluids are, their region of applicability, and the equations that govern their motion.

\subsection{What is a Fluid?}

In the classical description, a fluid is treated as a continuous medium, which assumes the existence of a separation between the microscopic and macroscopic scales of the system. In other words, it must be possible to choose a volume element large enough to contain many microscopic degrees of freedom, so that average thermodynamic quantities such as temperature $T$, pressure $P$, energy density $\epsilon$, and charge densities $n_X$ are well defined, yet still small enough that these quantities do not vary significantly within it. Furthermore, the microscopic processes responsible for the relaxation of the system must occur on scales smaller than those associated with macroscopic variations of the fluid, allowing each cell to be described by a state close to local thermodynamic equilibrium.

This formulation makes it possible to interpret the fluid as a collective state of matter, in which the evolution of each cell is determined not only by its local properties but also by the pressure gradients imposed by neighboring cells. In this way, microscopic interactions are effectively incorporated into the hydrodynamic equations, without the need to explicitly track the individual dynamics of the constituents of the system. In the context of relativistic heavy-ion collisions, this description does not necessarily require the system to be a weakly coupled gas with a well-defined mean free path in the classical sense. The essential requirement is that there exists a microscopic relaxation scale that is sufficiently small when compared to the macroscopic scales over which the hydrodynamic fields vary.

This validity condition can be quantified by means of, for example, the Knudsen number, defined as
\begin{equation}
K_n \equiv \frac{\ell_{\mathrm{micro}}}{L_{\mathrm{macro}}},
\label{eq:knudsen_number}
\end{equation}
where $\ell_{\mathrm{micro}}$ represents a typical microscopic scale of the system, such as the mean free path or a relaxation time, while $L_{\mathrm{macro}}$ represents the characteristic scale of spatial or temporal variation of the hydrodynamic fields. Schematically, this macroscopic scale can be associated with the gradients of the flow or of the thermodynamic variables, for example $L_{\mathrm{macro}}^{-1}\sim d_\mu u^\mu$ or $L_{\mathrm{macro}}^{-1}\sim |\nabla ^\mu P|/P$. The hydrodynamic description is expected to be appropriate when
\begin{equation}
K_n \ll 1,
\end{equation}
that is, when the microscopic scales are small compared to the scales over which the fluid varies. As $K_n$ increases, deviations from local equilibrium become more relevant, and dissipative corrections start to play an important role in the evolution of the system.

\subsection{Landau-Matching}

Assuming that the system at the end of the pre-equilibrium stage is sufficiently close to local thermodynamic equilibrium for hydrodynamics to apply, we decompose the energy-momentum tensor and conserved currents from the partonic cascade as follows:
\begin{align}
    &T^{\mu \nu} = \epsilon u^\mu u^\nu - (P + \Pi)\Delta^{\mu \nu} + \pi^{\mu \nu},\\ 
    &J_X^\mu = n_X u^\mu + V_X^\mu.
\end{align}
Here, $\Pi$ and $\pi^{\mu \nu}$ are the bulk pressure and shear tensor, respectively, while $V_X^\mu$ is the diffusion current associated with the conserved charge $X \in \{B,Q,S\}$. These are the dissipative contributions to the energy-momentum tensor and conserved currents. The tensor $\pi^{\mu \nu}$ is symmetric, traceless, and orthogonal to $u^\mu$, while $V_X^\mu$ is also orthogonal to $u^\mu$.

This form of $T^{\mu \nu}$ is known as the Landau choice \cite{landau1987fluid}. Under these conditions, it is possible to rewrite
\begin{equation}
    T^{\mu}_{\ \nu} u^\nu = \epsilon u^\mu,
\end{equation}
that is, as an eigenvalue and eigenvector problem, where we identify $u^\mu$ as the eigenvector of the matrix $T^{\mu}_{\ \nu} $ with eigenvalue $\epsilon$. We can obtain these quantities through a matrix diagonalization process.

The particle density and diffusion are obtained as
\begin{align}
    &n_X = u_\mu J^\mu_X,\\
    &V^\mu = J^\mu_X - n_X u^\mu,
\end{align}
and thus we can obtain the pressure $P =P(\epsilon, n_X)$ through an equation of state. The other quantities are obtained simply as
\begin{align}
    &\Pi = - P + \frac{\epsilon-T^{\mu}_{\ \mu} }{3},\\
    &\pi^{\mu \nu} = T^{\mu \nu} - \epsilon u^\mu u^\nu + (P + \Pi)\Delta^{\mu \nu}.
\end{align}
In this way, we determine all the information needed for the hydrodynamic evolution. In this work, we are considering only a nonzero baryon potential and will not include diffusion.

\subsection{Dynamic Equations}

We start from the conservation laws of the fluid energy tensor in the absence of sources
\begin{align}
    &d_\mu T^{\mu \nu} = 0, \label{eq:energystressconservation}\\
    &d_\mu J_X^\mu = 0
\end{align}
Applying $u^\mu$ and $\Delta^{\mu \nu}$ to \eqref{eq:energystressconservation} we obtain \cite{denicol2022microscopic}
\begin{align}
& D \epsilon + (\epsilon + P + \Pi)\theta - \pi^{\mu\nu} \sigma_{\mu \nu}
=0 \label{eq:energydensity}\\
&(\epsilon + P + \Pi) D u^\alpha
- \nabla^\alpha (P + \Pi)
+ \Delta^{\alpha}_{\ \nu} d_\mu \pi^{\mu\nu}
= 0\label{eq:motion}
\end{align}
These are the equations of motion of the fluid. For the evolution of the charge densities, the dynamical equation becomes
\begin{equation}
    Dn_X + n_X \theta + d_\mu V_X^\mu= 0.
    \label{eq:numberdensityeq}
\end{equation}
Using the thermodynamic relations \eqref{eq:gibbsduhem} and \eqref{eq:gibbsduhem2} together with the evolution equation for the energy density \eqref{eq:energydensity} and the equation for conserved charges \eqref{eq:numberdensityeq}, it is possible to obtain the so-called entropy production equation, where we define the entropy current $s^\mu$ as
\begin{equation}
    s^\mu = s u^\mu - \sum_X \frac{\mu_X}{T} V_X^\mu
\end{equation}
so that the dynamical equation for this current is
\begin{equation}
    d_\mu s^\mu
    = \frac{1}{T}\left(-\Pi \theta
    + \pi^{\mu \nu} \sigma_{\mu \nu}\right)
    - \sum_X V_X^\mu  \nabla_\mu \left( \frac{\mu_X}{T} \right).
\end{equation}
By the second law of thermodynamics, entropy production must necessarily be greater than zero, which imposes conditions on the terms in the previous equation to ensure that $d_\mu s^\mu \geq 0$.

In an isotropic fluid close to local equilibrium, the linear constitutive relations are fixed by symmetry. By Curie’s principle \cite{Koide:2007}, thermodynamic fluxes and forces with different tensorial terms do not couple linearly. Thus, at Navier–Stokes order, the dissipative sectors separate: 
\begin{itemize}
    \item $\Pi$, a scalar, is associated with the expansion scalar $\theta$.
    \item $V_X^\mu$, which are vectors, are related to the transverse gradients of the reduced chemical potentials $\alpha_X=\mu_X/T$.
    \item $\pi^{\mu \nu}$, which is symmetric, orthogonal to $u^\mu$, and traceless, is the shear tensor $\sigma^{\mu\nu}$.
\end{itemize}
 The first-order constitutive relations are
\begin{align}
    &\Pi = -\zeta \theta, \\
    &V_X^\mu = \kappa_{X}\nabla^\mu \alpha_X, \\
    &\pi^{\mu\nu} = 2\eta \sigma^{\mu\nu},
\end{align}
where $\zeta$, $\eta$, and $\kappa_{X}$ are called transport coefficients. We know this theory as the Navier–Stokes equations. This specific choice ensures the positivity of the entropy production equation once we have
\begin{equation}
    d_\mu s^\mu = \frac{1}{T}\left(\frac{\Pi^2}{\zeta} + \frac{1}{2\eta} \pi^{\mu \nu}\pi_{\mu \nu}\right) -  \sum_X \frac{1}{\kappa_{X}} V^\mu_X (V_X)_\mu.
\end{equation}
Since $V_X^\mu$ is orthogonal to fourvelocity, $V^\mu_X (V_X)_\mu$ is negative.

\subsection{Viscosity Equations}

Linear perturbations of the Navier–Stokes equations can give rise to waves that travel faster than the speed of light \cite{denicol2022microscopic}, rendering these hydrodynamic descriptions acausal. There are numerous methods to get around this. In this work, since we are not considering charge diffusion, we employed Israel-Stewart theory to implement the viscous hydrodynamic equations \cite{ISRAEL1979341, denicol2022microscopic}
\begin{align}
    &\tau_\Pi (\text D \Pi + \Pi\theta) + \Pi = -\zeta \theta,\label{bulk}\\
    &\tau_\pi \left(\Delta^{\mu \nu}_{\alpha \beta} \text D\pi^{\alpha \beta} + \frac{4}{3}\pi^{\mu \nu}\theta \right)+ \pi^{\mu \nu} = 2\eta \sigma^{\mu \nu}\label{shear},
\end{align}
where $\tau_\Pi$ and $\tau_\pi$ are the relaxation times.

\subsection{Ideal Hydrodynamics vs. Non-Ideal Hydrodynamics}

When $\Pi$, $\pi^{\mu\nu}$, and $V_X^\mu$ vanish, the system is in the ideal hydrodynamic regime. In this limit, the energy-momentum tensor is written solely in terms of the energy density, pressure, and fluid velocity, and the evolution does not produce entropy. Using the conservation equations and the local thermodynamic relations, the evolution equation for the energy density can be rewritten as
\begin{equation}
\partial_\mu(su^\mu)=0,
\end{equation}
showing that the entropy current is conserved. Therefore, in the absence of dissipative terms, the total entropy of the system remains constant during the hydrodynamic evolution.
In this same limit, the trace of the energy-momentum tensor is given by
\begin{equation}
T^\mu_{\ \mu}=\epsilon-3P.
\end{equation}
Thus, for a conformal system or for an ultrarelativistic ideal gas, in which $\epsilon=3P$, the energy-momentum tensor is traceless. However, as discussed in the construction of the lattice QCD equation of state, this relation does not hold, reflecting the breaking of conformal invariance.

This difference is particularly important in the hydrodynamization process. If the pre-equilibrium stage is described by approximately conformal dynamics, as in the case of ultrarelativistic particles for which $T^\mu_{\ \mu}\simeq 0$, the decomposition of this same energy-momentum tensor in terms of a nonconformal equation of state cannot, in general, be carried out with $\Pi=0$. In fact, in the hydrodynamic decomposition
\begin{equation}
T^{\mu\nu}=\epsilon u^\mu u^\nu-(P+\Pi)\Delta^{\mu\nu}+\pi^{\mu\nu},
\end{equation}
we have
\begin{equation}
T^\mu_{\ \mu}=\epsilon-3(P+\Pi),
\end{equation}
since $\pi^\mu_{\ \mu}=0$. Therefore, if the tensor coming from the pre-equilibrium stage is approximately traceless, matching to an equation of state with $\epsilon\neq 3P$ generates an initial bulk contribution approximately given by
\begin{equation}
\Pi \simeq \frac{\epsilon}{3}-P.
\end{equation}
In this way, even if the pre-equilibrium evolution is close to conformal dynamics, the transition to a hydrodynamic description with a realistic QCD equation of state naturally introduces non-ideal terms.



\section{Switching and Particlization }

During the hydrodynamic evolution, the fluid converts thermal energy into kinetic energy, accelerating and cooling down. As this occurs, the quark–gluon plasma undergoes reconfinement into hadrons, whose description is directly incorporated into the equation of state. The end of the hydrodynamic regime marks the beginning of a new stage characterized by hadronic scattering processes and decays, known as the afterburner. The transition from the hydrodynamic stage to the afterburner stage is called switching.

Switching occurs when each fluid cell cools down to a characteristic energy density $\epsilon_{sw}$, called the switching energy density. The value of $\epsilon_{sw}$ identifies points in the fluid hypervolume $\Delta$ that are selected and, taken together, form a hypersurface $\Sigma = \partial \Delta$. The particle distribution in momentum space for each species $i$ is described by the Cooper–Frye formula \cite{CooperFrye1974} in \eqref{eq:cff}, and the hypersurface integral is carried out over $\Sigma$, where we now rewrite the expression by factoring the constant $(2\pi)^3$ out of $f(x,p)$,
\begin{equation}
    p^0 \frac{d^3N_i}{d^3p} = \frac{1}{(2\pi)^3} \int_\Sigma \dd \sigma(x) \cdot p\, f_i(x,p)
\end{equation}
so that $f(x,p)$ is separated from this constant. In general, the distribution $f(x,p)$ that describes the particles is written as
\begin{equation}
    f(x,p) = f_{0i}(x,p) + \delta f_i(x,p).
    \label{eq:distviscouscorrections}
\end{equation}
The term $f_{0i}(x,p)$ is the equilibrium distribution, given in \eqref{eq:distideal}, with
\begin{equation}
    f_{0i}(x,p) = \frac{g_i}{\exp\left(\frac{u(x)\cdot p - B_i\mu_B(x)}{T}\right)+\theta}
\end{equation}
where $\theta = 1$ is associated with Fermi–Dirac statistics for baryons and $\theta = -1$ with Bose–Einstein statistics for mesons, $g_i$ denotes the degeneracy of species $i$, and $B_i$ its corresponding baryon number. In this context, $E(x,p) = u(x)\cdot p$ represents the energy of a particle with momentum $p$, as measured by an observer with four-velocity $u(x)$ comoving with the hypersurface $\Sigma$.

The term $\delta f_i(x,p)$ represents corrections due to the presence of viscosity in the fluid \cite{PhysRevC.68.034913}, that is, deviations from equilibrium. These corrections are calculated from the Boltzmann equation in \eqref{eq:boltzmann}, and in general they do not have exact solutions, only approximate ones. There are methods for computing these distributions, such as the Grad moment approximation, in particular the 14-moment approximation \cite{ISRAEL1979341,denicol2022microscopic}, a Chapman–Enskog expansion of the Boltzmann equation, often in the relaxation time approximation \cite{Jaiswal2013,DuslingSchaefer2012}, or solving the linearized Boltzmann equation to construct self-consistent freeze-out corrections \cite{MolnarWolff2017}. In this work, we adopt the 14-moment approximation. In this framework, we essentially have two corrections arising from the presence of shear and bulk,
\begin{equation}
    \delta f = \delta f_\pi + \delta f_\Pi,
\end{equation}
where the shear contribution is given by
\begin{equation}
    \delta f_{\pi i} = f_{0 i}\left(1 - g_i^{-1} \theta f_{0i}\right)\frac{p_\mu p_\nu \pi^{\mu \nu}}{2(\epsilon +P)T^2}
\end{equation}
and the bulk contribution is expressed as
\begin{equation}
    \delta f_{\Pi i} = f_{0i}\left(1 - g_i^{-1} \theta f_{0i}\right)\left( m^2 A_T +   E^2 A_E + A_B E  B_i\right)\Pi,
\end{equation}
where the coefficients $A_T$, $A_E$, and $A_B$ are presented in Appendix \ref{bulkcoef}. 

From the Cooper–Frye formula, it is possible to sample individual particles by means of Monte Carlo methods, generating lists of final particles that characterize the outcome of the hydrodynamic evolution. This procedure is known as particlization and is extremely useful in the afterburner stages.

\section{Afterburner}

Hadronic transport processes and decays occur by evolving the distribution
\(f(x,p)\), which results from the hydrodynamic evolution, through the Boltzmann
equation \eqref{eq:boltzmann}. The terms that appear in the collision term \(C[f]\) can be summarized as \cite{groot1980relativistic, CercignaniKremer2002,Weil:2016zrk}
\begin{equation}
C_i[f] =
C_i^{\mathrm{el.}}[f]
+
C_i^{\mathrm{inel.}}[f]
+
C_i^{\mathrm{res.}}[f]
+
C_i^{\mathrm{dec.}}[f]
+
C_i^{\mathrm{multi}}[f],
\label{eq:collisionterms}
\end{equation}
where \(i\) denotes the hadronic species under consideration. Under these terms, we have
\begin{itemize}
    \item \(C_i^{\mathrm{el.}}[f]\): corresponds to elastic scatterings,
    in which the particle species are preserved and only their
    momenta are changed. Such processes can be written as
    \begin{equation}
        a+b \rightarrow a+b .
    \end{equation}
    They are responsible for the kinetic redistribution of the system, modifying
    the momentum spectra without directly changing the abundances of the
    participating species.

    \item \(C_i^{\mathrm{inel.}}[f]\): describes inelastic binary
    scatterings, in which the particles in the final state may differ from the
    incoming particles. Schematically,
    \begin{equation}
        a+b \leftrightarrow c+d .
    \end{equation}
    These processes simultaneously affect the momentum distribution and the
    chemical composition of the hadronic gas, while respecting the conservation
    of energy-momentum and of the conserved quantum numbers.

    \item \(C_i^{\mathrm{res.}}[f]\): characterizes processes mediated by
    the formation of resonances. A typical channel is
    \begin{equation}
        a+b \leftrightarrow R ,
    \end{equation}
    where \(R\) represents an unstable resonance. After its formation, this
    resonance propagates as a particle until it decays or interacts again.
    These channels are central to low-energy hadronic dynamics, since a large
    part of the interactions between hadrons occurs via intermediate
    resonant states.

    \item \(C_i^{\mathrm{dec.}}[f]\): includes the decays of unstable
    particles and resonances. Examples of channels are
    \begin{equation}
        R \rightarrow a+b,
        \qquad
        R \rightarrow a+b+c .
    \end{equation}
    These terms act as loss terms for the unstable species \(R\) and
    as gain terms for the particles produced in the decay. The corresponding
    rates are determined by the partial decay widths and by the branching
    ratios of the allowed channels.

    \item \(C_i^{\mathrm{multi}}[f]\): gathers reactions with more than two
    particles in the initial or final state. In a generic form,
    \begin{equation}
        a_1+\cdots+a_n \leftrightarrow b_1+\cdots+b_m .
    \end{equation}
    This category encompasses processes such as \(2\leftrightarrow3\),
    \(2\leftrightarrow4\), \(2\leftrightarrow5\), \(3\leftrightarrow1\),
    \(3\leftrightarrow2\), \(4\leftrightarrow2\), and \(5\leftrightarrow2\).
    These channels are important for effectively describing multi-hadronic
 reactions, such as baryon–antibaryon annihilation, hadronic regeneration,
    and the production or absorption of light nuclei \cite{Staudenmaier2021}.
\end{itemize}

\section{Observables}

As in real collisions, the final stage of the theoretical description of the collision involves identifying the produced particles and computing the final observables. Experimentally, this process is associated with reconstructing the trajectories and identifying the species that reach the detector, together with the determination of their momenta. In the theoretical calculation, on the other hand, the species of each particle is defined directly from its microscopic properties. In the application of the Cooper–Frye formula, Eq.~\eqref{eq:cff}, the invariant distribution depends explicitly on the mass, quantum numbers, and degeneracy of each species. Equivalently, when particles are sampled using Monte Carlo methods, each produced hadron is identified by a standardized numerical code, usually following the \textit{Particle Data Group} \cite{ParticleDataGroup:2022} convention for particle identification in event generators and analysis packages. This convention makes it possible to distinguish particles from antiparticles, as well as to encode information associated with flavor content, spin, and other relevant quantum numbers.

Next, we will present observables that are relevant to our study, since reproducing them is essential to ensure the consistency of a hybrid simulation model. Only then will we be able to move on to the calculation of more complex quantities, which, in our case, are the HBT interferometry radii.

\subsection{Spectra Analysis}

\begin{figure}[!htbp]
\centering
\includegraphics[width=\linewidth]{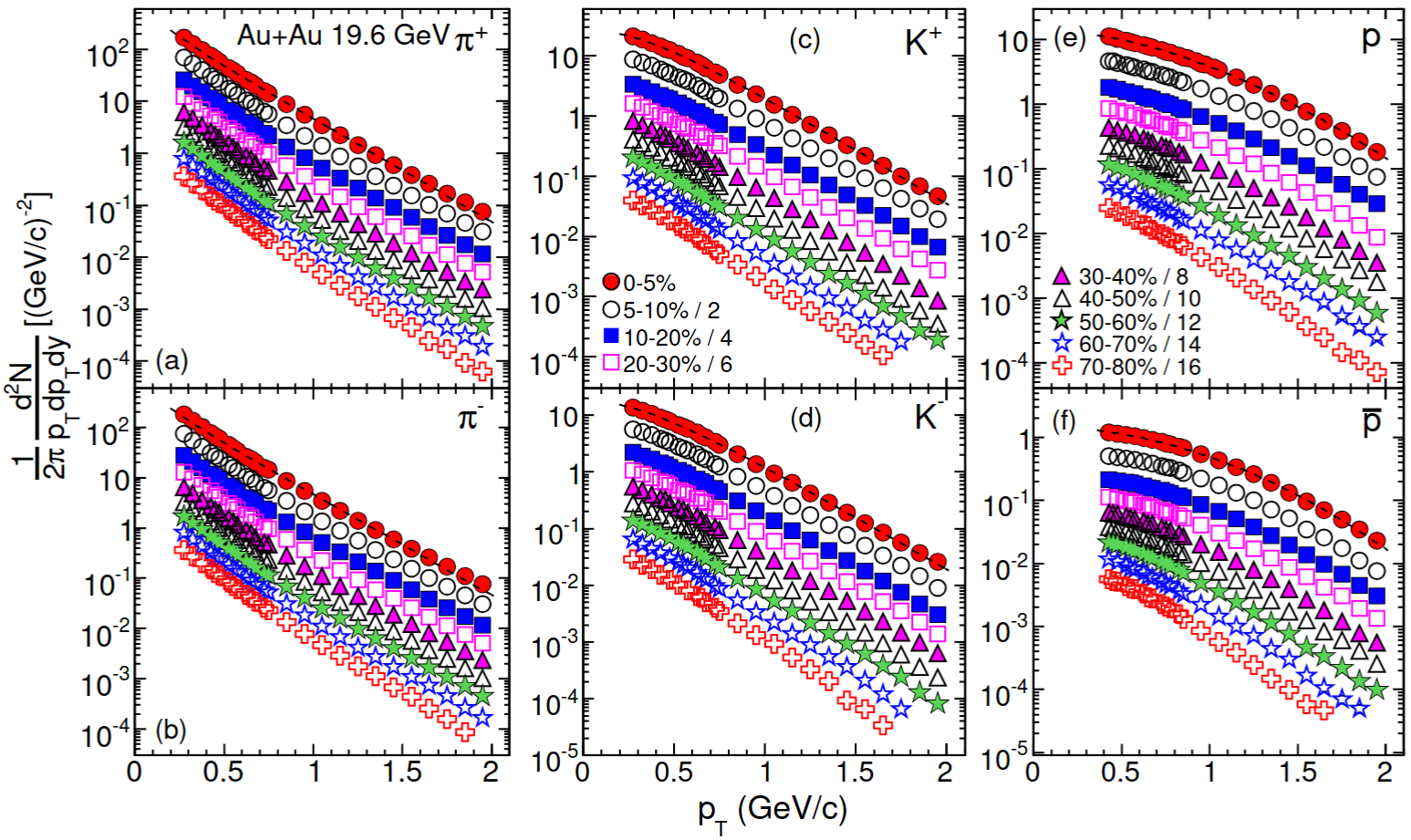}
\caption{Transverse momentum spectra for (a) $\pi^+$, (b) $\pi^-$, (c) K$^+$, (d) K$^-$, (e) $p$, and (f) $\bar p$ in Au+Au collisions at $\sqrt{s_{NN}}$ = 19.6 GeV for different centralities at mid-rapidity ($|y| < 0.1$, $dy = 0.2$). Taken from \cite{PhysRevC.96.044904}.}
\label{fig:dndpt19p6}
\end{figure}

\begin{figure}[!htbp]
\centering
\includegraphics[width=0.4\linewidth]{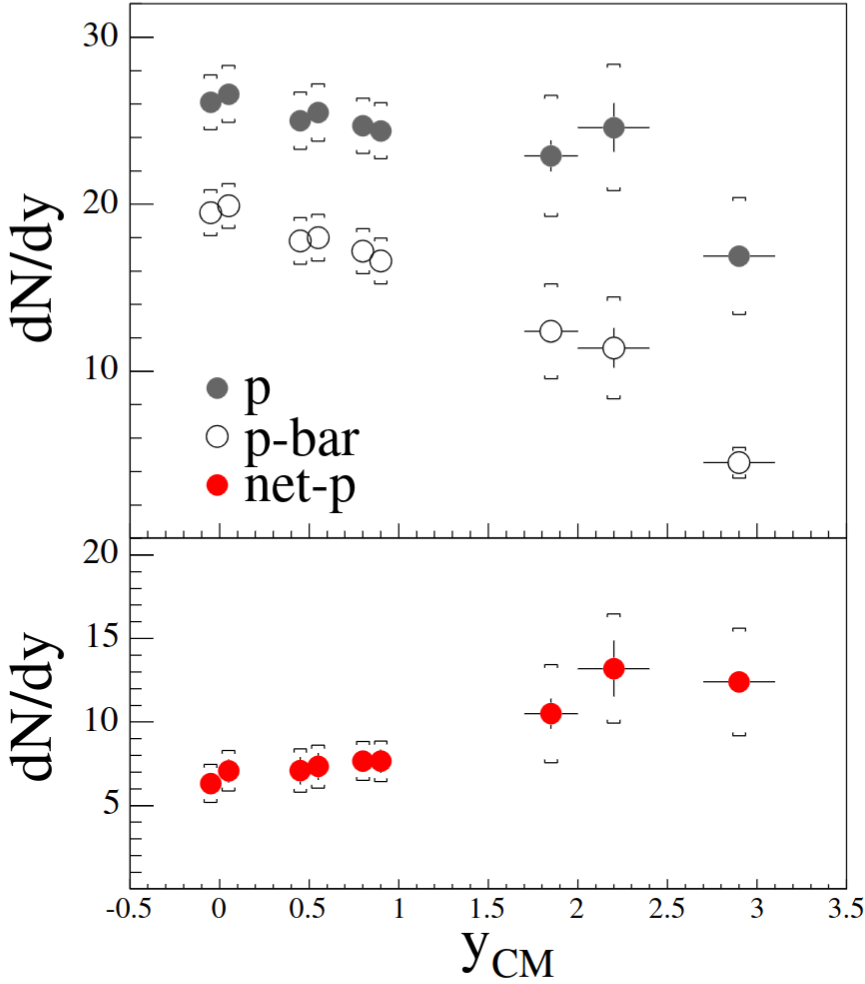}
\caption{Proton, anti-proton and net-proton rapidity distributions in Au+Au collisions at $\sqrt{s_{NN}}$ = 200 GeV. Taken from \cite{PhysRevLett.93.102301}.}
\label{fig:dndy200}
\end{figure}

With the particles properly identified, the most common observables calculated experimentally are the spectra, that is, histograms of the particles in terms of momentum variables such as rapidity $y$, pseudorapidity $\eta$, and transverse momentum $p_T$. These spectra are related to the invariant momentum distribution $p^0 d^3N/d^3p$ since the element $d^3p/p^0$ can be rewritten as
\begin{equation}
    \frac{d^3p}{p^0} = \frac{1}{p^0} \left|\frac{\partial (p^x, p^y, p^z) }{\partial (y, p_T, \phi) } \right|dp_T dy d\phi= p_T dp_T dy d\phi,
\end{equation}
so
\begin{equation}
    p^0\frac{d^3N}{d^3p} = \frac{1}{p_T}\frac{d^3 N}{dp_T dy d\phi}.
\end{equation}
By integrating this distribution over each pair of these variables, we obtain:
\begin{align}
    &\frac{dN}{dy} = \int dp_T p_T d\phi \left(  \frac{1}{p_T}\frac{d^3 N}{dp_T dy d\phi}\right) = \int dp_T p_T d\phi \left(p^0\frac{d^3N}{d^3p}\right),\label{eq:projections}\\
    &\frac{1}{p_T}\frac{dN}{dp_T} = \int dy d\phi \left(  \frac{1}{p_T}\frac{d^3 N}{dp_T dy d\phi}\right) = \int dy d\phi \left(p^0\frac{d^3N}{d^3p}\right),\\
    &\frac{dN}{d\phi} = \int dp_T p_T dy\left(  \frac{1}{p_T}\frac{d^3 N}{dp_T dy d\phi}\right) = \int dp_T p_T dy \left(p^0\frac{d^3N}{d^3p}\right) .
\end{align}
This shows that the histograms in $y$ and $p_T$ are indeed projections of the invariant distribution. We can see examples of experimental data from Au+Au collisions in the plots of Figures \ref{fig:dndpt19p6} and \ref{fig:dndy200}.

\begin{figure}[!htbp]
\centering
\includegraphics[width=0.5\linewidth]{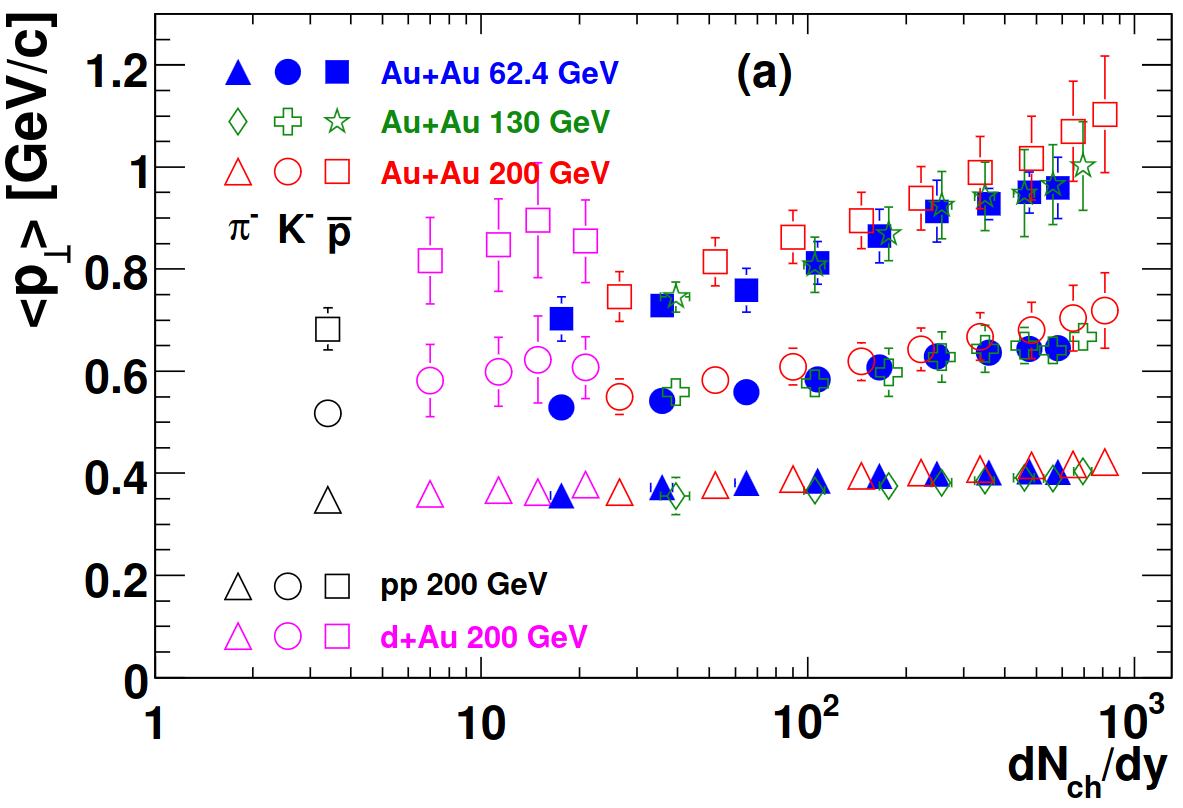}
\caption{Mean transverse momentum of charged particles as a function of rapidity for Au+Au collisions at 62.4, 130, and 200~GeV. Minimum-bias proton--proton and deuteron+Au collision data are also shown. Taken from \cite{PhysRevC.79.034909}.}
\label{fig:meanpt200}
\end{figure}

Other common observables are integrated observables, such as the calculation of the average number of charged particles.
\begin{equation}
    \langle N_{\rm ch}\rangle =  \sum_{\rm i = ch} N_i = \sum_{\rm i = ch}\int  dp_T p_T dy d\phi \left(p^0\frac{d^3N_i}{d^3p}\right),
\end{equation}
and of the mean transverse momentum of charged particles
\begin{equation}
    \langle p_T \rangle_{\rm ch} =  \frac{1}{\langle N_{\rm ch}\rangle}\sum_{\rm i = ch}\int  dp_T p_T dy d\phi \left(p^0\frac{d^3N_i}{d^3p}\right)p_T.
\end{equation}
These quantities are shown in Fig.~\ref{fig:meanpt200}.

\subsection{Event-by-Event Analysis}

In this work, we adopt an event-by-event (EbE) analysis. In this approach, each initial condition is evolved independently through the dynamical chain, generating a final hadron sample for each collision event. This procedure is important because individual initial conditions are not smooth profiles but contain event-by-event geometric fluctuations and granular structures, such as localized high-density regions, which can affect the subsequent hydrodynamic response. The observables are then computed for each event and averaged over the ensemble. For a generic observable $\mathcal{O}$, this procedure can be written as
\begin{equation}
    \langle \mathcal{O} \rangle
    =
    \frac{1}{N_{\rm ev}}
    \sum_{e=1}^{N_{\rm ev}}
    \mathcal{O}_e ,
\end{equation}
where $N_{\rm ev}$ is the number of events and $\mathcal{O}_e$ is the value of the observable calculated from the final particles of event $e$. The EbE procedure preserves the fluctuations and granularity of the initial state, allowing their hydrodynamic response to be reflected in the final observables, as commonly done in event-by-event hydrodynamic calculations~\cite{PhysRevLett.101.112301}.

\subsection{Anisotropic Flow}

A particularly relevant set of observables are the anisotropic flow coefficients. They play an important role because they make it possible to describe the emission geometry through the orientation of the momentum of the emitted particles. In this case, they can be used to characterize geometries compatible with a hydrodynamic evolution.

We define the expression
\begin{equation}
    v_n(p_T,y)e^{in\Psi_n(p_T,y)}
    =
    \frac{
    \int d\phi\, e^{in\phi}\,\frac{d^3N}{p_T dp_T dy d\phi}
    }{
    \int d\phi\,\frac{d^3N}{p_T dp_T dy d\phi}
    }
    \label{eq:complexflowvector}
\end{equation}
With this definition, the invariant distribution can be written as
\begin{equation}
    \frac{1}{p_T}\frac{d^3N}{dp_T\,dy\,d\phi}
    =
    \frac{1}{2\pi p_T}\frac{d^2N}{dp_T\,dy}
    \left[
        1
        +
        2\sum_{n=1}^{\infty}
        v_n(p_T,y)
        \cos\left(n[\phi-\Psi_n(p_T,y)]\right)
    \right],
    \label{eq:flowfourier}
\end{equation}
where \(v_n\) measures the strength of the harmonic of order \(n\), while \(\Psi_n\) defines its azimuthal orientation. In particular, \(v_2\) is called elliptic flow because it is associated with the elliptic component of the final momentum distribution \cite{Heinz_2009}. Meanwhile, \(v_3\) is related to triangular flow.


\subsubsection{Calculations with Cumulants}

We determine \(v_n\) using two-particle cumulants, which allow us to obtain them directly from azimuthal correlations, without the need to explicitly reconstruct the event plane. For an individual event, we define
\begin{equation}
    Q_n
    =
    \sum_{j=1}^{M} e^{in\phi_j},
    \label{eq:qvectorcumulant}
\end{equation}
where \(M=Q_0\) is the event multiplicity and $j$ is the summation index over the particles. The two-particle correlation can be written as
\begin{equation}
    \langle 2\rangle_n
    =
    \frac{
    Q_n Q_n^\ast - M
    }{
    M(M-1)
    }
    =
    \frac{1}{M(M-1)}
    \sum_{i\neq j}
    e^{in(\phi_i-\phi_j)}.
    \label{eq:twoparticleevent}
\end{equation}
The subtracted term \(M\) removes the \(i=j\) autocorrelations. Taking the average over events, we obtain
\begin{equation}
    c_n\{2\}
    =
    \left\langle\left\langle 2\right\rangle_n\right\rangle
    =
    \left\langle\left\langle
    \cos\left(n[\phi_1-\phi_2]\right)
    \right\rangle\right\rangle.
    \label{eq:twoparticlecumulant}
\end{equation}
When pair correlation is dominated by collective flow\cite{PhysRevC.83.044913},
\begin{equation}
    v_n\{2\}
    =
    \sqrt{c_n\{2\}}.
    \label{eq:vntwoparticle}
\end{equation}
For the differential flow, we correlate particles of interest in a given interval of \(p_T\) with the reference particles of the event. Defining
\begin{equation}
    p_n(p_T)
    =
    \sum_{a=1}^{m(p_T)}
    e^{in\phi_a},
    \label{eq:poivector}
\end{equation}
the differential correlation can be written as\cite{PhysRevC.64.054901}
\begin{equation}
    d_n\{2\}(p_T)
    =
    \left\langle
    \frac{
    p_n(p_T)Q_n^\ast - m(p_T)
    }{
    m(p_T)(M-1)
    }
    \right\rangle,
    \label{eq:differentialcumulant}
\end{equation}
so that
\begin{equation}
    v_n\{2\}(p_T)
    =
    \frac{
    d_n\{2\}(p_T)
    }{
    \sqrt{c_n\{2\}}
    }.
    \label{eq:vndifferential}
\end{equation}

\section{Further Developments}

In addition to the hydrodynamic description employed in this work, different extensions have been developed to make the modeling of heavy-ion collisions more consistent in regimes far from local equilibrium. Among them, second-order dissipative formulations, such as DNMR~\cite{Denicol:2012cn}, and first-order causal and stable formulations, such as BDNK~\cite{Bemfica:2020zjp,Kovtun:2019hdm}, stand out. More refined descriptions of the initial stage have also been explored, such as dynamical initialization, in which the energy-momentum tensor and conserved currents are continuously deposited into the hydrodynamic evolution, instead of being imposed at a single initial time~\cite{Shen:2017bsr}. Another important direction is the inclusion of core–corona components, as in EPOS-type models, which allow one to separate dense regions, treated hydrodynamically, from more dilute regions, in which particle production retains characteristics closer to those of elementary hadronic collisions~\cite{Werner:2007bf,Werner:2010aa}. These advances show that relativistic hydrodynamic modeling remains a very active field, in which different physical ingredients can be incorporated to reduce systematic uncertainties and strengthen the connection between experimental observables and the properties of strongly interacting matter.

\chapter{Hybrid Model for Heavy-Ion Collision}

In this chapter, we provide a description of the codes present in the hybrid model used for the heavy-ion collision simulation chain. In list form, they are:
\begin{enumerate}
    \item Pre-equilibrium and Initial Conditions: \texttt{AMPT} and \texttt{AMPT Smearer} \cite{PhysRevC.61.067901,PhysRevD.44.3501,PhysRevC.65.034904,ZHANG1998193}
    \item Hydrodynamic Evolution: \texttt{PRAIA} \cite{PRAIA}
    \item Particlization: \texttt{BQSSampler} \cite{pala2025nuclearconfectionery}
    \item Afterburner: \texttt{SMASH} \cite{Weil:2016zrk, Sciarra:2024smash}
\end{enumerate}
A flowchart of this description can be seen below in Fig. \ref{fig:fluxograma}.

\begin{figure}[H]
    \centering
    \includegraphics[width=1.\linewidth]{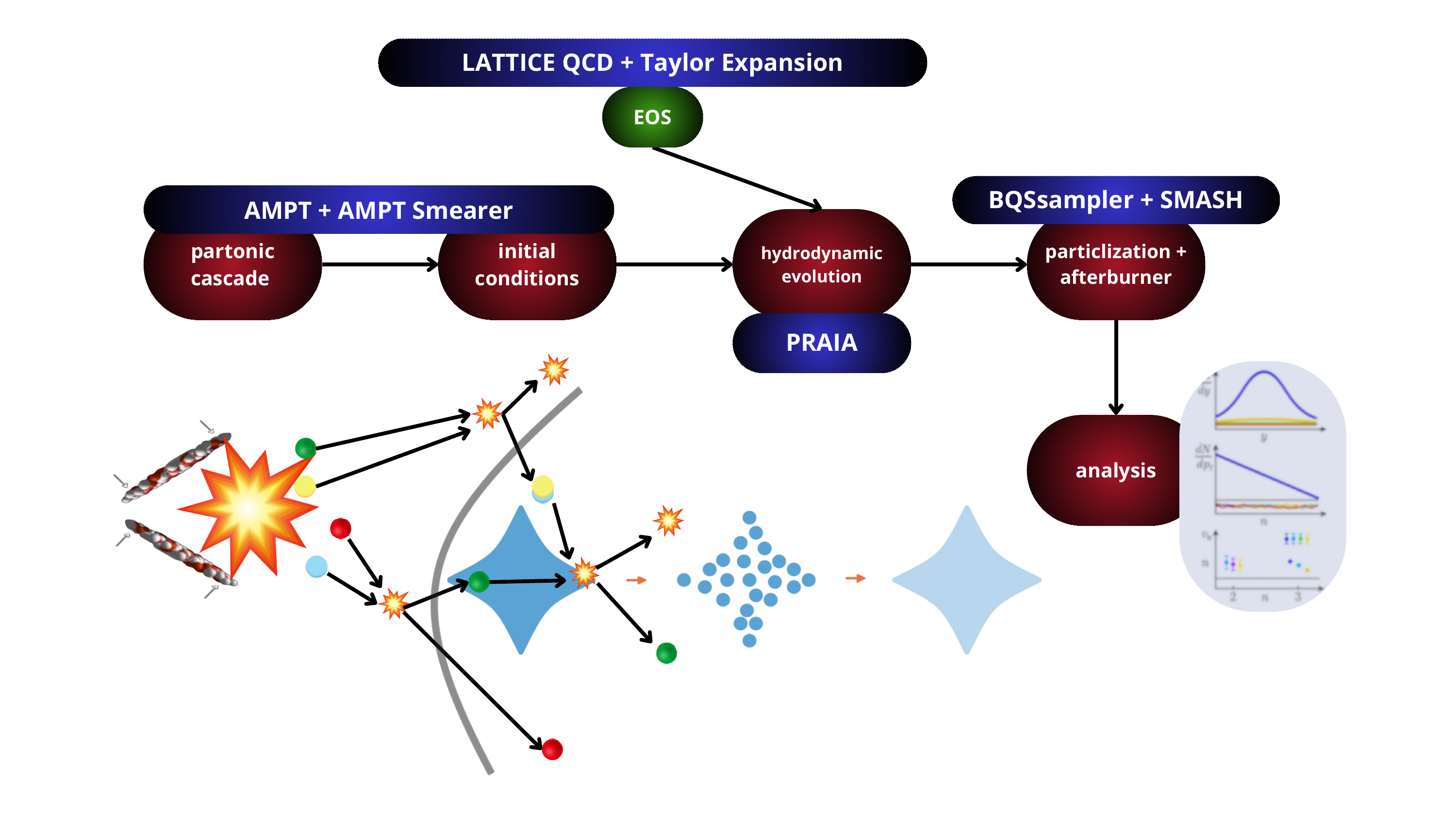}
    \caption{General flowchart of the simulation chain employed}
    \label{fig:fluxograma}
\end{figure}
\section{AMPT and AMPT Smearer}

\textit{A Multi-Phase Transport} -- \texttt{AMPT} -- is a (3+1)--D model developed to simulate nuclear collisions from the initial stage and pre-equilibrium partonic cascade up to the hadronic transport stage. The initial conditions are generated by the Heavy Ion Jet Interaction Generator (\texttt{HIJING}), which describes both hard processes through the production of minijet partons and soft processes through the excitation of color strings. The subsequent evolution of the partonic system is described by Zhang's Parton Cascade (\texttt{ZPC}), which models parton scatterings through a cascade approach based on binary collisions.

\texttt{AMPT} has two operation modes: the default mode and the string-melting mode. In the default mode, only the minijet partons produced by \texttt{HIJING} undergo the \texttt{ZPC} evolution. After the partonic interactions, these partons are recombined with their parent strings, which are subsequently hadronized through the Lund string fragmentation model. In the string-melting mode, the excited strings are instead converted into their constituent partons. These partons, together with the minijet partons, undergo the \texttt{ZPC} evolution before hadronization. Hadronization is then performed through a quark coalescence model.

In both modes, the resulting hadrons subsequently undergo hadronic interactions described by the relativistic transport model ART. In this work, the string-melting mode is employed, in which the system therefore evolves from the initial conditions generated by \texttt{HIJING}, through the partonic evolution described by \texttt{ZPC}, followed by quark-coalescence hadronization and the final hadronic transport stage described by ART.




For initial conditions and pre-equilibrium, \texttt{AMPT} is employed in the string-melting mode only up to the pre-equilibrium stage.  We evolve the partonic system up to a proper time $\tau_{\mathrm{switch}}$, and the resulting partonic configuration is then passed to the hydrodynamic evolution. Thus, the hadronization and hadronic transport stages described above are not included in the \texttt{AMPT} evolution used here.

\begin{figure}[H]
    \centering
    \includegraphics[width=0.3\linewidth]{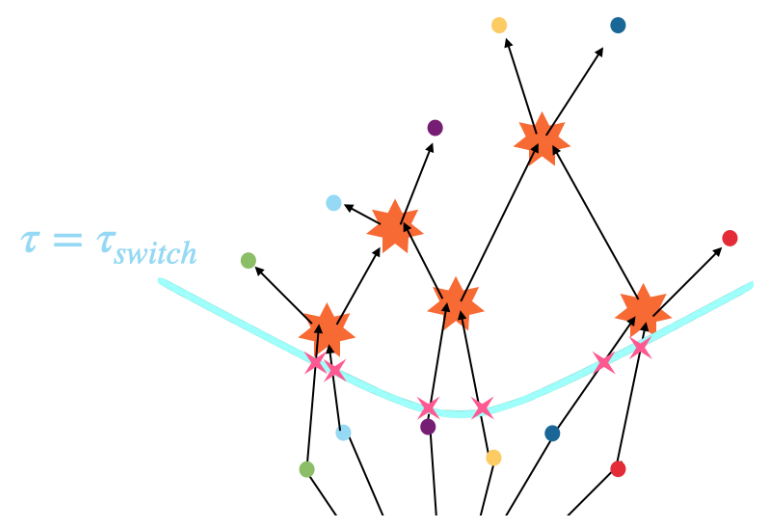}
    \caption{Illustration showing partons interacting and them being selected at the switching time. Figure produced by Kevin Pala and provided to the author.}
\end{figure}
Partons are discrete degrees of freedom characterized by their space-time position $x^\mu$, momentum $p^\mu$, and baryon charge. In order to initialize hydrodynamics, these discrete partons must be converted into continuous fields, namely the energy-momentum tensor $T^{\mu\nu}$ and the baryon density $\rho_b$. This is achieved through a smearing procedure \cite{PhysRevC.103.024903}, written as
\begin{align}
    &T^{\mu\nu}(x,y,\eta_s)=\sum_{i=\mathrm{partons}}\frac{p_i^\mu p_i^\nu}{p_i^0}\,\phi_i(x,y,\eta_s),\\
    &J_b^\mu(x,y,\eta_s)=\sum_{i=\mathrm{partons}}\frac{Q_i}{K}\frac{p_i^\mu}{p^0_i}\,\phi_i(x,y,\eta_s),
\end{align}

where $Q_i$ is the baryon charge of the $i$-th parton and $\phi_i$ is a Gaussian profile given by
\begin{equation}
\phi(x,y,\eta_s) =
\frac{K}{(2\pi)^{3/2}\sigma_r^2\tau_0\sigma_{\eta_s}}\exp\left[
-\frac{(x-x_i)^2+(y-y_i)^2}{2\sigma_r^2}
-\frac{(\eta_s-\eta_{s,i})^2}{2\sigma_{\eta_s}^2}
\right].
\end{equation}
The parameters $K$, $\sigma_r$, and $\sigma_{\eta_s}$ are chosen to reproduce the experimental data.

\begin{figure}[H]
    \centering
    \includegraphics[width=1.\linewidth]{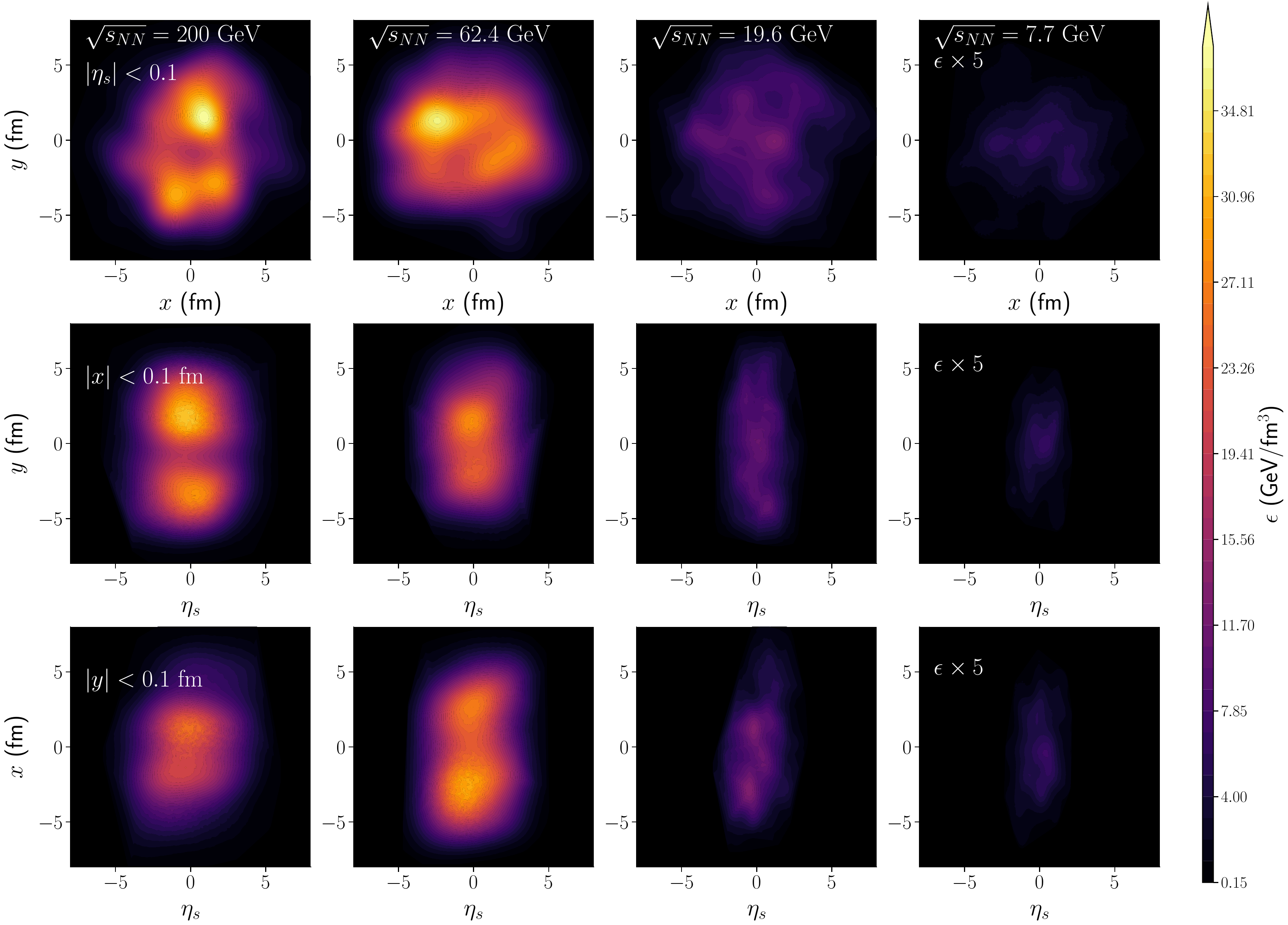}
    \caption{Initial energy density profiles obtained from the smeared \texttt{AMPT} partonic cascade after Landau matching for Au+Au collisions in the 0--5\% centrality class. The columns correspond to $\sqrt{s_{NN}}=200$, $62.4$, $19.6$, and $7.7$ GeV. The rows show, respectively, the $(x,y)$ plane for $|\eta_s|<0.1$, the $(\eta_s,y)$ plane for $|x|<0.1$ fm, and the $(\eta_s,x)$ plane for $|y|<0.1$ fm. A common color scale is used for all panels, fixed by the $200$ GeV profile. The $7.7$ GeV profiles were multiplied by a factor of $5$ only for visualization.}
    \label{fig:ampt2}
\end{figure}

In Fig.~\ref{fig:ampt2}, the energy density profiles are shown for several values of $\sqrt{s_{NN}}$ in the 0–5\% centrality window, derived through smearing and Landau matching within the \texttt{AMPT} model.

\section{PRAIA}

\begin{figure}[!htpb]
\centering
\begin{adjustbox}{max width=\textwidth, max totalheight=\textheight, keepaspectratio}
\begin{BVerbatim}[fontsize=\small]
#######################################################################################
///////////////////////////////////////////////////////////////////////////////////////
//                                                                                   //
// 8 888888888o   8 888888888o.            .8.           8 8888          .8.         //
// 8 8888    `88. 8 8888    `88.          .888.          8 8888         .888.        //
// 8 8888     `88 8 8888     `88         :88888.         8 8888        :88888.       //
// 8 8888     ,88 8 8888     ,88        . `88888.        8 8888       . `88888.      //
// 8 8888.   ,88' 8 8888.   ,88'       .8. `88888.       8 8888      .8. `88888.     //
// 8 888888888P'  8 888888888P'       .8`8. `88888.      8 8888     .8`8. `88888.    //
// 8 8888         8 8888`8b          .8' `8. `88888.     8 8888    .8' `8. `88888.   //
// 8 8888         8 8888 `8b.       .8'   `8. `88888.    8 8888   .8'   `8. `88888.  //
// 8 8888         8 8888   `8b.    .888888888. `88888.   8 8888  .888888888. `88888. //
// 8 8888         8 8888     `88. .8'       `8. `88888.  8 8888 .8'       `8. `88888.//
//                                                                                   //
///////////////////////////////////////////////////////////////////////////////////////
#######################################################################################
                                
                 /%#                       
                ###//                             
               %###//#                            
               ####//(/      /%(####%.                       
               &###////  /*#####///(                               
               .####*//%####(////(                                  
  .%#&####&#####(###/(##%#(/////                           
%//////////**#%####/###%###%###&###%                   @@@
     #%/%///(//###&##(##//////*/%#%#%&(         @ @@@@(@@(/@ @@@ 
            (%###%/// (*/%(/(//(///////%(          @@@@@@@@@@@ 
          /%%##(//*,   /*//                  @@@@@@@@@@@@(/(((((@@(((@@
         %####////#     (/(*            @@/@@((@@@@@@(@%((/(@((((/((@@@((%@ 
        .###/*/(#        (/*/         %@((@((((@@@/@/((@@(((/((@@/((((((/@((@((@@
        .##//(           .///      @@#@/((//@/(/@@(/(/(/(/(/((/@@/(((/((/(@/(/@@@@@  
         &,               #((/     @@((((((@/(((@//((@@(((((/(((@@((((@/((((((((@@@@
                          ////*        @@@@@@@@@@@@@@@@@@@@@@@@@@@@@@@@@@@@@@
                           (//*       @@..@@@@@&....%@,..@@@@@@@@@........%@
                          .*((/      @@.../@@@@..........@@@@@@@@@.........@@ 
                          /////      @@....@@@&..........@@@@@@@@@.........@@
                          (//*       @@....@@@@..........@@@@@@@@@,........@@
                         ///*/       @@@@@@@@@@(......../@@@@@@@@@&,.......@@ 
***************************************************************************************
***************************************************************************************
********* ..... .............          ....  ......*************************.....   ...
.....................................//*//#######....///////......******.........    ..
.#..,*,.#./*...............*//(.*./.../,.(((%%#  (#.((#/....   .          .      ......
.//////////*##.###(##(#./// ,...../...(%%((%%####/%%/#####(.,//...      ........,,*#///
*##/..######.....*/////././././/,/.//,//%%%#%%%%%%(###(%((*###((######.#((#.(*.#*  #///
.##########///..//./././,//////..///(////%%%#######%#%%%%(%%%%(%**##(####(##(#*.((/(,,(
#######################################################################################
                     RELATIVISTIC VISCOUS HYDR0DYNAMICS SIMULATOR                      
#######################################################################################
    Vinícius Franção, Jonatan Sola, Kevin Pala, Gabriel Denicol & Frédérique Grassi    
#######################################################################################
\end{BVerbatim}
\end{adjustbox}
\caption{Opening banner of the \textsc{PRAIA} code.}
\label{fig:praia-banner}
\end{figure}

To solve the equations of viscous relativistic hydrodynamics, we use the \texttt{PRAIA} code \cite{PRAIA}, developed by my research group. Initially conceived for exploring the QCD phase diagram, \texttt{PRAIA} was developed to simulate relativistic hydrodynamics and was later coupled with Maxwell’s equations for the study of resistive 3+1D magnetohydrodynamics in Milne coordinates. The code is written in \texttt{C++} and employs the Smoothed Particle Hydrodynamics (SPH) method to simulate fluid evolution.

Originally introduced in astrophysics~\cite{lucy1977numerical,gingold1977smoothed}, SPH has since been applied in several areas of computational fluid dynamics. In the context of heavy-ion collisions, SPH-based codes have been employed for more than two decades in hydrodynamic codes such as SPheRIO --- \textbf{S}moothed \textbf{P}article \textbf{h}ydrodynamic \textbf{e}volution of \textbf{R}elativistic heavy \textbf{IO}n collisions~\cite{aguiar2001smoothed,osada2001event,hama2005topics}, a Brazilian 3+1D ideal hydrodynamics code developed by T. Kodama, Y. Hama, T. Osada, and C. Aguiar that pioneered event-by-event simulations of high-energy heavy-ion collisions --- and v-USPhydro/CCAKE~\cite{noronha2013bulk,PhysRevC.90.034907,plumberg2025conservation,pala2025nuclearconfectionery} --- initially developed by J. Noronha-Hostler as a successor to SPheRIO, incorporating viscous effects and equations of state with three conserved charges (BQS).

Combined with fluctuating initial conditions generated by NeXuS~\cite{PhysRevC.65.054902}, SPheRIO formed the NeXSPheRIO framework, which was extensively employed in phenomenological studies of relativistic heavy-ion collisions. Its applications include the description of particle spectra, including identified hadrons~\cite{doi:10.1142/S0218301307007167,doi:10.1142/S0218301318500581}, anisotropic flow and correlations among flow harmonics~\cite{PhysRevLett.109.202302,PhysRevC.95.034901}, two-pion HBT interferometry~\cite{PhysRevLett.93.182301}, particle-number fluctuations and conserved-charge cumulants~\cite{PhysRevC.100.054905,PhysRevC.107.024910}, as well as investigations of different equations of state~\cite{doi:10.1142/S0218301307007167}. These studies also highlighted the importance of initial-state granularity and event-by-event fluctuations in shaping final-state observables~\cite{PhysRevLett.97.202302,PhysRevLett.101.112301,PhysRevC.85.024908}.

The main advantages of the SPH method include the natural preservation of conserved quantities and the absence of conventional advection algorithms. By representing the fluid as a set of interpolating particles, SPH also provides a fully Lagrangian description of hydrodynamics.

\subsection{Smoothed Particle Hydrodynamics}

Consider a quantity $A$ defined at spacetime points $x = (x^0, x^i)$, where $x^0$ is a temporal coordinate and $x^i$ are spatial coordinates (not necessarily Cartesian). The SPH interpolation is constructed as follows. We define a quantity $\sigma$ called the specific density of a conserved quantity $\nu$, that is,
\begin{equation}
    \sigma = \frac{\nu}{V},
\end{equation}
where $V$ is the volume. The corresponding density satisfies the continuity equation
\begin{equation}
    d_\mu (\sigma u^\mu) = \frac{1}{\sqrt{-\text{g}}} \partial_\mu (\sigma^* v^\mu) = 0
\end{equation}
where
\begin{align}
    &\sigma^* = \sqrt{-g} u^0 \sigma,\\
    &v^\mu = \frac{u^\mu}{u^0} = \frac{dx^\mu}{dx^0},
\end{align}
in which $\sigma^*$ is the so-called specific density in the computational reference of SPH. We define a normalized kernel function $W(x, h)$, where $h$ is the so-called smoothing length, with the properties
\begin{align}
    &\int d^3x' \, W(x', h) = 1, \\
    &\lim_{h \to 0} W(x, h) = \delta(x).
    \label{eq:kerneldefinition}
\end{align}
For a set of $N$ spacetime points $x_j$, we assign values $A_j = A(x_j)$, $\nu_j = \nu(x_j)$, and $\sigma_j = \sigma(x_j)$ for each point, such that to compute $A$ at another point $x_i$, we use
\begin{equation}
    A_i = \sum_{j=1}^N \frac{\nu_j}{\sigma^*_j} A_j W_{ij}
\end{equation}
where $W_{ij} = W(x_i - x_j, h)$. This allows the spatial gradients of $A$ to be computed directly on the kernel, since
\begin{equation}
    \frac{\partial A_i}{\partial x^k} = \sum^N_{j=1}\frac{\nu_j}{\sigma^*_j} A_j \frac{\partial W_{ij}}{\partial x^k}.
\end{equation}
and we can compute $\sigma^*$ directly as
\begin{equation}
    \sigma^*_i= \sum^N_{j=1}\nu_j W_{ij}
\end{equation}
These points are the SPH particles, which act as Lagrangian interpolation points for the fluid. All physical quantities are assigned to these particles, and the dynamical equations are then evolved for them.

\subsection{Equations of Motion}

The equations of motion are obtained from the conservation laws and are evolved as described below. For any dynamical tensor $A^{\alpha_1 \cdots \alpha_n}$, its comoving covariant derivative $DA^{\alpha_1 \cdots \alpha_n}$ can be written  as a linear function of the covariant derivative of the 4-velocity $Du^\beta$,
\begin{equation}
  DA^{\alpha_1 \cdots \alpha_n} = (M_A)^{\alpha_1 \cdots \alpha_n} _\beta Du^\beta + (F_A)^{\alpha_1 \cdots \alpha_n},
\end{equation}
where $(M_A)^{\alpha_1 \cdots \alpha_n} _\beta$ and $(F_A)^{\alpha_1 \cdots \alpha_n}$ are the matrix coefficients. Using these equations, we can write a general dynamic equation for the 4-velocity as
\begin{align}
    &M^\alpha_{\ \beta} Du^\beta = F^\alpha \text{, \ \ then } \\ 
    &Du^{\beta} = (M^{-1})^{\beta}_{\ \alpha} F^\alpha
\end{align}
since $M^\alpha_{\ \beta}$ is a matrix that depends on all $(M_A)^{\alpha_1 \cdots \alpha_n} _\beta $, and the same is true for $F^\alpha$. Inverting $M^\alpha_{\ \beta}$ gives $Du^\beta$, from which the comoving derivatives of the remaining dynamical variables are obtained. From this, we obtain the total derivatives of the dynamical quantities from the covariant derivatives by manipulating the Christoffel symbol terms. This provides the time derivatives used to advance the coupled equations numerically. Details on how we evolve the equation of motion and code tests are given in the Appendix.~(\ref{apdx:numericaleom}). 

\subsection{Hydrodynamic Tests}

\begin{figure}[!htbp]
    \centering
    \includegraphics[width=0.49\textwidth]{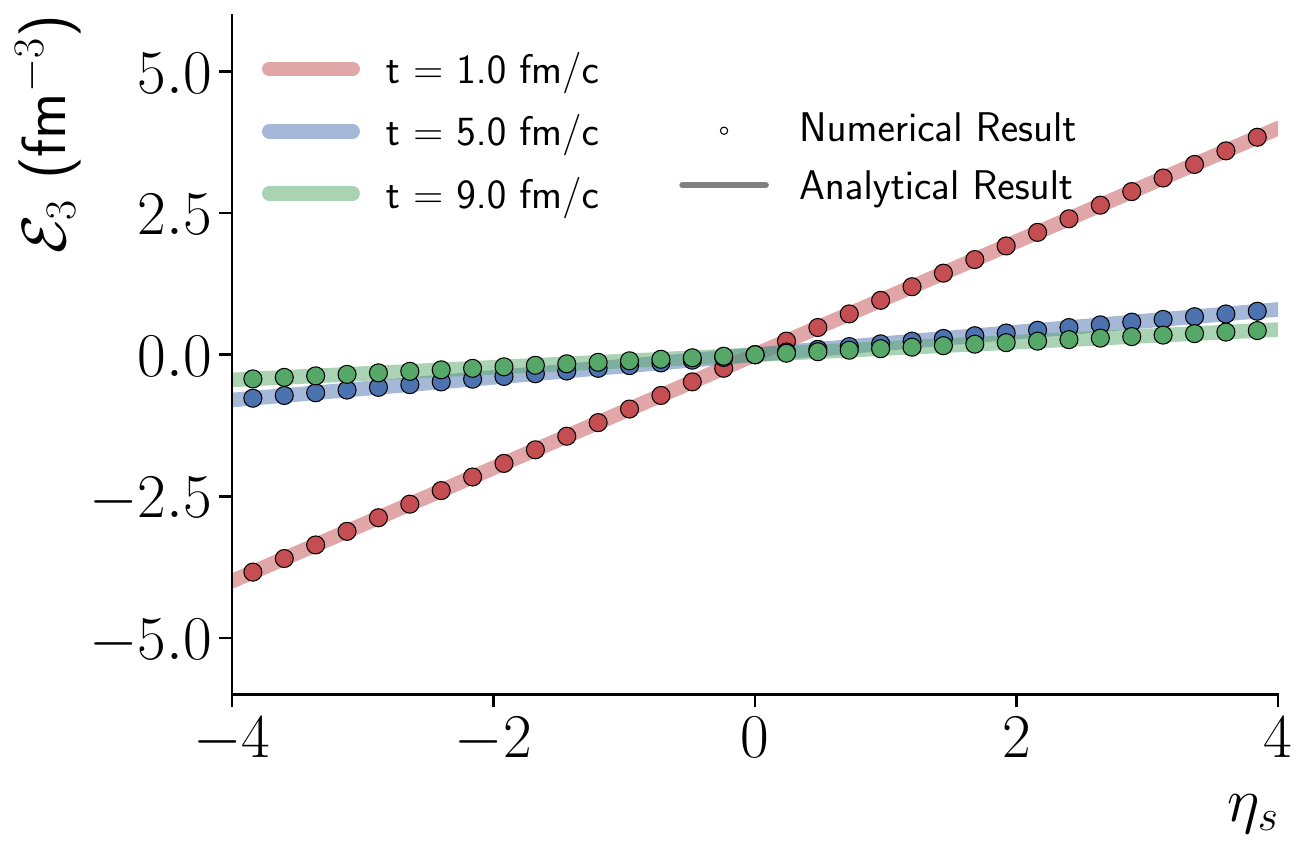}
    \includegraphics[width=0.49\textwidth]{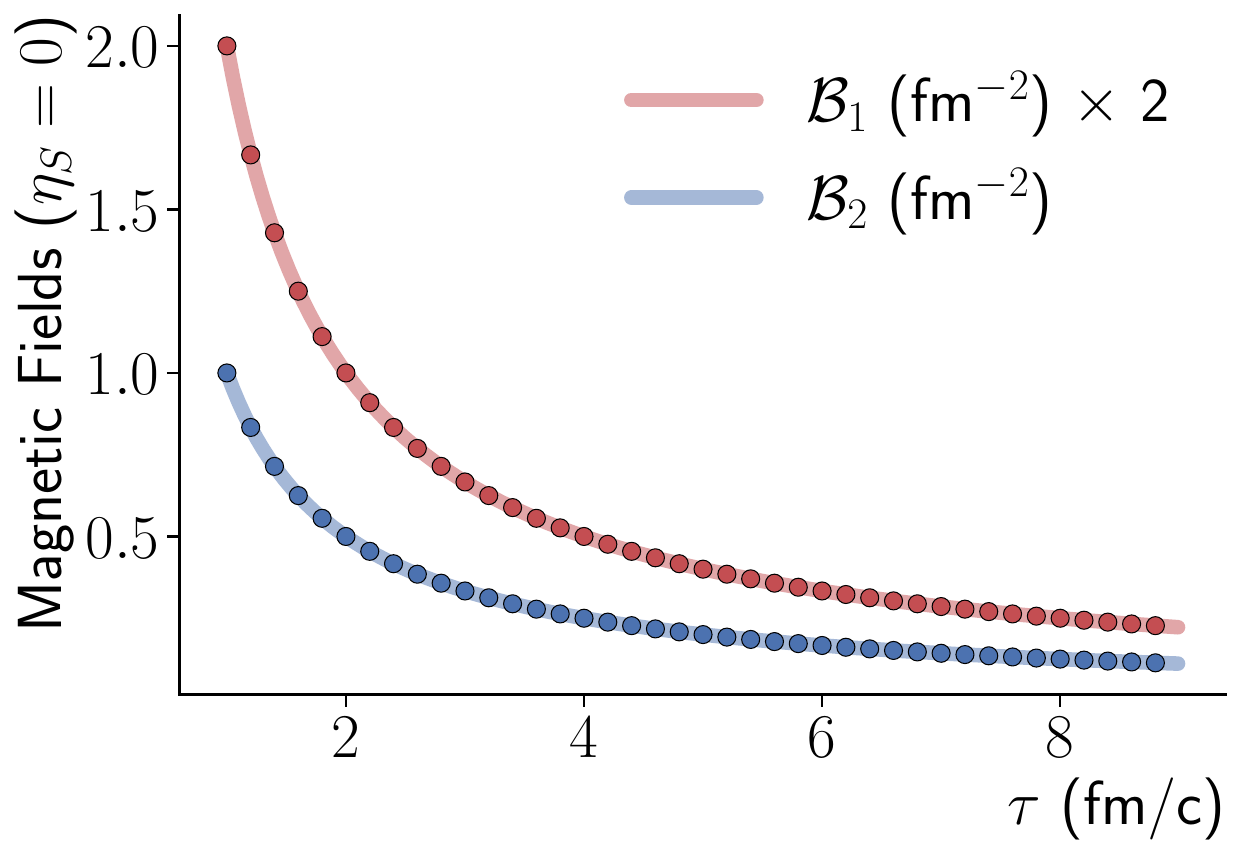}\\
    \includegraphics[width=0.49\textwidth]{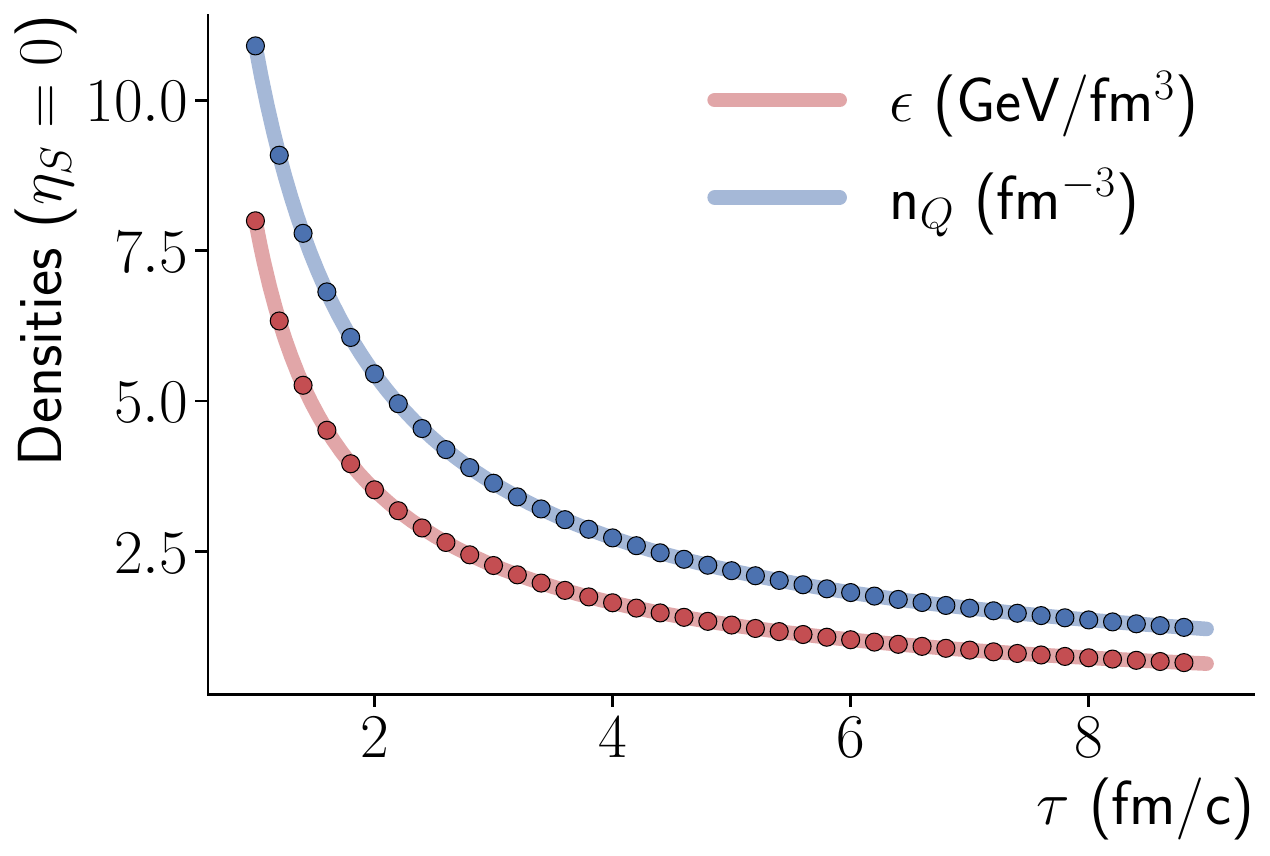}
    \includegraphics[width=0.49\textwidth]{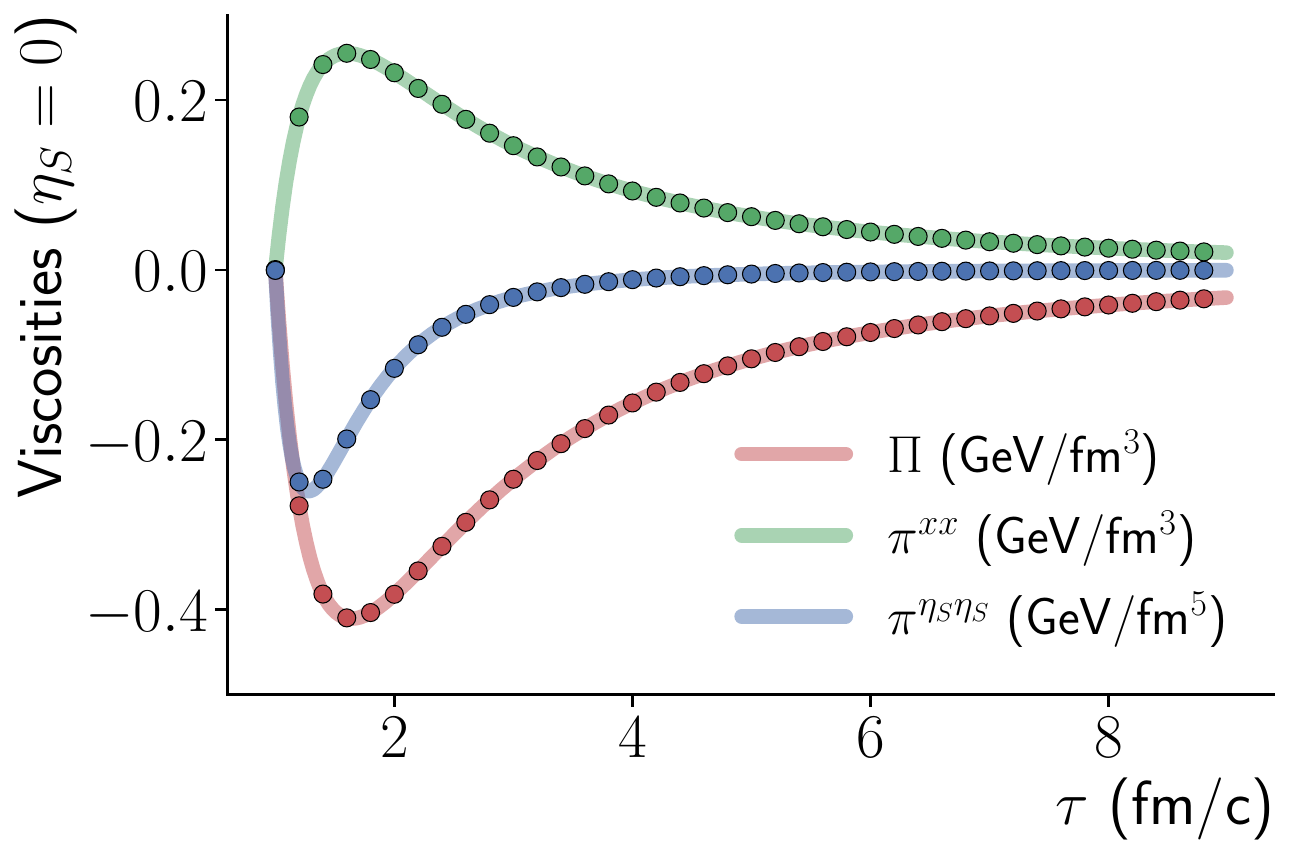}
    \caption{Time evolution of selected quantities using Bjorken flow background in the hydrodynamic and electromagnetic sectors using PRAIA. Taken from \cite{PRAIA}. $\mathcal E_3$, $\mathcal B_1$ and $\mathcal B_2$ are the electromagnetic fields.}
    \label{fig:hydrobjorken}
\end{figure}

\begin{figure}[!htbp]
    \centering
    \includegraphics[width=0.325\textwidth]{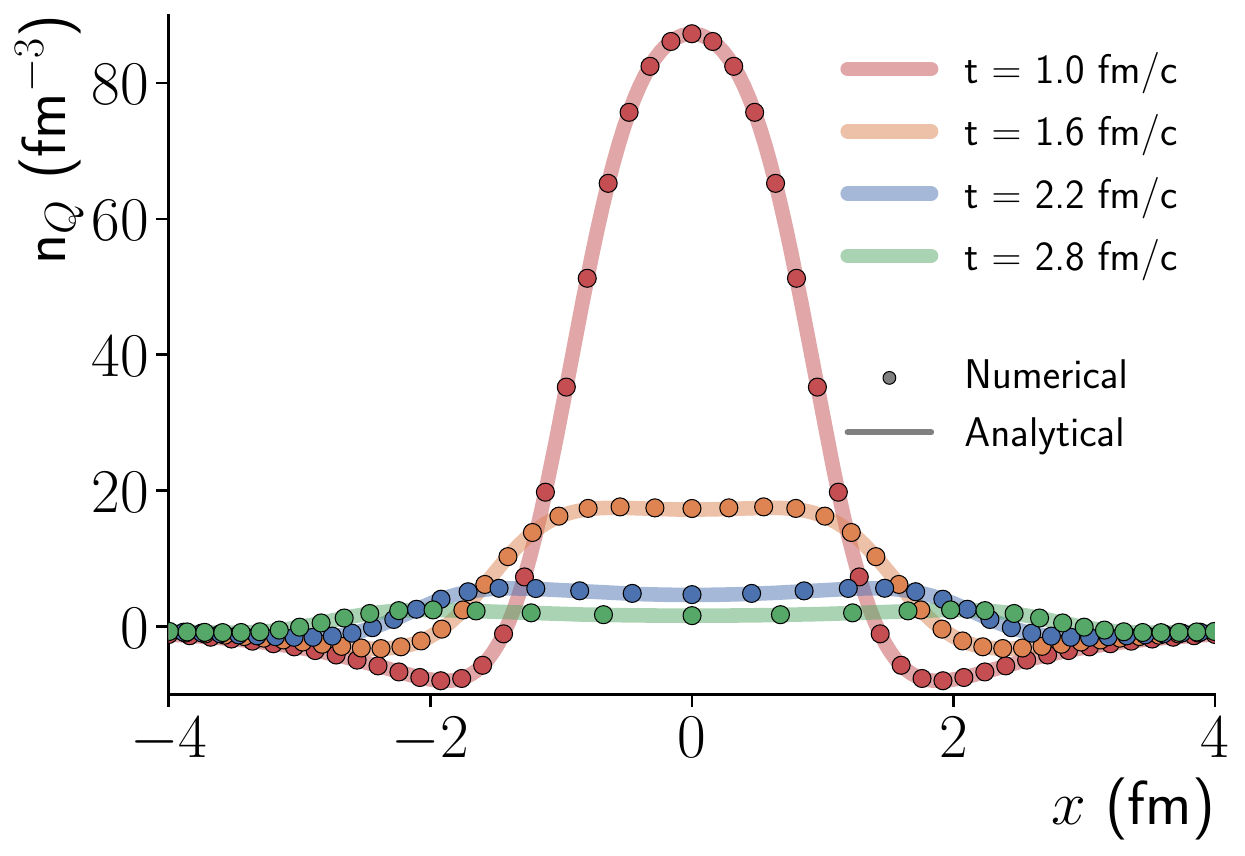}
    \includegraphics[width=0.325\textwidth]{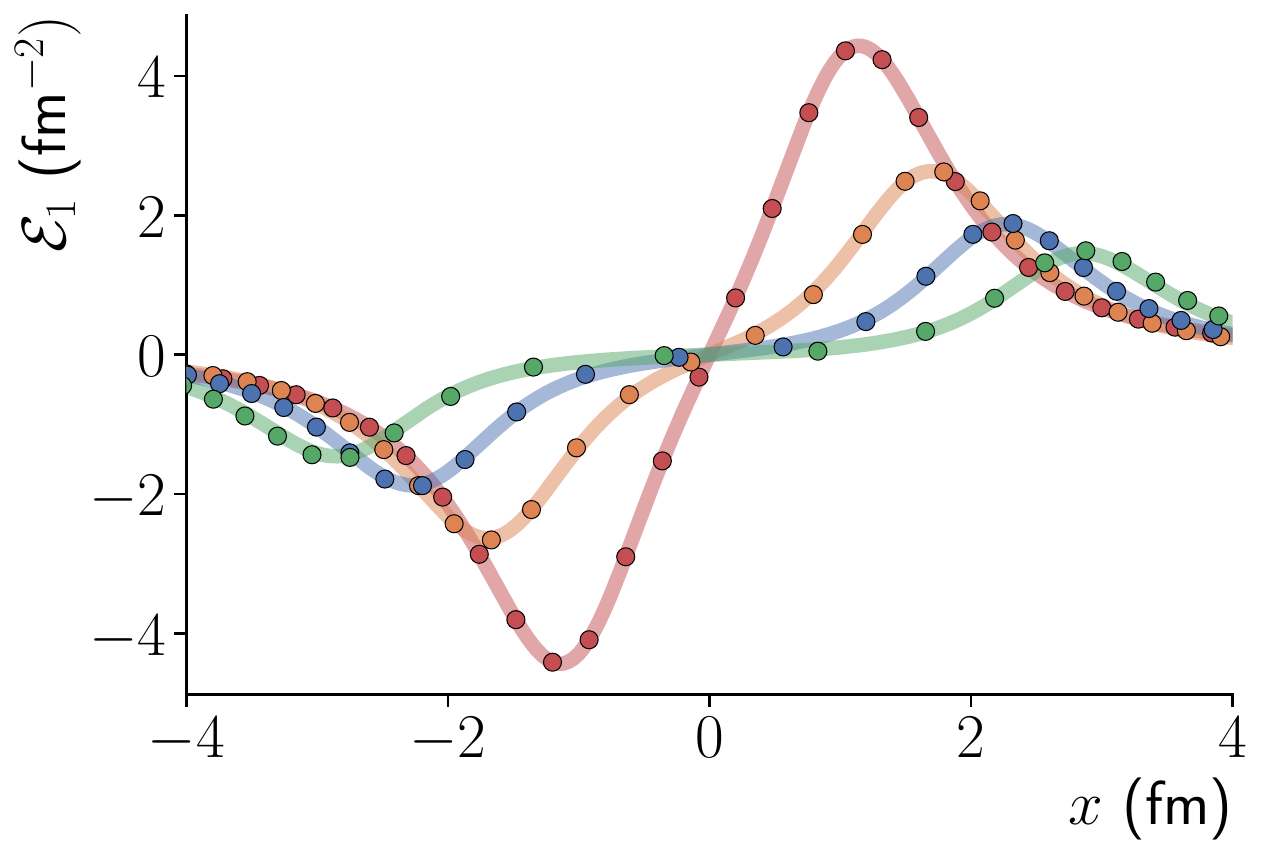}
    \includegraphics[width=0.325\textwidth]{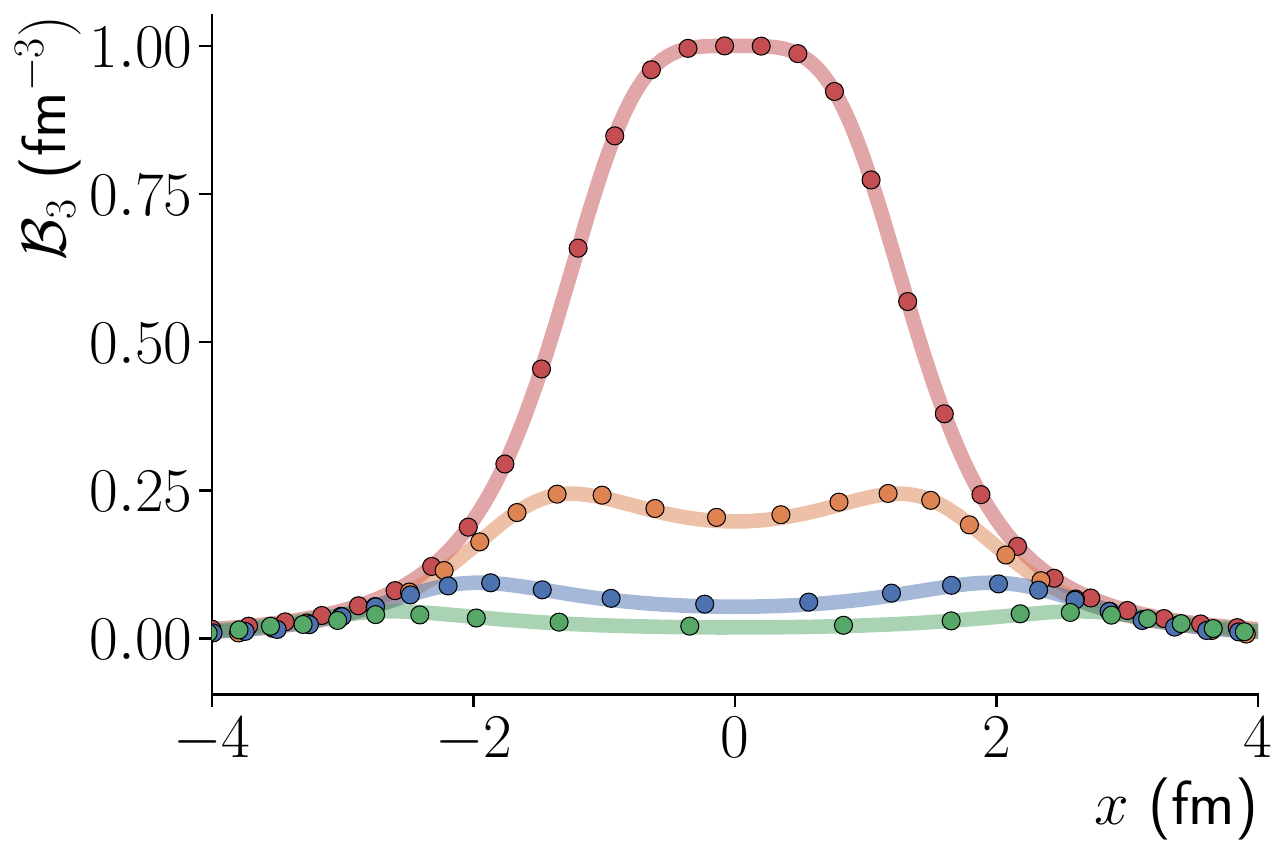}\\
    \includegraphics[width=0.325\textwidth]{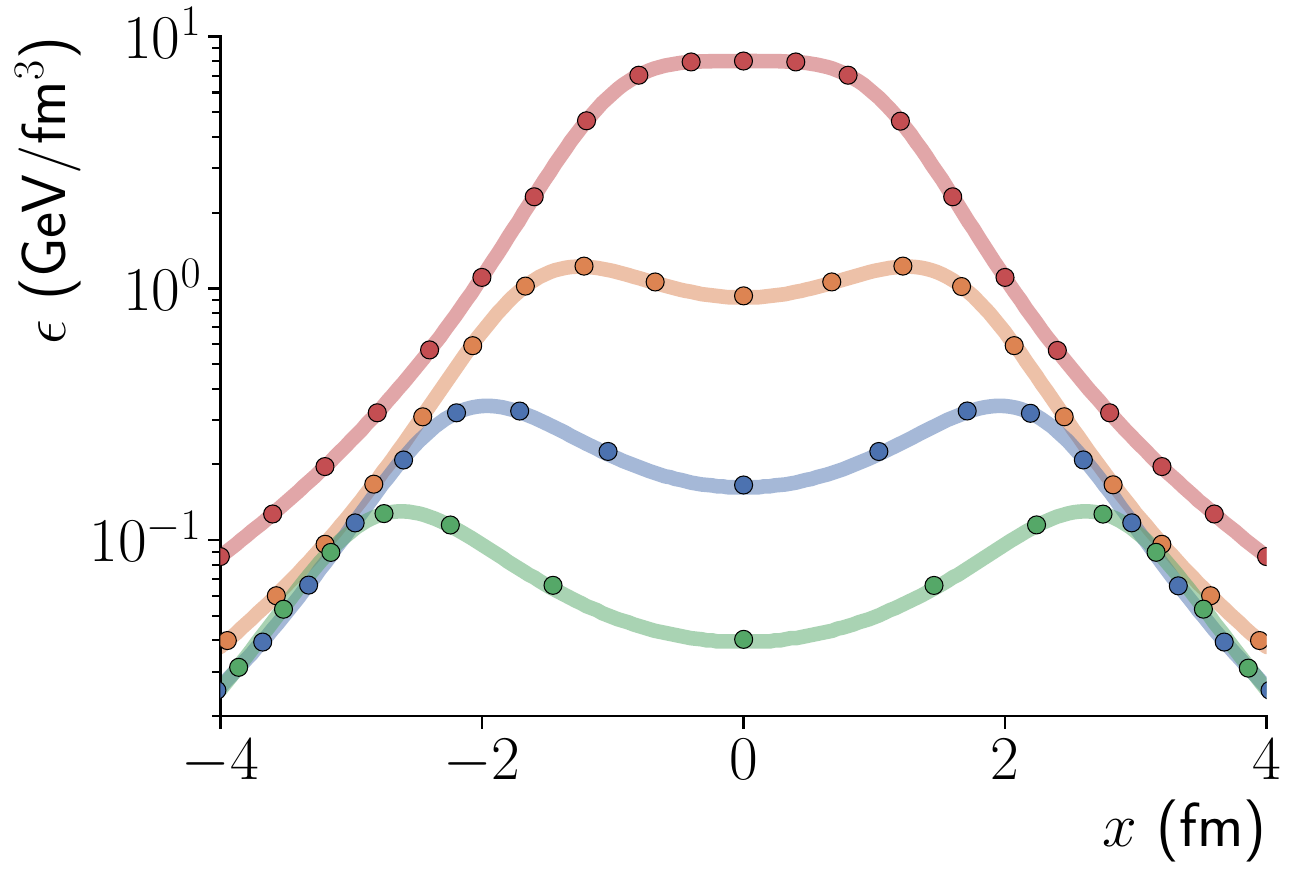}
    \includegraphics[width=0.325\textwidth]{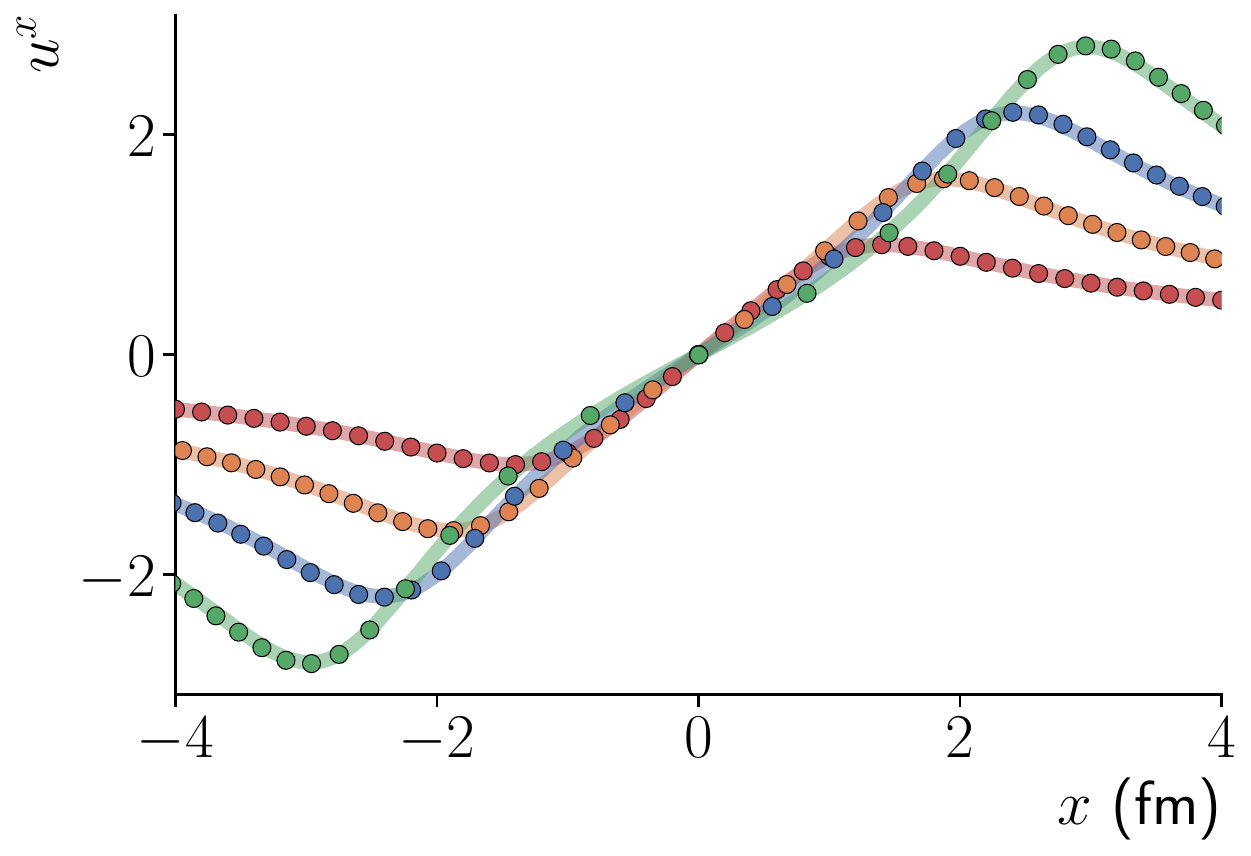}
    \includegraphics[width=0.325\textwidth]{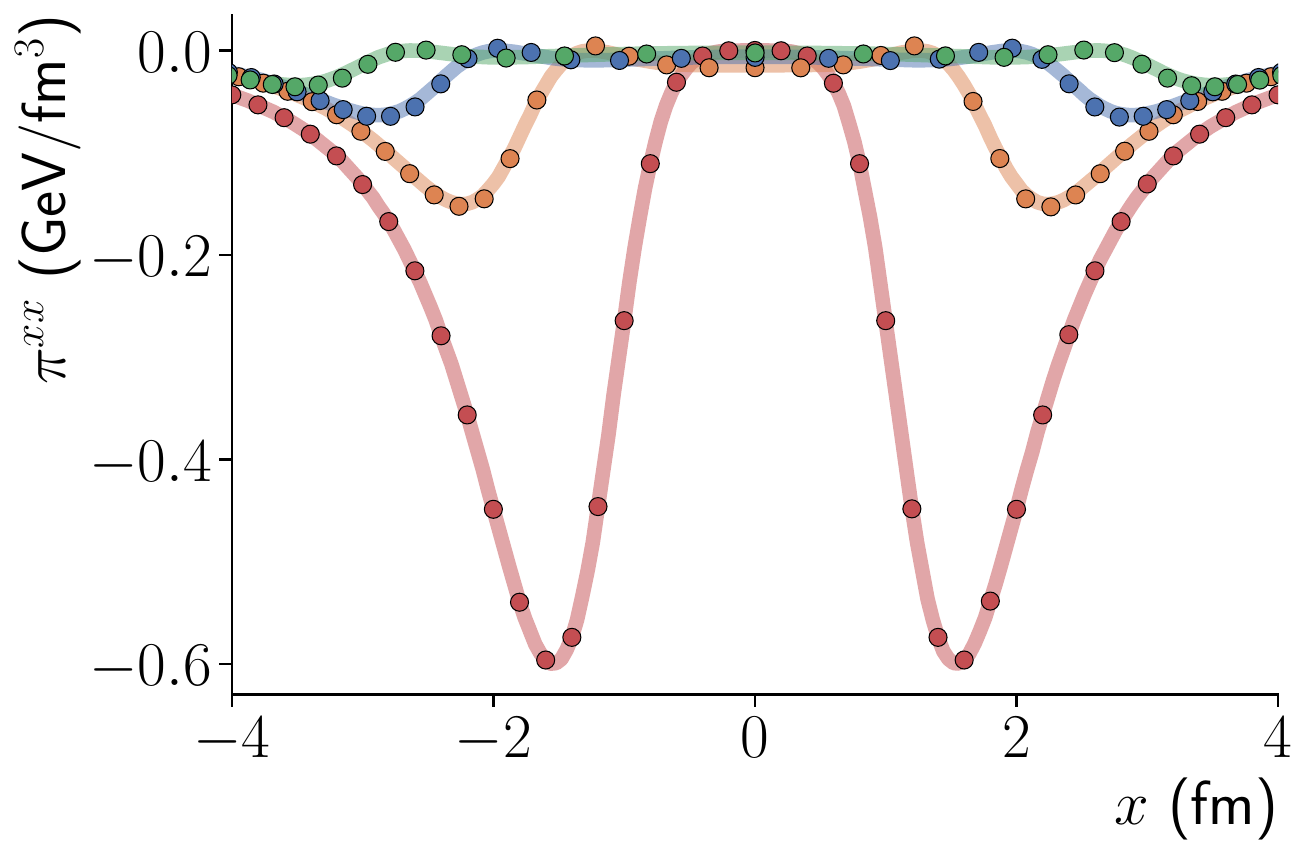}
    \caption{Time evolution of selected quantities on the Gubser-flow background in the hydrodynamic and electromagnetic sectors as functions of \(x\) using PRAIA. Taken from \cite{PRAIA}. $\mathcal E_1$ and $\mathcal B_3$ are the electromagnetic fields.}
    \label{fig:hydrogubser}
\end{figure}

\texttt{PRAIA} has been tested and validated by reproducing known analytical solutions of viscous hydrodynamics, such as the Bjorken model \cite{Bjorken1983,gubser2013complex} and the Gubser model \cite{gubser2010symmetry,PhysRevC.91.014903,v334-d32w}. More recently, we have developed analytical solutions for electromagnetic fields using this model \cite{PRAIA}, as seen in Figs. \ref{fig:hydrobjorken} and \ref{fig:hydrogubser}.

\subsection{Switching and Cooper-Frye Equations}

In the numerical hydrodynamic evolution, the SPH fluid particles do not necessarily reach the energy density $\epsilon_{\rm sw}$ exactly within a single time step. To determine the hydrodynamic fields precisely on the hypersurface defined by the constant switching energy density $\epsilon_{\rm sw}$, we linearly interpolate over the time interval during which the fluid cell cools below this value.

For a fluid represented by discrete SPH particles, the Cooper–Frye formula \eqref{eq:cff} in its discrete version takes the form
\begin{equation}
    p^0 \frac{d^3N}{d^3p} = \sum_{i} d \sigma \cdot p \, f(x,p)
    \label{eq:cooperfryediscrete}
\end{equation}
where $i$ is the index of the fluid particles that undergo switching. The computation of $d\sigma$ is explicit.
\begin{equation}
    d\sigma_\mu  = - \, \frac{\nu}{\sigma} \, \frac{\mathcal N _\mu}{|u\cdot \mathcal N|} = - \, \frac{\nu}{\sigma} \, \frac{N _\mu}{|u\cdot  N|}
\end{equation}
since the factor $\sqrt{N\cdot N}$ cancels out because $\mathcal N_\mu$ appears in both the numerator and the denominator. Using the expression in Eq. \eqref{eq:dsgimaA}, we use the approximation $d^3x \approx \nu/\sigma$ and introduce $ N^\mu = - \, \partial \epsilon_{\rm sw}/\partial x^\mu$. The minus sign appears because the fluid is in a cooling process. Since, in this case, the derivative during cooling is oriented toward the interior of the fluid, the minus sign compensates for this, making the resulting vector point outward. 

We can evaluate the spatial components of $N_\mu$ using the SPH method. For the temporal component, we start from
\begin{equation}
    D\epsilon = u^\mu \frac{\partial \epsilon}{\partial x^\mu} = u^0 \frac{\partial \epsilon}{\partial x^0} + u^i \frac{\partial \epsilon}{\partial x^i},
\end{equation}
which implies
\begin{equation}
    \frac{\partial \epsilon}{\partial x^0} = \frac{1}{u^0}\left( D\epsilon - u^i \frac{\partial \epsilon}{\partial x^i} \right),
\end{equation}
where $D\epsilon$ is known from the fluid dynamical equations.

\section{BQSSampler}

The \texttt{BQSSampler} is the module responsible for converting the switching hypersurface calculated in the previous subsection into a list of hadrons to be used later in the hadronic transport. From this information, the sampler generates particles event by event using a Monte Carlo procedure based on the Cooper–Frye prescription in \eqref{eq:cooperfryediscrete} with viscous corrections in Eq \eqref{eq:distviscouscorrections}.

The sampling procedure proceeds as follows. First, the code evaluates the expected number of particles, $\Delta N_{ij}$, associated with each hypersurface element $d\sigma_i$ and each particle species $j$, so that the resulting effective distribution is
\begin{align}
    &p^0 \frac{d^3N_{ij}}{d^3p} = \sigma_i \cdot p f_j(x,p),\\
    &\Delta N_{ij} = \int p_T dp_T dy d\phi \left(p^0 \frac{d^3N_{ij}}{d^3p}\right),
\end{align}
This contribution $\Delta N_{ij}$ is used as the parameter of a Poisson distribution,
\begin{equation}
    P(N_{ij}) = \exp(-\Delta N_{ij}) \frac{(\Delta N_{ij})^{N_{ij}}}{N_{ij}!}
\end{equation}
from which a sampling is performed via the Monte Carlo method to determine the effective number of times that a species $j$, originating from a hypersurface element $i$, will be produced. This selection is carried out by applying the Monte Carlo method again, now using the Cooper–Frye formula to determine the momenta $p$.

The position selected for emission coincides with that of hypersurface element $i$, plus a small spatial shift obtained by sampling from a continuous distribution with adjustable width. In this work, the value $0.01$ is adopted (in fm units for $(x,y)$). In this way, particles emitted from the same surface element do not reach \texttt{SMASH} with exactly identical positions, preventing numerical degeneracies in the collision criterion, in which pairs with zero spatial separation could be interpreted as immediate collisions at the initial instant.

In addition to the standard sampling mode, \texttt{BQSSampler} provides a specific structure for handling the conserved BQS charges, making it possible to control the net production of these charges at switching. In the configurations considered in this work, the sampler reads the hypersurface, performs Monte Carlo sampling with viscous corrections, and writes the final particle list in the \texttt{oscar.dat} format, which is then used as the initial condition for the hadronic evolution in \texttt{SMASH}.

\section{SMASH}

The final stage of the hybrid evolution is described with \texttt{SMASH} (Simulating Many Accelerated Strongly-interacting Hadrons), a microscopic hadronic transport approach. After the hydrodynamic fields are converted into particles on the particlization hypersurface, the system is no longer treated as a continuous fluid but as a dilute gas of interacting hadrons. 

\texttt{SMASH} solves the relativistic Boltzmann equation for the one-particle phase-space distribution of each hadron species,
\[
p^\mu \partial_\mu f_i(x,p) + m_i \frac{\partial\left[K^\mu f_i(x,p)\right]}{\partial p^\mu}
= C_i^{\rm coll},
\]
where \(f_i(x,p)\) is the distribution function of species \(i\), \(K^\mu\) denotes a possible mean-field force, and \(C_i^{\rm coll}\) is the collision term that may include the terms in Eq. \eqref{eq:collisionterms}. In the afterburner stage used in this work, the relevant dynamics are encoded in the collision and decay processes among hadrons. \texttt{SMASH} treats hadrons and resonances with vacuum properties, so that medium effects are not imposed through explicit in-medium spectral modifications but arise dynamically from the sequence of microscopic interactions.

The particle content is based on the hadronic states listed by the Particle Data Group, with resonance masses, widths, and decay channels implemented through vacuum spectral functions and branching ratios. The collision term includes elastic and inelastic reactions, resonance formation and decay, and conserves energy, momentum, baryon number, electric charge, and strangeness at each microscopic interaction.

In this work, \texttt{SMASH} is used as the hadronic afterburner following \texttt{BQSSampler}. For each hydrodynamic event, the sampled hadronic ensemble is propagated through \texttt{SMASH} with hadronic rescattering and decays. This stage is particularly important for femtoscopy because the two-pion correlation function is sensitive to the space-time coordinates at which the particles are effectively emitted.

\chapter{HBT Interferometry \textit{or} Femtoscopy}

Access to the geometrical structure of quantum systems is of fundamental importance for their physical characterization. In numerous physical systems, geometry emerges directly from the fundamental dynamics that control the system. Stellar objects, for instance, tend to acquire approximately spherical shapes as a consequence of the action of gravity on matter that undergoes nuclear fusion. The balance between gravitational attraction and internal pressure gradients provides a certain degree of stability to these systems, making it meaningful to assign macroscopic geometrical quantities to them, such as a stellar radius. A similar idea, although much more subtle and indirect, appears in the study of particles produced in high-energy nuclear collisions. The particles emitted at the end of the collision form a cloud with characteristic spatial and temporal structures. These structures are not arbitrary since they carry information about several stages of the collision process. Determining the geometry of the particle-emitting source provides a way to probe the space-time evolution of the matter created in nuclear collisions. 

Several observables carry information about the geometry and the collective dynamics of the system. However, since the system produced in such collisions is extremely small and short-lived, its space-time structure cannot be accessed through direct imaging. One way to investigate this structure is through Hanbury Brown–Twiss interferometry -- or femtoscopy --, which exploits quantum correlations to determine the space-time variances and covariances of the particle distribution produced in the collision, such as pions. In this work, femtoscopy serves as the primary method used to examine the space-time structure of the matter created in nuclear collisions and, thereby, to explore the QCD phase diagram. For this reason, an entire chapter is dedicated to this topic. The discussion starts with a historical overview of how HBT interferometry emerged and was later applied to the investigation of nuclear collisions. It then moves on to its formulation, in which femtoscopic observables are closely linked to the study of final-state particle distributions and to the search for indications of phase transitions in the QCD equation of state.

\section{A Historical Perspective}

\begin{figure}[!htbp]
    \centering
    \subfloat[]{
        \includegraphics[width=0.38\linewidth]{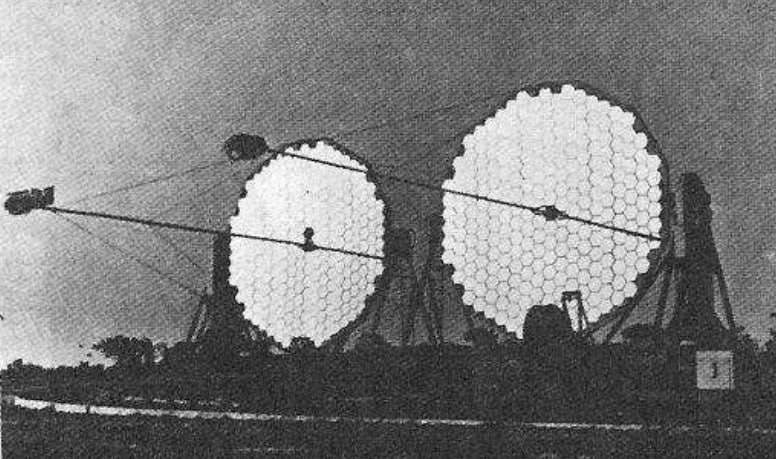}
    }
    \hfill
    \subfloat[]{
        \includegraphics[width=0.47\linewidth]{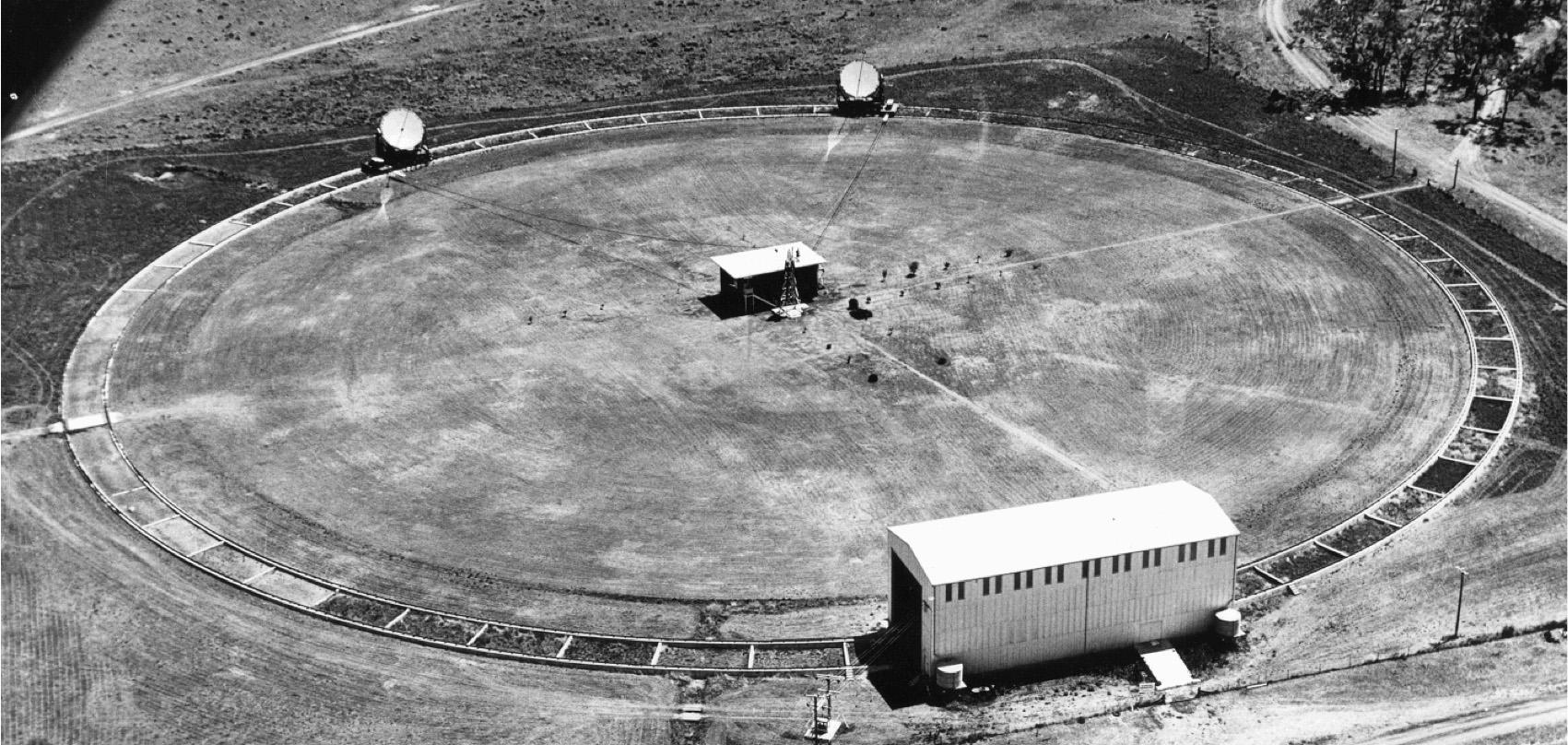}
    }
    \caption{
    Historical realization of the Hanbury Brown-Twiss intensity-interferometry method in astronomy. Panel (a) shows the two light collectors used in the Narrabri Stellar Intensity Interferometer, while panel (b) shows the circular track along which the collectors could be moved in order to vary the baseline between the detectors. 
    Instead of measuring optical amplitudes directly, the experiment measured correlations between intensity fluctuations, allowing the angular diameters of stars to be inferred from the loss of correlation with increasing detector separation. 
    Taken from Ref.~\cite{Padula2005HBT}.
    }
    \label{fig:narrabri_hbt}
\end{figure}

In the 1950s, Robert {\bf H}anbury {\bf B}rown and Richard Quentin {\bf T}wiss introduced a new method for measuring the angular diameter of stars, which later became known as {\it HBT interferometry} \cite{BrownTwiss1954} (Fig~\ref{fig:narrabri_hbt}). In contrast to conventional amplitude interferometry, their method was based on measuring correlations between intensity fluctuations recorded by two separate detectors. Since the strength of these correlations depends on the angular size of the emitting source, the technique provided an indirect way to extract stellar dimensions from the statistical properties of the detected radiation. The experimental apparatus was developed mainly by Hanbury Brown, whereas Twiss contributed to the theoretical and mathematical formulation of the method. After testing the principle in a laboratory experiment using blue light emitted by a mercury vapor lamp \cite{BrownTwiss1956Photon}, they applied the technique to the measurement of the angular diameter of Sirius \cite{BrownTwiss1956Sirius}.

Despite these successful measurements, the approach was at first met with skepticism by some members of the scientific community. The observation of correlations between signals detected at spatially separated points appeared counterintuitive when analyzed within a framework based on individual photons. Consequently, several objections were raised regarding its consistency with fundamental physical principles, particularly the conservation of energy \cite{SilvaFreire2013HBT}. This controversy was resolved with the development of a quantum interpretation of the effect. In this framework, the observed correlations arise from the quantum statistical properties of identical particles. In particular, photons, being bosons, exhibit an enhanced probability of joint detection, a phenomenon now associated with Bose-Einstein correlations. This interpretation provided the theoretical foundation required for the technique to become widely accepted and established its relevance beyond its original astronomical application \cite{Padula2005HBT}.

A closely related phenomenon was independently identified in high-energy physics by Gerson {\bf G}oldhaber, Sulamith {\bf G}oldhaber, Wonyong {\bf L}ee, and Abraham {\bf P}ais at the Lawrence Radiation Laboratory in Berkeley. Studying pion production, they observed nontrivial angular correlations between pairs of identical pions. They interpreted these correlations as arising from the same quantum mechanism as the Hanbury Brown-Twiss effect, that is, the bosonic nature of the emitted particles. In particle physics, this approach became known as the GGLP method \cite{GoldhaberGoldhaberLeePais1960}. In the same experimental program, the correlations in the opposite-charge pion channel were important to identifying the $\rho^0$ resonance through its decay mode $\rho^0 \to \pi^+\pi^-$.

The use of two-particle correlation techniques in high-energy collisions emerged independently of their original applications in astronomy. Since then, HBT interferometry has become a tool in the study of multiparticle production. In relativistic heavy-ion collisions, it is used to infer the space-time dimensions of the particle-emitting source formed after the expansion and cooling of strongly interacting matter \cite{LisaPrattSoltzWiedemann2005}. In this context, the technique is commonly referred to as {\it femtoscopy}, emphasizing its ability to probe length and time scales on the order of femtometers. Modern femtoscopic measurements probe not only the geometry of the particle-emitting source but also collective flow, resonance decay contributions, hadronic final-state interactions, and potential signals of phase-transition phenomena in the QCD equation of state.

\section{Physics behind the HBT Interferometry}

HBT interferometry is a method for deducing the space-time structure of a particle-emitting source from correlations in the momenta of particle pairs \cite{LisaPrattSoltzWiedemann2005,WiedemannHeinz1999}. These correlations originate from quantum statistics and are a direct consequence of the indistinguishability of identical particles. Specifically, the many-body wave function must be symmetric under the exchange of identical bosons and antisymmetric under the exchange of identical fermions \cite{GottfriedYan2003}. As a result, the two-particle probability distribution acquires interference contributions that raise the probability of detecting identical bosons with almost the same momenta, while reducing the corresponding probability for identical fermions.
 
We begin by explaining how the interference emerges at the level of two-particle amplitudes and then reformulate the result within a many-body framework appropriate for relativistic particle production. The link between this formal description and the experimentally accessible observables in heavy-ion femtoscopy will be elaborated in the subsequent sections.

\subsection{Two-particle Interference}

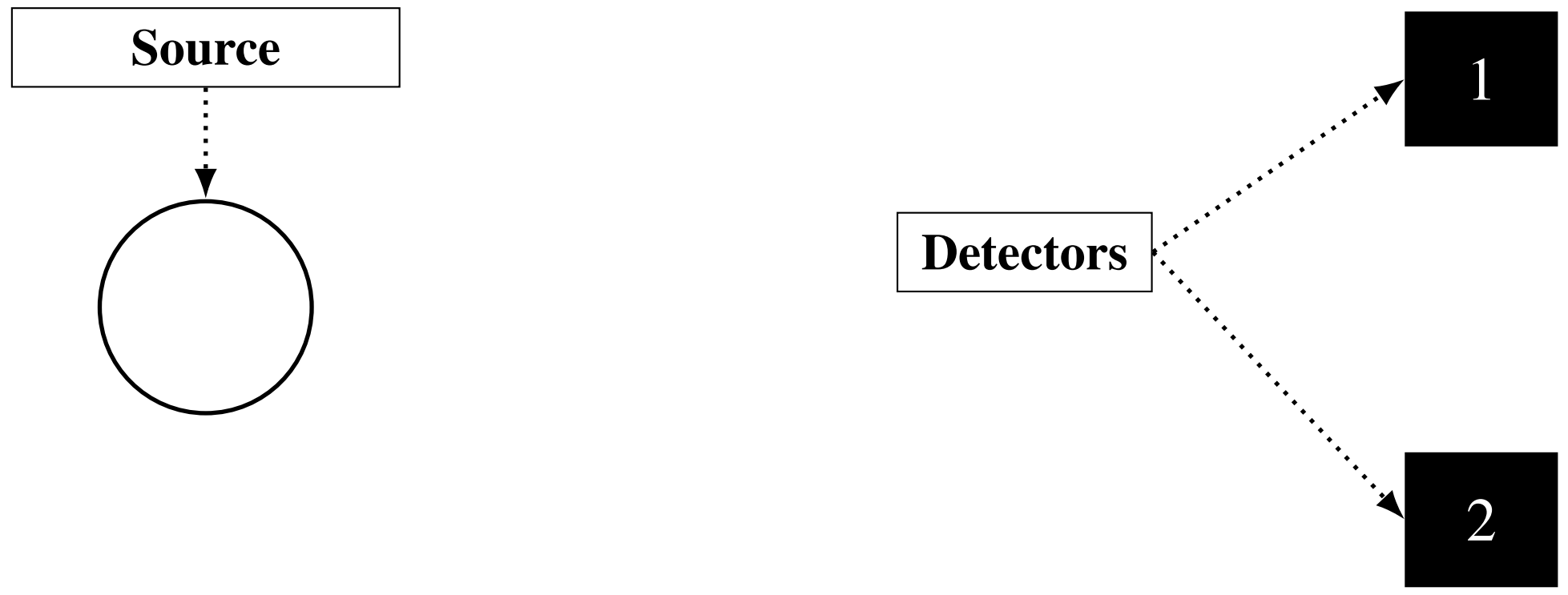
\begin{figure}[!htbp]
\centering
\begin{tikzpicture}[scale=1.3]

\node[labelbox] (sourcebox) at (0,2.2) {Source};
\node[sourcecircle] (source) at (0,0.55) {};

\draw[dottedarrow] (sourcebox.south) -- (source.north);

\node[labelbox, minimum width=2.1cm] (detlabel) at (5.2,0.9) {Detectors};

\node[detector] (d1) at (8.1,2.0) {1};
\node[detector] (d2) at (8.1,-0.8) {2};

\draw[dottedarrow] (detlabel.east) -- (d1.west);
\draw[dottedarrow] (detlabel.east) -- (d2.west);

\end{tikzpicture}
\caption{Schematic representation of an emitting source and two detectors.}
\label{fig:hbt_scheme}
\end{figure}

To illustrate the origin of this effect, consider a particle-emitting source and two detectors, as represented schematically in Fig.~\ref{fig:hbt_scheme}. Let \(x_i=(t_i,\vec{x}_i)\) denote the space-time point at which the particle \(i\) is emitted from the source, and let \(p_i=(p_i^0,\vec{p}_i)\) denote its asymptotic momentum measured by the detector. For a single particle, the probability amplitude associated with the emission of the source and the detection with momentum \(p_i\) may be denoted by \(\psi_i(x_i,p_i)\). The corresponding one-particle probability distribution is
\begin{equation}
    \mathcal{P}_{1i}(x_i,p_i)=|\psi_i(x_i,p_i)|^2 .
\end{equation}

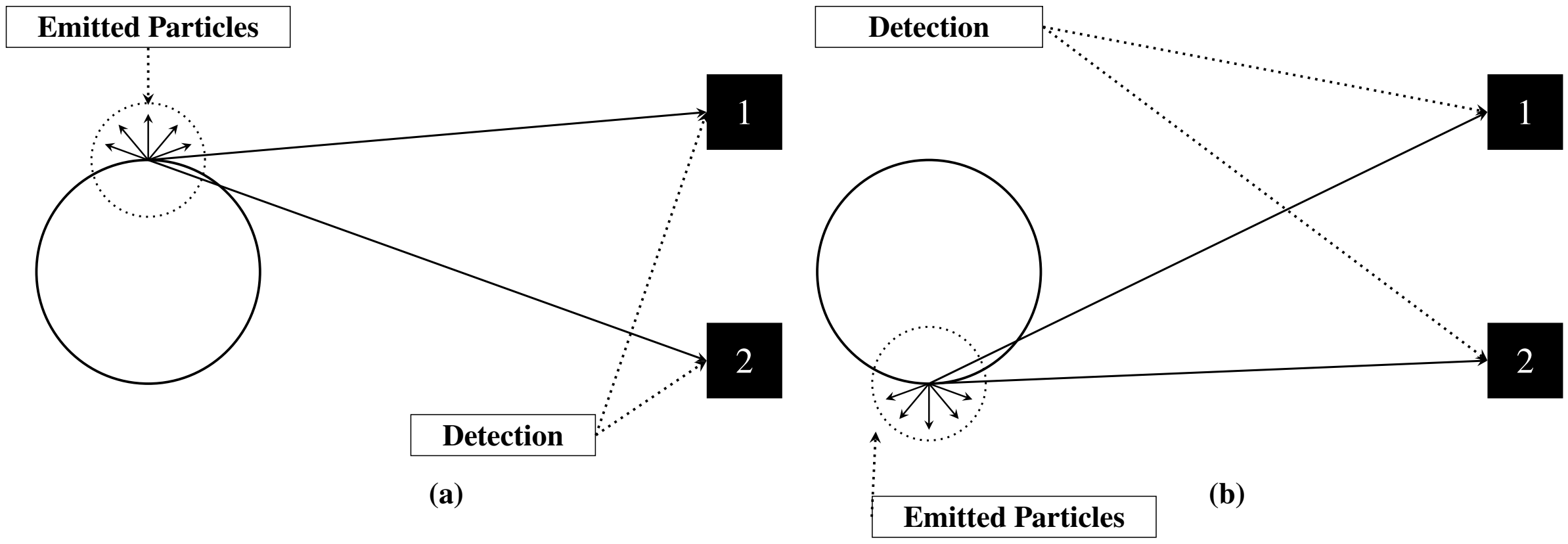
\begin{figure}[!htbp]
\centering
\resizebox{0.98\textwidth}{!}{%
\begin{tikzpicture}[
    labelbox/.style={
        draw,
        rectangle,
        minimum height=0.55cm,
        align=center,
        font=\bfseries
    },
    detector/.style={
        draw,
        rectangle,
        fill=black,
        minimum width=1.05cm,
        minimum height=1.05cm,
        text=white,
        font=\large
    },
    source/.style={
        draw,
        circle,
        minimum size=3.15cm,
        line width=1.0pt
    },
    solidarrow/.style={
        line width=0.8pt,
        ->,
        >=stealth
    },
    dottedarrow/.style={
        dotted,
        line width=1.0pt,
        ->,
        >=stealth
    },
    smallarrow/.style={
        line width=0.65pt,
        ->,
        >=stealth
    }
]


\begin{scope}[xshift=0cm]

\node[source] (sourceA) at (0,0) {};
\coordinate (emitA) at (0,1.575);

\node[labelbox, minimum width=4.0cm] (emitAbox) at (0,3.45) {Emitted Particles};
\draw[dottedarrow] (emitAbox.south) -- (0,2.35);

\draw[dotted, line width=0.8pt] (emitA) circle (0.80);

\draw[smallarrow] (emitA) -- ++(20:0.65);
\draw[smallarrow] (emitA) -- ++(50:0.65);
\draw[smallarrow] (emitA) -- ++(90:0.65);
\draw[smallarrow] (emitA) -- ++(130:0.65);
\draw[smallarrow] (emitA) -- ++(160:0.65);

\node[detector] (d1A) at (8.4,2.25) {1};
\node[detector] (d2A) at (8.4,-1.25) {2};

\draw[solidarrow] (emitA) -- (d1A.west);
\draw[solidarrow] (emitA) -- (d2A.west);

\node[labelbox, minimum width=2.6cm] (detAbox) at (5.0,-2.30) {Detection};

\draw[dottedarrow] (detAbox.east) -- (d1A.west);
\draw[dottedarrow] (detAbox.east) -- (d2A.west);

\node[font=\bfseries] at (4.2,-3.15) {(a)};

\end{scope}


\begin{scope}[xshift=11.0cm]

\node[source] (sourceB) at (0,0) {};
\coordinate (emitB) at (0,-1.575);

\node[labelbox, minimum width=3.2cm] (detBbox) at (0,3.45) {Detection};

\draw[dotted, line width=0.8pt] (emitB) circle (0.80);

\draw[smallarrow] (emitB) -- ++(-20:0.65);
\draw[smallarrow] (emitB) -- ++(-50:0.65);
\draw[smallarrow] (emitB) -- ++(-90:0.65);
\draw[smallarrow] (emitB) -- ++(-130:0.65);
\draw[smallarrow] (emitB) -- ++(-160:0.65);

\node[labelbox, minimum width=4.0cm] (emitBbox) at (1.2,-3.45) {Emitted Particles};
\draw[dottedarrow] (emitBbox.west) -- (-0.75,-2.25);

\node[detector] (d1B) at (8.4,2.25) {1};
\node[detector] (d2B) at (8.4,-1.25) {2};

\draw[solidarrow] (emitB) -- (d1B.west);
\draw[solidarrow] (emitB) -- (d2B.west);

\draw[dottedarrow] (detBbox.east) -- (d1B.west);
\draw[dottedarrow] (detBbox.east) -- (d2B.west);

\node[font=\bfseries] at (4.2,-3.15) {(b)};

\end{scope}

\end{tikzpicture}%
}
\caption{Schematic illustration of the two indistinguishable assignments contributing to the two-particle detection amplitude. Panel (a) emphasizes one assignment of emission points to detectors, while panel (b) represents the exchanged assignment.}
\label{fig:hbt_two_amplitudes}
\end{figure}

For a pair of identical particles, the two alternatives in which the particles are emitted from two points and detected with momenta \(p_1\) and \(p_2\) are physically indistinguishable. Therefore, the two-particle amplitude must be symmetrized for bosons and antisymmetrized for fermions. In a schematic first-quantized description, this amplitude can be written as
\begin{equation}
    \Psi(x_1,x_2,p_1,p_2)
    =
    \frac{
    \psi_1(x_1,p_1)\psi_2(x_2,p_2)
    \pm
    \psi_1(x_2,p_2)\psi_2(x_1,p_1)
    }{\sqrt{2}},
\end{equation}
where the plus sign applies to identical bosons and the minus sign applies to identical fermions. The probability of detecting the pair is, therefore,
\begin{equation}
    \mathcal{P}_2(x_1,p_1,x_2,p_2)
    =
    |\Psi(x_1,x_2,p_1,p_2)|^2 ,
\end{equation}
as we can see in Fig. \ref{fig:hbt_two_amplitudes}.

\begin{figure}[!htbp]
\centering
\resizebox{0.95\textwidth}{!}{%
\begin{tikzpicture}[
    detector/.style={
        draw,
        rectangle,
        fill=black,
        minimum width=1.45cm,
        minimum height=1.20cm
    },
    source/.style={
        draw,
        circle,
        minimum size=3.15cm,
        line width=1.0pt
    },
    solidarrow/.style={
        line width=0.85pt,
        ->,
        >=stealth
    },
    smallarrow/.style={
        line width=0.65pt,
        ->,
        >=stealth
    }
]

\node[source] (source) at (0,0) {};

\coordinate (emitA) at (0,1.575);
\coordinate (emitB) at (0,-1.575);

\draw[smallarrow] (emitA) -- ++(20:0.65);
\draw[smallarrow] (emitA) -- ++(50:0.65);
\draw[smallarrow] (emitA) -- ++(90:0.65);
\draw[smallarrow] (emitA) -- ++(130:0.65);
\draw[smallarrow] (emitA) -- ++(160:0.65);

\draw[smallarrow] (emitB) -- ++(-20:0.65);
\draw[smallarrow] (emitB) -- ++(-50:0.65);
\draw[smallarrow] (emitB) -- ++(-90:0.65);
\draw[smallarrow] (emitB) -- ++(-130:0.65);
\draw[smallarrow] (emitB) -- ++(-160:0.65);

\node[detector] (d1) at (10.0,2.15) {};
\node[detector] (d2) at (10.0,-2.15) {};

\draw[solidarrow] (emitA) -- (d1.west);
\draw[solidarrow] (emitA) -- (d2.west);

\draw[solidarrow] (emitB) -- (d1.west);
\draw[solidarrow] (emitB) -- (d2.west);

\end{tikzpicture}%
}
\caption{Schematic representation of two emission points on the source and the possible paths connecting them to two detectors.}
    \label{fig:hbt_pair_detection}
\end{figure}
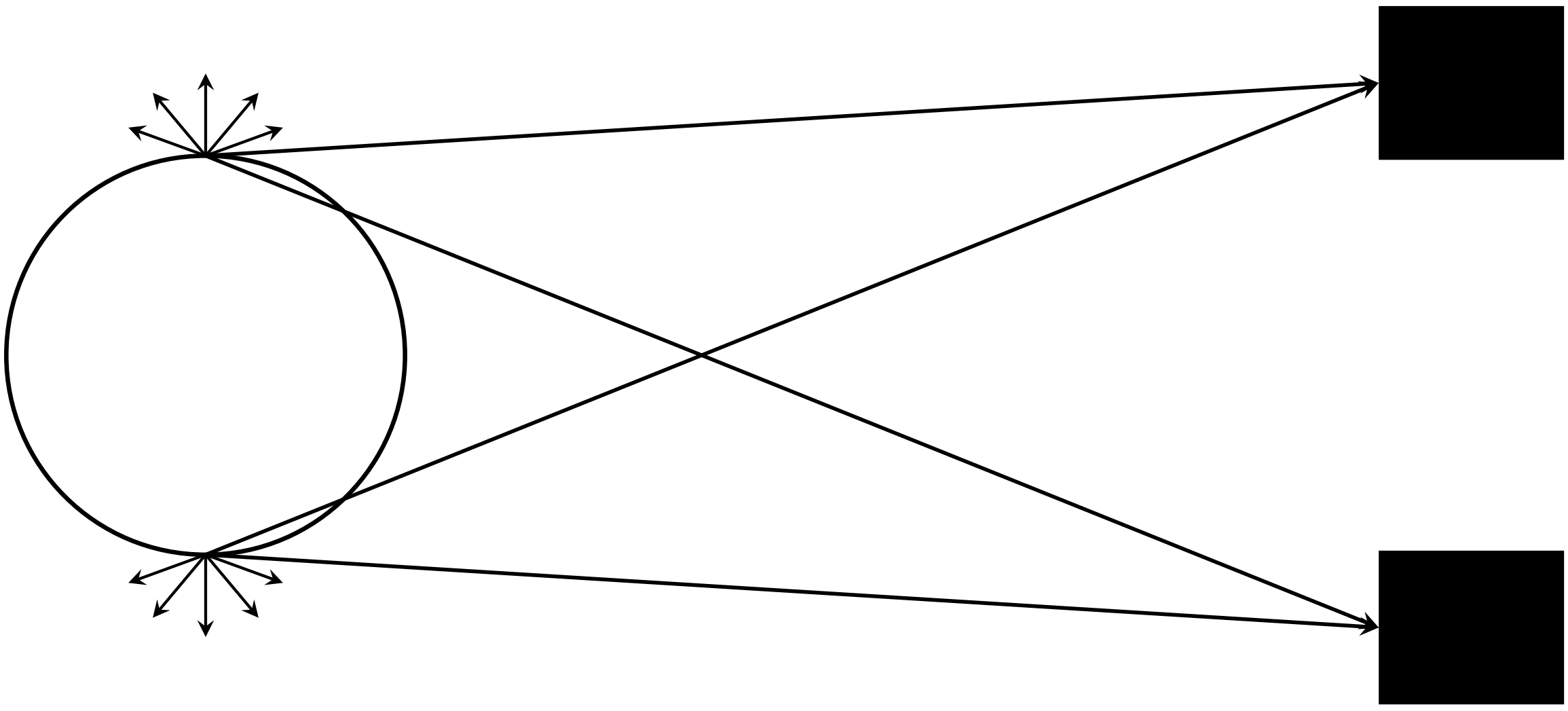

The interference term generated by the modulus squared of the symmetrized or antisymmetrized amplitude is the basic quantum-statistical origin of the HBT correlation. This is represented schematically in Fig.~\ref{fig:hbt_pair_detection}.

Since experimentally one measures the final momenta of the emitted particles, it is useful to integrate the emission points. The single-particle momentum distribution is then written as
\begin{equation}
    \mathcal{P}_{1i}(p_i)
    =
    \int d x_i \,
    \mathcal{P}_{1i}(x_i,p_i),
\end{equation}
while the two-particle momentum distribution is
\begin{equation}
    \mathcal{P}_2(p_1,p_2)
    =
    \int d x_1 \int d x_2 \,
    \mathcal{P}_2(x_1,p_1,x_2,p_2).
\end{equation}
The two-particle distribution can be written in a generic form.
\begin{equation}
    \mathcal{P}_2(p_1,p_2)
    =
    \mathcal{P}_{11}(p_1)
    \mathcal{P}_{12}(p_2)
    \left[1\pm\mathcal{R}(p_1,p_2)\right],
\end{equation}
where \(\mathcal{R}(p_1,p_2)\) encodes the correlation generated by the exchange symmetry of the amplitude of the two-particles. If \(\mathcal{R}(p_1,p_2)=0\), the two-particle distribution is factored into the product of the single-particle distributions, corresponding to the absence of correlations in this channel. The two-particle correlation function is, therefore, defined as
\begin{equation}
    \mathcal{C}_2(p_1,p_2)
    \equiv
    \frac{
    \mathcal{P}_2(p_1,p_2)
    }{
    \mathcal{P}_{11}(p_1)
    \mathcal{P}_{12}(p_2)
    }
    =
    1\pm\mathcal{R}(p_1,p_2).
    \label{eq:cf}
\end{equation}
For identical bosons, the plus sign leads to an enhancement of the correlation function, while for identical fermions, the minus sign leads to a suppression if $\mathcal R(p_1, p_2)$ is positive definite. This  construction is useful for understanding the physical origin of HBT interferometry, but the system produced in high-energy nuclear collisions is a relativistic many-body system, and its description requires a more general formalism. In what follows, the correlation function will therefore be reformulated in terms of statistical operators as well as creation and annihilation operators. This provides a more appropriate framework for connecting the quantum-statistical origin of the effect with the invariant particle distributions measured experimentally.

\subsection{Many-Body Formulation}

The particle system can be described by a statistical operator \(\hat\rho\). From it, one may construct the non-normalized reduced one-particle and two-particle statistical operators, denoted, respectively, by \(\hat\rho^{(1)}\) and \(\hat\rho^{(2)}\) \cite{GottfriedYan2003,MartinRothen2004}. Their matrix elements in momentum space are defined as
\begin{equation}
    \mel{p_1 r}{\hat \rho^{(1)}}{p_2 s}
    =
    \Tr\left\{
    \hat a^\dagger_{p_2 s}
    \hat a_{p_1 r}
    \hat \rho
    \right\}
    =
    \left\langle
    \hat a^\dagger_{p_2 s}
    \hat a_{p_1 r}
    \right\rangle ,
\end{equation}
and
\begin{equation}
\begin{split}
    \mel{p_1 r_1,p_2 r_2}{\hat \rho^{(2)}}{p_3 s_1,p_4 s_2}
    =
    \Tr\left\{
    \hat a^\dagger_{p_3 s_1}
    \hat a^\dagger_{p_4 s_2}
    \hat a_{p_2 r_2}
    \hat a_{p_1 r_1}
    \hat \rho
    \right\}
    =
    \left\langle
    \hat a^\dagger_{p_3 s_1}
    \hat a^\dagger_{p_4 s_2}
    \hat a_{p_2 r_2}
    \hat a_{p_1 r_1}
    \right\rangle .
\end{split}
\end{equation}
Here, \(\hat a^\dagger_{ps}\) and \(\hat a_{ps}\) are creation and annihilation operators associated with momentum \(p\) and additional quantum numbers \(s\), while \(\Tr\) denotes a trace over the full many-body Hilbert space. The total particle-number operator is defined as
\begin{equation}
    \hat N
    =
    \sum_{p,s}
    \hat a^\dagger_{ps}
    \hat a_{ps}.
\end{equation}
The traces of the reduced statistical operators satisfy
\begin{equation}
    \Tr_{(1)}\hat \rho^{(1)}
    =
    \sum_{p,s}
    \left\langle
    \hat a^\dagger_{ps}
    \hat a_{ps}
    \right\rangle
    =
    \langle \hat N \rangle ,
    \label{eq:partnumb}
\end{equation}
and
\begin{equation}
    \Tr_{(2)}\hat \rho^{(2)}
    =
    \langle \hat N^2-\hat N\rangle ,
\end{equation}
where \(\Tr_{(i)}\) denotes the trace over the \(i\)-particle subspace. Since these traces are not equal to unity, they must be included when defining normalized probability distributions. From this point on, the dependence on quantum numbers other than momentum will be omitted in order to simplify the notation. The one-particle probability distribution in momentum space is given by the diagonal matrix element of the one-particle reduced statistical operator,
\begin{equation}
    \mathcal P_1(p)
    =
    \frac{1}{\Tr_{(1)}\hat \rho^{(1)}}
    \mel{p}{\hat \rho^{(1)}}{p}
    =
    \frac{1}{\langle \hat N\rangle}
    \left\langle
    \hat a^\dagger_p
    \hat a_p
    \right\rangle .
    \label{eq:onepartdist}
\end{equation}
Similarly, the two-particle probability distribution in momentum space is obtained from the diagonal matrix element of the two-particle reduced statistical operator,
\begin{equation}
    \mathcal P_2(p_1,p_2)
    =
    \frac{1}{\Tr_{(2)}\hat \rho^{(2)}}
    \mel{p_1p_2}{\hat \rho^{(2)}}{p_1p_2}
    =
    \frac{1}{\langle \hat N^2-\hat N\rangle}
    \left\langle
    \hat a^\dagger_{p_1}
    \hat a^\dagger_{p_2}
    \hat a_{p_2}
    \hat a_{p_1}
    \right\rangle .
\end{equation}
The two-particle correlation function is then defined as
\begin{equation}
\begin{split}
    \mathcal C(p_1,p_2)
    &\equiv
    \frac{
    \mathcal P_2(p_1,p_2)
    }{
    \mathcal P_1(p_1)\mathcal P_1(p_2)
    }  =
    \frac{
    \langle \hat N\rangle^2
    }{
    \langle \hat N^2-\hat N\rangle
    }
    \frac{
    \left\langle
    \hat a^\dagger_{p_1}
    \hat a^\dagger_{p_2}
    \hat a_{p_2}
    \hat a_{p_1}
    \right\rangle
    }{
    \left\langle
    \hat a^\dagger_{p_1}
    \hat a_{p_1}
    \right\rangle
    \left\langle
    \hat a^\dagger_{p_2}
    \hat a_{p_2}
    \right\rangle
    } .
\end{split}
\end{equation}
For high-multiplicity systems, and neglecting multiplicity-fluctuation corrections, the prefactor
\(\langle \hat N\rangle^2/\langle \hat N^2-\hat N\rangle\) is often approximated by unity. In this approximation,
\begin{equation}
\begin{split}
    \mathcal C(p_1,p_2)
    \approx
    \frac{
    \left\langle
    \hat a^\dagger_{p_1}
    \hat a^\dagger_{p_2}
    \hat a_{p_2}
    \hat a_{p_1}
    \right\rangle
    }{
    \left\langle
    \hat a^\dagger_{p_1}
    \hat a_{p_1}
    \right\rangle
    \left\langle
    \hat a^\dagger_{p_2}
    \hat a_{p_2}
    \right\rangle
    } .
\end{split}
\end{equation}
For a chaotic source whose statistical operator is Gaussian in the creation and annihilation operators, the four-point function can be factorized by Wick's theorem \cite{BlaizotRipka1986}. This gives
\begin{equation}
\begin{split}
    &
    \left\langle
    \hat a^\dagger_{p_1}
    \hat a^\dagger_{p_2}
    \hat a_{p_2}
    \hat a_{p_1}
    \right\rangle
    =
    \left\langle
    \hat a^\dagger_{p_1}
    \hat a_{p_1}
    \right\rangle
    \left\langle
    \hat a^\dagger_{p_2}
    \hat a_{p_2}
    \right\rangle 
    \pm
    \left\langle
    \hat a^\dagger_{p_1}
    \hat a_{p_2}
    \right\rangle
    \left\langle
    \hat a^\dagger_{p_2}
    \hat a_{p_1}
    \right\rangle
    +
    \left\langle
    \hat a^\dagger_{p_1}
    \hat a^\dagger_{p_2}
    \right\rangle
    \left\langle
    \hat a_{p_2}
    \hat a_{p_1}
    \right\rangle .
\end{split}
\label{eq:wicktheo}
\end{equation}
The plus sign corresponds to bosons, while the minus sign corresponds to fermions. The last term contains anomalous expectation values. In the absence of anomalous pair correlations, namely when
\begin{equation}
    \left\langle
    \hat a^\dagger_{p_1}
    \hat a^\dagger_{p_2}
    \right\rangle
    =
    \left\langle
    \hat a_{p_2}
    \hat a_{p_1}
    \right\rangle
    =
    0,
\end{equation}
the correlation function reduces to
\begin{equation}
    \mathcal C(p_1,p_2)
    \approx
    1
    \pm
    \frac{
    \left\langle
    \hat a^\dagger_{p_1}
    \hat a_{p_2}
    \right\rangle
    \left\langle
    \hat a^\dagger_{p_2}
    \hat a_{p_1}
    \right\rangle
    }{
    \left\langle
    \hat a^\dagger_{p_1}
    \hat a_{p_1}
    \right\rangle
    \left\langle
    \hat a^\dagger_{p_2}
    \hat a_{p_2}
    \right\rangle
    } .
\end{equation}
Comparing this expression with Eq.~\eqref{eq:cf}, one identifies
\begin{equation}
    \mathcal R(p_1,p_2)
    \approx
    \frac{
    \left\langle
    \hat a^\dagger_{p_1}
    \hat a_{p_2}
    \right\rangle
    \left\langle
    \hat a^\dagger_{p_2}
    \hat a_{p_1}
    \right\rangle
    }{
    \left\langle
    \hat a^\dagger_{p_1}
    \hat a_{p_1}
    \right\rangle
    \left\langle
    \hat a^\dagger_{p_2}
    \hat a_{p_2}
    \right\rangle
    } .
\end{equation}
This result shows how the correlation term introduced phenomenologically in the first-quantized discussion emerges naturally from the many-body density-matrix formalism.

\subsection{Relativistic Invariant Distributions}

The previous discussion was formulated in terms of discrete momentum states. We now move to a continuous momentum-space description, which is more appropriate for relativistic particle distributions. In the continuum limit, the sums over momentum states and the corresponding creation and annihilation operators are replaced according to \cite{BurgessQFTNotes}
\begin{equation}
\begin{split}
    &\sum_p \longrightarrow \frac{V}{(2\pi)^3}\int  d^3 p,\\
    &\hat a_p \longrightarrow \hat a_p\sqrt{\frac{(2\pi)^3}{V}},
\end{split}
\end{equation}
where \(V\) is the normalization volume. With this convention, the expectation value of the total number of particles, defined in Eq.~\eqref{eq:partnumb}, becomes
\begin{equation}
    N
    =
    \langle \hat N\rangle
    =
    \int  d^3p\,
    \left\langle
    \hat a^\dagger_p
    \hat a_p
    \right\rangle .
\end{equation}
Therefore, the single-particle momentum distribution can be identified as
\begin{equation}
    \frac{ d^3N}{ d^3p}
    =
    \left\langle
    \hat a^\dagger_p
    \hat a_p
    \right\rangle .
    \label{eq:onepartdista}
\end{equation}
Comparing this expression with Eq.~\eqref{eq:onepartdist}, one sees that \( d^3N/ d^3p\) is proportional to the probability distribution of a single particle \(\mathcal P_1(p)\), with the proportionality factor given by the mean multiplicity. Analogously, the two-particle momentum distribution is defined as
\begin{equation}
    \frac{ d^6N}{ d^3p_1 d^3p_2}
    =
    \left\langle
    \hat a^\dagger_{p_1}
    \hat a^\dagger_{p_2}
    \hat a_{p_2}
    \hat a_{p_1}
    \right\rangle ,
    \label{eq:twopartdist}
\end{equation}
which is proportional to the probability distribution of particle pair \(\mathcal P_2(p_1,p_2)\).
The quantities introduced above are not written in a manifestly Lorentz-invariant form. Since \( d N\) is a scalar and \( d^3p/p^0\) is the Lorentz invariant momentum-space measure, the combination $p^0 d^3N / d^3p$ is Lorentz invariant, as we saw in the previous chapters. Similarly, for a pair of particles, the product of the two invariant measures implies that $p_1^0 p_2^0 d^6N /  d^3p_1 d^3p_2$ is also Lorentz invariant. We therefore define
\begin{equation}
    \mathcal{N}_1(p)
    \equiv
    p^0
    \frac{ d^3N}{ d^3p},
    \qquad
    \mathcal{N}_2(p_1,p_2)
    \equiv
    p_1^0p_2^0
    \frac{ d^6N}{ d^3p_1 d^3p_2}.
    \label{eq:invdist}
\end{equation}
Using the normalized probability distributions introduced previously, the two-particle correlation function may be written as
\begin{equation}
    \mathcal C(p_1,p_2)
    =
    \frac{\mathcal P_2(p_1,p_2)}
    {\mathcal P_1(p_1)\mathcal P_1(p_2)}
    =
    \frac{\langle\hat N\rangle^2}
    {\langle \hat N^2-\hat N\rangle}
    \frac{\mathcal{N}_2(p_1,p_2)}
    {\mathcal{N}_1(p_1)\mathcal{N}_1(p_2)} .
\end{equation}
Since \(\langle\hat N\rangle\), \(\langle \hat N^2\rangle\), \(\mathcal{N}_1\), and \(\mathcal{N}_2\) are Lorentz invariant quantities, the correlation function defined in this way is also Lorentz invariant. For high-multiplicity systems, and neglecting multiplicity-fluctuation corrections, one may again approximate $\langle \hat N^2-\hat N\rangle   \simeq \langle \hat N\rangle^2 $. In this approximation, the correlation function can be defined directly in terms of invariant distributions as
\begin{equation}
    C(p_1,p_2)
    \equiv
    \frac{
    \mathcal{N}_2(p_1,p_2)
    }{
    \mathcal{N}_1(p_1)\mathcal{N}_1(p_2)
    } .
    \label{eq:invcf}
\end{equation}

\subsection{Emission Function}

Now, let us define the Lorentz-invariant creation and annihilation operators \cite{Schwartz2014QFT},
\begin{equation}
    \hat \alpha^\dagger_p
    =
    \sqrt{p^0}\,\hat a^\dagger_p,
    \qquad
    \hat \alpha_p
    =
    \sqrt{p^0}\,\hat a_p .
\end{equation}
Using Eqs.~\eqref{eq:onepartdista}, \eqref{eq:twopartdist}, and \eqref{eq:invdist}, the invariant single-particle and two-particle distributions can be written as
\begin{equation}
    \mathcal{N}_1(p)
    =
    \left\langle
    \hat \alpha^\dagger_p
    \hat \alpha_p
    \right\rangle ,
    \qquad
    \mathcal{N}_2(p_1,p_2)
    =
    \left\langle
    \hat \alpha^\dagger_{p_1}
    \hat \alpha^\dagger_{p_2}
    \hat \alpha_{p_2}
    \hat \alpha_{p_1}
    \right\rangle .
    \label{eq:invdistdef}
\end{equation}
The off-diagonal expectation values of the creation and annihilation operators can be represented in terms of a Wigner function, or emission function, \(S(x,p)\) \cite{ChapmanHeinz1994,CurtrightFairlieZachos2014}. In this representation,
\begin{equation}
    \left\langle
    \hat \alpha^\dagger_{p_1}
    \hat \alpha_{p_2}
    \right\rangle
    =
    \int  d^4x\,
    S\left(x,\frac{p_1+p_2}{2}\right)
    \exp\left[i x\cdot(p_1-p_2)\right],
    \label{aadaggerdefinition}
\end{equation}
where \(i\) is the imaginary unit. In particular, for \(p_1=p_2=p\), one obtains the following
\begin{equation}
    \left\langle
    \hat \alpha^\dagger_p
    \hat \alpha_p
    \right\rangle
    =
    \int  d^4x\,S(x,p)
    =
    \mathcal{N}_1(p).
    \label{eq:wgnfunc} 
\end{equation}
The function \(S(x,p)\) describes the phase-space distribution of the source that emits particles with momentum \(p\) from the space-time point \(x\). It is convenient to introduce the mean and relative four-momenta of the pair
\begin{equation}
    k
    =
    \frac{p_1+p_2}{2},
    \qquad
    q
    =
    p_1-p_2.
    \label{eq:dynvar}
\end{equation}
The variable \(q\) measures the momentum difference between the two-particles, while \(k\) characterizes the momentum scale of the pair. For identical bosons emitted from a chaotic source, the quantum-statistical correlation is enhanced when the relative momentum is small, since the two final states become harder to distinguish. As \(|q|\) increases, this enhancement is reduced. However, in realistic systems, final-state interactions, resonance decays, collective flow, and possible non-femtoscopic correlations can modify this behavior \cite{KisielFlorkowskiBroniowskiPluta2006}. A simple example of an emission function is provided by the Cooper-Frye prescription. In this case, one may write
\begin{equation}
S(x,p)
=
\int_{\Sigma}
 d \sigma(x') \cdot p
f(x',p)
H(x-x'),
\label{eq:cfef}
\end{equation}
where \(H(x-x')\) is a function that determines the spatial distribution. If this kernel is normalized according to
\begin{equation}
\int  d^4x  H(x-x') = 1,
\label{eq:cfefdelta}
\end{equation}
then the invariant one-particle distribution obtained from the emission function reduces to the Cooper--Frye formula:
\begin{equation}
\begin{split}
    \int  d^4x S(x,p)
    &=
    \int_{\Sigma}
     d \sigma(x')\cdot p
    f(x',p)
    \int  d^4x H(x-x')\\
    &=
    \int_{\Sigma}
     d \sigma(x')\cdot p
    f(x',p)
    =
    \mathcal{N}_1(p).
\end{split}
\end{equation}
The usual localized Cooper-Frye emission is recovered by choosing
\begin{equation}
H(x-x') = \delta^{(4)}(x-x'),
\end{equation}
where $\delta^{(4)}$ is the four-dimensional Dirac delta function. More generally, one may replace the delta function with a smooth, normalized kernel, for example, a Gaussian profile. Thus, the kernel $H$ controls the space-time localization of the emission while preserving the Cooper-Frye spectrum, as long as it is properly normalized. By construction,
\begin{equation}
    S(x,p) = p^0\frac{d^7N}{d^3p\,d^4x}
\end{equation}
Up to this point, our derivation has been carried out in terms of Cartesian coordinates. For a generalized coordinate system, it suffices to take the invariant measure $d^4x \to \sqrt{-g}\, d^4x$ and use it in the functions where needed (such as in the delta). 

The definition in Eq.~\eqref{aadaggerdefinition} implies that
\begin{equation}
    \left\langle
    \hat \alpha^\dagger_{p_1}
    \hat \alpha_{p_2}
    \right\rangle
    =
    \left\langle
    \hat \alpha^\dagger_{p_2}
    \hat \alpha_{p_1}
    \right\rangle^*,
    \end{equation}
Combining this identity with Eqs.~\eqref{eq:invcf} and \eqref{eq:invdistdef}, and applying Wick's theorem as in Eq.~\eqref{eq:wicktheo}, we obtain the two-particle correlation function:
\begin{equation}
    C(p_1,p_2)
    =
    1
    \pm
    \frac{
    \left|
    \int  d^4x\,
    S(x,k)
    \exp(iq\cdot x)
    \right|^2
    }{
    \mathcal{N}_1(p_1)
    \mathcal{N}_1(p_2)
    } .
    \label{eq:corrfuncx}
\end{equation}
It is useful to define an effective normalized source distribution,
\begin{equation}
    \rho_{\rm eff}(x,p_1,p_2)
    =
    \frac{
    S(x,k)
    }{
    \sqrt{
    \mathcal{N}_1(p_1)
    \mathcal{N}_1(p_2)
    }
    } .
\end{equation}
The correlation function can then be written as
\begin{equation}
\begin{split}
    C(p_1,p_2)
    &=
    1
    \pm
    \left|
    \int  d^4x\,
    \rho_{\rm eff}(x,p_1,p_2)
    \exp(iq\cdot x)
    \right|^2 =
    1
    \pm
    \left|
    \tilde{\rho}_{\rm eff}(q;p_1,p_2)
    \right|^2 ,
\end{split}
\label{eq:corfunclas}
\end{equation}
where
\begin{equation}
    \tilde{\rho}_{\rm eff}(q;p_1,p_2)
    \equiv
    \int  d^4x\,
    \rho_{\rm eff}(x,p_1,p_2)
    \exp(iq\cdot x)
\end{equation}
is the Fourier transform of the effective source distribution. The notation keeps the dependence on \(p_1\) and \(p_2\), including the normalization in the denominator of \(\rho_{\rm eff}\).

\section{Femtoscopy in Heavy-Ion Collisions}
\label{sec:hbt_heavy_ion_collisions}

In relativistic heavy-ion collisions, the particle-emitting source is formed at the final stage of a complex dynamical evolution. Since this system is extremely small and short-lived, its space-time structure cannot be accessed through direct imaging. Interferometry provides indirect access through two-particle correlations. Their connection with the space-time distribution of emission points will be developed, leading to the so called HBT Radii observables.

Pion interferometry is the standard channel for this purpose. Pions are the most abundantly produced hadrons in relativistic heavy-ion collisions and, being spin-zero bosons, exhibit Bose-Einstein statistics for identical pairs. This makes identical pion correlations experimentally accessible and particularly suitable for femtoscopic analyses. Proton and baryon correlations can be useful, particularly at lower collision energies where baryon stopping plays an increasing role. However, in this study, we chose pion interferometry.

For the QCD phase diagram exploration, pion HBT does not probe baryon density directly, but it constrains how the equation of state and the dynamical evolution of the system shape the final emitting source. A softening of the equation of state, changes in the collective expansion rate, or modifications of the emission time distribution may leave traces in the two-pion correlation function. In this sense, pion interferometry provides information complementary to yields, spectra, anisotropic flow, and fluctuation observables. These quantities also carry information about the geometry and collective dynamics of the system, while femtoscopy gives more direct constraints on the space-time structure of the final state.

Using the Koonin–Pratt formalism, the relationship between the measured two-particle correlation function and the space–time distribution of emission points can be stated more directly, providing a natural link between experimentally extracted pair correlations and their interpretation in terms of femtoscopic source properties.

\subsection{\normalsize Koonin-Pratt Formalism}
For small relative momentum $q$, the dependence of the emission function in Eq.~\eqref{eq:corrfuncx} on \(q\) is typically much weaker than that of the phase factor \(\exp(iq\cdot x)\). This motivates the so-called smoothness approximation \cite{PrattCsorgoZimanyi1990,Pratt1997Smoothness}, in which one assumes
\begin{equation}
    S\left(x,k+\frac{q}{2}\right)
    S\left(y,k-\frac{q}{2}\right)
    \approx
    S(x,k)S(y,k).
    \label{eq:smoothapprox}
\end{equation}
Within this approximation, Eq.~\eqref{eq:corrfuncx} becomes
\begin{equation}
    C(k,q)
    =
    1
    \pm
    \frac{
    \left|
    \int  d^4x\,
    S(x,k)
    \exp(iq\cdot x)
    \right|^2
    }{
    \left(
    \int  d^4x\,S(x,k)
    \right)^2
    } .
    \label{eq:sacf}
\end{equation}
In this case $k \approx p_1 \approx p_2$, so $k\cdot k \approx m^2$ and
\begin{equation}
    k^0
    \approx
    E_k
    \equiv
    \sqrt{m^2+\vec{k}^{\,2}},
   \label{eq:onshellapp}
\end{equation}
where \(m\) is the mass of the emitted particles. This approximation is called on-shell. The modulus squared appearing in Eq.~\eqref{eq:sacf} can be rewritten as
\begin{equation}
\begin{split}
    \left|
    \int  d^4x\,
    S(x,k)
    \exp(iq\cdot x)
    \right|^2    &=
    \left[
    \int  d^4x\,
    S(x,k)
    \cos(q\cdot x)
    \right]^2
    +
    \left[
    \int  d^4x\,
    S(x,k)
    \sin(q\cdot x)
    \right]^2
    \\
    &=
    \int  d^4x
    \int  d^4y\,
    S(x,k)S(y,k)
    \cos\left[q\cdot(x-y)\right].
\end{split}
\end{equation}
Therefore,
\begin{equation}
    C(k,q)
    =
    1
    \pm
    \frac{
    \int  d^4x
    \int  d^4y\,
    S(x,k)S(y,k)
    \cos\left[q\cdot(x-y)\right]
    }{
    \left(
    \int  d^4x\,S(x,k)
    \right)^2
    } .
    \label{eq:sacf_cos}
\end{equation}
Equivalently, using the smoothness approximation in reverse to recover the dependence on \(p_1=k+q/2\) and \(p_2=k-q/2\) in the denominator of the correlation function, this expression may be written in the form
\begin{equation}
    C(k,q)
    \approx
    \frac{
    \int  d^4x
    \int  d^4y\,
    S\left(x,k+\frac{q}{2}\right)
    S\left(y,k-\frac{q}{2}\right)
    \left\{
    1
    \pm
    \cos\left[q\cdot(x-y)\right]
    \right\}
    }{
    \int  d^4x
    \int  d^4y\,
    S\left(x,k+\frac{q}{2}\right)
    S\left(y,k-\frac{q}{2}\right)
    } .
    \label{eq:sacf_cos_general}
\end{equation}

The factor in braces can be interpreted as the squared modulus of the properly symmetrized or antisymmetrized relative wave function of two identical free particles. Indeed, for two-particles emitted from the space-time points \(x\) and \(y\), one may write
\begin{equation}
\begin{split}
    &\Psi(q,x,y)
    =
    \frac{
    \exp\left[\frac{i}{2}q\cdot(x-y)\right]
    \pm
    \exp\left[-\frac{i}{2}q\cdot(x-y)\right]
    }{\sqrt{2}},
    \\
    &|\Psi(q,x,y)|^2
    =
    1
    \pm
    \cos\left[q\cdot(x-y)\right].
\end{split}
\label{eq:twopartwavefunction}
\end{equation}
The plus sign corresponds to identical bosons, while the minus sign corresponds to identical fermions. This observation motivates a more general form of the two-particle correlation function, commonly associated with the so-called Koonin-Pratt formulation \cite{Koonin1977, Pratt1984}. In this formulation, the free symmetrized wave function is replaced by a two-particle relative wave function \(\Psi(x,y,q,k)\), which may include not only quantum-statistical symmetrization but also final-state interactions between the emitted particles \cite{KisielFlorkowskiBroniowskiPluta2006,Florkowski2010}. One then writes
\begin{equation}
    C(k,q)
    \equiv
    \frac{
    \int  d^4x
    \int  d^4y\,
    S\left(x,k+\frac{q}{2}\right)
    S\left(y,k-\frac{q}{2}\right)
    |\Psi(x,y,q,k)|^2
    }{
    \int  d^4x
    \int  d^4y\,
    S\left(x,k+\frac{q}{2}\right)
    S\left(y,k-\frac{q}{2}\right)
    } .
    \label{eq:gencorfunc}
\end{equation}
This expression extends the purely quantum-statistical result by allowing the two-particle wave function to encode the interaction between the particles after emission. Such effects are particularly important when Coulomb or strong final-state interactions contribute to the measured correlation function. The same structure can be interpreted in terms of a two-particle phase-space distribution. In general, the invariant two-particle distribution can be written as \cite{YanoKoonin1978}
\begin{equation}
    \mathcal N_2(p_1,p_2)
    =
    \int  d^4x_1
    \int  d^4x_2\,
    S_2(p_1,p_2,x_1,x_2),
    \label{eq:phasespacetwopart}
\end{equation}
where \(S_2(p_1,p_2,x_1,x_2)\) plays the role of a two-particle Wigner distribution. Comparing this expression with Eq.~\eqref{eq:gencorfunc}, one identifies the approximate factorized form
\begin{equation}
    S_2(p_1,p_2,x_1,x_2)
    \approx
    S(x_1,p_1)
    S(x_2,p_2)
    |\Psi(x_1,x_2,q,k)|^2 .
    \label{eq:phasespacetwopartfunc}
\end{equation}
This form shows that a product of one-particle emission functions describes independent emission, while the two-particle wave function introduces the correlations generated by quantum statistics and by final-state interactions.

Finally, if the smoothness approximation is applied once again to the product of emission functions, Eq.~\eqref{eq:gencorfunc} reduces to
\begin{equation}
    C(k,q)
    \approx
    \frac{
    \int  d^4x
    \int  d^4y\,
    S(x,k)S(y,k)
    |\Psi(x,y,q,k)|^2
    }{
    \int  d^4x
    \int  d^4y\,
    S(x,k)S(y,k)
    } .
    \label{eq:gencorfunc2}
\end{equation}
This compact expression makes it clear that the measured correlation function can be interpreted as an average of the squared two-particle wave function over the space-time distribution of emission points.

\subsection{Correlation Functions from Monte Carlo Event Generators}

The final particle lists produced by hybrid simulations can be used to construct two-particle correlation functions. In experimental analyses, the correlation function is usually obtained as the ratio between a same-event pair distribution and a mixed-event reference distribution. For identical pions, for example, the numerator $A(\vec q,\vec k)$ is constructed by counting pairs of particles from the same event in bins of relative momentum and average pair momentum. The denominator $B(\vec q,\vec k)$ is constructed from pairs formed with particles taken from different events. This mixed-event distribution provides a reference that contains the same single-particle acceptance and phase-space effects, but not the femtoscopic pair correlation. The measured correlation function can then be written schematically as~\cite{LisaPrattSoltzWiedemann2005,WiedemannHeinz1999}
\begin{equation}
    \mathcal C(\vec q,\vec k)
    =
    \mathcal N
    \frac{A(\vec q,\vec k)}{B(\vec q,\vec k)}
    \xi(\vec q,\vec k),
    \label{eq:expcf}
\end{equation}
where $\mathcal N$ is a normalization factor and $\xi(\vec q,\vec k)$ represents possible correction factors or additional contributions introduced in the fitting procedure to account for effects such as particle purity, momentum resolution, final-state interactions, and non-femtoscopic correlations.

In simulations, the same physical object can be computed in different ways. One possibility is to evaluate the correlation function directly from the emission function on the particlization or switching hypersurface. This approach is close to the Cooper-Frye formulation, but it does not automatically include the subsequent hadronic evolution. Another possibility is to use the final particle lists generated after resonance decays and hadronic transport. In this case, the final particles define an effective discrete emission function, whose space-time points correspond to the points of production or the last interaction before the particles free flow to the detector~\cite{KisielFlorkowskiBroniowskiPluta2006,Florkowski2010,PlumbergHeinz2018}.

It is important to distinguish the effective one-particle source from the two-particle correlation weight. The Particlization and the final particle lists produced after decays and hadronic transport determine the space-time and momentum distribution of emitted particles, since they evolve and sample a one-particle distribution. However, they do not by themselves provide the quantum-statistical symmetrization of an identical-particle pair, nor the Coulomb or strong final-state interaction between the two-particles after emission. These effects enter the correlation function through the two-particle relative wave function, represented below by the factor $|\Psi|^2$. Thus, the product $S(x,p_1)S(y,p_2)$ describes the independent emission of the source, while $|\Psi(x,y,p_1,p_2)|^2$ provides the femtoscopic pair weight.

For a single event, the effective source can be represented as a discrete distribution. The invariant one-particle spectrum may be approximated by
\begin{equation}
    \mathcal N_1(p)
    =
    \int d^4x\,S(x,p)
    =
    p^0\frac{d^3N}{d^3p}
    \simeq
    \frac{p^0}{\Delta^3p}
    \sum_i
    \delta_\Delta(\vec p-\vec p_i),
    \label{eq:n1forevent}
\end{equation}
where the sum runs over particles of the species considered in the event, $\Delta^3p=\Delta p_x\Delta p_y\Delta p_z$ is the momentum-space bin volume, and
\begin{equation}
    \delta_\Delta(\vec p)
    =
    \begin{cases}
    1, & |p_x|\leq \Delta p_x/2,\ |p_y|\leq \Delta p_y/2,\ |p_z|\leq \Delta p_z/2,\\
    0, & \text{otherwise}.
    \end{cases}
    \label{eq:delta_bin}
\end{equation}
This expression states that the continuous emission function is approximated by a histogram over the final particles. The factor $p^0/\Delta^3p$ converts the particle count in a momentum bin into the invariant spectrum. The corresponding two-particle invariant distribution can be written as
\begin{equation}
\begin{split}
    \mathcal N_2(p_1,p_2)
    &=
    p_1^0p_2^0
    \frac{d^6N}{d^3p_1d^3p_2}
    \\
    &=
    \int d^4x\int d^4y\,
    S(x,p_1)S(y,p_2)
    |\Psi(x,y,p_1,p_2)|^2
    \\
    &\simeq
    \frac{p_1^0p_2^0}{\Delta^3p_1\Delta^3p_2}
    \sum_{i}
    \sum_{j\neq i}
    \delta_{\Delta_1}(\vec p_1-\vec p_i)
    \delta_{\Delta_2}(\vec p_2-\vec p_j)
    |\Psi(x_i,x_j,p_i,p_j)|^2 .
\end{split}
\label{eq:n2forevent}
\end{equation}
The restriction $i\neq j$ avoids constructing a pair with the same particle twice. In the absence of femtoscopic correlations, the corresponding uncorrelated pair distribution is obtained by replacing the squared two-particle wave function by unity. Therefore, for a single event, one may write the theoretical pair estimator
\begin{equation}
    C(p_1,p_2)
    =
    \frac{
    \sum_{i\neq j}
    \delta_{\Delta_1}(\vec p_1-\vec p_i)
    \delta_{\Delta_2}(\vec p_2-\vec p_j)
    |\Psi(x_i,x_j,p_i,p_j)|^2
    }{
    \sum_{i\neq j}
    \delta_{\Delta_1}(\vec p_1-\vec p_i)
    \delta_{\Delta_2}(\vec p_2-\vec p_j)
    } .
    \label{eq:cf_pair_estimator_p1p2}
\end{equation}
This expression should be interpreted as an average of the weights of the two-particles at the generated emission points. 

It is useful to rewrite $C$ in terms of $q$ and $k$. The transformation from $(\vec p_1,\vec p_2)$ to $(\vec k,\vec q)$ has a unit Jacobian 
\begin{equation}
    d^3p_1\,d^3p_2
    =
    \left|
    \frac{\partial(p_{1x},p_{1y},p_{1z},p_{2x},p_{2y},p_{2z})}
    {\partial(k_x,k_y,k_z,q_x,q_y,q_z)}
    \right|
    d^3k\,d^3q
    =
    d^3k\,d^3q ,
    \label{eq:jacobian_pkq}
\end{equation}
then we can replace $\delta(\vec p_1)\delta(\vec p_2)$ with $\delta(\vec q)\delta(\vec k)$. After binning directly in $\vec q$ and $\vec k$, the correlation function becomes
\begin{equation}
    C(k,q)
    =
    \frac{
    \sum_{i\neq j}
    \delta_\Delta(\vec q-\vec q_{ij})
    \delta_\Delta(\vec k-\vec k_{ij})
    |\Psi(x_i,x_j,q_{ij},k_{ij})|^2
    }{
    \sum_{i\neq j}
    \delta_\Delta(\vec q-\vec q_{ij})
    \delta_\Delta(\vec k-\vec k_{ij})
    },
    \label{eq:cf_pair_estimator_qk}
\end{equation}
where $q_{ij}=p_i-p_j$ and $k_{ij}=(p_i+p_j)/2$. For identical free bosons, one obtains \cite{Florkowski2010,KisielFlorkowskiBroniowskiPluta2006}
\begin{equation}
    C(k,q)
    =
    1+
    \frac{
    \sum_{i\neq j}
    \delta_\Delta(\vec q-\vec q_{ij})
    \delta_\Delta(\vec k-\vec k_{ij})
    \cos\left[q_{ij}\cdot(x_i-x_j)\right]
    }{
    \sum_{i\neq j}
    \delta_\Delta(\vec q-\vec q_{ij})
    \delta_\Delta(\vec k-\vec k_{ij})
    } .
    \label{eq:cfpeg}
\end{equation}
Hadronic transport and resonance decays modify the effective emission function by changing the final momenta and the last-emission points of the particles, however, the femtoscopic pair weight stemming from quantum statistics and mutual final-state interactions still needs to be provided at the correlation-function level.

Due to the number of pairs in a single event may be limited, it is usually more stable to accumulate numerator and denominator histograms over many events before taking their ratio. We define
\begin{equation}
\begin{split}
    \widetilde N_{\rm den}(k,q)
    &\equiv
    \sum_{i\neq j}
    \delta_\Delta(\vec q-\vec q_{ij})
    \delta_\Delta(\vec k-\vec k_{ij}),
    \\
    \widetilde N_{\rm num}(k,q)
    &\equiv
    \sum_{i\neq j}
    \delta_\Delta(\vec q-\vec q_{ij})
    \delta_\Delta(\vec k-\vec k_{ij})
    |\Psi(x_i,x_j,q_{ij},k_{ij})|^2 .
\end{split}
\label{eq:num_den_histograms}
\end{equation}
The event-averaged correlation function is then computed as the ratio of the accumulated numerator and denominator histograms,
\begin{equation}
    \mathcal C(k,q)
    =
    \frac{
    \left\langle \widetilde N_{\rm num}(k,q)\right\rangle_{\rm ev}
    }{
    \left\langle \widetilde N_{\rm den}(k,q)\right\rangle_{\rm ev}
    },
    \label{eq:cfav2}
\end{equation}
where $\langle\cdots\rangle_{\rm ev}$ denotes the accumulation over events. The ratio is taken only after accumulating the numerator and denominator histograms, rather than event by event, because events may contain different numbers of accepted pairs. In the theoretical calculation used here, the denominator is built from the same accepted pairs with unit weight, while the numerator contains the femtoscopic pair weight. Thus, the denominator represents the corresponding uncorrelated reference for the same pair sample.

\subsection{Kinematic Variables and Reference Frames}
\label{sec:kinematic_variables}
In Eq.~\eqref{eq:dynvar}, we introduced the average and relative four-momenta of a particle pair \(k^\mu=(k^0,\vec k)\) and \(q^\mu=(q^0,\vec q)\). These variables play a central role in femtoscopic analyses, since they separate the momentum scale of the pair from the momentum difference probed by the correlation function. Let us start with the definition again. Given a pair of particles with momenta $p_1^\mu = (p^0_1, \vec p_1)$, $p_2^\mu = (p^0_2, \vec p_2)$, we have
\begin{align}
    k &= \frac{p_1 + p_2}{2},\\
    q &= p_1 - p_2.
    \label{eq:dynvar1}
\end{align}
They are, respectively, the mean and relative momenta of the pair system. From this definition, it is straightforward to obtain
\begin{align}
    p_1 &= k + \frac{q}{2},\\
    p_2 &= k - \frac{q}{2}.
\label{eq:dynvar2}
\end{align}
When we take $p_1^2 = p_1\cdot p_1$ and $p_2^2 = p_2 \cdot p_2$, we get
\begin{align}
    m_1^2 &= k^2 + k\cdot q +\frac{q^2}{4},\\
    m_2^2 &= k^2 - k\cdot q +\frac{q^2}{4},
\end{align}
since, by relativistic properties, $p_i^2 = m_i^2$ where $m_i$ is the mass of each $i$ particle. By adding and subtracting the two equations above, we get
\begin{align}
    \frac{m_1^2 + m_2^2}{2} &= k^2 +\frac{q^2}{4},\\
    \frac{m_1^2 - m_2^2}{2} &=  k\cdot q.
\end{align}
In particular, if the particles are identical, that is, $m_1 = m_2 \equiv m$, we get
\begin{align}
    & k^2 +\frac{q^2}{4} = m^2,\\
    & k\cdot q =  0.
    \label{eq:systemeqkq}
\end{align}
For identical particles, the constraint $k\cdot q =0$ is exact and is also known as the {\it mass-shell constraint} \cite{WiedemannHeinz1999}. A corresponding relation may be used as an approximation in particle-antiparticle correlations when the two masses are taken to be equal. This second equation leads to
\begin{equation}
    q^0 = \frac{\vec k \cdot \vec q}{k^0} \equiv \vec \beta \cdot \vec q,
    \label{eq:q0}
\end{equation}
where we define the vector $\vec \beta$ as
\begin{equation}
    \vec \beta = \frac{\vec k}{k^0}.
    \label{eq:betanotation}
\end{equation}
This definition is interesting because, as we will see later, a Lorentz boost with velocity $\vec \beta$ takes the particle pair to its center of momentum frame. For $p_1 \cdot p_2$, using Eq.~\eqref{eq:dynvar2}, we have
\begin{equation}
    p_1 \cdot p_2 = k^2 - \frac{q^2}{4}.
\end{equation}
As one said, the Correlation function is Lorentz invariant, as we can see by the definition in Eq.~\eqref{eq:invcf}. On the other hand, in analyses that consider different frames of reference, we obtain projections of the correlation function in a way that is quite similar to that used to calculate the projections of the invariant distribution in the spectrum analysis presented in Eq. \eqref{eq:projections}. From this perspective, we can use numerous reference frames, but we will concentrate on three: the center-of-momentum frame, the Bertsch–Pratt coordinate system, and the Longitudinally Comoving System \cite{Florkowski2010}.

\subsubsection{ Collision Frame}

We will start from the definition of the collision axes to build the other coordinate systems. We call the axis along the collision the {\it longitudinal direction}, and we call the plane perpendicular to this axis the {\it transverse plane}
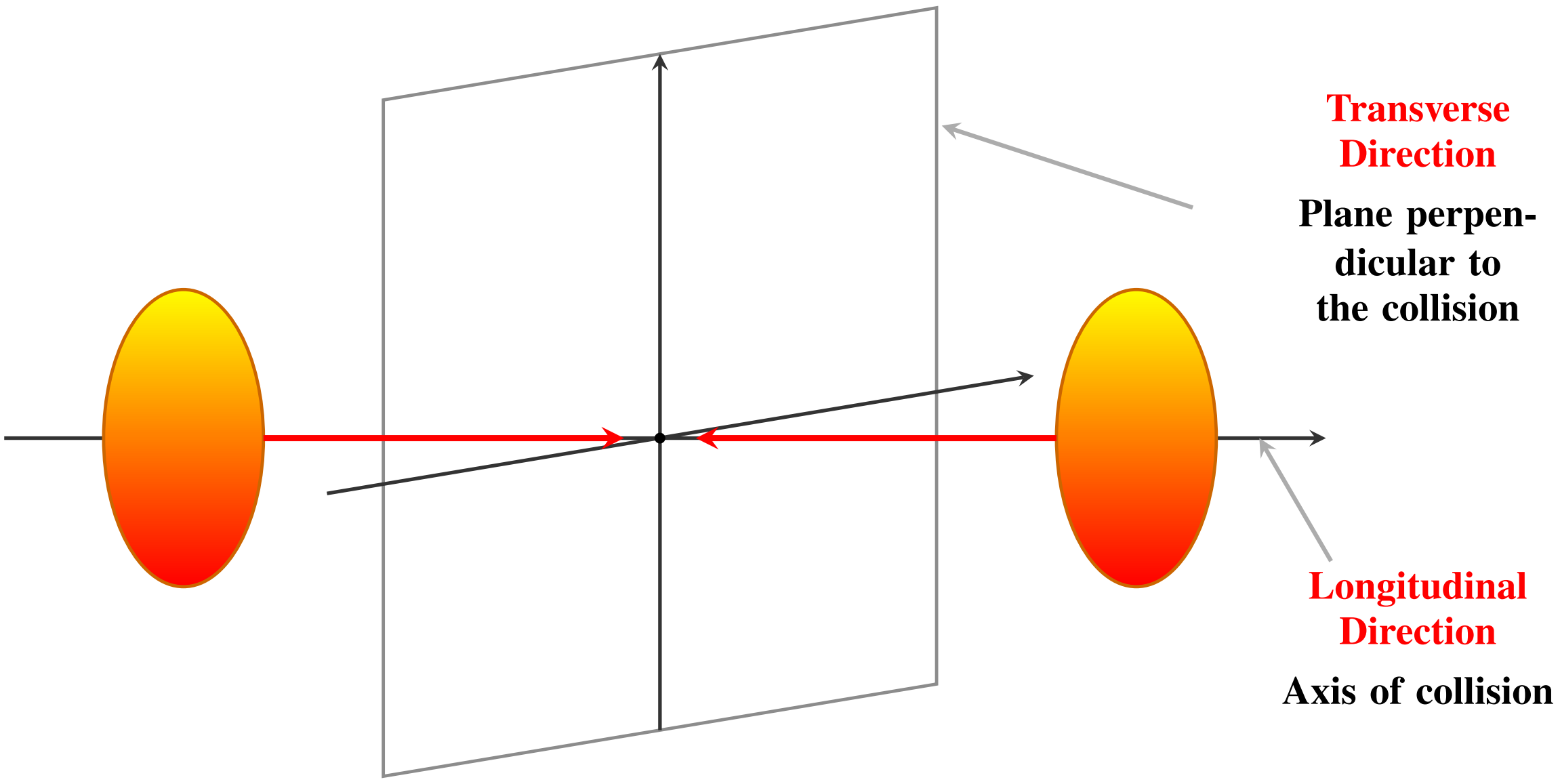
\begin{figure}[!htbp]
\centering
\begin{tikzpicture}[
    x=1cm,
    y=1cm,
    >=stealth,
    axis/.style={->, line width=1.0pt, black!80},
    beam/.style={->, red, line width=1.7pt},
    plane/.style={black!45, line width=0.9pt},
    callout/.style={->, gray!65, line width=1.2pt},
    physlabel/.style={
        anchor=west,
        align=center,
        text width=3.4cm,
        font=\small
    }
]

\coordinate (O) at (0,0);

\draw[axis] (-6.4,0) -- (6.5,0);

\coordinate (A) at (-2.7,-3.3);
\coordinate (B) at (-2.7, 3.3);
\coordinate (C) at ( 2.7, 4.2);
\coordinate (D) at ( 2.7,-2.4);

\draw[plane] (A) -- (B) -- (C) -- (D) -- cycle;

\draw[axis] (0,-2.85) -- (0,3.75);

\draw[axis] (-3.25,-0.54) -- (3.65,0.61);

\shade[
    draw=orange!80!black,
    line width=0.9pt,
    top color=yellow,
    middle color=orange,
    bottom color=red
] (-4.65,0) ellipse (0.78 and 1.45);

\shade[
    draw=orange!80!black,
    line width=0.9pt,
    top color=yellow,
    middle color=orange,
    bottom color=red
] (4.65,0) ellipse (0.78 and 1.45);

\draw[beam] (-3.87,0) -- (-0.35,0);
\draw[beam] (3.87,0) -- (0.35,0);

\fill[black] (O) circle (1.5pt);

\node[physlabel] (translabel) at (5.55,2.25)
{
    \textcolor{red}{\textbf{Transverse}}\\[-1pt]
    \textcolor{red}{\textbf{Direction}}\\[3pt]
    \textbf{Plane perpendicular to}\\[-1pt]
    \textbf{the collision}
};

\draw[callout] (5.20,2.25) -- (2.75,3.05);

\node[physlabel] (longlabel) at (5.55,-1.95)
{
    \textcolor{red}{\textbf{Longitudinal}}\\[-1pt]
    \textcolor{red}{\textbf{Direction}}\\[3pt]
    \textbf{Axis of collision}
};

\draw[callout] (6.55,-1.20) -- (5.85,0.0);

\end{tikzpicture}

\caption{Schematic representation of two nuclei moving along the longitudinal direction, identified with the beam axis. The transverse plane is perpendicular to this axis and contains the directions in which transverse dynamics are analyzed.}

    \label{fig:colisao}
\end{figure}
\noindent We use the convention in which the longitudinal axis is the $z$ axis, while the $x$ and $y$ axes span the transverse plane. The ordered basis $\{\hat x,\hat y,\hat z\}$ is chosen to be right-handed.

\subsubsection{\normalsize Center of Momentum Frame}

The center-of-momentum frame is defined as the reference frame in which the total $3-$momentum of the system is zero, that is, given a two-particle system with $4$-momenta $p_1$ and $p_2$,
\begin{equation}
    \vec p'_1 + \vec p'_2 = 0,
\end{equation}
where here the primes indicate that these variables are in the center-of-momentum frame. Since the mean momentum of the particle pair system is given by $k = (p_1 + p_2)/2$, we have
\begin{equation}
    \vec k' = 0.
\end{equation}
These definitions do not imply that the $0$th component of $p'_1$, $p'_2$, or $k'$ is zero. To transform from an arbitrary frame to the center-of-momentum frame, one uses a Lorentz boost with velocity
\begin{equation}
    \vec V = \frac{\vec k}{k^0} = \vec \beta
    \label{eq:boostboost}
\end{equation}
defined in the original frame. It is easy to prove this because the Lorentz transformation of $k$ is the same as that of $p_1$ and $p_2$, since $k$ is linearly dependent on $p_1$ and $p_2$. The transformation is\\
\begin{equation}
\begin{split}
   \vec{k}' &=  (\gamma - 1) ( \vec{k}\cdot\vec{V})\frac{\vec{V}}{\vec{V}^2}- \gamma \vec{V} k^0 + \vec{k} = \\
   &= \left(\frac{k^0}{\sqrt{k\cdot k}} -1 \right)(\vec k ^2)\frac{\vec k }{(\vec k ^2)} - \frac{k^0}{\sqrt{k \cdot k}}\vec k + \vec k = \\
   &= \frac{k^0}{\sqrt{k \cdot k}}\vec k - \vec k - \frac{k^0}{\sqrt{k \cdot k}}\vec k + \vec k =  0,
\end{split}
\end{equation}
where
\begin{equation}
    \gamma = \frac{1}{\sqrt{1-\vec{V}^2}} = \frac{k^0}{\sqrt{k\cdot k}}
\end{equation}
since $k^0 \geq 0$. We saw that $q\cdot k = 0$ for particles with the same mass. So, in the center of the momentum frame, 
\begin{equation}
    q'\cdot k' = q'^0 k'^0 - \vec q' \cdot \vec k' = q'^0 k'^0,
\end{equation}
which implies 
\begin{equation}
    q'^0 = 0,
\end{equation}
since $k'^0$ is arbitrary. This frame then has only 4 dynamic degrees of freedom (besides the particle positions) which are $\{k'^0, \vec{q}'\}$. In this case, $2k^0 = p^0_1 + p^0_2 = E_1 + E_2 \equiv E_T$ is the total energy of the particle system (this for any frame of reference), we just need to specify the total energy of the particle system and the relative $3-$momentum to specify the dynamic state of this system, which facilitates the HBT analysis. Since the scalar product $k\cdot k$ is invariant, we can calculate $k'^0$ as
\begin{equation*}
    k'^0 = \sqrt{k\cdot k} 
\end{equation*}
where, again, the non-prime quantities are evaluated in a frame different from the center-of-momentum frame. The Lorentz transformation of $\vec{q}$, since $q$ is also linearly dependent on $p_1$ and $p_2$, becomes
\begin{equation}
\begin{split}
    \vec{q}' &=  (\gamma - 1) ( \vec{q}\cdot\vec{V})\frac{\vec{V}}{\vec{V}^2}- \gamma \vec{V} q^0 + \vec{q}.
\end{split}
\end{equation}
In the non-relativistic limit $\vec V^2 \ll 1$, the center-of-momentum frame coincides with the center-of-mass frame. It is common to define 
\begin{equation}
    \vec q_{inv} \equiv \vec q'
\end{equation}
where
\begin{equation}
    q\cdot q = q' \cdot q' = - \vec q_{inv}^2
\end{equation}
so
\begin{equation}
    Q_{inv} = |\vec q_{inv}| = \sqrt{-q\cdot q}
\end{equation}

\subsubsection{\normalsize Bertsch–Pratt coordinate system}

\begin{figure}[!htbp] 
    \centering
   \includegraphics[width=\linewidth]{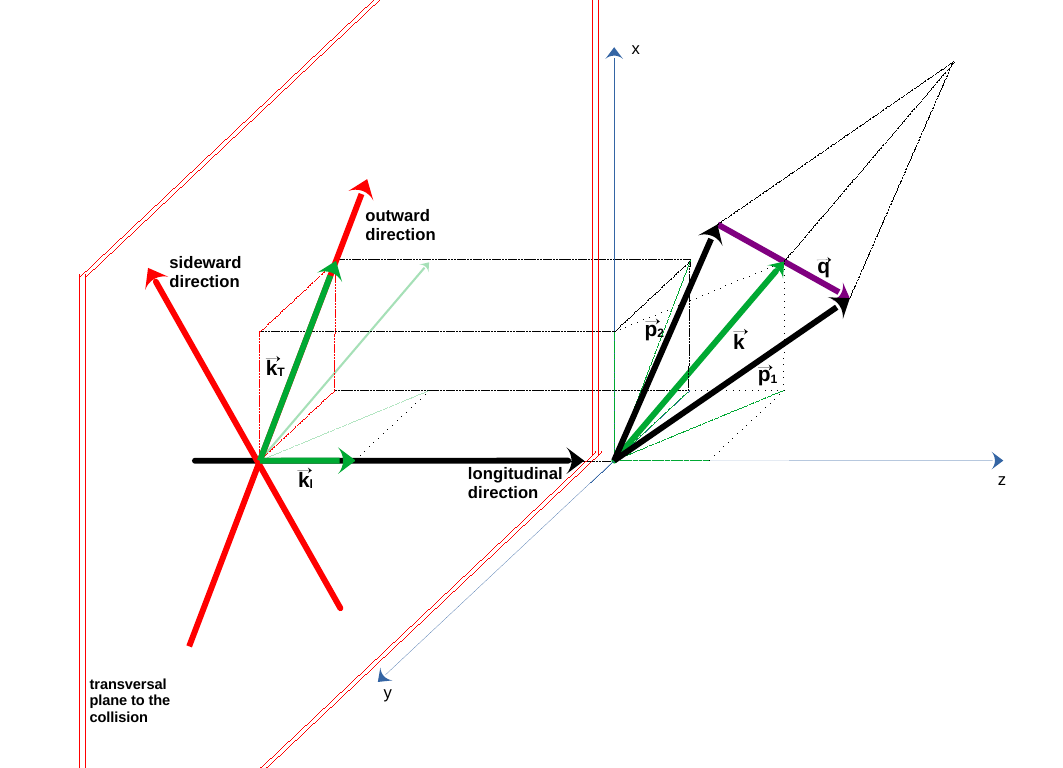}
   \caption{Bertsch–Pratt coordinate system scheme}
   \label{fig:oslsystem}
\end{figure}

Following the decomposition of $\vec q$ relative to $\vec K$ discussed by V. G. Grishin, G. I. Kopylov, and M. I. Podgoretskii~\cite{Padula2005HBT}, the Bertsch–Pratt coordinate system defines three spatial axes as follows. We first define the axis $\hat e_o$, called {\it outward}, parallel to the component of $\vec k$ transverse to the collision axis (see Fig.~\ref{fig:oslsystem}). The second axis, $\hat e_l$, lies along the longitudinal axis of the collision. The third axis, $\hat e_s$, called {\it sideward}, is defined by the cross product $\hat e_s \equiv \hat e_l \times \hat e_o$, so that the ordered triplet of unit vectors $\{\hat e_o , \hat e_s, \hat e_l\}$ is right-handed and forms a basis. This system is also known as the {\it out-side-long} (OSL) coordinate system.
\noindent We know that
\begin{equation}
    \vec k  = k_x \hat x + k_y \hat y + k_z \hat z \equiv \vec k_T+ k_z \hat z
\end{equation}
where $\hat x$ and $\hat y$ are any axes in the transverse collision plane and $\hat z$ is the collision axis. The transverse component of $\vec k$ can be written as
\begin{equation}
    \vec k_T = k_x \hat x + k_y \hat y \equiv k_T \hat e_o,
\end{equation}
so
\begin{equation}
    \hat e_o = \frac{\vec k_T}{k_T},
\end{equation}
where $k_T = |\vec k_T| = \sqrt{k_x^2 + k_y^2}$. The longitudinal component of $\vec k$ can be written as
\begin{equation}
    \vec k_l = k_z \hat z
\end{equation}
then it is trivial that
\begin{equation}
     \hat e_l = \hat z.
\end{equation}
Finally
\begin{equation}
    \hat e_s = \hat e_l \times \hat e_o = \frac{-k_y \hat x + k_x \hat y}{k_T}.
\end{equation}
This transformation can be rewritten in matrix form as
\begin{equation}
    \left(
    \begin{array}{c}
         \hat e_o\\
         \hat e_s\\
         \hat e_l
    \end{array}
    \right) = 
    \underbrace{\left( 
    \begin{array}{ccc}
         \frac{k_x}{k_T} & \frac{k_y}{k_T} & 0  \\
         -\frac{k_y}{k_T} & \frac{k_x}{k_T} & 0  \\
          0 & 0 & 1
    \end{array}
    \right)}_{\mathrm A}
    \left(
    \begin{array}{c}
         \hat x \\
         \hat y\\
         \hat z
    \end{array}
    \right).
    \label{eq:oslaxisdefinition}
\end{equation}
The matrix $A$ is an orthogonal matrix, that is, $A^T = A^{-1}$. Since $\det A=1$, it is a proper rotation matrix, so the Bertsch–Pratt coordinate system is a rotation of the coordinate system around the $z$ axis. Moreover, the corresponding four-dimensional block matrix is a Lorentz transformation because the Lorentz group includes spatial rotations \cite{DInvernoVickers2022}. Thus, we have
\begin{equation}
    \left(
    \begin{array}{c}
         \hat e_t\\
         \hat e_o\\
         \hat e_s\\
         \hat e_l
    \end{array}
    \right) = 
    \underbrace{\left( 
    \begin{array}{cccc}
        1& 0 & 0 & 0\\
        0&  &  & \\
        0&  & A &   \\
        0&  &   & 
    \end{array}
    \right)}_{ \Lambda^{\mu}_{\ \nu}}
    \left(
    \begin{array}{c}
         \hat t \\
         \hat x \\
         \hat y\\
         \hat z
    \end{array}
    \right).
\end{equation}
where $\Lambda^{\mu}_{\ \nu}$ is a Lorentz transformation itself. The components of $\vec q$ in this case become
\begin{align}
    q_o &= \Lambda^{o}_{\ \nu}q^\nu = \hat e_o\cdot\vec q  = \frac{q_x k_x + q_y k_y}{k_T},\\
    q_s &= \Lambda^{s}_{\ \nu}q^\nu = \hat e_s\cdot\vec q = \frac{q_y k_x - q_x k_y}{k_T},\\
    q_l &=  \Lambda^{l}_{\ \nu}q^\nu = \hat e_l\cdot\vec q  = q_z.
\end{align}
This system allows us to study HBT interferometry by separating the system with respect to the mean momentum of the particle pair into a longitudinal component and a transverse component only, and this will be reflected in the way we obtain the observable HBT radii. In this case, it is common to parameterize $k$ as
\begin{align}
    &k^\mu = (m_T \cosh(Y), k_T, 0, m_T\sinh(Y))\\
    &k\cdot k= (k^0)^2 - \vec k^2 = m_T^2 - k_T^2,
    \label{eq:kparametrization}
\end{align}
where given the parameters $(k_T, Y, q_o, q_s, q_l)$, it is possible to calculate $q^0$ and $m_T$. However, for these parameters, the set of equations in \eqref{eq:systemeqkq} leads to a complicated system of non-linear equations, so it is preferable to use the on-shell approximation in Eq.~\eqref{eq:onshellapp}, which simplifies the calculations.
\begin{equation}
    m_T \approx \sqrt{m^2 + k_T^2},
    \label{eq:mtapprox}
\end{equation}
and using Eq. \eqref{eq:q0}, it is simple to calculate $q^0$.

\subsubsection{\normalsize Longitudinally Comoving System}

The Longitudinally Comoving System (LCMS) is obtained by a boost along the longitudinal axis $z$. Using this system together with the OSL system, the mean momentum vector is, by definition,
\begin{equation}
    \vec k = (k_T, 0, 0).
\end{equation}
To go from the collision frame to this frame, the boost velocity along the $z$ axis is simply the $\beta_z$ component of the vector defined in Eq.~\eqref{eq:betanotation}, so that, given a $4-$vector $A = (A^t, \vec A)$ in the collision frame, we have 
\begin{equation}
    \left(
    \begin{array}{c}
         A'^t\\
         A^o\\
         A^s\\
         A'^l
    \end{array}
    \right) = 
    \left( 
    \begin{array}{cccc}
        \gamma_z& 0 & 0 & -\gamma_z \beta_z\\
        0& \frac{k_x}{k_T} & \frac{k_y}{k_T} & 0\\
        0& -\frac{k_y}{k_T} & \frac{k_x}{k_T} & 0\\
        -\gamma_z \beta_z& 0 &  0 & \gamma_z 
    \end{array}
    \right)
    \left(
     \begin{array}{c}
         A^t\\
         A^x\\
         A^y\\
         A^z
    \end{array}
    \right),
    \label{eq:osllcmsaxisdefinition}
\end{equation}
where $\gamma_z = 1/\sqrt{1-\beta_z^2}$. This system further reduces the number of variables analyzed and retains only the transverse component of the pair momentum in the average-momentum dependence, which will be useful for analyzing the observables obtained by HBT interferometry.

\subsection{Projected Correlation Functions and Frame Dependence}
\label{sec:projected_correlation_functions}

The correlation function introduced in the previous sections is, in its most complete form, a function of the two-particle momenta or, equivalently, of the mean and relative momenta of the pair. At this point, the invariant correlation function can be written as \(C(p_1,p_2)\) or \(C(k,q)\), and its scalar value does not depend on the Lorentz frame used to evaluate it. However, the correlation functions used in practical HBT analyzes are not the full six-dimensional object. They are projected estimators constructed after choosing a set of variables, cuts, bins, etc. The HBT radii are extracted not from the complete invariant function, but from ratios of projected pair distributions. For example, in a three-dimensional analysis using the OSL+LCMS decomposition and a given \(k_T\) bin, the numerator for a single event may be written schematically as
\begin{equation}
\small
    A_{\rm ev}(k_T; q_o,q_s,q_l)
    =
    \sum_{i\neq j}
    \delta_{\Delta k_T}(k_T-k_{T,ij})
    \delta_\Delta(q_o-q_{o,ij})
    \delta_\Delta(q_s-q_{s,ij})
    \delta_\Delta(q_l-q_{l,ij})
    W_{ij},
    \label{eq:projected_numerator_event}
\end{equation}
where \(W_{ij}\) is the squared wave function, and the corresponding denominator is
\begin{equation}
\small
    B_{\rm ev}(k_T;q_o,q_s,q_l)
    =
    \sum_{i\neq j}
    \delta_{\Delta k_T}(k_T-k_{T,ij})
    \delta_\Delta(q_o-q_{o,ij})
    \delta_\Delta(q_s-q_{s,ij})
    \delta_\Delta(q_l-q_{l,ij}),
    \label{eq:projected_denominator_event}
\end{equation}
or, in an experimental-like construction, an analogous mixed-event reference distribution. Both \(A_{\rm ev}\) and \(B_{\rm ev}\) are already projections of the pair distribution onto the chosen variables. For a given centrality class, the final correlation function is obtained only after accumulating these projected histograms over events. Thus the estimator is not the average of event-by-event ratios, but rather the ratio of the event-averaged numerator and denominator,
\begin{equation}
    C(k_T;q_o,q_s,q_l)
    =
    \frac{
    \left\langle
    A_{\rm ev}(k_T;q_o,q_s,q_l)
    \right\rangle_{\rm ev}
    }{
    \left\langle
    B_{\rm ev}(k_T;q_o,q_s,q_l)
    \right\rangle_{\rm ev}
    }.
    \label{eq:projected_correlation_ratio_of_averages}
\end{equation}
More generally, for a chosen set of projected variables \(\mathcal Q\), one may write
\begin{equation}
    C_{\rm proj}(\mathcal Q)
    =
    \frac{
    \left\langle
    \displaystyle
    \sum_{i\neq j}
    \delta_\Delta\!\left(\mathcal Q-\mathcal Q_{ij}\right)
    W_{ij}
    \right\rangle_{\rm ev}
    }{
    \left\langle
    \displaystyle
    \sum_{i\neq j}
    \delta_\Delta\!\left(\mathcal Q-\mathcal Q_{ij}\right)
    \right\rangle_{\rm ev}
    }.
    \label{eq:generic_projected_correlation}
\end{equation}
If \(\mathcal Q\) does not depend on the six free components, then the components are manifestly frame dependent. As a result, different pairs are, in general, considered in different local frames. This is exactly why the OSL+LCMS parameterization is helpful, as it reduces the degrees of freedom. At the same time, \(C(q_o,q_s,q_l;k_T)\) is not a Lorentz scalar in the same sense as the underlying \(C(p_1,p_2)\). These frame choices and projections do not represent an inconsistency of the femtoscopic formalism but reflect the fact that the measured correlation is obtained from a particular projection of the two-particle distribution. In high-energy collisions, the OSL+LCMS decomposition is helpful because it is defined with respect to the collision axis~\cite{WiedemannHeinz1999,LisaPrattSoltzWiedemann2005}. Experimental collaborations, such as STAR pion interferometry analyses\cite{AdamsSTAR2005} at RHIC, also explicitly use this decomposition~. 

For the calculations performed in this work, the Monte Carlo particles are generated in the collision frame. For each accepted pair, \(q\) and \(k\) are constructed and then transformed to the chosen analysis frame. The numerator and denominator are filled as projected histograms and averaged over events in the selected centrality class. The final correlation function is the ratio of these event-averaged projections. Therefore, the resulting correlation function should always be understood as an observable defined by that specific projection, binning, event class, and frame choice, rather than as a frame-independent object detached from the analysis procedure.

\section{Parameterizations and HBT Radii}
\label{sec:3.10}

We have presented a description of the method for calculating the correlation functions without, in principle, linking them to the observables that describe the physical dimensions characterizing the geometry of space-time in collisions. These observables, the HBT radii, quantify what we call the characteristic homogeneity lengths of the source for emitted particles. In practice, HBT radii are extracted by fitting two-particle correlation functions. In the present discussion, we use the correlation function in the form of Eq.~\eqref{eq:corfunclas}, where the effective distribution $\rho_{\rm eff}$ models the space-time distribution of the emitting source. We first consider a Gaussian source, which naturally leads to the standard definition of the HBT radii.

\subsection{Gaussian Emission and HBT Radii}

The use of a Gaussian distribution to model the emission source is historically important and convenient~\cite{WiedemannHeinz1999,LisaPrattSoltzWiedemann2005}. Although realistic sources produced in heavy-ion collisions are not exactly Gaussian, this approximation provides a simple connection between the curvature of the correlation function around $\vec q=0$ and the characteristic space-time widths of the region emitting particles with a given pair momentum. We define the four-dimensional multivariate Gaussian distribution as
\begin{equation}
    \mathcal G(x,B)
    =
    \frac{1}{\sqrt{(2\pi)^4\det B}}
    \exp\left[
    -\frac{1}{2}
    (x-\bar x)_\mu (B^{-1})^{\mu\nu}(x-\bar x)_\nu
    \right],
    \label{eq:gaussian_source}
\end{equation}
where $\bar x^\mu$ is the average emission point and $B^{\mu\nu}$ is the covariance matrix. With this normalization, $\mathcal G(x,B)$ integrates to unity. The effective source distribution can then be written as
\begin{equation}
    \rho_{\rm eff}(x,p_1,p_2)
    =
    \kappa(p_1,p_2)\mathcal G(x,B(p_1,p_2)),
    \label{eq:rho_eff_gaussian}
\end{equation}
where $\kappa$ is a residual normalization factor. Its Fourier transform is
\begin{equation}
\begin{split}
    \tilde\rho_{\rm eff}(k,q)
    &=
    \int d^4x\,\rho_{\rm eff}(x,p_1,p_2)\exp(iq\cdot x)
    \\
    &=
    \kappa(p_1,p_2)
    \exp\left(
    iq\cdot\bar x
    -\frac{1}{2}q_\mu B^{\mu\nu}(p_1,p_2)q_\nu
    \right).
\end{split}
\label{eq:rho_eff_gaussian_fourier}
\end{equation}
Therefore, using Eq.~\eqref{eq:corfunclas}, the correlation function becomes
\begin{equation}
    C(k,q)
    =
    1
    \pm
    \lambda(p_1,p_2)
    \exp\left(
    -q_\mu B^{\mu\nu}(p_1,p_2)q_\nu
    \right),
    \label{eq:foutransgauss}
\end{equation}
where $\lambda\equiv |\kappa|^2$. The physical meaning of this parameter will be discussed later. For a Gaussian source, the covariance matrix can be written as
\begin{equation}
    B^{\mu\nu}
    =
    \left\langle
    \Delta x^\mu \Delta x^\nu
    \right\rangle,
    \qquad
    \Delta x^\mu \equiv x^\mu-\bar x^\mu .
    \label{eq:covariance_matrix}
\end{equation}
Using the mass-shell constraint for identical particles, $q^0=\vec\beta\cdot\vec q$, with $\vec\beta=\vec k/k^0$, one obtains the following.
\begin{equation}
\begin{split}
    q_\mu B^{\mu\nu}q_\nu
    &=
    q_\mu
    \left\langle
    \Delta x^\mu \Delta x^\nu
    \right\rangle
    q_\nu
    \\
    &=
    (q^0)^2\left\langle \Delta t^2\right\rangle
    -2q^0q^i\left\langle \Delta t\,\Delta x_i\right\rangle
    +q^iq^j\left\langle \Delta x_i\Delta x_j\right\rangle
    \\
    &=
    q^iq^j
    \left\langle
    (\Delta x_i-\beta_i\Delta t)
    (\Delta x_j-\beta_j\Delta t)
    \right\rangle .
\end{split}
\label{eq:qBq_radii}
\end{equation}
Here, Latin indices denote spatial components and are summed over when repeated. No additional Minkowski metric contraction is implied in the last line. We therefore define
\begin{equation}
    R_{ij}^2
    \equiv
    \left\langle
    (\Delta x_i-\beta_i\Delta t)
    (\Delta x_j-\beta_j\Delta t)
    \right\rangle ,
    \label{eq:hbtradii}
\end{equation}
so the correlation function can then be written as
\begin{equation}
    C(k,q)
    =
    1
    \pm
    \lambda(p_1,p_2)
    \exp\left[
    -q^iR_{ij}^2(p_1,p_2)q^j
    \right].
    \label{eq:corfuncgenhbtradii}
\end{equation}
Using the smoothness approximation, the dependence on $p_1$ and $p_2$ is replaced by a dependence on the average pair momentum $k$, giving
\begin{equation}
    C(k,q)
    =
    1
    \pm
    \lambda(k)
    \exp\left[
    -q^iR_{ij}^2(k)q^j
    \right].
    \label{eq:corfuncgenhbtradii2}
\end{equation}
The parameters $R_{ij}^2(k)$ are called \textit{HBT radii}. Within the Gaussian approximation, these quantities are given by a combination of the second central moments of the effective emission function selected by the pair momentum $k$, and thus they probe homogeneity lengths rather than the source’s overall geometric extent. Equivalently, for the Gaussian form above, they can be obtained from the curvature of the correlation function around $\vec q=0$:
\begin{equation}
    R_{ij}^2
    =
    -\frac{1}{2\lambda}
    \left.
    \frac{\partial^2 C}{\partial q^i\partial q^j}
    \right|_{\vec q=0}.
    \label{eq:hbt_curvature}
\end{equation}

In a general three-dimensional analysis, the symmetric matrix $R_{ij}^2$ contains six independent components. Explicitly,
\begin{align}
    R_{11}^2 &= \left\langle\Delta x_1^2\right\rangle -2\beta_1\left\langle\Delta x_1\Delta t\right\rangle+\beta_1^2\left\langle\Delta t^2\right\rangle,\\
    R_{22}^2 &= \left\langle\Delta x_2^2\right\rangle -2\beta_2\left\langle\Delta x_2\Delta t\right\rangle+\beta_2^2\left\langle\Delta t^2\right\rangle,\\
    R_{33}^2 &= \left\langle\Delta x_3^2\right\rangle -2\beta_3\left\langle\Delta x_3\Delta t\right\rangle+\beta_3^2\left\langle\Delta t^2\right\rangle,\\
    R_{12}^2 &= R_{21}^2
    =
    \left\langle\Delta x_1\Delta x_2\right\rangle
    -\beta_2\left\langle\Delta x_1\Delta t\right\rangle
    -\beta_1\left\langle\Delta x_2\Delta t\right\rangle
    +\beta_1\beta_2\left\langle\Delta t^2\right\rangle,\\
    R_{13}^2 &= R_{31}^2
    =
    \left\langle\Delta x_1\Delta x_3\right\rangle
    -\beta_3\left\langle\Delta x_1\Delta t\right\rangle
    -\beta_1\left\langle\Delta x_3\Delta t\right\rangle
    +\beta_1\beta_3\left\langle\Delta t^2\right\rangle,\\
    R_{23}^2 &= R_{32}^2
    =
    \left\langle\Delta x_2\Delta x_3\right\rangle
    -\beta_3\left\langle\Delta x_2\Delta t\right\rangle
    -\beta_2\left\langle\Delta x_3\Delta t\right\rangle
    +\beta_2\beta_3\left\langle\Delta t^2\right\rangle .
\label{eq:hbt_radii_general_components}
\end{align}
The number of independent radii can be reduced by choosing a convenient frame and by imposing symmetries on the source. In the Bertsch-Pratt, or out-side-long, system, the outward direction is chosen parallel to the transverse pair momentum. Therefore, the sideward component of the pair velocity vanishes, $\beta_s=0$. In this basis, the diagonal radii are
\begin{align}
    R_o^2 &= \left\langle(\Delta x_o-\beta_T\Delta t)^2\right\rangle
    =
    \left\langle\Delta x_o^2\right\rangle
    -2\beta_T\left\langle\Delta x_o\Delta t\right\rangle
    +\beta_T^2\left\langle\Delta t^2\right\rangle,\\
    R_s^2 &= \left\langle\Delta x_s^2\right\rangle,\\
    R_l^2 &= \left\langle(\Delta x_l-\beta_l\Delta t)^2\right\rangle
    =
    \left\langle\Delta x_l^2\right\rangle
    -2\beta_l\left\langle\Delta x_l\Delta t\right\rangle
    +\beta_l^2\left\langle\Delta t^2\right\rangle.
\label{eq:hbt_radii_osl_general}
\end{align}
The corresponding cross terms are
\begin{align}
    R_{os}^2
    &=
    \left\langle
    (\Delta x_o-\beta_T\Delta t)\Delta x_s
    \right\rangle,\\
    R_{ol}^2
    &=
    \left\langle
    (\Delta x_o-\beta_T\Delta t)
    (\Delta x_l-\beta_l\Delta t)
    \right\rangle,\\
    R_{sl}^2
    &=
    \left\langle
    \Delta x_s
    (\Delta x_l-\beta_l\Delta t)
    \right\rangle .
\label{eq:hbt_cross_terms_osl}
\end{align}
In the LCMS, the longitudinal component of the pair velocity vanishes, $\beta_l=0$, while $\beta_o=\beta_T$ and $\beta_s=0$. Therefore,
\begin{align}
    R_o^2 &= 
    \left\langle\Delta x_o^2\right\rangle
    -2\beta_T\left\langle\Delta x_o\Delta t\right\rangle
    +\beta_T^2\left\langle\Delta t^2\right\rangle,\\
    R_s^2 &= \left\langle\Delta x_s^2\right\rangle,\\
    R_l^2 &= \left\langle\Delta x_l^2\right\rangle .
\label{eq:lcmsoslhbtradii}
\end{align}
For azimuthally integrated analyses of central collisions, one usually assumes reflection symmetries around the sideward and longitudinal directions. Under these symmetry assumptions, the cross terms vanish, and the Gaussian fit reduces to
\begin{equation}
    C(\vec q,k_T)
    =
    1
    \pm
    \lambda(k_T)
    \exp\left[
    -q_o^2R_o^2(k_T)
    -q_s^2R_s^2(k_T)
    -q_l^2R_l^2(k_T)
    \right].
    \label{cfobj}
\end{equation}
The radii $R_o$, $R_s$ and $R_l$ characterize the homogeneity lengths of the source in the outward, sideward and longitudinal directions, respectively. In particular, $R_o$ contains a temporal contribution through the pair velocity, while $R_s$ is insensitive to the emission duration in this approximation. 

In addition to the three-dimensional OSL analysis, one can also perform a one-dimensional
Gaussian fit as a function of the invariant relative momentum \(q_{\rm inv}\). If the effective relative source is approximately isotropic in the pair center-of-momentum frame, one may write schematically
\begin{equation}
    G(x)
    \propto
    \exp\left(
        -\frac{X_\mu X^\mu}{2R_{\rm inv}^2}
    \right),
\end{equation}
which leads to the one-dimensional Gaussian form
\begin{equation}
    C(q_{\rm inv})
    =
    1\pm\lambda\exp\left(-Q_{\rm inv}^2R_{\rm inv}^2\right).
    \label{eq:gaussian_qinv_1d}
\end{equation}
The parameter \(R_{\rm inv}\) should be understood as an effective femtoscopy scale associated with this projection. It is useful for compact comparisons and for analyses with limited statistics, but it does not retain the separate longitudinal, outward, and sideward information contained in the complete OSL parameterization~\cite{WiedemannHeinz1999,LisaPrattSoltzWiedemann2005,AdamsSTAR2005,AdamczykSTAR2015HBTBES}.
\begin{figure}[!htbp] 
    \centering
    \includegraphics[width=0.9\linewidth]{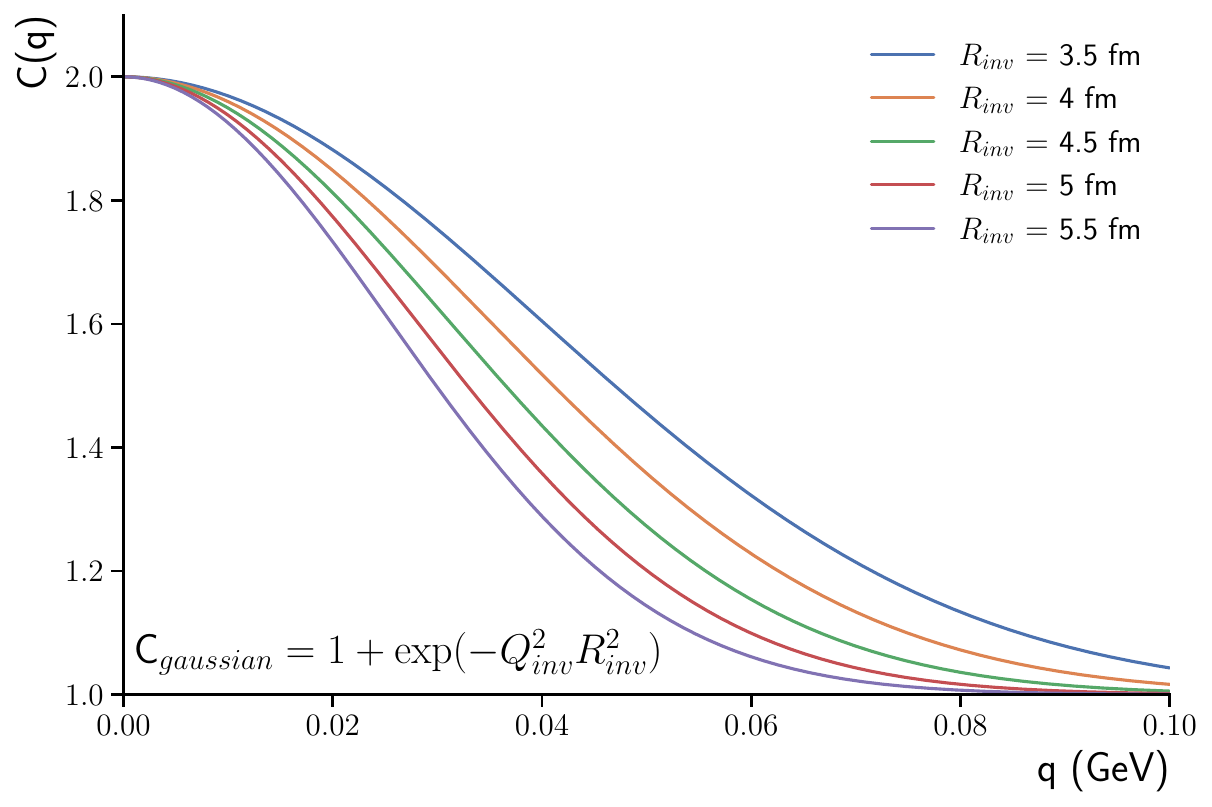}
    \caption{One-dimensional Gaussian correlation function $C(q)$ evaluated for several radius values $R_{inv}$ for bosons and $\lambda = 1$.}
\end{figure}

\subsubsection{The $\boldsymbol\lambda$ Parameter and Core-Halo Interpretation}

\begin{figure}[!htbp] 
    \centering
    \includegraphics[width=\linewidth]{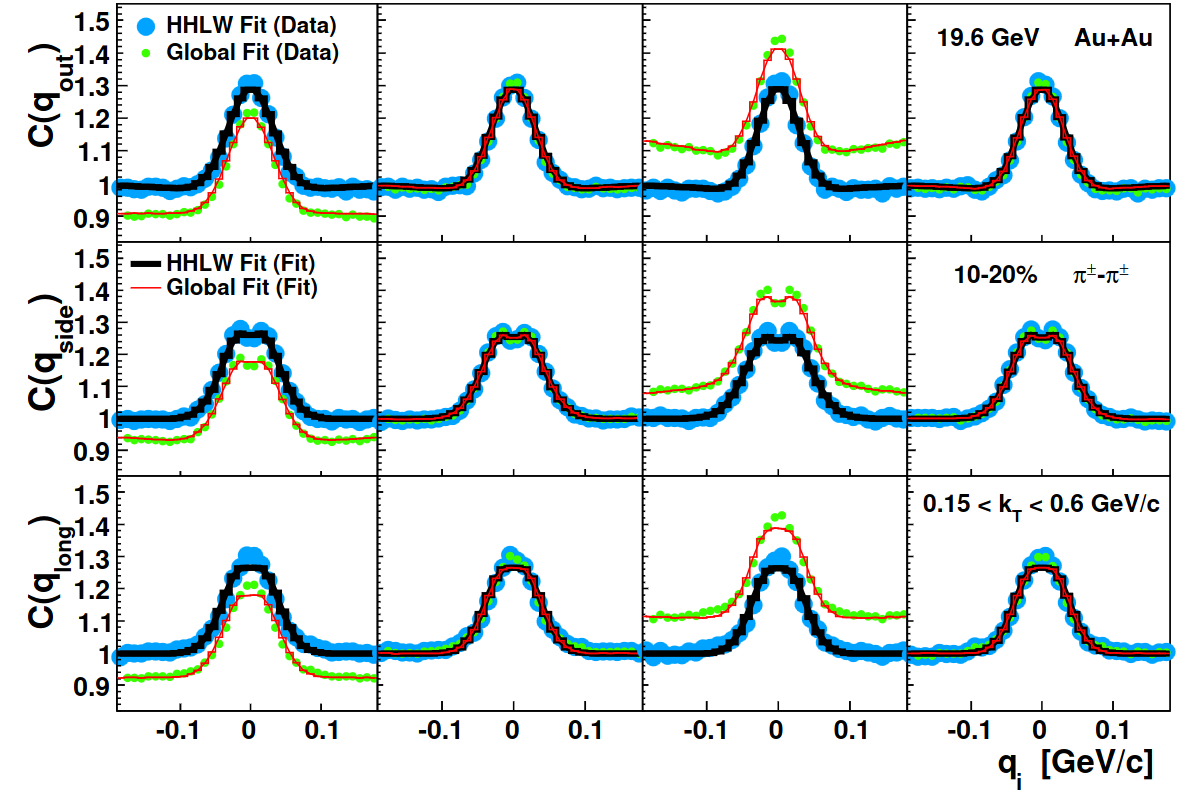}
    \caption{Experimental correlation functions with intercepts below the ideal value $C(0)=2$. Taken from Ref.~\cite{AdamczykSTAR2015HBTBES}.}
    \label{fig:lambda_intercept_example}
\end{figure}

For a completely chaotic Gaussian source without additional correlations and with perfect experimental resolution, the intercept of the identical-boson correlation function would be $C(\vec q=0)=2$, corresponding to $\lambda=1$. In experimental data analyses, the measured intercept is often smaller than this ideal value, as we can see in Fig. \ref{fig:lambda_intercept_example}. This reduction can be caused by long-lived resonance decays, partial coherence, final-state interactions, particle misidentification, finite momentum resolution, or binning effects. 

We say the parameter $\lambda$ controls the effective strength or intercept of the femtoscopic correlation. A standard way to interpret this reduction is the core–halo picture~\cite{WiedemannHeinz1999, 10.1007/BF02907008}. In this picture, the ``core'' is the compact part of the emitting source, formed by directly emitted pions and pions from short-lived resonances, whose space-time separations are resolved by the femtoscopic measurement. The ``halo'' is formed by pions from long-lived resonances or very late decays, whose characteristic length scales are much larger than the experimentally resolvable femtoscopic region. These long-distance contributions generate an extremely narrow correlation structure in relative momentum and therefore reduce the intercept extracted from a finite-resolution fit.

Writing the total pion yield at fixed $k_T$ as
\begin{equation}
    N_{\rm tot}(k_T)=N_{\rm core}(k_T)+N_{\rm halo}(k_T),
\end{equation}
one defines the core fraction as
\begin{equation}
    f_{\rm core}(k_T)=\frac{N_{\rm core}(k_T)}{N_{\rm tot}(k_T)}
    =
    \frac{N_{\rm core}(k_T)}
    {N_{\rm core}(k_T)+N_{\rm halo}(k_T)}.
\end{equation}
Since the Bose-Einstein enhancement is a two-particle correlation, the resolved correlation strength is proportional to the probability that both particles in the pair come from the core. Therefore, in the ideal chaotic and pure-source limit,
\begin{equation}
    \lambda(k_T)\simeq f_{\rm core}^2(k_T)
    =
    \left[
    \frac{N_{\rm core}(k_T)}
    {N_{\rm core}(k_T)+N_{\rm halo}(k_T)}
    \right]^2 .
\end{equation}
In practice, $\lambda$ should be regarded as an effective parameter: particle misidentification, finite momentum resolution, Coulomb treatment, residual non-Gaussian structure, pair cuts, and contamination from nonprimary pions can further reduce or modify the fitted value. For this reason, $\lambda$ is useful as a diagnostic of the effective correlation strength and source composition, but it is generally less direct as a geometric observable than the HBT radii.

\subsection{Bowler-Sinyukov Corrections}
\label{sec:bowler_sinyukov}

The Gaussian parameterization introduced above describes the Bose-Einstein enhancement associated with the symmetrization of identical bosons. For charged pion pairs, the measured small-\(q\) structure is also affected by the Coulomb final-state interaction, which can be incorporated through the Coulomb two-particle wave function in the Koonin-Pratt formalism~\cite{Sinyukov1998,Mrowczynski2009}. This effect is already contained, at the formal level, in Eq.~\eqref{eq:gencorfunc2} through the replacement of the free two-particle wave function by the interacting one. In actual experimental HBT fits, one usually does not evaluate the full Coulomb wave function pair by pair. Instead, the Coulomb effect is incorporated through an averaged correction factor \(K_{\rm Coul}\). Following the notation introduced in Sect.~\ref{sec:kinematic_variables}, let the prime variables denote the quantities evaluated in the center-of-momentum frame of the pair. In this frame $\vec p_1^{\,\prime} + \vec p_2^{\,\prime}=0$, $q^{\prime 0}=0$, $q_{\rm inv}=|\vec q^{\,\prime}|$. For identical particles, one may define
\begin{equation}
    \vec p^{\,\prime}
    \equiv
    \vec p_1^{\,\prime}
    =
    -\vec p_2^{\,\prime}
    =
    \frac{\vec q^{\,\prime}}{2},
    \qquad
    |\vec p^{\,\prime}|=\frac{q_{\rm inv}}{2}.
\end{equation}
The relative separation entering the two-particle wave function is written as
\begin{equation}
    \vec r^{\,\prime}
    =
    \vec x_1^{\,\prime}
    -
    \vec x_2^{\,\prime},
\end{equation}
where the positions are also evaluated in the pair center-of-momentum frame. For a pair of charged particles, the Coulomb scattering wave function may be written as~\cite{Mrowczynski2009}
\begin{equation}
    \psi_{\rm Coul}(\vec p^{\,\prime},\vec r^{\,\prime})
    =
    \sqrt{A_{\rm C}(\eta)}
    e^{i\vec p^{\,\prime}\cdot\vec r^{\,\prime}}
    F\left(
        -i\eta,1,
        i\left[
            |\vec p^{\,\prime}|\,|\vec r^{\,\prime}|
            -
            \vec p^{\,\prime}\cdot\vec r^{\,\prime}
        \right]
    \right),
    \label{eq:coulomb_wavefunction}
\end{equation}
where \(F(a,b,z)\equiv {}_1F_1(a;b;z)\) is the confluent hypergeometric function. The Coulomb parameter is
\begin{equation}
    \eta
    =
    \frac{\alpha_{\rm em}z_1z_2\mu}{|\vec p^{\,\prime}|}
    =
    \frac{1}{|\vec p^{\,\prime}|a_{\rm C}},
    \label{eq:coulomb_eta}
\end{equation}
where \(\alpha_{\rm em}\) is the fine-structure constant, \(\mu\) is the reduced mass, \(z_1\) and \(z_2\) are the particle charges in units of the elementary charge, and \(a_{\rm C}\) is the two-body Bohr radius. The factor
\begin{equation}
    A_{\rm C}(\eta)
    =
    \frac{2\pi\eta}{\exp(2\pi\eta)-1}
    \label{eq:gamow_factor}
\end{equation}
is the Gamow factor. For pairs of pions of the same sign, \(z_1z_2=+1\), \(\mu=m_\pi/2\), and the interaction is repulsive. Therefore, the pair probability is suppressed for small relative momentum. For an extended source, the Coulomb correction is obtained by averaging the squared Coulomb wave function over a relative source distribution~\cite{Sinyukov1998,Mrowczynski2009},
\begin{equation}
    K_{\rm Coul}(q_{\rm inv})
    =
    \int d^3r^{\,\prime}\,
    S_{\rm rel}(\vec r^{\,\prime})
    \left|
    \psi_{\rm Coul}(\vec p^{\,\prime},\vec r^{\,\prime})
    \right|^2,
    \label{eq:kcoul_definition}
\end{equation}
For a spherical Gaussian relative source, one commonly uses
\begin{equation}
    S_{\rm rel}(\vec r^{\,\prime};R_{\rm C})
    =
    \frac{1}{(4\pi R_{\rm C}^2)^{3/2}}
    \exp\left[
        -\frac{|\vec r^{\,\prime}|^2}{4R_{\rm C}^2}
    \right],
    \label{eq:relative_gaussian_source}
\end{equation}
so that
\begin{equation}
    K_{\rm Coul}(q_{\rm inv};R_{\rm C})
    =
    \frac{1}{(4\pi R_{\rm C}^2)^{3/2}}
    \int d^3r^{\,\prime}\,
    \exp\left[
        -\frac{|\vec r^{\,\prime}|^2}{4R_{\rm C}^2}
    \right]
    \left|
    \psi_{\rm Coul}(\vec p^{\,\prime},\vec r^{\,\prime})
    \right|^2 .
    \label{eq:kcoul_gaussian_source}
\end{equation}
In STAR pion interferometry analyses, for example, the correction factor \(K_{\rm Coul}\) is obtained by averaging \(|\psi_{\rm Coul}|^2\) over a spherical Gaussian source with a fixed Coulomb radius~\cite{AdamsSTAR2005,AdamczykSTAR2015HBTBES}. In the one-dimensional Gaussian case, the Bowler-Sinyukov form is
\begin{equation}
    C(q)
    =
    (1-\lambda)
    +
    \lambda K_{\rm Coul}(Q_{inv})
    \left(
    1+\exp\left[
        -Q_{inv}^2R_{inv}^2
    \right]
    \right),
    \label{eq:bowler_sinyukov_1d}
\end{equation}
In the OSL decomposition and in the LCMS, the corresponding three-dimensional parameterization used for identical pion pairs is
\begin{equation}
\small
    C(\vec q,k_T)
    =
    (1-\lambda)
    +
    \lambda K_{\rm Coul}(q_{\rm inv})
    \left(
    1+
    \exp\left[
    -q_o^2R_o^2(k_T)
    -q_s^2R_s^2(k_T)
    -q_l^2R_l^2(k_T)
    \right]
    \right).
    \label{eq:bowler_sinyukov_3d}
\end{equation}

\subsection{Lévy-Stable Emission}

The Gaussian parameterization introduced in Sect.~\ref{sec:3.10} provides a useful baseline for extracting HBT radii: within this approximation, the curvature of the correlation function near $\vec q=0$ is related to the second central moments of the effective emitting source. To describe possible deviations from this baseline, we now consider a more general parametrization that contains the Gaussian limit and lets the data determine the shape of the correlation function~\cite{CsorgoHegyiZajc2004,AdarePHENIX2018Levy}.

A natural generalization is given by Lévy-stable source distributions. Their relevance follows from the generalized central limit theorem~\cite{GnedenkoKolmogorov1954,Nolan2020Stable}. If a random variable is obtained as the sum of many independent elementary contributions with finite variance, the limiting distribution is Gaussian. If, on the other hand, the elementary contributions have sufficiently broad power-law tails, the limiting distribution belongs to the broader class of stable distributions. In this case, the Gaussian distribution is only the special member corresponding to the Lévy index of stability $\alpha=2$. For $0<\alpha<2$, the distribution remains stable under convolution after appropriate rescaling, but its tails are wider than the Gaussian ones.

In the context of femtoscopy, this provides a useful phenomenological framework. The effective emission function probed by HBT interferometry is not the full geometrical source but the source of particles with a given pair momentum. If this effective source is not well described by a Gaussian, the corresponding correlation peak need not be Gaussian either. A Lévy-stable source leads to a stretched-exponential form in relative momentum~\cite{CsorgoHegyiZajc2004},
\begin{equation}
    C(q) - 1 \propto \exp\left[-(qR)^\alpha\right],
\end{equation}
where $R$ is a Lévy scale parameter and $\alpha$ controls the shape of the correlation function. The Gaussian form is recovered for $\alpha=2$, while $\alpha=1$ corresponds to an exponential dependence in $q$, associated with a Cauchy-type source in the coordinate space. Smaller values of $\alpha$ correspond to increasingly long-tailed non-Gaussian source distributions.

To make the connection with the Gaussian case explicit, let $\rho(x)$ be a normalized four-dimensional effective source distribution. Its characteristic function is defined as
\begin{equation}
    \phi(Q)
    \equiv
    \left\langle \exp(-iQ\cdot x) \right\rangle
    =
    \int d^4x\, \rho(x)\,\exp(-iQ\cdot x).
\end{equation}
For a Gaussian source with mean $\bar x^\mu$ and covariance matrix $B^{\mu\nu}$, one obtains
\begin{equation}
    \phi_G(Q)
    =
    \exp\left[
        -iQ_\mu \bar x^\mu
        -\frac{1}{2}Q_\mu B^{\mu\nu}Q_\nu
    \right].
\end{equation}
The logarithm of the Gaussian characteristic function is, therefore, quadratic in $Q$. This quadratic structure is what allows the HBT radii, in the Gaussian approximation, to be related to second central moments of the effective emission region.

The symmetric Lévy-stable generalization replaces this quadratic dependence by a fractional power~\cite{CsorgoHegyiZajc2004,Nolan2020Stable},
\begin{equation}
    \phi_L(Q)
    =
    \exp\left[
        -iQ_\mu \bar x^\mu
        -\frac{1}{2}
        \left(Q_\mu B^{\mu\nu}Q_\nu\right)^{\alpha/2}
    \right],
    \qquad
    0<\alpha\leq 2.
\end{equation}
The parameter $\alpha$ is the Lévy index of stability. For $\alpha=2$, the Gaussian characteristic function is recovered. For $\alpha<2$, the corresponding ideal Lévy-stable distribution has power-law tails and does not possess a finite variance. Consequently, the matrix $B^{\mu\nu}$ should not be interpreted as a covariance matrix in the same sense as in the Gaussian case. Instead, it should be understood as a scale matrix that controls the width of the homogeneity region probed by the correlation function.

The inverse Fourier transform defines the corresponding Lévy-stable source,
\begin{equation}
    \mathcal{L}(x;B,\alpha)
    =
    \frac{1}{(2\pi)^4}
    \int d^4Q\,
    \exp(iQ\cdot x)
    \exp\left[
        -\frac{1}{2}
        \left(Q_\mu B^{\mu\nu}Q_\nu\right)^{\alpha/2}
    \right],
\end{equation}
where, for simplicity, the mean position has been set to zero. A nonzero mean only generates a phase in the Fourier transform and drops out of the squared modulus entering the correlation function.

We now write the effective source distribution as
\begin{equation}
    \rho_{\rm eff}(x,p_1,p_2)
    =
    \kappa(p_1,p_2)\,
    \mathcal{L}\left(x;B(p_1,p_2),\alpha\right),
\end{equation}
where $\kappa$ is a residual normalization factor. Using the smoothness approximation, the dependence on $p_1$ and $p_2$ is replaced by a dependence on the average pair momentum $k$. The Fourier transform then becomes the following.
\begin{equation}
    \widetilde{\rho}_{\rm eff}(k,q)
    =
    \kappa(k)
    \exp\left[
        -\frac{1}{2}
        \left(q_\mu B^{\mu\nu}(k)q_\nu\right)^{\alpha/2}
    \right].
\end{equation}
Substitution into the identical-particle correlation function gives
\begin{equation}
    C(k,q)
    =
    1
    \pm
    \lambda(k)
    \exp\left[
        -
        \left(q_\mu B^{\mu\nu}(k)q_\nu\right)^{\alpha/2}
    \right],
\end{equation}
with $\lambda(k)=|\kappa(k)|^2$. In the pion HBT analyses considered in this work, the bosonic sign is relevant.

\begin{figure}[!htbp] 
    \centering
    \includegraphics[width=\linewidth]{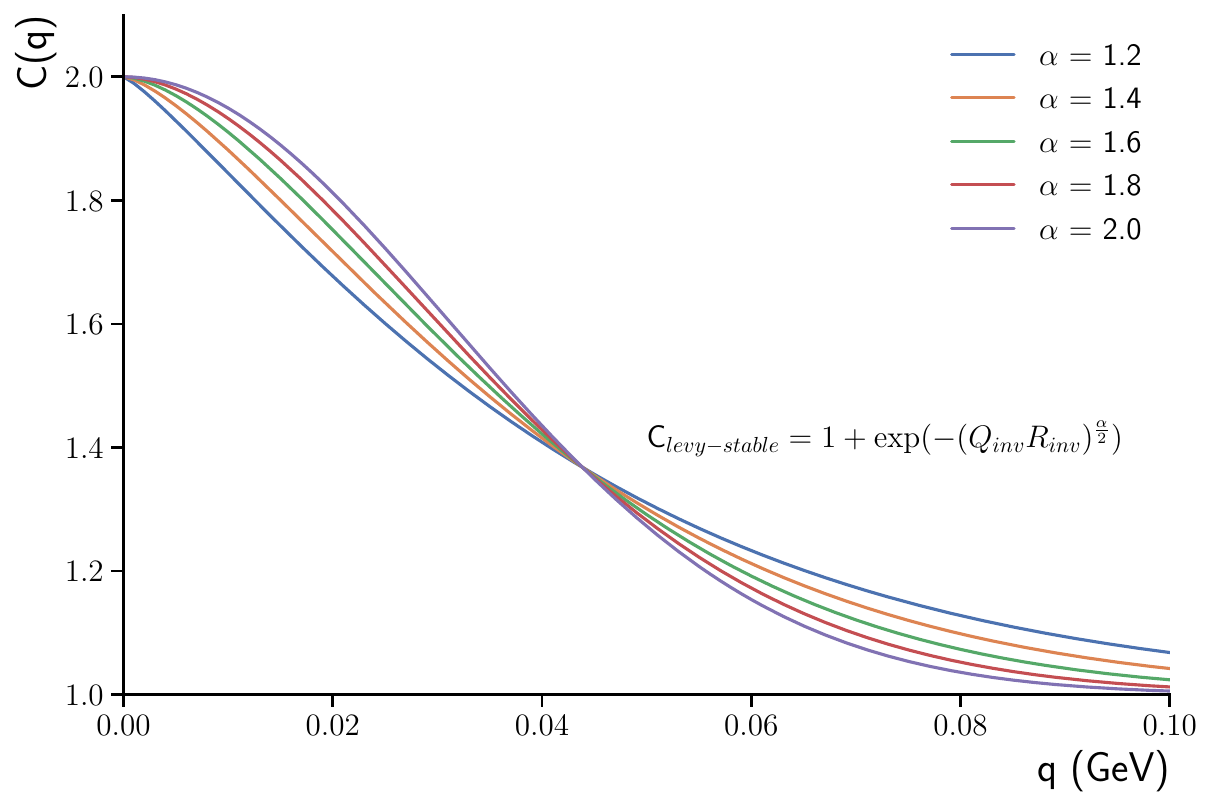}
    \caption{One-dimensional Lévy-Stable correlation function $C(q)$ evaluated for several $\alpha$ values with $R_{inv} = 4.5$ fm, for bosons and $\lambda = 1$.}
    \label{fig:levyexample}.
\end{figure}

Analogously to the Gaussian case, the on-shell condition allows us to express the four-dimensional quadratic form in terms of the spatial components of the relative momentum. We also replace $q_\mu B^{\mu\nu}(k)q_\nu$ by $q^i R_{ij}^2(k) q^j$. For $\alpha=2$, the quantities $R_{ij}^2$ coincide with the familiar Gaussian HBT radii defined through the second moments of the effective source. When $\alpha<2$, on the contrary, these parameters are viewed more appropriately as Lévy scale parameters rather than genuine variances. In this representation, the correlation function takes the form
\begin{equation}
    C(k,q)
    =
    1
    \pm
    \lambda(k)
    \exp\left[
        -
        \left(q^i R_{ij}^2(k)q^j\right)^{\alpha/2}
    \right].
\end{equation}
In the Bertsch-Pratt decomposition, combined with the LCMS, and for azimuthally integrated analyses where cross terms are neglected, this becomes
\begin{equation}
    C(\vec q,k_T)
    =
    1
    \pm
    \lambda(k_T)
    \exp\left\{
        -
        \left[
            q_o^2R_o^2(k_T)
            +
            q_s^2R_s^2(k_T)
            +
            q_l^2R_l^2(k_T)
        \right]^{\alpha/2}
    \right\}.
    \label{eq:levy_3d_hbt}
\end{equation}
For $\alpha=2$, Eq.~\eqref{eq:levy_3d_hbt} reduces to the usual Gaussian form. Thus, the Lévy-stable parametrization preserves the familiar femtoscopic scales $R_o$, $R_s$, and $R_l$, while introducing an additional source-shape parameter $\alpha$. This Lévy extension also applies to correlation functions of \(Q_{inv}\), as we can see plotted in Figure \ref{fig:levyexample}.

The physical interpretation of $\alpha$ has two complementary aspects. The first is conservative and purely phenomenological. In this interpretation, $\alpha$ quantifies the deviation of the effective source from a Gaussian shape. A value close to $\alpha=2$ indicates that the Gaussian approximation captures the shape of the measured correlation peak reasonably well. Values significantly below $2$ indicate that the correlation function has a stretched-exponential form and that the corresponding effective source contains non-Gaussian structures. In a realistic hybrid description, such deviations may arise from resonance decays, hadronic rescattering, finite emission duration, event-by-event fluctuations, anomalous diffusion, or the interplay between collective flow and freeze-out geometry. Therefore, even without invoking critical behavior, $\alpha$ provides an additional femtoscopic diagnostic of the full dynamical evolution~\cite{CsorgoHegyiZajc2004,AdarePHENIX2018Levy}.

The second interpretation is connected with the possible existence of a critical endpoint in the QCD phase diagram. Near a second-order phase transition, the correlation length grows, and the order-parameter correlation function develops a power-law behavior,
\begin{equation}
    \langle \phi(\vec r)\phi(0)\rangle
    -
    \langle \phi\rangle^2
    \sim
    \frac{1}{r^{d-2+\eta}},
\end{equation}
where \(r=|\vec r|\), \(d\) is the number of spatial dimensions, and \(\eta\) is a critical exponent. Since Lévy-stable distributions also describe scale-free, long-tailed structures, it has been proposed that the Lévy index extracted from Bose-Einstein or HBT correlations may become related to the critical exponent $\eta$ near a second-order QCD phase transition~\cite{CsorgoHegyiNovakZajc2005,CsorgoHegyiNovakZajc2006},
\begin{equation}
    \alpha_{\rm L\acute{e}vy}
    \sim
    \eta .
\end{equation}
In this scenario, the beam-energy dependence of $\alpha$ could provide a femtoscopic signature of the vicinity of the QCD critical endpoint.

This connection must be treated with care. The value of $\eta$ depends on the universality class assumed for the critical point. For the ordinary three-dimensional Ising universality class, the anomalous dimension is very small, with modern estimates giving approximately~\cite{ElShowk2014Ising}
\begin{equation}
    \eta_{\rm 3D\,Ising}
    \simeq
    0.036 .
\end{equation}
On the other hand, the femtoscopic literature discussing Lévy sources and critical behavior often emphasizes a larger value,
\begin{equation}
    \eta \simeq 0.50 \pm 0.05,
\end{equation}
associated with the three-dimensional random-field Ising model, together with a disconnected-susceptibility exponent
\begin{equation}
    \bar\eta \simeq 1.03 \pm 0.05 .
\end{equation}
Here, \(\bar\eta\) characterizes the scaling of the disconnected part of the order-parameter susceptibility in the random-field Ising framework, and should not be confused with the ordinary anomalous dimension \(\eta\). This is the origin of the commonly quoted expectation that a Lévy index approaching $\alpha_{\rm L\acute{e}vy}\sim 0.5$ could suggest critical behavior in this framework~\cite{Rieger1995RFIM,CsorgoHegyiNovakZajc2005,CsorgoHegyiNovakZajc2006}. Therefore, observation of $\alpha<2$ alone is not sufficient to identify the critical endpoint. Noncritical mechanisms can also generate non-Gaussian tails. A meaningful critical interpretation would require a systematic nonmonotonic behavior of $\alpha$ as a function of collision energy, centrality, transverse mass, and particle species, together with consistency with other observables sensitive to the QCD phase structure.

\begin{figure}[!htbp]
    \centering
    \includegraphics[width=\textwidth]{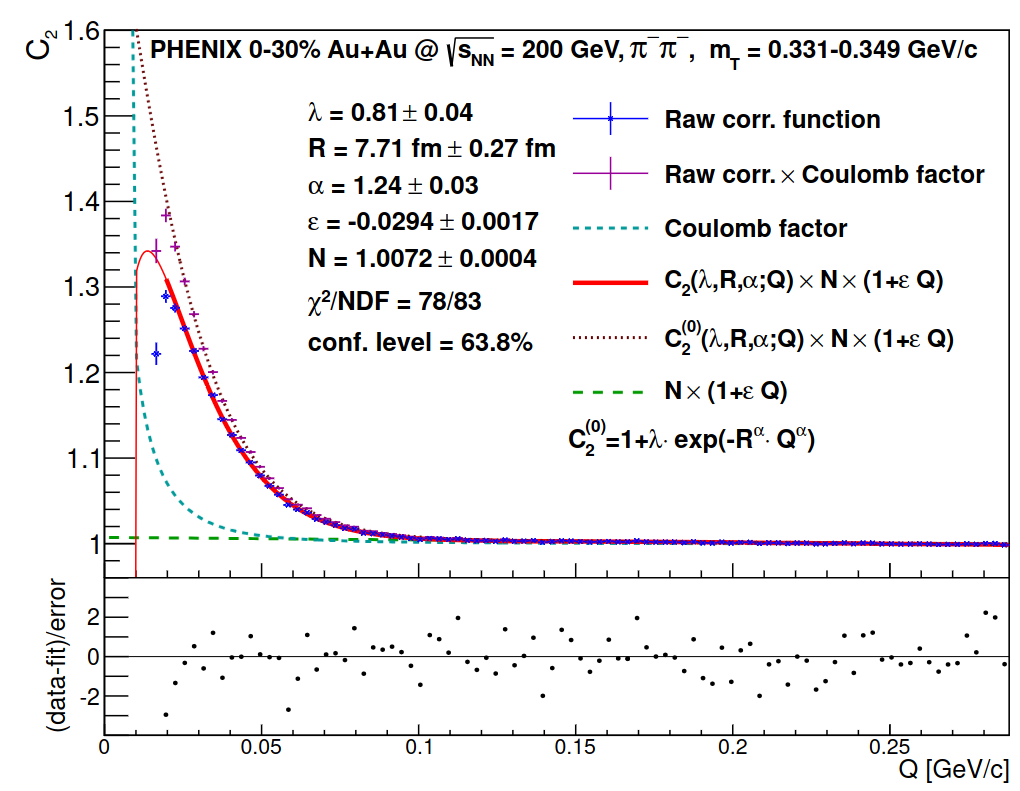}
    \caption{Example of a Lévy-stable fit to a two-pion Bose-Einstein correlation function measured by PHENIX in $0$-$30\%$ Au+Au collisions at $\sqrt{s_{NN}}=200$ GeV. The blue points show the raw $\pi^-\pi^-$ correlation function, while the red curve shows the fit including the Lévy-stable Bose-Einstein correlation, Coulomb correction, normalization, and residual long-range baseline. The corresponding uncorrected Lévy correlation component is also shown. This example illustrates how the parameters $\lambda$, $R$, and $\alpha$ are extracted from the measured correlation peak, with $\alpha<2$ indicating a non-Gaussian effective source shape. Taken from Ref.~\cite{AdarePHENIX2018Levy}.}
    \label{fig:phenix_levy_fit_example}
\end{figure}

Experimental measurements have shown that Lévy-stable forms can describe pion correlation functions in relativistic heavy-ion collisions. In particular, PHENIX analyses of two-pion Bose-Einstein correlations in \(\sqrt{s_{NN}}=200\) GeV Au+Au collisions found values of \(\alpha\) below the Gaussian limit \(\alpha=2\) (Fig.~\ref{fig:phenix_levy_fit_example}), indicating non-Gaussian source shapes, but not values close to the \(\alpha\sim0.5\) expectation associated with the critical scenario~\cite{AdarePHENIX2018Levy}. The example shown below should therefore be understood as a single-energy illustration of the method. A critical-point interpretation would require a beam-energy scan of \(\alpha\), looking for a systematic and nonmonotonic dependence on \(\sqrt{s_{NN}}\) rather than only a deviation from the Gaussian limit at one collision energy. This supports the usefulness of Lévy fits as a controlled extension of Gaussian femtoscopy, while also illustrating that a non-Gaussian source shape does not by itself constitute evidence for critical behavior.
\subsection{Analytical Estimates for Expanding Sources}

The interpretation of HBT radii as homogeneity lengths can be made more explicit through semi-analytical calculations based on hydrodynamic emission functions. A useful example is the analytical calculation of HBT radii for a longitudinally boost-invariant source, based on the Bjorken flow~\cite{Bjorken1983}.

First, let us consider a boost-invariant longitudinally expanding system with finite transverse support of radius \(R\). The hypersurface element is given by~\cite{PhysRevC.77.034911}
\begin{equation}
    d\sigma_\mu
    =
    r\tau
    \left(
    \cosh\eta_S,0,0,-\sinh\eta_S
    \right)
    dr\,d\phi\,d\eta_S .
\end{equation}
where \(r = \sqrt{x^2+y^2}\) is the radius in the transverse plane and \(\phi = \arctan(y/x)\) is the azimuthal angle. For a system emitting particles in the Bertsch–Pratt coordinate system + LCMS, the mean momentum of a particle pair can be described by the variable \(k = (m_T, k_T, 0, 0)\), where we use the approximation \eqref{eq:mtapprox}. Taking Eq.~\eqref{eq:cfef} with \(H\) given by a delta in Milne–cylindrical coordinates,
\begin{equation}
    H(x - x')
    =
    \frac{1}{\tau r}
    \delta(\tau-\tau')
    \delta(r-r')
    \delta(\phi-\phi')
    \delta(\eta_S-\eta'_S),
\end{equation}
and using a Boltzmann-type distribution \(f(x,k) = \exp(-u\cdot k/T)\), we obtain
\begin{equation}
    S(x,k)
    =
    m_T \cosh\eta_S
    \exp\left[-\frac{m_T\cosh\eta_S}{T}\right]
    \delta(\tau-\tau'),
\end{equation}
where the boost-invariant flow is given by~\cite{Bjorken1983}
\begin{equation}
    u^\mu
    =
    \left(
    \cosh\eta_S,0,0,\sinh\eta_S
    \right),
\end{equation}
and the integrals over the hypersurface vanish due to the delta functions. The complex integral in \eqref{eq:sacf} can be computed as (here using generalized coordinates)
\begin{equation}
    \int d^4x \, S(x,k)e^{iq\cdot x}
    =
    \tau m_T A(q_o,q_l) B(q_o,q_s),
\end{equation}
where in Milne cylindrical coordinates $\sqrt{-g} = \tau r$
\begin{align}
    A(q_o,q_l)
    &=
    \int_{-\infty}^{\infty}d\eta_S\,
    \cosh\eta_S\,
    \exp\left[-\frac{m_T}{T}\cosh\eta_S\right]
    \exp\left[
    i\tau\left(q_o\beta_T\cosh\eta_S-q_l\sinh\eta_S\right)
    \right],
    \\
    B(q_o,q_s)
    &=
    \int_0^R r\,dr
    \int_0^{2\pi}d\phi\,
    \exp\left[
    -ir\left(q_o\cos\phi+q_s\sin\phi\right)
    \right],
\end{align}
where $R$ is the radius of the colliding nuclei. These integrals can be calculated analytically. Let us start with the longitudinal part. We define $a \equiv \frac{m_T}{T}$, $b \equiv \tau q_l$, and $c \equiv \tau\beta_T q_o$, so that
\begin{equation}
    A(q_o,q_l)
    =
    \int_{-\infty}^{\infty}d\eta_S\,
    \cosh\eta_S\,
    \exp\left[-(a-ic)\cosh\eta_S-ib\sinh\eta_S\right].
\end{equation}
Using the integral representation of the modified Bessel function of the second kind \(K_\nu(z)\)~\cite{AbramowitzStegun1964,NISTHandbook2010,DLMF}, the longitudinal integral can be written as
\begin{equation}
    A(q_o,q_l)
    =
    2\frac{a-ic}{\rho}K_1(\rho),
    \label{eq:bjorken_A_result}
\end{equation}
where 
\begin{equation}
\rho = \sqrt{(a-ic)^2+b^2}.
\end{equation}
In the limit \(q_o = q_l = 0\), we obtain $A(0,0) = 2K_1(a)$. The transverse part can be written in terms of ordinary Bessel functions~\cite{AbramowitzStegun1964,NISTHandbook2010,DLMF}. Defining 
\begin{align}
    &q_T = \sqrt{q_o^2+q_s^2}, \\
    &z = q_T R,
\end{align}
we have
\begin{equation}
    q_o\cos\phi+q_s\sin\phi
    =
    q_T\cos(\phi-\alpha),
\end{equation}
where \(q_o=q_T\cos\alpha\) and \(q_s=q_T\sin\alpha\). Then,
\begin{equation}
    \int_0^{2\pi}d\phi\,
    \exp\left[-ir(q_o\cos\phi+q_s\sin\phi)\right]
    =
    2\pi J_0(q_T r),
\end{equation}
where \(J_\nu\) is the ordinary Bessel function of the first kind. Therefore,
\begin{align}
    B(q_o,q_s)
    &=
    2\pi\int_0^R r\,dr\,J_0(q_T r)
    \nonumber\\
    &=
    2\pi R\frac{J_1(q_T R)}{q_T}
    \nonumber\\
    &=
    \pi R^2\frac{2J_1(z)}{z}.
    \label{eq:bjorken_B_result}
\end{align}
At the limit \(q_T\rightarrow0\), like \(J_1(z)\simeq z/2\), it follows that
\begin{equation}
    B(0,0)
    =
    \pi R^2.
\end{equation}
On the other hand, the integral without the phase is obtained by taking \(q_o=q_s=q_l=0\):
\begin{align}
    \int d^4x\,S(x,k)
    &=
    \tau m_T A(0,0)B(0,0)
    \nonumber\\
    &=
    2\pi R^2\tau m_T K_1(a).
    \label{eq:bjorken_source_norm}
\end{align}
Thus, for identical bosons, Eq.~\eqref{eq:sacf} becomes

\begin{equation}
    C(k,q)
    =
    1+
    \left|
    \frac{a-ic}{\rho}
    \frac{K_1(\rho)}{K_1(a)}
    \right|^2
    \left[
    \frac{2J_1(z)}{z}
    \right]^2 .
    \label{eq:bjorken_correlation_exact}
\end{equation}
The usual Gaussian form of the HBT radii is obtained only in the limit of small relative momentum \(q\), that is,
\begin{align}
    &z\ll1,\\
    &|c|\ll1,\\
    &|b|\ll1.
\end{align}
In this limit~\cite{AbramowitzStegun1964,NISTHandbook2010,DLMF},
\begin{equation}
    \left[
    \frac{2J_1(z)}{z}
    \right]^2
    =
    1-\frac{z^2}{4}+\mathcal O(z^4)
    \simeq
    \exp\left[-\frac{z^2}{4}\right].
\end{equation}
The longitudinal-temporal part can be expanded by taking the logarithm of the first factor of the Eq.~\eqref{eq:bjorken_correlation_exact}. Up to second order in \(q\), we obtain\cite{AbramowitzStegun1964,NISTHandbook2010,DLMF}
\begin{equation}
    \ln\left|
    \frac{a-ic}{\rho}
    \frac{K_1(\rho)}{K_1(a)}
    \right|^2
    =
    -c^2\frac{d^2}{da^2}\ln K_1(a)
    -b^2\frac{K_2(a)}{aK_1(a)}
    +\mathcal O(q^4).
    \label{eq:bjorken_long_expansion}
\end{equation}
Thus
\begin{equation}
    C(k,q) \simeq 1+ \exp\left[-q_o^2R_o^2-q_s^2R_s^2-q_l^2R_l^2\right],
\end{equation}
where $\lambda = 1$ and
\begin{align}
    R_o^2
    &=
    \frac{R^2}{4}
    +
    \tau^2\beta_T^2
    \frac{d^2}{da^2}\ln K_1(a),
    \\
    R_s^2
    &=
    \frac{R^2}{4},
    \\
    R_l^2
    &=
    \tau^2
    \frac{K_2(a)}{aK_1(a)}.
\end{align}

\begin{figure}[!htbp] 
    \centering
    \includegraphics[width=\linewidth]{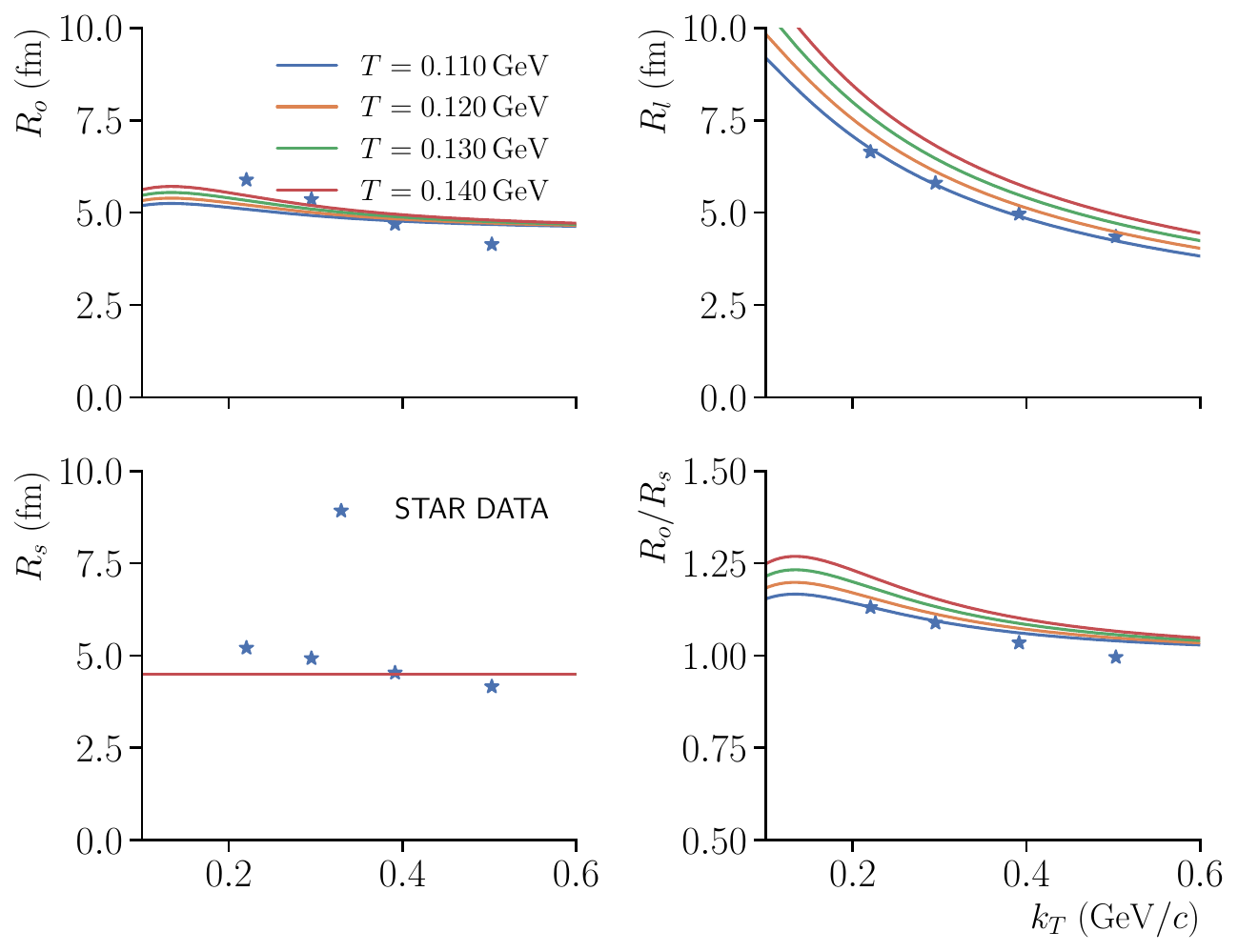}
    \caption{HBT radii from pion interferometry, calculated using the Bjorken model and adopting the parameters $R = 9$ fm and $\tau = 8$ fm/c for several temperatures $T$. Experimental data are from STAR for Au+Au collisions at 200 GeV in the $0\%$-$5\%$ centrality class. \cite{AdamczykSTAR2015HBTBES}.}
    \label{fig:analyticialhbtradii}
\end{figure}

These expressions show, in a controlled way, how the HBT radii emerge from the curvature of the Fourier transform of the emission function in the small-\(q\) limit. Even in this simple model, the radii do not necessarily coincide with the direct geometric widths of the system: \(R_s\) reflects the transverse extent perpendicular to the average pair momentum, \(R_l\) encodes the longitudinal structure of the boost-invariant source, and \(R_o\) also receives a contribution from the temporal dependence associated with \(t=\tau\cosh\eta\). This analytical example illustrates the relation between the Cooper–Frye emission function and the femtoscopic radii, while at the same time showing that the geometric interpretation of these radii depends on the assumptions made about the emitting source~\cite{WiedemannScottoHeinz1996,LisaPrattSoltzWiedemann2005,WiedemannHeinz1999}. In Figure \ref{fig:analyticialhbtradii}, we present a plot with the analytical expressions for the HBT radii.

\subsection{Emission-Time Sensitive Observable and QCD Phase Diagram Probe}

\begin{figure}[!htbp]
    \centering
    \includegraphics[width=0.55\textwidth]{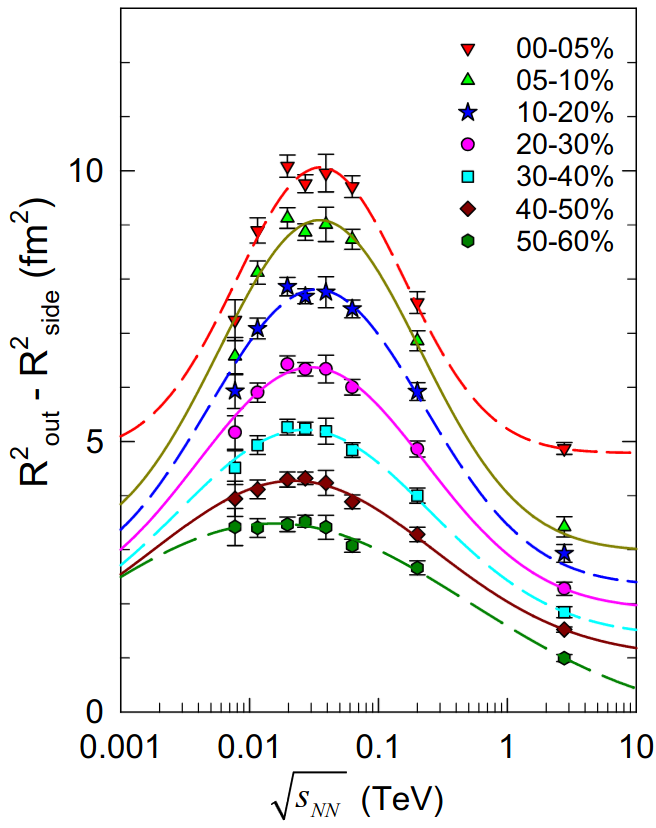}
    \caption{STAR-BES dependence of femtoscopic observables related to the emission duration, as discussed in Ref.~\cite{Lacey2015CEP}. Only statistical uncertainties are taken into account; systematic errors are not shown.}
    \label{fig:lacey_RoRs}
\end{figure}

As discussed in the previous sections, in the OSL decomposition combined with the LCMS, the outward and sideward radii can be written as
\begin{equation}
    R_o^2
    =
    \left\langle \Delta x_o^2 \right\rangle
    -2\beta_T\left\langle \Delta x_o\Delta t\right\rangle
    +\beta_T^2\left\langle \Delta t^2\right\rangle ,
    \qquad
    R_s^2
    =
    \left\langle \Delta x_s^2 \right\rangle .
\end{equation}
Therefore,
\begin{equation}
    R_o^2-R_s^2
    =
    \left(
    \left\langle \Delta x_o^2 \right\rangle
    -
    \left\langle \Delta x_s^2 \right\rangle
    \right)
    -2\beta_T\left\langle \Delta x_o\Delta t\right\rangle
    +\beta_T^2\left\langle \Delta t^2\right\rangle .
    \label{eq:RoRs_general}
\end{equation}
This relation shows that \(R_o^2-R_s^2\) is sensitive to the time structure of the emitting source. Under additional assumptions, such as similar outward and sideward variances and small position-time correlations, one obtains the commonly used estimate
\begin{equation}
    R_o^2-R_s^2
    \approx
    \beta_T^2\left\langle \Delta t^2\right\rangle .
    \label{eq:RoRs_time_approx}
\end{equation}
In this limit, the observable becomes sensitive to the duration of particle emission. Interest in this quantity arises from its link to the space-time dynamics of freeze-out and, indirectly, to the equation of state. A first-order phase transition, or any region where the equation of state softens significantly, can weaken pressure gradients, slow the expansion, and widen the distribution of emission times. This typically enhances the temporal contribution to $R_o^2$. In contrast, a smoother crossover usually results in a more uniform expansion and can yield a shorter effective emission duration. Consequently, the beam-energy dependence of $R_o^2 - R_s^2$ has been employed as a femtoscopic observable to search for the softening of the equation of state associated with a first-order phase transition or the presence of a QCD critical point. HBT radii extracted from identical charged-pion femtoscopy were systematically measured in Au+Au collisions across the Beam Energy Scan I (BES-I) energies~\cite{AdamczykSTAR2015HBTBES}. The nonmonotonic energy dependence of $R_o^2-R_s^2$, with a peak at $\sqrt{s_{NN}} = 19.6$ GeV, was proposed as a possible signature of a critical point~\cite{AdamczykSTAR2015HBTBES,Lacey2015CEP}, as shown in Fig.~\ref{fig:lacey_RoRs}.

Numerous theoretical and phenomenological studies have been carried out, and further work is ongoing, to describe HBT interferometry observables. Hybrid models have also been used to explore the QCD phase diagram and investigate how different equations of state affect the evolution of HBT radii~\cite{https://doi.org/10.1155/2016/9414239, Pratt2009ResolvingHBT, Soltz2013, Habich2015, Naboka2025, PhysRevC.96.024911, HUANG2026140423, PhysRevC.93.044913}.

\section{The HBT Puzzle and its Resolution}
\label{sec:hbt_puzzle}

The HBT puzzle emerged at the beginning of the RHIC program as a tension between the success of hydrodynamics in describing momentum-space observables and its difficulty in reproducing femtoscopic radii. Early hydrodynamic calculations were able to describe, at least qualitatively, transverse-momentum spectra and elliptic flow, suggesting that the matter produced at RHIC behaved as a strongly interacting fluid. The same calculations often predicted pion HBT radii incompatible with the measured two-particle correlations. This was a significant problem because the spectra and anisotropic flow constrain the momentum-space evolution of the system, while the HBT radii constrain the space-time distribution of the particles at decoupling.

\begin{figure}[!htbp]
    \centering
    \includegraphics[width=\linewidth]{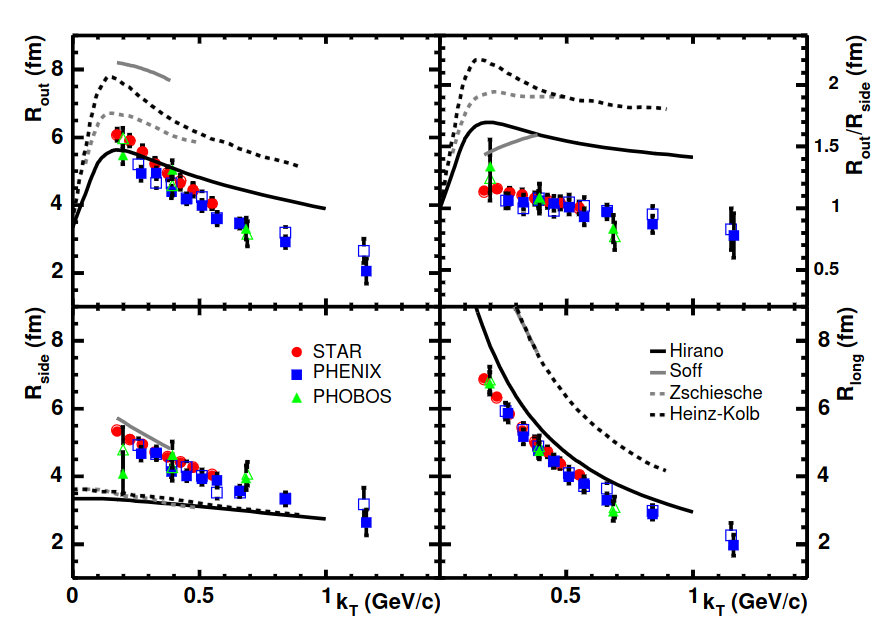}
    \caption{
    Illustration of the HBT puzzle at RHIC. 
    The figure compares pion HBT radii measured by STAR, PHENIX, and PHOBOS in central Au+Au collisions at top RHIC energies with several hydrodynamic calculations as functions of \(k_T\). 
    Early hydrodynamic models generally overpredicted \(R_l\) and \(R_o/R_s\), while the data showed \(R_o/R_s\sim 1\), revealing a tension between the space-time structure predicted by hydrodynamics and the femtoscopic measurements.
    Taken from Ref.~\cite{LisaPrattSoltzWiedemann2005}.
    }
    \label{fig:hbt_puzzle_lisa_review}
\end{figure}

The main tension involved the outward and sideward radii. In the OSL+LCMS decomposition, the outward radius contains both spatial and temporal contributions, whereas the sideward radius is mostly sensitive to the transverse homogeneity length perpendicular to the pair momentum. Early calculations with long-lived sources tended to produce values of \(R_o\) larger than \(R_s\). However, the RHIC data showed \(R_o/R_s\) close to unity. This indicated that the modeled emitting source was too large, too long-lived, or had incorrect space-time correlations. As discussed in the previous section, the difference $R_o^2-R_s^2$ is not a direct clock. It contains a spatial contribution, a position-time correlation term, and a term proportional to the variance of the emission time. 

An early semi-analytic approach used the Blast-Wave model of hydrodynamic expansion developed by Fabrice Retiere and Michael Lisa~\cite{PhysRevC.70.044907}. In this model, rapid, explosive expansion was proposed as a mechanism for reducing the transverse HBT radii $R_o$. Scott Pratt proposed a useful interpretation, arguing that the resolution of the puzzle did not come from a single missing ingredient, but from the combined effect of several corrections that made the evolution of the collision more explosive~\cite{Pratt2009ResolvingHBT}. Early transverse acceleration, a softening equation of state, viscosity, more realistic initial conditions, and a more complete treatment of the hadronic stage modified the predicted homogeneity regions in the direction indicated by the data. In this view, the resolution did not come from identifying a single dominant mechanism, but instead from a confluence of multiple effects acting in a coordinated way~\cite{Pratt2009LongSlow}.

\begin{itemize}

\item  Pre-equilibrium dynamics lower the tension by generating transverse flow before hydrodynamics begins. Early ideal-hydrodynamic studies often assumed negligible initial transverse velocity, delaying radial flow and implying larger sources and longer lifetimes. If pressure gradients act before local equilibrium, the system expands faster from the start, reducing homogeneity lengths and bringing calculated radii closer to experiment.

\item A pronounced first‑order phase transition leads to a soft mixed phase with weak pressure gradients, which slows the expansion and lengthens the system’s lifetime. This generally enhances \(R_l\) and the time-related contribution to \(R_o\). By contrast, a stiffer equation of state, compatible with a crossover at low baryon chemical potential, strengthens the pressure response and reduces the effective emission duration. As a result, the predicted \(R_o/R_s\) ratio shifts closer to the experimental values.

\item Viscous hydrodynamics alters the space-time evolution by changing longitudinal and transverse pressures, producing entropy, shaping radial and anisotropic flow, and modifying particlization distribution functions. These effects can reduce the overly large radii from ideal calculations and better reproduce spectra, flow, and femtoscopy simultaneously. Event-by-event initial conditions further help by replacing smooth geometries with fluctuating density profiles, which change local gradients and final homogeneity regions.

\item A sudden hydrodynamic kinetic freeze-out is too restrictive for heavy-ion collisions. After particlization, particles rescatter, decay, and decouple over a finite space-time region. Hadronic afterburners thus alter particle momenta and last-emission points relevant for HBT. Hybrid models combining hydrodynamics with hadronic transport yield a more realistic effective emission function than a single switching hypersurface.
\end{itemize}

\begin{figure}[!htbp]
    \centering
     \includegraphics[width=0.55\linewidth]{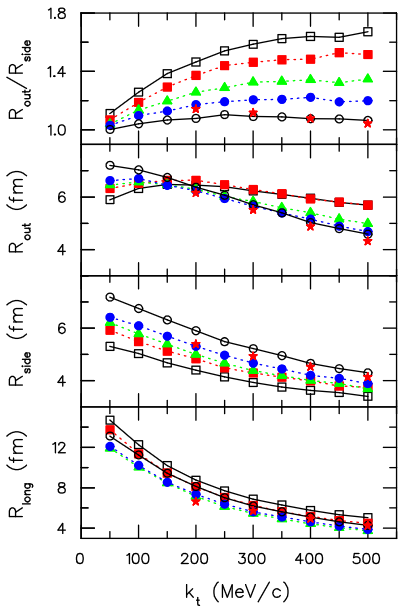}
\caption{Illustration of the progressive resolution of the HBT puzzle in relativistic heavy-ion collisions. The curves show model calculations obtained after adding successive physical ingredients to the dynamical description, including pre-equilibrium acceleration, a stiffer equation of state, viscosity, and a more realistic treatment of the late hadronic stage. The star symbols denote the experimental data points used for comparison. These ingredients reduce the predicted values of \(R_{\rm out}\), \(R_{\rm side}\), \(R_{\rm long}\), and especially the ratio \(R_{\rm out}/R_{\rm side}\), bringing the calculations closer to the measured RHIC femtoscopic radii. The result illustrates that the HBT puzzle was not solved by a single mechanism, but by the combined effect of several improvements that made the collision dynamics more explosive and the effective emitting source more realistic. Taken from Ref.~\cite{Pratt2009ResolvingHBT}.}
    \label{fig:pratt_hbt_puzzle_resolution}
\end{figure}

The modern view is therefore that the original HBT puzzle has been largely reduced by improving the full dynamical description of the collision (Fig.~\ref{fig:pratt_hbt_puzzle_resolution}). Nevertheless, femtoscopy still provides nontrivial constraints. A model may reproduce yields, spectra, and anisotropic flow while still producing an incorrect freeze-out geometry, emission duration, or resonance-decay contribution. Moreover, observables such as \(R_o^2-R_s^2\) remain sensitive not only to the equation of state but also to transverse flow, position-time correlations, hadronic rescattering, kinematic cuts, and final-state interaction corrections.

\section{Further Developments in Femtoscopy}

Although this work focuses mainly on pion HBT radii obtained from Gaussian and L\'evy-stable parameterizations, the scope of femtoscopy is considerably wider and encompasses several complementary approaches. One key extension is azimuthally sensitive HBT, where the radii are investigated as functions of the pair azimuthal angle relative to the event plane. This allows one to probe the anisotropic structure of the homogeneity regions and to relate the femtoscopic observables to the geometry and collective expansion dynamics of the source~\cite{LisaRetiereWells2000,AdamsSTAR2004AzHBT,LisaPrattSoltzWiedemann2005}.

Another direction is the study of final-state interactions beyond Bose-Einstein correlations of identical pions. For charged pions, Coulomb interactions are typically accounted for using correction schemes such as the Bowler-Sinyukov method. For other hadron pairs (especially kaons, protons, hyperons, and non-identical combinations), strong final-state interactions play a crucial role. In these systems, the correlation function serves not only to determine source sizes but also to constrain hadron–hadron interactions at low relative momentum. This has driven femtoscopic investigations of pion, kaon, proton, and strange-hadron pairs, as well as more recent studies involving charm hadrons~\cite{Lednicky2009,Kisiel2010NonIdentical,AcharyaALICE2020CommonBaryon,AcharyaALICE2022CharmScattering,AcharyaALICE2024CharmLight}.

Femtoscopy with non-identical particles offers further information because the pair wave function is not subject to symmetrization. Consequently, the correlation function may become sensitive to the mean relative emission shift between the two-particle species. Such measurements can help constrain the emission sequence arising from collective flow, resonance decays, and hadronic rescattering. Similarly, particle–antiparticle correlations can probe annihilation processes and strong-interaction channels that are inaccessible in identical-particle HBT~\cite{Lednicky2009,Kisiel2010NonIdentical}.

Another approach focuses on squeezed correlations between particles and antiparticles, or back-to-back configurations. These have been proposed as signatures of in-medium modifications of hadron properties. If the mass of a hadron is altered within the hot and dense medium and then reverts to its vacuum value at freeze-out, the mapping between the medium and asymptotic states can induce squeezed correlations between particles and antiparticles emitted with nearly opposite momenta. Although such effects have not yet become standard experimental observables, they exemplify how femtoscopy can be sensitive not only to geometry but also to potential in-medium changes of hadronic degrees of freedom~\cite{AsakawaCsorgoGyulassy1999,DudekPadula2010}.

Recent progress also extends femtoscopic techniques to small systems—high-multiplicity proton–proton and proton–nucleus collisions—as well as to heavy-flavor correlations. In these studies, the femtoscopic signal is used to test whether small collision systems develop collective space–time structures analogous to those in heavy-ion collisions and to constrain the interactions of heavy-flavor hadrons with the hadronic medium. In this broader perspective, femtoscopy goes beyond simply extracting pion source radii. It constitutes a set of correlation observables sensitive to source geometry, emission time scales, final-state interactions, hadronic rescattering, medium effects, and potential bound or resonant hadronic states~\cite{AcharyaALICE2020CommonBaryon,AcharyaALICE2022CharmScattering,AcharyaALICE2024CharmLight,TorresRincon2023DmesonFemtoscopy}.


\chapter{Femtoscopic Analysis of STAR BES using a Hybrid Model}

We performed simulations of $^{197}$Au + $^{197}$Au collisions at a set of center-of-mass energies corresponding to the RHIC Beam Energy Scan program, employing a hydrodynamic workflow based on the Smeared AMPT + PRAIA + BQSSampler + SMASH hybrid model in conjunction with the BEST EOS. From these simulations, we calculated selected observables for charged-hadron and identified-hadron transverse-momentum spectra, and we additionally extracted HBT radii characterizing pion–pion emission.

\begin{figure}[!htbp]
    \centering
    \includegraphics[width=\linewidth]{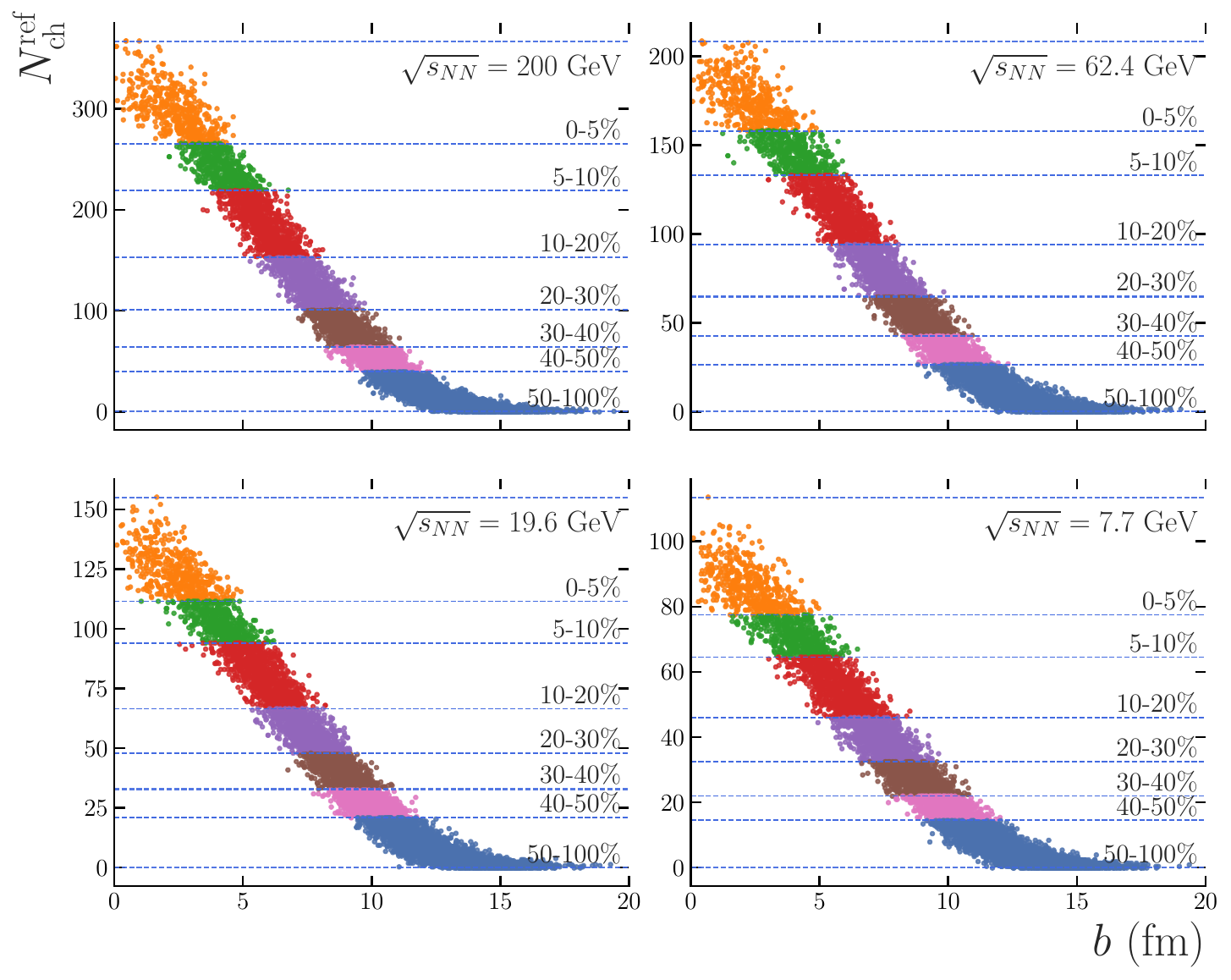}
    \caption{Centrality classification for minimum-bias $^{197}$Au+$^{197}$Au events at $\sqrt{s_{NN}}=200$, $62.4$, $19.6$ and $7.7$ GeV. Each point represents one AMPT event, shown as a function of the impact parameter $b$ and of the reference charged-particle multiplicity $N_{\rm ch}^{\rm ref}$. The dashed horizontal lines indicate the reference-multiplicity intervals used to define the centrality classes.}
    \label{fig:cent}
\end{figure}

\section{Parameter Choice and Centrality Classification}

The energies selected for the scan were $\sqrt{s_{NN}} =$ 200, 62.4, 19.6, and 7.7 GeV. The pre-equilibrium parameters were fixed as

\begin{table}[H]
    \centering
    \caption{Parameters used for Smeared-AMPT Pre-Equilibrium}
    \begin{tabular}{||r|c c c c||}
        $\sqrt{s_{NN}}$ (GeV) &  7.7 &  19.6 &  62.4 &  200  \\
        $\tau_0$ (fm/c) & 2 & 1.2 & 0.6 & 0.4 \\
        $K$ &2.3&2.9 & 2.8 & 2\\
        $\sigma_{\eta_S}$  &0.34&0.34&1.02&1.02\\
        $\sigma_r$ (fm) &0.62&0.68&1.02&1.02\\
    \end{tabular}
\end{table}

We generated a total of ten thousand events for each of the selected energies in minimum bias, and the centrality selection was performed based on the Reference Multiplicity method \cite{PhysRevC.107.024901}, where we integrate $dN_{\rm ch}/d\eta$ provided by the AMPT partonic cascade.  In Fig. \ref{fig:cent} we show the relation with the impact parameter $b$ for each of the events. In the most central bin (0-5\% centrality class), we randomly chose 100 events to simulate the centrality and performed an event-by-event analysis. 

With the smearing process and Landau matching, the hydrodynamic evolution was performed using PRAIA, and the parameters used for the simulations are shown in the table below.

\begin{table}[H]
    \centering
    \caption{Parameters used for Hydrodynamic Simulations}
    \begin{tabular}{||r|c||}
    $\eta T/\omega$ &  $0.08$\\
    $\tau_\pi \omega/\eta$ & $5$\\
    $\zeta T/\omega$ & $A(1/3-c_s^2)^2$\\
    $A$ & $8\pi/15$ \\
    $\tau_\Pi \omega/\zeta$ & $B(1/3-c_s^2)^2$\\
    $B$ & $1/15$\\
    $\epsilon_{SW}$ & $0.47 \  \text{GeV/fm}^{3}$\\
    \hline
    Time Step & $0.05 \text{ fm/c}$\\
    Integration Method & Runge-Kutta 2\\
    $h_{SPH}$ & 0.4\\
    \end{tabular}
\end{table}

For the kernel defined in Eq.~\eqref{eq:kerneldefinition}, the Cubic Spline function was adopted, as given by the following expression:

\begin{equation}
W(q) = \frac{1}{\pi h^3}
\begin{cases}
1 - \frac{3}{2}q^2 + \frac{3}{4}q^3, & 0 \leq q < 1, \\[2mm]
\frac{1}{4}(2-q)^3, & 1 \leq q < 2, \\[2mm]
0, & q \geq 2.
\end{cases}
\end{equation}

The configuration used for SMASH is the one presented below for results with active hadronic transport.
$ $\\
\begin{figure}[!htpb]
\centering
\begin{adjustbox}{max width=\textwidth, max totalheight=\textheight, keepaspectratio}
\begin{BVerbatim}[fontsize=\footnotesize]
Logging:
    default: INFO

General:
    Modus:  List  
    Time_Step_Mode: Fixed
    Delta_Time:    10
    End_Time:      1000
    Randomseed:    -1     
    JetScape xml config
    Nevents:       200    

Collision_Term:
    No_Collisions: False
    Force_Decays_At_End: True
    
Output:
    Output_Interval: 1000
    Particles:
        Format:          ["Oscar2013"]

Modi:
    List:
        File_Directory: "./"
        Filename: "oscar.dat"
\end{BVerbatim}
\end{adjustbox}
\caption{Configuration Used by Smash.}
\label{fig:SMASHCONFIGURATION}
\end{figure}

For runs with only decays, the flag \texttt{No\_Collisions} was set to \texttt{True}.

\section{Simulation Results for a Crossover-Only Equation of State}
We analyzed, for each configuration, the spectrum of particles produced in the simulations in pseudorapidity $\eta$ (for charged particles), as well as the transverse-momentum $p_T$ and rapidity $y$ spectra (for identified particles), and compared them with experimental data. The uncertainties of the results obtained in our simulations were calculated using the Jackknife method. \\

\subsection{Analysis of Charged-Particle and Identified-Particle Spectra}
\label{subsec:spectra}

\begin{figure}[!htbp]
    \centering
    \includegraphics[width=\linewidth]{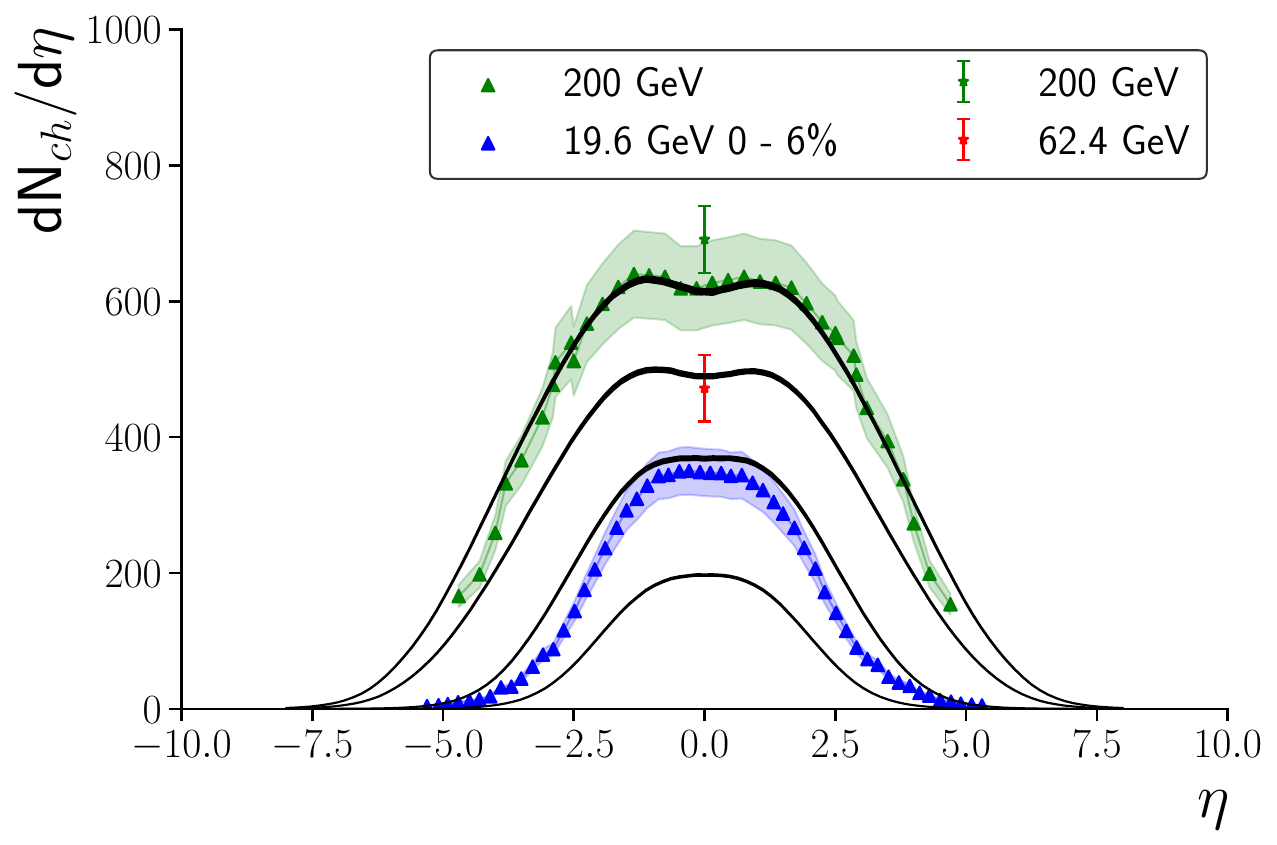}
    \caption{Pseudorapidity distribution of charged particles in $^{197}$Au+$^{197}$Au collisions in the $0$-$5\%$ centrality window for $\sqrt{s_{NN}}=200$, $62.4$, $19.6$ and $7.7$ GeV. The simulations are represented by the black curves and the colored points are the experimental data. Data at 200 GeV were obtained from BRAHMS (green triangles) \cite{PhysRevLett.88.202301} and STAR \cite{PhysRevC.79.034909} (green star). Data at 62.4 GeV were obtained from \cite{PhysRevC.79.034909} (red star). Data at 19.6 GeV were obtained from \cite{PhysRevLett.91.052303} (at this energy, the available experimental window was 0-6\%). No data were found for the 7.7 GeV energy. No cuts are applied in the calculation of this spectrum.}
    \label{fig:dndeta}
\end{figure}

The plot in Fig.~\ref{fig:dndeta} shows the pseudorapidity distribution of charged particles for the four simulated energies. For the energies for which experimental data are available, good agreement is observed between the simulation results and the experimental data. 

The curves reproduce both the normalization and the overall shape of the distributions for $\sqrt{s_{NN}}=200$, $62.4$, and $19.6$ GeV. For $19.6$ GeV, the comparison must be interpreted with some care since the available experimental data correspond to the $0$-$6\%$ centrality window, whereas the simulations were performed for $0$-$5\%$. For $\sqrt{s_{NN}}=7.7$ GeV, no experimental data for $dN_{\rm ch}/d\eta$ were found, so no quantitative conclusions can be drawn for this energy from this observable. The hierarchy of the simulated curves follows the expected behavior with collision energy: the charged-particle multiplicity decreases monotonically in the sequence $200$, $62.4$, $19.6$, and $7.7$ GeV.

\begin{figure}[!htbp]
    \centering
    \includegraphics[width=\linewidth]{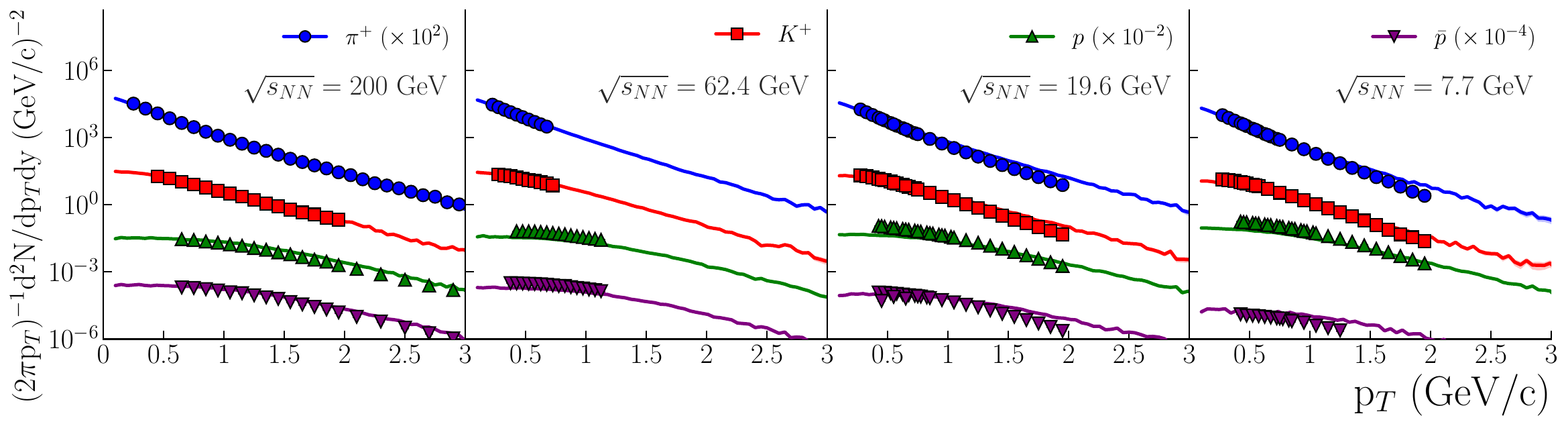}
    \caption{Transverse-momentum spectra for identified particles $\pi^+$, $K^+$, $p$ and $\bar p$ in the 0-5\% window for several $\sqrt{s_{NN}}$. The simulations are represented by full curves and experimental data are represented by points. Experimental data were obtained from \cite{PhysRevC.79.034909, PhysRevC.69.034909, PhysRevC.96.044904}. The momentum spectrum was obtained with a cut $|y| < 0.1$. }
    \label{fig:dndpt}
\end{figure}

\begin{figure}[!htbp]
    \centering
    \includegraphics[width=\linewidth]{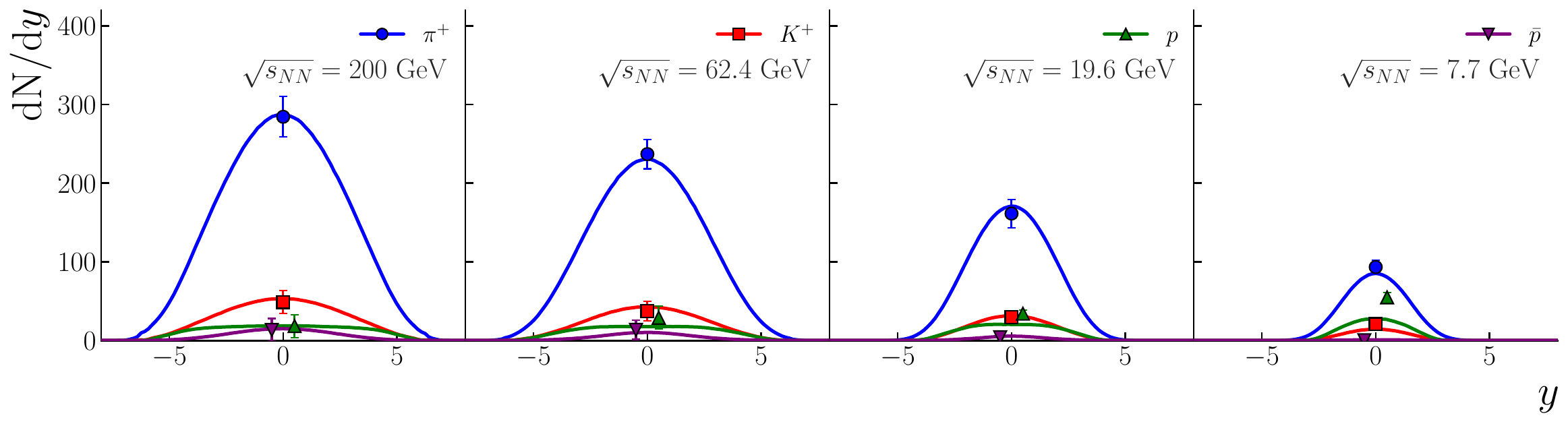}
    \caption{Rapidity spectra for identified particles $\pi^+$, $K^+$, $p$ and $\bar p$ in the 0-5\% window for several $\sqrt{s_{NN}}$.  The simulations are represented by full curves and experimental data are represented by points. Experimental data were obtained from \cite{PhysRevC.79.034909, PhysRevC.69.034909, PhysRevC.96.044904}. }
    \label{fig:dndy}
\end{figure}

The plots in Figs.~\ref{fig:dndpt} and~\ref{fig:dndy} present the transverse-momentum and rapidity spectra for identified particles. At $\sqrt{s_{NN}}=200$ and $62.4$ GeV, the description is quite satisfactory for all analyzed species. The $p_T$ spectra of $\pi^+$, $K^+$, $p$, and $\bar p$ are reproduced within the experimental uncertainties or are very close to them. The same behavior is observed for the rapidity distributions. The most visible exception occurs for protons at $62.4$ GeV, for which the simulated curve lies slightly above the experimental point at midrapidity. Even so, the difference remains compatible with a moderate discrepancy of the order of two error bars.

At $\sqrt{s_{NN}}=19.6$ GeV, the $p_T$ spectra of $\pi^+$ and $K^+$ remain well described up to approximately $p_T \sim 1.25\ \mathrm{GeV}/c$. A reasonable proximity between the simulated curves and the experimental data is also observed for protons and antiprotons, although the discrepancies are more evident than at $200$ and $62.4$ GeV. In the rapidity distributions, the agreement remains globally satisfactory, but proton production at midrapidity is overestimated by the simulation. At $\sqrt{s_{NN}}=7.7$ GeV, by contrast, the differences become systematic: although the $p_T$ spectra of $\pi^+$ and $K^+$ are still reasonably reproduced in the low- and intermediate-transverse-momentum region, the proton and antiproton curves deviate more significantly from the data. In addition, the rapidity distributions of all analyzed species are underestimated.

This behavior is consistent with the expectation that hydrodynamic hybrid models describe collisions at higher energies more robustly, where the formation of a dense medium close to local equilibrium is better justified. As the collision energy decreases, effects associated with baryonic dynamics become more important and may not be fully captured by the assumptions used in this simulation, even with an equation of state with finite baryon chemical potential. At $19.6$ GeV, a possible source of the discrepancies is the absence of baryon diffusion during the hydrodynamic evolution, which can directly affect the proton distribution and the shape of the rapidity distributions. At $7.7$ GeV, the limitations appear to be more severe and may indicate that the assumption of hydrodynamic initialization on a constant proper-time hypersurface, $\tau=\sqrt{t^2-z^2}$, becomes insufficient. In this regime, effects from dynamical initialization, finite energy-deposition times and a more complete description of longitudinal dynamics may be necessary to simultaneously reproduce the $p_T$ spectra and the rapidity distributions. Works including baryon-diffusion effects reproduce these data better at 19.6 and 7.7 GeV \cite{PhysRevC.105.034909}. 

\subsubsection{Semi-Peripheral Collisions}

Simulations were performed in the semi-peripheral $30$-$40\%$ centrality window, using the same parameters employed in the central $0$-$5\%$ analysis. The goal of this test was to verify whether the parametrization of the central-collision simulations also reproduced peripheral collisions. At this stage, we restricted the analysis to $\sqrt{s_{NN}}=200$, $62.4$ and $19.6$ GeV, focusing mainly on pion spectra. Each simulation set included 100 events.

\begin{figure}[!htbp]
    \centering
    \includegraphics[width=\linewidth]{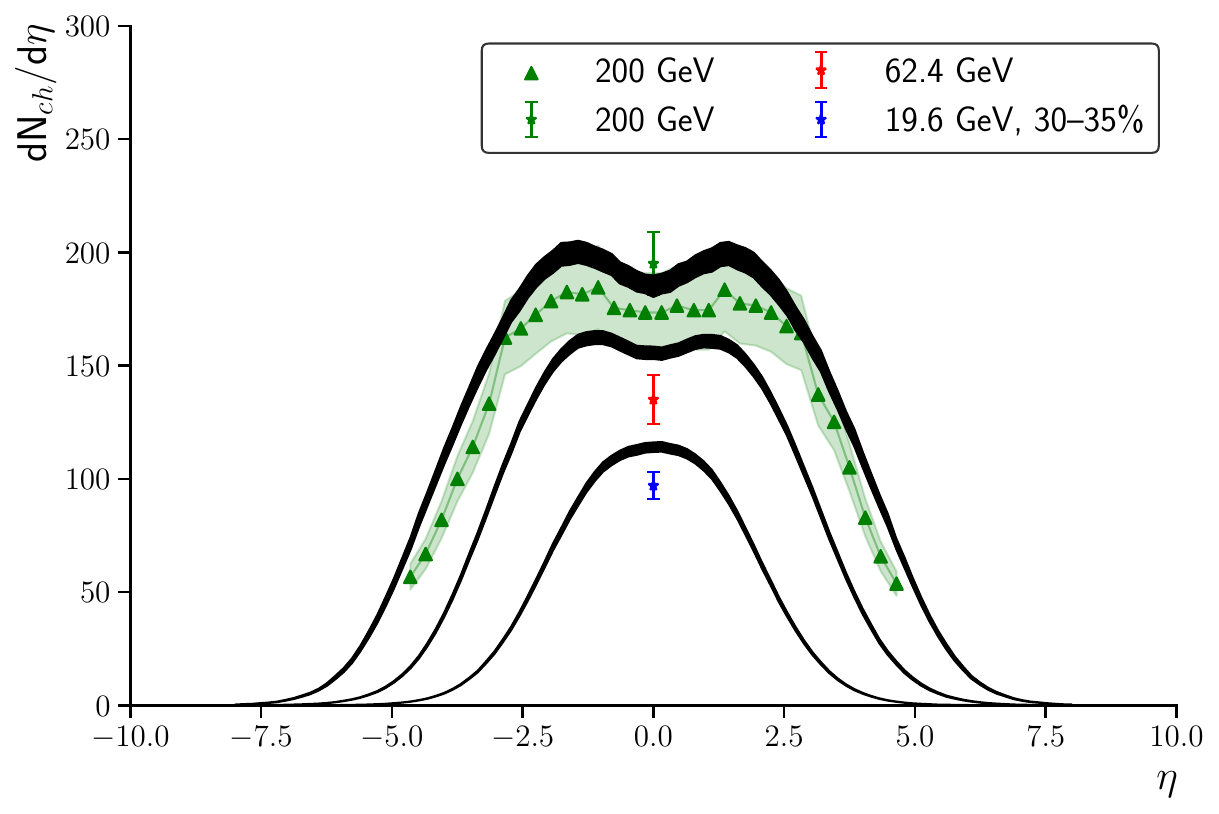}

    \caption{Pseudorapidity distribution of charged particles in $^{197}$Au+$^{197}$Au collisions in the $30$-$40\%$ centrality window for $\sqrt{s_{NN}}=200$, $62.4$ and $19.6$ GeV. The black curves represent the simulation results and the points correspond to the experimental data. For $\sqrt{s_{NN}}=19.6$ GeV, the available experimental data correspond to the $30$-$35\%$ centrality window. Experimental data are available in \cite{PhysRevC.79.034909, PhysRevC.69.034909} from BRAHMS (200 and 62.4 GeV) and ~\cite{PhysRevC.96.044904} from PHOBOS (19.6 GeV).}
    \label{fig:dndeta3040}
\end{figure}

The plot in Fig.~\ref{fig:dndeta3040} shows the pseudorapidity distribution of charged particles. Overall, the hierarchy with energy is preserved, with larger charged-particle production at $\sqrt{s_{NN}}=200$ GeV, followed by $62.4$ and $19.6$ GeV. However, unlike what was observed in the central window, the simulated curves do not reproduce the data with the same level of accuracy. Although the shape of the distributions is qualitatively compatible with the data, there is a tendency to overestimate the charged multiplicity at $200$ GeV and a more visible normalization difference at $62.4$ and $19.6$ GeV. For $19.6$ GeV, the comparison must also be interpreted with care, because the available experimental data correspond to the $30$-$35\%$ centrality window, whereas the simulation was performed for $30$-$40\%$.

This behavior is not unexpected. In hydrodynamic hybrid models, parameters adjusted to reproduce central collisions do not generally describe peripheral collisions automatically with the same accuracy~\cite{PhysRevC.100.014912}. Thus, a simultaneous quantitative description of different centralities may require a global calibration of the parameters, as in Bayesian analyses~\cite{PhysRevC.103.054904}. 

\begin{figure}[!htbp]
    \centering
    \includegraphics[width=0.49\linewidth]{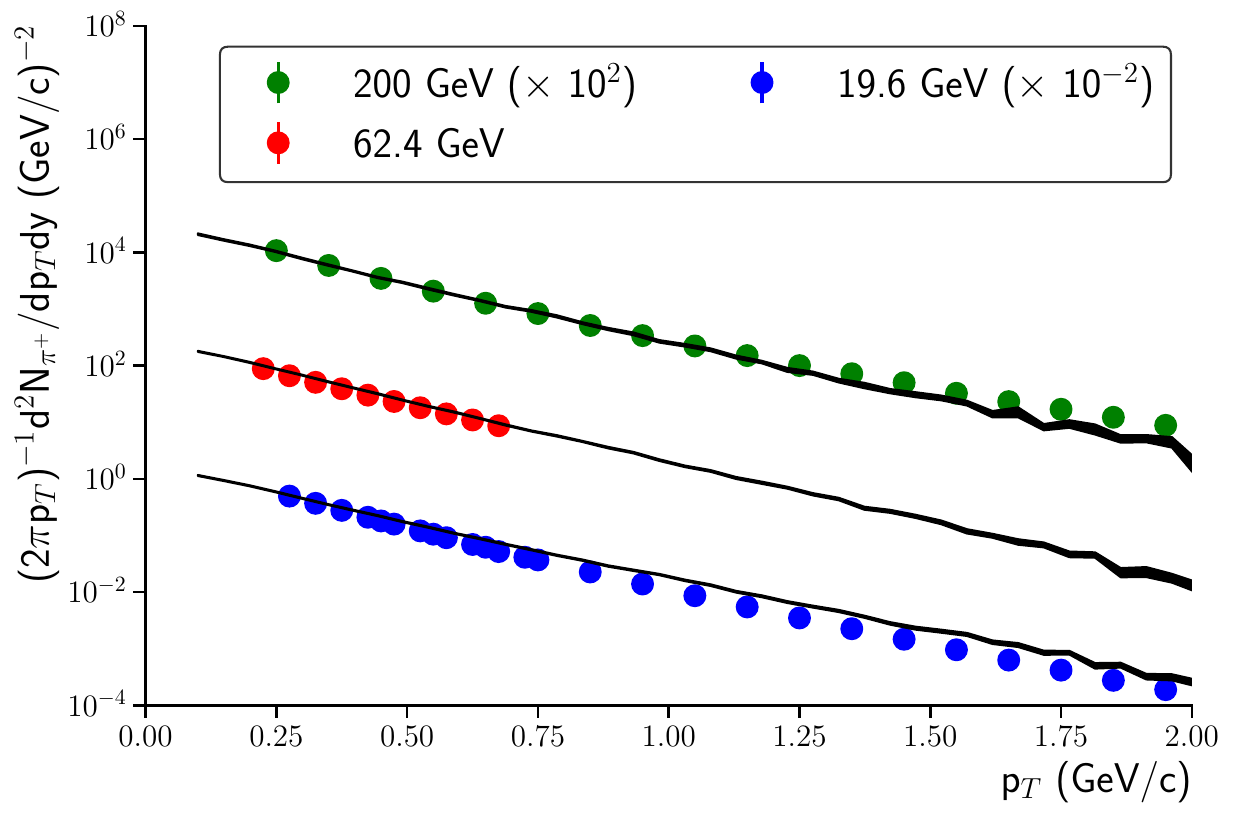}
    \includegraphics[width=0.49\linewidth]{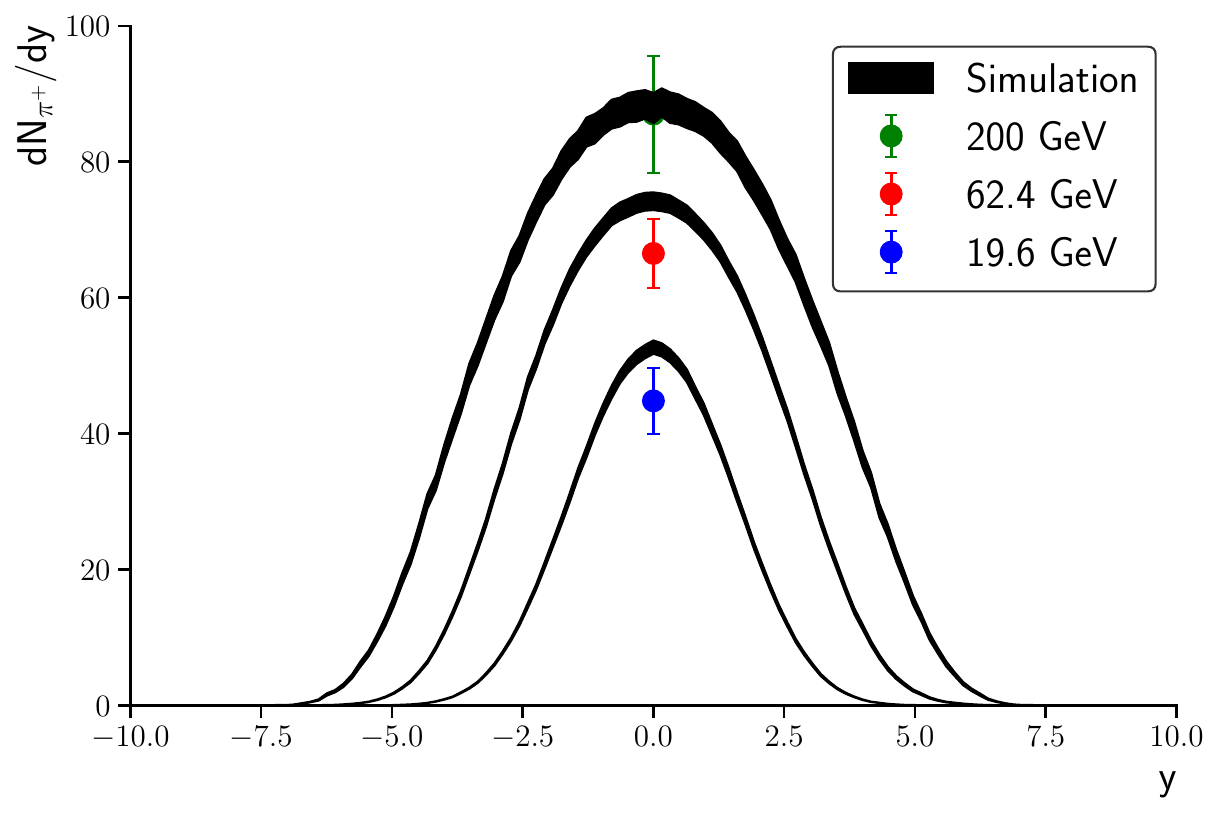}
    \caption{Transverse-momentum spectrum and rapidity distribution for positive pions in $^{197}$Au+$^{197}$Au collisions in the $30$-$40\%$ centrality window for $\sqrt{s_{NN}}=200$, $62.4$ and $19.6$ GeV. The black curves represent the simulation results and the points correspond to the experimental data. Experimental data are available in \cite{PhysRevLett.88.202301, PhysRevC.79.034909, PhysRevLett.91.052303}.}
    \label{fig:dndpt3040}
\end{figure}

The plots in Fig.~\ref{fig:dndpt3040} present the transverse-momentum spectra and rapidity distributions of $\pi^+$ for the same energies. Despite the differences observed in $dN_{\rm ch}/d\eta$, the pion spectra remain reasonably described. 

At $\sqrt{s_{NN}}=200$ and $62.4$ GeV, the simulated curves follow the experimental data well over a wide $p_T$ region, although small differences in slope and normalization appear. For $\sqrt{s_{NN}}=19.6$ GeV, the description is less precise, especially at higher $p_T$, but the low- and intermediate-$p_T$ regions still show satisfactory qualitative agreement.

The rapidity distributions of $\pi^+$ also show a reasonable description of the midrapidity normalization, mainly at $200$ GeV, while larger discrepancies appear at $62.4$ and $19.6$ GeV. Even so, considering that no retuning of the parameters was performed for this centrality, the results indicate that pion production remains satisfactory.

\subsection{Anisotropic Flow Analysis} 
\label{subsec:anisotropic}

\begin{figure}[!htbp]
\centering 
\includegraphics[width=\linewidth]{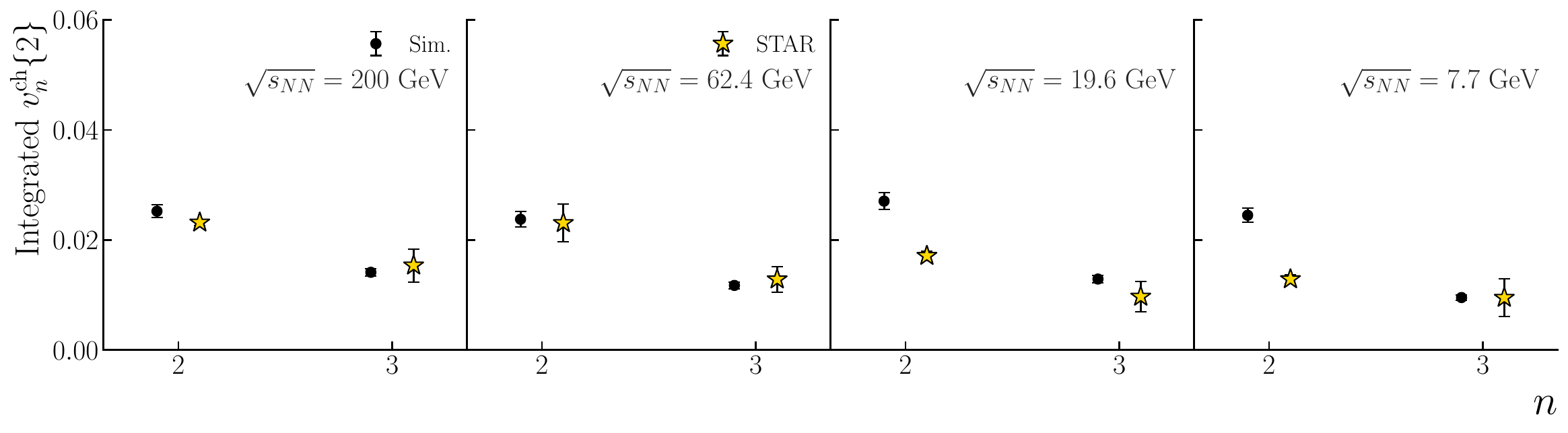}
\caption{Integrated anisotropic-flow coefficients of charged particles, $v_n^{\mathrm{ch}}\{2\}$, in $^{197}$Au+$^{197}$Au collisions at 0--5\% centrality, as a function of the harmonic order $n$ for different collision energies $\sqrt{s_{NN}}$. Simulation results are represented by black circles, while STAR measurements are shown as yellow stars. Experimental data for $v_2\{2\}$ are taken from~\cite{PhysRevC.86.014904} for $\sqrt{s_{NN}}=200$ and $62.4$ GeV and from~\cite{PhysRevC.86.054908} for $\sqrt{s_{NN}}=19.6$ and $7.7$ GeV. Data for $v_3\{2\}$ are taken from~\cite{PhysRevLett.116.112302}. } 
\label{fig:integratedvncharged} 
\end{figure}

As an additional check of the collective dynamics generated by the hybrid chain, we calculated the anisotropic-flow coefficients of charged particles and charged pions. This analysis is not intended as a complete flow calibration, but rather as a consistency test of the same event sample used later in the femtoscopic analysis. Figure~\ref{fig:integratedvncharged} shows the energy dependence of the integrated charged-particle coefficients $v_2^{\rm ch}\{2\}$ and $v_3^{\rm ch}\{2\}$. The simulation reproduces the overall scale of the STAR data. For $v_2^{\rm ch}\{2\}$, the agreement is better at the higher energies, while the lower-energy points show a tendency toward larger values in the simulation. This indicates that the model produces a reasonable elliptic response, although the conversion of the initial spatial anisotropy into momentum anisotropy may be somewhat too strong at the lowest energies. The triangular flow coefficient $v_3^{\rm ch}\{2\}$ is also reproduced at the correct order of magnitude and shows a weaker dependence on the collision energy, as expected for an observable more directly related to event-by-event fluctuations of the initial state. 

\subsection{Pion Interferometry Analysis Using a Gaussian Source}

\label{subsec:hbtanalysis}

We calculated the three-dimensional correlation functions $C(k_T,q_o,q_s,q_l)$~\eqref{cfobj} for identical same-sign pion pairs ($\pi^+\pi^+ +\pi^-\pi^-$), using the Bertsch-Pratt decomposition in the out-side-long (OSL) system and performing the longitudinal boost to the LCMS. From these correlation functions, we extracted the radii $R_o(k_T)$, $R_s(k_T)$, and $R_l(k_T)$, as well as the intensity parameter $\lambda(k_T)$ through a three-dimensional Gaussian fit.

\subsubsection{Pion Selection}

For the evaluation and fit of the correlation functions, single pions were selected using cuts in rapidity, $|y| < 0.5$, and transverse momentum, $0.15~{\rm GeV} < p_T < 0.8~{\rm GeV}$, in accordance with the STAR BES analysis~\cite{AdamczykSTAR2015HBTBES}. Moreover, because SMASH continues to propagate particles without interruption up to the final time of the hadronic simulation, the final positions recorded in the output do not, in general, coincide with the emission points relevant for the interferometry analysis. To address this, we reconstruct the emission points by back-propagating each particle from $t_{\rm final}$ to $t_{\rm emit}$ using the velocity $\vec{p}/p^0$. The resulting positions correspond to the space-time points of the particle’s last interaction or formation, i.e., the points that must be used in the phase $\cos(q \cdot (x_i - x_j))$ of the correlation function.

\subsubsection{HBT Radii}
\begin{figure}[!htbp]
    \centering
    \includegraphics[width=\linewidth]{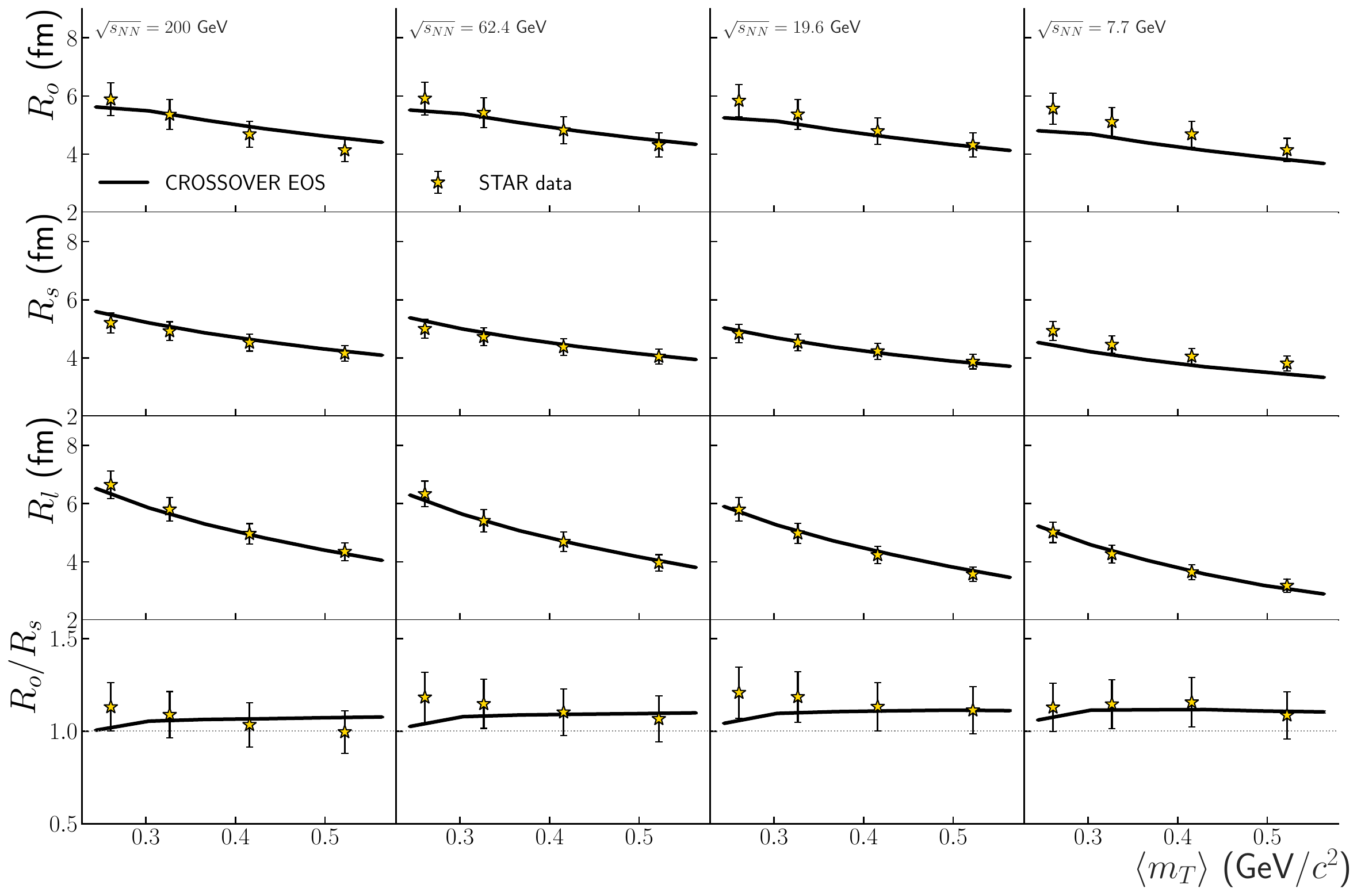}
    \caption{HBT radii $R_o$, $R_s$, $R_l$ and the ratio $R_o/R_s$ obtained from a Gaussian fit to the correlation functions for the different collision energies. Experimental data were taken from~\cite{AdamczykSTAR2015HBTBES}.}
    \label{fig:hbtradii}
\end{figure}

Results for HBT radii obtained from the correlation function fit are presented in Fig~\ref{fig:hbtradii} together with STAR BES data. The simulations provide a reasonable qualitative description of the STAR data. The results are given as a function of $m_T$, defined by equation~\eqref{eq:mtapprox}, in agreement with the experimental results~\cite{AdamczykSTAR2015HBTBES}. For most points in $m_T$, the simulated results remain close to the experimental values or compatible with the error bars, indicating that the space-time structure of the pion source is reasonably described by the hydrodynamic evolution followed by particlization, hadronic transport, and decays. 

At $\sqrt{s_{NN}}=200$ GeV, a globally satisfactory description of the three radii is observed. The radius $R_o$ follows the data within the experimental uncertainties over nearly the full $m_T$ range, remaining compatible, although at the edge of the error bar, in the intermediate-to-high $m_T$ region. The same qualitative behavior is observed for $R_s$ and $R_l$, which reproduce both the magnitude and the decreasing trend with $m_T$. Consequently, the ratio $R_o/R_s$ remains close to the data. At $\sqrt{s_{NN}}=62.4$ and $19.6$ GeV, the radii also remain close to the experimental data. At both energies, the simulated curves reproduce the decrease of $R_o$, $R_s$, and $R_l$ with increasing $m_T$, the expected signature of the collective expansion of the source. However, a small systematic trend appears in the low-$m_T$ sector, especially in $R_o$: the simulated curve lies slightly below the first experimental point compared to the rest of the curve. Since $R_s$ is described more stably, this reduction of $R_o$ at low $m_T$ propagates directly to the ratio $R_o/R_s$, producing a mild suppression of this ratio in that region. This effect does not compromise the global description, but it indicates that the part most sensitive to emission duration and to correlations between emission position and emission time may still be slightly underestimated by the model.

The case of $\sqrt{s_{NN}}=7.7$ GeV presents a discrepancy in $R_o$. Although $R_s$ and $R_l$ still reproduce the data, the $R_o$ curve lies below the experimental points over almost the full analyzed range. This suppression directly reduces the ratio $R_o/R_s$, which also appears systematically below the data. Since $R_o$ depends on the emission duration and on space-time correlations of the source (Eq.~\eqref{eq:lcmsoslhbtradii}), this result suggests that, at the lowest energy, the hybrid chain may be producing an effective pion source with a shorter emission duration or with non-negligible space-time correlations.

The energies $200$ and $7.7$ GeV show values close to unity, $\sum\chi^2/\sum{\rm ndf}=1.66$ and $1.04$, respectively, indicating good global compatibility according to this metric. At $62.4$ and $19.6$ GeV, the values are larger, $\sum\chi^2/\sum{\rm ndf}=10.87$ and $5.43$. These numbers do not necessarily imply a qualitative failure of the description because the visual deviations are small at several points. They indicate that moderate systematic discrepancies can be amplified when the experimental uncertainties are small and when the $\chi^2$ is summed over all radii and all $m_T$ points. Thus, despite the moderate systematic tensions reflected in the $\chi^2/\mathrm{ndf}$ values at 62.4 and 19.6 GeV, the visual comparison and the overall physical trends indicate a reasonable qualitative description of the femtoscopic observables, with the main discrepancies concentrated in $R_o$ and, consequently, in $R_o/R_s$.

\begin{figure}[!htbp]
    \centering
    \includegraphics[width=0.49\linewidth]{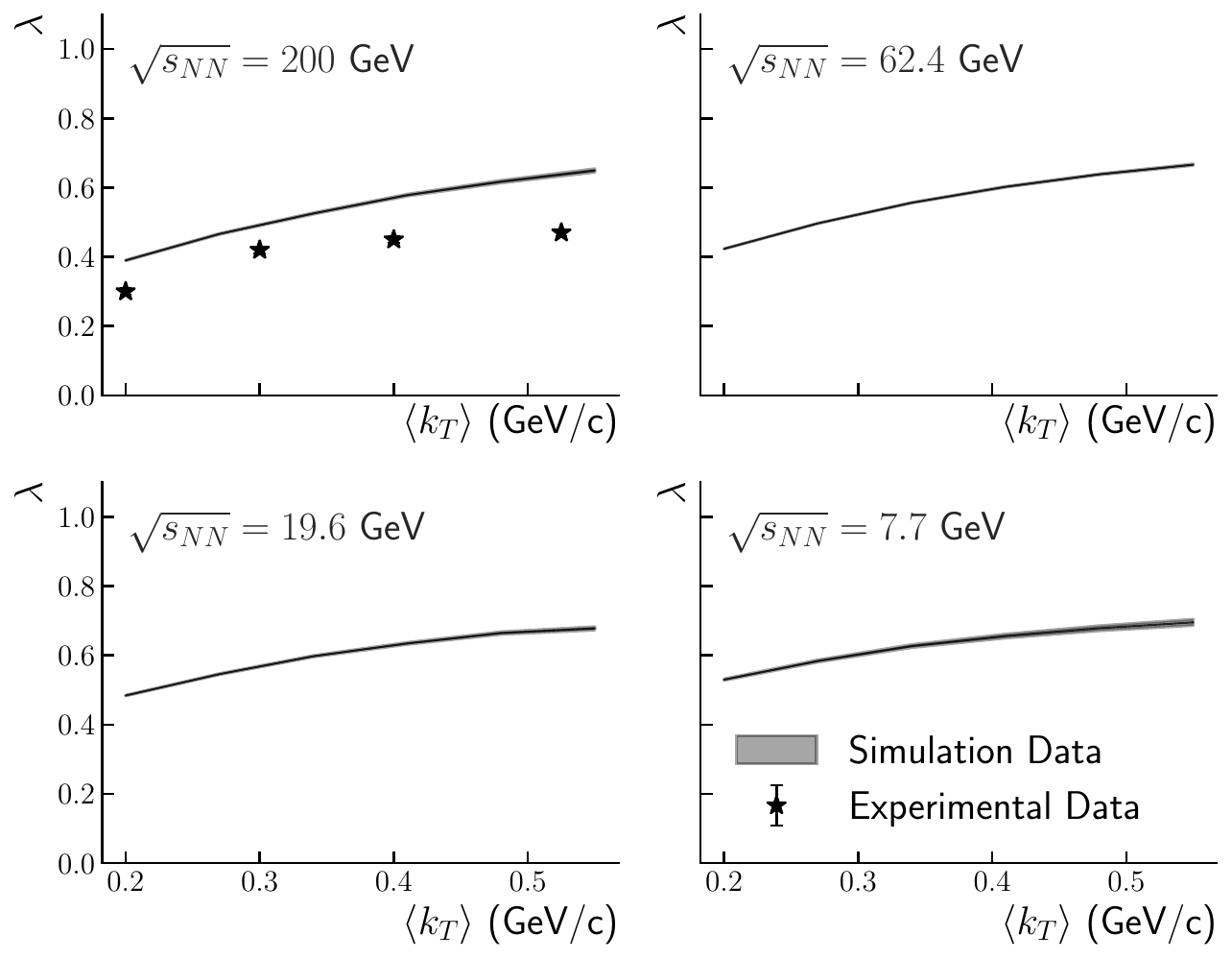}
    \includegraphics[width=0.49\linewidth]{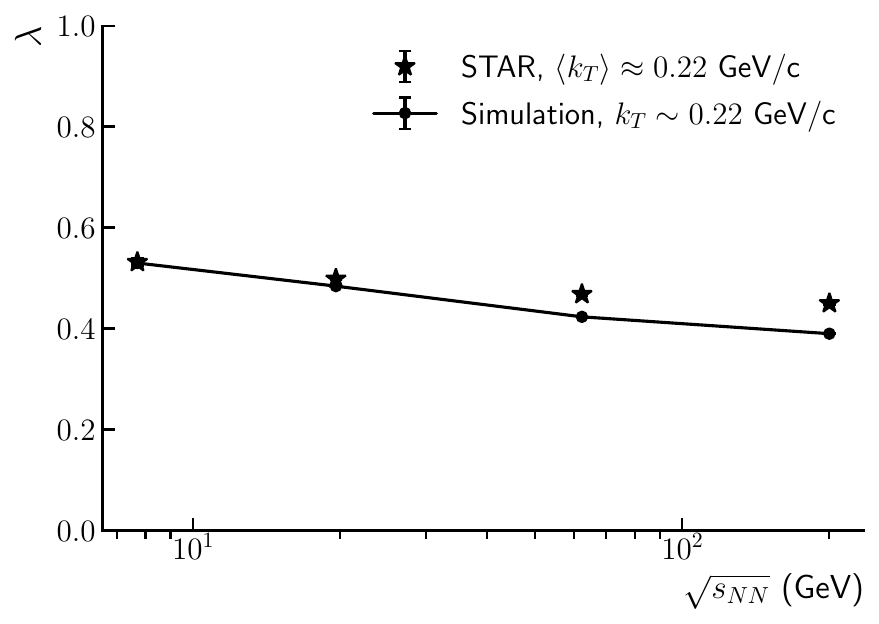}
    \caption{Extracted $\lambda$ parameter. The four panels on the left show $\lambda$ as a function of $\langle k_T\rangle$ for $\sqrt{s_{NN}}=200$, $62.4$, $19.6$ and $7.7$ GeV. The curves represent the simulation results, while the experimental points are from \cite{PhysRevC.71.044906}. The right panel shows the comparison at $\langle k_T\rangle \approx 0.22~\mathrm{GeV}/c^2$ between the simulation and the experimental data~\cite{AdamczykSTAR2015HBTBES}.}
    \label{fig:lambda-hbt}
\end{figure}

The plots in Fig.~\ref{fig:lambda-hbt} show the behavior of $\lambda$ for the four analyzed energies. In the left plot, the simulation results show a systematic monotonic increase of $\lambda$ as a function of $k_T$ at the four energies, with values on the order of $0.4$-$0.6$, compatible with the scale observed experimentally. Since experimental data are available only for $200$ GeV, we compare with them. The simulation results at this energy lie systematically above the experimental data, although they remain within the same order of magnitude. Unlike the HBT radii, the parameter $\lambda$ has a stronger sensitivity to details of the source composition, such as resonance decays, pion sample purity, and the cuts used in constructing the correlation function, among others~\cite{AdamczykSTAR2015HBTBES,PhysRevC.71.044906}. In addition, the systematic error assigned to this parameter is not provided. Since the HBT radii have large systematic errors, a similar behavior should be expected for the uncertainties assigned to $\lambda$.

 In the right panel, we compare the value of $\lambda$ as a function of $\sqrt{s_{NN}}$ for $k_T = 0.22$ GeV/c$^2$ with the experimental data~\cite{AdamczykSTAR2015HBTBES}. The comparison shows qualitative compatibility between the simulation and the data: the order of magnitude is correctly reproduced, and the variation between energies remains moderate, as observed experimentally. The best agreement occurs at the lower energies, while at $62.4$ and $200$ GeV, the simulation tends to produce values slightly lower than the experimental points. The restricted analysis at this energy \cite{PhysRevC.71.044906} reported smaller values of $\lambda$ than those later reported in the Beam Energy Scan analysis \cite{AdamczykSTAR2015HBTBES}. The STAR BES study itself observes that the value of $\lambda$ at $200$ GeV is larger than the earlier result and associates this difference with the implementation of an anti-electron cut, which reduces sample contamination. Therefore, the discrepancy between the simulation and the $200$ GeV data should be understood within this sensitivity of $\lambda$ to sample purity and not necessarily as a difference in the geometrical structure of the source.

\begin{figure}[!htbp]
    \centering
    \includegraphics[width=\linewidth]{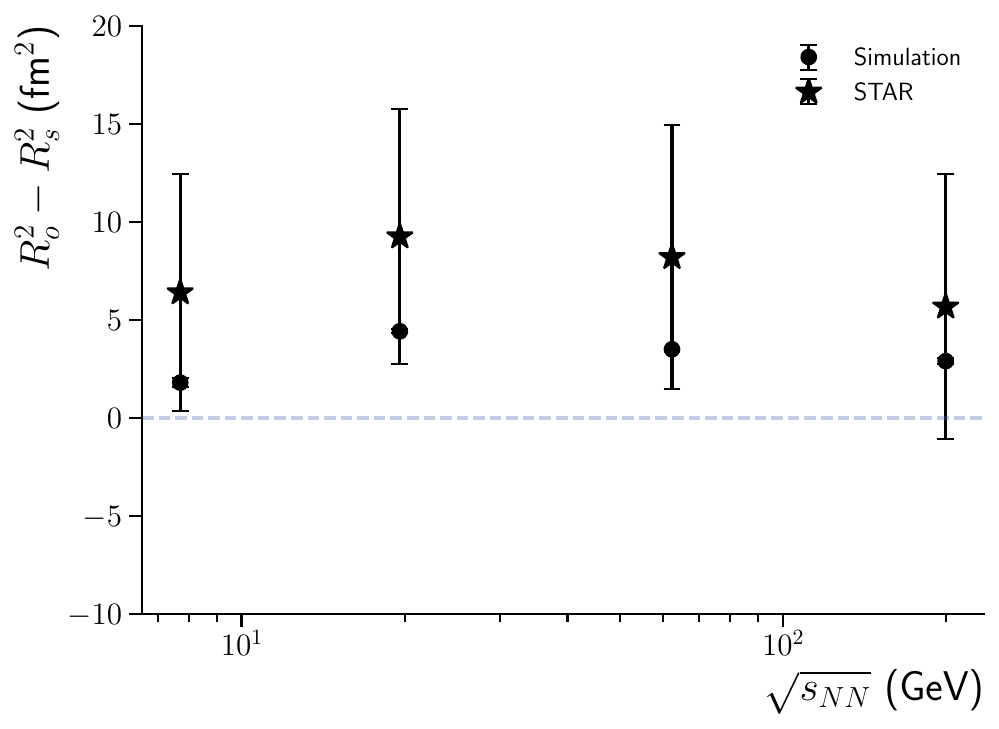}
    \caption{Comparison of $R_o^2-R_s^2$ as a function of $\sqrt{s_{NN}}$. The experimental points are obtained from~\cite{AdamczykSTAR2015HBTBES}. Although the simulated values lie systematically below the experimental central points, they remain compatible with the data within the uncertainties.}
    \label{fig:ro2-rs2-energy}
\end{figure}

Finally, we analyzed the observable $R_o^2-R_s^2$. Figure~\ref{fig:ro2-rs2-energy} shows the comparison between the simulation results and the STAR BES data for central collisions at the considered energies \cite{AdamczykSTAR2015HBTBES}. Overall, the values obtained with the hybrid chain lie systematically below the experimental central points but are compatible with the data within the error bars. This result indicates that, even using an equation of state containing only a crossover transition, the simulation is able to reproduce the experimental scale of this observable within the available uncertainties.

The experimental error bars of $R_o^2-R_s^2$ become significantly large due to the propagation of the uncertainties of the radii $R_o$ and $R_s$, especially when systematic contributions are included (around 7 to 10\% of the radii values~\cite{AdamczykSTAR2015HBTBES}). Since $R_o^2-R_s^2$ is obtained from the difference of two observables of similar magnitude, small relative uncertainties in each radius can lead to a large absolute uncertainty in the quadratic difference. Consequently, the experimental data itself become compatible, at some energies, with values close to zero or even negative.

This point is important because it limits the direct interpretation of $R_o^2-R_s^2$ as a measure of the emission duration. Although this observable contains a contribution associated with the temporal extent of the source (Eq.~\eqref{eq:RoRs_general}), this interpretation depends on additional assumptions, such as symmetric transverse hydrodynamic expansion or the neglect of space-time correlations between the emission position and the emission time. The compatibility of the data with negative values suggests that such approximations may not be adequate in general: the geometrical difference between the out and side directions may not be negligible, and the correlation $\langle \Delta x_o \Delta t'\rangle$ can contribute significantly to the final value of $R_o^2-R_s^2$.

The large uncertainties make it difficult to assign a trend to the experimental curve. In particular, any relation as a function of $\sqrt{s_{NN}}$ must be interpreted with care, since the error bars allow for different trends compatible with the data. This limitation directly affects interpretations that associate $R_o^2-R_s^2$ with predominantly temporal signals~\cite{Lacey2015CEP}. The fact that the results obtained here remain compatible with the experimental data within the uncertainties indicates that this observable, by itself, does not provide conclusive evidence for the presence of critical structure. 

\subsubsection{Calculation on Direct Emission and on Only Decays}

\begin{figure}[!htbp]
    \centering
    \includegraphics[width=\linewidth]{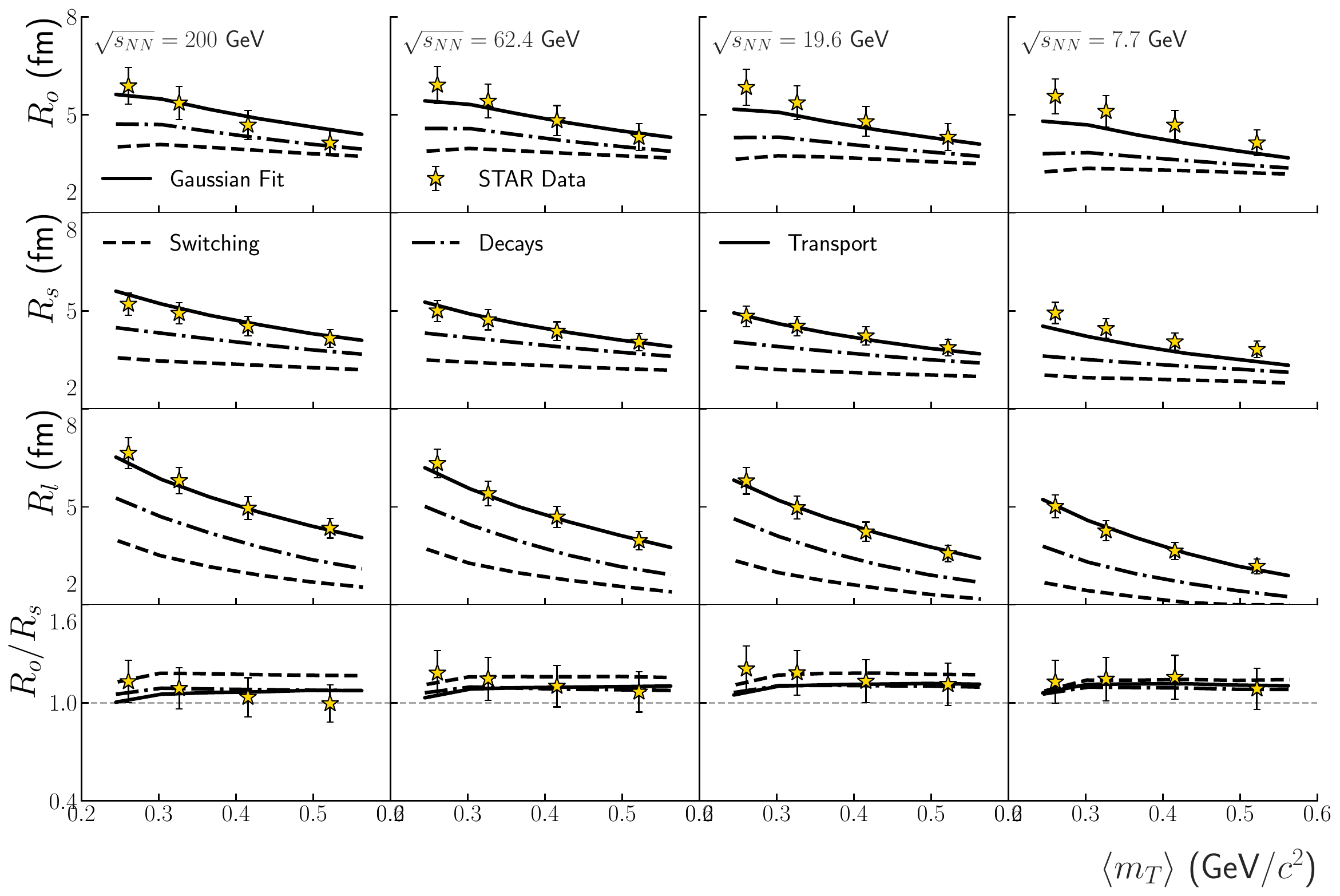}
    \caption{HBT radii $R_o$, $R_s$, $R_l$ and the ratio $R_o/R_s$ obtained from a Gaussian fit to the correlation functions for the different collision energies. The results are shown at two stages: immediately after switching (dashed lines) and including only the decays (dash-dotted lines). We also present results for the complete evolution in hadronic transport for comparison.}
    \label{fig:comparisionhbt}
\end{figure}

We also obtained HBT radii by performing a Gaussian fit to the correlation functions in two additional cases: (1) directly after switching and (2) with decays only, as opposed to full transport. They are shown in Fig. \ref{fig:comparisionhbt}.

It is possible to observe an ordering relationship in which the inclusion of decays and the full use of transport systematically increase the HBT radii, whereas, for the ratio $R_o/R_s$, both decays and full transport lower the curves and bring them closer to the data.

\subsection{Pion Interferometry Analysis Using a Lévy Source}

\begin{figure}[!htbp]
    \centering
    \includegraphics[width=\linewidth]{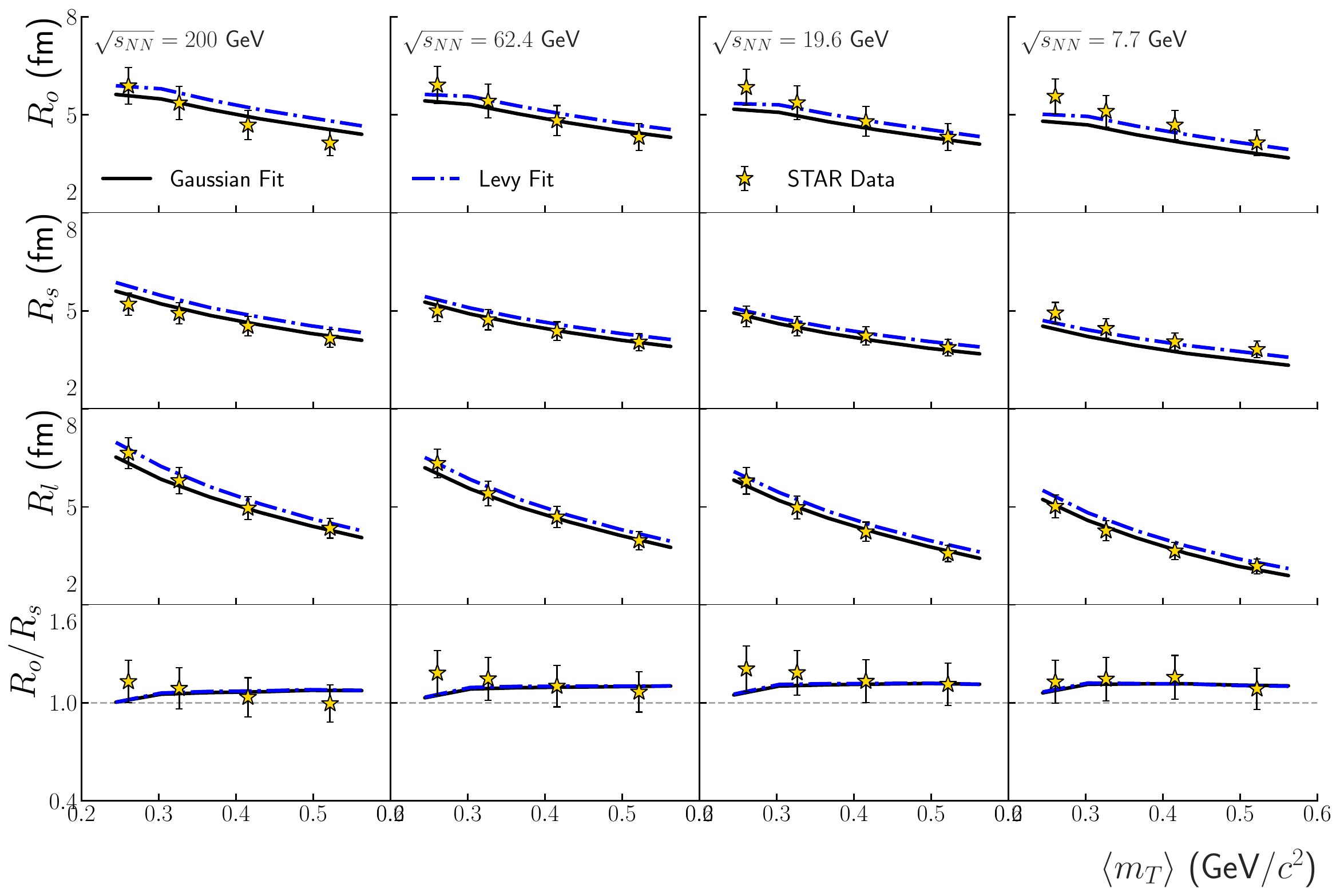}
    \caption{HBT radii $R_o$, $R_s$, $R_l$ and the ratio $R_o/R_s$ obtained from a Lévy fit (curves in blue) to the correlation functions for the different collision energies.}
    \label{fig:hbtradii2}
\end{figure}

We use the Lévy fit of Eq. \eqref{eq:levy_3d_hbt} to the three-dimensional correlation functions $C(k_T, q_o, q_s, q_l)$ to obtain the HBT radii and the factor $\alpha$ for the characterization of the source. The plots in Fig. \ref{fig:hbtradii2} show the radii obtained with the Lévy fit compared with the radii obtained with the Gaussian fit and the experimental data.

\begin{figure}[!htbp]
    \centering
    \includegraphics[width=\linewidth]{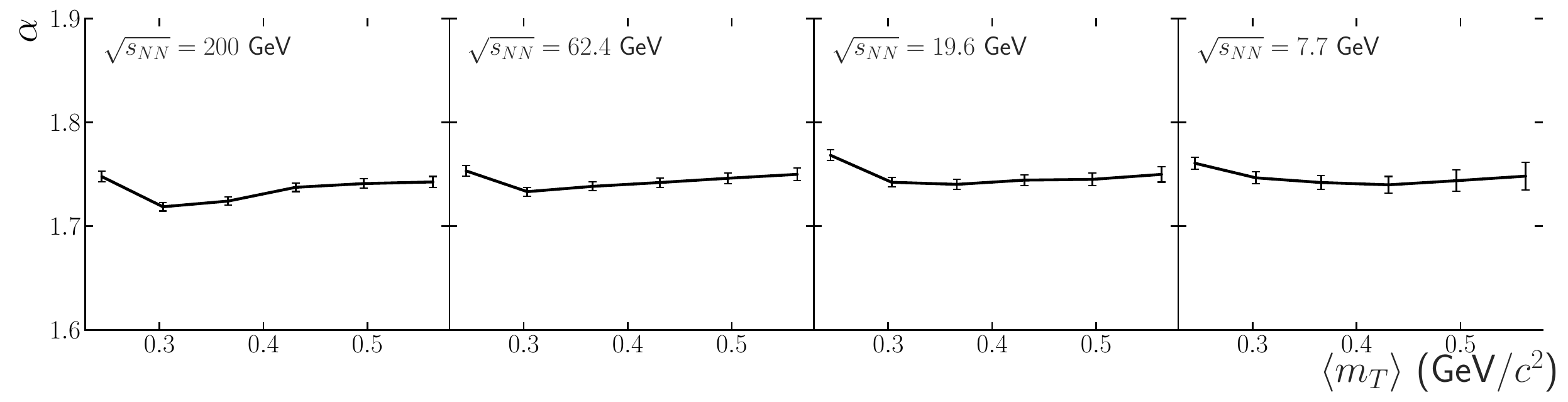}
    \caption{Values of $\alpha$ obtained from the Lévy fit.}
    \label{fig:alpha1}
\end{figure}

In general, the radii obtained with the Lévy fit show an increase in all HBT radii while preserving the ratio $R_o/R_s$. The plot in Fig. \ref{fig:alpha1} shows the values of $\alpha$ obtained as a function of $\langle m_T \rangle$. One may say that the results obtained with the Lévy fit are close to the experimental data. However, since those data were obtained using a Gaussian fit, it is not possible to state that they describe the same pion source.

Most extracted $\alpha$ values fall in the range $1.7$–$1.8$, i.e., below $2$, which suggests that with the Lévy fit, the inferred source properties deviate from a Gaussian description. This additional parameter allows the shape of the source distribution to deviate from the Gaussian form and to be determined directly from the fit.

\begin{figure}[!htbp]
    \centering
    \includegraphics[width=\linewidth]{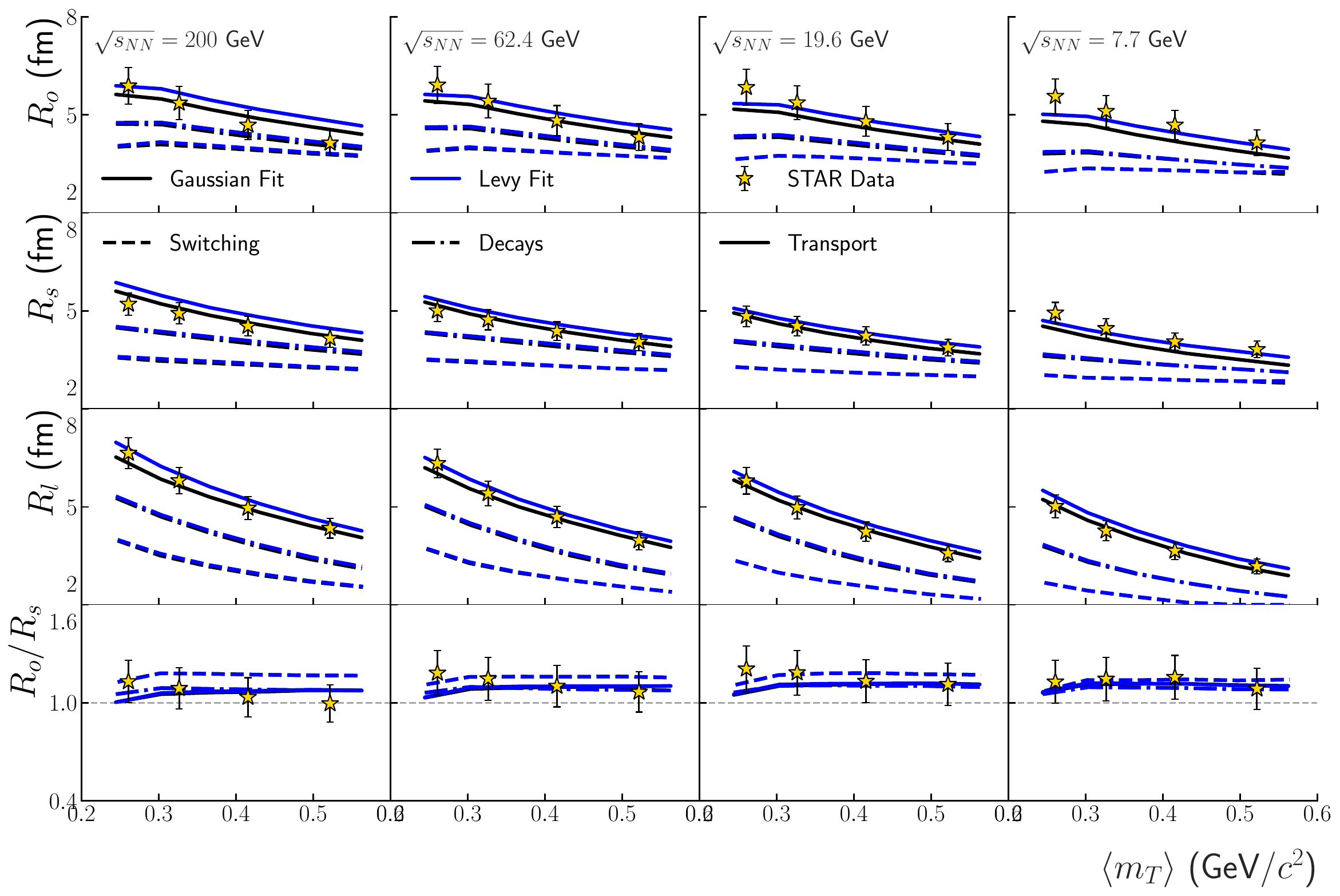}
    \caption{HBT radii $R_o$, $R_s$, $R_l$ and the ratio $R_o/R_s$ obtained from a Lévy fit to the correlation functions for the different collision energies (blue curves). The results are shown at two stages: immediately after switching (dashed lines) and including only the decays (dashdotted lines). We also present results for the full hadronic transport evolution (full lines) and, for comparison, the outcomes of the Gaussian fits (black curves).}
    \label{fig:comparision2}
\end{figure}

\begin{figure}[!htbp]
    \centering
    \includegraphics[width=\linewidth]{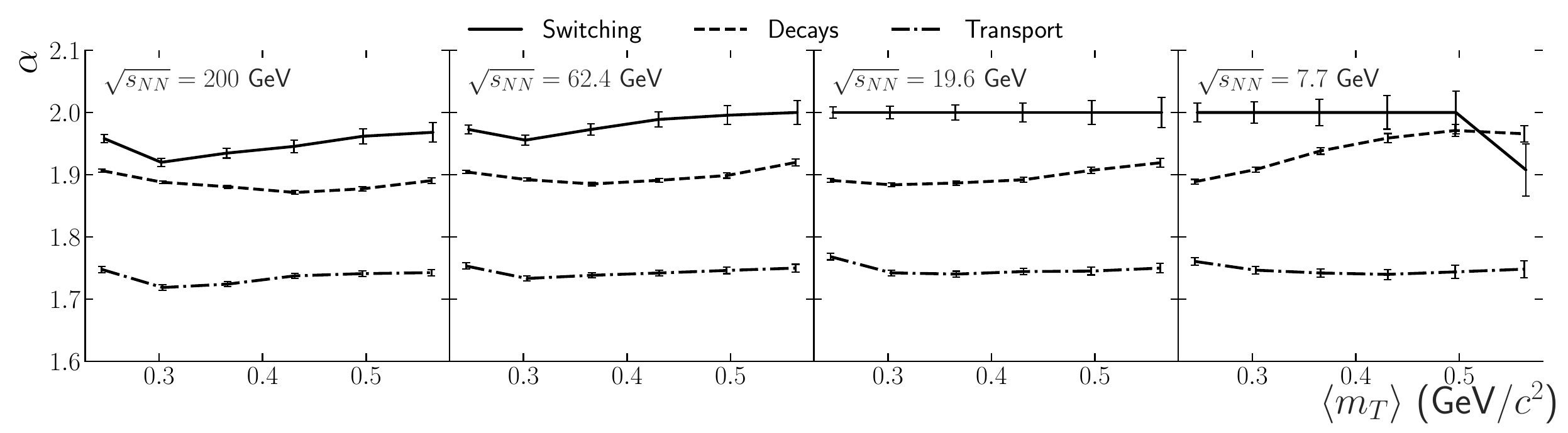}
    \caption{Values of $\alpha$ obtained with the Lévy fit. The results are shown at two stages: immediately after switching (dashed lines) and including only the decays (dashdotted lines). We also present results for the full hadronic transport evolution (full lines).}
    \label{fig:alphacomparision}
\end{figure}

An analysis using a Lévy fit was also performed to obtain the radii (Fig. \ref{fig:comparision2}) and the parameter $\alpha$ (Fig. \ref{fig:alphacomparision}) directly after the switching and with only decays.

The radii obtained with the Lévy fit do not differ from those obtained with the Gaussian fit right after the switching and with decays only. For collision energies of 19.6 GeV, 62.4 GeV, and 200 GeV, the values of $\alpha$, however, appear closer to $2$ right after the switching, which shows that the sources are more nearly Gaussian, while the decays reduce the value of $\alpha$ to approximately $1.9$. For 7.7 GeV, the analysis works well for low values of $\langle m_T \rangle$, but deviates from the other energies for higher values.

\section{Simulation Results for a CEP Equation of State}

Using an equation of state with a critical point and a first-order phase transition, as described in Chap. \ref{chap:2}, we jointly implement a critical behavior in the bulk transport coefficient \cite{PhysRevC.95.034902} in Eq. \eqref{bulk}.
\begin{equation}
    \frac{\zeta T}{\epsilon + P} \to \frac{\zeta' T}{\epsilon + P} = A\left(\frac{1}{3} - c_s^2\right)\left[1+ \left(\frac{\xi}{\xi_0}\right)^3\right]
\end{equation}
where the dependence on the correlation length $\xi$ introduces critical-point effects into this coefficient. The exponent $3$ is a result of the universality class of the Ising model~\cite{PhysRevC.95.034902}.

\begin{figure}[!htbp]
    \centering
    \includegraphics[width=0.95\linewidth]{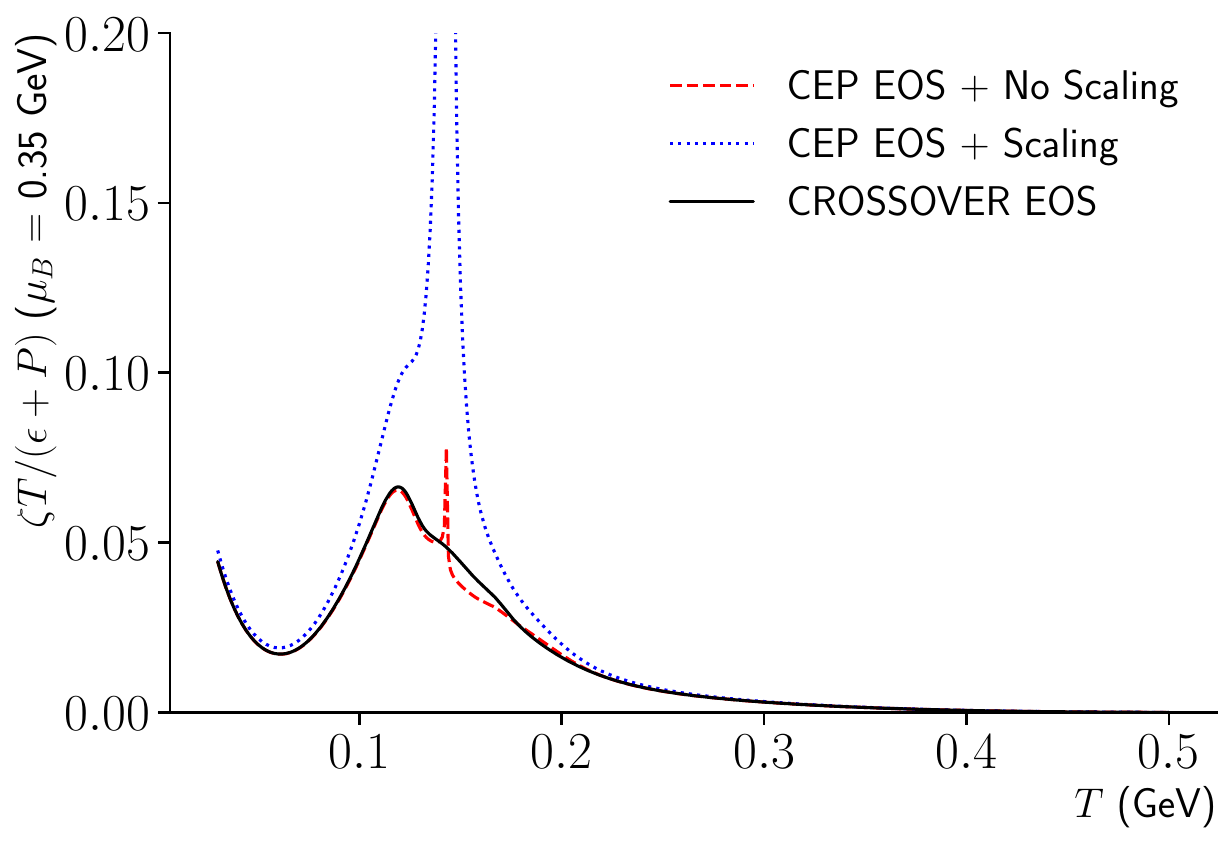}
    \caption{Comparison between the different behaviors of $\zeta T/(\epsilon + P)$ for an equation of state with a critical point, with and without scaling, as a function of $T$ for $\mu_B = 350$ MeV. The dotted blue curve represents the behavior of the coefficient with scaling, while the dashed red curve corresponds to the case without scaling, both for the equation of state with a critical point. The black curve, in turn, shows the coefficient for the equation of state with crossover-only.} 
    \label{fig:criticalscaling}.
\end{figure}

 We do not apply a correction to the shear coefficient, since the dependence of the critical correction is $\sim 1 + (\xi/\xi_0)^{0.05}$, where the critical exponent $0.05$ is too small \cite{PhysRevD.100.074017}. In Fig.~\ref{fig:criticalscaling} we can see the dependence of the critical coefficient as a function of $T$ for $\mu_B = 350$ MeV. Since the critical point is located at $(T_{\rm CEP},\mu_{B,\rm CEP})\simeq(143,350)$, we can observe a divergent behavior of the coefficient both with and without scaling for the equation of state with a critical point. However, the effect is much more pronounced with scaling. One sets $\xi_0 = 1$.

\subsection{Phase-Space Trajectories}

\begin{figure}[!htbp]
    \centering
    \includegraphics[width=\linewidth]{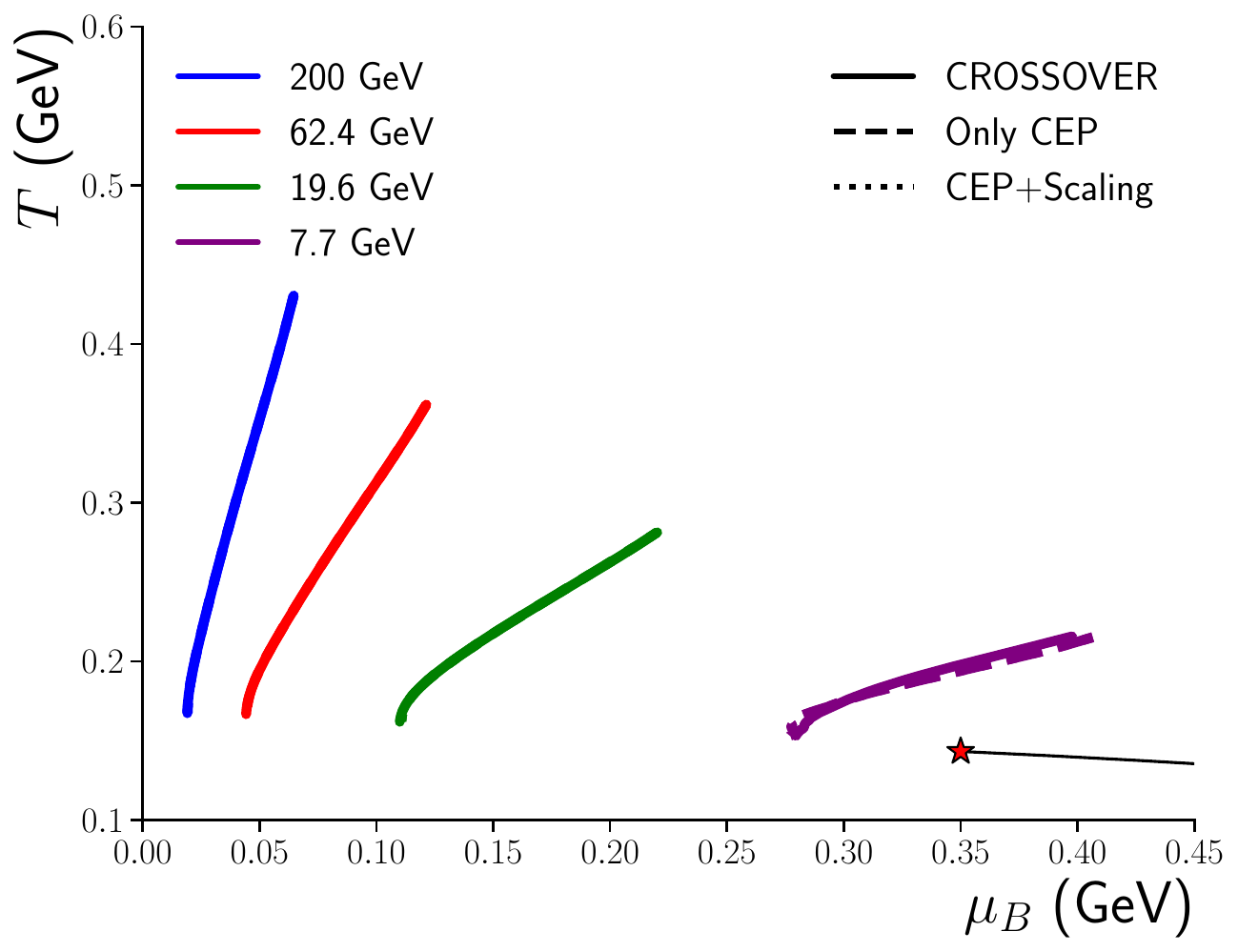}
    \caption{Phase-space $T \times \mu_B$ trajectories of a point at the center of the fluid in the average hydrodynamic evolution for each of the simulated collision energies. The red star marks the location of the critical point, and the short black line denotes the first-order phase transition.} 
    \label{fig:criticalscalingtrajectories}
\end{figure}

We reran the simulations using the equation of state that includes a critical point, both with and without applying critical scaling. Initially, we present the trajectories in the phase diagram at the center of the fluid, averaged over the hydrodynamic evolution for each collision energy, as shown in Fig.~\ref{fig:criticalscalingtrajectories}.

For the average central trajectories shown here, incorporating the critical point and first-order phase transition produces no visible modification at energies above $19.6$ GeV, while only a slight effect is seen at $7.7$ GeV. Even though this is a trajectory at the center of the fluid, this gives us some insight into what to expect, since the observables, in general, are calculated as averages over the entire domain. All curves pass through the crossover region of the equation of state.

\subsection{Identified-Particle Spectra and Anisotropic Flow Analysis}

\begin{figure}[!htbp]
    \centering
    \includegraphics[width=\linewidth]{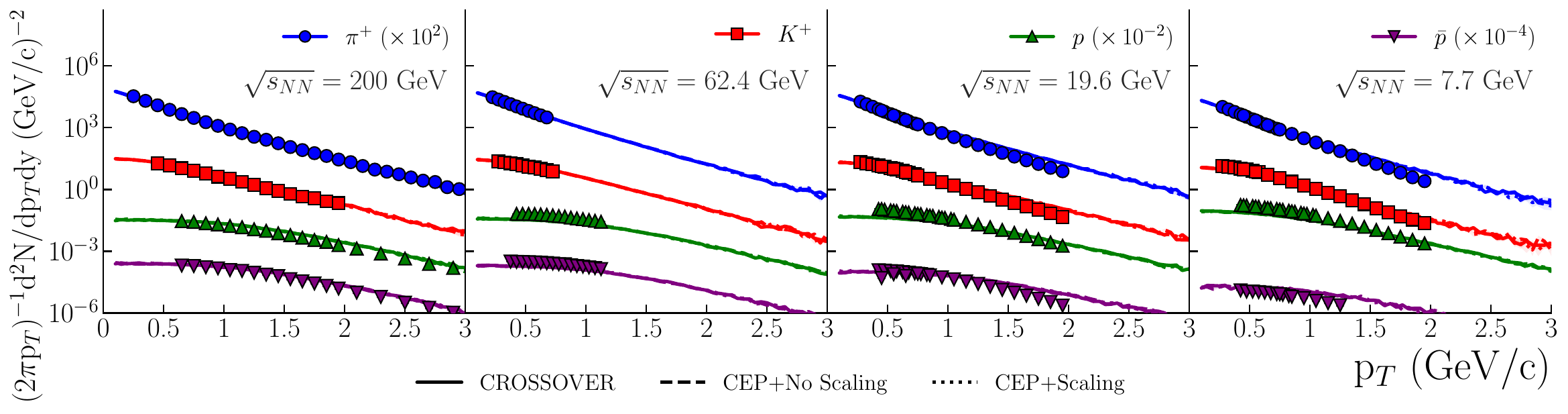}
    \caption{Momentum spectra calculated as in \ref{fig:dndpt} using an equation of state with a critical point and critical scaling, in comparison with the crossover-only equation of state.}
    \label{fig:dndpt2}
\end{figure}

\begin{figure}[!htbp]
    \centering
    \includegraphics[width=\linewidth]{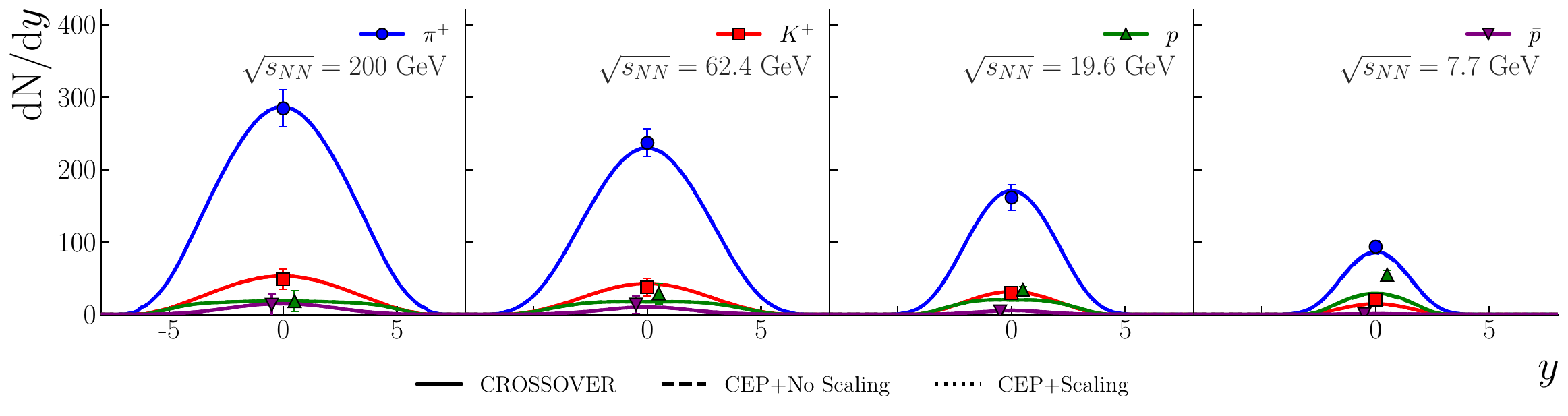}
    \caption{Rapidity spectra calculated as in \ref{fig:dndy} using an equation of state with a critical point and critical scaling, compared with the crossover-only equation of state.}
    \label{fig:dndy2}
\end{figure}

\begin{figure}[!htbp]
\centering 
    \includegraphics[width=\linewidth]{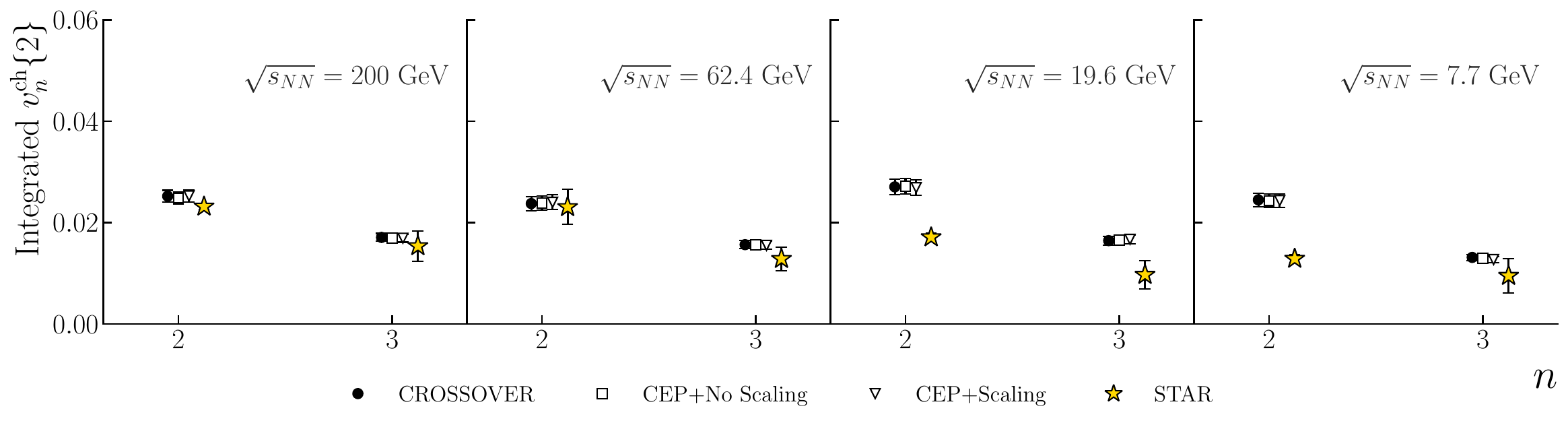}
\caption{Integrated anisotropic flow calculated as in \ref{fig:integratedvncharged} using an equation of state with a critical point and critical scaling, compared with the crossover-only equation of state.} 
\label{fig:integratedvncharged2} 
\end{figure}

We recalculated all spectrum observables (Figs. \ref{fig:dndpt2} and \ref{fig:dndy2}) and anisotropic flow (Fig. \ref{fig:integratedvncharged2}), as done in Subsections \ref{subsec:spectra} and \ref{subsec:anisotropic}, using the equation of state with a critical point and critical scaling.

As expected, no substantial changes were observed between the results obtained with the critical point and critical scaling and those derived from the crossover-only equation of state.

\subsection{Pion Interferometry Analysis Using a Gaussian Source}
\begin{figure}[!htbp]
    \centering
    \includegraphics[width=\linewidth]{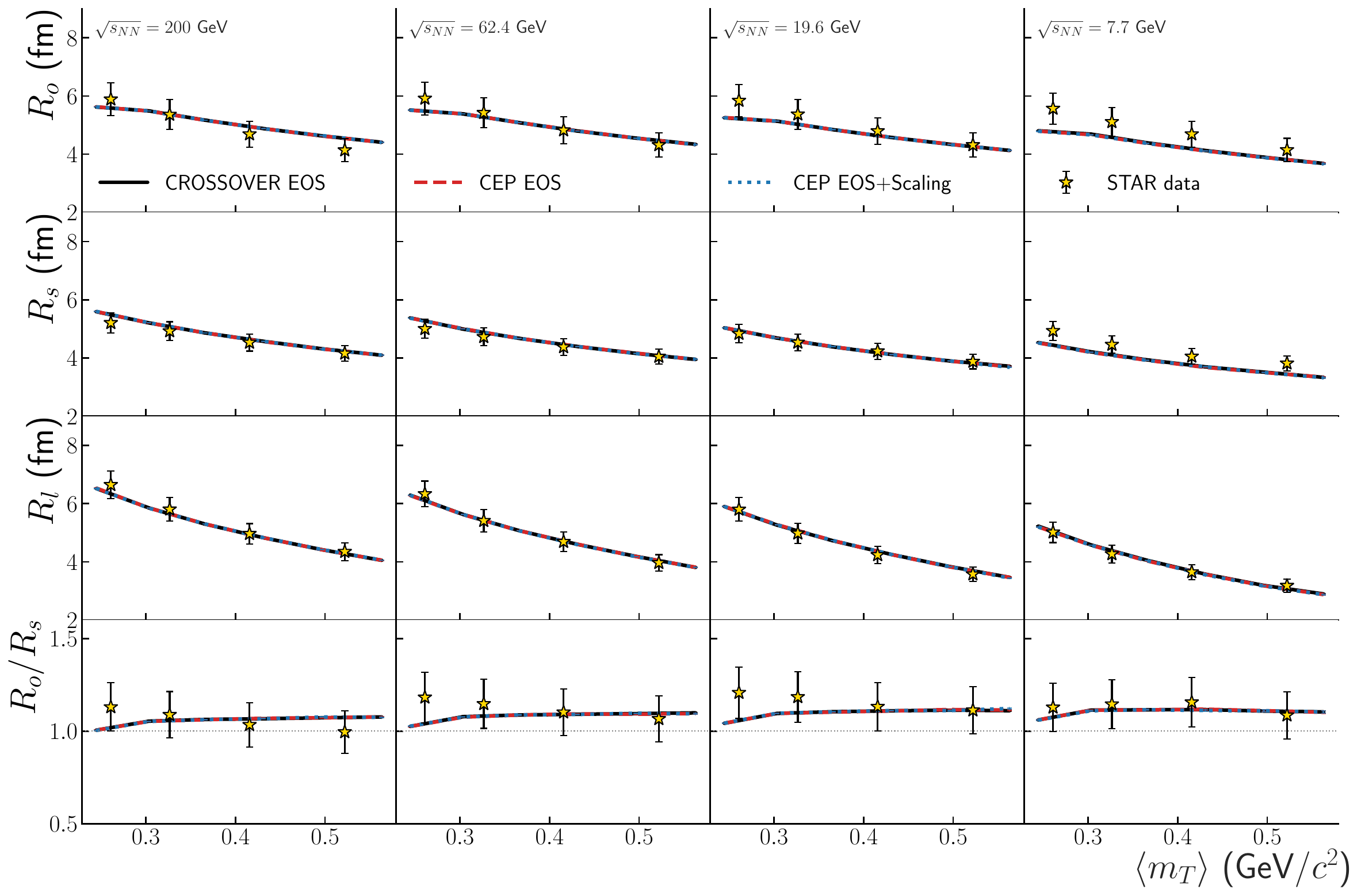}
    \caption{HBT radii extracted from a Gaussian fit using an equation of state with a critical point and critical scaling, compared with the crossover-only equation of state.} 
    \label{fig:hbtradii1critical}
\end{figure}

We recalculated the HBT radii, as done in Section \ref{subsec:hbtanalysis}, for the simulations, employing the equation of state with a critical point and critical scaling. In Fig. \ref{fig:hbtradii1critical} we show the results in comparison with the simulations without a critical point and with the experimental data.

Simulations employing an equation of state that includes a critical point showed no modification of the HBT radii, even when critical scaling was incorporated into the bulk transport coefficient, in agreement with the phase-space trajectories in Fig. \ref{fig:criticalscalingtrajectories}.

\begin{figure}[!htbp]
    \centering
    \includegraphics[width=\linewidth]{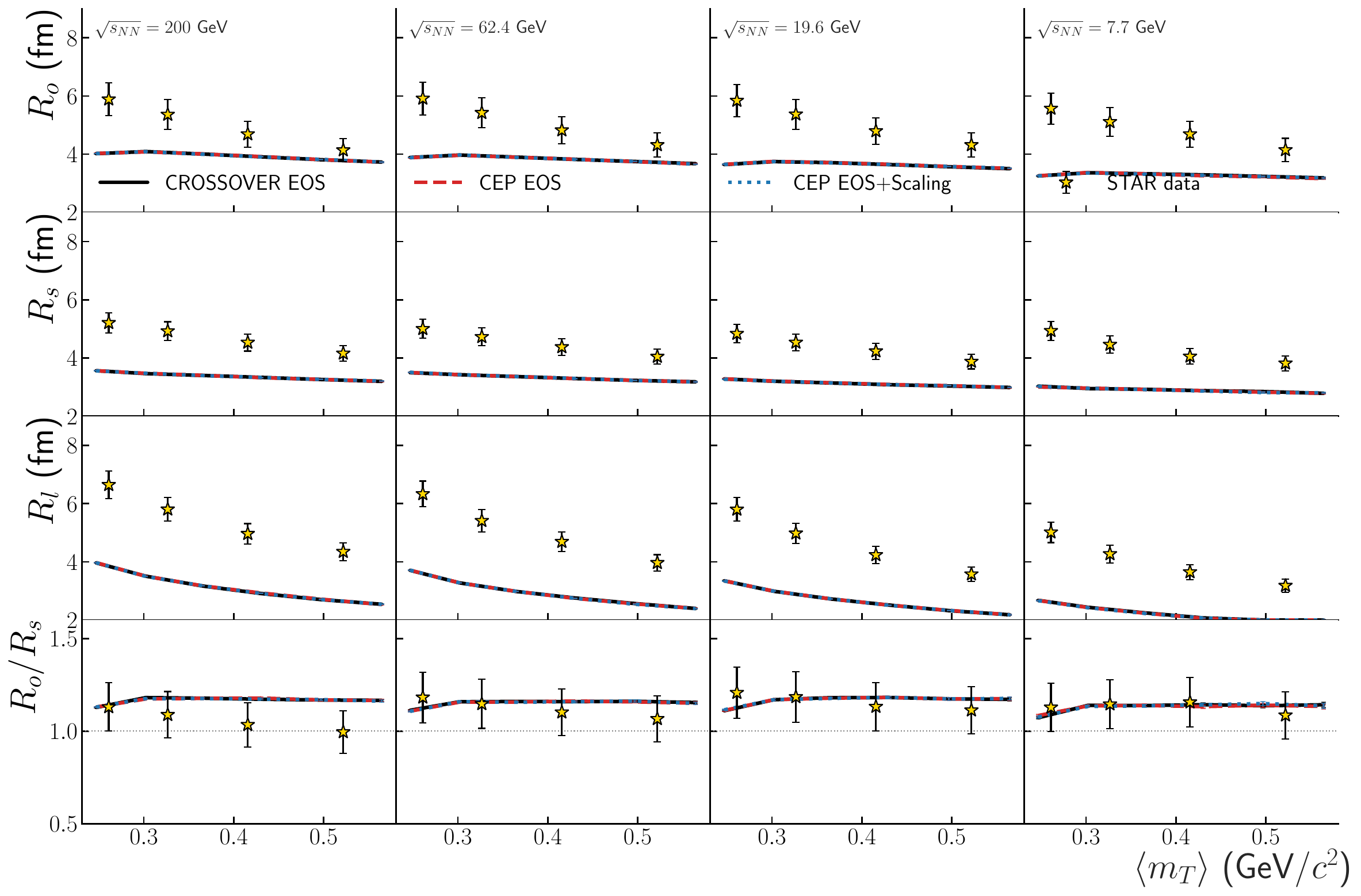}
    \caption{HBT radii obtained from a Gaussian fit are taken directly from switching using an equation of state that includes a critical point and critical scaling, compared with the crossover-only equation of state.} 
    \label{fig:hbtradii1criticalswitching}
\end{figure}

\begin{figure}[!htbp]
    \centering
    \includegraphics[width=\linewidth]{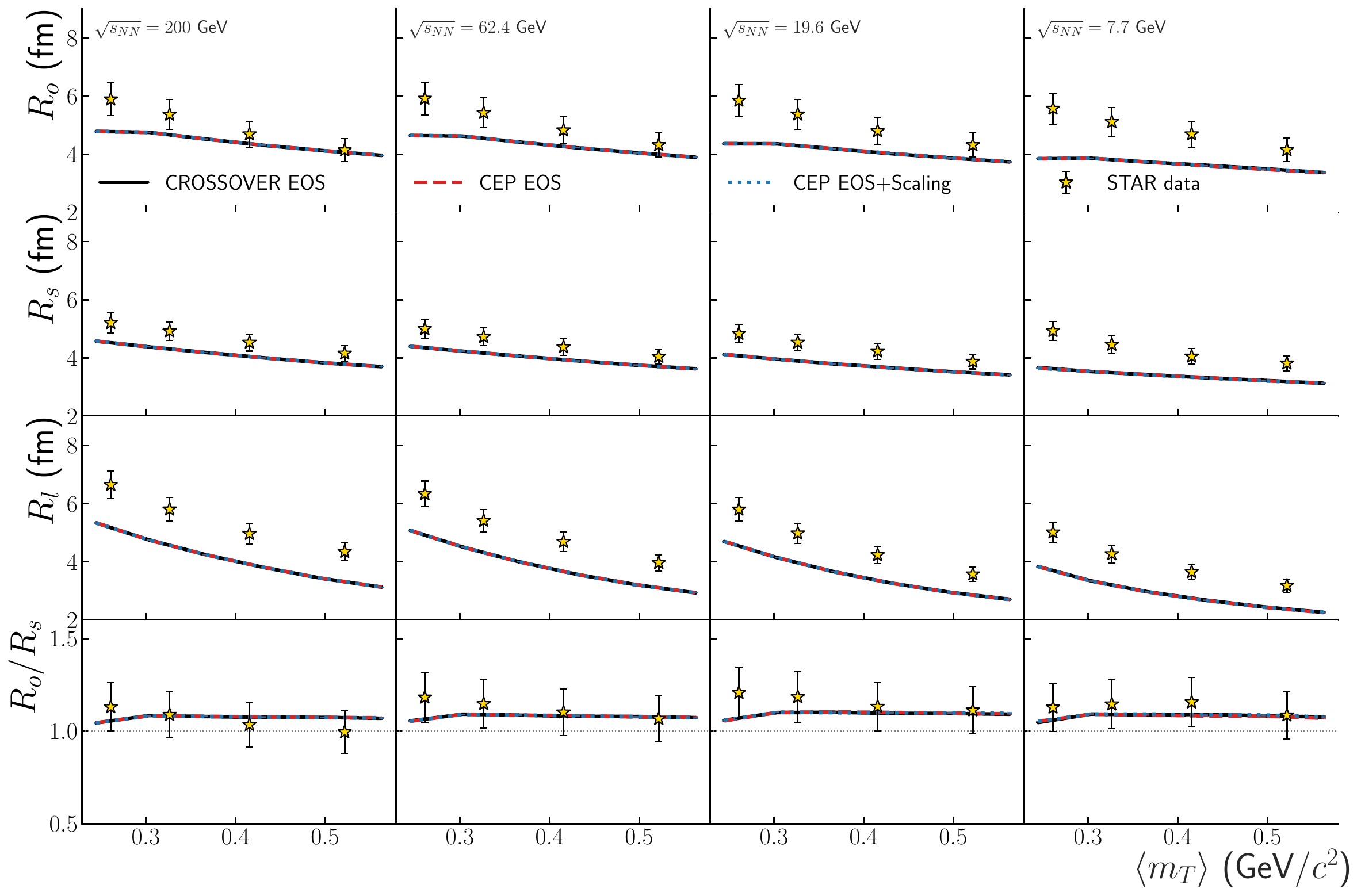}
    \caption{HBT radii extracted from a Gaussian fit from only decays using an equation of state with a critical point and critical scaling, compared with the crossover-only equation of state.} 
    \label{fig:hbtradii1criticaldecays}
\end{figure}

We also repeated this calculation, comparing the HBT radii obtained when considering direct emission at switching (Fig. \ref{fig:hbtradii1criticalswitching}), including those originating exclusively from decays (Fig. \ref{fig:hbtradii1criticaldecays}). No changes in the behavior of the radii were identified at these stages, indicating that the effects of the critical point and critical scaling are no longer manifesting in the hydrodynamic evolution itself. This weakens the hypothesis that full transport is exclusively responsible for eliminating these differences.

\subsection{Pion Interferometry Analysis Using a Lévy Source}

\begin{figure}[!htbp]
    \centering
    \includegraphics[width=\linewidth]{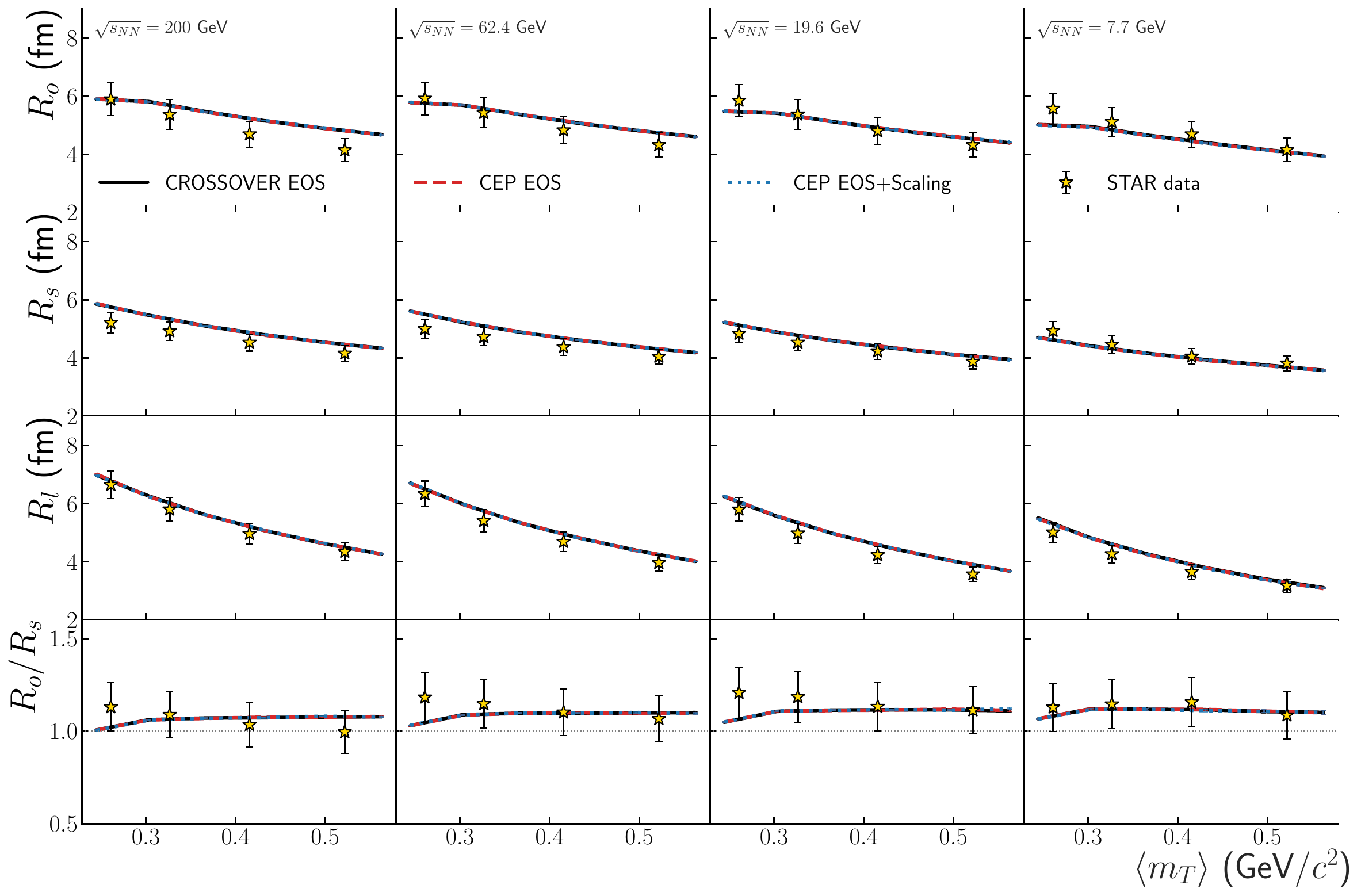}
    \caption{HBT radii extracted from a Lévy fit using an equation of state with a critical point and critical scaling, compared with the crossover-only equation of state.} 
    \label{fig:hbtradii2critical}
\end{figure}

\begin{figure}[!htbp]
    \centering
    \includegraphics[width=\linewidth]{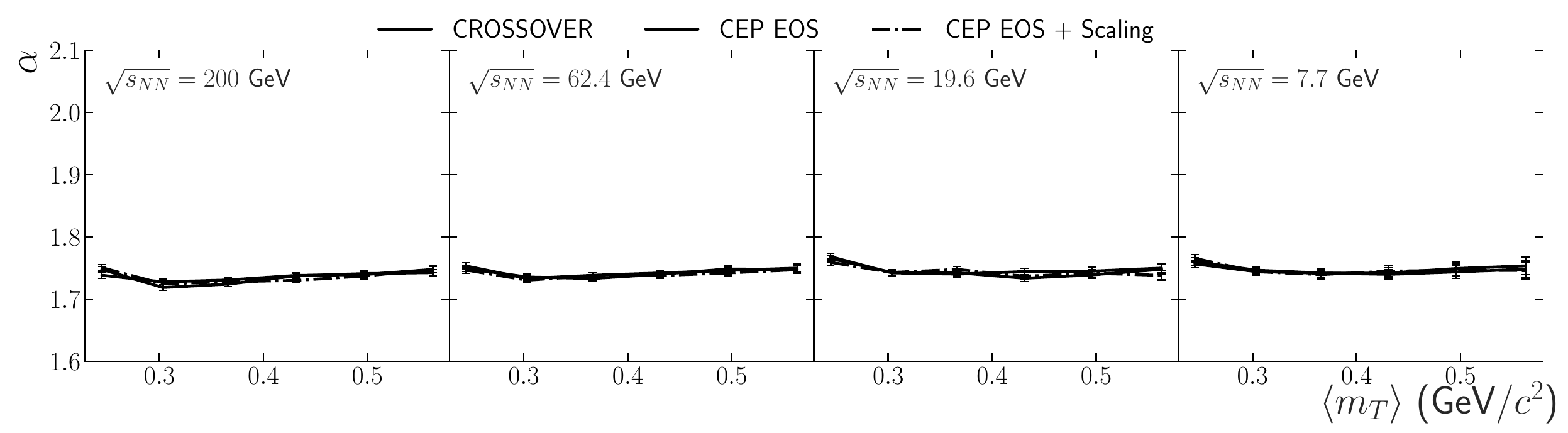}
    \caption{Values of $\alpha$ obtained with the Lévy fit using an equation of state with a critical point and critical scaling, compared with the crossover-only equation of state.}
    \label{fig:alphacomparision2}
\end{figure}

We obtained the HBT radii using the Lévy fit for simulations that employ the equation of state with a critical point and the critical scaling of the bulk transport coefficient together with the parameter $\alpha$. The results are presented in Figures \ref{fig:hbtradii2critical} and \ref{fig:alphacomparision2}.

\begin{figure}[!htbp]
    \centering
    \includegraphics[width=\linewidth]{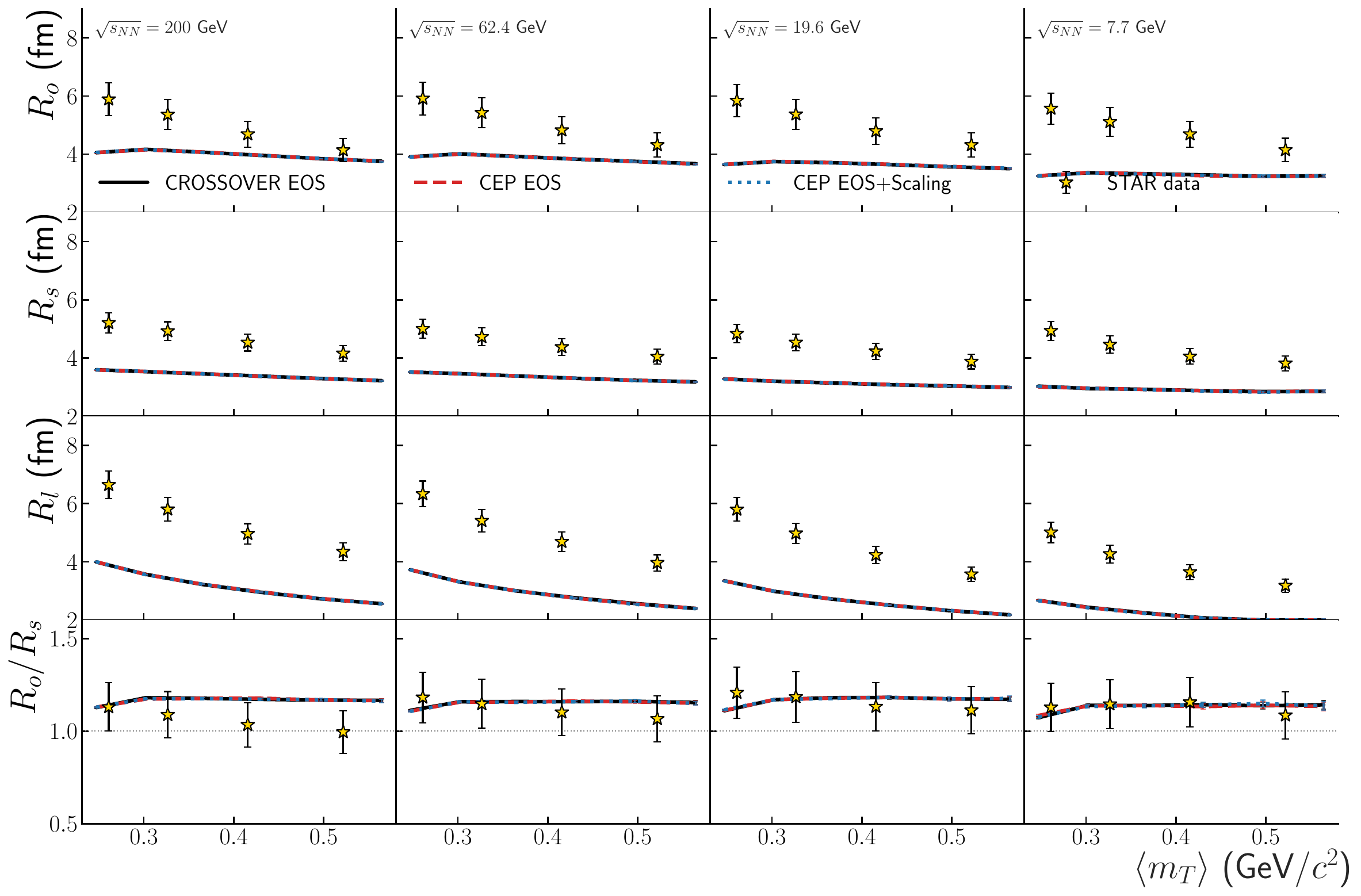}
    \caption{HBT radii obtained from a Lévy fit are taken directly from switching using an equation of state that includes a critical point and critical scaling, compared with the crossover-only equation of state.} 
    \label{fig:hbtradii2criticalswitching}
\end{figure}

\begin{figure}[!htbp]
    \centering
    \includegraphics[width=\linewidth]{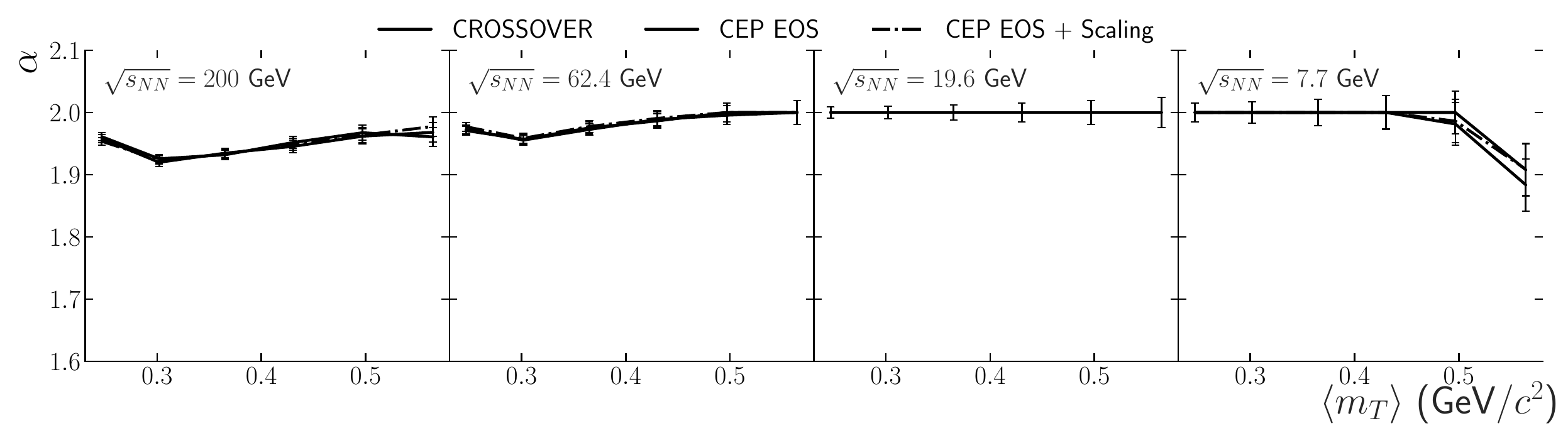}
    \caption{Values of $\alpha$ obtained with the Lévy fit directly from switching using an equation of state with a critical point and critical scaling, compared with the crossover-only equation of state.}
    \label{fig:alphacomparision2switching}
\end{figure}

\begin{figure}[!htbp]
    \centering
    \includegraphics[width=\linewidth]{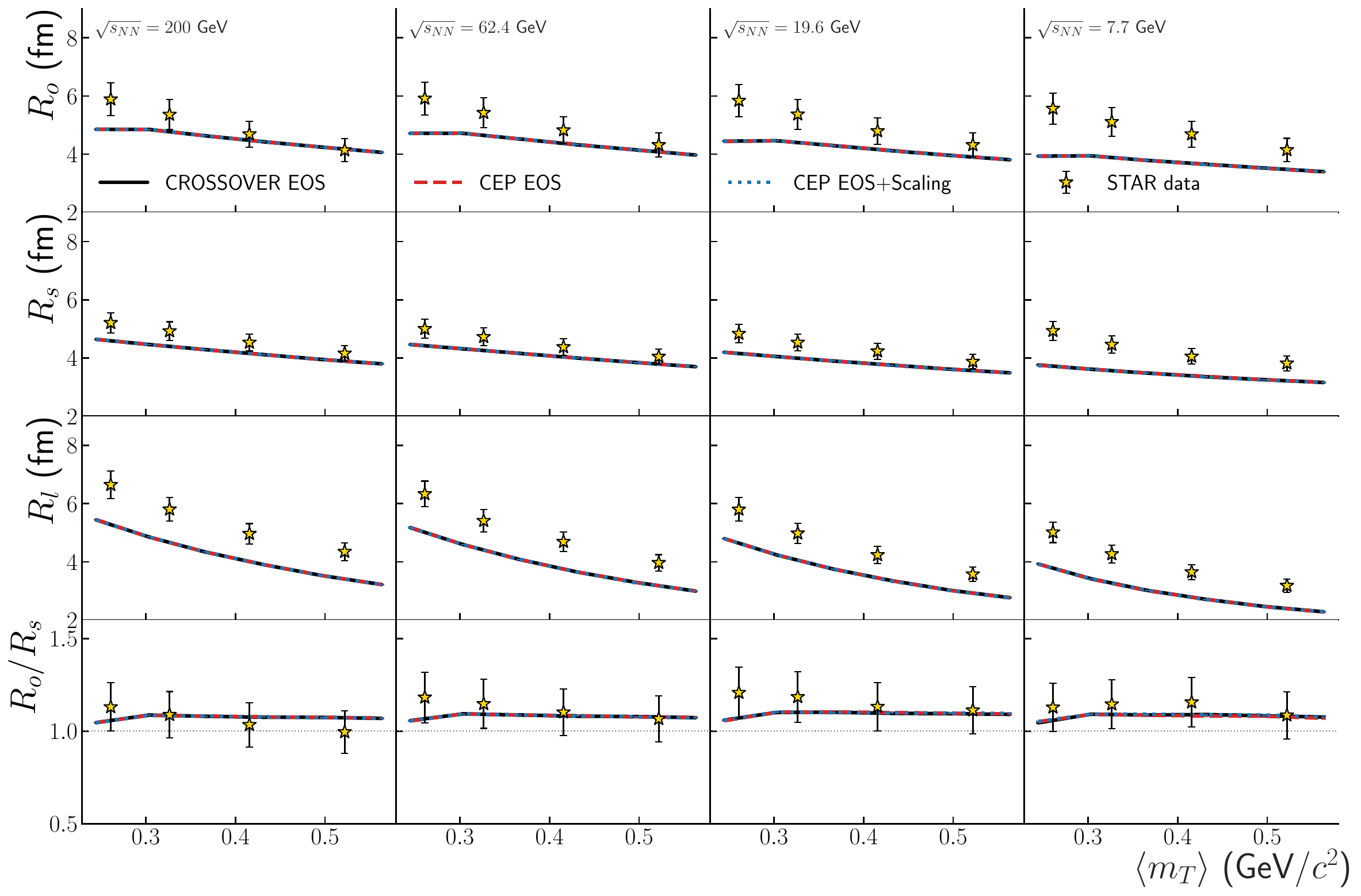}
    \caption{HBT radii extracted from a Lévy fit from only decays using an equation of state with a critical point and critical scaling, compared with the crossover-only equation of state.} 
    \label{fig:hbtradii2criticaldecays}
\end{figure}

\begin{figure}[!htbp]
    \centering
    \includegraphics[width=\linewidth]{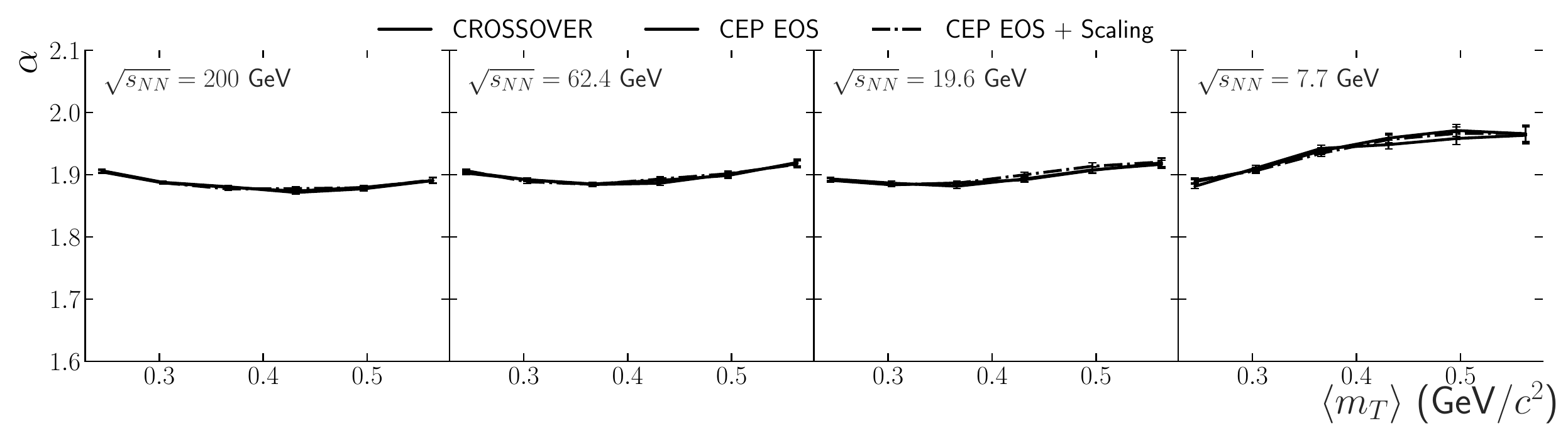}
    \caption{Values of $\alpha$ obtained with the Lévy fit from only decays using an equation of state with a critical point and critical scaling, compared with the crossover-only equation of state.}
    \label{fig:alphacomparision2decays}
\end{figure}

Once again, no significant changes were observed among the extracted radii. We present the results obtained directly from switching (Figures \ref{fig:hbtradii2criticalswitching} and \ref{fig:alphacomparision2switching}) and from only decays (Figures \ref{fig:hbtradii2criticaldecays} and \ref{fig:alphacomparision2decays}).


It is observed that the value of $\alpha$ does not exhibit significant changes between simulations using the crossover-only equation of state and simulations using an equation of state with a critical point, suggesting that the reduction of this parameter still arises predominantly from the intrinsic emission dynamics rather than from the incorporation of critical effects in the equation of state. This result, by itself, does not constitute evidence for the absence of a critical point, but it indicates that, for the collision energies analyzed, the pion emission channel does not display measurable sensitivity to such effects in the hybrid model employed. 

\pagebreak
$ $\\
\pagebreak

\section{HBT Radii Analysis Using Direct Calculation}

Since the HBT radii extracted by fitting the correlation functions do not provide direct access to the space-time variances of the emitting source, we directly calculated the moments of the emission distribution of final pions. This procedure is not new in HBT interferometry, \cite{Zi-Wei-Lin_2004, PhysRevLett.89.152301, molnar2003decoupling, PhysRevLett.92.052301, PhysRevLett.86.3981}, and we use this construction as a way to decompose the observable $R_o^2-R_s^2$ into its contributions.

To obtain these variances and covariances using the particle pairs, we proceed as follows. Let $f$ denote a quantity assigned to each particle. For every event $e$ within a specified $k_T$ interval, we introduce the non normalized pairwise average of the fluctuation $\delta f$ together with its corresponding normalization
\begin{align}
    &\langle \delta f \rangle_{\rm pair,e} = \sum_{i}\sum_{j\neq i} \delta_{\Delta k_T}\left(k_T - \frac{|\vec{p_{T,i}} - \vec{p_{T,j}}|}{2}\right)(f_{i} - f_j),\\
    &\langle 1 \rangle_{\rm pair,e} = \sum_{i}\sum_{j\neq i} \delta_{\Delta k_T}\left(k_T - \frac{|\vec{p_{T,i}} - \vec{p_{T,j}}|}{2}\right),
\end{align}
in correspondence with the distribution of pairs of the Eq.~\eqref{eq:gencorfunc2} after integrating over $q$ and written as a function of $k_T$ in Eq.~\eqref{eq:generic_projected_correlation}. We define the event average as
\begin{equation}
  \langle f \rangle_{\rm pair} =  \dfrac{\sum_e\langle f \rangle_{\rm pair,e}}{\sum_e\langle 1 \rangle_{\rm pair,e}}.
\end{equation}
For each accepted pair, we boost the relative separation
$\delta x_{ij}=x_i-x_j$ to the pair Bertsch-Pratt + LCMS coordinates, defining the pair differences $\delta x_o$, $\delta x_s$, $\delta x_l$, and $\delta t'$, and generically calling them $\delta x_A$. The corresponding single-particle source variances are obtained from the pair moments for each event $e$
\begin{align}
    &\left\langle \Delta x_A^2 \right\rangle
    =
    \frac{1}{2}
    \left\langle \delta x_A ^2 \right\rangle_{\rm pair},\\
    &\left\langle \Delta x_A \Delta x_B \right\rangle
    =
    \frac{1}{2}
    \left\langle \delta x_A \delta x_B \right\rangle_{\rm pair}.
    \label{eq:variancescovariances}
\end{align}
The factor $1/2$ accounts for the fact that the variance of the difference between two particles emitted from the same source is twice the variance of the single-particle distribution (details are given in the Appendix~\ref{apdx:moments}). 

\begin{figure}[!htbp]
    \centering
    \includegraphics[width=\linewidth]{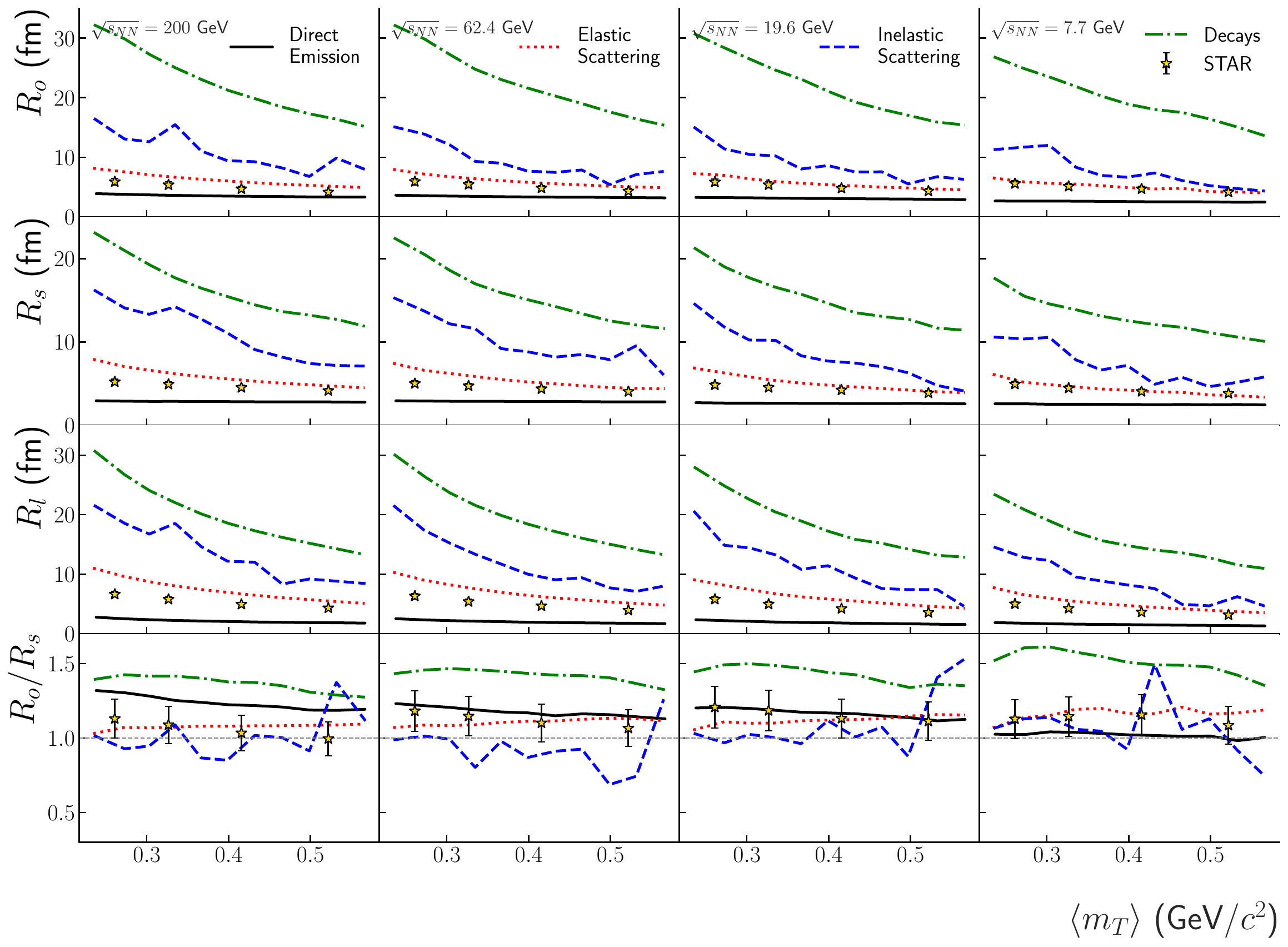}
    \caption{HBT radii of pions obtained through direct calculation for the types of pions emitted after the transport phase and decays. Experimental data are included for comparison.}
    \label{fig:hbtradiipionstype}
\end{figure}

The pair selection followed the procedure outlined in Section~\ref{subsec:hbtanalysis}: we imposed the experimental cuts and backpropagated the SMASH simulation from its final time to the time of the last interaction for each produced pion. For pions produced in Au+Au collisions, using the hybrid chain with the crossover-only equation of state, we calculated the HBT radii using the variances and covariances, as defined in Eq.~\eqref{eq:lcmsoslhbtradii}, where the pion pairs were classified according to their last interaction in the hadronic transport process: direct pions (originating from the switching and not interacting after emission), pions from elastic collisions, pions from inelastic collisions, and pions from decays. The curves in Fig.~\ref{fig:hbtradiipionstype} present the results of this calculation.

It can be seen that the radii computed for direct pions are smaller than the experimental values, while the radii for all other interaction types are consistently larger than the experimental data, as indicated by the relation below:
\begin{equation*}
    \fbox{
    \text{Direct Emission} < \text{Elastic Scattering} < \text{Inelastic Scattering} < \text{Decays}.
    }
\end{equation*}

This result is not surprising, since the effect of increasing radii due, for example, to decays and resonances is well known~\cite{ PhysRevC.56.3265, PhysRevC.56.R610}. An analogous trend is observed in the ratio $Ro/Rs$, but the growth behavior is different: the ratio from direct pions is larger than that of elastically scattered pions, except for the noisier data associated with pions from inelastic scatterings. For 7.7 GeV, the trend in the ratio $Ro/Rs$ is reversed. None of the pion contributions results in a source size compatible with the experimental data. In fact, apart from the direct contributions, the estimates remain systematically above the data. Additionally, the pions from decays lead to substantially higher radius values than the data, exceeding the value of $R_o$ by more than a factor of 3. In the context of the direct calculation, the sum of the contributions to the HBT radii does not, in general, imply a reduction of those radii. In particular, for the calculation of $R_s^2$ and $R_l^2$ in Eq.\eqref{eq:lcmsoslhbtradii}, the inclusion of new pairs in \eqref{eq:variancescovariances} necessarily generates positive contributions, so that, when considering pairs of the type direct $\pi^\pm$ + non-direct $\pi^\pm$, the estimated values of the radii tend to increase.

\subsection{Using Kinematic Cuts for Core Determination}

In extracting the HBT radii by fitting the correlation functions, the parameter $\lambda$ effectively incorporates the reduction in correlation strength associated with halo contributions. Thus, the radii obtained from the fit mainly characterize the resolvable part of the pion source, i.e., the core. In the direct calculation, there is no parameter like $\lambda$ that plays the role of absorbing the halo contributions, so the isolation of the core must be performed in some other way. As we saw previously, the core is not characterized by pion pairs originating from a particular type of hadronic transport interaction. This motivates the introduction of some kind of pion cut that makes it possible to isolate the resolvable part of the source, in order to obtain HBT radii compatible with the experimental homogeneity regions and with those obtained by the Gaussian fit to the curves in Fig.~\ref{fig:hbtradii}.

\begin{figure}[!htbp]
    \centering
    \includegraphics[width=\linewidth]{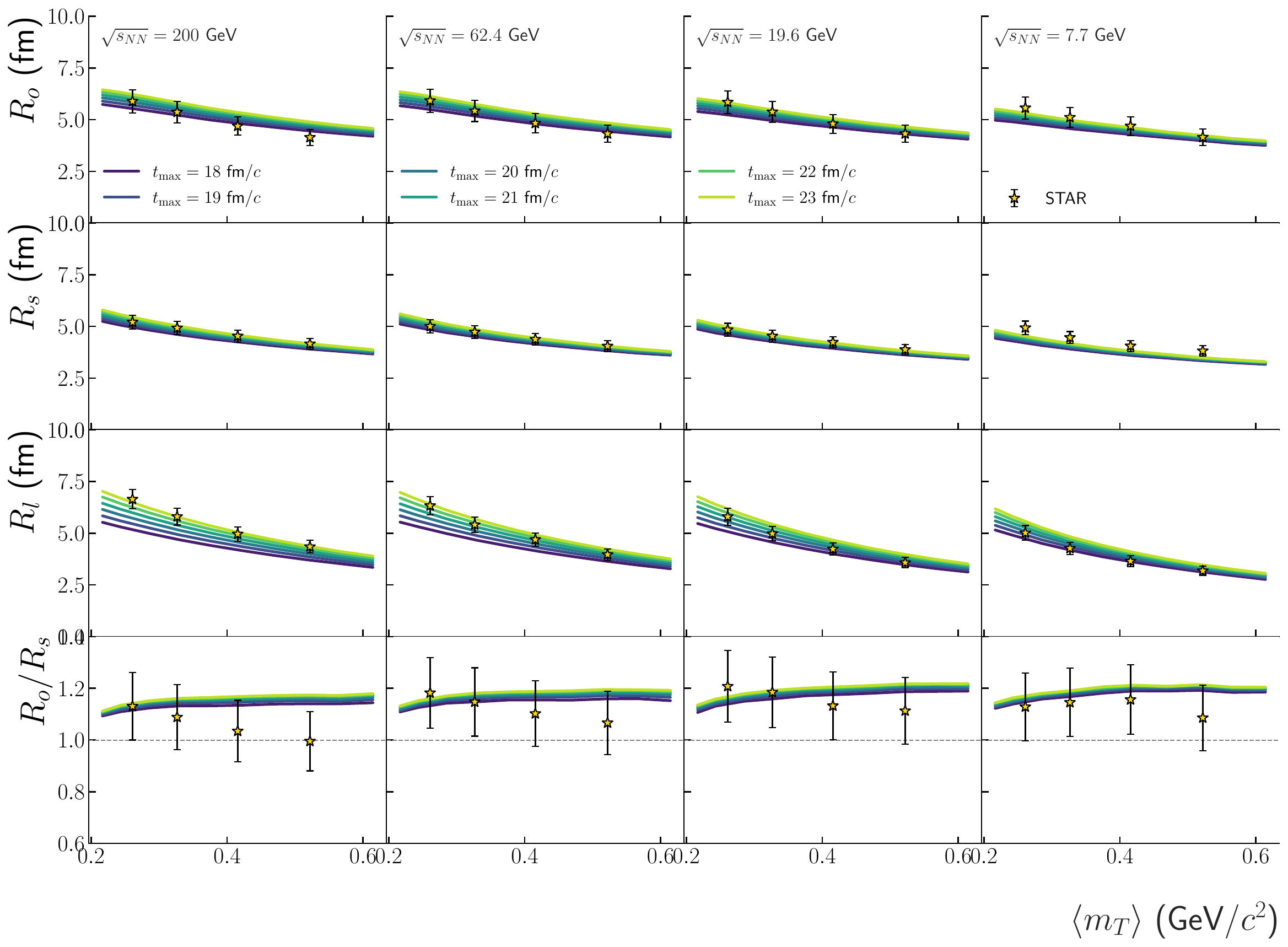}
    \caption{HBT radii were obtained via direct calculation for BES energies, considering different values of $t_{max}$ in the range between 18 and 23 fm/c. Experimental data are presented for comparison. }
    \label{fig:hbtradiitimecut}
\end{figure}

There are several ways to introduce pion cuts. The first one we tested is tied to each individual pion, in particular to the last interaction undergone by each one. In the calculation of correlation functions, the relevant times for femtoscopic interactions are those at which pions undergo kinetic freeze-out, that is, when they stop interacting and fly to the detector. We therefore initially tested a cut selecting pions whose final interaction occurred before a $t_{max}$. The physical justification behind this cut is that pions that interact for a long time extend the tails of the Gaussian distribution, which increases the distribution variances and, consequently, the radii. We chose values of $t_{max}$ ranging from $18$ fm/c to $23$ fm/c and then calculated the HBT radii using the variance- and covariance-dependent expressions, including all produced pions whose last interaction occurred before $t_{max}$. The results are shown in Fig.~\ref{fig:hbtradiitimecut}. There is a non-monotonic growth relationship for all radii, including the ratio $R_o/R_l$, as this time increases. The difference between the curves is more pronounced in $R_l$ and smaller in $R_s$. In general, the $R_o/R_l$ curves obtained with the direct calculation lay above the data.
\begin{figure}[!htbp]
    \centering
    \includegraphics[width=\linewidth]{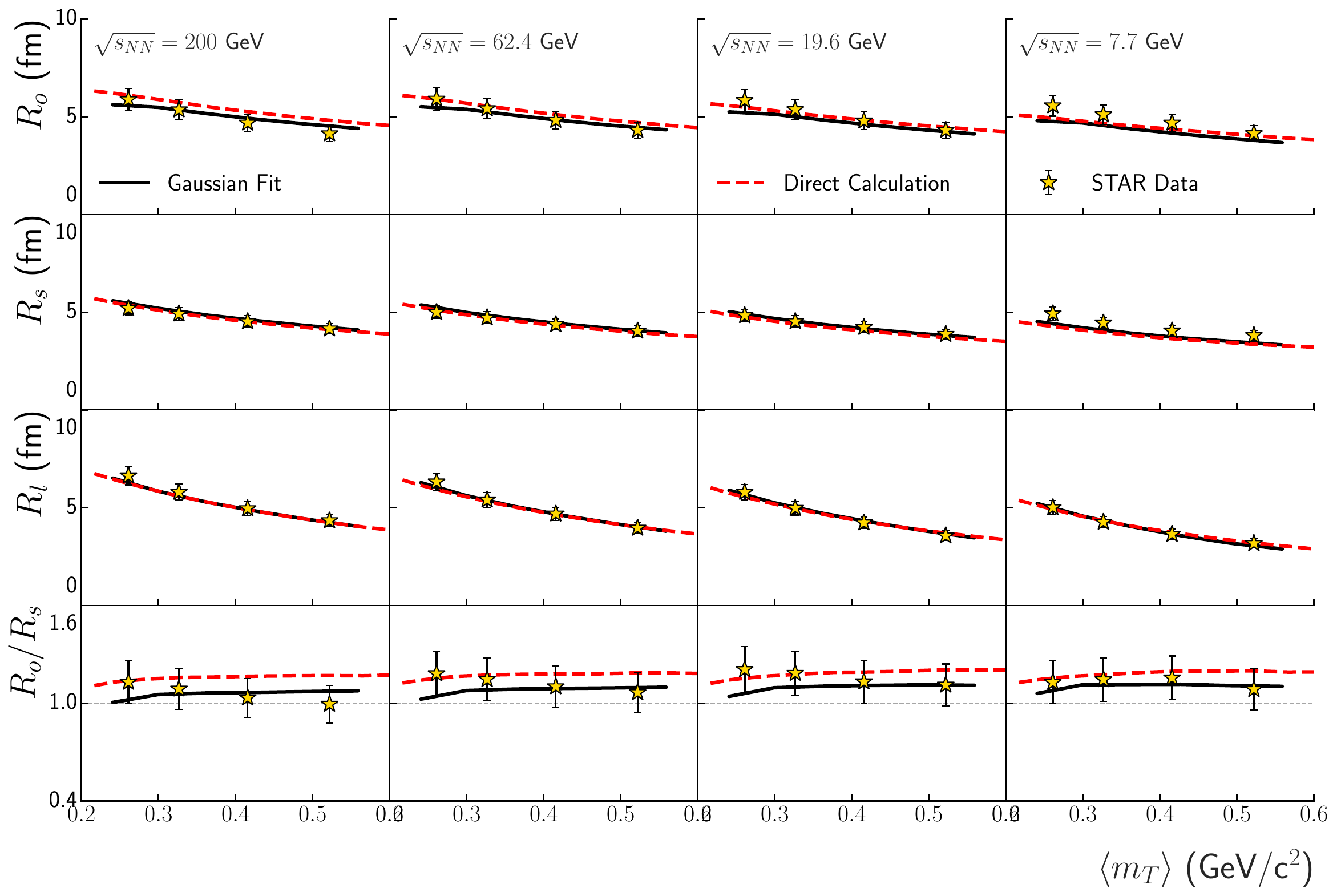}
    \caption{Comparison between the radii obtained via direct calculation and the cuts at $t_{max}$ that reproduce the radii obtained with a Gaussian fit. Experimental data are presented for comparison. }
    \label{fig:hbtradiitimecut2}
\end{figure}

In the curves of Fig.~\ref{fig:hbtradiitimecut2}, we show that for certain collision energies the pion HBT radii are well approximated by specific values of $t_{max}$, as shown in the table below:

\begin{table}[H]
    \centering
    \begin{tabular}{||r|c c c c||}
        $\sqrt{s_{NN}}$ (GeV) &  7.7 &  19.6 &  62.4 &  200  \\
        $t_{max}$ (fm/c) & 19 & 20 & 21 & 22
    \end{tabular}
\end{table}

It is found that selecting these parameters adequately matches the values of $R_s$ and $R_l$ obtained from the Gaussian fit. In contrast, the parameter $R_o$ remains above the value estimated by this fit and, consequently, above the experimental data. This pattern suggests the need to incorporate an additional pion selection (cut) criterion, less restrictive in nature, so as to predominantly affect the determination of the radius $R_o$.

A cutoff criterion associated with the transverse components $out$ and $side$ was proposed, in order to tie it to the pion-pair system rather than to each pion individually, as is the case for $t_{max}$. Each pion pair defines a reference frame in the center-of-momentum system, as discussed in Section~\ref{sec:kinematic_variables}. Combining these two aspects, the transverse distance in the pair’s center-of-momentum frame, $r_T^*$, is introduced: the boost given by \eqref{eq:boostboost} is applied to the transverse separations to obtain $\Delta x^*$ and $\Delta y^*$, such that
\begin{equation}
    (r_T^*)^2 = (\Delta x^*)^2 + (\Delta y^*)^2.
\end{equation}

\begin{figure}[!htbp]
    \centering
    \includegraphics[width=\linewidth]{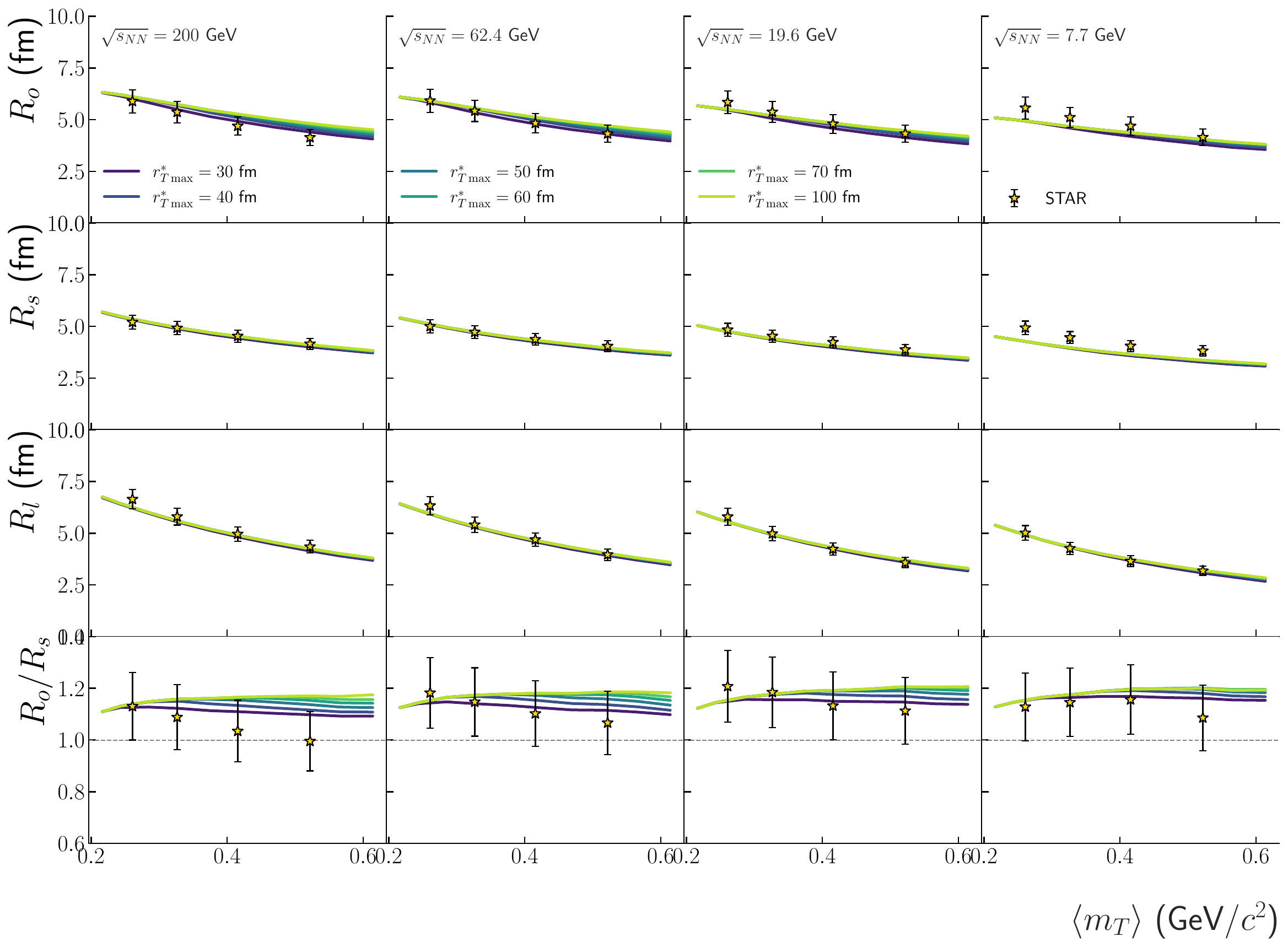}
    \caption{HBT radii were obtained via direct calculation for BES energies, considering different values of $r^*_{T,max}$ in the range between 30 and 100 fm. Experimental data are presented for comparison. }
    \label{fig:hbtradiispacecut}
\end{figure}

We chose values of $r^*_{T,max}$ ranging from $30$ fm to $100$ fm and then calculated the HBT radii using the variance and covariance dependent expressions, including all produced pion pairs whose $r^*_{T} < r^*_{T,max}$. The results are shown in Fig.~\ref{fig:hbtradiispacecut}, where we continue to use the $t^*_{max}$ values from the previous table that best describe the results of the Gaussian fit.

Applying this cut did not result in visible changes in the radii $R_s$ and $R_l$, but it induced a reduction in the values of $R_o$ in the high-$m_T$ regions, associated with a decrease in $r^*_{T,\max}$. As a consequence, a decrease was observed in the $R_o/R_s$ curves.

From the analytical calculations~\ref{fig:analyticialhbtradii}, one observes that the HBT radii $R_l$ are properly described by the boost invariance predicted in the Bjorken model. The direct calculation of the variances reproduces this behavior only when the time cut is applied, which can be interpreted as a consequence of the temporal hydrodynamic evolution occurring along hypersurfaces of constant proper time, $\tau=\sqrt{t^2-z^2}$, in analogy with the Bjorken scenario itself. Thus, the dependence of $R_l$ constitutes a signature of boost-invariance symmetry, taking the experimental cuts into account.

Imposing an upper limit on $r_T^*$ can introduce a cut on pion pairs that are very widely separated in the transverse plane, which contribute to the broadening of the tails of the Gaussian distribution in that direction but do not exhibit comparable separations in the longitudinal direction. In the absence of this cut, such contributions tend to yield larger values of $R_o$, although $R_s$ remains satisfactorily reproduced. Adopting a limit on $r_T^*$ defined in the pair center-of-mass frame ensures, in particular, that the relevant separation scale is determined by the particle-pair system itself.

\begin{figure}[!htbp]
    \centering
    \includegraphics[width=\linewidth]{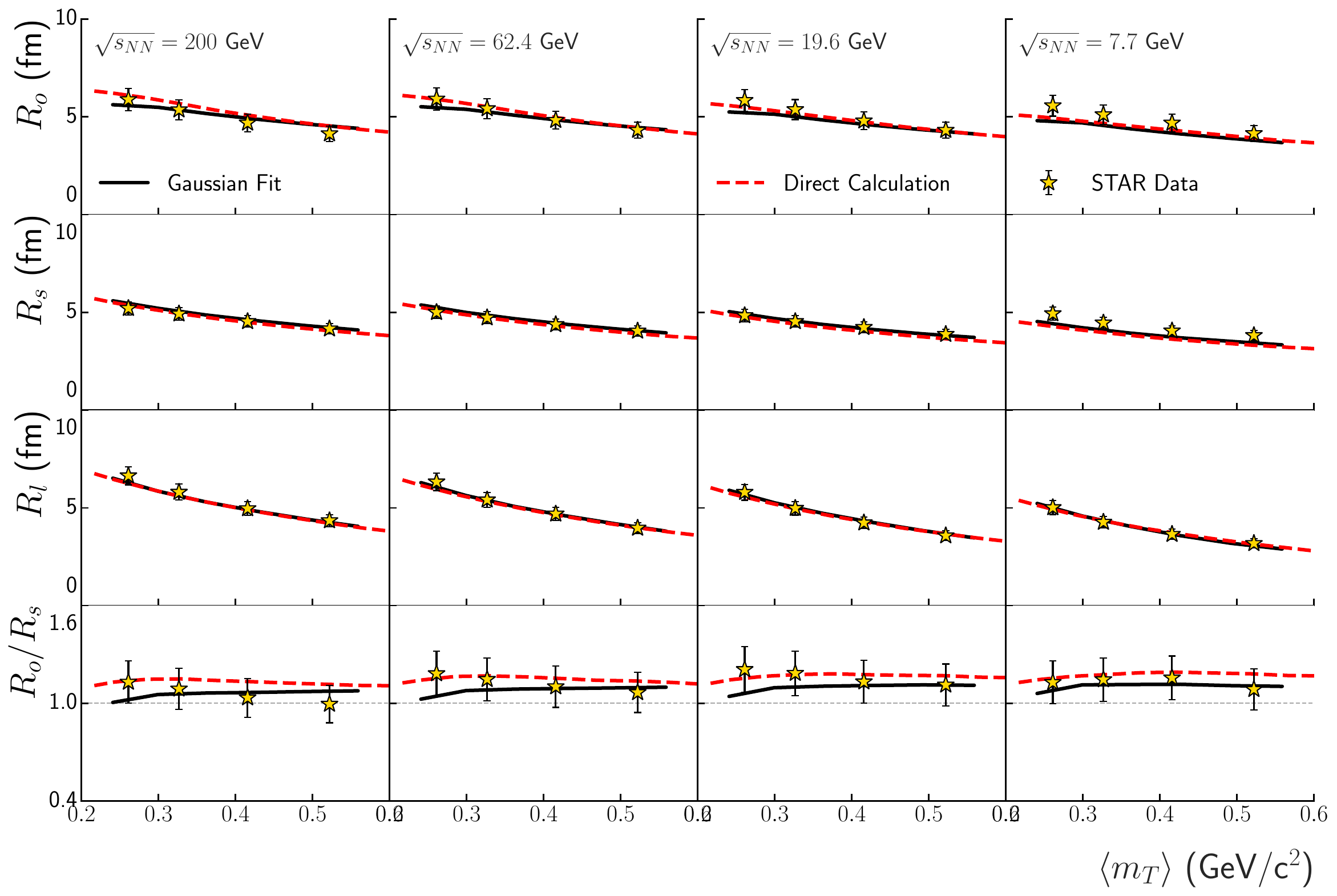}
    \caption{Comparison between the radii obtained via direct calculation with the cuts at $t_{max}$ and $r^*_{T,max}$ that reproduce the radii obtained with a Gaussian fit. Experimental data are presented for comparison.  }
    \label{fig:hbtradiispacecut2}
\end{figure}

Here, for every collision energy, applying a cut of $r^*_{T,\max}=40$ fm allowed us to lower the $R_o$ curves so that they match the Gaussian fit, as shown in Fig.~\ref{fig:hbtradiispacecut2}. It is possible to note that there is a slight tension between the direct calculation method and the HBT radii obtained with the Gaussian fit at low $m_T$ for $R_o$. This discrepancy may be related to the limitations of the Gaussian fit or even to particle statistics. However, compared with the experimental data themselves, the result remains satisfactory. This means that, for lower values of $m_T$, the ratio $R_o/R_s$ obtained with the direct calculation reproduces the experimental data better than the ratio obtained with the Gaussian fit.

Up to this point, we have been able to reproduce, through direct calculation, the HBT radii obtained from the Gaussian fit to the correlation functions. We observed this agreement at all simulated collision energies, which provides evidence that the HBT radii extracted from the fit capture the behavior of the space–time variances of the pion-emitting regions in the simulations. However, it is important to emphasize that these cuts are not decisive for classifying pions that the literature describes as belonging to the core and the halo. They serve only as a tool to isolate these produced regions of homogeneity in the simulations.

\subsection{Characterizing Source Moments}

\begin{figure}[!htbp]
    \centering
    \includegraphics[width=\linewidth]{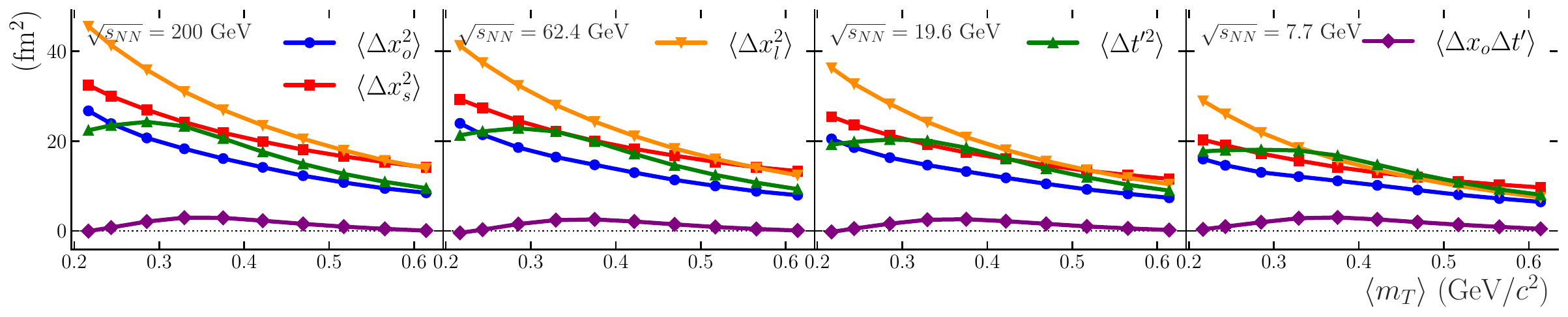}
    \caption{Variances and covariances, obtained by direct calculation using the cuts defined for reproducing the HBT radii using the Bertsch–Pratt+LCMS reference frame.}
    \label{fig:variances}
\end{figure}

Using the cuts that best fit the radii, we plotted the variances and covariances involved in the calculations of the HBT radii in Eq.~\eqref{eq:lcmsoslhbtradii} in Fig.~\ref{fig:variances}.

We know that $R_l^2 = \langle \Delta x_l^2 \rangle$ and $R_s^2 = \langle \Delta x_s^2 \rangle$, so there is not much we can say once these parameters already reproduce the Gaussian fit and the data. The most interesting point that arises here concerns the components that make up $R_o^2$. Note that, over almost the entire range of $m_T$, $\langle \Delta x_s^2 \rangle > \langle \Delta x_o^2 \rangle$; they only coincide for high values of $m_T$. This probably stems from the fact that the $out$ axis is defined along the direction of emission of the particle pair, whereas the $side$ axis is perpendicular to the transverse plane of the collision. Variances are measures of deviation from the mean, and this asymmetry can be explained by considering that, in the $out$ direction, there is a greater accumulation of pair density, precisely because it is the pair emission axis, which makes the distribution in that direction denser and closer to the mean. On the other hand, in the $side$ direction, which is perpendicular to the emission in the transverse plane, the concentration of pairs is smaller, which makes the distribution sparse and increases the variance.

The term $\langle \Delta x_o \Delta t' \rangle$ is not zero over the entire $m_T$ range considered. However, its magnitude remains significantly smaller when compared with the other variances. It is observed that it starts from zero, reaches a maximum at $m_T \approx 0.35$ GeV, and then decreases again until it vanishes. In turn, the temporal term $\langle \Delta t'^2 \rangle$ exhibits higher values, of the same order of magnitude as the purely spatial variances. This term increases until it reaches a maximum at $m_T \approx 0.29$ and then begins to decrease, displaying behavior qualitatively similar to that of the spatial variances, which, in turn, show an approximately monotonic decrease over $m_T$.

It is observed that, in the context of the hybrid model considered, the approximations $\langle \Delta x_o^2 \rangle\approx\langle\Delta x_s^2 \rangle$ and $\langle \Delta x_o \Delta t'\rangle \approx 0$ are not satisfied. This result is particularly interesting, since it shows that, for the simulation results presented here, $R_o^2 - R_s^2$ may not be dominated by the temporal term $\langle \Delta t'^2 \rangle$, as expressed in Eq.~\eqref{eq:RoRs_time_approx}. We can verify this by examining the curves of these terms, as in the plot of Fig.~\ref{fig:ro2-rs2-energy}.

\subsection{Probing $R_o^2 - R_s^2$ as a Time Emission Parameter}

\begin{figure}[!htpb]
    \centering
    \includegraphics[width=0.49\linewidth]{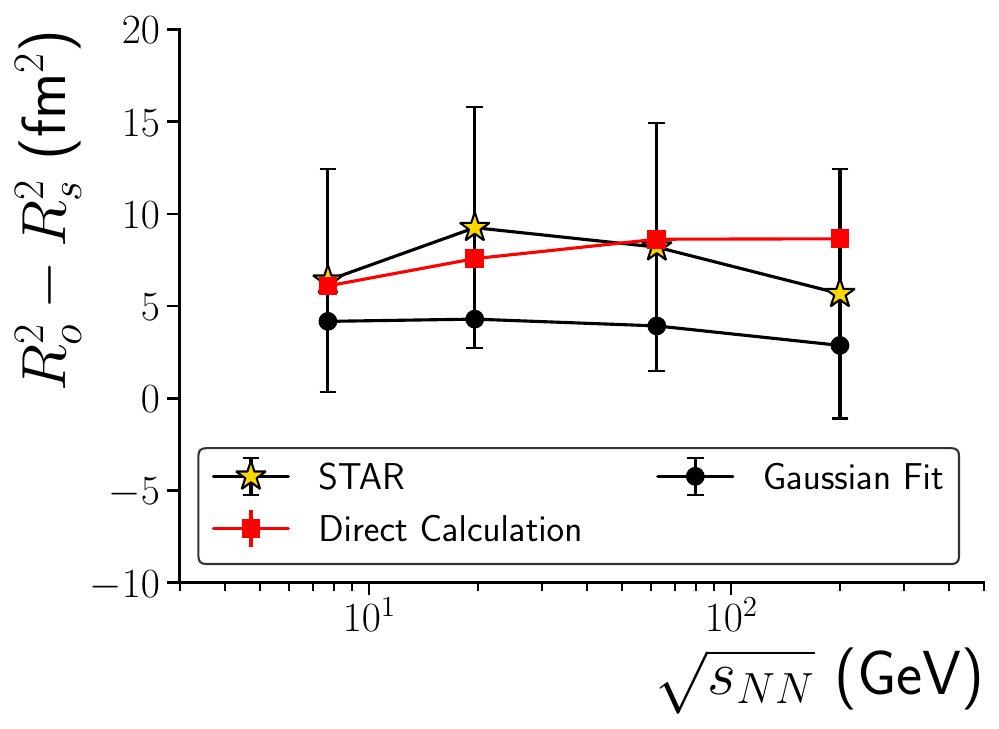}
    \includegraphics[width=0.49\linewidth]{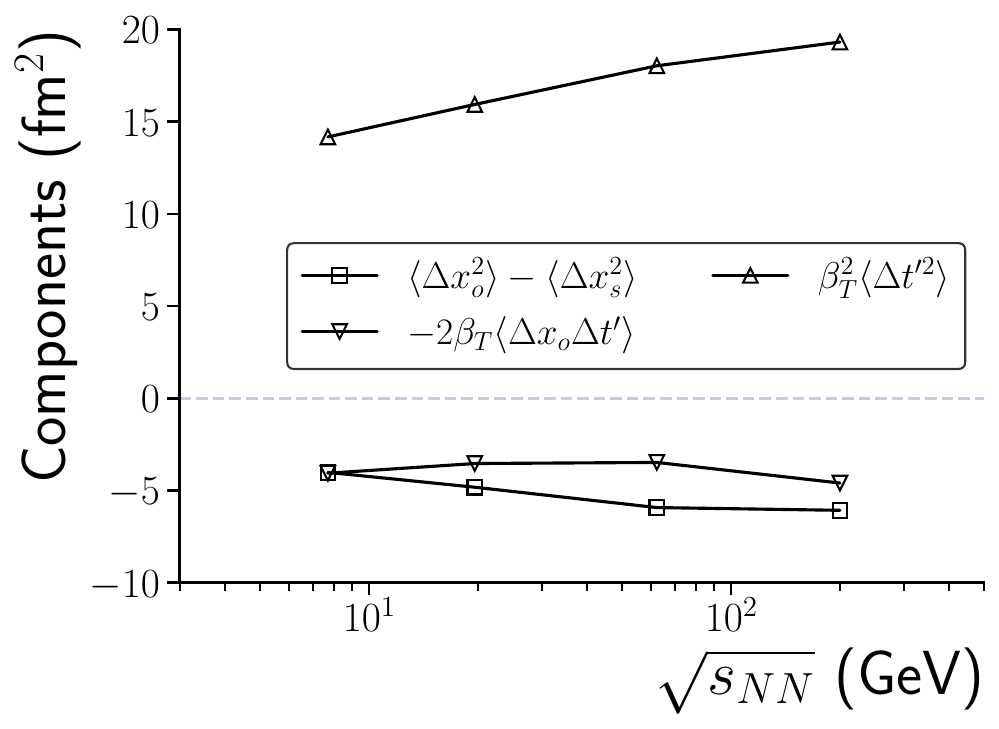}
    \caption{Energy dependence of $R_o^2-R_s^2$ and of its direct source decomposition. The left panel compares the STAR data with the values obtained from the Gaussian fit of the correlation function and from the direct calculation of source variances. The right panel shows the corresponding direct contributions $\langle\Delta x_o^2\rangle-\langle\Delta x_s^2\rangle$, $-2\beta_T\langle\Delta x_o\Delta t'\rangle$, and $\beta_T^2\langle\Delta t'^2\rangle$, evaluated in the same momentum bin. The final value of $R_o^2-R_s^2$ results from a cancellation between the positive duration term and the spatial and space-time correlation terms.}
    \label{fig:ro2-rs2-energy2}
\end{figure}

In the graph on the left of fig.~\ref{fig:ro2-rs2-energy2}, we present the curves of $R_o^2-R_s^2$ as a function of $\sqrt{s_{NN}}$ with $k_T \approx 0.3$, and on the right, we show the curves for $\langle \Delta x_o^2 \rangle - \langle \Delta x_s^2 \rangle$, $-2\langle \beta_T \Delta x_o \Delta t' \rangle$, and $\langle \beta_T^2 \Delta t'^2 \rangle$ to understand the behavior of these terms.

Both $\langle \Delta x_o^2 \rangle - \langle \Delta x_s^2 \rangle$ and $-2\langle \beta_T \Delta x_o \Delta t' \rangle$ take negative values, of the order of $\sim -5$ fm$^2$, while $\langle \beta_T^2 \Delta t'^2 \rangle$ takes positive values that increase with $\sqrt{s_{NN}}$, from $\sim 15$ fm$^2$ to $\sim 20$ fm$^2$. The combination of these terms results in an $R_o^2-R_s^2$ that increases with $\sqrt{s_{NN}}$, ranging between $5$ and $10$ fm$^2$. Thus, under the conditions of the hybrid model employed, the approximations made in the literature concerning the temporal dominance of the $\langle \beta_T^2 \Delta t'^2 \rangle$ term over the others do not hold. This makes the interpretation of $R_o^2-R_s^2$ as an observable providing a direct measure of the emission time more difficult.

On the other hand, the direct calculation does not reproduce the behavior of $R_o^2-R_s^2$ obtained from the Gaussian fit. The results are systematically higher and exhibit a monotonically increasing behavior as a function of $\sqrt{s_{NN}}$, in contrast to the non-monotonic behavior shown by the experimental data, although they remain within the error bars.

\pagebreak

\chapter{Conclusions}

In this work, we investigate whether an equation of state containing a critical point and a first-order transition can produce observable effects on the HBT radii of pions produced in Au+Au collisions at energies in the STAR-BES program. To this end, we employ a hybrid framework consisting of AMPT-based initial conditions, 3+1D viscous relativistic hydrodynamic evolution at finite baryon chemical potential, particlization via the Cooper–Frye prescription, and hadronic evolution with SMASH. The HBT radii $R_o$, $R_s$, and $R_l$ were extracted by fitting correlation functions of identical pions.

Initially, we used an equation of state containing only a smooth crossover-type transition. The spectra of charged and identified particles were described reasonably well, especially at the higher and intermediate BES energies. Discrepancies at $\sqrt{s_{NN}}=7.7~\mathrm{GeV}$ may reflect challenges in modeling energy deposition, longitudinal dynamics, and the evolution of conserved charges at low collision energies. The anisotropic flow analysis showed that the model reproduces the overall scale of $v_2$ and $v_3$, providing a consistent description of the collective dynamics observed experimentally, with larger discrepancies observed at $\sqrt{s_{NN}}=19.6$ GeV and 7.7 GeV.

The HBT radii obtained with a Gaussian fit reproduce the scale and the $m_T$ dependence observed in the STAR data. The main discrepancy appears in $R_o$ and, consequently, in $R_o/R_s$, particularly at the lowest energy analyzed. This indicates that, although the space-time evolution of the pion source is described reasonably well, there are still limitations in the detailed description of the dynamics, especially at low energies. The analysis at different stages of the evolution showed that decays already produce systematic modifications in the HBT radii, increasing their values and reducing $R_o/R_s$, while full hadronic transport introduces additional effects. Thus, the dynamics after switching play an important role in determining the femtoscopic properties of the source. Fits with a Lévy source yielded radii similar to those obtained with the Gaussian parametrization, but resulted in values of $\alpha<2$, indicating deviations from a Gaussian source shape.

Next, we repeated the analysis using an equation of state featuring a critical point and a first-order transition, both without and with the inclusion of critical scaling in the bulk-viscosity coefficient. Despite the modifications introduced in the hydrodynamic evolution in the critical region, no substantial differences were observed in the particle spectra, anisotropic flow, or HBT radii relative to the crossover case. The same behavior was observed when the HBT radii were calculated directly at switching and after including only the decays. Thus, within the model employed and for the energies investigated, the presence of the critical point and the critical scaling incorporated into the bulk-viscosity coefficient did not yield a clearly identifiable signature in the final observables studied.

This result does not necessarily imply the absence of critical effects in the collision, but it indicates that their manifestation in the analyzed observables may be diminished during the system’s evolution and, in particular, by the integration of effects throughout the hydrodynamic and hadronic dynamics. The absence of a visible modification in the HBT radii also shows that introducing a critical structure into the equation of state, even when accompanied by a critical-scaling parameterization in a transport coefficient, is not sufficient, by itself, to produce an evident femtoscopic signature.

As next steps, it is important to investigate a more complete description of non-equilibrium critical dynamics, including the dynamical evolution of the correlation length and of critical modes coupled to hydrodynamics. It will also be relevant to extend the study to different centrality classes and to perform a more systematic energy scan in order to explore different trajectories in the phase diagram and increase the likelihood of probing the critical region. Extending the analysis to additional centrality classes could also allow us to examine how the HBT radii vary with particle multiplicity and compare this behavior with experimental results. Finally, a more detailed investigation of the impact of switching, decays, and hadronic transport on the spatiotemporal structure of the source may help identify which femtoscopic observables are most sensitive to critical dynamics and under which conditions this sensitivity can be preserved up to the final state.

\begin{singlespace}
\printbibliography[heading=bibintoc,title={References}]

@article{NeddermeyerAnderson1937,
  author  = {Neddermeyer, Seth H. and Anderson, Carl D.},
  title   = {Note on the Nature of Cosmic-Ray Particles},
  journal = {Physical Review},
  volume  = {51},
  number  = {10},
  pages   = {884--886},
  year    = {1937},
  doi     = {10.1103/PhysRev.51.884}
}

@article{StreetStevenson1937,
  author  = {Street, J. C. and Stevenson, E. C.},
  title   = {New Evidence for the Existence of a Particle of Mass Intermediate Between the Proton and Electron},
  journal = {Physical Review},
  volume  = {52},
  pages   = {1003--1004},
  year    = {1937},
  doi     = {10.1103/PhysRev.52.1003}
}

@article{RochesterButler1947,
  author  = {Rochester, G. D. and Butler, C. C.},
  title   = {Evidence for the Existence of New Unstable Elementary Particles},
  journal = {Nature},
  volume  = {160},
  pages   = {855--857},
  year    = {1947},
  doi     = {10.1038/160855a0}
}

@techreport{GellMann1961,
  author      = {Gell-Mann, Murray},
  title       = {The Eightfold Way: A Theory of Strong Interaction Symmetry},
  institution = {California Institute of Technology, Synchrotron Laboratory},
  number      = {CTSL-20},
  year        = {1961},
  doi         = {10.2172/4008239}
}

@article{Neeman1961,
  author  = {Ne'eman, Y.},
  title   = {Derivation of Strong Interactions from a Gauge Invariance},
  journal = {Nuclear Physics},
  volume  = {26},
  number  = {2},
  pages   = {222--229},
  year    = {1961},
  doi     = {10.1016/0029-5582(61)90134-1}
}

@article{GellMann1964,
  author  = {Gell-Mann, Murray},
  title   = {A Schematic Model of Baryons and Mesons},
  journal = {Physics Letters},
  volume  = {8},
  number  = {3},
  pages   = {214--215},
  year    = {1964},
  doi     = {10.1016/S0031-9163(64)92001-3}
}

@techreport{Zweig1964a,
  author      = {Zweig, George},
  title       = {An SU(3) Model for Strong Interaction Symmetry and Its Breaking},
  institution = {CERN},
  number      = {CERN-TH-401},
  year        = {1964},
  note        = {Version I}
}

@techreport{Zweig1964b,
  author      = {Zweig, George},
  title       = {An SU(3) Model for Strong Interaction Symmetry and Its Breaking. II},
  institution = {CERN},
  number      = {CERN-TH-412},
  year        = {1964},
  note        = {Version II}
}

@article{BrownTwiss1954,
  author  = {Brown, R. Hanbury and Twiss, R. Q.},
  title   = {A New Type of Interferometer for Use in Radio Astronomy},
  journal = {Philosophical Magazine},
  volume  = {45},
  number  = {366},
  pages   = {663--682},
  year    = {1954},
  doi     = {10.1080/14786440708520475}
}

@article{BrownTwiss1956Photon,
  author  = {Brown, R. Hanbury and Twiss, R. Q.},
  title   = {Correlation between Photons in Two Coherent Beams of Light},
  journal = {Nature},
  volume  = {177},
  number  = {4497},
  pages   = {27--29},
  year    = {1956},
  doi     = {10.1038/177027a0}
}

@article{BrownTwiss1956Sirius,
  author  = {Brown, R. Hanbury and Twiss, R. Q.},
  title   = {A Test of a New Type of Stellar Interferometer on Sirius},
  journal = {Nature},
  volume  = {178},
  number  = {4541},
  pages   = {1046--1048},
  year    = {1956},
  doi     = {10.1038/1781046a0}
}

@article{SilvaFreire2013HBT,
  author  = {Silva, Indianara and Freire, Olival},
  title   = {The Concept of the Photon in Question: The Controversy Surrounding the Hanbury Brown--Twiss Effect circa 1956--1958},
  journal = {Historical Studies in the Natural Sciences},
  volume  = {43},
  number  = {4},
  pages   = {453--491},
  year    = {2013},
  doi     = {10.1525/hsns.2013.43.4.453}
}

@article{Padula2005HBT,
  author  = {Padula, Sandra S.},
  title   = {{HBT} Interferometry: Historical Perspective},
  journal = {Brazilian Journal of Physics},
  volume  = {35},
  pages   = {70--99},
  year    = {2005},
  doi     = {10.1590/S0103-97332005000100005}
}

@article{GoldhaberGoldhaberLeePais1960,
  author  = {Goldhaber, G. and Goldhaber, S. and Lee, W. and Pais, A.},
  title   = {Influence of Bose--Einstein Statistics on the Antiproton-Proton Annihilation Process},
  journal = {Physical Review},
  volume  = {120},
  number  = {1},
  pages   = {300--312},
  year    = {1960},
  doi     = {10.1103/PhysRev.120.300}
}

@article{LisaPrattSoltzWiedemann2005,
  author  = {Lisa, Michael Annan and Pratt, Scott and Soltz, Ron and Wiedemann, Urs},
  title   = {Femtoscopy in Relativistic Heavy Ion Collisions: Two Decades of Progress},
  journal = {Annual Review of Nuclear and Particle Science},
  volume  = {55},
  pages   = {357--402},
  year    = {2005},
  doi     = {10.1146/annurev.nucl.55.090704.151533},
  eprint  = {nucl-ex/0505014},
  archivePrefix = {arXiv}
}

@article{WiedemannHeinz1999,
  author  = {Wiedemann, Urs Achim and Heinz, Ulrich},
  title   = {Particle Interferometry for Relativistic Heavy-Ion Collisions},
  journal = {Physics Reports},
  volume  = {319},
  number  = {4-5},
  pages   = {145--230},
  year    = {1999},
  doi     = {10.1016/S0370-1573(99)00032-0},
  eprint  = {nucl-th/9901094},
  archivePrefix = {arXiv}
}

@book{GottfriedYan2003,
  author    = {Gottfried, Kurt and Yan, Tung-Mow},
  title     = {Quantum Mechanics: Fundamentals},
  publisher = {Springer},
  year      = {2003},
  doi       = {10.1007/978-0-387-21623-2}
}

@book{MartinRothen2004,
  author    = {Martin, Philippe A. and Rothen, Fran{\c{c}}ois},
  title     = {Many-Body Problems and Quantum Field Theory: An Introduction},
  publisher = {Springer},
  year      = {2004},
  doi       = {10.1007/978-3-662-08490-8}
}

@book{BlaizotRipka1986,
  author    = {Blaizot, Jean-Paul and Ripka, Georges},
  title     = {Quantum Theory of Finite Systems},
  publisher = {MIT Press},
  address   = {Cambridge, MA},
  year      = {1986},
  isbn      = {9780262022149}
}

@misc{BurgessQFTNotes,
  author       = {Burgess, C. P.},
  title        = {From Schr{\"o}dinger to Feynman: A Hands-On Guide to Quantum Field Theory},
  howpublished = {Class notes},
  url          = {https://physics.mcmaster.ca/~cburgess/cburgess/?page_id=630},
  note         = {Accessed as online lecture notes}
}

@book{DInvernoVickers2022,
  author    = {D'Inverno, Ray and Vickers, James},
  title     = {Introducing Einstein's Relativity: A Deeper Understanding},
  publisher = {Oxford University Press},
  year      = {2022},
  doi       = {10.1093/oso/9780198862024.001.0001}
}

@book{Schwartz2014QFT,
  author    = {Schwartz, Matthew D.},
  title     = {Quantum Field Theory and the Standard Model},
  publisher = {Cambridge University Press},
  year      = {2014},
  doi       = {10.1017/CBO9781139540940}
}

@article{ChapmanHeinz1994,
  author  = {Chapman, S. and Heinz, U.},
  title   = {{HBT} Correlators: Current Formalism versus Wigner Function Formulation},
  journal = {Physics Letters B},
  volume  = {340},
  number  = {4},
  pages   = {250--253},
  year    = {1994},
  doi     = {10.1016/0370-2693(94)01277-6}
}

@book{CurtrightFairlieZachos2014,
  author    = {Curtright, Thomas L. and Fairlie, David B. and Zachos, Cosmas K.},
  title     = {A Concise Treatise on Quantum Mechanics in Phase Space},
  publisher = {World Scientific},
  year      = {2014},
  doi       = {10.1142/8870}
}

@article{KisielFlorkowskiBroniowskiPluta2006,
  author  = {Kisiel, Adam and Florkowski, Wojciech and Broniowski, Wojciech and Pluta, Jan},
  title   = {Femtoscopy in Hydro-Inspired Models with Resonances},
  journal = {Physical Review C},
  volume  = {73},
  pages   = {064902},
  year    = {2006},
  doi     = {10.1103/PhysRevC.73.064902},
  eprint  = {nucl-th/0602039},
  archivePrefix = {arXiv}
}

@article{CooperFrye1974,
  author  = {Cooper, Fred and Frye, Graham},
  title   = {Single-Particle Distribution in the Hydrodynamic and Statistical Thermodynamic Models of Multiparticle Production},
  journal = {Physical Review D},
  volume  = {10},
  number  = {1},
  pages   = {186--189},
  year    = {1974},
  doi     = {10.1103/PhysRevD.10.186}
}

@article{PrattCsorgoZimanyi1990,
  author  = {Pratt, S. and Cs{\"o}rg{\H{o}}, T. and Zim{\'a}nyi, J.},
  title   = {Detailed Predictions for Two-Pion Correlations in Ultrarelativistic Heavy-Ion Collisions},
  journal = {Physical Review C},
  volume  = {42},
  number  = {6},
  pages   = {2646--2652},
  year    = {1990},
  doi     = {10.1103/PhysRevC.42.2646}
}

@article{Pratt1997Smoothness,
  author  = {Pratt, Scott},
  title   = {Validity of the Smoothness Assumption for Calculating Two-Boson Correlations in High-Energy Collisions},
  journal = {Physical Review C},
  volume  = {56},
  number  = {2},
  pages   = {1095--1098},
  year    = {1997},
  doi     = {10.1103/PhysRevC.56.1095}
}

@book{Florkowski2010,
  author    = {Florkowski, Wojciech},
  title     = {Phenomenology of Ultra-Relativistic Heavy-Ion Collisions},
  publisher = {World Scientific},
  year      = {2010},
  doi       = {10.1142/7396}
}

@article{YanoKoonin1978,
  author  = {Yano, F. B. and Koonin, S. E.},
  title   = {Determining Pion Source Parameters in Relativistic Heavy-Ion Collisions},
  journal = {Physics Letters B},
  volume  = {78},
  number  = {5},
  pages   = {556--559},
  year    = {1978},
  doi     = {10.1016/0370-2693(78)90638-X}
}

@article{PlumbergHeinz2018,
  author  = {Plumberg, Christopher and Heinz, Ulrich},
  title   = {Hanbury Brown--Twiss Correlation Functions and Radii from Event-by-Event Hydrodynamics},
  journal = {Physical Review C},
  volume  = {98},
  pages   = {034910},
  year    = {2018},
  doi     = {10.1103/PhysRevC.98.034910},
  eprint  = {1711.11112},
  archivePrefix = {arXiv},
  primaryClass = {nucl-th}
}

@article{AdamsSTAR2005,
  author  = {Adams, J. and others},
  collaboration = {STAR Collaboration},
  title   = {Pion Interferometry in {Au+Au} Collisions at $\sqrt{s_{NN}}=200$ {GeV}},
  journal = {Physical Review C},
  volume  = {71},
  pages   = {044906},
  year    = {2005},
  doi     = {10.1103/PhysRevC.71.044906},
  eprint  = {nucl-ex/0411036},
  archivePrefix = {arXiv}
}

@article{Lacey2015CEP,
  author  = {Lacey, Roy A.},
  title   = {Indications for a Critical End Point in the Phase Diagram for Hot and Dense Nuclear Matter},
  journal = {Physical Review Letters},
  volume  = {114},
  pages   = {142301},
  year    = {2015},
  doi     = {10.1103/PhysRevLett.114.142301},
  eprint  = {1411.7931},
  archivePrefix = {arXiv},
  primaryClass = {nucl-ex}
}

@article{Pratt2009ResolvingHBT,
  author  = {Pratt, Scott},
  title   = {Resolving the Hanbury Brown--Twiss Puzzle in Relativistic Heavy Ion Collision},
  journal = {Physical Review Letters},
  volume  = {102},
  pages   = {232301},
  year    = {2009},
  doi     = {10.1103/PhysRevLett.102.232301},
  eprint  = {0811.3363},
  archivePrefix = {arXiv},
  primaryClass = {nucl-th}
}

@article{Pratt2009LongSlow,
  author        = {Pratt, Scott},
  title         = {The Long Slow Death of the {HBT} Puzzle},
  journal       = {Nuclear Physics A},
  volume        = {830},
  pages         = {51c--57c},
  year          = {2009},
  doi           = {10.1016/j.nuclphysa.2009.09.004},
  eprint        = {0907.1094},
  archivePrefix = {arXiv},
  primaryClass  = {nucl-th}
}

@article{WiedemannScottoHeinz1996,
  author  = {Wiedemann, Urs Achim and Scotto, Pierre and Heinz, Ulrich},
  title   = {Transverse Momentum Dependence of Hanbury-Brown--Twiss Correlation Radii},
  journal = {Physical Review C},
  volume  = {53},
  pages   = {918--931},
  year    = {1996},
  doi     = {10.1103/PhysRevC.53.918},
  eprint  = {nucl-th/9508040},
  archivePrefix = {arXiv}
}

@article{Koonin1977,
  author  = {Koonin, Steven E.},
  title   = {Proton pictures of high-energy nuclear collisions},
  journal = {Physics Letters B},
  volume  = {70},
  number  = {1},
  pages   = {43--47},
  year    = {1977},
  doi     = {10.1016/0370-2693(77)90340-9}
}

@article{Pratt1984,
  author  = {Pratt, Scott},
  title   = {Pion Interferometry for Exploding Sources},
  journal = {Physical Review Letters},
  volume  = {53},
  pages   = {1219--1221},
  year    = {1984},
  doi     = {10.1103/PhysRevLett.53.1219}
}

@article{AdamczykSTAR2015HBTBES,
  author        = {Adamczyk, L. and others},
  collaboration = {STAR Collaboration},
  title         = {Beam energy dependent two-pion interferometry and the freeze-out eccentricity of pions in heavy ion collisions at STAR},
  journal       = {Physical Review C},
  volume        = {92},
  number        = {1},
  pages         = {014904},
  year          = {2015},
  doi           = {10.1103/PhysRevC.92.014904},
  eprint        = {1403.4972},
  archivePrefix = {arXiv},
  primaryClass  = {nucl-ex}
}

@article{Sinyukov1998,
  author  = {Sinyukov, Yu. and Lednicky, R. and Akkelin, S. V. and Pluta, J. and Erazmus, B.},
  title   = {Coulomb corrections for interferometry analysis of expanding hadron systems},
  journal = {Physics Letters B},
  volume  = {432},
  pages   = {248--257},
  year    = {1998},
  doi     = {10.1016/S0370-2693(98)00653-4}
}

@article{Mrowczynski2009,
  author        = {Maj, Radoslaw and Mrowczynski, Stanislaw},
  title         = {Coulomb effects in femtoscopy},
  journal       = {Physical Review C},
  volume        = {80},
  pages         = {034907},
  year          = {2009},
  doi           = {10.1103/PhysRevC.80.034907},
  eprint        = {0903.0111},
  archivePrefix = {arXiv},
  primaryClass  = {nucl-th}
}

@article{CsorgoHegyiZajc2004,
  author        = {Cs{\"o}rg{\H{o}}, T. and Hegyi, S. and Zajc, W. A.},
  title         = {Bose--Einstein correlations for L{\'e}vy stable source distributions},
  journal       = {European Physical Journal C},
  volume        = {36},
  pages         = {67--78},
  year          = {2004},
  doi           = {10.1140/epjc/s2004-01870-9},
  eprint        = {nucl-th/0310042},
  archivePrefix = {arXiv}
}

@book{GnedenkoKolmogorov1954,
  author    = {Gnedenko, B. V. and Kolmogorov, A. N.},
  title     = {Limit Distributions for Sums of Independent Random Variables},
  publisher = {Addison-Wesley},
  address   = {Cambridge, MA},
  year      = {1954}
}

@book{Nolan2020Stable,
  author    = {Nolan, John P.},
  title     = {Univariate Stable Distributions: Models for Heavy Tailed Data},
  publisher = {Springer},
  address   = {Cham},
  year      = {2020},
  doi       = {10.1007/978-3-030-52915-4}
}

@article{CsorgoHegyiNovakZajc2005,
  author        = {Cs{\"o}rg{\H{o}}, T. and Hegyi, S. and Nov{\'a}k, T. and Zajc, W. A.},
  title         = {Bose--Einstein or HBT correlations and the anomalous dimension of QCD},
  journal       = {Acta Physica Polonica B},
  volume        = {36},
  pages         = {329--337},
  year          = {2005},
  eprint        = {hep-ph/0412243},
  archivePrefix = {arXiv}
}

@inproceedings{CsorgoHegyiNovakZajc2006,
  author        = {Cs{\"o}rg{\H{o}}, T. and Hegyi, S. and Nov{\'a}k, T. and Zajc, W. A.},
  title         = {Bose--Einstein or HBT correlation signature of a second order QCD phase transition},
  booktitle     = {Particle Production Spanning MeV and TeV Energies},
  series        = {AIP Conference Proceedings},
  volume        = {828},
  pages         = {525--532},
  year          = {2006},
  doi           = {10.1063/1.2197460},
  eprint        = {nucl-th/0512060},
  archivePrefix = {arXiv}
}

@article{ElShowk2014Ising,
  author        = {El-Showk, Sheer and Paulos, Miguel F. and Poland, David and Rychkov, Slava and Simmons-Duffin, David and Vichi, Alessandro},
  title         = {Solving the 3D Ising Model with the Conformal Bootstrap II. c-Minimization and Precise Critical Exponents},
  journal       = {Journal of Statistical Physics},
  volume        = {157},
  pages         = {869--914},
  year          = {2014},
  doi           = {10.1007/s10955-014-1042-7},
  eprint        = {1403.4545},
  archivePrefix = {arXiv},
  primaryClass  = {hep-th}
}

@article{Rieger1995RFIM,
  author        = {Rieger, Heiko},
  title         = {Critical behavior of the three-dimensional random-field Ising model: Two-exponent scaling and discontinuous transition},
  journal       = {Physical Review B},
  volume        = {52},
  pages         = {6659--6667},
  year          = {1995},
  doi           = {10.1103/PhysRevB.52.6659},
  eprint        = {cond-mat/9503041},
  archivePrefix = {arXiv}
}

@article{AdarePHENIX2018Levy,
  author        = {Adare, A. and others},
  collaboration = {PHENIX Collaboration},
  title         = {L{\'e}vy-stable two-pion Bose--Einstein correlations in $\sqrt{s_{NN}}=200$ GeV Au+Au collisions},
  journal       = {Physical Review C},
  volume        = {97},
  pages         = {064911},
  year          = {2018},
  doi           = {10.1103/PhysRevC.97.064911},
  eprint        = {1709.05649},
  archivePrefix = {arXiv},
  primaryClass  = {nucl-ex}
}

@article{Soltz2013,
  author        = {Soltz, R. A. and Garishvili, I. and Cheng, M. and Abelev, B. and Glenn, A. and Newby, J. and Klay, J. L. and Lacey, R.},
  title         = {Constraining the Initial Temperature and Shear Viscosity in a Hybrid Hydrodynamic Model of $\sqrt{s_{NN}}=200$ {GeV} {Au+Au} Collisions Using Pion Spectra, Elliptic Flow, and Femtoscopic Radii},
  journal       = {Physical Review C},
  volume        = {87},
  pages         = {044901},
  year          = {2013},
  doi           = {10.1103/PhysRevC.87.044901},
  eprint        = {1208.0897},
  archivePrefix = {arXiv},
  primaryClass  = {nucl-th}
}

@article{LisaRetiereWells2000,
  author        = {Lisa, M. A. and Retiere, F. and Wells, R. M.},
  title         = {Tilted pion sources from azimuthally sensitive {HBT} interferometry},
  journal       = {Physics Letters B},
  volume        = {496},
  pages         = {1--8},
  year          = {2000},
  doi           = {10.1016/S0370-2693(00)00952-7},
  eprint        = {nucl-th/0007022},
  archivePrefix = {arXiv}
}

@article{AdamsSTAR2004AzHBT,
  author        = {Adams, J. and others},
  collaboration = {STAR Collaboration},
  title         = {Azimuthally Sensitive Hanbury Brown--Twiss Interferometry in {Au+Au} Collisions at $\sqrt{s_{NN}}=200$ {GeV}},
  journal       = {Physical Review Letters},
  volume        = {93},
  pages         = {012301},
  year          = {2004},
  doi           = {10.1103/PhysRevLett.93.012301},
  eprint        = {nucl-ex/0312009},
  archivePrefix = {arXiv}
}

@article{Lednicky2009,
  author        = {Lednick{\'y}, R.},
  title         = {Finite-size effects on two-particle production in continuous and discrete spectrum},
  journal       = {Physics of Particles and Nuclei},
  volume        = {40},
  pages         = {307--352},
  year          = {2009},
  doi           = {10.1134/S1063779609030034},
  eprint        = {nucl-th/0501065},
  archivePrefix = {arXiv}
}

@article{Kisiel2010NonIdentical,
  author        = {Kisiel, Adam},
  title         = {Nonidentical-particle femtoscopy at $\sqrt{s_{NN}}=200$ {GeV} in hydrodynamics with statistical hadronization},
  journal       = {Physical Review C},
  volume        = {81},
  pages         = {064906},
  year          = {2010},
  doi           = {10.1103/PhysRevC.81.064906},
  eprint        = {0909.5349},
  archivePrefix = {arXiv},
  primaryClass  = {nucl-th}
}

@article{AsakawaCsorgoGyulassy1999,
  author        = {Asakawa, M. and Cs{\"o}rg{\H{o}}, T. and Gyulassy, M.},
  title         = {Squeezed Correlations and Spectra for Mass-Shifted Bosons},
  journal       = {Physical Review Letters},
  volume        = {83},
  pages         = {4013--4016},
  year          = {1999},
  doi           = {10.1103/PhysRevLett.83.4013},
  eprint        = {nucl-th/9810034},
  archivePrefix = {arXiv}
}

@article{DudekPadula2010,
  author        = {Dudek, Danuce M. and Padula, Sandra S.},
  title         = {Squeezed $K^+K^-$ correlations in high energy heavy ion collisions},
  journal       = {Physical Review C},
  volume        = {82},
  pages         = {034905},
  year          = {2010},
  doi           = {10.1103/PhysRevC.82.034905},
  eprint        = {1006.5899},
  archivePrefix = {arXiv},
  primaryClass  = {nucl-th}
}

@article{AcharyaALICE2020CommonBaryon,
  author        = {Acharya, S. and others},
  collaboration = {ALICE Collaboration},
  title         = {Search for a common baryon source in high-multiplicity pp collisions at the {LHC}},
  journal       = {Physics Letters B},
  volume        = {811},
  pages         = {135849},
  year          = {2020},
  doi           = {10.1016/j.physletb.2020.135849},
  eprint        = {2004.08018},
  archivePrefix = {arXiv},
  primaryClass  = {nucl-ex}
}

@article{AcharyaALICE2022CharmScattering,
  author        = {Acharya, S. and others},
  collaboration = {ALICE Collaboration},
  title         = {First study of the two-body scattering involving charm hadrons},
  journal       = {Physical Review D},
  volume        = {106},
  pages         = {052010},
  year          = {2022},
  doi           = {10.1103/PhysRevD.106.052010},
  eprint        = {2201.05352},
  archivePrefix = {arXiv},
  primaryClass  = {nucl-ex}
}

@article{AcharyaALICE2024CharmLight,
  author        = {Acharya, S. and others},
  collaboration = {ALICE Collaboration},
  title         = {Studying the interaction between charm and light-flavor mesons},
  journal       = {Physical Review D},
  volume        = {110},
  pages         = {032004},
  year          = {2024},
  doi           = {10.1103/PhysRevD.110.032004},
  eprint        = {2401.13541},
  archivePrefix = {arXiv},
  primaryClass  = {nucl-ex}
}

@article{TorresRincon2023DmesonFemtoscopy,
  author        = {Torres-Rincon, Juan M. and Tolos, Laura},
  title         = {Femtoscopy of $D$ mesons and light mesons upon unitarized effective field theories},
  journal       = {Physical Review D},
  volume        = {108},
  pages         = {096008},
  year          = {2023},
  doi           = {10.1103/PhysRevD.108.096008},
  eprint        = {2307.02102},
  archivePrefix = {arXiv},
  primaryClass  = {hep-ph}
}

@book{denicol2022microscopic,
  title={Microscopic foundations of relativistic fluid dynamics},
  author={Denicol, Gabriel S and Rischke, Dirk H},
  year={2022},
  publisher={Springer},
  doi={10.1007/978-3-030-82077-0}
}

@article{Chodos1974MITBag,
  author        = {Chodos, A. and Jaffe, R. L. and Johnson, K. and Thorn, C. B. and Weisskopf, V. F.},
  title         = {New Extended Model of Hadrons},
  journal       = {Physical Review D},
  volume        = {9},
  pages         = {3471--3495},
  year          = {1974},
  doi           = {10.1103/PhysRevD.9.3471}
}

@article{Aoki2006Nature,
  author        = {Aoki, Y. and Endrodi, G. and Fodor, Z. and Katz, S. D. and Szabo, K. K.},
  title         = {The Order of the Quantum Chromodynamics Transition Predicted by the Standard Model of Particle Physics},
  journal       = {Nature},
  volume        = {443},
  pages         = {675--678},
  year          = {2006},
  doi           = {10.1038/nature05120},
  eprint        = {hep-lat/0611014},
  archivePrefix = {arXiv}
}

@article{Bazavov2014EoS,
  author        = {Bazavov, A. and Bhattacharya, T. and DeTar, C. and Ding, H.-T. and Gottlieb, S. and Gupta, R. and Hegde, P. and Heller, U. M. and Karsch, F. and Laermann, E. and Levkova, L. and Mukherjee, S. and Petreczky, P. and Schmidt, C. and Schroeder, C. and Soltz, R. A. and Soeldner, W. and Sugar, R. and Wagner, M. and Vranas, P.},
  title         = {The Equation of State in (2+1)-Flavor QCD},
  journal       = {Physical Review D},
  volume        = {90},
  pages         = {094503},
  year          = {2014},
  doi           = {10.1103/PhysRevD.90.094503},
  eprint        = {1407.6387},
  archivePrefix = {arXiv},
  primaryClass  = {hep-lat}
}

@article{Borsanyi2014EoS,
  author        = {Borsanyi, S. and Fodor, Z. and Hoelbling, C. and Katz, S. D. and Krieg, S. and Szabo, K. K.},
  title         = {Full Result for the QCD Equation of State with 2+1 Flavors},
  journal       = {Physics Letters B},
  volume        = {730},
  pages         = {99--104},
  year          = {2014},
  doi           = {10.1016/j.physletb.2014.01.007},
  eprint        = {1309.5258},
  archivePrefix = {arXiv},
  primaryClass  = {hep-lat}
}

@article{Allton2002Taylor,
  author        = {Allton, C. R. and Ejiri, S. and Hands, S. J. and Kaczmarek, O. and Karsch, F. and Laermann, E. and Schmidt, C. and Scorzato, L.},
  title         = {The QCD Thermal Phase Transition in the Presence of a Small Chemical Potential},
  journal       = {Physical Review D},
  volume        = {66},
  pages         = {074507},
  year          = {2002},
  doi           = {10.1103/PhysRevD.66.074507},
  eprint        = {hep-lat/0204010},
  archivePrefix = {arXiv}
}

@article{Allton2003EoS,
  author        = {Allton, C. R. and Ejiri, S. and Hands, S. J. and Kaczmarek, O. and Karsch, F. and Laermann, E. and Schmidt, C.},
  title         = {Equation of State for Two Flavor QCD at Nonzero Chemical Potential},
  journal       = {Physical Review D},
  volume        = {68},
  pages         = {014507},
  year          = {2003},
  doi           = {10.1103/PhysRevD.68.014507},
  eprint        = {hep-lat/0305007},
  archivePrefix = {arXiv}
}

@article{Borsanyi2021FiniteMuB,
  author        = {Borsanyi, S. and Fodor, Z. and Guenther, J. N. and Kara, R. and Katz, S. D. and Parotto, P. and Pasztor, A. and Ratti, C. and Szabo, K. K.},
  title         = {Lattice QCD Equation of State at Finite Chemical Potential from an Alternative Expansion Scheme},
  journal       = {Physical Review Letters},
  volume        = {126},
  pages         = {232001},
  year          = {2021},
  doi           = {10.1103/PhysRevLett.126.232001},
  eprint        = {2102.06660},
  archivePrefix = {arXiv},
  primaryClass  = {hep-lat}
}

@article{Mondal2022FiniteMuB,
  author        = {Mondal, S. and Mukherjee, S. and Hegde, P.},
  title         = {Lattice QCD Equation of State for Nonvanishing Chemical Potential by Resumming Taylor Expansions},
  journal       = {Physical Review Letters},
  volume        = {128},
  pages         = {022001},
  year          = {2022},
  doi           = {10.1103/PhysRevLett.128.022001},
  eprint        = {2106.03165},
  archivePrefix = {arXiv},
  primaryClass  = {hep-lat}
}

@article{DeGrand1975Bag,
  author        = {DeGrand, T. A. and Jaffe, R. L. and Johnson, K. and Kiskis, J.},
  title         = {Masses and Other Parameters of the Light Hadrons},
  journal       = {Physical Review D},
  volume        = {12},
  pages         = {2060--2076},
  year          = {1975},
  doi           = {10.1103/PhysRevD.12.2060}
}

@article{PhysRevLett.88.202301,
  title = {Pseudorapidity Distributions of Charged Particles from $\mathrm{Au}+\mathrm{Au}$ Collisions at the Maximum RHIC Energy, $\sqrt{{\mathit{s}}_{\mathit{NN}}}\phantom{\rule{0ex}{0ex}}=\phantom{\rule{0ex}{0ex}}200\mathrm{GeV}$},
  author = {Bearden, I. G. and Beavis, D. and Besliu, C. and Blyakhman, Y. and Budick, B. and B\o{}ggild, H. and Chasman, C. and Christensen, C. H. and Christiansen, P. and Cibor, J. and Debbe, R. and Enger, E. and Gaardh\o{}je, J. J. and Hagel, K. and Hansen, O. and Holm, A. and Holme, A. K. and Ito, H. and Jakobsen, E. and Jipa, A. and J\o{}rdre, J. I. and Jundt, F. and J\o{}rgensen, C. E. and Karabowicz, R. and Keutgen, T. and Kim, E. J. and Kozik, T. and Larsen, T. M. and Lee, J. H. and Lee, Y. K. and L\o{}vh\o{}iden, G. and Majka, Z. and Makeev, A. and McBreen, B. and Mikelsen, M. and Murray, M. and Natowitz, J. and Nielsen, B. S. and Norris, J. and Olchanski, K. and Olness, J. and Ouerdane, D. and P\l{}aneta, R. and Rami, F. and Ristea, C. and R\"ohrich, D. and Samset, B. H. and Sandberg, D. and Sanders, S. J. and Sheetz, R. A. and Staszel, P. and Thorsteinsen, T. F. and Tveter, T. S. and Videb\ae{}k, F. and Wada, R. and Wieloch, A. and Zgura, I. S.},
  collaboration = {BRAHMS Collaboration},
  journal = {Phys. Rev. Lett.},
  volume = {88},
  issue = {20},
  pages = {202301},
  numpages = {4},
  year = {2002},
  month = {May},
  publisher = {American Physical Society},
  doi = {10.1103/PhysRevLett.88.202301}
}

@article{PhysRevC.79.034909,
  title = {Systematic measurements of identified particle spectra in $\mathit{pp}$, $d+\mathrm{Au}$, and $\mathrm{Au}+\mathrm{Au}$ collisions at the STAR detector},
  author = {Abelev, B. I. and Aggarwal, M. M. and Ahammed, Z. and Anderson, B. D. and Arkhipkin, D. and Averichev, G. S. and Bai, Y. and Balewski, J. and Barannikova, O. and Barnby, L. S. and Baudot, J. and Baumgart, S. and Beavis, D. R. and Bellwied, R. and Benedosso, F. and Betts, R. R. and Bhardwaj, S. and Bhasin, A. and Bhati, A. K. and Bichsel, H. and Bielcik, J. and Bielcikova, J. and Biritz, B. and Bland, L. C. and Bombara, M. and Bonner, B. E. and Botje, M. and Bouchet, J. and Braidot, E. and Brandin, A. V. and Bruna, E. and Bueltmann, S. and Burton, T. P. and Bystersky, M. and Cai, X. Z. and Caines, H. and S\'anchez, M. Calder\'on de la Barca and Callner, J. and Catu, O. and Cebra, D. and Cendejas, R. and Cervantes, M. C. and Chajecki, Z. and Chaloupka, P. and Chattopadhyay, S. and Chen, H. F. and Chen, J. H. and Chen, J. Y. and Cheng, J. and Cherney, M. and Chikanian, A. and Choi, K. E. and Christie, W. and Chung, S. U. and Clarke, R. F. and Codrington, M. J. M. and Coffin, J. P. and Cormier, T. M. and Cosentino, M. R. and Cramer, J. G. and Crawford, H. J. and Das, D. and Dash, S. and Daugherity, M. and Silva, C. De and Dedovich, T. G. and DePhillips, M. and Derevschikov, A. A. and de Souza, R. Derradi and Didenko, L. and Djawotho, P. and Dogra, S. M. and Dong, X. and Drachenberg, J. L. and Draper, J. E. and Du, F. and Dunlop, J. C. and Mazumdar, M. R. Dutta and Edwards, W. R. and Efimov, L. G. and Elhalhuli, E. and Elnimr, M. and Emelianov, V. and Engelage, J. and Eppley, G. and Erazmus, B. and Estienne, M. and Eun, L. and Fachini, P. and Fatemi, R. and Fedorisin, J. and Feng, A. and Filip, P. and Finch, E. and Fine, V. and Fisyak, Y. and Gagliardi, C. A. and Gaillard, L. and Gangadharan, D. R. and Ganti, M. S. and Garcia-Solis, E. and Ghazikhanian, V. and Ghosh, P. and Gorbunov, Y. N. and Gordon, A. and Grebenyuk, O. and Grosnick, D. and Grube, B. and Guertin, S. M. and Guimaraes, K. S. F. F. and Gupta, A. and Gupta, N. and Guryn, W. and Haag, B. and Hallman, T. J. and Hamed, A. and Harris, J. W. and He, W. and Heinz, M. and Heppelmann, S. and Hippolyte, B. and Hirsch, A. and Hjort, E. and Hoffman, A. M. and Hoffmann, G. W. and Hofman, D. J. and Hollis, R. S. and Huang, H. Z. and Humanic, T. J. and Igo, G. and Iordanova, A. and Jacobs, P. and Jacobs, W. W. and Jakl, P. and Jin, F. and Jones, P. G. and Joseph, J. and Judd, E. G. and Kabana, S. and Kajimoto, K. and Kang, K. and Kapitan, J. and Kaplan, M. and Keane, D. and Kechechyan, A. and Kettler, D. and Khodyrev, V. Yu. and Kiryluk, J. and Kisiel, A. and Klein, S. R. and Knospe, A. G. and Kocoloski, A. and Koetke, D. D. and Kopytine, M. and Kotchenda, L. and Kouchpil, V. and Kravtsov, P. and Kravtsov, V. I. and Krueger, K. and Krus, M. and Kuhn, C. and Kumar, L. and Kurnadi, P. and Lamont, M. A. C. and Landgraf, J. M. and LaPointe, S. and Lauret, J. and Lebedev, A. and Lednicky, R. and Lee, C.-H. and LeVine, M. J. and Li, C. and Li, Y. and Lin, G. and Lin, X. and Lindenbaum, S. J. and Lisa, M. A. and Liu, F. and Liu, H. and Liu, J. and Liu, L. and Ljubicic, T. and Llope, W. J. and Longacre, R. S. and Love, W. A. and Lu, Y. and Ludlam, T. and Lynn, D. and Ma, G. L. and Ma, Y. G. and Mahapatra, D. P. and Majka, R. and Mall, O. I. and Mangotra, L. K. and Manweiler, R. and Margetis, S. and Markert, C. and Matis, H. S. and Matulenko, Yu. A. and McShane, T. S. and Meschanin, A. and Millane, J. and Miller, M. L. and Minaev, N. G. and Mioduszewski, S. and Mischke, A. and Mitchell, J. and Mohanty, B. and Molnar, L. and Morozov, D. A. and Munhoz, M. G. and Nandi, B. K. and Nattrass, C. and Nayak, T. K. and Nelson, J. M. and Nepali, C. and Netrakanti, P. K. and Ng, M. J. and Nogach, L. V. and Nurushev, S. B. and Odyniec, G. and Ogawa, A. and Okada, H. and Okorokov, V. and Olson, D. and Pachr, M. and Page, B. S. and Pal, S. K. and Pandit, Y. and Panebratsev, Y. and Pawlak, T. and Peitzmann, T. and Perevoztchikov, V. and Perkins, C. and Peryt, W. and Phatak, S. C. and Planinic, M. and Pluta, J. and Poljak, N. and Poskanzer, A. M. and Potukuchi, B. V. K. S. and Prindle, D. and Pruneau, C. and Pruthi, N. K. and Putschke, J. and Raniwala, R. and Raniwala, S. and Ray, R. L. and Reed, R. and Ridiger, A. and Ritter, H. G. and Roberts, J. B. and Rogachevskiy, O. V. and Romero, J. L. and Rose, A. and Roy, C. and Ruan, L. and Russcher, M. J. and Rykov, V. and Sahoo, R. and Sakrejda, I. and Sakuma, T. and Salur, S. and Sandweiss, J. and Sarsour, M. and Schambach, J. and Scharenberg, R. P. and Schmitz, N. and Seger, J. and Selyuzhenkov, I. and Seyboth, P. and Shabetai, A. and Shahaliev, E. and Shao, M. and Sharma, M. and Shi, S. S. and Shi, X.-H. and Sichtermann, E. P. and Simon, F. and Singaraju, R. N. and Skoby, M. J. and Smirnov, N. and Snellings, R. and Sorensen, P. and Sowinski, J. and Spinka, H. M. and Srivastava, B. and Stadnik, A. and Stanislaus, T. D. S. and Staszak, D. and Strikhanov, M. and Stringfellow, B. and Suaide, A. A. P. and Suarez, M. C. and Subba, N. L. and Sumbera, M. and Sun, X. M. and Sun, Y. and Sun, Z. and Surrow, B. and Symons, T. J. M. and de Toledo, A. Szanto and Takahashi, J. and Tang, A. H. and Tang, Z. and Tarnowsky, T. and Thein, D. and Thomas, J. H. and Tian, J. and Timmins, A. R. and Timoshenko, S. and Tlusty, D. and Tokarev, M. and Tram, V. N. and Trattner, A. L. and Trentalange, S. and Tribble, R. E. and Tsai, O. D. and Ulery, J. and Ullrich, T. and Underwood, D. G. and Buren, G. Van and van Leeuwen, M. and Molen, A. M. Vander and Vanfossen, J. A. and Varma, R. and Vasconcelos, G. M. S. and Vasilevski, I. M. and Vasiliev, A. N. and Videbaek, F. and Vigdor, S. E. and Viyogi, Y. P. and Vokal, S. and Voloshin, S. A. and Wada, M. and Waggoner, W. T. and Wang, F. and Wang, G. and Wang, J. S. and Wang, Q. and Wang, X. and Wang, X. L. and Wang, Y. and Webb, J. C. and Westfall, G. D. and Whitten, C. and Wieman, H. and Wissink, S. W. and Witt, R. and Wu, Y. and Xu, N. and Xu, Q. H. and Xu, Y. and Xu, Z. and Yepes, P. and Yoo, I.-K. and Yue, Q. and Zawisza, M. and Zbroszczyk, H. and Zhan, W. and Zhang, H. and Zhang, S. and Zhang, W. M. and Zhang, Y. and Zhang, Z. P. and Zhao, Y. and Zhong, C. and Zhou, J. and Zoulkarneev, R. and Zoulkarneeva, Y. and Zuo, J. X.},
  collaboration = {STAR Collaboration},
  journal = {Phys. Rev. C},
  volume = {79},
  issue = {3},
  pages = {034909},
  numpages = {58},
  year = {2009},
  month = {Mar},
  publisher = {American Physical Society},
  doi = {10.1103/PhysRevC.79.034909}
}

@article{PhysRevLett.91.052303,
  title = {Significance of the Fragmentation Region in Ultrarelativistic Heavy-Ion Collisions},
  author = {Back, B. B. and Baker, M. D. and Barton, D. S. and Betts, R. R. and Ballintijn, M. and Bickley, A. A. and Bindel, R. and Budzanowski, A. and Busza, W. and Carroll, A. and Decowski, M. P. and Garc\'{\i}a, E. and George, N. and Gulbrandsen, K. and Gushue, S. and Halliwell, C. and Hamblen, J. and Heintzelman, G. A. and Henderson, C. and Hofman, D. J. and Hollis, R. S. and Ho\l{}y\ifmmode \acute{n}\else \'{n}\fi{}ski, R. and Holzman, B. and Iordanova, A. and Johnson, E. and Kane, J. L. and Katzy, J. and Khan, N. and Kucewicz, W. and Kulinich, P. and Kuo, C. M. and Lin, W. T. and Manly, S. and McLeod, D. and Micha\l{}owski, J. and Mignerey, A. C. and Nouicer, R. and Olszewski, A. and Pak, R. and Park, I. C. and Pernegger, H. and Reed, C. and Remsberg, L. P. and Reuter, M. and Roland, C. and Roland, G. and Rosenberg, L. and Sagerer, J. and Sarin, P. and Sawicki, P. and Skulski, W. and Steadman, S. G. and Steinberg, P. and Stephans, G. S. F. and Stodulski, M. and Sukhanov, A. and Tang, J.-L. and Teng, R. and Trzupek, A. and Vale, C. and van Nieuwenhuizen, G. J. and Verdier, R. and Wadsworth, B. and Wolfs, F. L. H. and Wosiek, B. and Wo\ifmmode \acute{z}\else \'{z}\fi{}niak, K. and Wuosmaa, A. H. and Wys\l{}ouch, B.},
  journal = {Phys. Rev. Lett.},
  volume = {91},
  issue = {5},
  pages = {052303},
  numpages = {4},
  year = {2003},
  month = {Aug},
  publisher = {American Physical Society},
  doi = {10.1103/PhysRevLett.91.052303}
}

@article{PhysRevC.69.034909,
  title = {Identified charged particle spectra and yields in $\text{Au}+\text{Au}$ collisions at $\sqrt{{\mathbit{s}}_{\mathbit{N}\mathbit{N}}}=200\phantom{\rule{0.3em}{0ex}}\text{GeV}$},
  author = {Adler, S. S. and Afanasiev, S. and Aidala, C. and Ajitanand, N. N. and Akiba, Y. and Alexander, J. and Amirikas, R. and Aphecetche, L. and Aronson, S. H. and Averbeck, R. and Awes, T. C. and Azmoun, R. and Babintsev, V. and Baldisseri, A. and Barish, K. N. and Barnes, P. D. and Bassalleck, B. and Bathe, S. and Batsouli, S. and Baublis, V. and Bazilevsky, A. and Belikov, S. and Berdnikov, Y. and Bhagavatula, S. and Boissevain, J. G. and Borel, H. and Borenstein, S. and Brooks, M. L. and Brown, D. S. and Bruner, N. and Bucher, D. and Buesching, H. and Bumazhnov, V. and Bunce, G. and Burward-Hoy, J. M. and Butsyk, S. and Camard, X. and Chai, J.-S. and Chand, P. and Chang, W. C. and Chernichenko, S. and Chi, C. Y. and Chiba, J. and Chiu, M. and Choi, I. J. and Choi, J. and Choudhury, R. K. and Chujo, T. and Cianciolo, V. and Cobigo, Y. and Cole, B. A. and Constantin, P. and d'Enterria, D. G. and David, G. and Delagrange, H. and Denisov, A. and Deshpande, A. and Desmond, E. J. and Dietzsch, O. and Drapier, O. and Drees, A. and du Rietz, R. and Durum, A. and Dutta, D. and Efremenko, Y. V. and El Chenawi, K. and Enokizono, A. and En`yo, H. and Esumi, S. and Ewell, L. and Fields, D. E. and Fleuret, F. and Fokin, S. L. and Fox, B. D. and Fraenkel, Z. and Frantz, J. E. and Franz, A. and Frawley, A. D. and Fung, S.-Y. and Garpman, S. and Ghosh, T. K. and Glenn, A. and Gogiberidze, G. and Gonin, M. and Gosset, J. and Goto, Y. and Granier de Cassagnac, R. and Grau, N. and Greene, S. V. and Grosse Perdekamp, M. and Guryn, W. and Gustafsson, H.-\AA{}. and Hachiya, T. and Haggerty, J. S. and Hamagaki, H. and Hansen, A. G. and Hartouni, E. P. and Harvey, M. and Hayano, R. and He, X. and Heffner, M. and Hemmick, T. K. and Heuser, J. M. and Hibino, M. and Hill, J. C. and Holzmann, W. and Homma, K. and Hong, B. and Hoover, A. and Ichihara, T. and Ikonnikov, V. V. and Imai, K. and Isenhower, D. and Ishihara, M. and Issah, M. and Isupov, A. and Jacak, B. V. and Jang, W. Y. and Jeong, Y. and Jia, J. and Jinnouchi, O. and Johnson, B. M. and Johnson, S. C. and Joo, K. S. and Jouan, D. and Kametani, S. and Kamihara, N. and Kang, J. H. and Kapoor, S. S. and Katou, K. and Kelly, S. and Khachaturov, B. and Khanzadeev, A. and Kikuchi, J. and Kim, D. H. and Kim, D. J. and Kim, D. W. and Kim, E. and Kim, G.-B. and Kim, H. J. and Kistenev, E. and Kiyomichi, A. and Kiyoyama, K. and Klein-Boesing, C. and Kobayashi, H. and Kochenda, L. and Kochetkov, V. and Koehler, D. and Kohama, T. and Kopytine, M. and Kotchetkov, D. and Kozlov, A. and Kroon, P. J. and Kuberg, C. H. and Kurita, K. and Kuroki, Y. and Kweon, M. J. and Kwon, Y. and Kyle, G. S. and Lacey, R. and Ladygin, V. and Lajoie, J. G. and Lebedev, A. and Leckey, S. and Lee, D. M. and Lee, S. and Leitch, M. J. and Li, X. H. and Lim, H. and Litvinenko, A. and Liu, M. X. and Liu, Y. and Maguire, C. F. and Makdisi, Y. I. and Malakhov, A. and Manko, V. I. and Mao, Y. and Martinez, G. and Marx, M. D. and Masui, H. and Matathias, F. and Matsumoto, T. and McGaughey, P. L. and Melnikov, E. and Messer, F. and Miake, Y. and Milan, J. and Miller, T. E. and Milov, A. and Mioduszewski, S. and Mischke, R. E. and Mishra, G. C. and Mitchell, J. T. and Mohanty, A. K. and Morrison, D. P. and Moss, J. M. and M\"uhlbacher, F. and Mukhopadhyay, D. and Muniruzzaman, M. and Murata, J. and Nagamiya, S. and Nagle, J. L. and Nakamura, T. and Nandi, B. K. and Nara, M. and Newby, J. and Nilsson, P. and Nyanin, A. S. and Nystrand, J. and O'Brien, E. and Ogilvie, C. A. and Ohnishi, H. and Ojha, I. D. and Okada, K. and Ono, M. and Onuchin, V. and Oskarsson, A. and Otterlund, I. and Oyama, K. and Ozawa, K. and Pal, D. and Palounek, A. P.T. and Pantuev, V. S. and Papavassiliou, V. and Park, J. and Parmar, A. and Pate, S. F. and Peitzmann, T. and Peng, J.-C. and Peresedov, V. and Pinkenburg, C. and Pisani, R. P. and Plasil, F. and Purschke, M. L. and Purwar, A. K. and Rak, J. and Ravinovich, I. and Read, K. F. and Reuter, M. and Reygers, K. and Riabov, V. and Riabov, Y. and Roche, G. and Romana, A. and Rosati, M. and Rosnet, P. and Ryu, S. S. and Sadler, M. E. and Saito, N. and Sakaguchi, T. and Sakai, M. and Sakai, S. and Samsonov, V. and Sanfratello, L. and Santo, R. and Sato, H. D. and Sato, S. and Sawada, S. and Schutz, Y. and Semenov, V. and Seto, R. and Shaw, M. R. and Shea, T. K. and Shibata, T.-A. and Shigaki, K. and Shiina, T. and Silva, C. L. and Silvermyr, D. and Sim, K. S. and Singh, C. P. and Singh, V. and Sivertz, M. and Soldatov, A. and Soltz, R. A. and Sondheim, W. E. and Sorensen, S. P. and Sourikova, I. V. and Staley, F. and Stankus, P. W. and Stenlund, E. and Stepanov, M. and Ster, A. and Stoll, S. P. and Sugitate, T. and Sullivan, J. P. and Takagui, E. M. and Taketani, A. and Tamai, M. and Tanaka, K. H. and Tanaka, Y. and Tanida, K. and Tannenbaum, M. J. and Tarj\'an, P. and Tepe, J. D. and Thomas, T. L. and Tojo, J. and Torii, H. and Towell, R. S. and Tserruya, I. and Tsuruoka, H. and Tuli, S. K. and Tydesj\"o, H. and Tyurin, N. and van Hecke, H. W. and Velkovska, J. and Velkovsky, M. and Villatte, L. and Vinogradov, A. A. and Volkov, M. A. and Vznuzdaev, E. and Wang, X. R. and Watanabe, Y. and White, S. N. and Wohn, F. K. and Woody, C. L. and Xie, W. and Yang, Y. and Yanovich, A. and Yokkaichi, S. and Young, G. R. and Yushmanov, I. E. and Zajc, W. A. and Zhang, C. and Zhou, S. and Zhou, S. J. and Zolin, L.},
  collaboration = {PHENIX Collaboration},
  journal = {Phys. Rev. C},
  volume = {69},
  issue = {3},
  pages = {034909},
  numpages = {32},
  year = {2004},
  month = {Mar},
  publisher = {American Physical Society},
  doi = {10.1103/PhysRevC.69.034909}
}

@article{PhysRevC.96.044904,
  title = {Bulk properties of the medium produced in relativistic heavy-ion collisions from the beam energy scan program},
  author = {Adamczyk, L. and Adkins, J. K. and Agakishiev, G. and Aggarwal, M. M. and Ahammed, Z. and Ajitanand, N. N. and Alekseev, I. and Anderson, D. M. and Aoyama, R. and Aparin, A. and Arkhipkin, D. and Aschenauer, E. C. and Ashraf, M. U. and Attri, A. and Averichev, G. S. and Bai, X. and Bairathi, V. and Behera, A. and Bellwied, R. and Bhasin, A. and Bhati, A. K. and Bhattarai, P. and Bielcik, J. and Bielcikova, J. and Bland, L. C. and Bordyuzhin, I. G. and Bouchet, J. and Brandenburg, J. D. and Brandin, A. V. and Brown, D. and Bunzarov, I. and Butterworth, J. and Caines, H. and Calder\'on de la Barca S\'anchez, M. and Campbell, J. M. and Cebra, D. and Chakaberia, I. and Chaloupka, P. and Chang, Z. and Chankova-Bunzarova, N. and Chatterjee, A. and Chattopadhyay, S. and Chen, X. and Chen, J. H. and Chen, X. and Cheng, J. and Cherney, M. and Christie, W. and Contin, G. and Crawford, H. J. and Das, S. and De Silva, L. C. and Debbe, R. R. and Dedovich, T. G. and Deng, J. and Derevschikov, A. A. and Didenko, L. and Dilks, C. and Dong, X. and Drachenberg, J. L. and Draper, J. E. and Dunkelberger, L. E. and Dunlop, J. C. and Efimov, L. G. and Elsey, N. and Engelage, J. and Eppley, G. and Esha, R. and Esumi, S. and Evdokimov, O. and Ewigleben, J. and Eyser, O. and Fatemi, R. and Fazio, S. and Federic, P. and Federicova, P. and Fedorisin, J. and Feng, Z. and Filip, P. and Finch, E. and Fisyak, Y. and Flores, C. E. and Fulek, L. and Gagliardi, C. A. and Garand, D. and Geurts, F. and Gibson, A. and Girard, M. and Grosnick, D. and Gunarathne, D. S. and Guo, Y. and Gupta, A. and Gupta, S. and Guryn, W. and Hamad, A. I. and Hamed, A. and Harlenderova, A. and Harris, J. W. and He, L. and Heppelmann, S. and Heppelmann, S. and Hirsch, A. and Hoffmann, G. W. and Horvat, S. and Huang, T. and Huang, B. and Huang, X. and Huang, H. Z. and Humanic, T. J. and Huo, P. and Igo, G. and Jacobs, W. W. and Jentsch, A. and Jia, J. and Jiang, K. and Jowzaee, S. and Judd, E. G. and Kabana, S. and Kalinkin, D. and Kang, K. and Kauder, K. and Ke, H. W. and Keane, D. and Kechechyan, A. and Khan, Z. and Kiko\l{}a, D. P. and Kisel, I. and Kisiel, A. and Kochenda, L. and Kocmanek, M. and Kollegger, T. and Kosarzewski, L. K. and Kraishan, A. F. and Kravtsov, P. and Krueger, K. and Kulathunga, N. and Kumar, L. and Kvapil, J. and Kwasizur, J. H. and Lacey, R. and Landgraf, J. M. and Landry, K. D. and Lauret, J. and Lebedev, A. and Lednicky, R. and Lee, J. H. and Li, X. and Li, C. and Li, W. and Li, Y. and Lidrych, J. and Lin, T. and Lisa, M. A. and Liu, H. and Liu, P. and Liu, Y. and Liu, F. and Ljubicic, T. and Llope, W. J. and Lomnitz, M. and Longacre, R. S. and Luo, S. and Luo, X. and Ma, G. L. and Ma, L. and Ma, Y. G. and Ma, R. and Magdy, N. and Majka, R. and Mallick, D. and Margetis, S. and Markert, C. and Matis, H. S. and Meehan, K. and Mei, J. C. and Miller, Z. W. and Minaev, N. G. and Mioduszewski, S. and Mishra, D. and Mizuno, S. and Mohanty, B. and Mondal, M. M. and Morozov, D. A. and Mustafa, M. K. and Nasim, Md. and Nayak, T. K. and Nelson, J. M. and Nie, M. and Nigmatkulov, G. and Niida, T. and Nogach, L. V. and Nonaka, T. and Nurushev, S. B. and Odyniec, G. and Ogawa, A. and Oh, K. and Okorokov, V. A. and Olvitt, D. and Page, B. S. and Pak, R. and Pandit, Y. and Panebratsev, Y. and Pawlik, B. and Pei, H. and Perkins, C. and Pile, P. and Pluta, J. and Poniatowska, K. and Porter, J. and Posik, M. and Poskanzer, A. M. and Pruthi, N. K. and Przybycien, M. and Putschke, J. and Qiu, H. and Quintero, A. and Ramachandran, S. and Ray, R. L. and Reed, R. and Rehbein, M. J. and Ritter, H. G. and Roberts, J. B. and Rogachevskiy, O. V. and Romero, J. L. and Roth, J. D. and Ruan, L. and Rusnak, J. and Rusnakova, O. and Sahoo, N. R. and Sahu, P. K. and Salur, S. and Sandweiss, J. and Saur, M. and Schambach, J. and Schmah, A. M. and Schmidke, W. B. and Schmitz, N. and Schweid, B. R. and Seger, J. and Sergeeva, M. and Seyboth, P. and Shah, N. and Shahaliev, E. and Shanmuganathan, P. V. and Shao, M. and Sharma, A. and Sharma, M. K. and Shen, W. Q. and Shi, Z. and Shi, S. S. and Shou, Q. Y. and Sichtermann, E. P. and Sikora, R. and Simko, M. and Singha, S. and Skoby, M. J. and Smirnov, N. and Smirnov, D. and Solyst, W. and Song, L. and Sorensen, P. and Spinka, H. M. and Srivastava, B. and Stanislaus, T. D. S. and Strikhanov, M. and Stringfellow, B. and Sugiura, T. and Sumbera, M. and Summa, B. and Sun, Y. and Sun, X. M. and Sun, X. and Surrow, B. and Svirida, D. N. and Tang, A. H. and Tang, Z. and Taranenko, A. and Tarnowsky, T. and Tawfik, A. and Th\"ader, J. and Thomas, J. H. and Timmins, A. R. and Tlusty, D. and Todoroki, T. and Tokarev, M. and Trentalange, S. and Tribble, R. E. and Tribedy, P. and Tripathy, S. K. and Trzeciak, B. A. and Tsai, O. D. and Ullrich, T. and Underwood, D. G. and Upsal, I. and Van Buren, G. and van Nieuwenhuizen, G. and Vasiliev, A. N. and Videb\ae{}k, F. and Vokal, S. and Voloshin, S. A. and Vossen, A. and Wang, G. and Wang, Y. and Wang, F. and Wang, Y. and Webb, J. C. and Webb, G. and Wen, L. and Westfall, G. D. and Wieman, H. and Wissink, S. W. and Witt, R. and Wu, Y. and Xiao, Z. G. and Xie, W. and Xie, G. and Xu, J. and Xu, N. and Xu, Q. H. and Xu, Y. F. and Xu, Z. and Yang, Y. and Yang, Q. and Yang, C. and Yang, S. and Ye, Z. and Ye, Z. and Yi, L. and Yip, K. and Yoo, I.-K. and Yu, N. and Zbroszczyk, H. and Zha, W. and Zhang, Z. and Zhang, X. P. and Zhang, J. B. and Zhang, S. and Zhang, J. and Zhang, Y. and Zhang, J. and Zhang, S. and Zhao, J. and Zhong, C. and Zhou, L. and Zhou, C. and Zhu, X. and Zhu, Z. and Zyzak, M.},
  collaboration = {STAR Collaboration},
  journal = {Phys. Rev. C},
  volume = {96},
  issue = {4},
  pages = {044904},
  numpages = {33},
  year = {2017},
  month = {Oct},
  publisher = {American Physical Society},
  doi = {10.1103/PhysRevC.96.044904}
}

@article{PhysRevC.105.034909,
  title = {$(3+1)$-D viscous hydrodynamics at finite net baryon density: Identified particle spectra, anisotropic flows, and flow fluctuations across energies relevant to the beam-energy scan at RHIC},
  author = {Wu, Xiang-Yu and Qin, Guang-You and Pang, Long-Gang and Wang, Xin-Nian},
  journal = {Phys. Rev. C},
  volume = {105},
  issue = {3},
  pages = {034909},
  numpages = {10},
  year = {2022},
  month = {Mar},
  publisher = {American Physical Society},
  doi = {10.1103/PhysRevC.105.034909}
}

@article{PhysRevC.100.014912,
  title = {Bulk quantities in nuclear collisions from running-coupling ${k}_{T}$ factorization and hybrid simulations},
  author = {Giannini, Andre V. and Grassi, Fr\'ed\'erique and Luzum, Matthew},
  journal = {Phys. Rev. C},
  volume = {100},
  issue = {1},
  pages = {014912},
  numpages = {10},
  year = {2019},
  month = {Jul},
  publisher = {American Physical Society},
  doi = {10.1103/PhysRevC.100.014912}
}

@article{PhysRevC.103.054904,
  title = {Multisystem Bayesian constraints on the transport coefficients of QCD matter},
  author = {Everett, D. and Ke, W. and Paquet, J.-F. and Vujanovic, G. and Bass, S. A. and Du, L. and Gale, C. and Heffernan, M. and Heinz, U. and Liyanage, D. and Luzum, M. and Majumder, A. and McNelis, M. and Shen, C. and Xu, Y. and Angerami, A. and Cao, S. and Chen, Y. and Coleman, J. and Cunqueiro, L. and Dai, T. and Ehlers, R. and Elfner, H. and Fan, W. and Fries, R. J. and Garza, F. and He, Y. and Jacak, B. V. and Jacobs, P. M. and Jeon, S. and Kim, B. and Kordell, M. and Kumar, A. and Mak, S. and Mulligan, J. and Nattrass, C. and Oliinychenko, D. and Park, C. and Putschke, J. H. and Roland, G. and Schenke, B. and Schwiebert, L. and Silva, A. and Sirimanna, C. and Soltz, R. A. and Tachibana, Y. and Wang, X.-N. and Wolpert, R. L.},
  collaboration = {JETSCAPE Collaboration},
  journal = {Phys. Rev. C},
  volume = {103},
  issue = {5},
  pages = {054904},
  numpages = {51},
  year = {2021},
  month = {May},
  publisher = {American Physical Society},
  doi = {10.1103/PhysRevC.103.054904}
}

@article{PhysRevC.71.044906,
  title = {Pion interferometry in Au+Au collisions at $\sqrt{{s}_{NN}}=200\phantom{\rule{0.3em}{0ex}}\mathrm{GeV}$},
  author = {Adams, J. and Aggarwal, M. M. and Ahammed, Z. and Amonett, J. and Anderson, B. D. and Arkhipkin, D. and Averichev, G. S. and Badyal, S. K. and Bai, Y. and Balewski, J. and Barannikova, O. and Barnby, L. S. and Baudot, J. and Bekele, S. and Belaga, V. V. and Bellwied, R. and Berger, J. and Bezverkhny, B. I. and Bharadwaj, S. and Bhasin, A. and Bhati, A. K. and Bhatia, V. S. and Bichsel, H. and Billmeier, A. and Bland, L. C. and Blyth, C. O. and Bonner, B. E. and Botje, M. and Boucham, A. and Brandin, A. V. and Bravar, A. and Bystersky, M. and Cadman, R. V. and Cai, X. Z. and Caines, H. and S\'anchez, M. Calder\'on de la Barca and Castillo, J. and Cebra, D. and Chajecki, Z. and Chaloupka, P. and Chattopadhyay, S. and Chen, H. F. and Chen, Y. and Cheng, J. and Cherney, M. and Chikanian, A. and Christie, W. and Coffin, J. P. and Cormier, T. M. and Cramer, J. G. and Crawford, H. J. and Das, D. and Das, S. and Moura, M. M. de and Derevschikov, A. A. and Didenko, L. and Dietel, T. and Dogra, S. M. and Dong, W. J. and Dong, X. and Draper, J. E. and Du, F. and Dubey, A. K. and Dunin, V. B. and Dunlop, J. C. and Mazumdar, M. R. Dutta and Eckardt, V. and Edwards, W. R. and Efimov, L. G. and Emelianov, V. and Engelage, J. and Eppley, G. and Erazmus, B. and Estienne, M. and Fachini, P. and Faivre, J. and Fatemi, R. and Fedorisin, J. and Filimonov, K. and Filip, P. and Finch, E. and Fine, V. and Fisyak, Y. and Fomenko, K. and Fu, J. and Gagliardi, C. A. and Gans, J. and Ganti, M. S. and Gaudichet, L. and Geurts, F. and Ghazikhanian, V. and Ghosh, P. and Gonzalez, J. E. and Grachov, O. and Grebenyuk, O. and Grosnick, D. and Guertin, S. M. and Guo, Y. and Gupta, A. and Gutierrez, T. D. and Hallman, T. J. and Hamed, A. and Hardtke, D. and Harris, J. W. and Heinz, M. and Henry, T. W. and Hepplemann, S. and Hippolyte, B. and Hirsch, A. and Hjort, E. and Hoffmann, G. W. and Huang, H. Z. and Huang, S. L. and Hughes, E. W. and Humanic, T. J. and Igo, G. and Ishihara, A. and Jacobs, P. and Jacobs, W. W. and Janik, M. and Jiang, H. and Jones, P. G. and Judd, E. G. and Kabana, S. and Kang, K. and Kaplan, M. and Keane, D. and Khodyrev, V. Yu. and Kiryluk, J. and Kisiel, A. and Kislov, E. M. and Klay, J. and Klein, S. R. and Klyachko, A. and Koetke, D. D. and Kollegger, T. and Kopytine, M. and Kotchenda, L. and Kramer, M. and Kravtsov, P. and Kravtsov, V. I. and Krueger, K. and Kuhn, C. and Kulikov, A. I. and Kumar, A. and Kutuev, R. Kh. and Kuznetsov, A. A. and Lamont, M. A. C. and Landgraf, J. M. and Lange, S. and Laue, F. and Lauret, J. and Lebedev, A. and Lednicky, R. and Lehocka, S. and LeVine, M. J. and Li, C. and Li, Q. and Li, Y. and Lin, G. and Lindenbaum, S. J. and Lisa, M. A. and Liu, F. and Liu, L. and Liu, Q. J. and Liu, Z. and Ljubicic, T. and Llope, W. J. and Long, H. and Longacre, R. S. and Noriega, M. L\'opez and Love, W. A. and Lu, Y. and Ludlam, T. and Lynn, D. and Ma, G. L. and Ma, J. G. and Ma, Y. G. and Magestro, D. and Mahajan, S. and Mahapatra, D. P. and Majka, R. and Mangotra, L. K. and Manweiler, R. and Margetis, S. and Markert, C. and Martin, L. and Marx, J. N. and Matis, H. S. and Matulenko, Yu. A. and McClain, C. J. and McShane, T. S. and Meissner, F. and Melnick, Yu. and Meschanin, A. and Miller, M. L. and Minaev, N. G. and Mironov, C. and Mischke, A. and Mishra, D. K. and Mitchell, J. and Mohanty, B. and Molnar, L. and Moore, C. F. and Morozov, D. A. and Munhoz, M. G. and Nandi, B. K. and Nayak, S. K. and Nayak, T. K. and Nelson, J. M. and Netrakanti, P. K. and Nikitin, V. A. and Nogach, L. V. and Nurushev, S. B. and Odyniec, G. and Ogawa, A. and Okorokov, V. and Oldenburg, M. and Olson, D. and Pal, S. K. and Panebratsev, Y. and Panitkin, S. Y. and Pavlinov, A. I. and Pawlak, T. and Peitzmann, T. and Perevoztchikov, V. and Perkins, C. and Peryt, W. and Petrov, V. A. and Phatak, S. C. and Picha, R. and Planinic, M. and Pluta, J. and Porile, N. and Porter, J. and Poskanzer, A. M. and Potekhin, M. and Potrebenikova, E. and Potukuchi, B. V. K. S. and Prindle, D. and Pruneau, C. and Putschke, J. and Rakness, G. and Raniwala, R. and Raniwala, S. and Ravel, O. and Ray, R. L. and Razin, S. V. and Reichhold, D. and Reid, J. G. and Renault, G. and Retiere, F. and Ridiger, A. and Ritter, H. G. and Roberts, J. B. and Rogachevskiy, O. V. and Romero, J. L. and Rose, A. and Roy, C. and Ruan, L. and Sahoo, R. and Sakrejda, I. and Salur, S. and Sandweiss, J. and Savin, I. and Sazhin, P. S. and Schambach, J. and Scharenberg, R. P. and Schmitz, N. and Schweda, K. and Seger, J. and Seyboth, P. and Shahaliev, E. and Shao, M. and Shao, W. and Sharma, M. and Shen, W. Q. and Shestermanov, K. E. and Shimanskiy, S. S. and Sichtermann, E and Simon, F. and Singaraju, R. N. and Skoro, G. and Smirnov, N. and Snellings, R. and Sood, G. and Sorensen, P. and Sowinski, J. and Speltz, J. and Spinka, H. M. and Srivastava, B. and Stadnik, A. and Stanislaus, T. D. S. and Stock, R. and Stolpovsky, A. and Strikhanov, M. and Stringfellow, B. and Suaide, A. A. P. and Sugarbaker, E. and Suire, C. and Sumbera, M. and Surrow, B. and Symons, T. J. M. and Toledo, A. Szanto de and Szarwas, P. and Tai, A. and Takahashi, J. and Tang, A. H. and Tarnowsky, T. and Thein, D. and Thomas, J. H. and Timoshenko, S. and Tokarev, M. and Trainor, T. A. and Trentalange, S. and Tribble, R. E. and Tsai, O. D. and Ulery, J. and Ullrich, T. and Underwood, D. G. and Urkinbaev, A. and Buren, G. Van and Leeuwen, M. van and Molen, A. M. Vander and Varma, R. and Vasilevski, I. M. and Vasiliev, A. N. and Vernet, R. and Vigdor, S. E. and Viyogi, Y. P. and Vokal, S. and Voloshin, S. A. and Vznuzdaev, M. and Waggoner, W. T. and Wang, F. and Wang, G. and Wang, G. and Wang, X. L. and Wang, Y. and Wang, Y. and Wang, Z. M. and Ward, H. and Watson, J. W. and Webb, J. C. and Wells, R. and Westfall, G. D. and Wetzler, A. and Jr., C. Whitten and Wieman, H. and Wissink, S. W. and Witt, R. and Wood, J. and Wu, J. and Xu, N. and Xu, Z. and Xu, Z. Z. and Yamamoto, E. and Yepes, P. and Yurevich, V. I. and Zanevsky, Y. V. and Zhang, H. and Zhang, W. M. and Zhang, Z. P. and Zolnierczuk, P. A and Zoulkarneev, R. and Zoulkarneeva, Y. and Zubarev, A. N.},
  collaboration = {STAR Collaboration},
  journal = {Phys. Rev. C},
  volume = {71},
  issue = {4},
  pages = {044906},
  numpages = {21},
  year = {2005},
  month = {Apr},
  publisher = {American Physical Society},
  doi = {10.1103/PhysRevC.71.044906}
}

@article{Zi-Wei-Lin_2004,
doi = {10.1088/0954-3899/30/1/031},
url = {https://doi.org/10.1088/0954-3899/30/1/031},
year = {2003},
month = {dec},
publisher = {},
volume = {30},
number = {1},
pages = {S263},
author = {Zi-Wei Lin and C M Ko},
title = {Kaon interferometry at RHIC from the AMPT model*},
journal = {Journal of Physics G: Nuclear and Particle Physics}
}

@article{PhysRevLett.89.152301,
  title = {Partonic Effects on Pion Interferometry at the Relativistic Heavy-Ion Collider},
  author = {Lin, Zi-wei and Ko, C. M. and Pal, Subrata},
  journal = {Phys. Rev. Lett.},
  volume = {89},
  issue = {15},
  pages = {152301},
  numpages = {4},
  year = {2002},
  month = {Sep},
  publisher = {American Physical Society},
  doi = {10.1103/PhysRevLett.89.152301}
}

@article{molnar2003decoupling,
  title={The decoupling problem at RHIC},
  author={Molnar, Denes and Gyulassy, Miklos},
  journal={Acta Physica Hungarica A) Heavy Ion Physics},
  volume={18},
  number={1},
  pages={69--78},
  year={2003},
  publisher={Springer},
  doi={10.1556/APH.18.2003.1.8}
}

@article{PhysRevLett.92.052301,
  title = {Differential Freeze-Out and Pion Interferometry at RHIC from Covariant Transport Theory},
  author = {Moln\'ar, D\'enes and Gyulassy, Miklos},
  journal = {Phys. Rev. Lett.},
  volume = {92},
  issue = {5},
  pages = {052301},
  numpages = {4},
  year = {2004},
  month = {Feb},
  publisher = {American Physical Society},
  doi = {10.1103/PhysRevLett.92.052301}
}

@article{PhysRevLett.86.3981,
  title = {Pion Interferometry at RHIC: Probing a Thermalized Quark-Gluon Plasma?},
  author = {Soff, Sven and Bass, Steffen A. and Dumitru, Adrian},
  journal = {Phys. Rev. Lett.},
  volume = {86},
  issue = {18},
  pages = {3981--3984},
  numpages = {0},
  year = {2001},
  month = {Apr},
  publisher = {American Physical Society},
  doi = {10.1103/PhysRevLett.86.3981}
}

@article{Weil:2016zrk,
    author = {Weil, J. and Steinberg, V. and Staudenmaier, J. and Pang, L. G. and Oliinychenko, D. and Mohs, J. and Kretz, M. and Kehrenberg, T. and Goldschmidt, A. and B{\"a}uchle, B. and Auvinen, J. and Attems, M. and Petersen, H.},
    title = {Particle production and equilibrium properties within a new hadron transport approach for heavy-ion collisions},
    journal = {Phys. Rev. C},
    volume = {94},
    number = {5},
    pages = {054905},
    year = {2016},
    doi = {10.1103/PhysRevC.94.054905},
    eprint = {1606.06642},
    archivePrefix = {arXiv},
    primaryClass = {nucl-th}
}

@article{Sciarra:2024smash,
    author = {Sciarra, Alessandro and Elfner, Hannah},
    title = {SMASH as an event generator for heavy-ion collisions},
    journal = {Front. Phys.},
    volume = {12},
    pages = {1502621},
    year = {2024},
    doi = {10.3389/fphy.2024.1502621},
    eprint = {2409.20024},
    archivePrefix = {arXiv},
    primaryClass = {nucl-th}
}

@article{pala2025nuclearconfectionery,
  title={NuclearConfectionery: Multi-stage Simulation Framework for Modeling Relativistic Heavy-ion Collisions},
  author={Pala, Kevin P and Virk, Surkhab Kaur and Almaalol, Dekrayat and Danhoni, Isabella and Yao, Nanxi and Long, Isaac and Serenone, Willian and Mart{\'\i}n, Jordi Salinas San and Yared, Alayna A and Plumberg, Christopher and others},
  journal={arXiv preprint arXiv:2511.22852},
  year={2025},
  doi={10.48550/arXiv.2511.22852}
}

@article{PhysRevC.61.067901,
  title = {Multiphase transport model for relativistic nuclear collisions},
  author = {Zhang, Bin and Ko, C. M. and Li, Bao-An and Lin, Ziwei},
  journal = {Phys. Rev. C},
  volume = {61},
  issue = {6},
  pages = {067901},
  numpages = {4},
  year = {2000},
  month = {May},
  publisher = {American Physical Society},
  doi = {10.1103/PhysRevC.61.067901}
}

@article{PhysRevD.44.3501,
  title = {hijing: A Monte Carlo model for multiple jet production in $\mathrm{pp}$, $\mathrm{pA}$, and $\mathrm{AA}$ collisions},
  author = {Wang, Xin-Nian and Gyulassy, Miklos},
  journal = {Phys. Rev. D},
  volume = {44},
  issue = {11},
  pages = {3501--3516},
  numpages = {0},
  year = {1991},
  month = {Dec},
  publisher = {American Physical Society},
  doi = {10.1103/PhysRevD.44.3501}
}

@article{PhysRevC.65.034904,
  title = {Partonic effects on the elliptic flow at relativistic heavy ion collisions},
  author = {Lin, Zi-wei and Ko, C. M.},
  journal = {Phys. Rev. C},
  volume = {65},
  issue = {3},
  pages = {034904},
  numpages = {6},
  year = {2002},
  month = {Feb},
  publisher = {American Physical Society},
  doi = {10.1103/PhysRevC.65.034904}
}

@article{ZHANG1998193,
title = {ZPC 1.0.1: a parton cascade for ultrarelativistic heavy ion collisions},
journal = {Computer Physics Communications},
volume = {109},
number = {2},
pages = {193-206},
year = {1998},
issn = {0010-4655},
doi = {https://doi.org/10.1016/S0010-4655(98)00010-1},
author = {Bin Zhang}
}

@article{PhysRevC.103.024903,
  title = {Hydrodynamic study of hyperon spin polarization in relativistic heavy ion collisions},
  author = {Fu, Baochi and Xu, Kai and Huang, Xu-Guang and Song, Huichao},
  journal = {Phys. Rev. C},
  volume = {103},
  issue = {2},
  pages = {024903},
  numpages = {11},
  year = {2021},
  month = {Feb},
  publisher = {American Physical Society},
  doi = {10.1103/PhysRevC.103.024903},
  url = {https://link.aps.org/doi/10.1103/PhysRevC.103.024903}
}

@article{ISRAEL1979341,
title = {Transient relativistic thermodynamics and kinetic theory},
journal = {Annals of Physics},
volume = {118},
number = {2},
pages = {341-372},
year = {1979},
issn = {0003-4916},
doi = {10.1016/0003-4916(79)90130-1},
author = {W. Israel and J. M. Stewart}
}

@book{leal2007advanced,
  title={Advanced transport phenomena: fluid mechanics and convective transport processes},
  author={Leal, L Gary},
  volume={7},
  year={2007},
  publisher={Cambridge university press},
  doi={10.1017/CBO9780511800245}
}

@book{groot1980relativistic,
  title={Relativistic kinetic theory: principles and applications},
  author={Groot, Sybren Ruurds and Leeuwen, Willem Andries and van Weert, Christianus Gerardus and Weert, Christianus Gerardus},
  year={1980},
  publisher={North Holland},
  isbn      ={9780444854537}
}

@book{liboff2003kinetic,
  title={Kinetic theory: classical, quantum, and relativistic descriptions},
  author={Liboff, Richard L},
  year={2003},
  publisher={Springer},
  doi={doi: 10.1007/b97467}
}

@book{Landau1980,
    author    = {Landau, L. D. and Lifshitz, E. M.},
    title     = {Statistical Physics, Part 1},
    series    = {Course of Theoretical Physics},
    volume    = {5},
    edition   = {3},
    publisher = {Pergamon Press},
    year      = {1980},
    isbn      = {9780080230399}
}

@article{PhysRevD.10.186,
  title = {Single-particle distribution in the hydrodynamic and statistical thermodynamic models of multiparticle production},
  author = {Cooper, Fred and Frye, Graham},
  journal = {Phys. Rev. D},
  volume = {10},
  issue = {1},
  pages = {186--189},
  numpages = {0},
  year = {1974},
  month = {Jul},
  publisher = {American Physical Society},
  doi = {10.1103/PhysRevD.10.186},
  url = {https://link.aps.org/doi/10.1103/PhysRevD.10.186}
}

@book{DineenMultivariateCalculusGeometry,
  author    = {Dineen, Se{\'a}n},
  title     = {Multivariate Calculus and Geometry},
  edition   = {3},
  series    = {Springer Undergraduate Mathematics Series},
  publisher = {Springer London},
  year      = {2014},
  doi       = {10.1007/978-1-4471-6419-7},
  isbn      = {9781447164197}
}

@incollection{Poisson2004Hypersurfaces,
  author    = {Poisson, Eric},
  title     = {Hypersurfaces},
  booktitle = {A Relativist's Toolkit: The Mathematics of Black-Hole Mechanics},
  publisher = {Cambridge University Press},
  year      = {2004},
  pages     = {59--117},
  doi       = {10.1017/CBO9780511606601.005}
}

@book{Gourgoulhon2012,
  author    = {Gourgoulhon, {\'E}ric},
  title     = {3+1 Formalism in General Relativity: Bases of Numerical Relativity},
  series    = {Lecture Notes in Physics},
  volume    = {846},
  publisher = {Springer},
  year      = {2012},
  doi       = {10.1007/978-3-642-24525-1}
}

@article{w87d-ps6f,
  title = {Pushing the hybrid approach to low beam energies with dynamic initial conditions from hadronic transport},
  author = {G\'oes-Hirayama, Renan and Egger, Joscha and Paul\'{\i}nyov\'a, Zuzana and Karpenko, Iurii and Elfner, Hannah},
  journal = {Phys. Rev. C},
  volume = {113},
  issue = {4},
  pages = {044906},
  numpages = {14},
  year = {2026},
  month = {Apr},
  publisher = {American Physical Society},
  doi = {10.1103/w87d-ps6f},
  url = {https://link.aps.org/doi/10.1103/w87d-ps6f}
}

@book{Kardar2007,
  author    = {Kardar, Mehran},
  title     = {Statistical Physics of Fields},
  publisher = {Cambridge University Press},
  address   = {Cambridge},
  year      = {2007},
  isbn      = {9780521873413},
  doi       = {10.1017/CBO9780511815881}
}

@article{hama2005topics,
title={Topics on hydrodynamic model of nucleus-nucleus collisions},
author={Hama, Yogiro and Kodama, Takashi and Socolowski Jr, O},
journal={Brazilian Journal of Physics},
volume={35},
pages={24--51},
year={2005},
publisher={SciELO Brasil},
doi={10.1590/S0103-97332005000100003}
}

@article{PhysRevC.106.014909,
  title = {Embedding a critical point in a hadron to quark-gluon crossover equation of state},
  author = {Kapusta, J. I. and Welle, T. and Plumberg, C.},
  journal = {Phys. Rev. C},
  volume = {106},
  issue = {1},
  pages = {014909},
  numpages = {5},
  year = {2022},
  month = {Jul},
  publisher = {American Physical Society},
  doi = {10.1103/PhysRevC.106.014909}
}

@article{PhysRevC.101.034901,
  title = {QCD equation of state matched to lattice data and exhibiting a critical point singularity},
  author = {Parotto, Paolo and Bluhm, Marcus and Mroczek, Debora and Nahrgang, Marlene and Noronha-Hostler, J. and Rajagopal, Krishna and Ratti, Claudia and Sch\"afer, Thomas and Stephanov, Mikhail},
  journal = {Phys. Rev. C},
  volume = {101},
  issue = {3},
  pages = {034901},
  numpages = {18},
  year = {2020},
  month = {Mar},
  publisher = {American Physical Society},
  doi = {10.1103/PhysRevC.101.034901},
}

@article{PhysRevD.110.094006,
  title = {Bayesian location of the QCD critical point from a holographic perspective},
  author = {Hippert, Mauricio and Grefa, Joaquin and Manning, T. Andrew and Noronha, Jorge and Noronha-Hostler, Jacquelyn and Vazquez, Israel Portillo and Ratti, Claudia and Rougemont, Romulo and Trujillo, Michael},
  journal = {Phys. Rev. D},
  volume = {110},
  issue = {9},
  pages = {094006},
  numpages = {14},
  year = {2024},
  month = {Nov},
  publisher = {American Physical Society},
  doi = {10.1103/PhysRevD.110.094006}
}

@article{PhysRevLett.81.4816,
  title = {Signatures of the Tricritical Point in QCD},
  author = {Stephanov, M. and Rajagopal, K. and Shuryak, E.},
  journal = {Phys. Rev. Lett.},
  volume = {81},
  issue = {22},
  pages = {4816--4819},
  numpages = {0},
  year = {1998},
  month = {Nov},
  publisher = {American Physical Society},
  doi = {10.1103/PhysRevLett.81.4816}
}

@article{HAMA2006169,
title = {3D Relativistic Hydrodynamic Computations Using Lattice-QCD-Inspired Equations of State},
journal = {Nuclear Physics A},
volume = {774},
pages = {169-178},
year = {2006},
note = {QUARK MATTER 2005},
issn = {0375-9474},
doi = {https://doi.org/10.1016/j.nuclphysa.2006.06.024},
author = {Yogiro Hama and Rone P.G. Andrade and Frédérique Grassi and Otávio Socolowski and Takeshi Kodama and Bernardo Tavares and Sandra S. Padula},
}

@article{Ising1925,
  author  = {Ising, Ernst},
  title   = {{Beitrag zur Theorie des Ferromagnetismus}},
  journal = {Zeitschrift f{\"u}r Physik},
  volume  = {31},
  number  = {1},
  pages   = {253--258},
  year    = {1925},
  doi     = {10.1007/BF02980577}
}

@book{Goldenfeld1992,
  author    = {Goldenfeld, Nigel},
  title     = {Lectures on Phase Transitions and the Renormalization Group},
  series    = {Frontiers in Physics},
  volume    = {85},
  publisher = {Westview Press},
  address   = {Boulder, CO},
  year      = {1992},
  isbn      = {9780201554090},
  doi       = {10.1201/9780429493492}
}

@book{Cardy1996,
  author    = {Cardy, John},
  title     = {Scaling and Renormalization in Statistical Physics},
  series    = {Cambridge Lecture Notes in Physics},
  volume    = {5},
  publisher = {Cambridge University Press},
  address   = {Cambridge},
  year      = {1996},
  isbn      = {9780521499590},
  doi       = {10.1017/CBO9781316036440}
}

@article{PelissettoVicari2002,
  author        = {Pelissetto, Andrea and Vicari, Ettore},
  title         = {Critical phenomena and renormalization-group theory},
  journal       = {Physics Reports},
  volume        = {368},
  number        = {6},
  pages         = {549--727},
  year          = {2002},
  doi           = {10.1016/S0370-1573(02)00219-3},
  archivePrefix = {arXiv},
  eprint        = {cond-mat/0012164}
}

@article{Hasenbusch2010,
  author        = {Hasenbusch, Martin},
  title         = {A finite size scaling study of lattice models in the three-dimensional {Ising} universality class},
  journal       = {Physical Review B},
  volume        = {82},
  number        = {17},
  pages         = {174433},
  year          = {2010},
  doi           = {10.1103/PhysRevB.82.174433},
  archivePrefix = {arXiv},
  eprint        = {1004.4486},
  primaryClass  = {cond-mat.stat-mech}
}

@article{Kos2016,
  author        = {Kos, Filip and Poland, David and Simmons-Duffin, David and Vichi, Alessandro},
  title         = {Precision islands in the {Ising} and {$O(N)$} models},
  journal       = {Journal of High Energy Physics},
  volume        = {2016},
  number        = {8},
  pages         = {036},
  year          = {2016},
  doi           = {10.1007/JHEP08(2016)036},
  archivePrefix = {arXiv},
  eprint        = {1603.04436},
  primaryClass  = {hep-th}
}

@article{karthein2021strangeness,
  title={Strangeness-neutral equation of state for QCD with a critical point},
  author={Karthein, JM and Mroczek, D and Nava Acuna, AR and Noronha-Hostler, J and Parotto, P and Price, DRP and Ratti, C},
  journal={The European Physical Journal Plus},
  volume={136},
  number={6},
  pages={621},
  year={2021},
  publisher={Springer},
  doi={10.1140/epjp/s13360-021-01615-5}
}

@article{BELLWIED2015559,
title = {The QCD phase diagram from analytic continuation},
journal = {Physics Letters B},
volume = {751},
pages = {559-564},
year = {2015},
issn = {0370-2693},
doi = {10.1016/j.physletb.2015.11.011}
}

@article{PhysRevD.93.014507,
  title = {Critical line of $2+1$ flavor QCD: Toward the continuum limit},
  author = {Cea, Paolo and Cosmai, Leonardo and Papa, Alessandro},
  journal = {Phys. Rev. D},
  volume = {93},
  issue = {1},
  pages = {014507},
  numpages = {9},
  year = {2016},
  month = {Jan},
  publisher = {American Physical Society},
  doi = {10.1103/PhysRevD.93.014507}
}

@article{PhysRevD.92.054503,
  title = {Curvature of the chiral pseudocritical line in QCD: Continuum extrapolated results},
  author = {Bonati, Claudio and D'Elia, Massimo and Mariti, Marco and Mesiti, Michele and Negro, Francesco and Sanfilippo, Francesco},
  journal = {Phys. Rev. D},
  volume = {92},
  issue = {5},
  pages = {054503},
  numpages = {11},
  year = {2015},
  month = {Sep},
  publisher = {American Physical Society},
  doi = {10.1103/PhysRevD.92.054503}
}

@misc{stephanov2006qcdphasediagramoverview,
      title={QCD phase diagram: an overview}, 
      author={M. A. Stephanov},
      year={2006},
      eprint={hep-lat/0701002},
      archivePrefix={arXiv},
      primaryClass={hep-lat}
}

@article{GIORDANO2021121986,
title = {Towards a reliable lower bound on the location of the critical endpoint},
journal = {Nuclear Physics A},
volume = {1005},
pages = {121986},
year = {2021},
note = {The 28th International Conference on Ultra-relativistic Nucleus-Nucleus Collisions: Quark Matter 2019},
issn = {0375-9474},
doi = {10.1016/j.nuclphysa.2020.121986}
}

@misc{WikimediaWaterPhaseDiagram,
  author       = {{Cmglee and Manlleus}},
  title        = {{Phase diagram of water (simplified)}},
  howpublished = {Wikimedia Commons},
  year         = {2026},
  note         = {Original work by Cmglee; simplified version by Manlleus. Licensed under CC BY-SA 3.0 and GFDL. Accessed: 2026-06-07},
  url          = {https://commons.wikimedia.org/wiki/File:Phase_diagram_of_water_simplified.svg}
}

@misc{WikimediaIsingLattice,
  author       = {{Ta2o}},
  title        = {{2D ising model on lattice}},
  howpublished = {Wikimedia Commons},
  year         = {2025},
  note         = {Own work. Licensed under CC BY 4.0. Accessed: 2026-06-07},
  url          = {https://commons.wikimedia.org/wiki/File:2D_ising_model_on_lattice.svg}
}

@article{Gorenstein1999ExcludedVolume,
  author        = {Gorenstein, M. I. and Kostyuk, A. P. and Krivenko, Ya. D.},
  title         = {Van der Waals excluded volume model of multicomponent hadron gas},
  journal       = {Journal of Physics G: Nuclear and Particle Physics},
  volume        = {25},
  pages         = {L75--L83},
  year          = {1999},
  doi           = {10.1088/0954-3899/25/9/101},
  archivePrefix = {arXiv},
  eprint        = {nucl-th/9906068}
}

@article{lucy1977numerical,
title={A numerical approach to the testing of the fission hypothesis},
author={Lucy, Leon B},
journal={Astronomical Journal, vol. 82, Dec. 1977, p. 1013-1024.},
volume={82},
pages={1013--1024},
year={1977},
doi={10.1086/112164}
}

@article{gingold1977smoothed,
title={Smoothed particle hydrodynamics: theory and application to non-spherical stars},
author={Gingold, Robert A and Monaghan, Joseph J},
journal={Monthly notices of the royal astronomical society},
volume={181},
number={3},
pages={375--389},
year={1977},
publisher={Oxford University Press Oxford, UK},
doi={10.1093/mnras/181.3.375}
}

@article{aguiar2001smoothed,
title={Smoothed particle hydrodynamics for relativistic heavy-ioncollisions},
author={Aguiar, Carlos E and Kodama, Takeshi and Osada, Takeshi and Hama, Yogiro},
journal={Journal of Physics G: Nuclear and Particle Physics},
volume={27},
number={1},
pages={75},
year={2001},
publisher={IOP Publishing},
doi={10.1088/0954-3899/27/1/306}
}

@incollection{osada2001event,
title={Event-by-event analysis of ultra-relativistic heavy-ion collisions in smoothed particle hydrodynamics},
author={Osada, Takeshi and Aguiar, Carlos Eduardo and Hama, Yogiro and Kodama, Takeshi},
booktitle={Relativistic Aspects of Nuclear Physics},
pages={174--195},
year={2001},
publisher={World Scientific},
doi={10.1142/9789812799814_0013}
}

@article{noronha2013bulk,
  title = {Bulk viscosity effects in event-by-event relativistic hydrodynamics},
  author = {Noronha-Hostler, Jacquelyn and Denicol, Gabriel S. and Noronha, Jorge and Andrade, Rone P. G. and Grassi, Fr\'ed\'erique},
  journal = {Phys. Rev. C},
  volume = {88},
  issue = {4},
  pages = {044916},
  numpages = {17},
  year = {2013},
  month = {Oct},
  publisher = {American Physical Society},
  doi = {10.1103/PhysRevC.88.044916}
}

@article{PhysRevC.90.034907,
  title = {Bulk viscosity-driven suppression of shear viscosity effects on the flow harmonics at energies available at the BNL Relativistic Heavy Ion Collider},
  author = {Noronha-Hostler, Jacquelyn and Noronha, Jorge and Grassi, Fr\'ed\'erique},
  journal = {Phys. Rev. C},
  volume = {90},
  issue = {3},
  pages = {034907},
  numpages = {17},
  year = {2014},
  month = {Sep},
  publisher = {American Physical Society},
  doi = {10.1103/PhysRevC.90.034907}
}

@article{plumberg2025conservation,
title={Conservation of B, S, and Q charges in relativistic viscous hydrodynamics solved with smoothed particle hydrodynamics},
author={Plumberg, Christopher and Almaalol, Dekrayat and Dore, Travis and Mroczek, D{\'e}bora and Mart{\'\i}n, Jordi Salinas San and Serenone, Willian M and Spychalla, Lydia and Carzon, Patrick and Sievert, Matt and Gardim, Fernando G and others},
journal={Physical Review C},
volume={111},
number={4},
pages={044905},
year={2025},
publisher={APS},
doi={10.1103/PhysRevC.111.044905}
}

@article{Dore:2022qyz,
    author = {Dore, Travis and Karthein, Jamie M. and Long, Isaac and Mroczek, Debora and Noronha-Hostler, Jacquelyn and Parotto, Paolo and Ratti, Claudia and Yamauchi, Yukari},
    title = {Critical lensing and kurtosis near a critical point in the QCD phase diagram in and out-of-equilibrium},
    journal = {Phys. Rev. D},
    volume = {106},
    number = {9},
    pages = {094024},
    year = {2022},
    doi = {10.1103/PhysRevD.106.094024},
    eprint = {2207.04086},
    archivePrefix = {arXiv},
    primaryClass = {nucl-th}
}

@article{Noronha-Hostler:2019ayj,
    author = {Noronha-Hostler, Jacquelyn and Parotto, Paolo and Ratti, Claudia and Stafford, Jamie M.},
    title = {Lattice-based equation of state at finite baryon number, electric charge and strangeness chemical potentials},
    journal = {Phys. Rev. C},
    volume = {100},
    number = {6},
    pages = {064910},
    year = {2019},
    doi = {10.1103/PhysRevC.100.064910},
    eprint = {1902.06723},
    archivePrefix = {arXiv},
    primaryClass = {hep-ph}
}

@article{Abuali:2025texs4d,
    author = {Abuali, Ahmed and Borsanyi, Szabolcs and Fodor, Zoltan and Jahan, Johannes and Kahangirwe, Micheal and Parotto, Paolo and Pasztor, Attila and Ratti, Claudia and Shah, Hitansh and Trabulsi, Seth A.},
    title = {New 4D lattice QCD equation of state: Extended density coverage from a generalized {$T^\prime$} expansion},
    journal = {Phys. Rev. D},
    volume = {112},
    number = {5},
    pages = {054502},
    year = {2025},
    doi = {10.1103/2dmh-26yh},
    eprint = {2504.01881},
    archivePrefix = {arXiv},
    primaryClass = {hep-lat}
}

@book{AbramowitzStegun1964,
  editor    = {Abramowitz, Milton and Stegun, Irene A.},
  title     = {Handbook of Mathematical Functions with Formulas, Graphs, and Mathematical Tables},
  series    = {Applied Mathematics Series},
  volume    = {55},
  publisher = {National Bureau of Standards},
  address   = {Washington, D.C.},
  year      = {1964}
}

@book{NISTHandbook2010,
  editor    = {Olver, Frank W. J. and Lozier, Daniel W. and Boisvert, Ronald F. and Clark, Charles W.},
  title     = {NIST Handbook of Mathematical Functions},
  publisher = {Cambridge University Press},
  address   = {New York},
  year      = {2010}
}

@misc{DLMF,
  author       = {{NIST Digital Library of Mathematical Functions}},
  title        = {{NIST Digital Library of Mathematical Functions}},
  howpublished = {\url{https://dlmf.nist.gov/}},
  note         = {Release 1.2.6; F. W. J. Olver, A. B. Olde Daalhuis, D. W. Lozier, B. I. Schneider, R. F. Boisvert, C. W. Clark, B. R. Miller, B. V. Saunders, H. S. Cohl, and M. A. McClain, eds.},
  year         = {2026}
}

@article{PhysRevC.77.034911,
  title = {Viscous evolution of the rapidity distribution of matter created in relativistic heavy-ion collisions},
  author = {Bo\ifmmode \dot{z}\else \.{z}\fi{}ek, Piotr},
  journal = {Phys. Rev. C},
  volume = {77},
  issue = {3},
  pages = {034911},
  numpages = {11},
  year = {2008},
  month = {Mar},
  publisher = {American Physical Society},
  doi = {10.1103/PhysRevC.77.034911},
  url = {https://link.aps.org/doi/10.1103/PhysRevC.77.034911}
}

@article{Bjorken1983,
  author = {Bjorken, J. D.},
  title = {Highly Relativistic Nucleus-Nucleus Collisions: The Central Rapidity Region},
  journal = {Phys. Rev. D},
  volume = {27},
  pages = {140--151},
  year = {1983},
  doi = {10.1103/PhysRevD.27.140}
}

@article{PhysRevC.107.024901,
  title = {Pion, kaon, and (anti)proton production in $\text{U}+\text{U}$ collisions at $\sqrt{{s}_{NN}}=193$ GeV measured with the STAR detector},
  author = {Abdallah, M. S. and Aboona, B. E. and Adam, J. and Adams, J. R. and Adkins, J. K. and Agakishiev, G. and Aggarwal, I. and Aggarwal, M. M. and Ahammed, Z. and Aitbaev, A. and Alekseev, I. and Anderson, D. M. and Aparin, A. and Atchison, J. and Averichev, G. S. and Bairathi, V. and Baker, W. and Ball Cap, J. G. and Barish, K. and Bhagat, P. and Bhasin, A. and Bhatta, S. and Bordyuzhin, I. G. and Brandenburg, J. D. and Brandin, A. V. and Cai, X. Z. and Caines, H. and Calder\'on de la Barca S\'anchez, M. and Cebra, D. and Chakaberia, I. and Chan, B. K. and Chang, Z. and Chattopadhyay, S. and Chen, D. and Chen, J. and Chen, J. H. and Chen, X. and Chen, Z. and Cheng, J. and Choudhury, S. and Christie, W. and Chu, X. and Crawford, H. J. and Daugherity, M. and Dedovich, T. G. and Deppner, I. M. and Derevschikov, A. A. and Dhamija, A. and Di Carlo, L. and Didenko, L. and Dixit, P. and Dong, X. and Drachenberg, J. L. and Duckworth, E. and Dunlop, J. C. and Engelage, J. and Eppley, G. and Esumi, S. and Evdokimov, O. and Ewigleben, A. and Eyser, O. and Fatemi, R. and Fawzi, F. M. and Fazio, S. and Feng, C. J. and Feng, Y. and Finch, E. and Fisyak, Y. and Francisco, A. and Fu, C. and Geurts, F. and Ghimire, N. and Gibson, A. and Gopal, K. and Gou, X. and Grosnick, D. and Gupta, A. and Hamed, A. and Han, Y. and Harasty, M. D. and Harris, J. W. and Harrison, H. and He, S. and He, W. and He, X. H. and He, Y. and Heppelmann, S. and Hoffman, E. and Hu, C. and Hu, Q. and Hu, Y. and Huang, H. and Huang, H. Z. and Huang, S. L. and Huang, T. and Huang, X. and Huang, Y. and Humanic, T. J. and Isenhower, D. and Isshiki, M. and Jacobs, W. W. and Jena, C. and Ji, Y. and Jia, J. and Jiang, K. and Jin, C. and Ju, X. and Judd, E. G. and Kabana, S. and Kabir, M. L. and Kalinkin, D. and Kang, K. and Kapukchyan, D. and Kauder, K. and Ke, H. W. and Keane, D. and Kechechyan, A. and Kelsey, M. and Kimelman, B. and Kiselev, A. and Knospe, A. G. and Ko, H. S. and Kochenda, L. and Korobitsin, A. A. and Kravtsov, P. and Kumar, L. and Kumar, S. and Kunnawalkam Elayavalli, R. and Kwasizur, J. H. and Lacey, R. and Lan, S. and Landgraf, J. M. and Lebedev, A. and Lednicky, R. and Lee, J. H. and Leung, Y. H. and Lewis, N. and Li, C. and Li, C. and Li, W. and Li, W. and Li, X. and Li, Y. and Li, Y. and Li, Z. and Liang, X. and Liang, Y. and Lin, T. and Lin, Y. and Liu, F. and Liu, H. and Liu, H. and Liu, T. and Liu, X. and Liu, Y. and Ljubicic, T. and Llope, W. J. and Longacre, R. S. and Loyd, E. and Lu, T. and Lukow, N. S. and Luo, X. F. and Ma, L. and Ma, R. and Ma, Y. G. and Magdy, N. and Mallick, D. and Manukhov, S. L. and Margetis, S. and Matis, H. S. and Mazer, J. A. and McNamara, G. and Minaev, N. G. and Mishra, D. and Mohanty, B. and Mondal, M. M. and Mooney, I. and Morozov, D. A. and Nagy, M. I. and Nain, A. S. and Nam, J. D. and Nasim, Md. and Nayak, K. and Neff, D. and Nelson, J. M. and Nemes, D. B. and Nie, M. and Nigmatkulov, G. and Niida, T. and Nishitani, R. and Nogach, L. V. and Nonaka, T. and Nunes, A. S. and Odyniec, G. and Ogawa, A. and Oh, S. and Okorokov, V. A. and Okubo, K. and Page, B. S. and Pak, R. and Pan, J. and Pandav, A. and Pandey, A. K. and Panebratsev, Y. and Parfenov, P. and Paul, A. and Perkins, C. and Pokhrel, B. R. and Porter, J. and Posik, M. and Pruthi, N. K. and Putschke, J. and Qin, Z. and Qiu, H. and Quintero, A. and Racz, C. and Radhakrishnan, S. K. and Raha, N. and Ray, R. L. and Ritter, H. G. and Rogachevsky, O. V. and Romero, J. L. and Roy, D. and Roy Chowdhury, P. and Ruan, L. and Sahoo, A. K. and Sahoo, N. R. and Sako, H. and Salur, S. and Samigullin, E. and Sato, S. and Schmidke, W. B. and Schmitz, N. and Seger, J. and Sergeeva, M. and Seto, R. and Seyboth, P. and Shah, N. and Shahaliev, E. and Shanmuganathan, P. V. and Shao, M. and Shao, T. and Sharma, R. and Sheikh, A. I. and Shen, D. Y. and Shen, K. and Shi, S. S. and Shi, Y. and Shou, Q. Y. and Sichtermann, E. P. and Singh, J. and Singha, S. and Sinha, P. and Skoby, M. J. and S\"ohngen, Y. and Solyst, W. and Song, Y. and Srivastava, B. and Stanislaus, T. D. S. and Stewart, D. J. and Strikhanov, M. and Stringfellow, B. and Sun, C. and Sun, X. M. and Sun, X. and Sun, Y. and Sun, Y. and Surrow, B. and Svirida, D. N. and Sweger, Z. W. and Tang, A. H. and Tang, Z. and Taranenko, A. and Tarnowsky, T. and Thomas, J. H. and Tlusty, D. and Todoroki, T. and Tokarev, M. and Tomkiel, C. A. and Trentalange, S. and Tribble, R. E. and Tribedy, P. and Tripathy, S. K. and Tsai, O. D. and Tsang, C. Y. and Tu, Z. and Ullrich, T. and Underwood, D. G. and Upsal, I. and Van Buren, G. and Vasiliev, A. N. and Verkest, V. and Videb\ae{}k, F. and Vokal, S. and Voloshin, S. A. and Wang, F. and Wang, G. and Wang, J. S. and Wang, P. and Wang, X. and Wang, Y. and Wang, Y. and Wang, Z. and Webb, J. C. and Weidenkaff, P. C. and Westfall, G. D. and Wielanek, D. and Wieman, H. and Wissink, S. W. and Wu, J. and Wu, J. and Wu, Y. and Xi, B. and Xiao, Z. G. and Xie, G. and Xie, W. and Xu, H. and Xu, N. and Xu, Q. H. and Xu, Y. and Xu, Z. and Xu, Z. and Yan, G. and Yan, Z. and Yang, C. and Yang, Q. and Yang, S. and Yang, Y. and Ye, Z. and Ye, Z. and Yi, L. and Yip, K. and Yu, Y. and Zha, W. and Zhang, C. and Zhang, D. and Zhang, J. and Zhang, S. and Zhang, S. and Zhang, Y. and Zhang, Y. and Zhang, Y. and Zhang, Z. J. and Zhang, Z. and Zhang, Z. and Zhao, F. and Zhao, J. and Zhao, M. and Zhou, C. and Zhou, J. and Zhou, Y. and Zhu, X. and Zurek, M. and Zyzak, M.},
  collaboration = {STAR Collaboration},
  journal = {Phys. Rev. C},
  volume = {107},
  issue = {2},
  pages = {024901},
  numpages = {17},
  year = {2023},
  month = {Feb},
  publisher = {American Physical Society},
  doi = {10.1103/PhysRevC.107.024901},
}

@article{PhysRevC.68.034913,
  title = {Effect of shear viscosity on spectra, elliptic flow, and Hanbury Brown--Twiss radii},
  author = {Teaney, Derek},
  journal = {Phys. Rev. C},
  volume = {68},
  issue = {3},
  pages = {034913},
  numpages = {11},
  year = {2003},
  month = {Sep},
  publisher = {American Physical Society},
  doi = {10.1103/PhysRevC.68.034913}
}

@article{Jaiswal2013,
  author        = {Jaiswal, Amaresh},
  title         = {Relativistic dissipative hydrodynamics from kinetic theory with relaxation time approximation},
  journal       = {Physical Review C},
  volume        = {87},
  pages         = {051901},
  year          = {2013},
  doi           = {10.1103/PhysRevC.87.051901},
  eprint        = {1302.6311},
  archivePrefix = {arXiv},
  primaryClass  = {nucl-th}
}

@article{DuslingSchaefer2012,
  author        = {Dusling, Kevin and Schaefer, Thomas},
  title         = {Bulk viscosity, particle spectra and flow in heavy-ion collisions},
  journal       = {Physical Review C},
  volume        = {85},
  pages         = {044909},
  year          = {2012},
  doi           = {10.1103/PhysRevC.85.044909},
  eprint        = {1109.5181},
  archivePrefix = {arXiv},
  primaryClass  = {hep-ph}
}

@article{MolnarWolff2017,
  author        = {Molnar, Denes and Wolff, Zack},
  title         = {Self-consistent conversion of a viscous fluid to particles},
  journal       = {Physical Review C},
  volume        = {95},
  pages         = {024903},
  year          = {2017},
  doi           = {10.1103/PhysRevC.95.024903},
  eprint        = {1404.7850},
  archivePrefix = {arXiv},
  primaryClass  = {nucl-th}
}

@book{CercignaniKremer2002,
  author    = {Cercignani, Carlo and Kremer, Gilberto M.},
  title     = {The Relativistic Boltzmann Equation: Theory and Applications},
  publisher = {Birkh{\"a}user},
  address   = {Basel},
  year      = {2002},
  doi       = {10.1007/978-3-0348-8165-4}
}

@article{Staudenmaier2021,
  author        = {Staudenmaier, Jan and Oliinychenko, Dmytro and Torres-Rincon, Juan M. and Elfner, Hannah},
  title         = {{``Smashing more than two''}: Deuteron production in relativistic heavy ion collisions via stochastic multi-particle reactions},
  journal       = {Physical Review C},
  volume        = {104},
  pages         = {034908},
  year          = {2021},
  doi           = {10.1103/PhysRevC.104.034908},
  eprint        = {2106.14287},
  archivePrefix = {arXiv},
  primaryClass  = {nucl-th}
}

@article{kincses2025levy,
  title={L{\'e}vy walk of pions in heavy-ion collisions},
  author={Kincses, D{\'a}niel and Nagy, M{\'a}rton and Csan{\'a}d, M{\'a}t{\'e}},
  journal={Communications Physics},
  volume={8},
  number={1},
  pages={55},
  year={2025},
  publisher={Nature Publishing Group UK London},
  doi={doi.org/10.1038/s42005-025-01973-x}
}

@article{aaij2022centrality,
  title={Centrality determination in heavy-ion collisions with the LHCb detector},
  author={Aaij, Roel and Abell{\'a}n Beteta, Carlos and Ackernley, Thomas and Adeva, Bernardo and Adinolfi, Marco and Afsharnia, Hossein and Aidala, Christine Angela and Aiola, Salvatore and Ajaltouni, Ziad and Akar, Simon and others},
  journal={Journal of Instrumentation},
  volume={17},
  number={05},
  pages={P05009},
  year={2022},
  publisher={IoP Publishing},
  doi={10.1088/1748-0221/17/05/P05009}
}

@article{deJager1974,
  author  = {C. W. de Jager and H. de Vries and C. de Vries},
  title   = {Nuclear charge- and magnetization-density-distribution parameters from elastic electron scattering},
  journal = {Atomic Data and Nuclear Data Tables},
  volume  = {14},
  number  = {5--6},
  pages   = {479--508},
  year    = {1974},
  doi     = {10.1016/S0092-640X(74)80002-1}
}

@article{deVries1987,
  author  = {H. de Vries and C. W. de Jager and C. de Vries},
  title   = {Nuclear charge-density-distribution parameters from elastic electron scattering},
  journal = {Atomic Data and Nuclear Data Tables},
  volume  = {36},
  pages   = {495--536},
  year    = {1987},
  doi     = {10.1016/0092-640X(87)90013-1}
}

@article{WoodsSaxon1954,
  author  = {R. D. Woods and D. S. Saxon},
  title   = {Diffuse Surface Optical Model for Nucleon-Nuclei Scattering},
  journal = {Physical Review},
  volume  = {95},
  pages   = {577--578},
  year    = {1954},
  doi     = {10.1103/PhysRev.95.577}
}

@article{PhysRevC.110.064907,
  title = {Impact of the pre-equilibrium stage for the determination of nuclear geometry in high-energy isobar collisions},
  author = {Gardim, Fernando G. and Giannini, Andr\'e V. and Grassi, Fr\'ed\'erique and Pala, Kevin P. and Serenone, Willian M.},
  journal = {Phys. Rev. C},
  volume = {110},
  issue = {6},
  pages = {064907},
  numpages = {12},
  year = {2024},
  month = {Dec},
  publisher = {American Physical Society},
  doi = {10.1103/PhysRevC.110.064907}
}

@article{PhysRevLett.112.112501,
  title = {Signatures of $\ensuremath{\alpha}$ Clustering in Light Nuclei from Relativistic Nuclear Collisions},
  author = {Broniowski, Wojciech and Ruiz Arriola, Enrique},
  journal = {Phys. Rev. Lett.},
  volume = {112},
  issue = {11},
  pages = {112501},
  numpages = {5},
  year = {2014},
  month = {Mar},
  publisher = {American Physical Society},
  doi = {10.1103/PhysRevLett.112.112501}
}

@article{Miller:2007ri,
    author = {Miller, Michael L. and Reygers, Klaus and Sanders, Stephen J. and Steinberg, Peter},
    title = {Glauber Modeling in High-Energy Nuclear Collisions},
    journal = {Ann. Rev. Nucl. Part. Sci.},
    volume = {57},
    pages = {205--243},
    year = {2007},
    doi = {10.1146/annurev.nucl.57.090506.123020},
    eprint = {nucl-ex/0701025},
    archivePrefix = {arXiv}
}

@article{Alver:2008aq,
    author = {Alver, B. and Baker, M. and Loizides, C. and Steinberg, P.},
    title = {The PHOBOS Glauber Monte Carlo},
    eprint = {0805.4411},
    archivePrefix = {arXiv},
    primaryClass = {nucl-ex},
    year = {2008},
    doi={https://doi.org/10.48550/arXiv.0805.4411}
}

@misc{BraunMunzingerSchmah2021,
    author       = {Braun-Munzinger, Peter and Schmah, Alexander},
    title        = {Quark-Gluon Plasma Physics: Basics of Nucleon-Nucleon and Nucleus-Nucleus Collisions},
    howpublished = {Lecture slides, Heidelberg University},
    year         = {2021},
    note         = {Slide 32, accessed June 23, 2026},
    url          = {https://www.physi.uni-heidelberg.de/~schmah/QGP_SS2021/qgp2021_04_basics.pdf}
}

@article{Vredevoogd:2008id,
  author        = {Vredevoogd, Joshua and Pratt, Scott},
  title         = {Universal Flow in the First Stage of Relativistic Heavy Ion Collisions},
  journal       = {Physical Review C},
  volume        = {79},
  pages         = {044915},
  year          = {2009},
  doi           = {10.1103/PhysRevC.79.044915},
  eprint        = {0810.4325},
  archivePrefix = {arXiv},
  primaryClass  = {nucl-th}
}

@article{Liu:2015nwa,
  author        = {Liu, Jia and Shen, Chun and Heinz, Ulrich W.},
  title         = {Pre-equilibrium evolution effects on heavy-ion collision observables},
  journal       = {Physical Review C},
  volume        = {91},
  number        = {6},
  pages         = {064906},
  year          = {2015},
  doi           = {10.1103/PhysRevC.91.064906},
  eprint        = {1504.02160},
  archivePrefix = {arXiv},
  primaryClass  = {nucl-th}
}

@article{McLerran:1993ni,
  author        = {McLerran, Larry D. and Venugopalan, Raju},
  title         = {Computing quark and gluon distribution functions for very large nuclei},
  journal       = {Physical Review D},
  volume        = {49},
  pages         = {2233--2241},
  year          = {1994},
  doi           = {10.1103/PhysRevD.49.2233},
  eprint        = {hep-ph/9309289},
  archivePrefix = {arXiv}
}

@article{Schenke:2012wb,
  author        = {Schenke, Bjoern and Tribedy, Prithwish and Venugopalan, Raju},
  title         = {Fluctuating Glasma initial conditions and flow in heavy ion collisions},
  journal       = {Physical Review Letters},
  volume        = {108},
  pages         = {252301},
  year          = {2012},
  doi           = {10.1103/PhysRevLett.108.252301},
  eprint        = {1202.6646},
  archivePrefix = {arXiv},
  primaryClass  = {nucl-th}
}

@article{Gale:2012rq,
  author        = {Gale, Charles and Jeon, Sangyong and Schenke, Bjoern and Tribedy, Prithwish and Venugopalan, Raju},
  title         = {Event-by-event anisotropic flow in heavy-ion collisions from combined Yang-Mills and viscous fluid dynamics},
  journal       = {Physical Review Letters},
  volume        = {110},
  pages         = {012302},
  year          = {2013},
  doi           = {10.1103/PhysRevLett.110.012302},
  eprint        = {1209.6330},
  archivePrefix = {arXiv},
  primaryClass  = {nucl-th}
}

@article{Arnold:2002zm,
  author        = {Arnold, Peter Brockway and Moore, Guy D. and Yaffe, Laurence G.},
  title         = {Effective kinetic theory for high temperature gauge theories},
  journal       = {Journal of High Energy Physics},
  volume        = {2003},
  number        = {01},
  pages         = {030},
  year          = {2003},
  doi           = {10.1088/1126-6708/2003/01/030},
  eprint        = {hep-ph/0209353},
  archivePrefix = {arXiv}
}

@article{Kurkela:2018wud,
  author        = {Kurkela, Aleksi and Mazeliauskas, Aleksas and Paquet, Jean-Fran{\c{c}}ois and Schlichting, S{\"o}ren and Teaney, Derek},
  title         = {Effective kinetic description of event-by-event pre-equilibrium dynamics in high-energy heavy-ion collisions},
  journal       = {Physical Review C},
  volume        = {99},
  number        = {3},
  pages         = {034910},
  year          = {2019},
  doi           = {10.1103/PhysRevC.99.034910},
  eprint        = {1805.00961},
  archivePrefix = {arXiv},
  primaryClass  = {hep-ph}
}

@article{Kurkela:2018vqr,
  author        = {Kurkela, Aleksi and Mazeliauskas, Aleksas and Paquet, Jean-Fran{\c{c}}ois and Schlichting, S{\"o}ren and Teaney, Derek},
  title         = {Matching the Nonequilibrium Initial Stage of Heavy Ion Collisions to Hydrodynamics with QCD Kinetic Theory},
  journal       = {Physical Review Letters},
  volume        = {122},
  number        = {12},
  pages         = {122302},
  year          = {2019},
  doi           = {10.1103/PhysRevLett.122.302},
  eprint        = {1805.01604},
  archivePrefix = {arXiv},
  primaryClass  = {hep-ph}
}

@article{Lin:2004en,
  author        = {Lin, Zi-Wei and Ko, Che Ming and Li, Bao-An and Zhang, Bin and Pal, Subrata},
  title         = {A multi-phase transport model for relativistic heavy ion collisions},
  journal       = {Physical Review C},
  volume        = {72},
  pages         = {064901},
  year          = {2005},
  doi           = {10.1103/PhysRevC.72.064901},
  eprint        = {nucl-th/0411110},
  archivePrefix = {arXiv}
}

@article{Petersen:2008dd,
  author        = {Petersen, Hannah and Steinheimer, Jan and Burau, Gerhard and Bleicher, Marcus and St{\"o}cker, Horst},
  title         = {A fully integrated transport approach to heavy ion reactions with an intermediate hydrodynamic stage},
  journal       = {Physical Review C},
  volume        = {78},
  pages         = {044901},
  year          = {2008},
  doi           = {10.1103/PhysRevC.78.044901},
  eprint        = {0806.1695},
  archivePrefix = {arXiv},
  primaryClass  = {nucl-th}
}

@article{Oliinychenko:2015lva,
  author        = {Oliinychenko, Dmytro and Petersen, Hannah},
  title         = {Deviations of the energy-momentum tensor from equilibrium in the initial state for hydrodynamics from transport approaches},
  journal       = {Physical Review C},
  volume        = {93},
  number        = {3},
  pages         = {034905},
  year          = {2016},
  doi           = {10.1103/PhysRevC.93.034905},
  eprint        = {1508.04378},
  archivePrefix = {arXiv},
  primaryClass  = {nucl-th}
}

@book{landau1987fluid,
  title={Fluid Mechanics},
  author={Landau, Lev Davidovich and Lifshitz, Evgenii Mikhailovich},
  volume={6},
  year={1987},
  publisher={Elsevier},
  doi={https://doi.org/10.1016/C2013-0-03650-X},
  isbn={978-0-08-029142-0}
}

@article{Koide:2007,
  author  = {Koide, T. and Denicol, G. S. and Mota, Ph. and Kodama, T.},
  title   = {Relativistic dissipative hydrodynamics: A minimal causal theory},
  journal = {Physical Review C},
  volume  = {75},
  pages   = {034909},
  year    = {2007},
  doi     = {10.1103/PhysRevC.75.034909},
  eprint  = {hep-ph/0609117},
  archivePrefix = {arXiv}
}

@article{ParticleDataGroup:2022,
  author        = {Workman, R. L. and others},
  collaboration = {Particle Data Group},
  title         = {Review of Particle Physics},
  journal       = {Progress of Theoretical and Experimental Physics},
  volume        = {2022},
  pages         = {083C01},
  year          = {2022},
  doi           = {10.1093/ptep/ptac097},
  note          = {See also the review ``Monte Carlo Particle Numbering Scheme''}
}

@article{PhysRevLett.93.102301,
  title = {Nuclear Stopping in $\mathrm{A}\mathrm{u}+\mathrm{A}\mathrm{u}$ Collisions at $\sqrt{{s}_{NN}}=200\text{ }\text{ }\mathrm{G}\mathrm{e}\mathrm{V}$},
  author = {Bearden, I. G. and Beavis, D. and Besliu, C. and Budick, B. and B\o{}ggild, H. and Chasman, C. and Christensen, C. H. and Christiansen, P. and Cibor, J. and Debbe, R. and Enger, E. and Gaardh\o{}je, J. J. and Germinario, M. and Hagel, K. and Hansen, O. and Holm, A. and Holme, A. K. and Ito, H. and Jipa, A. and Jundt, F. and J\o{}rdre, J. I. and J\o{}rgensen, C. E. and Karabowicz, R. and Kim, E. J. and Kozik, T. and Larsen, T. M. and Lee, J. H. and Lee, Y. K. and L\o{}vh\o{}iden, G. and Majka, Z. and Makeev, A. and Mikelsen, M. and Murray, M. and Natowitz, J. and Nielsen, B. S. and Norris, J. and Olchanski, K. and Ouerdane, D. and P\l{}aneta, R. and Rami, F. and Ristea, C. and R\"ohrich, D. and Samset, B. H. and Sandberg, D. and Sanders, S. J. and Scheetz, R. A. and Staszel, P. and Tveter, T. S. and Videb\ae{}k, F. and Wada, R. and Yin, Z. and Zgura, I. S.},
  collaboration = {BRAHMS Collaboration},
  journal = {Phys. Rev. Lett.},
  volume = {93},
  issue = {10},
  pages = {102301},
  numpages = {5},
  year = {2004},
  month = {Aug},
  publisher = {American Physical Society},
  doi = {10.1103/PhysRevLett.93.102301}
}

@article{PhysRevLett.101.112301,
  title = {Importance of Granular Structure in the Initial Conditions for the Elliptic Flow},
  author = {Andrade, R. P. G. and Grassi, F. and Hama, Y. and Kodama, T. and Qian, W. L.},
  journal = {Phys. Rev. Lett.},
  volume = {101},
  issue = {11},
  pages = {112301},
  numpages = {4},
  year = {2008},
  month = {Sep},
  publisher = {American Physical Society},
  doi = {10.1103/PhysRevLett.101.112301}
}

@misc{NSWWSTARTrigger,
  author       = {{Nuclear Science and Technology}},
  title        = {{STAR Experiment: Trigger system}},
  year         = {2026},
  howpublished = {\url{https://nsww.org/projects/bnl/star/trigger-system-e.php}},
  note         = {Educational material, Fig. 2. Accessed: 2026-06-25}
}

@phdthesis{Goes-Hirayama:2026ies,
    author = "G{\'o}es-Hirayama, Renan Nobuyuki",
    title = "{Dynamical production of dileptons within a hybrid approach}",
    doi = "10.21248/gups.97813",
    school = "Goethe U., Frankfurt (main)",
    year = "2026"
}

@article{Heinz_2009,
doi = {10.1088/1751-8113/42/21/214003},
year = {2009},
month = {may},
publisher = {},
volume = {42},
number = {21},
pages = {214003},
author = {Heinz, Ulrich},
title = {The strongly coupled quark–gluon plasma created at RHIC*},
journal = {Journal of Physics A: Mathematical and Theoretical},
}

@article{PhysRevC.83.044913,
  title = {Flow analysis with cumulants: Direct calculations},
  author = {Bilandzic, Ante and Snellings, Raimond and Voloshin, Sergei},
  journal = {Phys. Rev. C},
  volume = {83},
  issue = {4},
  pages = {044913},
  numpages = {11},
  year = {2011},
  month = {Apr},
  publisher = {American Physical Society},
  doi = {10.1103/PhysRevC.83.044913}
}

@article{PhysRevC.64.054901,
  title = {Flow analysis from multiparticle azimuthal correlations},
  author = {Borghini, Nicolas and Dinh, Phuong Mai and Ollitrault, Jean-Yves},
  journal = {Phys. Rev. C},
  volume = {64},
  issue = {5},
  pages = {054901},
  numpages = {23},
  year = {2001},
  month = {Sep},
  publisher = {American Physical Society},
  doi = {10.1103/PhysRevC.64.054901},
  url = {https://link.aps.org/doi/10.1103/PhysRevC.64.054901}
}

@article{Denicol:2012cn,
  author = {Denicol, Gabriel S. and Niemi, Harri and Moln{\'a}r, Etele and Rischke, Dirk H.},
  title = {Derivation of transient relativistic fluid dynamics from the Boltzmann equation},
  journal = {Phys. Rev. D},
  volume = {85},
  pages = {114047},
  year = {2012},
  doi = {10.1103/PhysRevD.85.114047},
  eprint = {1202.4551},
  archivePrefix = {arXiv},
  primaryClass = {nucl-th},
  note = {Erratum: Phys. Rev. D 91, 039902 (2015)}
}

@article{Bemfica:2020zjp,
  author = {Bemfica, F{\'a}bio S. and Disconzi, Marcelo M. and Noronha, Jorge},
  title = {First-Order General-Relativistic Viscous Fluid Dynamics},
  journal = {Phys. Rev. X},
  volume = {12},
  number = {2},
  pages = {021044},
  year = {2022},
  doi = {10.1103/PhysRevX.12.021044},
  eprint = {2009.11388},
  archivePrefix = {arXiv},
  primaryClass = {gr-qc}
}

@article{Kovtun:2019hdm,
  author = {Kovtun, Pavel},
  title = {First-order relativistic hydrodynamics is stable},
  journal = {JHEP},
  volume = {10},
  pages = {034},
  year = {2019},
  doi = {10.1007/JHEP10(2019)034},
  eprint = {1907.08191},
  archivePrefix = {arXiv},
  primaryClass = {hep-th}
}

@article{Shen:2017bsr,
  author = {Shen, Chun and Schenke, Bj{\"o}rn},
  title = {Dynamical initial state model for relativistic heavy-ion collisions},
  journal = {Phys. Rev. C},
  volume = {97},
  number = {2},
  pages = {024907},
  year = {2018},
  doi = {10.1103/PhysRevC.97.024907},
  eprint = {1710.00881},
  archivePrefix = {arXiv},
  primaryClass = {nucl-th}
}

@article{Werner:2007bf,
  author = {Werner, Klaus},
  title = {Core-corona separation in ultra-relativistic heavy ion collisions},
  journal = {Phys. Rev. Lett.},
  volume = {98},
  pages = {152301},
  year = {2007},
  doi = {10.1103/PhysRevLett.98.152301},
  eprint = {0704.1270},
  archivePrefix = {arXiv},
  primaryClass = {nucl-th}
}

@article{Werner:2010aa,
  author = {Werner, Klaus and Karpenko, Iu. and Pierog, T. and Bleicher, Marcus and Mikhailov, Kirill},
  title = {Event-by-event simulation of the three-dimensional hydrodynamic evolution from flux tube initial conditions in ultrarelativistic heavy ion collisions},
  journal = {Phys. Rev. C},
  volume = {82},
  pages = {044904},
  year = {2010},
  doi = {10.1103/PhysRevC.82.044904},
  eprint = {1004.0805},
  archivePrefix = {arXiv},
  primaryClass = {nucl-th}
}

@software{PRAIA,
    author  = {{USPHydro Collaboration}},
    title   = {{PRAIA}: Relativistic Hydrodynamics Code},
    howpublished = {\url{https://github.com/USPHydro/PRAIAv2}},
    note    = {Source code available on GitHub},
    urldate = {2026-10-22}
}

@article{gubser2013complex,
  title={Complex deformations of Bjorken flow},
  author={Gubser, Steven S},
  journal={Physical Review C—Nuclear Physics},
  volume={87},
  number={1},
  pages={014909},
  year={2013},
  publisher={APS},
doi={10.1103/PhysRevC.87.014909}
}

@article{gubser2010symmetry,
title={Symmetry constraints on generalizations of Bjorken flow},
author={Gubser, Steven S},
journal={Physical Review D—Particles, Fields, Gravitation, and Cosmology},
volume={82},
number={8},
pages={085027},
year={2010},
publisher={APS},
doi={10.1103/PhysRevD.82.085027}
}

@article{PhysRevC.91.014903,
  title = {Solutions of conformal Israel-Stewart relativistic viscous fluid dynamics},
  author = {Marrochio, Hugo and Noronha, Jorge and Denicol, Gabriel S. and Luzum, Matthew and Jeon, Sangyong and Gale, Charles},
  journal = {Phys. Rev. C},
  volume = {91},
  issue = {1},
  pages = {014903},
  numpages = {12},
  year = {2015},
  month = {Jan},
  publisher = {American Physical Society},
  doi = {10.1103/PhysRevC.91.014903}
}

@article{v334-d32w,
  title = {Viscous Gubser flow with conserved charges to benchmark fluid simulations},
  author = {Ingles, Kevin and Salinas San Mart\'{\i}n, Jordi and Serenone, Willian and Noronha-Hostler, Jacquelyn},
  journal = {Phys. Rev. C},
  volume = {113},
  issue = {2},
  pages = {024915},
  numpages = {12},
  year = {2026},
  month = {Feb},
  publisher = {American Physical Society},
  doi = {10.1103/v334-d32w}
}

@article{Yukawa1935,
  author  = {Yukawa, Hideki},
  title   = {On the Interaction of Elementary Particles. I},
  journal = {Proceedings of the Physico-Mathematical Society of Japan. 3rd Series},
  volume  = {17},
  pages   = {48--57},
  year    = {1935},
  doi     = {10.11429/ppmsj1919.17.0_48}
}

@article{Lattes1947Nature,
  author  = {Lattes, C. M. G. and Muirhead, H. and Occhialini, G. P. S. and Powell, C. F.},
  title   = {Processes Involving Charged Mesons},
  journal = {Nature},
  volume  = {159},
  pages   = {694--697},
  year    = {1947},
  doi     = {10.1038/159694a0}
}

@misc{NobelYukawa1949,
  author       = {{Nobel Prize Outreach}},
  title        = {The Nobel Prize in Physics 1949},
  howpublished = {\url{https://www.nobelprize.org/prizes/physics/1949/summary/}},
  note         = {Accessed: 2026-06-26}
}

@misc{NobelPowell1950,
  author       = {{Nobel Prize Outreach}},
  title        = {The Nobel Prize in Physics 1950},
  howpublished = {\url{https://www.nobelprize.org/prizes/physics/1950/summary/}},
  note         = {Accessed: 2026-06-26}
}

@article{Schrodinger1926,
  author  = {Schr{\"o}dinger, E.},
  title   = {Quantisierung als Eigenwertproblem},
  journal = {Annalen der Physik},
  volume  = {384},
  number  = {4},
  pages   = {361--376},
  year    = {1926},
  doi     = {10.1002/andp.19263840404}
}

@article{Dirac1928,
  author  = {Dirac, P. A. M.},
  title   = {The Quantum Theory of the Electron},
  journal = {Proceedings of the Royal Society of London. Series A},
  volume  = {117},
  number  = {778},
  pages   = {610--624},
  year    = {1928},
  doi     = {10.1098/rspa.1928.0023}
}

@article{Anderson1933,
  author  = {Anderson, Carl D.},
  title   = {The Positive Electron},
  journal = {Physical Review},
  volume  = {43},
  number  = {6},
  pages   = {491--494},
  year    = {1933},
  doi     = {10.1103/PhysRev.43.491}
}

@article{Darwin1928,
  author  = {Darwin, C. G.},
  title   = {The Wave Equations of the Electron},
  journal = {Proceedings of the Royal Society of London. Series A},
  volume  = {118},
  number  = {780},
  pages   = {654--680},
  year    = {1928},
  doi     = {10.1098/rspa.1928.0076}
}

@article{Thomson1897,
  author  = {Thomson, J. J.},
  title   = {Cathode Rays},
  journal = {The London, Edinburgh, and Dublin Philosophical Magazine and Journal of Science},
  volume  = {44},
  number  = {269},
  pages   = {293--316},
  year    = {1897},
  doi     = {10.1080/14786449708621070}
}

@article{GeigerMarsden1909,
  author  = {Geiger, H. and Marsden, E.},
  title   = {On a Diffuse Reflection of the $\alpha$-Particles},
  journal = {Proceedings of the Royal Society of London. Series A},
  volume  = {82},
  number  = {557},
  pages   = {495--500},
  year    = {1909},
  doi     = {10.1098/rspa.1909.0054}
}

@article{Rutherford1911,
  author  = {Rutherford, E.},
  title   = {The Scattering of $\alpha$ and $\beta$ Particles by Matter and the Structure of the Atom},
  journal = {The London, Edinburgh, and Dublin Philosophical Magazine and Journal of Science},
  volume  = {21},
  number  = {125},
  pages   = {669--688},
  year    = {1911},
  doi     = {10.1080/14786440508637080}
}

@article{CockcroftWalton1932a,
  author  = {Cockcroft, J. D. and Walton, E. T. S.},
  title   = {Artificial Production of Fast Protons},
  journal = {Nature},
  volume  = {129},
  pages   = {242},
  year    = {1932},
  doi     = {10.1038/129242a0}
}

@article{CockcroftWalton1932b,
  author  = {Cockcroft, J. D. and Walton, E. T. S.},
  title   = {Disintegration of Lithium by Swift Protons},
  journal = {Nature},
  volume  = {129},
  pages   = {649},
  year    = {1932},
  doi     = {10.1038/129649a0}
}

@article{LawrenceLivingston1932,
  author  = {Lawrence, E. O. and Livingston, M. S.},
  title   = {The Production of High Speed Light Ions without the Use of High Voltages},
  journal = {Physical Review},
  volume  = {40},
  pages   = {19--35},
  year    = {1932},
  doi     = {10.1103/PhysRev.40.19}
}

@misc{GeigerMarsdenApparatusWikimedia,
  author       = {Kurzon},
  title        = {Geiger-Marsden experiment.svg},
  year         = {2014},
  howpublished = {Wikimedia Commons},
  note         = {Licensed under CC BY-SA 3.0. Accessed: 2026-06-27}
}

@article{Arnison1983W,
  author  = {Arnison, G. and others},
  title   = {Experimental Observation of Isolated Large Transverse Energy Electrons with Associated Missing Energy at $\sqrt{s}=540$ GeV},
  journal = {Physics Letters B},
  volume  = {122},
  pages   = {103--116},
  year    = {1983},
  doi     = {10.1016/0370-2693(83)91177-2}
}

@article{Arnison1983Z,
  author  = {Arnison, G. and others},
  title   = {Experimental Observation of Lepton Pairs of Invariant Mass Around 95 GeV/$c^2$ at the CERN SPS Collider},
  journal = {Physics Letters B},
  volume  = {126},
  pages   = {398--410},
  year    = {1983},
  doi     = {10.1016/0370-2693(83)90188-0}
}

@article{CDF1995Top,
  author  = {Abe, F. and others},
  collaboration = {CDF Collaboration},
  title   = {Observation of Top Quark Production in $\bar{p}p$ Collisions with the Collider Detector at Fermilab},
  journal = {Physical Review Letters},
  volume  = {74},
  pages   = {2626--2631},
  year    = {1995},
  doi     = {10.1103/PhysRevLett.74.2626}
}

@article{DZero1995Top,
  author  = {Abachi, S. and others},
  collaboration = {D0 Collaboration},
  title   = {Observation of the Top Quark},
  journal = {Physical Review Letters},
  volume  = {74},
  pages   = {2632--2637},
  year    = {1995},
  doi     = {10.1103/PhysRevLett.74.2632}
}

@article{ATLAS2012Higgs,
  author  = {Aad, G. and others},
  collaboration = {ATLAS Collaboration},
  title   = {Observation of a New Particle in the Search for the Standard Model Higgs Boson with the ATLAS Detector at the LHC},
  journal = {Physics Letters B},
  volume  = {716},
  pages   = {1--29},
  year    = {2012},
  doi     = {10.1016/j.physletb.2012.08.020}
}

@article{CMS2012Higgs,
  author  = {Chatrchyan, S. and others},
  collaboration = {CMS Collaboration},
  title   = {Observation of a New Boson at a Mass of 125 GeV with the CMS Experiment at the LHC},
  journal = {Physics Letters B},
  volume  = {716},
  pages   = {30--61},
  year    = {2012},
  doi     = {10.1016/j.physletb.2012.08.021}
}

@article{Adams2005STAR,
  author  = {Adams, J. and others},
  collaboration = {STAR Collaboration},
  title   = {Experimental and Theoretical Challenges in the Search for the Quark-Gluon Plasma: The STAR Collaboration's Critical Assessment of the Evidence from RHIC Collisions},
  journal = {Nuclear Physics A},
  volume  = {757},
  pages   = {102--183},
  year    = {2005},
  doi     = {10.1016/j.nuclphysa.2005.03.085}
}

@article{Adcox2005PHENIX,
  author  = {Adcox, K. and others},
  collaboration = {PHENIX Collaboration},
  title   = {Formation of Dense Partonic Matter in Relativistic Nucleus--Nucleus Collisions at RHIC: Experimental Evaluation by the PHENIX Collaboration},
  journal = {Nuclear Physics A},
  volume  = {757},
  pages   = {184--283},
  year    = {2005},
  doi     = {10.1016/j.nuclphysa.2005.03.086}
}

@article{Caines2017,
  author  = {Caines, Helen},
  title   = {The Search for Critical Behavior and Other Features of the QCD Phase Diagram -- Current Status and Future Prospects},
  journal = {Nuclear Physics A},
  volume  = {967},
  pages   = {121--128},
  year    = {2017},
  doi     = {10.1016/j.nuclphysa.2017.05.116}
}

@article{Yang2017BESII,
  author  = {Yang, Chi},
  collaboration = {STAR Collaboration},
  title   = {The STAR Beam Energy Scan Phase II Physics and Upgrades},
  journal = {Nuclear Physics A},
  volume  = {967},
  pages   = {800--803},
  year    = {2017},
  doi     = {10.1016/j.nuclphysa.2017.05.042}
}

@article{aa07,
  author  = {Belen'kji, S. Z. and Landau, L. D.},
  title   = {Hydrodynamic Theory of Multiple Production of Particles},
  journal = {Il Nuovo Cimento},
  volume  = {3},
  number  = {1 Suppl.},
  pages   = {15--31},
  year    = {1956},
  doi     = {10.1007/BF02745507},
  url     = {https://doi.org/10.1007/BF02745507}
}

@article{aa08,
  author  = {Fermi, E.},
  title   = {High Energy Nuclear Events},
  journal = {Progress of Theoretical Physics},
  volume  = {5},
  number  = {4},
  pages   = {570--583},
  year    = {1950},
  month   = jul,
  doi     = {10.1143/ptp/5.4.570},
  url     = {https://doi.org/10.1143/ptp/5.4.570}
}

@article{aa09,
  author  = {Fermi, E.},
  title   = {Angular Distribution of the Pions Produced in High Energy Nuclear Collisions},
  journal = {Physical Review},
  volume  = {81},
  pages   = {683},
  year    = {1951},
  month   = mar,
  doi     = {10.1103/PhysRev.81.683},
  url     = {https://doi.org/10.1103/PhysRev.81.683}
}

@article{Arsene2005BRAHMS,
  author        = {Arsene, I. and others},
  collaboration = {BRAHMS Collaboration},
  title         = {Quark-Gluon Plasma and Color Glass Condensate at RHIC? The Perspective from the BRAHMS Experiment},
  journal       = {Nuclear Physics A},
  volume        = {757},
  pages         = {1--27},
  year          = {2005},
  doi           = {10.1016/j.nuclphysa.2005.02.130}
}

@article{Back2005PHOBOS,
  author        = {Back, B. B. and others},
  collaboration = {PHOBOS Collaboration},
  title         = {The PHOBOS Perspective on Discoveries at RHIC},
  journal       = {Nuclear Physics A},
  volume        = {757},
  pages         = {28--101},
  year          = {2005},
  doi           = {10.1016/j.nuclphysa.2005.03.084}
}

@article{Teaney2001,
  author  = {Teaney, D. and Lauret, J. and Shuryak, E. V.},
  title   = {Flow at the SPS and RHIC as a Quark-Gluon Plasma Signature},
  journal = {Physical Review Letters},
  volume  = {86},
  pages   = {4783--4786},
  year    = {2001},
  doi     = {10.1103/PhysRevLett.86.4783}
}

@article{Kolb2000,
  author  = {Kolb, P. F. and Huovinen, P. and Heinz, U. and Heiselberg, H.},
  title   = {Elliptic Flow at SPS and RHIC: From Kinetic Transport to Hydrodynamics},
  journal = {Physics Letters B},
  volume  = {500},
  pages   = {232--240},
  year    = {2001},
  doi     = {10.1016/S0370-2693(01)00079-X}
}

@article{Huovinen2001,
  author  = {Huovinen, P. and Kolb, P. F. and Heinz, U. and Ruuskanen, P. V. and Voloshin, S. A.},
  title   = {Radial and Elliptic Flow at RHIC: Further Predictions},
  journal = {Physics Letters B},
  volume  = {503},
  pages   = {58--64},
  year    = {2001},
  doi     = {10.1016/S0370-2693(01)00219-2}
}

@misc{wismer_thomson_tube_2000,
  title        = {Thomson's gas discharge tube},
  author       = {{Science History Institute Archives}},
  year         = {2000},
  howpublished = {Robert K. Wismer Collection of Chemical Museum Slides, Box 1, Object 2006.501.016},
  note         = {Science History Institute Archives, Philadelphia. Accessed: 2026-06-29},
  url          = {https://sciencehistory.libraryhost.com/repositories/3/archival_objects/51122}
}

@article{pdg2025,
  author        = {{Particle Data Group}},
  title         = {Review of Particle Physics},
  journal       = {Physical Review D},
  volume        = {110},
  pages         = {030001},
  year          = {2024},
  note          = {2025 update}
}

@misc{wiki_standard_model,
  author       = {{Wikipedia contributors}},
  title        = {Standard Model},
  year         = {2023},
  howpublished = {\url{https://en.wikipedia.org/wiki/Standard_Model}},
  note         = {Accessed: 07 June 2023}
}

@article{PhysRevC.95.034902,
  title = {Phenomenological consequences of enhanced bulk viscosity near the QCD critical point},
  author = {Monnai, Akihiko and Mukherjee, Swagato and Yin, Yi},
  journal = {Phys. Rev. C},
  volume = {95},
  issue = {3},
  pages = {034902},
  numpages = {8},
  year = {2017},
  month = {Mar},
  publisher = {American Physical Society},
  doi = {10.1103/PhysRevC.95.034902}
}

@article{PhysRevD.100.074017,
  title = {Critical behavior of the bulk viscosity in QCD},
  author = {Martinez, M. and Sch\"afer, T. and Skokov, V.},
  journal = {Phys. Rev. D},
  volume = {100},
  issue = {7},
  pages = {074017},
  numpages = {9},
  year = {2019},
  month = {Oct},
  publisher = {American Physical Society},
  doi = {10.1103/PhysRevD.100.074017}
}

@article{doi:10.1142/S0217751X05027965,
author = {Stephanov, M. A.},
title = {QCD PHASE DIAGRAM AND THE CRITICAL POINT},
journal = {International Journal of Modern Physics A},
volume = {20},
number = {19},
pages = {4387-4392},
year = {2005},
doi = {10.1142/S0217751X05027965}
}

@article{PhysRevC.56.3265,
  title = {Resonance contributions to Hanbury-Brown--Twiss correlation radii},
  author = {Wiedemann, Urs Achim and Heinz, Ulrich},
  journal = {Phys. Rev. C},
  volume = {56},
  issue = {6},
  pages = {3265--3286},
  numpages = {0},
  year = {1997},
  month = {Dec},
  publisher = {American Physical Society},
  doi = {10.1103/PhysRevC.56.3265},
}

@article{PhysRevC.56.R610,
  title = {New class of Hanbury-Brown--Twiss parameters},
  author = {Wiedemann, Urs Achim and Heinz, Ulrich},
  journal = {Phys. Rev. C},
  volume = {56},
  issue = {2},
  pages = {R610(R)--R613(R)},
  numpages = {0},
  year = {1997},
  month = {Aug},
  publisher = {American Physical Society},
  doi = {10.1103/PhysRevC.56.R610},
}

@article{10.1007/BF02907008,
  author  = {Cs{\"o}rg{\H{o}}, T. and L{\"o}rstad, B. and Zim{\'a}nyi, J.},
  title   = {Bose-Einstein correlations for systems with large halo},
  journal = {Zeitschrift f{\"u}r Physik C Particles and Fields},
  year    = {1996},
  volume  = {71},
  number  = {3},
  pages   = {491--497},
  doi     = {10.1007/BF02907008},
  url     = {https://doi.org/10.1007/BF02907008}
}

@article{PhysRevC.107.054911,
  title = {Mapping out the thermodynamic stability of a QCD equation of state with a critical point using active learning},
  author = {Mroczek, D. and Hjorth-Jensen, M. and Noronha-Hostler, J. and Parotto, P. and Ratti, C. and Vilalta, R.},
  journal = {Phys. Rev. C},
  volume = {107},
  issue = {5},
  pages = {054911},
  numpages = {14},
  year = {2023},
  month = {May},
  publisher = {American Physical Society},
  doi = {10.1103/PhysRevC.107.054911},
  url = {https://link.aps.org/doi/10.1103/PhysRevC.107.054911}
}

@article{PhysRevC.86.014904,
  title = {Energy and system-size dependence of two- and four-particle ${v}_{2}$ measurements in heavy-ion collisions at $\sqrt{{s}_{NN}}=62.4$ and 200 GeV and their implications on flow fluctuations and nonflow},
  author = {Agakishiev, G. and Aggarwal, M. M. and Ahammed, Z. and Alakhverdyants, A. V. and Alekseev, I. and Alford, J. and Anderson, B. D. and Anson, C. D. and Arkhipkin, D. and Averichev, G. S. and Balewski, J. and Beavis, D. R. and Behera, N. K. and Bellwied, R. and Betancourt, M. J. and Betts, R. R. and Bhasin, A. and Bhati, A. K. and Bichsel, H. and Bielcik, J. and Bielcikova, J. and Bland, L. C. and Bordyuzhin, I. G. and Borowski, W. and Bouchet, J. and Braidot, E. and Brandin, A. V. and Bridgeman, A. and Brovko, S. G. and Bruna, E. and Bueltmann, S. and Bunzarov, I. and Burton, T. P. and Cai, X. Z. and Caines, H. and Calder{\'o}n de la Barca S{\'a}nchez, M. and Cebra, D. and Cendejas, R. and Cervantes, M. C. and Chaloupka, P. and Chattopadhyay, S. and Chen, H. F. and Chen, J. H. and Chen, J. Y. and Chen, L. and Cheng, J. and Cherney, M. and Chikanian, A. and Choi, K. E. and Christie, W. and Chung, P. and Codrington, M. J. M. and Corliss, R. and Cramer, J. G. and Crawford, H. J. and Cui, X. and Davila Leyva, A. and De Silva, L. C. and Debbe, R. R. and Dedovich, T. G. and Deng, J. and Derevschikov, A. A. and Derradi de Souza, R. and Didenko, L. and Djawotho, P. and Dong, X. and Drachenberg, J. L. and Draper, J. E. and Du, C. M. and Dunlop, J. C. and Efimov, L. G. and Elnimr, M. and Engelage, J. and Eppley, G. and Estienne, M. and Eun, L. and Evdokimov, O. and Fatemi, R. and Fedorisin, J. and Fersch, R. G. and Filip, P. and Finch, E. and Fine, V. and Fisyak, Y. and Gagliardi, C. A. and Gangadharan, D. R. and Geurts, F. and Ghosh, P. and Gorbunov, Y. N. and Gordon, A. and Grebenyuk, O. G. and Grosnick, D. and Gupta, A. and Gupta, S. and Guryn, W. and Haag, B. and Hajkova, O. and Hamed, A. and Han, L-X. and Harris, J. W. and Hays-Wehle, J. P. and Heinz, M. and Heppelmann, S. and Hirsch, A. and Hoffmann, G. W. and Hofman, D. J. and Huang, B. and Huang, H. Z. and Humanic, T. J. and Huo, L. and Igo, G. and Jacobs, W. W. and Jena, C. and Jin, F. and Joseph, J. and Judd, E. G. and Kabana, S. and Kang, K. and Kapitan, J. and Kauder, K. and Ke, H. W. and Keane, D. and Kechechyan, A. and Kettler, D. and Kikola, D. P. and Kiryluk, J. and Kisiel, A. and Kizka, V. and Klein, S. R. and Knospe, A. G. and Koetke, D. D. and Kollegger, T. and Konzer, J. and Koralt, I. and Koroleva, L. and Korsch, W. and Kotchenda, L. and Kouchpil, V. and Kravtsov, P. and Krueger, K. and Krus, M. and Kumar, L. and Lamont, M. A. C. and Landgraf, J. M. and LaPointe, S. and Lauret, J. and Lebedev, A. and Lednicky, R. and Lee, J. H. and Leight, W. and LeVine, M. J. and Li, C. and Li, L. and Li, N. and Li, W. and Li, X. and Li, X. and Li, Y. and Li, Z. M. and Lima, L. M. and Lisa, M. A. and Liu, F. and Liu, H. and Liu, J. and Ljubicic, T. and Llope, W. J. and Longacre, R. S. and Lu, Y. and Lukashov, E. V. and Luo, X. and Ma, G. L. and Ma, Y. G. and Mahapatra, D. P. and Majka, R. and Mall, O. I. and Manweiler, R. and Margetis, S. and Markert, C. and Masui, H. and Matis, H. S. and McDonald, D. and McShane, T. S. and Meschanin, A. and Milner, R. and Minaev, N. G. and Mioduszewski, S. and Mitrovski, M. K. and Mohammed, Y. and Mohanty, B. and Mondal, M. M. and Morozov, B. and Morozov, D. A. and Munhoz, M. G. and Mustafa, M. K. and Naglis, M. and Nandi, B. K. and Nayak, T. K. and Nogach, L. V. and Nurushev, S. B. and Odyniec, G. and Ogawa, A. and Oh, K. and Ohlson, A. and Okorokov, V. and Oldag, E. W. and Oliveira, R. A. N. and Olson, D. and Pachr, M. and Page, B. S. and Pal, S. K. and Pandit, Y. and Panebratsev, Y. and Pawlak, T. and Pei, H. and Peitzmann, T. and Perkins, C. and Peryt, W. and Pile, P. and Planinic, M. and Pluta, J. and Plyku, D. and Poljak, N. and Porter, J. and Poskanzer, A. M. and Powell, C. B. and Prindle, D. and Pruneau, C. and Pruthi, N. K. and Pujahari, P. R. and Putschke, J. and Qiu, H. and Raniwala, R. and Raniwala, S. and Redwine, R. and Reed, R. and Ritter, H. G. and Roberts, J. B. and Rogachevskiy, O. V. and Romero, J. L. and Ruan, L. and Rusnak, J. and Sahoo, N. R. and Sakrejda, I. and Salur, S. and Sandweiss, J. and Sangaline, E. and Sarkar, A. and Schambach, J. and Scharenberg, R. P. and Schaub, J. and Schmah, A. M. and Schmitz, N. and Schuster, T. R. and Seele, J. and Seger, J. and Selyuzhenkov, I. and Seyboth, P. and Shah, N. and Shahaliev, E. and Shao, M. and Sharma, B. and Sharma, M. and Shi, S. S. and Shou, Q. Y. and Sichtermann, E. P. and Simon, F. and Singaraju, R. N. and Skoby, M. J. and Smirnov, N. and Solanki, D. and Sorensen, P. and de Souza, U. G. and Spinka, H. M. and Srivastava, B. and Stanislaus, T. D. S. and Steadman, S. G. and Stevens, J. R. and Stock, R. and Strikhanov, M. and Stringfellow, B. and Suaide, A. A. P. and Suarez, M. C. and Subba, N. L. and Sumbera, M. and Sun, X. M. and Sun, Y. and Sun, Z. and Surrow, B. and Svirida, D. N. and Symons, T. J. M. and Szanto de Toledo, A. and Takahashi, J. and Tang, A. H. and Tang, Z. and Tarini, L. H. and Tarnowsky, T. and Thein, D. and Thomas, J. H. and Tian, J. and Timmins, A. R. and Tlusty, D. and Tokarev, M. and Trentalange, S. and Tribble, R. E. and Tribedy, P. and Trzeciak, B. A. and Tsai, O. D. and Ullrich, T. and Underwood, D. G. and Van Buren, G. and van Nieuwenhuizen, G. and Vanfossen, J. A. and Varma, R. and Vasconcelos, G. M. S. and Vasiliev, A. N. and Videb{\ae{}}k, F. and Viyogi, Y. P. and Vokal, S. and Voloshin, S. A. and Wada, M. and Walker, M. and Wang, F. and Wang, G. and Wang, H. and Wang, J. S. and Wang, Q. and Wang, X. L. and Wang, Y. and Webb, G. and Webb, J. C. and Westfall, G. D. and Whitten, C. and Wieman, H. and Wissink, S. W. and Witt, R. and Witzke, W. and Wu, Y. F. and Xiao, Z. and Xie, W. and Xu, H. and Xu, N. and Xu, Q. H. and Xu, W. and Xu, Y. and Xu, Z. and Xue, L. and Yang, Y. and Yang, Y. and Yepes, P. and Yip, K. and Yoo, I-K. and Zawisza, M. and Zbroszczyk, H. and Zhan, W. and Zhang, J. B. and Zhang, S. and Zhang, W. M. and Zhang, X. P. and Zhang, Y. and Zhang, Z. P. and Zhao, F. and Zhao, J. and Zhong, C. and Zhu, X. and Zhu, Y. H. and Zoulkarneeva, Y.},
  collaboration = {STAR Collaboration},
  journal = {Phys. Rev. C},
  volume = {86},
  issue = {1},
  pages = {014904},
  numpages = {15},
  year = {2012},
  month = {Jul},
  publisher = {American Physical Society},
  doi = {10.1103/PhysRevC.86.014904},
  url = {https://link.aps.org/doi/10.1103/PhysRevC.86.014904}
}

@article{PhysRevC.86.054908,
  title = {Inclusive charged hadron elliptic flow in Au $+$ Au collisions at $\sqrt{{s}_{NN}}=7.7$--39 GeV},
  author = {Adamczyk, L. and Agakishiev, G. and Aggarwal, M. M. and Ahammed, Z. and Alakhverdyants, A. V. and Alekseev, I. and Alford, J. and Anderson, B. D. and Anson, C. D. and Arkhipkin, D. and Aschenauer, E. and Averichev, G. S. and Balewski, J. and Banerjee, A. and Barnovska, Z. and Beavis, D. R. and Bellwied, R. and Betancourt, M. J. and Betts, R. R. and Bhasin, A. and Bhati, A. K. and Bichsel, H. and Bielcik, J. and Bielcikova, J. and Bland, L. C. and Bordyuzhin, I. G. and Borowski, W. and Bouchet, J. and Brandin, A. V. and Brovko, S. G. and Bruna, E. and B{\"u}ltmann, S. and Bunzarov, I. and Burton, T. P. and Butterworth, J. and Cai, X. Z. and Caines, H. and Calder{\'o}n de la Barca S{\'a}nchez, M. and Cebra, D. and Cendejas, R. and Cervantes, M. C. and Chaloupka, P. and Chang, Z. and Chattopadhyay, S. and Chen, H. F. and Chen, J. H. and Chen, J. Y. and Chen, L. and Cheng, J. and Cherney, M. and Chikanian, A. and Christie, W. and Chung, P. and Chwastowski, J. and Codrington, M. J. M. and Corliss, R. and Cramer, J. G. and Crawford, H. J. and Cui, X. and Davila Leyva, A. and De Silva, L. C. and Debbe, R. R. and Dedovich, T. G. and Deng, J. and Derradi de Souza, R. and Dhamija, S. and Didenko, L. and Ding, F. and Dion, A. and Djawotho, P. and Dong, X. and Drachenberg, J. L. and Draper, J. E. and Du, C. M. and Dunkelberger, L. E. and Dunlop, J. C. and Efimov, L. G. and Elnimr, M. and Engelage, J. and Eppley, G. and Eun, L. and Evdokimov, O. and Fatemi, R. and Fazio, S. and Fedorisin, J. and Fersch, R. G. and Filip, P. and Finch, E. and Fisyak, Y. and Gagliardi, C. A. and Gangadharan, D. R. and Geurts, F. and Gibson, A. and Gliske, S. and Gorbunov, Y. N. and Grebenyuk, O. G. and Grosnick, D. and Gupta, S. and Guryn, W. and Haag, B. and Hajkova, O. and Hamed, A. and Han, L-X. and Harris, J. W. and Hays-Wehle, J. P. and Heppelmann, S. and Hirsch, A. and Hoffmann, G. W. and Hofman, D. J. and Horvat, S. and Huang, B. and Huang, H. Z. and Huck, P. and Humanic, T. J. and Huo, L. and Igo, G. and Jacobs, W. W. and Jena, C. and Joseph, J. and Judd, E. G. and Kabana, S. and Kang, K. and Kapitan, J. and Kauder, K. and Ke, H. W. and Keane, D. and Kechechyan, A. and Kesich, A. and Kettler, D. and Kikola, D. P. and Kiryluk, J. and Kisel, I. and Kisiel, A. and Kizka, V. and Klein, S. R. and Koetke, D. D. and Kollegger, T. and Konzer, J. and Koralt, I. and Koroleva, L. and Korsch, W. and Kotchenda, L. and Kravtsov, P. and Krueger, K. and Kulakov, I. and Kumar, L. and Lamont, M. A. C. and Landgraf, J. M. and LaPointe, S. and Lauret, J. and Lebedev, A. and Lednicky, R. and Lee, J. H. and Leight, W. and LeVine, M. J. and Li, C. and Li, L. and Li, W. and Li, X. and Li, X. and Li, Y. and Li, Z. M. and Lima, L. M. and Lisa, M. A. and Liu, F. and Ljubicic, T. and Llope, W. J. and Longacre, R. S. and Lu, Y. and Luo, X. and Luszczak, A. and Ma, G. L. and Ma, Y. G. and Madagodagettige Don, D. M. M. D. and Mahapatra, D. P. and Majka, R. and Mall, O. I. and Margetis, S. and Markert, C. and Masui, H. and Matis, H. S. and McDonald, D. and McShane, T. S. and Mioduszewski, S. and Mitrovski, M. K. and Mohammed, Y. and Mohanty, B. and Mondal, M. M. and Morozov, B. and Munhoz, M. G. and Mustafa, M. K. and Naglis, M. and Nandi, B. K. and Nasim, Md. and Nayak, T. K. and Nelson, J. M. and Nogach, L. V. and Novak, J. and Odyniec, G. and Ogawa, A. and Oh, K. and Ohlson, A. and Okorokov, V. and Oldag, E. W. and Oliveira, R. A. N. and Olson, D. and Ostrowski, P. and Pachr, M. and Page, B. S. and Pal, S. K. and Pan, Y. X. and Pandit, Y. and Panebratsev, Y. and Pawlak, T. and Pawlik, B. and Pei, H. and Perkins, C. and Peryt, W. and Pile, P. and Planinic, M. and Pluta, J. and Plyku, D. and Poljak, N. and Porter, J. and Poskanzer, A. M. and Powell, C. B. and Prindle, D. and Pruneau, C. and Pruthi, N. K. and Przybycien, M. and Pujahari, P. R. and Putschke, J. and Qiu, H. and Raniwala, R. and Raniwala, S. and Ray, R. L. and Redwine, R. and Reed, R. and Riley, C. K. and Ritter, H. G. and Roberts, J. B. and Rogachevskiy, O. V. and Romero, J. L. and Ross, J. F. and Ruan, L. and Rusnak, J. and Sahoo, N. R. and Sakrejda, I. and Salur, S. and Sandacz, A. and Sandweiss, J. and Sangaline, E. and Sarkar, A. and Schambach, J. and Scharenberg, R. P. and Schmah, A. M. and Schmidke, B. and Schmitz, N. and Schuster, T. R. and Seele, J. and Seger, J. and Seyboth, P. and Shah, N. and Shahaliev, E. and Shao, M. and Sharma, B. and Sharma, M. and Shi, S. S. and Shou, Q. Y. and Sichtermann, E. P. and Singaraju, R. N. and Skoby, M. J. and Smirnov, D. and Smirnov, N. and Solanki, D. and Sorensen, P. and deSouza, U. G. and Spinka, H. M. and Srivastava, B. and Stanislaus, T. D. S. and Steadman, S. G. and Stephans, G. S. F. and Stevens, J. R. and Stock, R. and Strikhanov, M. and Stringfellow, B. and Suaide, A. A. P. and Suarez, M. C. and Sumbera, M. and Sun, X. M. and Sun, Y. and Sun, Z. and Surrow, B. and Svirida, D. N. and Symons, T. J. M. and Szanto de Toledo, A. and Takahashi, J. and Tang, A. H. and Tang, Z. and Tarini, L. H. and Tarnowsky, T. and Thein, D. and Thomas, J. H. and Tian, J. and Timmins, A. R. and Tlusty, D. and Tokarev, M. and Trentalange, S. and Tribble, R. E. and Tribedy, P. and Trzeciak, B. A. and Tsai, O. D. and Turnau, J. and Ullrich, T. and Underwood, D. G. and Van Buren, G. and van Nieuwenhuizen, G. and Vanfossen, J. A. and Varma, R. and Vasconcelos, G. M. S. and Videb{\ae{}}k, F. and Viyogi, Y. P. and Vokal, S. and Voloshin, S. A. and Vossen, A. and Wada, M. and Wang, F. and Wang, G. and Wang, H. and Wang, J. S. and Wang, Q. and Wang, X. L. and Wang, Y. and Webb, G. and Webb, J. C. and Westfall, G. D. and Whitten, C. and Wieman, H. and Wissink, S. W. and Witt, R. and Witzke, W. and Wu, Y. F. and Xiao, Z. and Xie, W. and Xin, K. and Xu, H. and Xu, N. and Xu, Q. H. and Xu, W. and Xu, Y. and Xu, Z. and Xue, L. and Yang, Y. and Yang, Y. and Yepes, P. and Yi, Y. and Yip, K. and Yoo, I-K. and Zawisza, M. and Zbroszczyk, H. and Zhang, J. B. and Zhang, S. and Zhang, W. M. and Zhang, X. P. and Zhang, Y. and Zhang, Z. P. and Zhao, F. and Zhao, J. and Zhong, C. and Zhu, X. and Zhu, Y. H. and Zoulkarneeva, Y. and Zyzak, M.},
  collaboration = {STAR Collaboration},
  journal = {Phys. Rev. C},
  volume = {86},
  issue = {5},
  pages = {054908},
  numpages = {16},
  year = {2012},
  month = {Nov},
  publisher = {American Physical Society},
  doi = {10.1103/PhysRevC.86.054908},
  url = {https://link.aps.org/doi/10.1103/PhysRevC.86.054908}
}

@article{PhysRevLett.116.112302,
  title = {Beam Energy Dependence of the Third Harmonic of Azimuthal Correlations in $\mathrm{Au}+\mathrm{Au}$ Collisions at RHIC},
  author = {Adamczyk, L. and Adkins, J. K. and Agakishiev, G. and Aggarwal, M. M. and Ahammed, Z. and Alekseev, I. and Aparin, A. and Arkhipkin, D. and Aschenauer, E. C. and Attri, A. and Averichev, G. S. and Bai, X. and Bairathi, V. and Bellwied, R. and Bhasin, A. and Bhati, A. K. and Bhattarai, P. and Bielcik, J. and Bielcikova, J. and Bland, L. C. and Bordyuzhin, I. G. and Bouchet, J. and Brandenburg, J. D. and Brandin, A. V. and Bunzarov, I. and Butterworth, J. and Caines, H. and Calder{\'o}n de la Barca S{\'a}nchez, M. and Campbell, J. M. and Cebra, D. and Chakaberia, I. and Chaloupka, P. and Chang, Z. and Chatterjee, A. and Chattopadhyay, S. and Chen, J. H. and Chen, X. and Cheng, J. and Cherney, M. and Christie, W. and Contin, G. and Crawford, H. J. and Das, S. and De Silva, L. C. and Debbe, R. R. and Dedovich, T. G. and Deng, J. and Derevschikov, A. A. and di Ruzza, B. and Didenko, L. and Dilks, C. and Dong, X. and Drachenberg, J. L. and Draper, J. E. and Du, C. M. and Dunkelberger, L. E. and Dunlop, J. C. and Efimov, L. G. and Engelage, J. and Eppley, G. and Esha, R. and Evdokimov, O. and Eyser, O. and Fatemi, R. and Fazio, S. and Federic, P. and Fedorisin, J. and Feng, Z. and Filip, P. and Fisyak, Y. and Flores, C. E. and Fulek, L. and Gagliardi, C. A. and Garand, D. and Geurts, F. and Gibson, A. and Girard, M. and Greiner, L. and Grosnick, D. and Gunarathne, D. S. and Guo, Y. and Gupta, S. and Gupta, A. and Guryn, W. and Hamad, A. I. and Hamed, A. and Haque, R. and Harris, J. W. and He, L. and Heppelmann, S. and Heppelmann, S. and Hirsch, A. and Hoffmann, G. W. and Horvat, S. and Huang, T. and Huang, X. and Huang, B. and Huang, H. Z. and Huck, P. and Humanic, T. J. and Igo, G. and Jacobs, W. W. and Jang, H. and Jentsch, A. and Jia, J. and Jiang, K. and Judd, E. G. and Kabana, S. and Kalinkin, D. and Kang, K. and Kauder, K. and Ke, H. W. and Keane, D. and Kechechyan, A. and Khan, Z. H. and Kiko{\l{}}a, D. P. and Kisel, I. and Kisiel, A. and Kochenda, L. and Koetke, D. D. and Kosarzewski, L. K. and Kraishan, A. F. and Kravtsov, P. and Krueger, K. and Kumar, L. and Lamont, M. A. C. and Landgraf, J. M. and Landry, K. D. and Lauret, J. and Lebedev, A. and Lednicky, R. and Lee, J. H. and Li, X. and Li, C. and Li, X. and Li, Y. and Li, W. and Lin, T. and Lisa, M. A. and Liu, F. and Ljubicic, T. and Llope, W. J. and Lomnitz, M. and Longacre, R. S. and Luo, X. and Ma, R. and Ma, G. L. and Ma, Y. G. and Ma, L. and Magdy, N. and Majka, R. and Manion, A. and Margetis, S. and Markert, C. and Matis, H. S. and McDonald, D. and McKinzie, S. and Meehan, K. and Mei, J. C. and Minaev, N. G. and Mioduszewski, S. and Mishra, D. and Mohanty, B. and Mondal, M. M. and Morozov, D. A. and Mustafa, M. K. and Nandi, B. K. and Nasim, Md. and Nayak, T. K. and Nigmatkulov, G. and Niida, T. and Nogach, L. V. and Noh, S. Y. and Novak, J. and Nurushev, S. B. and Odyniec, G. and Ogawa, A. and Oh, K. and Okorokov, V. A. and Olvitt, D. and Page, B. S. and Pak, R. and Pan, Y. X. and Pandit, Y. and Panebratsev, Y. and Pawlik, B. and Pei, H. and Perkins, C. and Pile, P. and Pluta, J. and Poniatowska, K. and Porter, J. and Posik, M. and Poskanzer, A. M. and Pruthi, N. K. and Putschke, J. and Qiu, H. and Quintero, A. and Ramachandran, S. and Raniwala, S. and Raniwala, R. and Ray, R. L. and Ritter, H. G. and Roberts, J. B. and Rogachevskiy, O. V. and Romero, J. L. and Ruan, L. and Rusnak, J. and Rusnakova, O. and Sahoo, N. R. and Sahu, P. K. and Sakrejda, I. and Salur, S. and Sandweiss, J. and Sarkar, A. and Schambach, J. and Scharenberg, R. P. and Schmah, A. M. and Schmidke, W. B. and Schmitz, N. and Seger, J. and Seyboth, P. and Shah, N. and Shahaliev, E. and Shanmuganathan, P. V. and Shao, M. and Sharma, A. and Sharma, B. and Sharma, M. K. and Shen, W. Q. and Shi, Z. and Shi, S. S. and Shou, Q. Y. and Sichtermann, E. P. and Sikora, R. and Simko, M. and Singha, S. and Skoby, M. J. and Smirnov, N. and Smirnov, D. and Solyst, W. and Song, L. and Sorensen, P. and Spinka, H. M. and Srivastava, B. and Stanislaus, T. D. S. and Stepanov, M. and Stock, R. and Strikhanov, M. and Stringfellow, B. and Sumbera, M. and Summa, B. and Sun, Z. and Sun, X. M. and Sun, Y. and Surrow, B. and Svirida, D. N. and Tang, Z. and Tang, A. H. and Tarnowsky, T. and Tawfik, A. and Th{\"a}der, J. and Thomas, J. H. and Timmins, A. R. and Tlusty, D. and Todoroki, T. and Tokarev, M. and Trentalange, S. and Tribble, R. E. and Tribedy, P. and Tripathy, S. K. and Tsai, O. D. and Ullrich, T. and Underwood, D. G. and Upsal, I. and Van Buren, G. and van Nieuwenhuizen, G. and Vandenbroucke, M. and Varma, R. and Vasiliev, A. N. and Vertesi, R. and Videb{\ae{}}k, F. and Vokal, S. and Voloshin, S. A. and Vossen, A. and Wang, F. and Wang, G. and Wang, J. S. and Wang, H. and Wang, Y. and Wang, Y. and Webb, G. and Webb, J. C. and Wen, L. and Westfall, G. D. and Wieman, H. and Wissink, S. W. and Witt, R. and Wu, Y. and Xiao, Z. G. and Xie, W. and Xie, G. and Xin, K. and Xu, Y. F. and Xu, Q. H. and Xu, N. and Xu, H. and Xu, Z. and Xu, J. and Yang, S. and Yang, Y. and Yang, Y. and Yang, C. and Yang, Y. and Yang, Q. and Ye, Z. and Ye, Z. and Yepes, P. and Yi, L. and Yip, K. and Yoo, I.-K. and Yu, N. and Zbroszczyk, H. and Zha, W. and Zhang, X. P. and Zhang, Y. and Zhang, J. and Zhang, J. and Zhang, S. and Zhang, S. and Zhang, Z. and Zhang, J. B. and Zhao, J. and Zhong, C. and Zhou, L. and Zhu, X. and Zoulkarneeva, Y. and Zyzak, M.},
  collaboration = {STAR Collaboration},
  journal = {Phys. Rev. Lett.},
  volume = {116},
  issue = {11},
  pages = {112302},
  numpages = {9},
  year = {2016},
  month = {Mar},
  publisher = {American Physical Society},
  doi = {10.1103/PhysRevLett.116.112302},
  url = {https://link.aps.org/doi/10.1103/PhysRevLett.116.112302}
}

@article{PhysRevC.65.054902,
  title = {Initial condition for quark-gluon plasma evolution},
  author = {Drescher, H. J. and Liu, F. M. and Ostapchenko, S. and Pierog, T. and Werner, K.},
  journal = {Phys. Rev. C},
  volume = {65},
  issue = {5},
  pages = {054902},
  numpages = {9},
  year = {2002},
  month = {Apr},
  publisher = {American Physical Society},
  doi = {10.1103/PhysRevC.65.054902},
  url = {https://link.aps.org/doi/10.1103/PhysRevC.65.054902}
}

@article{PhysRevC.70.044907,
  title = {Observable implications of geometrical and dynamical aspects of freeze-out in heavy ion collisions},
  author = {Reti{\`e}re, Fabrice and Lisa, Michael Annan},
  journal = {Phys. Rev. C},
  volume = {70},
  issue = {4},
  pages = {044907},
  numpages = {33},
  year = {2004},
  month = {Oct},
  publisher = {American Physical Society},
  doi = {10.1103/PhysRevC.70.044907},
  url = {https://link.aps.org/doi/10.1103/PhysRevC.70.044907}
}

@article{doi:10.1142/S0218301318500581,
author = {Dudek, Danuce M. and Qian, Wei-Liang and Wu, Chen and Socolowski, Ot{\'a}vio and Padula, Sandra S. and Krein, Gast{\~a}o and Hama, Yogiro and Kodama, Takeshi},
title = {Effects of equation of state on hydrodynamic expansion, spectra, flow harmonics and two-pion interferometry},
journal = {International Journal of Modern Physics E},
volume = {27},
number = {07},
pages = {1850058},
year = {2018},
doi = {10.1142/S0218301318500581},
url = { https://doi.org/10.1142/S0218301318500581},
eprint = {https://doi.org/10.1142/S0218301318500581}
}

@article{doi:10.1142/S0218301307007167,
author = {QIAN, WEI-LIANG and ANDRADE, RONE and GRASSI, FR{\'E}D{\'E}RIQUE and SOCOLOWSKI, OTAVIO and KODAMA, TAKESHI and HAMA, YOGIRO},
title = {EFFECT OF CHEMICAL FREEZE OUT ON IDENTIFIED PARTICLE SPECTRA AT 200 AGeV Au-Au COLLISIONS AT RHIC USING SPheRIO},
journal = {International Journal of Modern Physics E},
volume = {16},
number = {07n08},
pages = {1877-1882},
year = {2007},
doi = {10.1142/S0218301307007167},
}

@article{PhysRevLett.109.202302,
  title = {Anisotropic Flow in Event-by-Event Ideal Hydrodynamic Simulations of $\sqrt{{s}_{NN}}\mathbf{=}200\text{ }\text{ }\mathrm{GeV}$ $\mathrm{Au}\mathbf{+}\mathrm{Au}$ Collisions},
  author = {Gardim, Fernando G. and Grassi, Fr{\'e}d{\'e}rique and Luzum, Matthew and Ollitrault, Jean-Yves},
  journal = {Phys. Rev. Lett.},
  volume = {109},
  issue = {20},
  pages = {202302},
  numpages = {5},
  year = {2012},
  month = {Nov},
  publisher = {American Physical Society},
  doi = {10.1103/PhysRevLett.109.202302},
  url = {https://link.aps.org/doi/10.1103/PhysRevLett.109.202302}
}

@article{PhysRevC.95.034901,
  title = {Hydrodynamic predictions for mixed harmonic correlations in 200 GeV Au+Au collisions},
  author = {Gardim, Fernando G. and Grassi, Frederique and Luzum, Matthew and Noronha-Hostler, Jacquelyn},
  journal = {Phys. Rev. C},
  volume = {95},
  issue = {3},
  pages = {034901},
  numpages = {8},
  year = {2017},
  month = {Mar},
  publisher = {American Physical Society},
  doi = {10.1103/PhysRevC.95.034901},
  url = {https://link.aps.org/doi/10.1103/PhysRevC.95.034901}
}

@article{PhysRevLett.93.182301,
  title = {Fluctuations of the Initial Conditions and the Continuous Emission in the Hydrodynamical Description of Two-Pion Interferometry},
  author = {Socolowski, O. and Grassi, F. and Hama, Y. and Kodama, T.},
  journal = {Phys. Rev. Lett.},
  volume = {93},
  issue = {18},
  pages = {182301},
  numpages = {4},
  year = {2004},
  month = {Oct},
  publisher = {American Physical Society},
  doi = {10.1103/PhysRevLett.93.182301},
  url = {https://link.aps.org/doi/10.1103/PhysRevLett.93.182301}
}

@article{PhysRevC.100.054905,
  title = {${p}_{T}$-dependent particle number fluctuations from principal-component analyses in hydrodynamic simulations of heavy-ion collisions},
  author = {Gardim, Fernando G. and Grassi, Fr{\'e}d{\'e}rique and Ishida, Pedro and Luzum, Matthew and Ollitrault, Jean-Yves},
  journal = {Phys. Rev. C},
  volume = {100},
  issue = {5},
  pages = {054905},
  numpages = {6},
  year = {2019},
  month = {Nov},
  publisher = {American Physical Society},
  doi = {10.1103/PhysRevC.100.054905},
  url = {https://link.aps.org/doi/10.1103/PhysRevC.100.054905}
}

@article{PhysRevC.107.024910,
  title = {Cumulants of conserved charges in hydrodynamic simulations},
  author = {Hirayama, Renan and Grassi, Fr{\'e}d{\'e}rique and Serenone, Willian Matioli and Ollitrault, Jean-Yves},
  journal = {Phys. Rev. C},
  volume = {107},
  issue = {2},
  pages = {024910},
  numpages = {10},
  year = {2023},
  month = {Feb},
  publisher = {American Physical Society},
  doi = {10.1103/PhysRevC.107.024910},
  url = {https://link.aps.org/doi/10.1103/PhysRevC.107.024910}
}

@article{PhysRevLett.97.202302,
  title = {Examining the Necessity to Include Event-By-Event Fluctuations in Experimental Evaluations of Elliptical Flow},
  author = {Andrade, R. and Grassi, F. and Hama, Y. and Kodama, T. and Socolowski, O.},
  journal = {Phys. Rev. Lett.},
  volume = {97},
  issue = {20},
  pages = {202302},
  numpages = {4},
  year = {2006},
  month = {Nov},
  publisher = {American Physical Society},
  doi = {10.1103/PhysRevLett.97.202302},
  url = {https://link.aps.org/doi/10.1103/PhysRevLett.97.202302}
}

@article{PhysRevC.85.024908,
  title = {Mapping the hydrodynamic response to the initial geometry in heavy-ion collisions},
  author = {Gardim, Fernando G. and Grassi, Fr{\'e}d{\'e}rique and Luzum, Matthew and Ollitrault, Jean-Yves},
  journal = {Phys. Rev. C},
  volume = {85},
  issue = {2},
  pages = {024908},
  numpages = {7},
  year = {2012},
  month = {Feb},
  publisher = {American Physical Society},
  doi = {10.1103/PhysRevC.85.024908},
  url = {https://link.aps.org/doi/10.1103/PhysRevC.85.024908}
}

@article{Habich2015,
    author  = {Habich, M. and Nagle, J. L. and Romatschke, P.},
    title   = {Particle spectra and HBT radii for simulated central nuclear collisions of C+C, Al+Al, Cu+Cu, Au+Au, and Pb+Pb from {$\sqrt{s}=62.4$--$2760$} GeV},
    journal = {Eur. Phys. J. C},
    volume  = {75},
    number  = {1},
    pages   = {15},
    year    = {2015},
    doi     = {10.1140/epjc/s10052-014-3206-7}
}

@misc{Naboka2025,
    author        = {Naboka, Vladislav and Sinyukov, Yuri and Adzhymambetov, Musfer and Zbroszczyk, Hanna},
    title         = {Interferometry Radii at RHIC BES Energies within the integrated HydroKinetic Model},
    year          = {2025},
    eprint        = {2506.19101},
    archivePrefix = {arXiv},
    primaryClass  = {hep-ph},
    doi           = {10.48550/arXiv.2506.19101}
}

@article{PhysRevC.96.024911,
  title = {Correlation femtoscopy study at energies available at the JINR Nuclotron-based Ion Collider fAcility and the BNL Relativistic Heavy Ion Collider within a viscous hydrodynamic plus cascade model},
  author = {Batyuk, P. and Karpenko, Iu. and Lednicky, R. and Malinina, L. and Mikhaylov, K. and Rogachevsky, O. and Wielanek, D.},
  journal = {Phys. Rev. C},
  volume = {96},
  issue = {2},
  pages = {024911},
  numpages = {10},
  year = {2017},
  month = {Aug},
  publisher = {American Physical Society},
  doi = {10.1103/PhysRevC.96.024911},
  url = {https://link.aps.org/doi/10.1103/PhysRevC.96.024911}
}

@article{HUANG2026140423,
title = {Excitation function of femtoscopic Lévy source parameters of pion pairs in EPOS4},
journal = {Physics Letters B},
volume = {876},
pages = {140423},
year = {2026},
issn = {0370-2693},
doi = {https://doi.org/10.1016/j.physletb.2026.140423},
url = {https://www.sciencedirect.com/science/article/pii/S0370269326002777},
author = {Yan Huang and Mátyás Molnár and Dániel Kincses and Máté Csanád},
}

@article{PhysRevC.93.044913,
  title = {Hydrodynamic simulations of relativistic heavy-ion collisions with different lattice quantum chromodynamics calculations of the equation of state},
  author = {Moreland, J. Scott and Soltz, Ron A.},
  journal = {Phys. Rev. C},
  volume = {93},
  issue = {4},
  pages = {044913},
  numpages = {10},
  year = {2016},
  month = {Apr},
  publisher = {American Physical Society},
  doi = {10.1103/PhysRevC.93.044913},
  url = {https://link.aps.org/doi/10.1103/PhysRevC.93.044913}
}

@article{https://doi.org/10.1155/2016/9414239,
author = {Zhang, S. and Ma, Y. G. and Chen, J. H. and Zhong, C.},
title = {Beam Energy Dependence of Hanbury-Brown-Twiss Radii from a Blast-Wave Model},
journal = {Advances in High Energy Physics},
volume = {2016},
number = {1},
pages = {9414239},
doi = {https://doi.org/10.1155/2016/9414239},
url = {https://onlinelibrary.wiley.com/doi/abs/10.1155/2016/9414239},
year = {2016}
}
\end{singlespace}

\appendix

\chapter{Hypersurface Element}
\label{chap:surfacelemente}

If this hypersurface can be described as a level set of an arbitrary function $A(x^\mu) = A_0$, where $A_0$ is a constant value, we can define the vector $N_\mu$ orthogonal to the hypersurface as \cite{DineenMultivariateCalculusGeometry, Poisson2004Hypersurfaces,Gourgoulhon2012}
\begin{equation}
N_\mu = \pm\frac{\partial A}{\partial x^\mu}\bigg|_{A = A_0},
\end{equation}
where $\pm$ fixes the orientation of this surface. Therefore, there is a proportionality relation $d\sigma_\mu \propto N_\mu$. For a non-null hypersurface, defining the norm
\begin{equation}
d\Sigma^2 = d\sigma \cdot d\sigma,
\end{equation}
then
\begin{align}
\frac{N_\mu}{\sqrt{|N \cdot N|}} = \frac{d\sigma_\mu}{\sqrt{|d\Sigma^2|}} \equiv \mathcal N_\mu,
\end{align}
For an observer comoving with this hypersurface, described by a $4-$velocity $\mathcal U^\mu(x)$, we can define the comoving volume element $d^3x$ of this hypersurface as \cite{Poisson2004Hypersurfaces,Gourgoulhon2012}
\begin{equation}
d^3x = \left|\mathcal U(x) \cdot d \sigma (x)\right|,
\end{equation}
such that, when $U = (1,0,0,0)$, using the definition of $d\sigma$ in Eq.~\eqref{eq:surfaceelement}, the expression above reduces to the identity
\begin{equation}
d^3x =\left|\frac{\partial(x^1, x^2, x^3)}{\partial(y^1, y^2, y^3)}\right|d^3y.
\end{equation}
From these definitions, it follows directly that
\begin{equation}
\sqrt{|d\Sigma^2|}  = \frac{d^3x}{|\mathcal U \cdot \mathcal N|},
\end{equation}
so that we can rewrite
\begin{equation}
d\sigma_\mu = \frac{d^3x }{|\mathcal U \cdot \mathcal N|}\mathcal N_\mu
\label{eq:dsgimaA}
\end{equation}
and this expression is very relevant for the numerical implementation.

\chapter{Bulk Coefficients}
\label{bulkcoef}

The coefficients \(A_T\), \(A_E\), and \(A_B\) are obtained by imposing the Landau matching conditions on the generalized Grad ansatz. For \(\mu_Q = \mu_S = 0\), they are given by
\begin{align}
A_T &= -\frac{U}{\Delta}, \\
A_E &= \frac{X}{\Delta}, \\
A_B &= -\left(M_{20}^{-1}\right)^{Bb}
\left(\mathcal{B}_{10}^{b} A_T + N_{30}^{b} A_E\right),
\end{align}
where there is an implicit sum over \(b \in \{B,Q,S\}\). The auxiliary quantities are
\begin{align}
\Delta &= XY - UW, \\
X &= \mathcal{A}_{20}
 - N_{30}^{a}\left(M_{20}^{-1}\right)^{ab}\mathcal{B}_{10}^{b}, \\
Y &= J_{41}
 - N_{31}^{a}\left(M_{20}^{-1}\right)^{ab}N_{30}^{b}, \\
U &= J_{40}
 - N_{30}^{a}\left(M_{20}^{-1}\right)^{ab}N_{30}^{b}, \\
W &= \mathcal{A}_{22}
 - N_{31}^{a}\left(M_{20}^{-1}\right)^{ab}\mathcal{B}_{10}^{b}.
\end{align}
The thermodynamic moments appearing in these expressions are defined by
\begin{align}
J_{kq} &= \sum_i J_{kq}^{(i)}, \\
N_{kq}^{a} &= \sum_i q_i^a J_{kq}^{(i)}, \\
M_{kq}^{ab} &= \sum_i q_i^a q_i^b J_{kq}^{(i)}, \\
\mathcal{B}_{10}^{a} &= \sum_i q_i^a m_i^2 J_{10}^{(i)}, \\
\mathcal{A}_{20} &= \sum_i m_i^2 J_{20}^{(i)}, \\
\mathcal{A}_{22} &= \sum_i m_i^2 J_{21}^{(i)}.
\end{align}
Here \(q_i^a = (B_i, Q_i, S_i)\), with \(B_i\), \(Q_i\), and \(S_i\) being, respectively, the baryon number, electric charge, and strangeness of species \(i\). Even when taking \(\mu_Q = \mu_S = 0\), the particle charges \(Q_i\) and \(S_i\) enter the sums above.

The microscopic integrals are
\begin{equation}
J_{kq}^{(i)}
=
\frac{g_i}{2\pi^2(\hbar c)^3}
\frac{T^{k+2}}{(2q+1)!!}
\int_0^\infty d\bar p\,
\bar p^{2q+2}
\bar E_i^{k-2q-1}
f_{0i}\left(1+\theta_i f_{0i}\right),
\end{equation}
with
\begin{equation}
f_{0i}
=
\frac{1}
{\exp\!\left[\bar E_i - B_i\mu_B/T\right] + \theta_i},
\qquad
\bar E_i = \sqrt{\bar p^2 + \left(\frac{m_i}{T}\right)^2}.
\end{equation}
In this convention, \(\theta_i = -1\) for bosons, \(\theta_i = +1\) for fermions, and \(\theta_i = 0\) in the Boltzmann limit.
\chapter{Equations of Motion and Evolution}

\label{apdx:numericaleom}
\textit{Special thanks to my research colleague Jonatan Sola for his work in developing the theoretical framework presented below.}\\
$ $\\
\noindent Given a tensor $T^{\mu \nu}$ expressed by a conservation law
\begin{equation}
    \small d_\mu T^{\mu \nu} = S^\nu,
    \label{eq:energystresstensorconservation}
\end{equation}
where $S^\nu$ is a source term, we can expand the left-hand side as
\begin{equation}
    \small d_\mu T^{\mu \nu} = \frac{1}{\sqrt{-g}}\partial_\mu(\sqrt{-g}T^{\mu \nu}) + \Gamma^{\nu}_{\mu \beta} T^{\mu \beta}  = \frac{1}{\sqrt{-g}}\left( \partial_0(\sqrt{-g}T^{0 \nu}) + \partial_i(\sqrt{-g}T^{i \nu})\right)+ \Gamma^{\nu}_{\mu \beta} T^{\mu \beta} ,
\end{equation}
and
\begin{equation}
\small
\begin{split}
\partial_0(\sqrt{-g}T^{0 \nu}) + \partial_i(\sqrt{-g}T^{i \nu}) &= \partial_0(\sqrt{-g}T^{0 \nu}) + \partial_i(\sqrt{-g}T^{i \nu}) +  v^i\partial_i(\sqrt{-g}T^{0 \nu}) -  v^i\partial_i(\sqrt{-g}T^{0 \nu}) =\\
&= \dv{(\sqrt{-g}T^{0 \nu})}{x^0} +\partial_i(\sqrt{-g}T^{i \nu}) - v^i\partial_i(\sqrt{-g}T^{0 \nu})  =\\
&= \dv{(\sqrt{-g}T^{0 \nu})}{x^0} +\partial_i(\sqrt{-g}T^{i \nu}) - \partial_i(v^i\sqrt{-g}T^{0 \nu}) + \sqrt{-g} T^{0 \nu}\partial_i v^i.
\end{split}
\end{equation}
Using the continuity equation for the specific density $\sigma = 1/V$, where $V$ is the volume, we can write:
\begin{equation}
\small
\begin{split}
    d_\mu (\sigma u^\mu) &= 0\\
    \frac{1}{\sqrt{-g}} \partial_\mu ( \sqrt{-g} \sigma u ^\mu) &= 0  \\
    \partial_\mu ( \sigma^* v ^\mu) & = 0\\
     \partial_0( \sigma^* ) + \partial_i (\sigma v^i) & = 0\\
     \partial_0( \sigma^* ) + v^i\partial_i (\sigma^* ) +\sigma^*\partial_i ( v^i) & = 0\\
     \dv{\sigma^*}{x^0} = -\sigma^* \partial_i v^i\\
     \therefore \partial_i v^i = -\frac{1}{\sigma^*} \dv{\sigma^*}{x^0} ,
\end{split}
\end{equation}
where $v^\mu = u^\mu/u^0$ and $\sigma^* \equiv \sqrt{-g}u^0\sigma$, so
\begin{equation}
\small
\begin{split}
    \partial_0(\sqrt{-g}T^{0 \nu}) + \partial_i(\sqrt{-g}T^{i \nu}) &= \dv{(\sqrt{-g}T^{0 \nu})}{x^0} +\partial_i(\sqrt{-g}T^{i \nu}) - \partial_i(v^i\sqrt{-g}T^{0 \nu}) - \frac{\sqrt{-g} T^{0 \nu}}{\sigma^*} \dv{\sigma^*}{x^0} ,
\end{split}
\end{equation}
then eq. \ref{eq:energystresstensorconservation} becomes
\begin{equation}
\small
\begin{split}
   \dv{(\sqrt{-g}T^{0 \nu})}{x^0} - \partial_i(\sqrt{-g}(T^{0 \nu}v^i - T^{i\nu})) - \frac{\sqrt{-g} T^{0 \nu}}{\sigma^*} \dv{\sigma^*}{x^0}  &= -\sqrt{-g} \Gamma^{\nu}_{\mu \beta} T^{\mu \beta} + \sqrt{-g}S^\nu\\
   \sqrt{-g}\dv{T^{0 \nu}}{x^0} + T^{0 \nu}\dv{\sqrt{-g} }{x^0} -\partial_i(\sqrt{-g}(T^{0 \nu}v^i - T^{i\nu})) - \frac{\sqrt{-g} T^{0 \nu}}{\sigma^*} \dv{\sigma^*}{x^0}  &= - \sqrt{-g} \Gamma^{\nu}_{\mu \beta} T^{\mu \beta} + \sqrt{-g}S^\nu\\
    \sqrt{-g} \left(\dv{T^{0 \nu}}{x^0} - T^{0 \nu} \left(\frac{1}{\sigma^*}\dv{\sigma^*}{x^0} - \frac{1}{\sqrt{-g}}\dv{\sqrt{-g}}{x^0}   \right)  \right) -\partial_i(\sqrt{-g}(T^{0 \nu}v^i - T^{i\nu})) &= - \sqrt{-g} \Gamma^{\nu}_{\mu \beta} T^{\mu \beta} + \sqrt{-g}S^\nu.
\end{split}
\end{equation}
Here, we define the quantity $\tilde T^{\mu \nu}$ as
\begin{equation}
    \tilde T^{\mu \nu} = \frac{\sqrt{-g}}{\sigma^*}T^{\mu \nu},
\end{equation}
so that
\begin{equation}
\begin{split}
      &\dv{\tilde T^{0 \nu}}{x^0} = \frac{T^{0 \nu}}{\sigma^*}\dv{\sqrt{-g}}{x^0} - \frac{\sqrt{-g}T^{0 \nu}}{(\sigma^*)^2}\dv{\sigma^*}{x^0} + \frac{\sqrt{-g}}{\sigma^*}\dv{T^{0 \nu}}{x^0}, \\
      &\therefore \dv{T^{0 \nu}}{x^0} - T^{0 \nu} \left(\frac{1}{\sigma^*}\dv{\sigma^*}{x^0} - \frac{1}{\sqrt{-g}}\dv{\sqrt{-g}}{x^0}   \right) = \frac{\sigma^*}{\sqrt{-g}}\dv{\tilde T^{0 \nu}}{x^0} , 
\end{split}
\end{equation}
and eq. \ref{eq:energystresstensorconservation} becomes
\begin{equation}
    \sigma^*\dv{\tilde T^{0 \nu}}{x^0} = \partial_i(\sqrt{-g}(T^{0 \nu}v^i - T^{i\nu})) -\sqrt{-g} \Gamma^{\nu}_{\mu \beta} T^{\mu \beta} + \sqrt{-g}S^\nu,
\end{equation}
so that
\begin{equation}
    \dv{\tilde T^{0 \nu}}{x^0} = \frac{1}{\sigma^*}\partial_i(\sqrt{-g}(T^{0 \nu}v^i - T^{i\nu})) - \Gamma^{\nu}_{\mu \beta} \tilde T^{\mu \beta} + \tilde S^\nu,
\end{equation}
where $\tilde S^\nu = \sqrt{-g} S^\nu/\sigma^*$. Note that this quantity does not depend on time derivatives of other quantities - it depends exclusively on the computation of spatial gradients (calculated via the SPH method) and on the tensor quantities themselves. We then extend this to the quantity in terms of the comoving derivative $D$, as this simplifies the use of Christoffel symbol terms during intermediate calculations. We only need to handle them explicitly when computing time derivatives.
\begin{equation}
\begin{split}
    DT^{0\nu} &= u^0 \dv{T^{0 \nu}}{x^0} + u^\lambda\Gamma^{0}_{\lambda \beta} T^{\beta \nu} + u^\lambda\Gamma^{\nu}_{\lambda \beta} T^{0\beta} \\
    &=\frac{\sigma^* u^0}{\sqrt{-g}} \dv{\tilde T^{0 \nu}}{x^0} +  u^0T^{0 \nu} \left(\frac{1}{\sigma^*}\dv{\sigma^*}{x^0} - \frac{1}{\sqrt{-g}}\dv{\sqrt{-g}}{x^0}\right)+ u^\lambda\Gamma^{0}_{\lambda \beta} T^{\beta \nu} + u^\lambda\Gamma^{\nu}_{\lambda \beta} T^{0\beta} ,
\end{split}
\label{eq:DT0ISPH}
\end{equation}
and one more time, there are not time derivative terms in this equation. Usually in SPH based codes, when we consider the dynamical equations derived from a conservation law $d_\mu T^{\mu \nu} = S^\nu$, we project it along directions parallel and perpendicular to the $4-$velocity to obtain the thermodynamic equations and equations of motion. This traditional approach often made deriving the equations of motion extremely difficult. In our reformulation, we retain the projection of the conservation law along the $4-$velocity direction, but replace the orthogonal projection with the derivative $DT^{0i}$. Given the form of $T^{\mu \nu}$ and the left-hand side $DT^{0i} = \cdots$ expressed through either directly computable terms or SPH-evaluated quantities, these become simpler. Here, $T^{\mu \nu}$ is not necessarily the energy-stress tensor $ - $ this equation structure holds for any tensor obeying conservation law \ref{eq:energystresstensorconservation}, such as the electromagnetic tensor. For a viscous fluid
\begin{equation}
T^{\mu \nu} = \epsilon  u^\mu u^\nu- (P+\Pi) \Delta^{\mu \nu} + \pi^{\mu \nu},
\end{equation}
then
\begin{equation}
    DT^{0i} = D\epsilon u^0 u^i - (DP + D\Pi) \Delta^{0i}+ (\epsilon+P+\Pi) (u^i Du^0 + u^0 Du ^i) + D\pi^{0 i}.
    \label{eq:DT0I}
\end{equation}
Since $DT^{0i}$ is already provided by Equation \ref{eq:DT0ISPH}, we obtain a direct connection with the derivative terms on the left-hand side of Equation \ref{eq:DT0I}, which simplifies the organization of these terms as
\begin{equation}
\begin{split}
    D\epsilon u^0 u^i - (DP + D\Pi) \Delta^{0i}+ (\epsilon+P+\Pi) (u^i Du^0 + u^0 Du ^i) + D\pi^{0 i} = \\
   = \frac{\sigma^* u^0}{\sqrt{-g}} \dv{\tilde T^{0 \nu}}{x^0} +  u^0T^{0 \nu} \left(\frac{1}{\sigma^*}\dv{\sigma^*}{x^0} - \frac{1}{\sqrt{-g}}\dv{\sqrt{-g}}{x^0}\right)+ u^\lambda\Gamma^{0}_{\lambda \beta} T^{\beta \nu} + u^\lambda\Gamma^{\nu}_{\lambda \beta} T^{0\beta} .
\end{split}
\end{equation}
Similarly, if a $4-$vector $N^\mu$ obeys a conservation law
\begin{equation}
    d_\mu N^\mu = Q,
    \label{eq:chargeconserv}
\end{equation}
we can express
\begin{equation}
    \dv{\tilde N^0}{x^0} = \frac{1}{\sigma^*}\partial_i(\sqrt{-g}(N^0v^i - N^i)) + \tilde Q,
\end{equation}
where $\tilde N^\mu = \sqrt{-g}N^\mu/\sigma^*$ and $\tilde Q = \sqrt{-g}Q/\sigma^*$. Consequently, the dynamical equation for $DN^0$ becomes
\begin{equation}
\begin{split}
    DN^0 = \frac{\sigma^* u^0}{\sqrt{-g}} \dv{\tilde N^0}{x^0} + u^0N^0 \left(\frac{1}{\sigma^*}\dv{\sigma^*}{x^0} - \frac{1}{\sqrt{-g}}\dv{\sqrt{-g}}{x^0}   \right) + u^\lambda\Gamma^{0}_{\lambda \beta} N^\beta .
\end{split}
\label{eq:dn0}
\end{equation}
This formulation allows us to incorporate the evolution of conserved charges into our simulations. For example, if $N^\mu = nu^\mu$ and we want to evolve $n$, we only need to consider
\begin{equation}
    DN^0 = Dn u^0 + nDu^0,
\label{eq:dn1}
\end{equation}
and use the left-hand side of \ref{eq:dn0}, where we only have spatial gradients computed via SPH along with the terms involving $N$ itself and the source term $Q$.

\subsection*{Matrix Coefficient Method}
Our main goal is to write an equation of motion in the form
\begin{equation}
    M^\alpha_{\ \beta} Du^\beta = F^\alpha
    \label{eq:motioneq1}
\end{equation}
because the inversion of this linear system is simply
\begin{equation}
    Du^\beta = (M^{-1})^\beta_{\ \alpha} F^\alpha
    \label{eq:motioneq2}
\end{equation}
allowing us to obtain the $4$-accelerations
\begin{equation}
    Du^\alpha = u^0 \dv{u^\alpha}{x^0} + u^\lambda \Gamma^\alpha_{\lambda \beta}u^\beta 
    \ \ \ \ \  \Rightarrow \ \ \ \ \  
    \dv{u^\alpha}{x^0} = \frac{Du^\alpha - u^\lambda \Gamma^\alpha_{\lambda \beta}u^\beta}{u^0},
\end{equation}
and thus we obtain the evolution of the $4$-velocities and the positions:
\begin{equation}
    \dv{x^\mu}{x^0} = \frac{u^\mu}{u^0}.
\end{equation}
In general, it is possible to write any dynamical equation as a linear expression of the covariant $4$-acceleration. That is, given a quantity $A$, we can write its derivative as
\begin{equation}
\begin{split}
    &DA = (M_A)_\alpha Du^\alpha + F_A \\
    &DA^{\mu} = (M_A)^\mu_\alpha Du^\alpha + (F_A)^\mu\\
    &DA^{\mu \nu} = (M_A)^{\mu \nu}_\alpha Du^\alpha + (F_A)^{\mu \nu}\\
    &\vdots
\end{split}
\end{equation}
where the terms $(M_A)^{\cdots}_\alpha$ and $(F_A)^{\cdots}$ are coefficient matrices resulting from the linearity of the dynamical equations. Once the comoving derivative $DA^{\cdots}$ is computed, we must recover $\dd A^{\cdots} / \dd x^0$ to obtain the temporal derivative. This is done directly from the definition (as shown above for the 4-velocities), since
\begin{equation}
    DA^{\cdots} = u^0 \dv{A^{\cdots}}{x^0} + \text{ Christoffel symbol terms},
\end{equation}
which implies
\begin{equation}
    \dv{A^{\cdots}}{x^0} = \frac{DA^{\cdots} - \text{ Christoffel symbol terms}}{u^0},
\end{equation}
thus resolving the issue of time evolution. The equations implemented for viscous hydrodynamics are:\\
$ $\\
For the expansion term $\theta = d_\mu u^\mu$, we start from the definition of the specific density equation and obtain
\begin{equation}
\small
\begin{split}
    \theta &= - \frac{u^0}{\sigma}\dv{\sigma}{x^0} = -\frac{1}{\sigma}\dv{}{\xi}\left(\frac{\sigma^*}{u^0\sqrt{-g}}\right) 
    = -\frac{u^0}{\sigma^*}\dv{\sigma^*}{x^0} + \dv{u^0}{x^0} + \frac{u^0}{\sqrt{-g}}\dv{\sqrt{-g}}{x^0}\\
    &= -\frac{u^0}{\sigma^*}\dv{\sigma^*}{x^0} + \frac{u^0}{\sqrt{-g}}\dv{\sqrt{-g}}{x^0} + \frac{Du^0 - u^\lambda \Gamma^0_{\lambda \beta}u^\beta}{u^0}. 
\end{split}
\end{equation}
so that, using $Du^0 = \delta^0_{\ \alpha} Du^\alpha$,
\begin{equation}
\small
\begin{split}
    &\theta = (M_\theta)_\alpha Du^\alpha + F_\theta,\\
    &(M_\theta)_\alpha = \frac{\delta^{0}_{\ \alpha}}{u^0}, \quad 
    F_\theta = -\frac{u^0}{\sigma^*}\dv{\sigma^*}{x^0} + \frac{u^0}{\sqrt{-g}}\dv{\sqrt{-g}}{x^0} - \frac{u^\lambda \Gamma^0_{\lambda \beta}u^\beta}{u^0}. 
\end{split}
\end{equation}
For the bulk viscosity, from Eq.~\eqref{bulk}, we have
\begin{equation}
\small
\begin{split}
    &D\Pi = (M_\Pi)_\alpha Du^\alpha + F_\Pi,\\
    &(M_\Pi)_\alpha = -\left(\Pi + \frac{\zeta}{\tau_\Pi} \right)(M_\theta)_\alpha, \quad 
    F_\Pi = -\left(\frac{\Pi}{\tau_\Pi} + \left(\Pi + \frac{\zeta}{\tau_\Pi} \right)F_\theta\right).
\end{split}
\end{equation}
For the shear viscosity, from Eq.~\eqref{shear}, we also have
\begin{equation}
\small
\begin{split}
    &D\pi^{\mu \nu} = (M_\pi)^{\mu \nu}_\alpha Du^\alpha + (F_\pi)^{\mu \nu},\\
    & (M_\pi)^{ij}_\alpha = -\left( u^i \pi_\alpha^{\ j} + u^j \pi_\alpha^{\ i} + \frac{4}{3}\frac{\pi^{ij}\delta^0_{\ \alpha}}{u^0} + \frac{2\eta}{\tau_\pi} \left(\frac{u^i\delta^j_{\ \alpha} + u^j \delta^i_{\ \alpha}}{2} + \frac{\Delta^{ij}}{3}(M_\theta)_\alpha \right)\right),\\
    &(F_\pi)^{ij} = -\left(\frac{4}{3}\pi^{ij}F_\theta + \frac{\pi^{ij}}{\tau_\pi} - \frac{2\eta}{\tau_\pi}\left(\frac{d^i u^j + d^j u^i}{2} - \frac{\Delta^{ij}}{3}F_\theta\right) \right),\\
    &(M_\pi)^{0i}_\alpha = -\frac{\left((M_\pi)^{ij}_\alpha u_j + \pi^{i}_{\ \alpha} \right)}{u_0}, \quad (F_\pi)^{0i} = - \frac{(F_\pi)^{ij}u_j}{u_0},\\
    &(M_\pi)^{00}_\alpha = -g_{ij} (M_\pi)^{ij}_\alpha, \quad (F_\pi)^{00} = -g_{ij}(F_\pi)^{ij}.
\end{split}
\end{equation}
Since the shear tensor is orthogonal to the $4$-velocity, $\pi^{\mu \nu}u_\nu = 0$, symmetric, $\pi^{\mu \nu} = \pi^{\nu \mu}$, and traceless, $\pi^\mu_{\ \mu} = 0$, we evolve only the components $\pi^{11}$, $\pi^{21}$, $\pi^{22}$, $\pi^{31}$, $\pi^{32}$, and $\pi^{33}$, and all other components and their dynamical evolution are obtained from these properties. For the energy density, from Eq.~\eqref{eq:energydensity}, we have
\begin{equation}
\begin{split}
    &D\epsilon = (M_\epsilon)_\alpha Du^\alpha + F_\epsilon,\\
    &(M_\epsilon)_\alpha = -\left( (\epsilon + P + \Pi)(M_\theta)_\alpha - \frac{\pi^0_{\ \alpha}}{u^0} \right), \quad
    F_\epsilon = - \left((\epsilon + P + \Pi)F_\theta + \frac{\pi^0_{\ \alpha} u^i d_i u^\alpha}{u^0} - \pi^i_{\ \alpha} d_i u^\alpha \right).
\end{split}
\end{equation}
Since we are only considering baryon charge without diffusion, we have from \eqref{eq:dn0} and \eqref{eq:dn1}
\begin{align}
    &Dn_B = (M_{n_B})_\alpha Du^\alpha + (F_{n_B}),\\
    &(M_{n_B})_\alpha = -\frac{n_B}{u^0} \delta^0_{\ \alpha} , \quad 
    (F_{n_B}) = DJ_B^0  .    
\end{align}
For the pressure, since we evolve $\epsilon$ and $n_B$, we get 
\begin{equation}
    DP = \pdv{P}{\epsilon} D\epsilon + \pdv{P}{n_B}Dn_B,
\end{equation}
where $\partial P/\partial \epsilon$ and $\partial P/\partial n_B$ are given by the equation of state, and thus
\begin{equation}
\begin{split}
    &DP = (M_P)_\alpha Du^\alpha + F_P,\\
    &(M_P)_\alpha = \pdv{p}{\epsilon}\left(M_\epsilon\right)_\alpha+\pdv{p}{n_B}\left(M_{n_B}\right)_\alpha, \quad F_P = \pdv{p}{\epsilon}F_\epsilon+\pdv{p}{n_B}F_{n_B}.
\end{split}
\end{equation}
For the full equation of motion, we take the energy-momentum tensor equations \eqref{eq:DT0I} and \eqref{eq:DT0ISPH}, along with the property $u_\alpha Du^\alpha = 0$, and obtain
\begin{equation}
\begin{split}
    & M^\mu_{\ \alpha} Du^\alpha = F^\mu,\\
    & M^i_{\ \alpha} = \left((M_\epsilon)_\alpha + (M_P)_\alpha + (M_\Pi)_\alpha\right) u^0 u^i + (\epsilon + p + \Pi)(u^i \delta^0_{\ \alpha} + u^0 \delta^i_{\ \alpha}) + (M_\pi)^{0i}_\alpha,\\
    & F^{i} = - (F_\epsilon + F_P + F_\Pi) u^0 u^i - (F_\pi)^{0i} + DT^{0i},\\
    & M^0_{\ \alpha} = u_\alpha, \quad F^0 = 0,
\end{split}
\label{eq:motioneqmatrix}
\end{equation}
where we have already considered a diagonal metric. Here we find the form of the matrix $M$ that we aimed for in Eq.~\eqref{eq:motioneq1}, and by inverting it to obtain Eq.~\eqref{eq:motioneq2}, the dynamics of relativistic viscous hydrodynamics is fully determined.

\chapter{Pair moments and source variances}
\label{apdx:moments}
Let \(x_1\) and \(x_2\) be two independent emissions from the same single-particle source. This independence follows from the factorized two-particle emission function used in the direct moment calculation,
\begin{equation}
    S_2(x_1,x_2;K)=S_1(x_1,K)S_1(x_2,K).
\end{equation}
For any source coordinate \(x_A\), with \(A=o,s,l,t'\), we define the relative separation
\begin{equation}
    \delta x_A=x_{A,1}-x_{A,2}.
\end{equation}
Since particles 1 and 2 are sampled from the same single-particle source,
\begin{equation}
    \left\langle \delta x_A \right\rangle_{\rm pair}
    =
    \left\langle x_{A,1}\right\rangle
    -
    \left\langle x_{A,2}\right\rangle
    =0.
\end{equation}
Therefore, the pair mixed moment is equal to the covariance of the relative separations,
\begin{equation}
    \left\langle \delta x_A \delta x_B \right\rangle_{\rm pair}
    =
    {\rm Cov}(\delta x_A,\delta x_B).
\end{equation}
Using \(\delta x_A=x_{A,1}-x_{A,2}\) and
\(\delta x_B=x_{B,1}-x_{B,2}\), one obtains 
\begin{equation}
\begin{split}
    {\rm Cov}(\delta x_A,\delta x_B)
    &=
    {\rm Cov}(x_{A,1}-x_{A,2},x_{B,1}-x_{B,2})
    \\
    &=
    {\rm Cov}(x_{A,1},x_{B,1})
    +
    {\rm Cov}(x_{A,2},x_{B,2})
    \\
    &\quad
    -
    {\rm Cov}(x_{A,1},x_{B,2})
    -
    {\rm Cov}(x_{A,2},x_{B,1}).
\end{split}
\end{equation}
\normalsize The last two terms vanish because they involve correlations between different particles, which are independent samples of the factorized source. The first two terms are equal because both particles are sampled from the same source:
\begin{equation}
    {\rm Cov}(x_{A,1},x_{B,1})
    =
    {\rm Cov}(x_{A,2},x_{B,2})
    =
    \left\langle \Delta x_A \Delta x_B \right\rangle .
\end{equation}
Thus,
\begin{equation}
    \left\langle \delta x_A\delta x_B\right\rangle_{\rm pair}
    =
    2\left\langle \Delta x_A\Delta x_B\right\rangle ,
\end{equation}
or equivalently,
\begin{equation}
    \left\langle \Delta x_A\Delta x_B\right\rangle
    =
    \frac{1}{2}
    \left\langle \delta x_A\delta x_B\right\rangle_{\rm pair}.
\end{equation}
The diagonal case \(A=B\) gives the usual relation between the pair relative variance and the single-particle source variance,
\begin{equation}
    \left\langle \Delta x_A^2\right\rangle
    =
    \frac{1}{2}
    \left\langle \delta x_A^2\right\rangle_{\rm pair}.
\end{equation}
The factor \(1/2\) therefore follows directly from the factorization of the two-particle source into independent single-particle emission functions. If the pair distribution were weighted by the two-particle wave function or by final-state interactions, the effective two-particle distribution would no longer be simply \(S_1(x_1,K)S_1(x_2,K)\), and this identity would not necessarily hold in the same form.

\end{document}